\documentclass{article}
\input{LatexFiles/preamble.tex}

\title{Modelling non-stationary extremal dependence through a geometric approach}
\author[1,2*]{C. J. R. Murphy-Barltrop}
\author[3]{J. L. Wadsworth}
\author[4,5]{M. de Carvalho}
\author[6]{B. D. Youngman}
\affil[1]{Technische Universität Dresden, Institut Für Mathematische Stochastik, Helmholtzstraße 10, 01069 Dresden, Germany}
\affil[2]{Center for Scalable Data Analytics and Artificial Intelligence (ScaDS.AI) Dresden/Leipzig, Germany}
\affil[3]{School of Mathematical Sciences, Lancaster University LA1 4YF, UK}
\affil[4]{School of Mathematics, University of Edinburgh EH9 3FD, UK}
\affil[5]{CIDMA---Department of Mathematics, University of Aveiro, Portugal}
\affil[6]{Department of Mathematics and Statistics, University of Exeter EX4 4QE, UK}
\affil[*]{Correspondence to: callum.murphy-barltrop@tu-dresden.de}
\date{\today}

\begin{document}

\maketitle

\begin{abstract}
    Non-stationary extremal dependence, whereby the relationship between the extremes of multiple variables evolves over time, is commonly observed in many environmental and financial data sets. However, most multivariate extreme value models are only suited to stationary data. A recent approach to multivariate extreme value modelling uses a geometric framework, whereby extremal dependence features are inferred through the limiting shapes of scaled sample clouds. This framework can capture a wide range of dependence structures, and a variety of inference procedures have been proposed in the stationary setting. In this work, we first extend the geometric framework to the non-stationary setting and outline assumptions to ensure the necessary convergence conditions hold. We then introduce a flexible, semi-parametric modelling framework for obtaining estimates of limit sets in the non-stationary setting. Through rigorous simulation studies, we demonstrate that our proposed framework can capture a wide range of dependence forms and is robust to different model formulations. We illustrate the proposed methods on financial returns data and present several practical uses.
\end{abstract}

\noindent%

\doublespacing

\section{Introduction} \label{sec:intro}

\subsection{Extremal dependence} \label{subsec:ex_dep}

Understanding the relationship between the extremes of multiple variables, termed extremal dependence, is an important area of research, with applications in environmental science \citep{Brunner2016}, engineering \citep{Tendijck2021}, actuarial science \citep{Quinn2019} and finance \citep{Nolde2021}. Typically, modelling procedures aim to estimate fixed coefficients or parameters related to the extremal dependence structure, which are subsequently used to infer extremal statistics. This field of study, known as multivariate extreme value theory, helps practitioners to carry out robust joint risk assessments across multiple variables, or locations, relevant to a given application.  

In the bivariate setting, there exist two distinct subclasses of extremal dependence. Given variables $X_i \sim F_i, i=1,2$, define the conditional probability
\begin{equation*}
    \chi(u) := \Pr(F_2(X_2)>u \mid F_1(X_1)>u) \in [0,1],
\end{equation*}
and set $\chi = \lim _{u \rightarrow 1^-} \chi(u) \in [0,1]$, where this limit exists \citep[e.g.,][]{Coles1999}. This dependence measure provides a summary of joint tail behaviour, and when $\chi > 0$ and $\chi = 0$, we say that the random vector $\boldsymbol{X} := (X_1,X_2)$ exhibits asymptotic dependence (AD) and asymptotic independence (AI), respectively. Notably, classical extremal dependence modelling procedures, which are developed under the framework of regular variation, have been shown to perform poorly for data exhibiting AI \citep{Ledford1997,Heffernan2004}. Since AI is commonly observed in many relevant data sets, this is significant drawback of classical approaches, and many recent works have argued against their use when AI is present \citep[e.g.,][]{Opitz2016,Murphy-Barltrop2024c,Huser2024}. In this paper, we pioneer the development of non-stationary extremal dependence modelling via a geometric approach based on limits sets \citep{Nolde2014,Nolde2022}. The framework can accommodate both AD and AI structures, but the modelling choices we subsequently make are better suited to AI. For simplicity, we also restrict attention to the bivariate setting throughout, noting that many of the introduced techniques can, in principle at least, be extended to higher dimensions; see Section~\ref{sec:discussion} for further discussion. 

\subsection{Non-stationary extremal dependence} \label{subsec:ns_dep}

Most modelling frameworks for extremal dependence make the simplifying assumption that data are independent and identically distributed (IID). However, many real-world data sets exhibit non-stationarity, whereby the distribution of the data is not fixed in time. Non-stationarity can be present in two distinct forms: within the marginal distributions $F_1,F_2$, and within the (extremal) dependence structure of $\boldsymbol{X}$. We restrict attention to the latter form, noting that stationary margins can be obtained by fitting non-stationary marginal models and transforming the data via the probability integral transform to a standard marginal scale. Such transformations do not alter the dependence structure of $\boldsymbol{X}$, owing to Sklar's theorem \citep{Sklar1959}. However, as recently demonstrated in \citet{Kakampakou2024}, selecting the appropriate marginal model for non-stationary data is not straightforward, and poor estimates of the marginal distribution can strongly affect the representation of extremal dependence; see also \citet{Murphy-Barltrop2024b}. 

In the context of extremal dependence, non-stationarity can result in dependence coefficients or parameters that are not fixed in time, requiring modelling procedures that can capture trends in these quantities. Let $\boldsymbol{X}_t:= (X_{1,t},X_{2,t})$ denote a continuous, non-stationary bivariate process on standard stationary margins indexed by $t$, where $t$ typically represents time, and $\boldsymbol{Z}_t$ is a continuous, $p$-dimensional covariate process (the predictor variables) that directly influences $\boldsymbol{X}_t$ (the response vector). Our interest lies in understanding, and modelling, the relationship between the extremal dependence structure of $\boldsymbol{X}_t$ and the covariate process $\boldsymbol{Z}_t$. Relatively few approaches have been proposed in this context, even though non-stationary dependence is a common feature across many relevant data sets \citep[e.g.,][]{Jonathan2014a, Castro-Camilo2018, Murphy-Barltrop2024b}. We note here that typically we only observe any given non-stationary process for a discrete number of time points; we therefore let $\{1,\hdots,T \}$ denote a finite set of arbitrary, fixed indices. 

The majority of the available models for non-stationary dependence are targeted at data structures exhibiting AD \citep[see, e.g.,][]{Carvalho2014,Mhalla2017,Castro-Camilo2018,Mhalla2019}, and will therefore perform poorly for data exhibiting AI. In the AI setting, \citet{Jonathan2014a}, \citet{Guerrero2021} and \citet{Talento2025} consider non-stationary extensions of the conditional extremes model proposed by \citet{Heffernan2004}, while \citet{Mhalla2019}, \citet{Murphy-Barltrop2024b} and \citet{Andre2023} propose non-stationary extensions, and inference techniques, for a quantity termed the angular dependence function \citep{Wadsworth2013}. Alongside these approaches, \citet{Mhalla2019} and \citet{Lee2024} model coefficients of extremal dependence in the non-stationary setting; in particular, the limit $\chi$ and the coefficient of tail dependence $\eta \in (0,1]$, first proposed by \citet{Ledford1996}, which quantifies the dependence amongst asymptotically independent data structures; see Appendix~\ref{sec:appen_led_tawn} for a formal definition.

The available approaches for modelling non-stationary extremal dependence offer limited utility. For example, coefficients such as $\eta$ and $\chi$ merely summarise the extremal dependence, and provide no information outside of the region where all variables are jointly large (i.e., the top right corner of the positive quadrant $\RR^2_+$). Furthermore, the \citet{Heffernan2004} and \citet{Wadsworth2013} modelling frameworks suffer from various drawbacks; the former requires one to fit several separate models, which can result in contradictory conclusions \citep{Liu2014,Simpson2024a}, while the latter can only be used for specific probability calculations (e.g., joint survivor sets). These limitations, combined with the sparsity of existing approaches, motivate novel developments for modelling non-stationary extremal dependence in the AI setting. 

\subsection{Limit sets} 
\label{subsec:limit_sets}

An increasingly popular modelling tool for multivariate extremes involves the use of limit sets, which arise from the asymptotic convergence of scaled sample clouds for suitably chosen scaling sequences and marginal distributions, to infer extremal dependence properties and perform inference. We briefly summarise this convergence property below. Let $\boldsymbol{X}= (X_1,X_2)$ be a continuous, stationary random vector with marginal distribution functions asymptotically equal to a von Mises function, i.e., light-tailed margins \citep{Nolde2022}. Consider the set $C_n := {\{ \boldsymbol{X}_i/r_n\}^n_{i=1}}$ denoting $n$ independent copies of $\boldsymbol{X}$ scaled by a suitably chosen sequence $r_n>0$ satisfying $r_n \to \infty$ as $n \to \infty$. Under mild conditions, $C_n$ converges in probability, as $n \to \infty$, onto a deterministic set $\mathcal{G} := \{\boldsymbol{x}: g(\boldsymbol{x}) \leq 1 \}$ \citep{Davis1988,Kinoshita1991,Balkema2010}. The set $\mathcal{G}$ is known as the \textit{limit set}, and $g:\RR^2 \mapsto \RR_+$ is termed the \textit{gauge function} of $\mathcal{G}$ .

\begin{figure}[h]
    \centering
    \includegraphics[width=\linewidth]{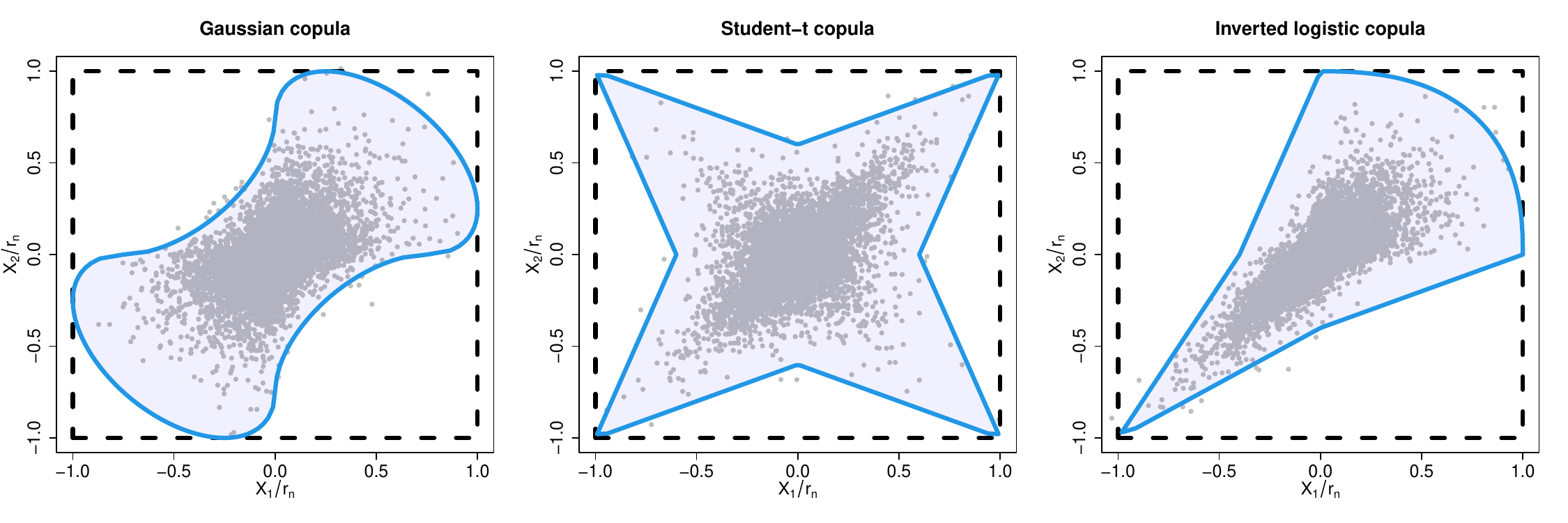}
    \caption{Theoretical limit and boundary sets, denoted as blue shaded regions and thick lines, respectively, for Gaussian (left), student-$t$ (centre) and inverted logistic (right) copulas. }
    \label{fig:theoretical_limit_sets}
\end{figure}

The shape of $\mathcal{G}$ is affected by the margins and dependence structure of $\boldsymbol{X}$. To focus on dependence only, a common approach is to standardize margins. We adopt standard Laplace margins in this article. The theoretical limit sets for three copulas on Laplace margins are illustrated in Figure~\ref{fig:theoretical_limit_sets}, alongside the scaled sample clouds $C_n$ for a large, but finite, sample size $n$. One can observe that even at the finite level, the scaled observations lie approximately within the interior $\mathcal{G}$. Further background on limit sets and conditions for convergence of $C_n$ are discussed in Section~\ref{subsec:ns_limit_sets}.

Many recent works have introduced techniques for estimating $\mathcal{G}$. These approaches can be divided into two categories; those that only target limit set estimation, and those that use the geometric framework as a vehicle for full joint tail inference. For the former, \citet{Simpson2024a} and \citet{Majumder2024} introduce semi-parametric techniques using generalised additive models (GAMs) and B\'ezier splines, respectively, to approximate $\mathcal{G}$. Note that we also employ GAMs and splines for our proposed modelling approach, although in an entirely different context (i.e., non-stationary extremal dependence) requiring more complex formulations. For the latter category, \citet{Wadsworth2024} propose a parametric copula-based approach, \citet{Campbell2024} present a semi-parametric piecewise linear model, \citet{Papastathopoulos2025} introduce a latent Gaussian model, and \citet{Murphy-Barltrop2024d} put forward a deep learning approach. In each of these works, the authors perform rigorous simulation studies, and demonstrate that the proposed techniques are able to accurately capture a wide variety of extremal dependence structures encompassing both dependence classes. 

It is clear that modelling multivariate extremes through limit sets represents an important line of research. Furthermore, existing estimation approaches for $\mathcal{G}$ have clearly demonstrated many advantages of the geometric extremes framework; namely, robustness across difference dependence structures, self-consistent conclusions, and applicability in higher dimensions than many competing approaches. However, to date, existing techniques have exclusively focused on the case when data are identically distributed, and thus cannot be applied for data exhibiting non-stationary dependence.

\subsection{Outline} \label{subsec:overview}
In this article, we introduce a novel non-stationary framework for modelling extremal dependence via limit sets. Our approach can be applied to data exhibiting both AD and AI, with more of a focus placed on the latter class. Moreover, inference from our approach is not limited to summary dependence measures or specific joint tail probabilities: for any index $t$, our model can be used for simulation from the joint tail of $\boldsymbol{X}_t$, for probability estimation in general joint tail regions, and for approximating bivariate risk measures. 

In Section~\ref{sec:ns_limit_sets}, we outline a natural extension of the geometric extremes framework to the non-stationary setting, and detail our asymptotic assumptions. In Section~\ref{sec:methodology}, we explain our GAM framework for estimating the non-stationary counterpart of the gauge function (equivalently, the limit set). In Section~\ref{sec:sim_study}  we apply our framework to a range of simulated data sets, demonstrating the proposed methodology can approximate non-stationary boundary sets across many dependence structures. In Section~\ref{sec:case_study}, we apply our model to financial data, with diagnostics indicating good performance. We conclude in Section~\ref{sec:discussion} with a discussion and outlook on future work.  

\section{Limit sets and their non-stationary extensions} 
\label{sec:ns_limit_sets}
\subsection{Background on limit sets} \label{subsec:limit_sets_conditions}

The limit set $\mathcal{G}$ of the scaled sample cloud $C_n$ was introduced in Section~\ref{subsec:limit_sets}. When $\boldsymbol{X}$ has standard Laplace margins, a sufficient condition for the convergence of $C_n$ onto $\mathcal{G} = \{\boldsymbol{x} \in \mathbb{R}^2: g(\boldsymbol{x})\leq 1\}$ to occur is that the joint density, $f_{\boldsymbol{X}}(\cdot)$, of $\boldsymbol{X}$ exists, and that 
\begin{equation} \label{eqn:gauge}
    -\log f_{\boldsymbol{X}}(u\boldsymbol{x}) \sim ug(\boldsymbol{x}), \; \; \boldsymbol{x} \in \RR^2, \; u \to \infty,
\end{equation}
for a continuous $g(\cdot)$ \citep{Balkema2010, Nolde2022}. The limit set $\mathcal{G}$ is star-shaped at $\boldsymbol{0}:=(0,0)$ and compact \citep{Kinoshita1991}. It is clear from equation~\eqref{eqn:gauge} that the gauge function $g(\cdot)$ is $1$-homogenous, i.e., $g(c\boldsymbol{x}) = cg(\boldsymbol{x})$ for any $c > 0$, $\boldsymbol{x}\in \RR^2$, linking to the star-shaped property. In standard Laplace margins, $\mathcal{G} \subset [-1,1]^2$, with the corresponding limit set possessing componentwise maxima and minima $\boldsymbol{1}$ and $-\boldsymbol{1}$, respectively. Note that for any random vector exhibiting AD, such as the student-$t$ copula, we have $\boldsymbol{1} \in \mathcal{G}$ (see Figure~\ref{fig:theoretical_limit_sets}). We refer the reader to \citet{Nolde2022} for a detailed discussion and theoretical treatment of limit sets. 

Following \citet{Nolde2014}, we also define the \textit{boundary} set $\partial \mathcal{G} := \{\boldsymbol{x}: g(\boldsymbol{x}) = 1 \} \subset \mathcal{G}$, noting this set possesses the same componentwise maxima and minima as $\mathcal{G}$. The boundary sets are also illustrated for the three copula examples in Figure~\ref{fig:theoretical_limit_sets}. We remark that one only needs to know $\partial \mathcal{G}$ in order to specify $\mathcal{G}$; hence no knowledge is lost through just considering the boundary set. We therefore use the term `limit set' to refer to both $\partial \mathcal{G}$ and $\mathcal{G}$ throughout this work. 

Recent works have demonstrated that the limit set is directly related to extremal dependence structure of $\boldsymbol{X}$ \citep{Nolde2014,Nolde2022}. Specifically, $\partial \mathcal{G}$ links several representations for multivariate extremes, immediately providing coefficients and parameters associated with the models proposed in \citet{Ledford1996}, \citet{Heffernan2004}, \citet{Wadsworth2013} and \citet{Simpson2020}. For example, taking $\eta \in (0,1]$, we have that $\eta := \max \{s \in [0,1] : [s,\infty)^2 \cap \partial \mathcal{G} \neq \emptyset \}$. In this sense, the study of deterministic geometric shapes is equivalent to the study of extremal dependence, and we refer to this topic as \textit{geometric extremes} herein.  Furthermore, in a practical setting, estimates of $\partial \mathcal{G}$ can be used to obtain estimates of risk measures, including, but not limit to, joint tail probabilities \citep{Wadsworth2024}, return curves \citep{Murphy-Barltrop2024c}, and return level sets \citep{Papastathopoulos2025}. Thus, knowledge of $\partial \mathcal{G}$ provides not only a great deal of information about the extremal dependence structure, but also supplies practitioners with a framework for performing inference on multivariate extremes. 

A brief review of statistical estimation techniques for $\mathcal{G}$ was given in Section~\ref{subsec:limit_sets}. When performing statistical estimation in the geometric extremes framework, it is helpful to decompose a random vector into angular and radial components. However, the method by which one performs this decomposition is ambiguous, owing to the theoretically infinite number of angular-radial systems that are available; see \citet{Mackay2023} for further discussion. As such, we define angular radial systems in a somewhat general manner, and leave the details of our chosen system(s) to Section~\ref{sec:methodology}.

Let $\| \cdot \|:\RR^2 \to \RR_+$ denote any valid norm. Given $\boldsymbol{X} \in \RR^2\setminus \boldsymbol{0}$, define radial $R$ and angular $\boldsymbol{W}$ variables by $\boldsymbol{X}\mapsto(R,\boldsymbol{W}):=(\|\boldsymbol{X}\|,\boldsymbol{X}/\|\boldsymbol{X}\|)$, such that $R >0$ and $\boldsymbol{W}\in\mathcal{S}^{1}$, where $\mathcal{S}^{1} := \{ \boldsymbol{x} \in \RR^{2}: \|\boldsymbol{x}\| = 1\}$ denotes the (closed) unit ball of $\|\cdot\|$. Clearly $\boldsymbol{X} = R\boldsymbol{W}$, implying the behaviour of $\boldsymbol{X}$ is directly related to the joint distribution of $(R,\boldsymbol{W})$. It is trivial to show that the mapping $t:\RR^2\setminus \boldsymbol{0} \mapsto \RR_+ \times \mathcal{S}^{1}$, where $t(\boldsymbol{x}) := (\| \boldsymbol{x} \|, \boldsymbol{x}/\| \boldsymbol{x} \|)$, is bijective; thus, no information is lost through considering $(R,\boldsymbol{W})$. Owing to the star-shaped property of $\mathcal{G}$ and the homogeneity of $g(\cdot)$, one can then reformulate the boundary set in terms of the angular component; specifically, we have $\partial \mathcal{G} = \{  \boldsymbol{w}/g(\boldsymbol{w}) : \boldsymbol{w} \in \mathcal{S}^{1} \}$. This implies that $\partial \mathcal{G}$ can be specified by only evaluating the gauge function on $\mathcal{S}^{1}$.

All of the existing limit set estimation approaches make model assumptions about the tail of $(R \mid \boldsymbol{W} = \boldsymbol{w})$, $\boldsymbol{w} \in \mathcal{S}^{1}$, from which the limit set can be approximated. Treating the angle as a covariate, or predictor variable, this modelling strategy is similar to a non-stationary univariate extreme value analysis, which is also the foundation of the recently-developed SPAR model for multivariate extremes \citep{Mackay2023}, and for which a wide range of modelling approaches are available \citep[see, e.g.,][]{Murphy-Barltrop2024,Mackay2024}.

\subsection{Non-stationary limit sets}
\label{subsec:ns_limit_sets}

Let $\boldsymbol{X}_t$ denote a non-stationary, bivariate process with stationary Laplace margins. We choose this marginal scale since it offers a more complete picture of extremal dependence compared to single tailed marginal distributions, especially in the case of negative dependence; see \citet{Murphy-Barltrop2024d} for a detailed discussion. To extend the geometric extremes framework to the non-stationary setting, we first require additional notation and assumptions. First, we assume the process $\boldsymbol{X}_t$ is conditionally stationary, i.e., given $\boldsymbol{z}_t \in \RR^p$, the distribution of $(\boldsymbol{X}_t \mid \boldsymbol{Z}_t = \boldsymbol{z}_t)$ is stationary, independent of $t$ \citep[e.g.,][]{Caires2005}. We also assume the process is conditionally independent in the sense that $(\boldsymbol{X}_t \mid \boldsymbol{Z}_t) \indep (\boldsymbol{X}_{t'} \mid \boldsymbol{Z}_{t'})$ for any $t \neq t'$. Furthermore, for all $t \in \{1,\hdots,T \}$ and any covariate realisation $\boldsymbol{z}_t \in \RR^p$, we assume that the joint density of $(\boldsymbol{X}_t \mid \boldsymbol{Z}_t = \boldsymbol{z}_t)$, denoted $f_{\boldsymbol{X}_t\mid \boldsymbol{Z}_t}(\cdot \mid \boldsymbol{z}_t)$, exists and satisfies  
\begin{equation} \label{eqn:ns_gauge}
    -\log f_{\boldsymbol{X}_t\mid \boldsymbol{Z}_t}(u\boldsymbol{x} \mid \boldsymbol{z}_t) \sim ug_{\boldsymbol{z}_t}(\boldsymbol{x}), \; \; \boldsymbol{x} \in \RR^2, \; u \to \infty,
\end{equation}
where $g_{\boldsymbol{z}_t}(\cdot)$ is termed the \textit{non-stationary gauge function}. Letting $C^{\boldsymbol{z}_t}_n := {\{ \boldsymbol{X}^{i}_t/r_n\}^n_{i=1}}$ be $n$ independent copies of the conditional variable $(\boldsymbol{X}_t \mid \boldsymbol{Z}_t = \boldsymbol{z}_t)$ and combining these assumptions, we obtain the following lemma, the proof of which follows directly from Proposition 2.2 of \citet{Nolde2022}.
\begin{lemma}
    Given a conditionally stationary and conditionally independent process $\boldsymbol{X}_t\mid \boldsymbol{Z}_t$ with a density $f_{\boldsymbol{X}_t\mid \boldsymbol{Z}_t}(\cdot \mid \boldsymbol{z}_t)$ satisfying equation~\eqref{eqn:ns_gauge}, we have that the scaled sample cloud $C^{\boldsymbol{z}_t}_n$ converges in probability onto $\mathcal{G}_{\boldsymbol{z}_t} := \{\boldsymbol{x}_t: g_{\boldsymbol{z}_t}(\boldsymbol{x}) \leq 1 \} \subset [-1,1]^2$. 
\end{lemma}
 
As in Section~\ref{subsec:limit_sets_conditions}, we define the boundary set $\partial \mathcal{G}_{\boldsymbol{z}_t} := \{\boldsymbol{x}_t: g_{\boldsymbol{z}_t}(\boldsymbol{x}_t) = 1 \} \subset \mathcal{G}_{\boldsymbol{z}_t}$, which we term the \textit{non-stationary limit set} henceforth. We acknowledge that in practical application, it is very unlikely that one will observe repeated observations from the variable $(\boldsymbol{X}_t \mid \boldsymbol{Z}_t = \boldsymbol{z}_t)$. Consequently, our estimation strategy, outlined in Section~\ref{sec:methodology}, takes a global approach to estimate $\partial \mathcal{G}_{\boldsymbol{z}_t}$, whereby information is pooled over time to account for the limited information. 

We emphasise here that the framework outlined in this section is more general than many existing approaches for modelling non-stationary extremal dependence detailed in Section~\ref{subsec:ns_dep}. In particular, knowledge of $\partial \mathcal{G}_{\boldsymbol{z}_t}$ allows us to estimate a wide range of dependence coefficients, parameters and risk measures. For instance, taking the coefficient of tail dependence $\eta \in (0,1]$, a non-stationary counterpart is given by $\eta_{\boldsymbol{z}_t} := \max \{s \in [0,1] : [s,\infty)^2 \cap \partial \mathcal{G}_{\boldsymbol{z}_t} \neq \emptyset \}$. Such quantities are useful for summarising trends in extremal dependence and performing probability estimation. As another example use case, consider the following: given any $p \in (0,1)$ close to 1, $\partial \mathcal{G}_{\boldsymbol{z}_t}$ allows us to specify a set $\mathcal{A}_{\boldsymbol{z_t}}^p$ satisfying $\Pr[\boldsymbol{X}_t \in \mathcal{A}_{\boldsymbol{z_t}}^p \mid \boldsymbol{Z}_t = \boldsymbol{z}_t ] = p$. Such sets are known as environmental contours \citep{Haver2004} or return level sets \citep{Papastathopoulos2025}. In the geometric extremes setting, this set is centred at $\mathbf{0}$ and is computed using high quantiles at a fixed probability level from the conditional variables $(R \mid \boldsymbol{W} = \boldsymbol{w}), \boldsymbol{w} \in \sphere$, thus ensuring an equal probability of exceedance in any angular direction. We term $\mathcal{A}_{\boldsymbol{z_t}}^p$ a \textit{non-stationary return level set}, and remark that such sets provide interpretable and intuitive summaries of joint extremal risks that are used to aid with design analysis in practice \citep{Mackay2021}; see Section~\ref{sec:case_study} for further discussion. 

Finally, the framework outlined in this section can be used for full tail inference on $(\boldsymbol{X}_t \mid \boldsymbol{Z}_t = \boldsymbol{z}_t)$, including, but not limited to, simulation. This represents a significant practical advantage since many applications of multivariate extreme value theory require a means of efficiently simulating synthetic data in the joint tail, e.g., flood risk mitigation \citep{Keef2013a} and catastrophe modelling  \citep{Quinn2019}. Moreover, simulated data allows one to estimate non-stationary tail probabilities for $(\boldsymbol{X}_t \mid \boldsymbol{Z}_t = \boldsymbol{z}_t)$; such probabilities are often used to aid with structural engineering \citep{Jonathan2014}, portfolio management \citep{Nolde2021} and environmental planning \citep{Gouldby2017}.

\section{Modelling and learning from data} \label{sec:methodology}

Having outlined a non-stationary framework for limit sets in Section~\ref{sec:ns_limit_sets}, we now turn our attention to estimation of the non-stationary gauge function. As illustrated in Section~\ref{subsec:limit_sets_conditions}, we are only required to estimate $g_{\boldsymbol{z}_t}(\cdot)$ on any unit ball to obtain the boundary set, and we therefore restrict our estimation domain to $\sphere$. Following Section~\ref{subsec:limit_sets_conditions}, we define $R_t := \|\boldsymbol{X}_t\|$ and $\boldsymbol{W}_t:=\boldsymbol{X}_t/\|\boldsymbol{X}_t\|$, and turn our interest to understanding the tail of $(R_t \mid \boldsymbol{W}_t = \boldsymbol{w}_t, \boldsymbol{Z}_t = \boldsymbol{z}_t)$ for any given $t \in \{1,\hdots,T \}$. Naturally we wish to avoid making strong assumptions about the joint tail of $(R_t \mid \boldsymbol{W}_t = \boldsymbol{w}_t, \boldsymbol{Z}_t = \boldsymbol{z}_t)$ while simultaneously avoiding over-fitting. Furthermore, we desire an estimator that is global in the sense that we do not restrict attention to subregions of $\sphere$, or $\{1,\hdots,T\}$, during inference. Such estimators allow one to pool information over the observation period, accounting for the limited number of data points available at any time point. 

\subsection{A simplified representation} \label{subsec:simple_rep}

To simplify the representation of the joint tail for our estimation procedure, we take two additional steps. First, we define the polar angular variable $\Phi_t := \atantwo (X_{2,t},X_{1,t}) \in [0,2\pi)$, where $\atantwo$ denotes the 2-argument arctangent function. Given any norm $\| \cdot \|$, one can identify any point $\boldsymbol{w} \in \sphere$ with an angle $\phi \in [0,2\pi)$: in particular, the transformation $v:[0,2\pi) \mapsto \sphere$ given by $\boldsymbol{v}(\phi) = (\cos(\phi),\sin(\phi))/\| (\cos(\phi),\sin(\phi))\|$ is one-to-one. Consequently, we can represent the unit ball for any norm via the polar angular interval $[0,2\pi)$, and no information is lost by considering $\Phi_t$ in place of $\boldsymbol{W}_t$. Furthermore, it is more straightforward to define functions on the interval $[0,2\pi)$ compared to $\sphere$, owing to the fact the former requires only a univariate periodic formulation to be valid. We therefore reformulate the non-stationary gauge function $g_{\boldsymbol{z}_t}(\cdot)$ via $m(\phi,\boldsymbol{z}_t):=g_{\boldsymbol{z}_t}(\boldsymbol{v}(\phi))$, $\phi \in [0,2\pi)$. We opt for the notation $m(\phi,\boldsymbol{z}_t)$ over $m_{\boldsymbol{z}_t}(\phi)$ to emphasise the fact that, under our modelling formulation, we treat both $\phi$ and $\boldsymbol{z}_t$ as predictor variables in our framework; see Section~\ref{subsec:GAMs_overview}. Observe that under this reformulation, we have $\partial \mathcal{G}_{\boldsymbol{z}_t} = \{  \boldsymbol{v}(\phi)/m(\phi,\boldsymbol{z}_t) : \phi \in [0,2\pi) \}$; see Section \ref{subsec:limit_sets_conditions}.

To further simplify the formulation, we treat the discrete index $t$ as a continuous covariate (i.e., we set $\boldsymbol{z}_t = t$), and explicitly assume that any relevant or unobserved covariates vary smoothly as a function of $t$. Combined, these steps allow us to simplify the formulation of the framework introduced in Section~\ref{sec:ns_limit_sets} and henceforth, we use $t$ and $\Phi_t$ in place of $\boldsymbol{z}_t$ and $\boldsymbol{W}_t$, respectively. 

\subsection{Modelling the conditional radial tails} \label{subsec:cond_radii_tail}
To approximate limit sets in the stationary setting, \citet{Wadsworth2024} study the tail behaviour of the variable $(R\mid \Phi = \phi)$. While the authors did not consider the non-stationary setting, their derivations extend to this case. In particular, if equation~\eqref{eqn:ns_gauge} holds, we have that $f_{R_t\mid \Phi_t, t}(r\mid\phi,t) \propto r^{d-1}\exp\{-rm(\phi,t)[1+o(1)]\}$, $r \to \infty$, where $f_{R_t\mid \Phi_t,t}(\cdot\mid \phi,t)$ denotes the density function $(R_t \mid \Phi_t = \phi, t)$. In addition, their arguments imply that for many dependence structures
\begin{equation} \label{eqn:ns_gamma_kernel}
    f_{R_t\mid \Phi_t, t}(r\mid\phi,t) \propto r^{d-1}\exp\{-rm(\phi,t)\}[1+o(1)], \hspace{1em} r \to \infty,
\end{equation}
 giving asymptotic equivalence of equation~\eqref{eqn:ns_gamma_kernel} with a gamma kernel. This motivates the following modelling assumption: 
\begin{equation} \label{eqn:trunc_gamma_assum}
    (R_t \mid \Phi_t = \phi, t, R_t>r^\tau(\phi,t) ) \sim \text{truncGamma}(d,m(\phi,t)),
\end{equation}
where `truncGamma' is shorthand for the \textit{truncated gamma} distribution with shape and rate parameters $d$ and $m(\phi,t)$, respectively, and $r^\tau(\phi,t)$ is the $\tau$-quantile of $(R_t \mid \Phi_t = \phi, t)$ for some $\tau \in (0,1)$ close to 1, i.e., $\Pr (R_t \leq r^\tau(\phi,t) \mid \Phi_t = \phi, t) = \tau$. Selection of $\tau$ is discussed in Section~\ref{sec:sim_study}. 

For our framework, we assume that the assumption outlined in equation~\eqref{eqn:trunc_gamma_assum} holds for all $t$ and $\phi \in [0,2\pi)$. Empirical evidence from many approaches indicates that this assumption is flexible enough to capture a wide range of dependence structures \citep{Wadsworth2024,Majumder2024,Murphy-Barltrop2024d}. This resulting inference procedure involves two-steps: estimation of the non-stationary quantile function $r^\tau(\phi,t)$, and estimation of the non-stationary gauge function $m(\phi,t)$. We represent both of these functions using GAMs, an overview of which is provided in Section~\ref{subsec:GAMs_overview}.

We remark that one can also leave the gamma shape parameter of equation \eqref{eqn:trunc_gamma_assum} as a free parameter to estimate, as has been done in several existing limit set estimation approaches \citep[e.g.,][]{Wadsworth2024,Murphy-Barltrop2024d}. Fixing this quantity to $d$ corresponds to assuming that the asymptotic form of equation \eqref{eqn:ns_gamma_kernel} holds exactly at some finite level. In unreported results, we found that including the shape parameter in the non-stationary framework did not improve the quality of limit set estimates, and in some cases even reduced their quality. We therefore chose to fix the shape at $d$, which has the added benefit of reducing parameter variability.

\subsection{Modelling covariate interactions with GAMs} \label{subsec:GAMs_overview}

GAMs \citep{Wood2017} provide a flexible, semi-parametric framework for capturing complex functional forms without requiring rigid modelling assumptions. Typically, GAMs are used in practice to capture the relationship between a response variable and a set of predictor variables; the resulting fitted model can subsequently be used for prediction and interpolation across the joint domain of the predictor variables. Unlike classical regression techniques, which tend to assume simple linear relationships between variables, GAMs exploit smooth functional forms to capture non-linear patterns, alongside complex interactions that may arise between dependent predictor variables. 

When fitting GAMs in practice, the optimal choice of smoothing functions and tuning parameters is context dependent, and fine tuning is typically required to ensure the model accurately represents the underlying data structure. Consequently, we only describe GAMs in the context of our proposed framework. As noted in Section~\ref{subsec:cond_radii_tail}, we wish to estimate the functions $r^\tau(\phi,t)$ and $m(\phi,t)$ for all $t$ and $\phi \in [0,2\pi)$. Thus, under the GAM framework, we view $t$ and $\phi$ as predictor variables. 

We propose the following GAM formulations  
\begin{equation} \label{eqn:GAM_form}
    \log[r^\tau(\phi,t)] = \beta_0 + s_{tp}(t, \phi), \; \; \;     \log[m(\phi,t)] = \beta^*_0 + s_{tp}^*(t, \phi),    
\end{equation}
where $\beta_0,\beta^*_0 \in \RR$ are intercept terms and 
$s_{tp}(\cdot, \cdot), s_{tp}^*(\cdot, \cdot)$ are smooth tensor product splines capturing the interaction between $t$ and $\phi$. Tensor splines are formed by taking products of univariate smooth spline functions defined on the predictor variables. Each univariate spline is formed of a piecewise combination of flexible polynomial functions, and the points at which these polynomials are connected are called \textit{knots}. In our case, we use cubic and cyclic cubic splines for $t$ and $\phi$, respectively, where latter spline ensures the periodicity over the polar angular variable. Cubic splines posses many desirable properties, such as optimality in terms of smoothness, high flexibility and computational efficiency \citep{Wood2017}. Note that more complex modelling frameworks involving adding univariate smooth splines to the tensor splines of equation~\eqref{eqn:GAM_form} did not lead to noticeable improvements within our modelling framework. An illustration of tensor product spline basis functions are provided in Appendix~\ref{sec:appen_additional_figures}.  

Identical GAM formulations are taken for $r^\tau(\phi,t)$ and $m(\phi,t)$; this is due to the fact, as $\tau \to 1$, these functions are approximately inversely proportional \citep{Wadsworth2024} and thus will vary in a similar manner over the predictor variable domain(s). Let $\boldsymbol{\beta}$ and $\boldsymbol{\beta}^*$ denote the spline coefficients associated with the tensor product functions
$s_{tp}(\cdot, \cdot)$ and $s_{tp}^*(\cdot, \cdot)$, respectively. Denote the number of knots (i.e., the \textit{basis dimensions} of the univariate splines) as $\kappa_t$ and $\kappa_{\phi}$ for $t$ and $\phi$, respectively. Note that the basis dimensions correspond to the flexibility of the model, and their selection typically represents a trade-off; higher dimensions results in greater flexibility but increased parameter variability. In practice, the function in equation~\eqref{eqn:GAM_form} is also penalised to avoid over-fitting, which commonly occurs for models with large parameter sets. We return to this discussion in Section~\ref{subsec:model_formulation}. The resulting parameter sets are given by $\boldsymbol{\theta} := (\beta_0,\boldsymbol{\beta})$ and $\boldsymbol{\theta}^* := (\beta^*_0,\boldsymbol{\beta}^*)$, with tuning parameters $\kappa_t$ and $\kappa_{\phi}$ and joint predictor domain $\mathcal{D}:=\{1,\hdots,T \} \times [0,2\pi)$. We now consider estimation of $\boldsymbol{\theta}$ and $\boldsymbol{\theta}^*$. 

\subsection{Quantile regression for the conditional radial variable} \label{subsec:quant_reg}
Estimation of $r^\tau(\phi,t)$ represents a quantile regression problem, for which a wide range of approaches are available \citep{Koenker2017}. These techniques are typically based on the pinball loss function, which, when minimised with respect to a parameter set, returns quantile estimates for any probability level $\tau$. Many recent works have incorporated GAM formulations in the context of quantile regression, allowing for flexible semi-parametric inference \citep[e.g.,][]{Yu2001,Oh2011,Koenker2011,Youngman2019}. However, there are several issues with these approaches. First, the standard pinball function is piecewise linear and is consequently difficult to optimise, leading to difficulty when incorporating complex model formulations with large parameter vectors, such as \eqref{eqn:GAM_form}. Furthermore, some existing approaches require one to manually select `smoothing' parameters (i.e., parameters to mitigate overfitting for highly parametrised models), even though this selection is not straightforward and is crucial for balancing the trade-off between model fit and complexity. Moreover, to estimate GAM parameters, many approaches adopt an asymmetric Laplace model as a working assumption, even though such distributional assumptions are problematic and can result in poorly calibrated, inaccurate quantile estimates \citep{Fasiolo2021}. 

In this work, we employ the approach introduced in \citet{Fasiolo2021}, which overcomes the drawbacks of existing techniques. They introduce a novel technique for fitting GAM-based quantile regression models and demonstrate that the resulting estimator is well calibrated, stable and efficient. They also introduce a novel loss function which does not suffer from the optimisation issues faced by the pinball function. This novel loss function is penalised to impose smoothness on the resulting splines (i.e., avoid overfitting), resulting in penalty parameters. Penalty parameter estimates are obtained via a Bayesian approach, with a so-called `learning rate' which determines the relative weights of the loss and penalty functions. Selection of basis dimensions is discussed in Section~\ref{subsec:model_formulation}, and for every $t$ and $\phi \in [0,2\pi)$, we denote the corresponding quantile estimate by $\hat{r}^\tau(\phi,t)$. 

\subsection{Restricted maximum likelihood estimation} \label{subsec:reml} 
With quantile estimates obtained over $\mathcal{D}$, what remains is to estimate the non-stationary gauge function via the model in \eqref{eqn:trunc_gamma_assum}. For this, we employ the framework of \citet{Youngman2019} to estimate the parameter vector $\boldsymbol{\theta}^*$. Taking the log-likelihood function associated with equation~\eqref{eqn:trunc_gamma_assum}, this approach employs a technique known as restricted maximum likelihood, whereby the log-likelihood function is penalised to avoid over-fitting, thus allowing for flexible GAM formulations in the likelihood function \citep{Wood2011,Wood2016}. In a similar manner to Section~\ref{subsec:quant_reg}, this results in penalty parameters that also need to be estimated as part of the inference procedure. Under the restricted maximum likelihood framework, this is achieved by making distributional assumptions on the smoothing parameters, then maximising the resulting likelihoods. For rigorous mathematical details and guidance on model fitting, see \citet{Wood2017} and \citet{Youngman2019}. 

Given $t$ and $\phi \in [0,2\pi)$, we denote the corresponding non-stationary gauge function estimate by $\hat{m}(\phi,t)$, with the corresponding non-stationary boundary set estimate given by $\widehat{\partial \mathcal{G}}_t = \{  v(\phi)/\hat{m}(\phi,t) : \phi \in [0,2\pi) \}$. Owing to the continuity of the function in equation~\eqref{eqn:GAM_form}, this limit set associated with this estimate is always compact \citep{Murphy-Barltrop2024d}. However, there is no guarantee that $\widehat{\partial \mathcal{G}}_t \in [-1,1]^2$, or that the boundary set estimate possesses componentwise maxima and minima $\boldsymbol{1}$ and $-\boldsymbol{1}$, respectively. See Sections~\ref{sec:case_study} and \ref{sec:discussion} for further discussion. 

Thus far, we have provided a general framework for inference on non-stationary boundary sets, but we have been intentionally ambiguous regarding specific features of the model. In particular, we have not yet specified a norm $\| \cdot\|$, the quantile level $\tau$, or the basis dimensions $\kappa_t$ and $\kappa_{\phi}$; these are important quantities for model inference, and as it turns out, the appropriate values vary depending on the form(s) of extremal dependence. Furthermore, one must also consider the effect of the observation window $T$ and account for sampling variability for any given level of information. Selection of these features is discussed in Sections \ref{sec:sim_study} and \ref{sec:case_study}.  

\subsection{Model checking} \label{subsec:model_checking}

To assess model fits, we introduce two diagnostic tools. Example applications of each diagnostic discussed in this section can be found in Section~\ref{sec:case_study}. Since we do not observe repeated observations for any time point $t$, one cannot compute scaled sample clouds and compare these to fixed boundary set estimates (see, e.g., Figure~\ref{fig:theoretical_limit_sets}), as is common for many stationary approaches. Therefore, a more nuanced approach is required for assessing model fits in the non-stationary setting. 

Firstly, we adopt the goodness-of-fit metric proposed by \citet{Wadsworth2024}, where all threshold exceeding observations are transformed to a standard exponential scale using the model of equation~\eqref{eqn:trunc_gamma_assum}. These transformed observations can be compared to theoretical quantiles via a quantile-quantile (QQ) plot, providing a global diagnostic for model fits over both the polar angle and time variables. Moreover, assuming the transformed data are IID, confidence bounds can be added to these plots by exploiting the fact the $k$-th order statistic of a uniform distribution theoretically follows a $\text{Beta}(k, n + 1 - k)$ distribution \citep{David2003}. 

Next, we consider the return level set diagnostic proposed in \citet{Murphy-Barltrop2024d}. Introduced in the stationary setting, this model computes the empirical probability of points inside an estimated return level set $\hat{\mathcal{A}}^p_t$ for some $p$ close to one, and compares this value to the nominal level. Repeating this procedure over a range of probability levels, one obtains a QQ plot representing the nominal and estimated probabilities. This diagnostic is easily extended to the non-stationary setting by averaging return level set probabilities over time. Assuming unbiased estimation, we have that 
\begin{equation}\label{eqn:rl_set_prob}
    \frac{1}{T} \sum_{t=1}^T\Pr(\boldsymbol{X}_t \in \mathcal{A}_{t}^p) =  \sum_{t=1}^T \frac{p}{T} = p
\end{equation}
Equation~\eqref{eqn:rl_set_prob} can be approximated empirically via the technique described in \citet{Murphy-Barltrop2024d}. In particular, given observations $\boldsymbol{x}_t, \; t=1,\hdots,T$ with $T$ large, we set $\hat{p} = (1/T)\sum_{t=1}^T \mathbbm{1}( \|\boldsymbol{x}_t \| \leq \hat{r}^p( \atantwo (x_{2,t},x_{1,t}) ,t))$, where $\mathbbm{1}(\cdot)$ is the indicator function and $\hat{r}^p(\cdot,\cdot)$ denotes the estimated $p$-quantile function of $(R_t \mid \Phi_t = \phi, t)$ (i.e., the radial values of $\hat{\mathcal{A}}^p_t$) derived from the truncated gamma model fit. The corresponding diagnostic describes how well the fitted model captures the observed structure in the data, and indicates the suitability of the model for inferring risk measures. 

\section{Simulation study} \label{sec:sim_study}

In this section, we explore the performance of the framework introduced in Sections~\ref{sec:ns_limit_sets} and \ref{sec:methodology}. Section~\ref{subsec:sim_examples} introduces a range of simulated examples exhibiting non-stationary dependence. In Section~\ref{subsec:performance_metrics}, we discuss metrics for assessing model performance over $\mathcal{D}$. Section~\ref{subsec:gam_smooth_learn} discusses estimation of the penalty parameters from the frameworks introduced in Sections~\ref{subsec:quant_reg} and \ref{subsec:reml}. Model formulation is discussed in Section~\ref{subsec:model_formulation} and in Section~\ref{subsec:sim_study_results}, we present results for each of the simulated examples. To estimate the quantile and gauge functions, we employ the \texttt{qgam} \citep{Fasiolo2021b} and \texttt{evgam} \citep{Youngman2022} software packages, respectively, in the \texttt{R} programming language. Example code for fitting our framework is freely available at \url{https://github.com/callumbarltrop/NSGE}, and wrapper functions will soon be incorporated into the Github version of \texttt{evgam}; see \url{https://github.com/byoungman/evgam}. 

\subsection{Simulated examples of non-stationary dependence structures} \label{subsec:sim_examples}
Given an observation window $T$, we define a range of non-stationary structures in terms of the time covariate $t$. All samples are simulated on standard Laplace margins. The first two examples are obtained using the bivariate Gaussian copula, for which dependence is characterised by the correlation coefficient $\rho \in [-1,1]$. For the first case, we set $\rho(t) := 0.2 + 0.6(t-1)/(T-1)$, giving $\rho(1) = 0.2$ and $\rho(T) = 0.8$, i.e., moving from weak to strong positive dependence. For the second example, we define the functions $a(t) := 0.45(t-1)/(T-1)$ and $b(t):= 0.5\sin(2.5\pi(t-1)/(T-1))$, and set $\rho_2(t) = a(t) + b(t)$; this corresponds to a harmonic, increasing correlation function. The dependence trend is complex in this case, moving from positive, to negative, then back to positive dependence. An illustrative figure of $\rho_2(t)$ over $t$ is provided in Appendix~\ref{sec:appen_additional_figures}.  

For the third example, we use the inverted bivariate extreme value copula with the logistic family \citep{Ledford1997}. Dependence is classified through the parameter $\alpha \in (0,1)$, with positive dependence increasing as $\alpha$ approaches $0$. We set $\alpha(t):= 0.3 + 0.4(t-1)/(T-1)$, resulting in a trend that moves from strong to weak positive dependence. 

For the fourth example, we consider the student-$t$ copula family, for which dependence is quantified via a parameter $\sigma \in [-1,1]$ representing the degree of collinearity in the data, alongside the degrees of freedom parameter $\nu > 0$. Lower values of $\nu$ correspond with stronger tail dependence. In this case, we fix $\sigma = 0.5$ for all $t$ and set $\nu(t) := 0.5 + 1.5(t-1)/(T-1)$; this corresponds to a trend whereby the collinearity parameter remains constant over time, but the tail dependence gradually decreases. 

Finally, for the fifth example, we consider the copula model introduced by \citet{Huser2019}. This model was first proposed on one-tailed margins, and consequently we reformulate it slightly to define it with standard Laplace margins. A version of this model with asymmetric tails has also been considered by \citet{Gong2022a}. Let $S$ denote a standard Laplace variable and $\boldsymbol{V}$ denote a bivariate Gaussian copula on standard Laplace margins with correlation coefficient $\rho$, with $S$ and $\boldsymbol{V}$ independent from each other. Given any $\delta > 0$, define the random vector 
\begin{equation}\label{eqn:hw_model}
    \boldsymbol{Y} = \begin{cases}
            \delta S + \boldsymbol{V}, & \delta < 1, \\
            S + \boldsymbol{V}/\delta, & \delta \geq 1.
        \end{cases}
\end{equation}
Note that $\boldsymbol{Y}$ does not posses standard Laplace margins, and additional marginal transformation involving the parameter $\delta$ is required to obtain standardised margins. The model in equation~\eqref{eqn:hw_model} can be interpreted as follows: for $\delta < 1$, $\boldsymbol{V}$ is heavier tailed than $\delta S$, inducing AI, while for $\delta \geq 1$, $S$ dominates $\boldsymbol{V}/\delta$, resulting in AD. One can observe that a Gaussian copula is recovered as we let $\delta$ approach $0$. The transition between the two dependence classes occurs at $\delta = 1$, corresponding to an interior point of the parameter space; this is in contrast to most extreme value models, for which this transition occurs at the boundary of the parameter space(s) \citep{Wadsworth2017}. Consequently, we fix $\rho = 0.5$ and set $\delta(t):= 0.7 + 1.8(t-1)/(T-1)$, corresponding to a smooth transition between AI and AD. 

\begin{figure}[h]
    \centering
    \includegraphics[width=\linewidth]{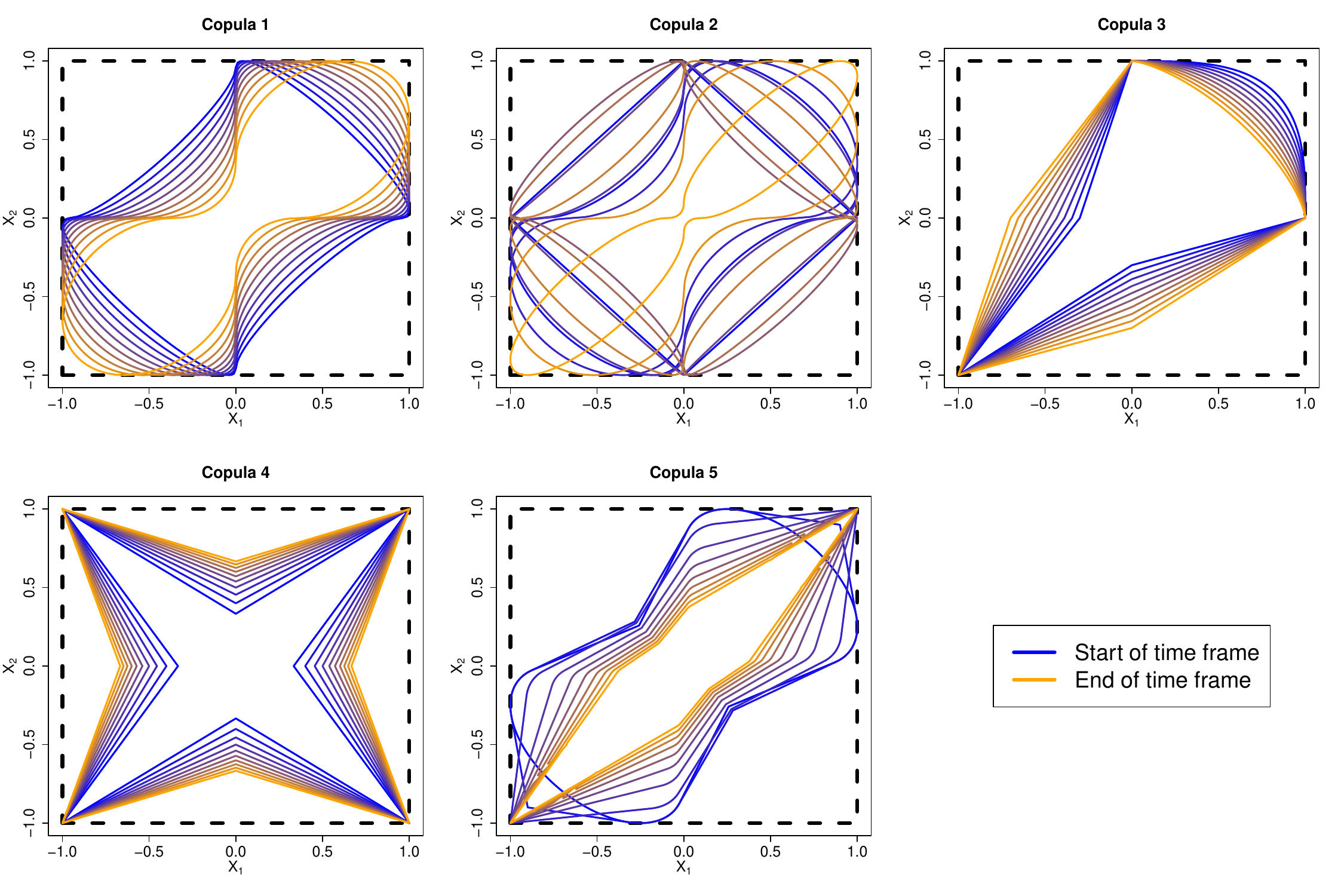}
    \caption{Boundary sets for each copula example at a finite, increasing set of equally spaced time indices. The boundary sets at the start and end of the observation interval are illustrated in blue and orange, respectively, with a colour transition used to visualise the trend in dependence.}
    \label{fig:ns_boundary_set}
\end{figure}

We simulate $200$ samples from each of the copula structures outlined in this section. This allows us to assess both the bias and variability of our proposed modelling framework across the different dependence structures. For each copula example, illustrations of the resulting boundary sets over time are provided in Figure~\ref{fig:ns_boundary_set}. We note that the cases where `pointiness' is observed in some, or all, of the quadrant corners (e.g., the student-$t$) correspond to copulas possessing strong tail dependence within such regions.  

\subsection{Performance metrics} \label{subsec:performance_metrics}
The overall goal of our modelling framework is to estimate non-stationary boundary sets, and consequently we assess performance in terms of set estimates. However, owing to the additional time dimension, assessing performance is not straightforward in the non-stationary setting. For example, one could plot all two-dimensional boundary set estimates on a time axes alongside the true sets, but the corresponding three-dimensional plot would be difficult to interpret, and it may be unclear in which regions the model performs well. Furthermore, many performance metrics proposed in the stationary setting, such as mean integrated squared error of the gauge function over $[0,2\pi)$ \citep{Murphy-Barltrop2024} or bias in extremal coefficients \citep{Simpson2024a,Majumder2024}, cannot be easily generalised to the non-stationary setting. 

Consequently, we follow \citet{Murphy-Barltrop2024b}, who propose fixing either the time index or the polar angle, then evaluating performance visually over the domain of the other component. Firstly, given a set of fixed time points $\{1, \lfloor T/2 \rfloor, T \}$, corresponding to the start, middle and end of the observation window, respectively, we compute estimates of the non-stationary boundary set at each $t \in \{1, \lfloor T/2 \rfloor, T \}$ and plot these alongside the true boundary sets. This allows us to visually assess what  boundary set shapes can be represented by the framework proposed in Section~\ref{sec:methodology}. Evaluating the joint tail at the boundaries of the observation window also provides a good indicator of the overall model performance, since there is less pooled information available at these points. Secondly, we fix angles $\{\pi/4, 3\pi/4, 5\pi/4, 7\pi/4 \}$, corresponding to the main diagonals in all four quadrants. For each $\phi \in \{\pi/4, 3\pi/4, 5\pi/4, 7\pi/4 \}$, we estimate the corresponding Cartesian point on the boundary set, denoted $\hat{\boldsymbol{x}}^{\phi}_t := v(\phi)/\hat{m}(\phi,t) \in \widehat{\partial \mathcal{G}}_t$, for every $t$. We then compute the Euclidean distance of $\hat{\boldsymbol{x}}^{\phi}_t$, i.e., $\hat{r}^{\phi}_t := \sqrt{ (\hat{x}^{\phi}_{1,t})^2 + (\hat{x}^{\phi}_{2,t})^2 }$, and plot this estimate against $t$, alongside the corresponding true values. This allows us to visually assess how well our framework is capturing the simulated dependence trends in different regions. Note that we evaluate the Euclidean distance of the Cartesian point to allow for comparison between different norm definitions of the radial component; plotting the estimates of $\hat{m}(\phi,t)$ for different norm definitions, for instance, would be meaningless, since the coordinates at which the gauge function is evaluated are dependent on the choice of norm. 

\subsection{Selecting GAM smoothing parameters and learning rates} \label{subsec:gam_smooth_learn}

We note that both steps of our modelling procedure, as outlined in Sections~\ref{subsec:quant_reg} and \ref{subsec:reml}, require smoothing parameters for penalising loss functions used for optimisation. Such parameters can be estimated automatically by making distributional assumptions directly on the smoothing parameters \citep{Wood2003}. Moreover, for the Bayesian quantile regression framework, a learning rate parameter must be estimated prior to sampling from the posteriors of the GAM parameters. Estimation of, or obtaining samples from, the smoothing parameters is computationally expensive. Moreover, across different data sets from the same copula and sample size, it is unlikely the same smoothing parameters will be estimated twice, implying one cannot easily compare GAM coefficient estimates for different copula samples. Therefore, prior to running our large simulation study, we opt to fix model formulations to speed up computation and ensure comparability across sampling iterations. 

For each of the possible combinations of model tuning parameters and features (see Section~\ref{subsec:reml}), we fit our modelling framework to $10$ unique samples from each copula and use the procedures outlined in \citet{Fasiolo2021} and \citet{Youngman2019} to automatically select the relevant parameters. For the Bayesian quantile regression procedure, we compute the median learning rate across the $10$ samples and then fix the learning rate to this estimate for the remainder of the simulation study. Note that being in Bayesian framework, this still requires one to sample from the posteriors of the GAM parameter distributions, and there is no guarantee such distributions will be the same across samples. For the frequentist REML framework, we compute the median smoothing parameters for the tensor product spline and then fix the smoothing parameter for any given combination to these medians. This ensures the same penalised likelihood function is used for each setup. 

\subsection{Selecting a model formulation} \label{subsec:model_formulation}
As noted in Section~\ref{subsec:reml}, our proposed framework requires us to specify a range of tuning parameters and features. Trying to select all of these terms simultaneously would be very difficult, and it is unlikely that the same model formulation would be optimal across all simulated example. Consequently, we select each model component separately, loosely assuming the optimal choices of each model feature can be selected independently of others. 

\begin{figure}[h]
    \centering
    \includegraphics[width=\linewidth]{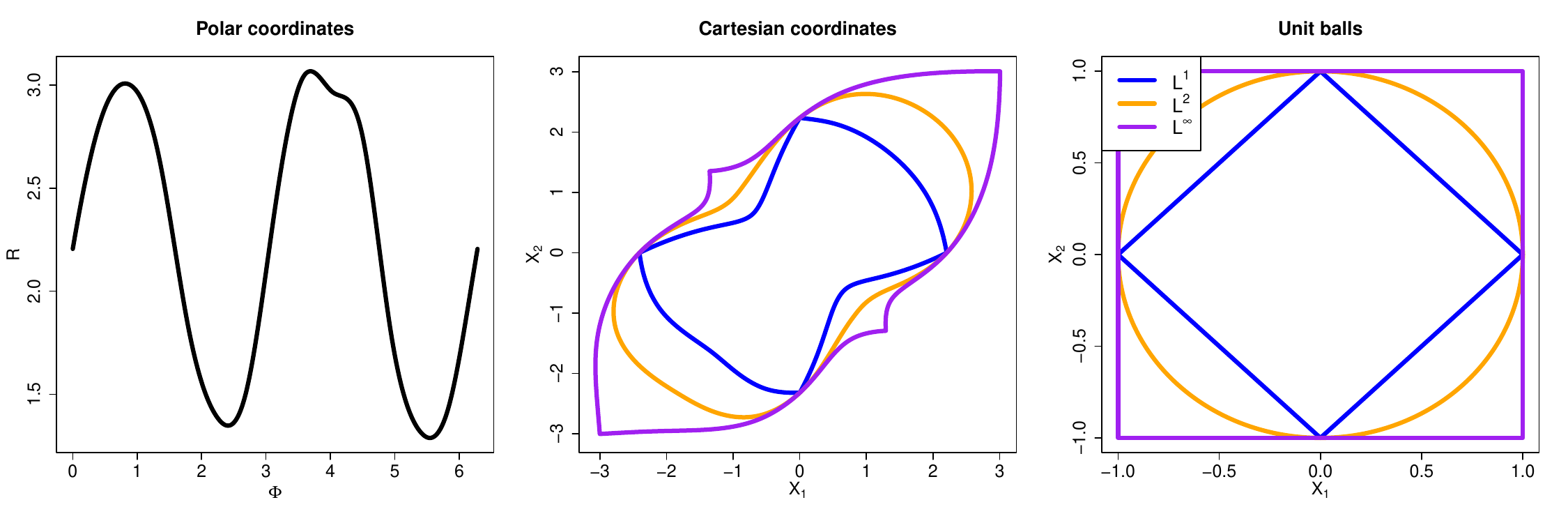}
    \caption{Plot illustrating the affect of norm choice on Cartesian set constructions. The left and centre panels illustrate the spline functions on polar and Cartesian scales, respectively. The right panel illustrates the unit balls for each choice of norm. }
    \label{fig:different_norms}
\end{figure}

For the norm function $\|\cdot \|$, we consider three popular choices: $\|\boldsymbol{x} \|_1 = |x_1| + |x_2|$, $\|\boldsymbol{x} \|_2 = \sqrt{x^2_1 + x^2_2}$ and $\|\boldsymbol{x} \|_{\infty} = \max\{|x_1|,|x_2|\}$, corresponding to the $L^1$, $L^2$ and $L^{\infty}$ norms, respectively. As observed in \citet{Murphy-Barltrop2024}, the unit ball associated with each norm directly influences the shape of the estimated boundary set. To see this, consider Figure~\ref{fig:different_norms}: in this plot, we have defined a cyclic cubic spline in polar coordinates, and for each norm, the corresponding Cartesian coordinates have been computed. While the spline is smooth on the polar scale, the resulting shapes in Cartesian coordinates possess noticeable kinks for the $L^1$ and $L^{\infty}$ norms. This occurs due to the shapes of corresponding unit balls, which are also illustrated in right panel of Figure~\ref{fig:different_norms}. Consequently, certain norm choices may be more appropriate for scaled sample clouds which appear particularly `pointy' within certain regions. We stress here that in practice, it is unlikely the same spline would be estimated across all three norms, and Figure~\ref{fig:different_norms} is meant to merely illustrate the difference in shapes when splines are transformed to the Cartesian scale. 

For the quantile level, we consider $\tau \in \{0.5,0.6,0.7,0.8,0.9\}$. This choice represents a typical bias-variance trade-off that is often observed in extreme value theory \citep[see, e.g.,][]{Coles2001}, and for the framework proposed in Section~\ref{sec:methodology}, this equates to assessing the rate of convergence of equation~\eqref{eqn:trunc_gamma_assum} to the true asymptotic tail distribution of the conditional radii variables. We wish to select $\tau$ as high as possible without compromising the robustness or reliability of the modelling framework. 

For the time basis dimension, we consider $\kappa_t \in \{5,10,15 \}$ and define spline knots at equally spaced points along the observation window, i.e., for $j = 1,\hdots,\kappa_t$, set $k^t_j := 1 + (T-1)(j-1)/(\kappa_t - 1)$. As noted in \citet{Wood2017}, the exact location of spline knots is not so important, so long as the basis dimension is sufficiently high enough to provide adequate flexibility. For the polar angular basis dimension, we consider $\kappa_{\phi} \in \{9,17\}$ and again define knots at equally spaced points on the angular window, i.e., for $j = 1,\hdots,\kappa_{\phi}$, set $k^{\phi}_j := 2\pi(j-1)/(\kappa_{\phi} - 1)$. These choices are practically motivated by the fact such basis dimensions result in knots that include the axes and principal diagonals in $\RR^2$; these regions are often where the boundary set exhibits key features related to extremal dependence properties. As such, it is important that our modelling framework can capture the behaviour around these regions. An illustrative figure of the polar angular knot locations is provided in Appendix~\ref{sec:appen_additional_figures}, and we refer to \citet{Simpson2024} for further discussion. 

To begin, we consider the effect of the quantile level $\tau$. For this, we first fix the basis dimensions to their maximal values, i.e., $\kappa_t = 15$, $\kappa_{\phi} = 17$, so as to provide maximal flexibility, and just consider the $L^2$ norm. For each example in Section~\ref{subsec:sim_examples}, we also set $T = 25,000$ to provide a large, but not unreasonable, sample size. Note that smaller $T$ values are subsequently considered once all tuning parameters have been selected. For each copula example, we then apply the framework introduced in Section~\ref{sec:methodology} across $200$ examples and compute the performance metrics discussed in Section~\ref{subsec:performance_metrics}. These results are given in Appendix~\ref{subsec:appen_tau}. Encouragingly, the model appears remarkably robust to the choice of $\tau$, accurately capturing the range of dependence structures over $\mathcal{D}$. Naturally there is higher variability at higher $\tau$ levels, as one would expect, though we also observe slightly better coverage for such values. Consequently, we fix $\tau = 0.8$ for the remainder of this article; this value appears to offer a reasonable trade-off between bias and variance. 

Next, we assess the effect of the basis dimensions $\kappa_t$ and $\kappa_{\phi}$. For the former, we fix $\kappa_{\phi} = 17$ and again restrict attention to the $L^2$ norm, then vary $\kappa_t \in \{ 5,10,15\}$. Using the performance metrics described in Section~\ref{subsec:performance_metrics}, the results are given in Appendix~\ref{subsec:appen_basis_dimension}. For most of the copula examples, there appears to be little difference among the basis dimensions. However, for the complex dependence trend exhibited by copula 2, $\kappa_t = 5$ does not appear sufficiently flexible to capture the underlying trends. Maintaining the same setup with $\kappa_t = 15$ and allowing $\kappa_{\phi}$ to vary in $\{ 9,17\}$, we evaluate the effect of basis dimension for the angular component, with the results again given in Appendix~\ref{subsec:appen_basis_dimension}. It is clear that the additional flexibility arising from setting $\kappa_{\phi} = 17$ allows the framework to better capture the complex variety of boundary set shapes. Taking these combined results into account, we fix $\kappa_t = 10$ and $\kappa_{\phi} = 17$ for the remainder of the article.

Finally, we consider the impact of the norm choice on the modelling framework. For this, we restrict attention to the first and fourth copula examples, where the underlying boundary sets are smooth and pointy respectively. With all other components of the model specified, we vary the norm across the three choices (i.e., $L^1$, $L^2$ and $L^{\infty}$). These results are given in Appendix~\ref{subsec:appen_norm}.  It is clear that the choice of norm does not appear to significantly alter the rate of convergence for the modelling framework described in equation~\eqref{eqn:trunc_gamma_assum}, but unsurprisingly affects the types of Cartesian shapes that can be represented by the framework. Visually, it appears that the $L^2$ and $L^{\infty}$ norms would be preferable for the first and fourth examples, respectively. However, without knowledge of theoretical boundary sets, it is not clear how one should select the norm choice in practice; we refer to Section~\ref{sec:discussion} for further discussion. Consequently, for the remainder of this simulation study, we restrict attention to the $L^2$ norm, acknowledging that this will impose smoothness on the resulting boundary set estimates. 

\subsection{Results} \label{subsec:sim_study_results}

\begin{figure}[h]
    \centering
    \begin{subfigure}{0.32\textwidth}
        \includegraphics[width=\linewidth]{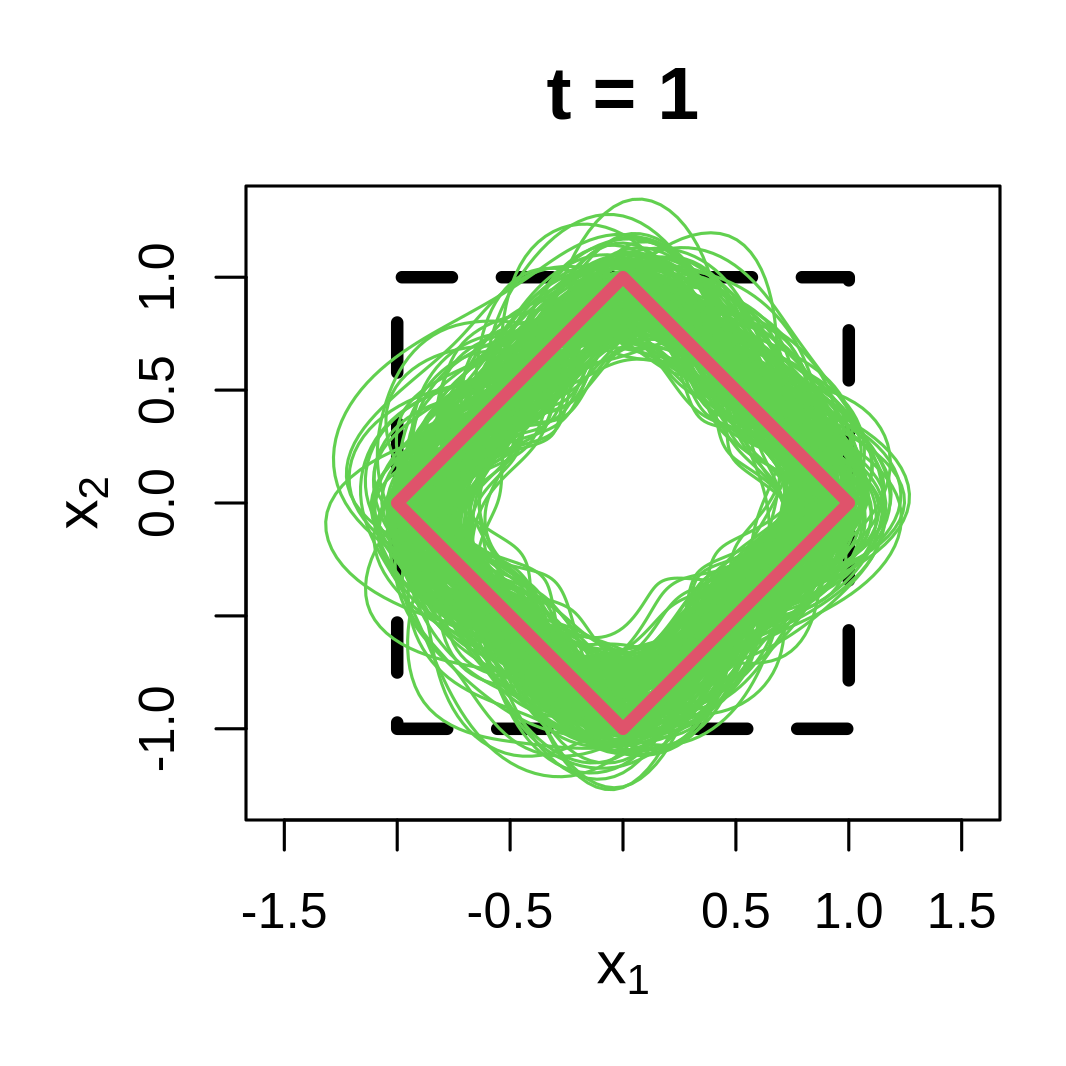}
    \end{subfigure}
    \begin{subfigure}{0.32\textwidth}
        \includegraphics[width=\linewidth]{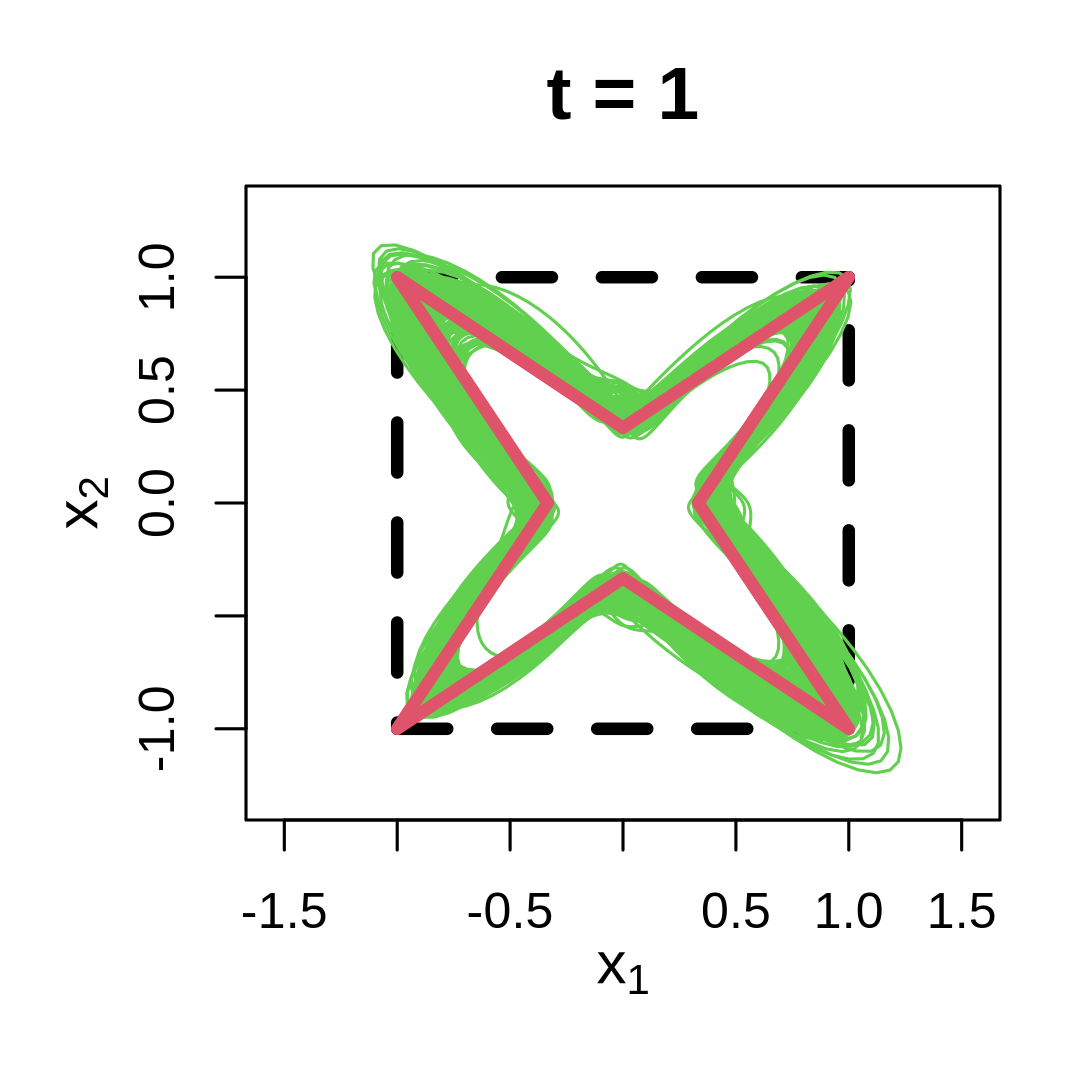}
    \end{subfigure}
    \begin{subfigure}{0.32\textwidth}
        \includegraphics[width=\linewidth]{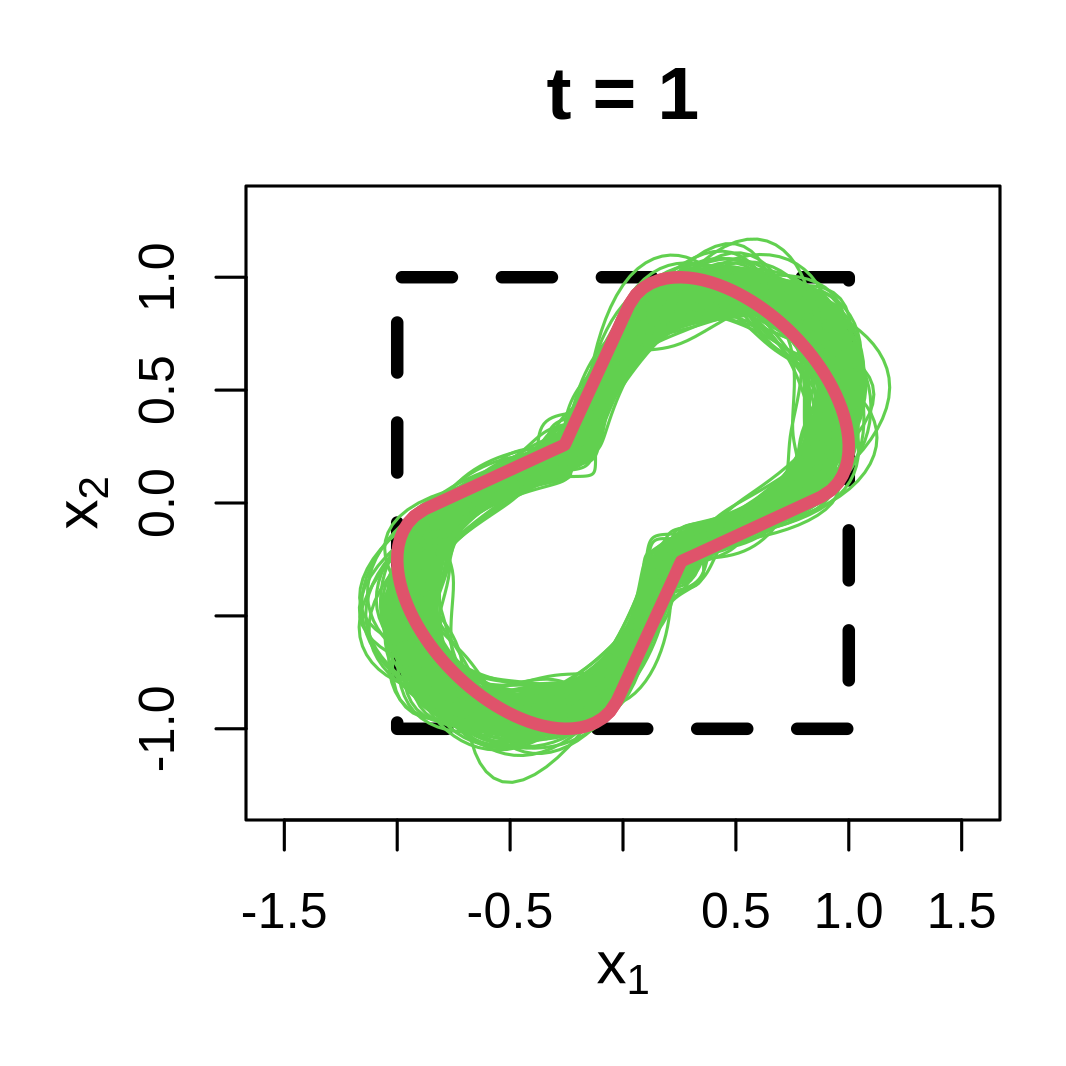}
    \end{subfigure}

    \begin{subfigure}{0.32\textwidth}
        \includegraphics[width=\linewidth]{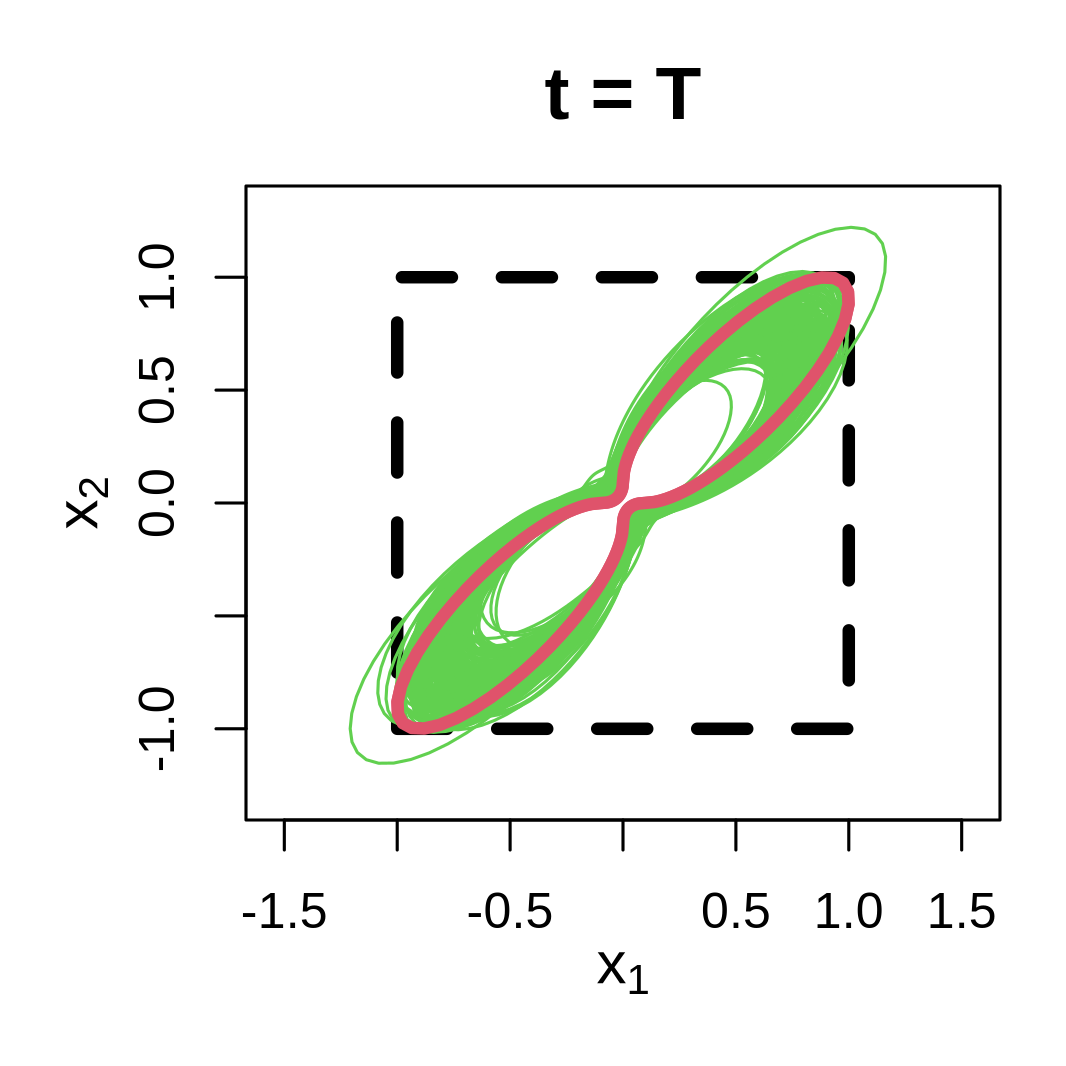}
    \end{subfigure}
    \begin{subfigure}{0.32\textwidth}
        \includegraphics[width=\linewidth]{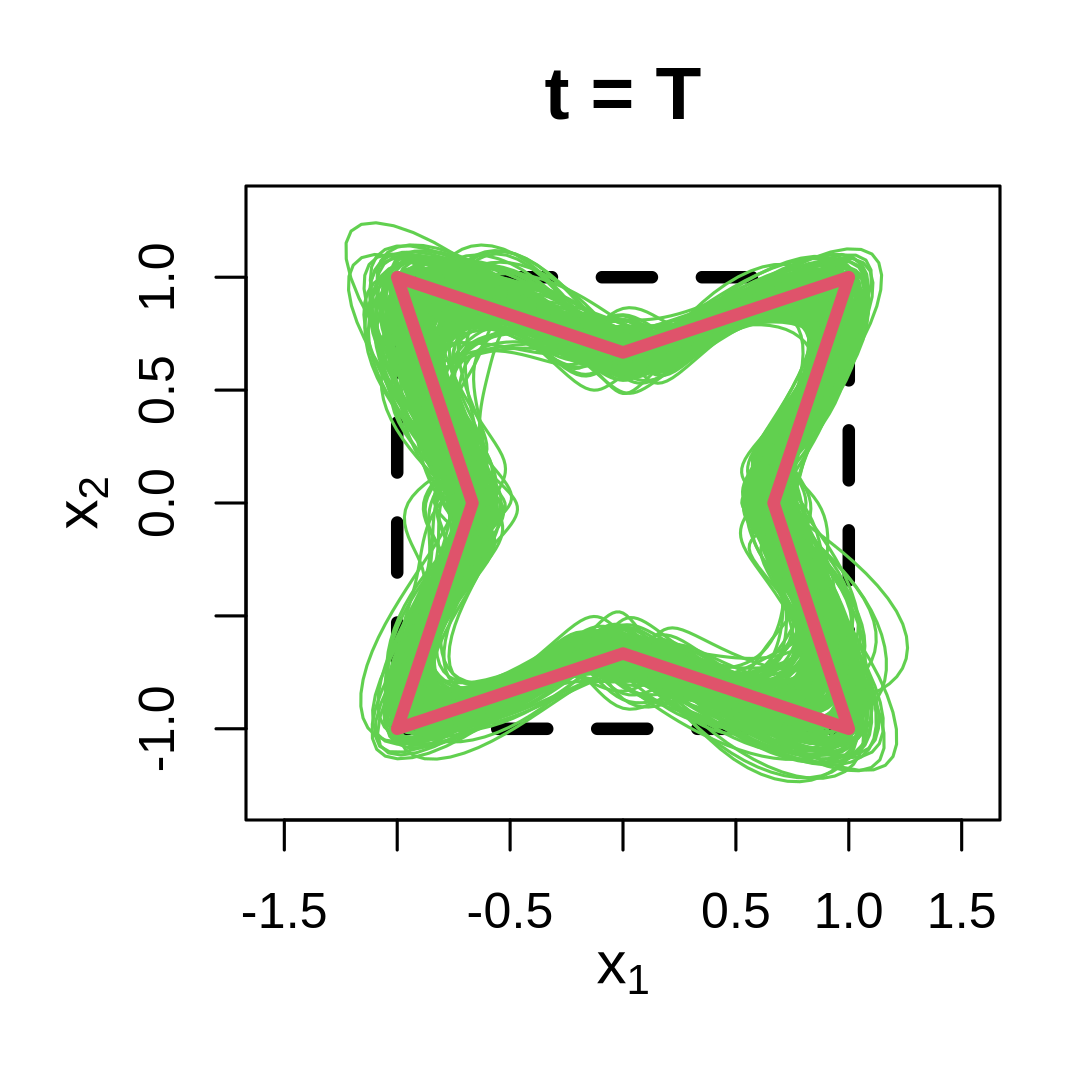}
    \end{subfigure}
    \begin{subfigure}{0.32\textwidth}
        \includegraphics[width=\linewidth]{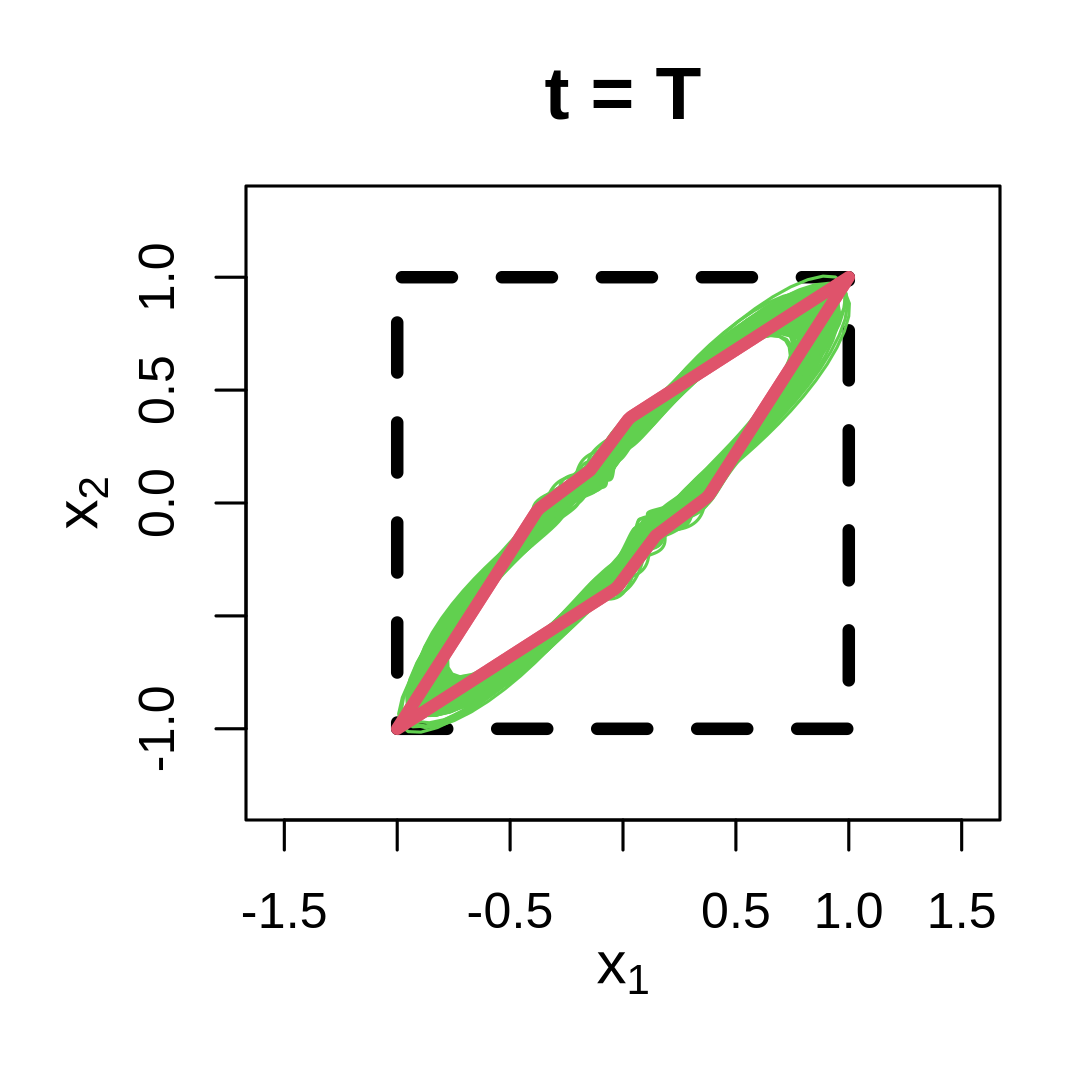}
    \end{subfigure}
    
    \caption{Boundary set estimates for $t=1$ (top row) and $t=T$ (bottom row). The left, centre and right columns correspond to the second, fourth and fifth copula examples, respectively. The green lines correspond to the resampled set estimates, while the red solid and black dotted lines denote the true boundary set and the region $[-1,1]^2$, respectively.}
    \label{fig:final_res_bs}
\end{figure}

Using the selected tuning parameters and model features, we present our simulation results. Fixing $T=25,000$, Figure~\ref{fig:final_res_bs} illustrates the boundary set estimates, along with the theoretical boundary sets, for $t=1$ and $t = T$ with the second, fourth and fifth copula examples; the results for the remaining copulas and time points are given in Appendix~\ref{subsec:appen_final_results}. As can be observed, the estimated boundary sets capture the structures of the theoretical sets at the start and end of the time frame.   

Setting $\phi=\pi/4$ and $\phi = 3\pi/4$ and again considering the second, fourth and fifth copula examples, the estimated radii of the corresponding boundary set points over time are illustrated against the truth in Figure~\ref{fig:final_res_radii}, with the figures for the remaining angles and copulas given in Appendix~\ref{subsec:appen_final_results}. The framework appears to capture the dependence trends in most cases, especially for the Gaussian copulas. However, the model can fail to capture specific dependence features: for example, the abrupt transition between AI and AD for the copula described by equation~\eqref{eqn:hw_model}, and the `pointiness' of the student-$t$ copula in the corners. Given the limited amount of knowledge available generally within the non-stationary setting, estimating dependence features exactly is seldom possible, and we already know that the $L^2$ norm will not be able to represent the pointedness that arises in certain quadrants. On the whole, however, the slight bias in radii observed at certain angles is secondary to the fact that the estimated Cartesian sets provide reasonable approximations of the true limit sets for each of the copula examples.

\begin{figure}[h]
    \centering
    \begin{subfigure}{0.32\textwidth}
        \includegraphics[width=\linewidth]{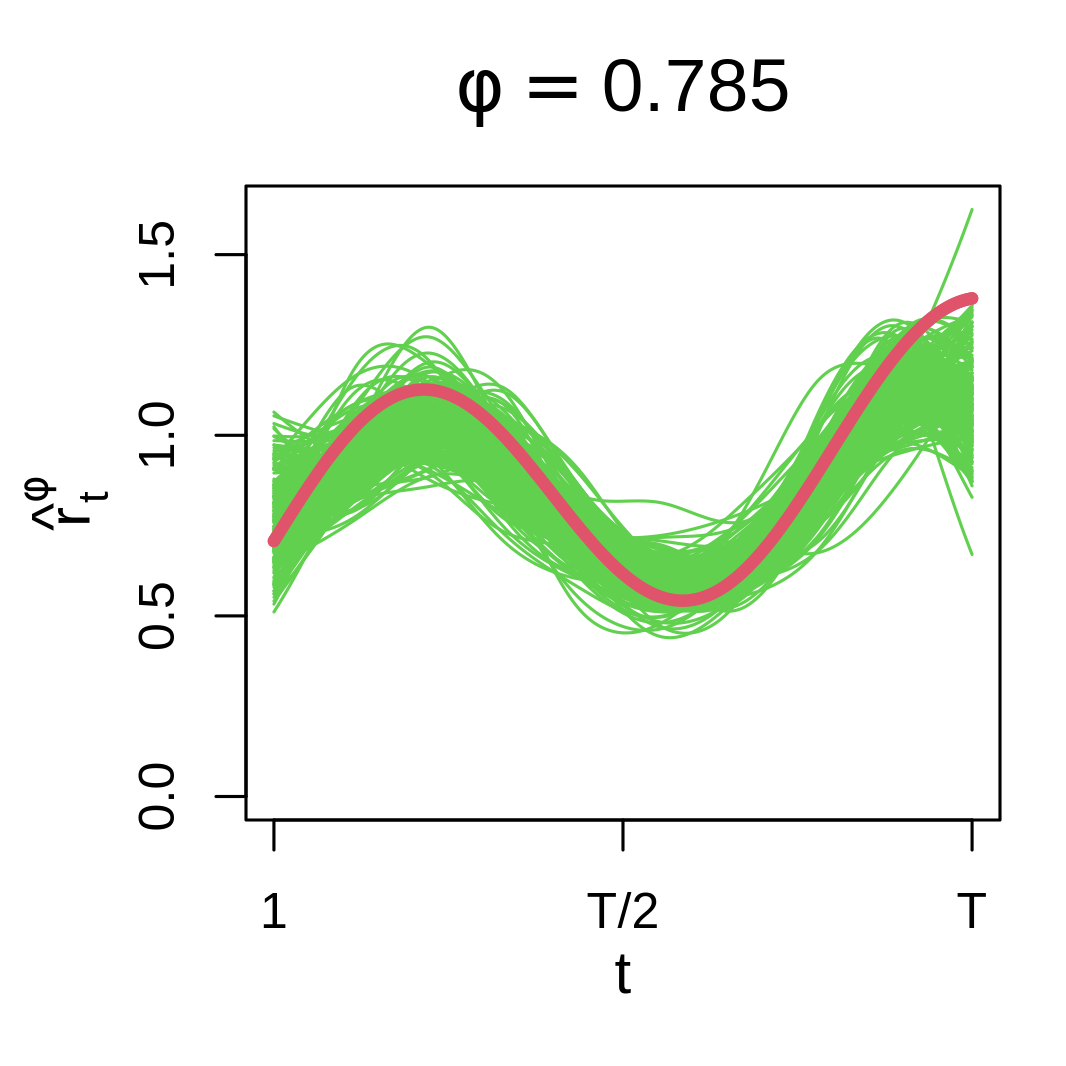}
    \end{subfigure}
    \begin{subfigure}{0.32\textwidth}
        \includegraphics[width=\linewidth]{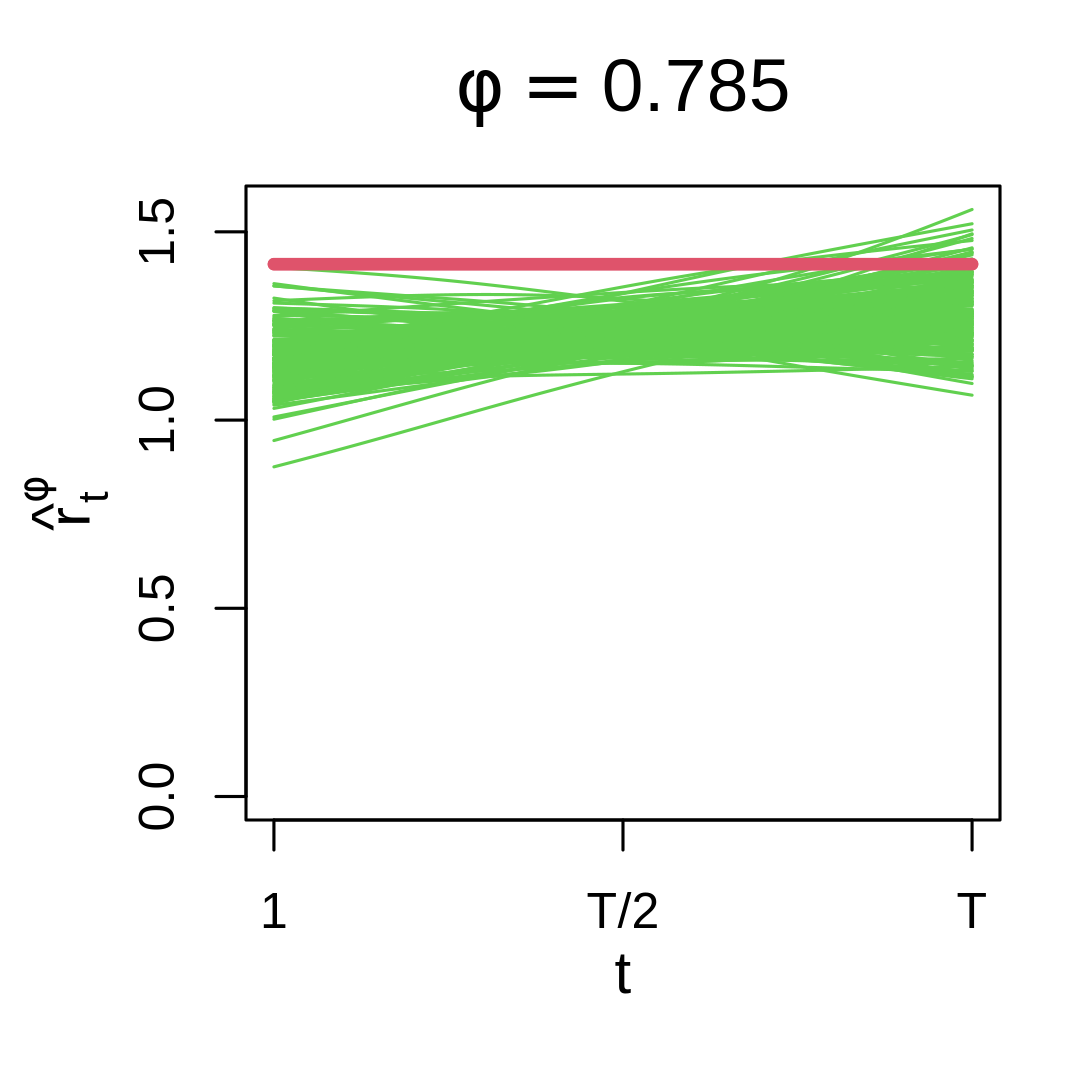}
    \end{subfigure}
    \begin{subfigure}{0.32\textwidth}
        \includegraphics[width=\linewidth]{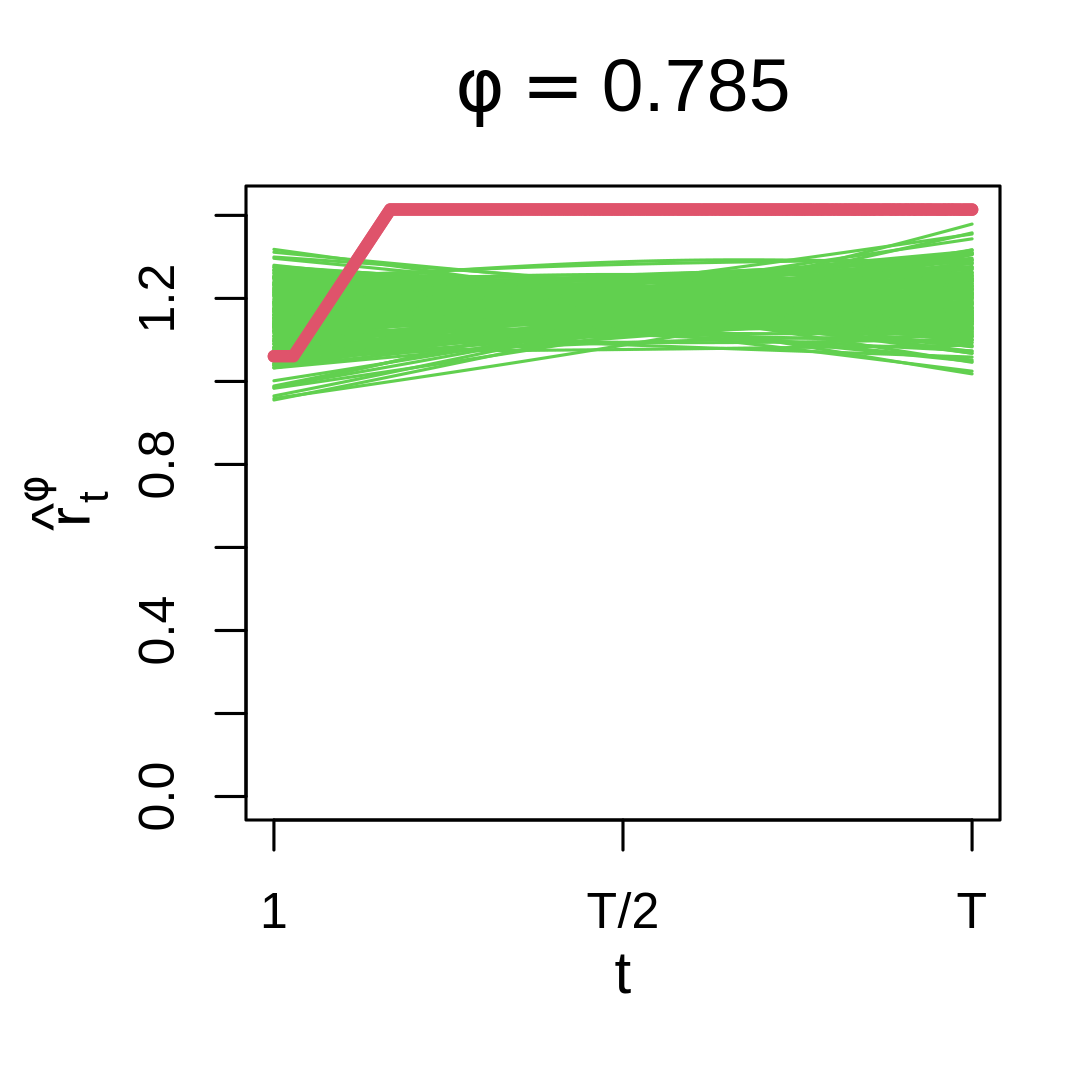}
    \end{subfigure}

    \begin{subfigure}{0.32\textwidth}
        \includegraphics[width=\linewidth]{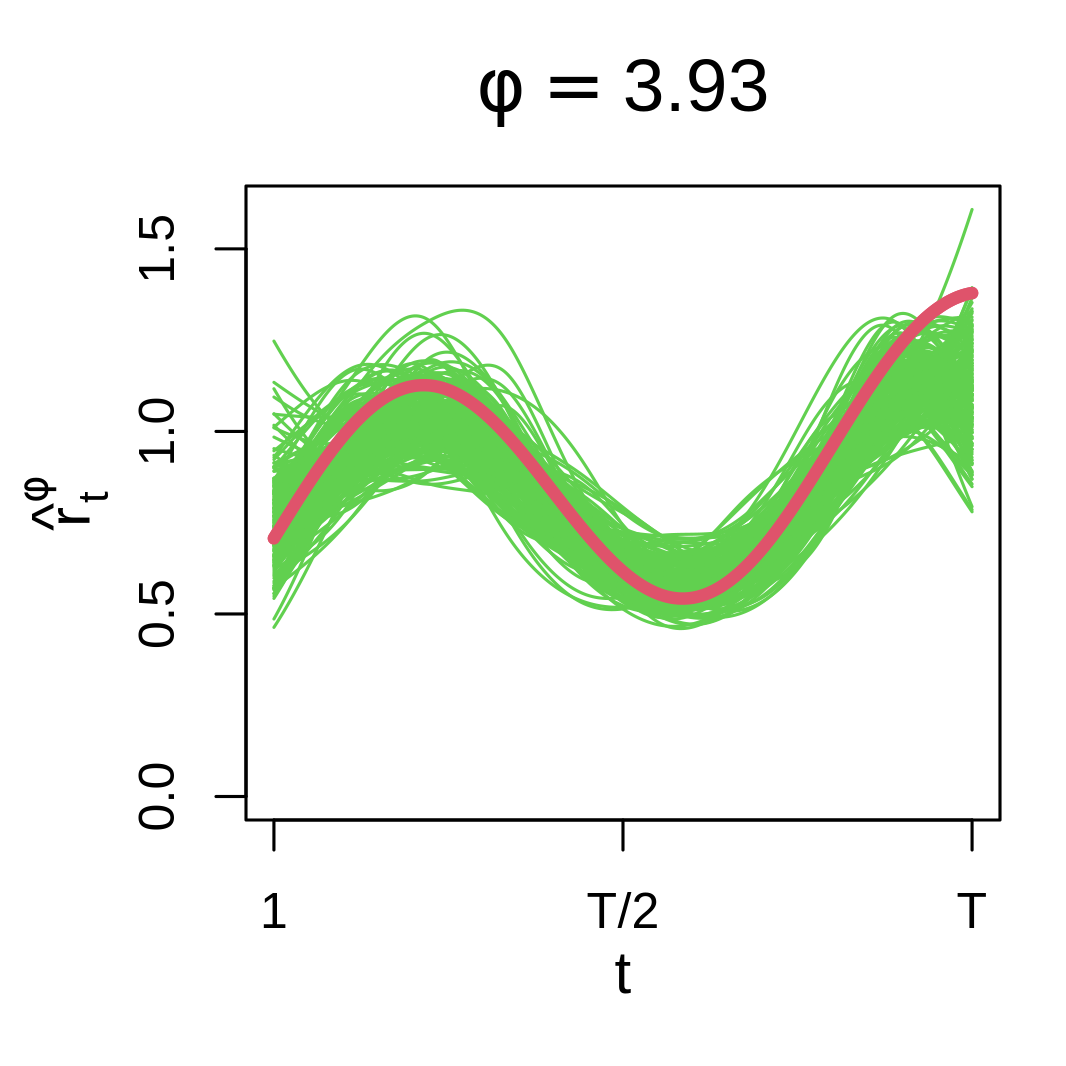}
    \end{subfigure}
    \begin{subfigure}{0.32\textwidth}
        \includegraphics[width=\linewidth]{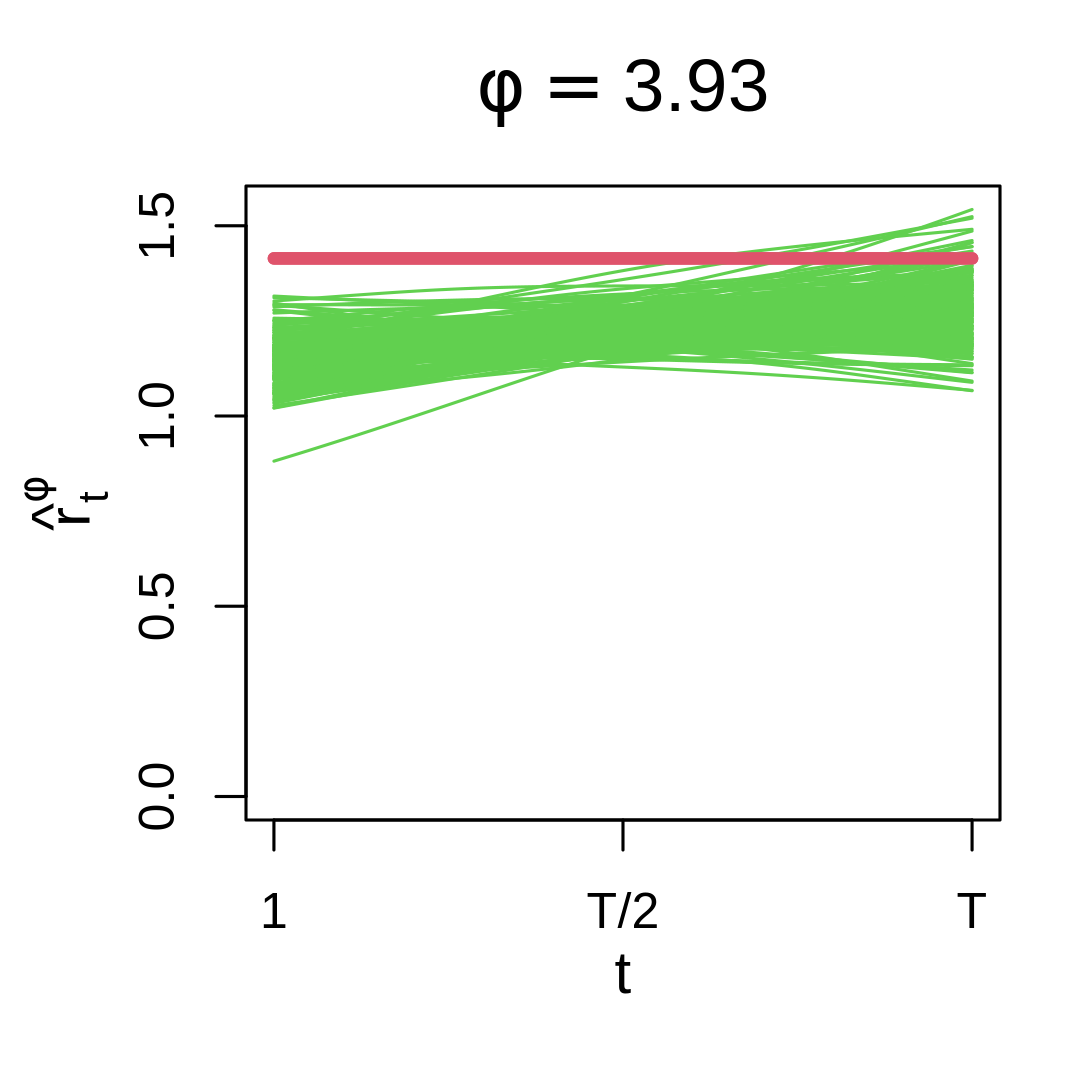}
    \end{subfigure}
    \begin{subfigure}{0.32\textwidth}
        \includegraphics[width=\linewidth]{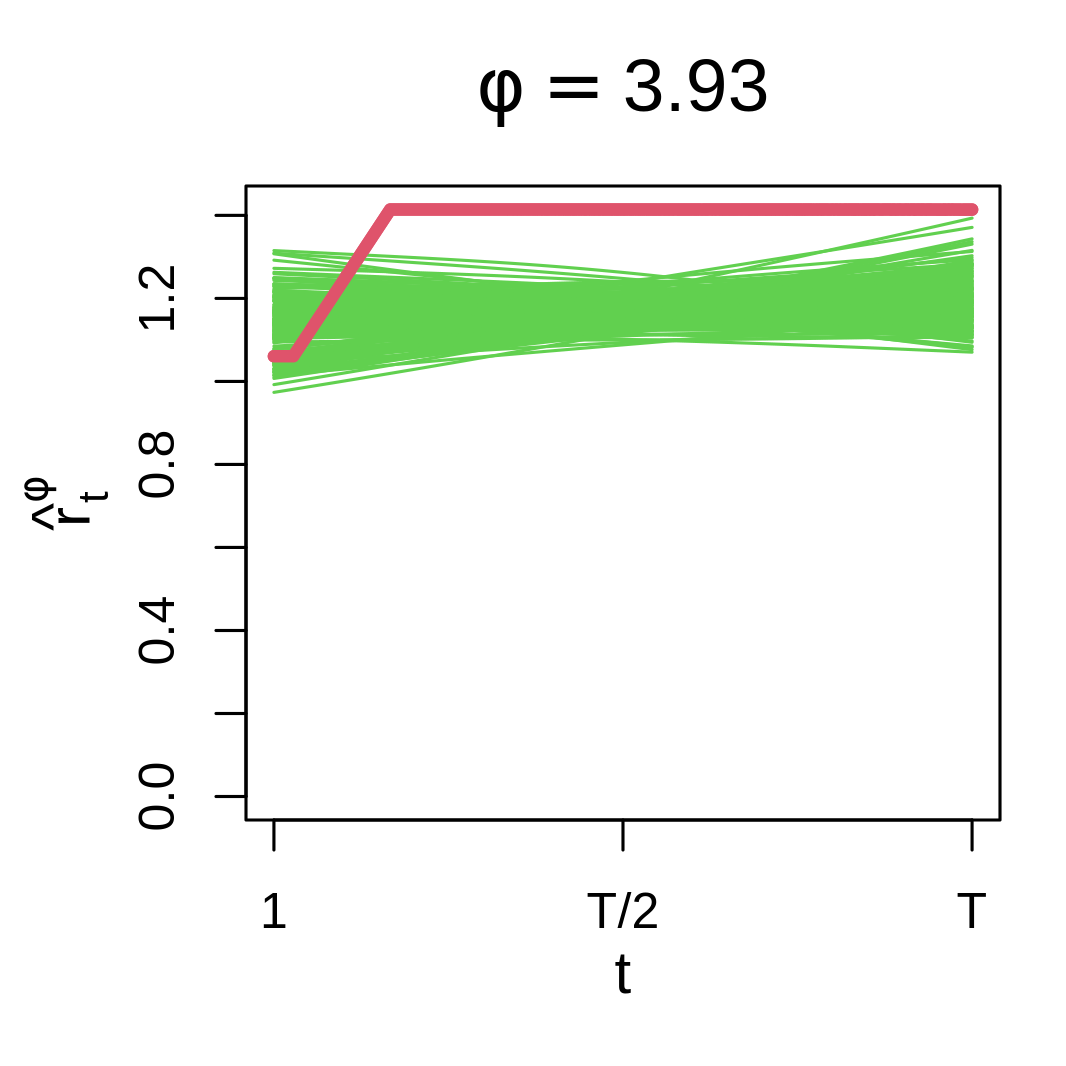}
    \end{subfigure}
    
    \caption{Boundary set radii estimates over time for $\phi=\pi/4$ (top row) and $\phi = 3\pi/4$ (bottom row). The left, centre and right columns correspond to the second, fourth and fifth copula examples, respectively. The legend is identical to that of Figure~\ref{fig:final_res_bs}.}
    \label{fig:final_res_radii}
\end{figure}

Finally, we vary the observation period $T \in \{5,000, \; 10,000, \; 25,000 \}$ and assess the corresponding boundary set estimates. These results are given in Appendix~\ref{subsec:appen_ss}. Unsurprisingly, increasing the sample size improves bias and reduces variability. However, even for $T=5,000$, the model still appears able to approximate the shapes of the true boundary sets, and thus is somewhat robust to lower sample sizes.   

Overall, our proposed framework is able to accurately capture a wide range of extremal dependence trends, demonstrating the robustness and versatility of the model. Furthermore, the choices of tuning parameters and features generally appear to have little overall influence on the model fits, provided an adequate amount of flexibility is permitted. 

\section{Case study} \label{sec:case_study}
\subsection{Financial context and data description} \label{subsec:case_overview}

We now apply our modelling framework to the NASDAQ stock market. Specifically, we consider data from Apple, Microsoft, Amazon, Google, and Nvidia; these firms are all part of the `Magnificent Seven' (M7), a term which refers to a group of seven highly influential big tech companies. As of September 9th, 2025, these firms constitute the five largest companies (in terms of market capitalisation) in the world. We do not consider the remaining M7 companies (Meta and Tesla) due to limited observational periods of the corresponding time series. In recent years, M7 companies have significantly impacted the U.S. stock market, driving a substantial portion of its growth. Their dominance stems from their leadership in key sectors like consumer electronics, cloud computing, social media, artificial intelligence, and electric vehicles. See \citet{Peprah2024} and \citet{de2025} for further discussion on M7 and on extreme losses of `big tech' stocks, respectively. Owing to the fact these companies began publicly training on different days, the observational dates vary depending on which stocks are being considered. The listing days range from December 12th, 1980 (Apple) to January 22nd, 1999 (Nvidia). Daily prices data was accessed from NASDAQ on August 25th, 2025, resulting in time series ending on August 22nd, 2025. 

For each company, we take the daily opening price $P_t$ and compute \textit{log-returns} $Q_t := \log( P_{t}/P_{t-1} )$, for all $t > 1$. This quantity represents a mathematically convenient way of representing price changes in financial markets \citep{Tsay2010}. To avoid computational issues, we remove any time indices $t$ for which $Q_t = 0$ for any stock; this corresponds to two-day periods where there was no change in the opening price of at least one stock. Including such values would result in mixed discrete-continuous time series, which would violate the assumptions of our modelling framework.

In the context of financial risk management, many stakeholders are interested in understanding the joint tails of log-returns across different assets. For example, one may wish to analyse whether an extreme increase or decrease in one stock price will also result in similar changes across other stocks. This motivates the use of multivariate extreme value theory to evaluate joint tail behaviour in regions where variables are simultaneously extreme---either in the lower or upper tails. Considering the bivariate setting, this context represents an application for which tail behaviour across all quadrants is of interest; hence, statistical techniques restricted to a single quadrant (i.e., where both variables are simultaneously large) offer less utility in this setting.  

In Section~\ref{subsec:case_marg_models}, we describe our procedure for modelling the marginal distributions of each time series in order to transform to standard Laplace margins. We apply our modelling framework in Section~\ref{subsec:case_fitting}; the diagnostics subsequently presented in Section~\ref{subsec:case_diag} indicate reasonable model fits across all pairs. Finally, in Section~\ref{subsec:case_sim_ret_level}, we summarise dependence trends over the observation period and introduce further practical use cases for our proposed framework. 

\subsection{Marginal modelling of returns} \label{subsec:case_marg_models}

To apply the framework introduced in Section~\ref{subsec:ns_limit_sets}, we first require marginally IID observations on standard Laplace margins. This requires careful consideration for the log-returns data in question. In particular, log-returns series are unlikely to be independent; see \citet{McNeil2015} for detailed discussion. By computing the autocorrelation function (ACF) of $Q_t$ and $Q^2_t$ at different lags for each stock, it is clear that our data exhibit non-negligible temporal dependence; see Appendix~\ref{sec:appen_add_case_figs} for the corresponding plots.

To account for the observed temporal dependence, we model the log-returns using a GARCH(1,1) process and compute the standardised residuals, which we then take to be independent. For this, we assume $Q_t = \mu +  \sigma_t \varepsilon_t$, where $\mu \in \RR$ denotes a mean term, $\varepsilon_t$ denotes the residual process (with zero mean and unit variance), and the conditional volatility $\sigma_t^2$ satisfies
\[
\sigma_t^2 = c + \alpha Q_{t-1}^2 + \beta \sigma_{t-1}^2,
\]
with parameters $c > 0$, $\alpha, \beta \geq 0$ and $\alpha + \beta < 1$. After fitting the model, the standardised residuals $\hat{\varepsilon}_t = (Q_t - \hat{\mu}) / \hat{\sigma}_t$, which we term the \textit{filtered log-returns}, should be approximately independent and identically distributed if the GARCH model is well specified. This works because the GARCH filter removes serial dependence in the conditional variance, capturing volatility clustering and leaving residuals that behave more like white noise \citep{Engle1982,Bollerslev1986}. 

The plotted time series of the filtered log-returns are given in Appendix~\ref{sec:appen_add_case_figs}, alongside ACF plots for $\hat{\varepsilon}_t$ and $\hat{\varepsilon}^2_t$. The latter plots indicate the the standardised residuals are independent in time. Furthermore, the plotted series do not seem to suggest any obvious periodicity or long term trends are present, suggesting the data may exhibit marginal stationarity. We additionally tested this property by applying Augmented Dickey--Fuller \citep{Dickey1979} and Kwiatkowski--Phillips--Schmidt--Shin \citep[KPSS,][]{Kwiatkowski1992} tests to each filtered time series. In almost all cases, both tests at a $5\%$ significance level indicated a stationarity assumption was reasonable for the data; the only exception was the KPSS test on the Apple time series for the Apple--Google pairing. Given the same test indicated stationarity for the remaining Apple series, and the computed $p$-value was just below the $\alpha=0.05$ significance level, it is therefore reasonable to assume that this is a type 1 error that can be ignored. We henceforth assume that the filtered log-returns are marginally IID.

We now employ standard techniques to model the individual marginal distributions. In particular, we employ the semi-parametric approach of \citet{Coles1991}, where a generalised Pareto distribution (GPD) is fitted to the tails while the body is modelled empirically. In the case of returns data, we are interested in understanding both the extreme high and low values. Letting $\mathcal{E}$ denote a filtered log-returns series with observations $\{\varepsilon_1,\hdots,\varepsilon_T \}$ and given some $\alpha$ close to $0$, define $l$ and $h$ to be the empirical $\alpha$ and $(1-\alpha)$ quantiles, respectively, of $\mathcal{E}$. We use the following marginal model
\begin{equation*} 
    \hat{F}_\mathcal{E}(\varepsilon) = \begin{cases} 1 - \alpha\{1 + \xi_1(\varepsilon-h)/\sigma_1 \}_+^{-1/\xi_1}, \hspace{1em} &\text{for} \; \varepsilon > h, \\  \Tilde{F}_\mathcal{E}(\varepsilon), \hspace{1em}  &\text{for} \; l \leq \varepsilon \leq h, \\
    \alpha\{1 + \xi_2(l-\varepsilon)/\sigma_2 \}_+^{-1/\xi_2}, \hspace{1em} &\text{for} \; \varepsilon < l,
    \end{cases}
\end{equation*}
where $(\sigma_1,\xi_1)$ and $(\sigma_2,\xi_2)$ are the estimated GPD parameters from the conditional tail variables $(\mathcal{E} - h) \mid \mathcal{E} > h$ and $(l - \mathcal{E}) \mid l > \mathcal{E}$, respectively, $\{x \}_+ := \max\{0,x\}$, and $\Tilde{F}_\mathcal{E}$ is an empirical rank transform given by 
$\Tilde{F}_\mathcal{E}(\varepsilon) = \sum_{t=1}^T \mathbbm{1}(\varepsilon_{t} \leq \varepsilon)/(T+1)$.

The selection of $\alpha$ is equivalent to selecting a threshold for the GPD, which represents a bias-variance trade-off in practice. For this, we applied a leading automated threshold selection technique \citep{Murphy2024} which outperforms many existing approaches. However, we found the model fits from the \citet{Murphy2024} approach to be inadequate for several time series. Therefore, we instead manually tested a range of $\alpha$ values and eventually fixed $\alpha = 0.03$; this appeared to give reasonable model fits for all of the tails. The resulting GPD QQ plots are given in Appendix~\ref{sec:appen_add_case_figs}. 

With marginal models specified, the probability integral transform is applied to obtain each series on standard Laplace margins. As an additional check, the series are split into non-overlapping rolling windows, and maximum likelihood techniques are used to estimate the location and scale parameter of the Laplace distribution for each window. These rolling window estimates remain close to the true Laplace parameters (i.e., zero location, unit scale) in all cases, indicating the margins have been successfully standardised. These plots are given in Appendix~\ref{sec:appen_add_case_figs}. 

\subsection{Modelling time-varying extremal dependence of returns} \label{subsec:case_fitting}

Time-varying joint modelling in stock markets has been recognised as fundamental since at least \cite{patton2006}, and we now use the framework proposed in Section~\ref{sec:methodology} to this end. Since our model is bivariate, there are a total of 10 pairs of variables we could consider. For the sake of brevity, we restrict our analysis to consider the extremal dependence between Apple and all of the remaining stocks. Apple has been listed for longer than any of the remaining companies, thus representing the most `mature' company, and the stock movements of Apple are often seen as an indicator of broader market trends, particularly within the technology sector. The size of observation windows varies from $T = 6,598$ for the Apple--Nvidia pair to $T = 9,432$ for Apple--Microsoft. As demonstrated in Section~\ref{sec:sim_study}, such sample sizes are sufficient for applying the modelling framework proposed in Section~\ref{sec:methodology}.

\begin{figure}[H]
    \centering
     \begin{subfigure}[b]{.25\textwidth}
        
        \includegraphics[width=\textwidth]{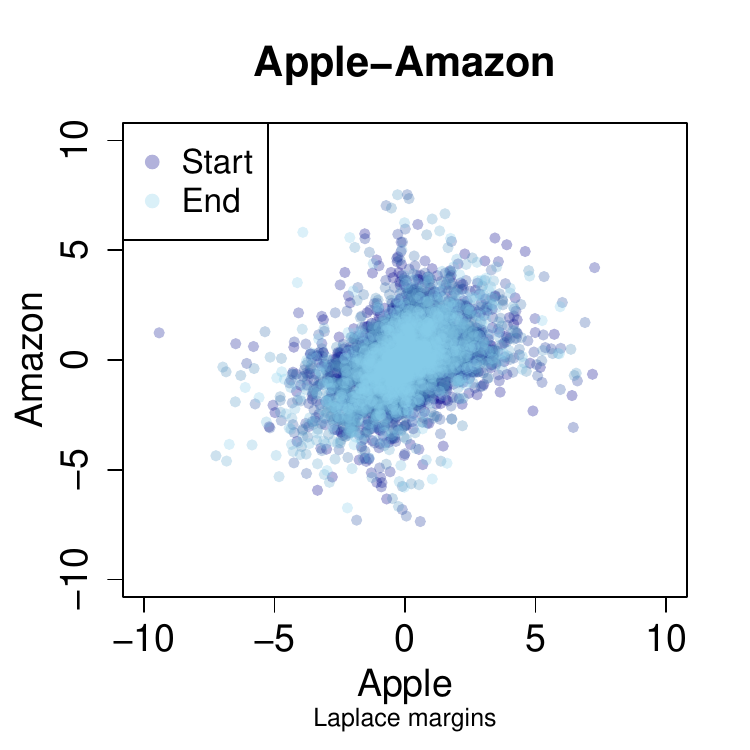}
    \end{subfigure}%
    \begin{subfigure}[b]{.25\textwidth}
        
        \includegraphics[width=\textwidth]{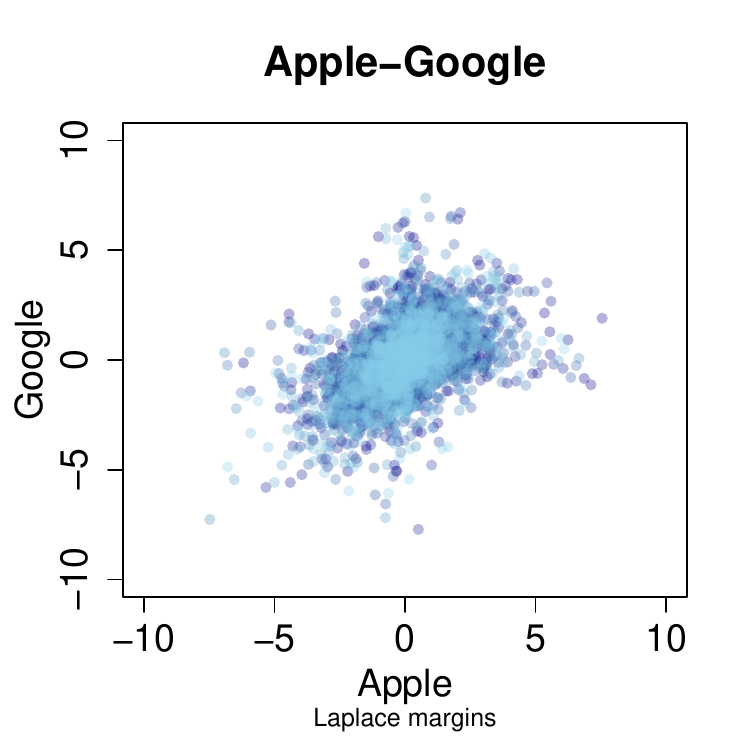}
    \end{subfigure}%
      \begin{subfigure}[b]{.25\textwidth}
        
        \includegraphics[width=\textwidth]{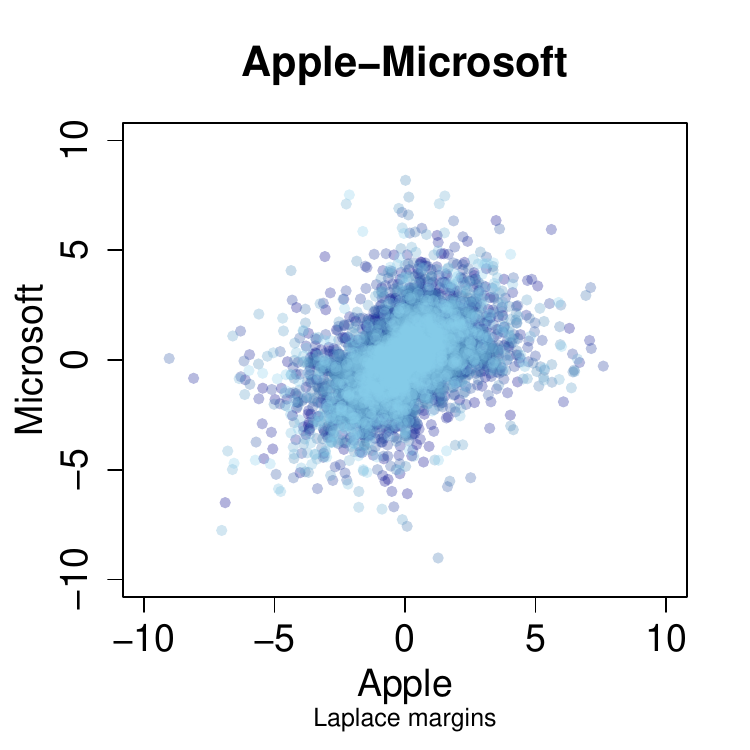}
    \end{subfigure}%
      \begin{subfigure}[b]{.25\textwidth}
        
        \includegraphics[width=\textwidth]{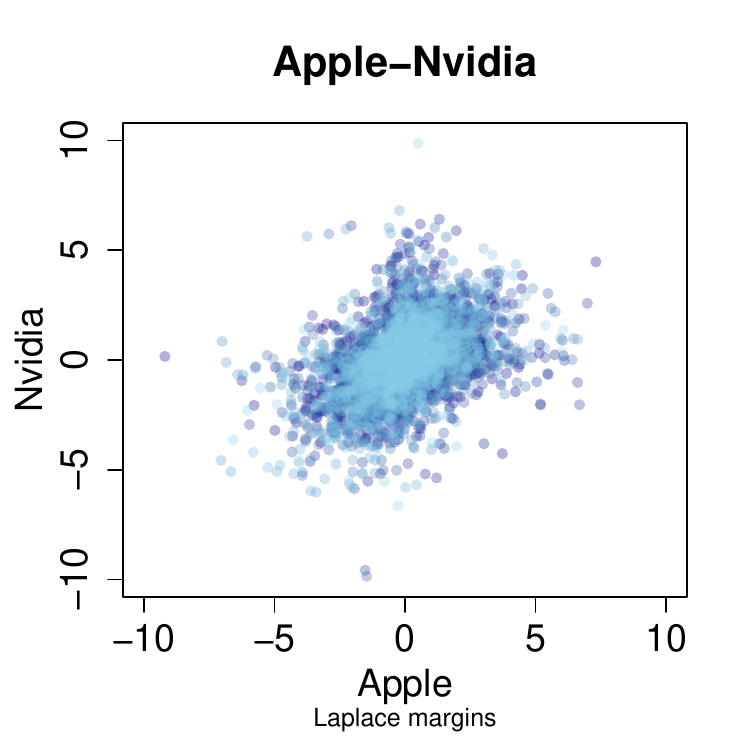}
    \end{subfigure}%
   
    \caption{Bivariate scatterplots of each data set on Laplace margins. A colour gradient is used to illustrate the changing joint behaviour over time, with observations at the start and end of the time periods illustrated in dark and light blue, respectively.}
    \label{fig:scatterplots}
\end{figure}

Figure~\ref{fig:scatterplots} illustrates bivariate scatterplots of each bivariate data set on Laplace margins, processed as described in Section~\ref{subsec:case_marg_models}. A colour gradient is used to illustrate the change in joint behaviour over time, suggesting subtle variations in dependence. Ideally, one would wish to formally test for this feature; however, there are no best practices established for such testing within the literature, and we found a previous technique, which involves computing dependence coefficients across overlapping rolling windows \citep{Castro-Camilo2018,Murphy-Barltrop2024b}, to be unreliable and misleading. In particular, for simulated IID data, the rolling windows approach could suggest spurious trends when using point estimates together with pointwise confidence intervals. We refer to Section~\ref{sec:discussion} for further discussion.

\begin{figure}[H]
    \centering
     \begin{subfigure}[b]{.25\textwidth}
        
        \includegraphics[width=\textwidth]{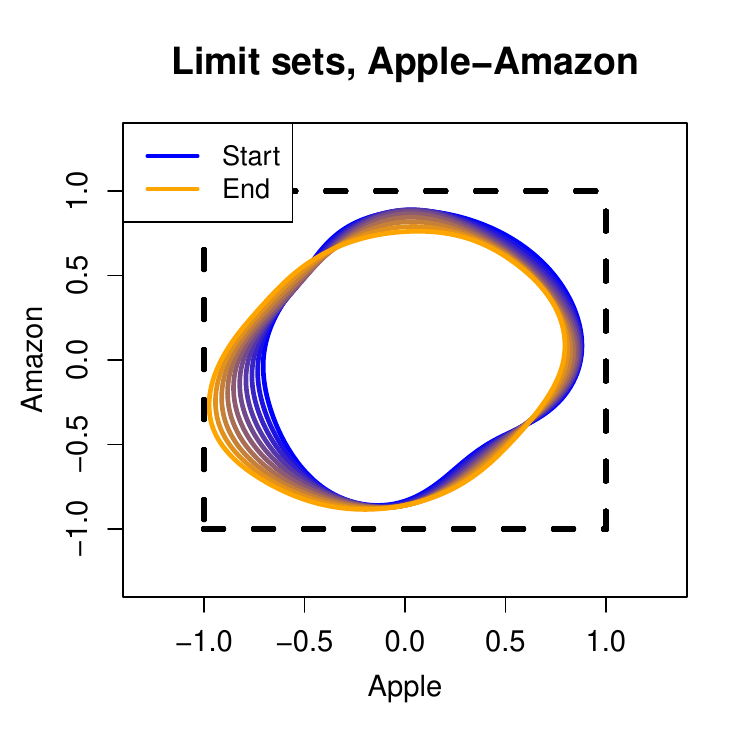}
    \end{subfigure}%
    \begin{subfigure}[b]{.25\textwidth}
        
        \includegraphics[width=\textwidth]{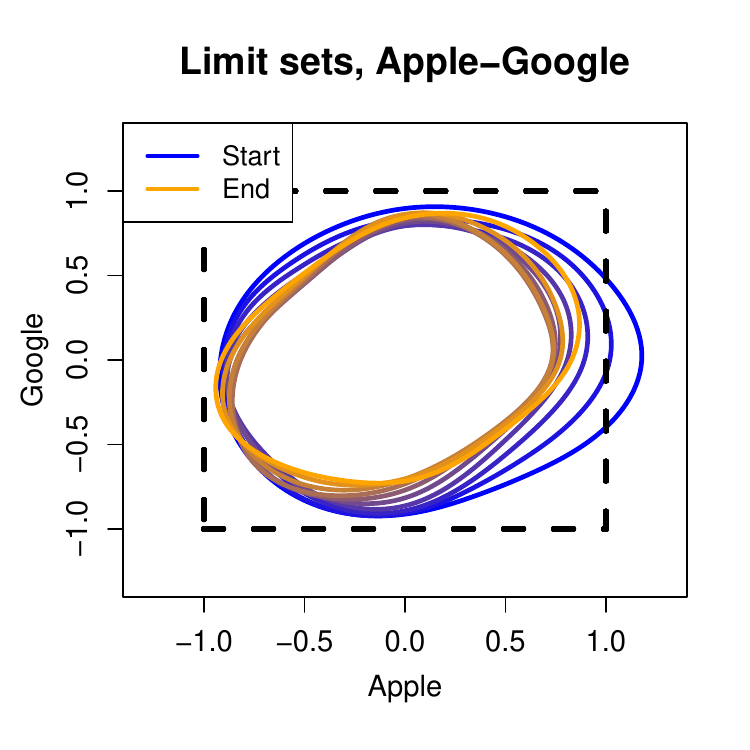}
    \end{subfigure}%
      \begin{subfigure}[b]{.25\textwidth}
        
        \includegraphics[width=\textwidth]{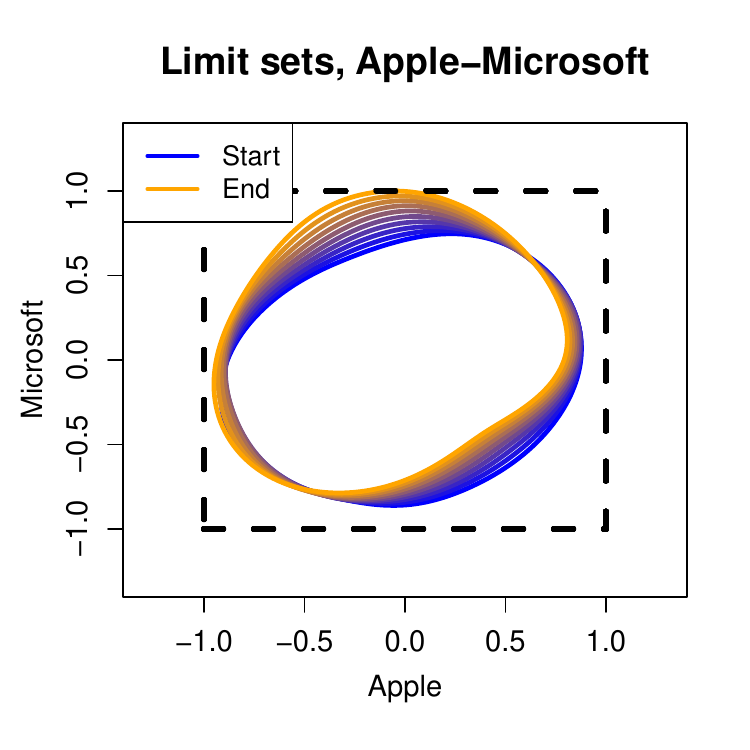}
    \end{subfigure}%
      \begin{subfigure}[b]{.25\textwidth}
        
        \includegraphics[width=\textwidth]{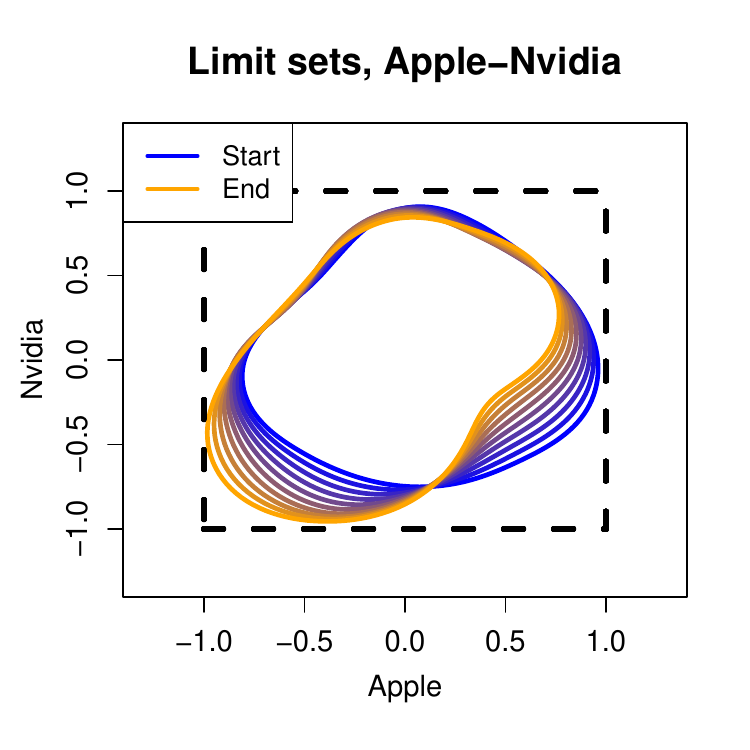}
    \end{subfigure}%
   
    \caption{Non-stationary limit set estimates $\widehat{\partial \mathcal{G}}_t$ over time for each of the filtered log-returns pairs'. The colour scale is used to illustrate the variation in time, with the blue and orange sets corresponding to the start and end of the observation periods, respectively.}
    \label{fig:all_boundary_sets}
\end{figure}

To apply the framework, the same tuning parameters and formulations as suggested in Section~\ref{subsec:model_formulation}, combined with the $L^2$ norm, are used for model fitting. Figure~\ref{fig:all_boundary_sets} illustrates the time-varying limit sets for each pair of stocks; the corresponding threshold quantile sets, i.e., $\widehat{\mathcal{R}}^{\tau}_t = \{  v(\phi)\hat{r^{\tau}}(\phi,t) : \phi \in [0,2\pi) \}$ are given in Appendix~\ref{sec:appen_add_case_figs}. These estimates suggest a range of potentially complex, non-linear extremal dependence trends are present for each pair, with the type of trend varying over quadrants. For example, within the positive quadrant, the extremal dependence varies moderately for the Apple--Google pair but only very slightly for the remaining pairs. Our analysis also reveals the dynamics governing joint extreme losses (i.e., third quadrant). For example, the pairs Apple--Amazon and Apple--Nvidia, shown in Figure~\ref{fig:all_boundary_sets}, exhibit strengthening extremal dependence in the lower-left (third) quadrant, suggesting that joint crashes between these stocks have become more frequent in recent periods. These instances illustrate the benefits of applying the geometric approach on Laplace margins: one can evaluate the dependence structure across \emph{all} quadrants simultaneously, providing a more complete picture of trends in extremal dependence. 

\subsection{Diagnostics} \label{subsec:case_diag}

\begin{figure}[H]
    \centering
     \begin{subfigure}[b]{.25\textwidth}
        
        \includegraphics[width=\textwidth]{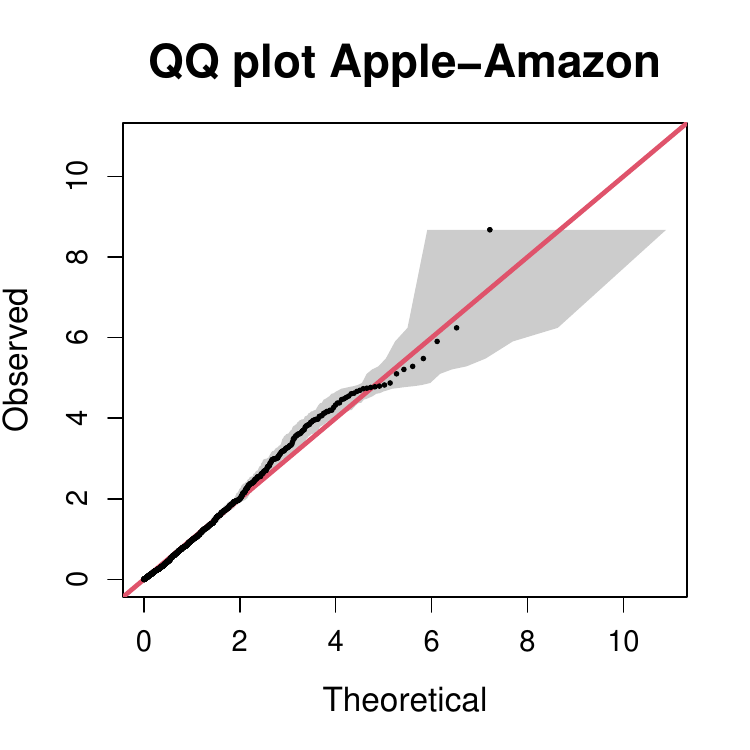}
    \end{subfigure}%
    \begin{subfigure}[b]{.25\textwidth}
        
        \includegraphics[width=\textwidth]{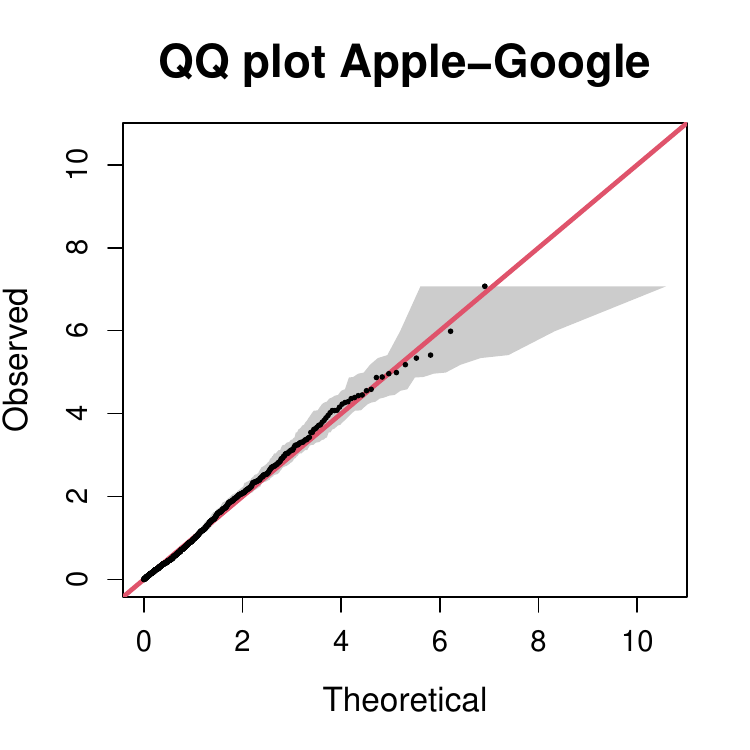}
    \end{subfigure}%
      \begin{subfigure}[b]{.25\textwidth}
        
        \includegraphics[width=\textwidth]{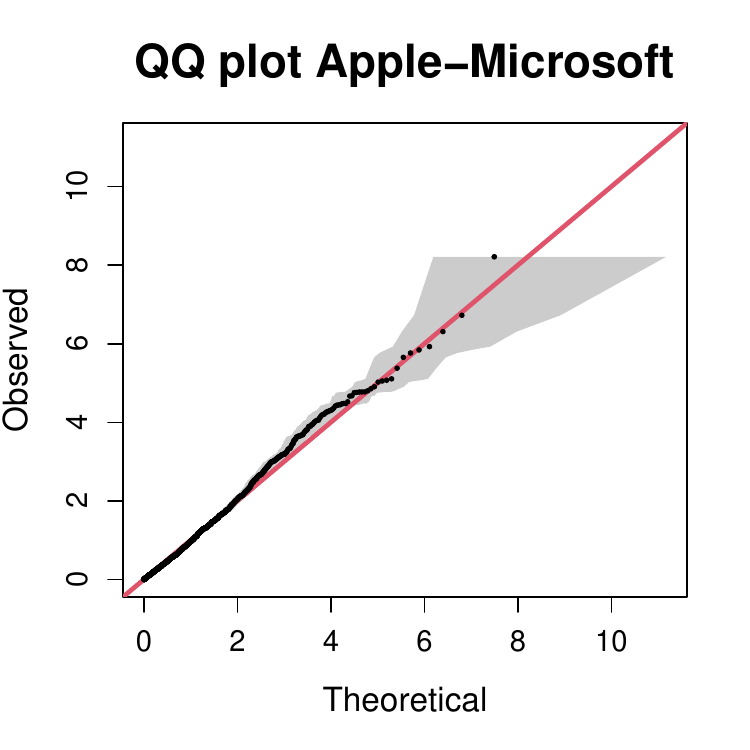}
    \end{subfigure}%
      \begin{subfigure}[b]{.25\textwidth}
        
        \includegraphics[width=\textwidth]{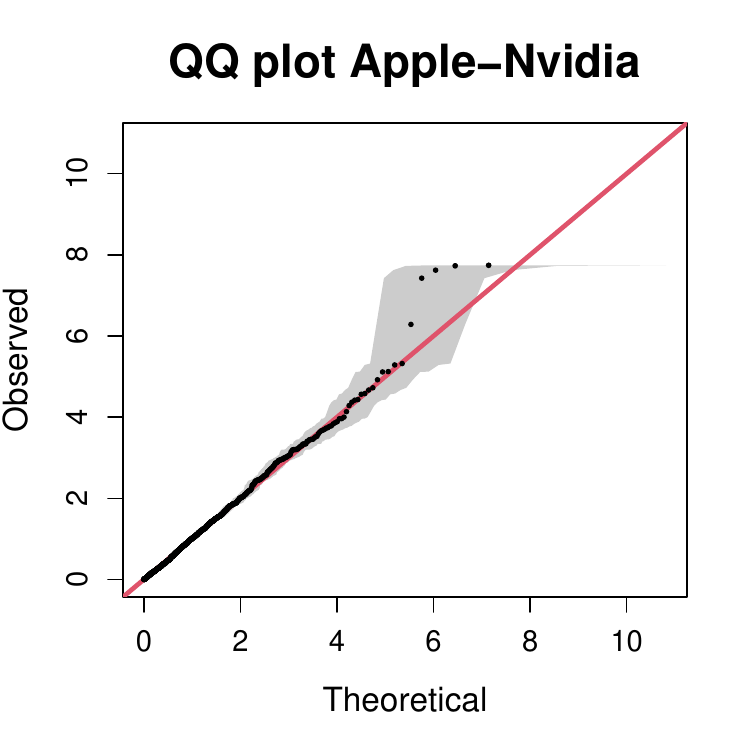}
    \end{subfigure}%
   
    \caption{Truncated gamma QQ plot diagnostic for each pair of stocks. The red lines denote equity, and the grey shaded regions denote 95\% pointwise confidence intervals. }
    \label{fig:qq_plots_all}
\end{figure}

To assess the models fits, we apply the diagnostics introduced in Section~\ref{subsec:model_checking}. Firstly, QQ plots for the truncated gamma model fits were computed with 95\% confidence intervals; Figure~\ref{fig:qq_plots_all} illustrates these plots for all pairs. One can observe good agreement between the observed and theoretical quantiles in all cases, with only a slight deviation in the tail for the Apple--Nvidia pairing. These plots indicate the truncated gamma model assumption of equation~\eqref{eqn:trunc_gamma_assum} is appropriate for capturing the radial tails across all pairs. 

The return level set diagnostic is also computed for each model fit. Specifically, we define a set of 200 equally spaced probabilities from 0.8 to 0.99, and evaluate the average empirical probabilities over time. Letting $p$ and $\hat{p}$ be the true and empirical probabilities, respectively, we compare the pairs $(-\log(1-p),-\log(1-\hat{p}))$; this allows us to evaluate performance for reasonably extreme return level set probabilities. The corresponding probability plots are given in Appendix~\ref{sec:appen_add_case_figs}. In all cases, we observe very good agreement between the pairs, indicating the estimated return level sets (and model fits) correctly represent the structure in the data.  

\subsection{Simulation and evaluating joint risk} \label{subsec:case_sim_ret_level}
\subsubsection*{Extremal dependence across financial episodes}
We now demonstrate further potential use cases for our framework. To begin, we summarise the trends in dependence via estimates of coefficients. As detailed in Appendix~\ref{sec:appen_led_tawn}, coefficients of tail dependence, which summarise the form of joint tail behaviour, can be defined separately for each quadrant. As noted in Sections~\ref{sec:ns_limit_sets}, these coefficients can be approximated directly from the estimated limit sets using the theoretical results introduced in \citet{Nolde2014} and \citet{Nolde2022}. For each pair, the time-varying $\eta_t$ estimates in quadrants 1 and 3 (i.e., the joint upper and lower tails, respectively) are shown in Figure~\ref{fig:eta_ests_quads_1_3}, with the remaining quadrants considered in Appendix~\ref{sec:appen_add_case_figs}. Such coefficients provide interpretable summaries of the extremal dependence behaviour over time for different stock market dynamics. For example, these estimates suggest that the probability of joint high or low days for Apple and Nvidia returns are increasing throughout the time period, whereas the same probabilities for Apple and Google vary non-linearly over the observation period. It is also clear that the estimated trends can be very slight in some cases, such as the Apple and Microsoft pairing. 

\begin{figure}[H]
    \centering
     \begin{subfigure}[b]{.25\textwidth}
        
        \includegraphics[width=\textwidth]{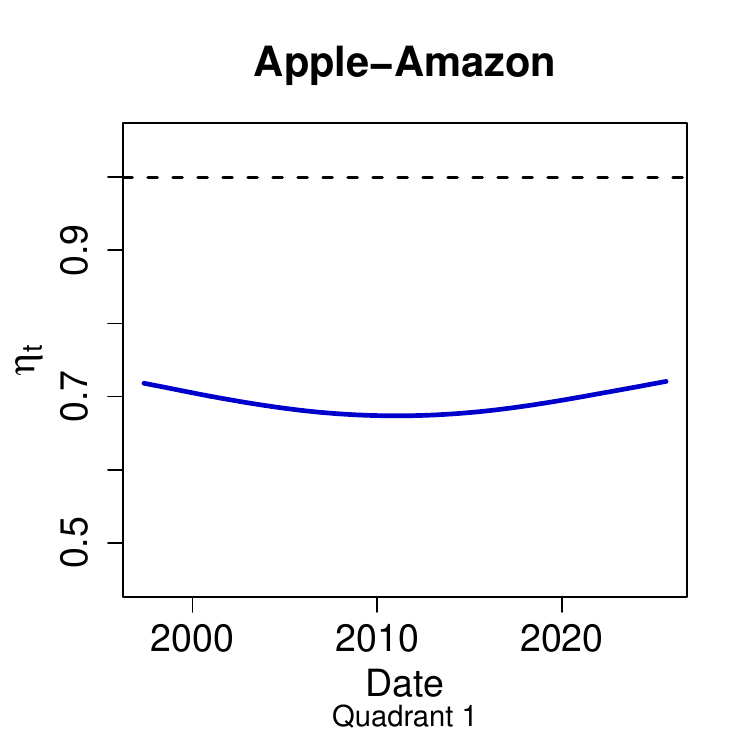}
    \end{subfigure}%
    \begin{subfigure}[b]{.25\textwidth}
        
        \includegraphics[width=\textwidth]{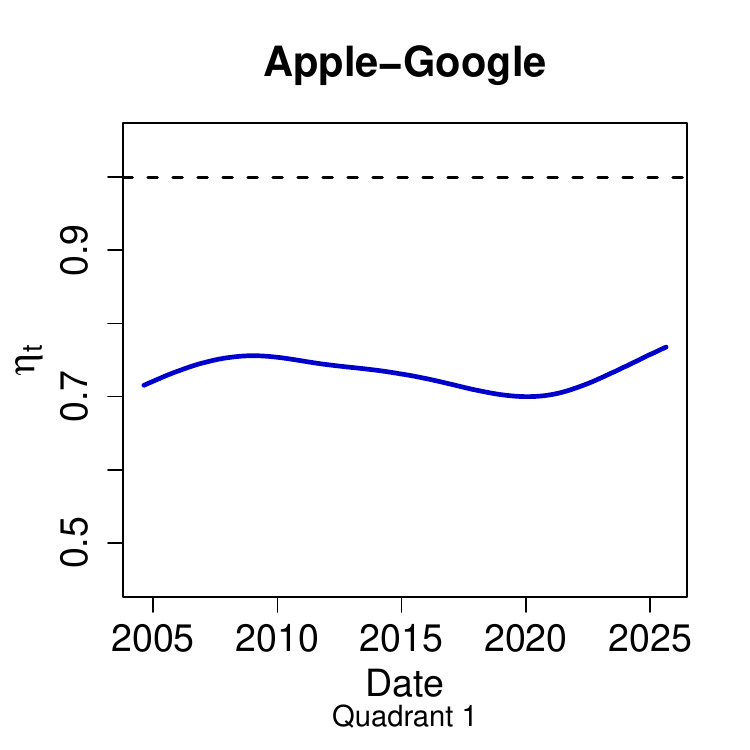}
    \end{subfigure}%
      \begin{subfigure}[b]{.25\textwidth}
        
        \includegraphics[width=\textwidth]{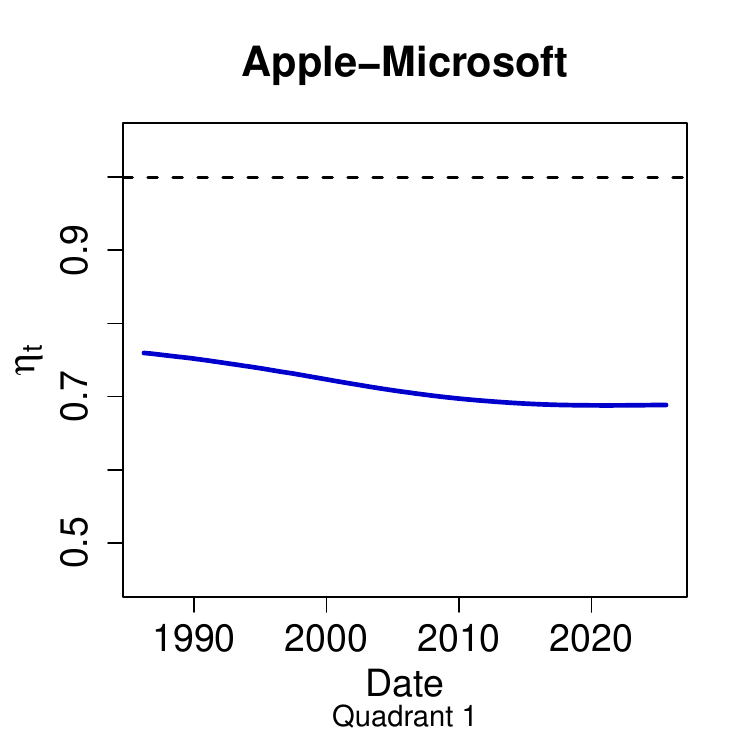}
    \end{subfigure}%
      \begin{subfigure}[b]{.25\textwidth}
        
        \includegraphics[width=\textwidth]{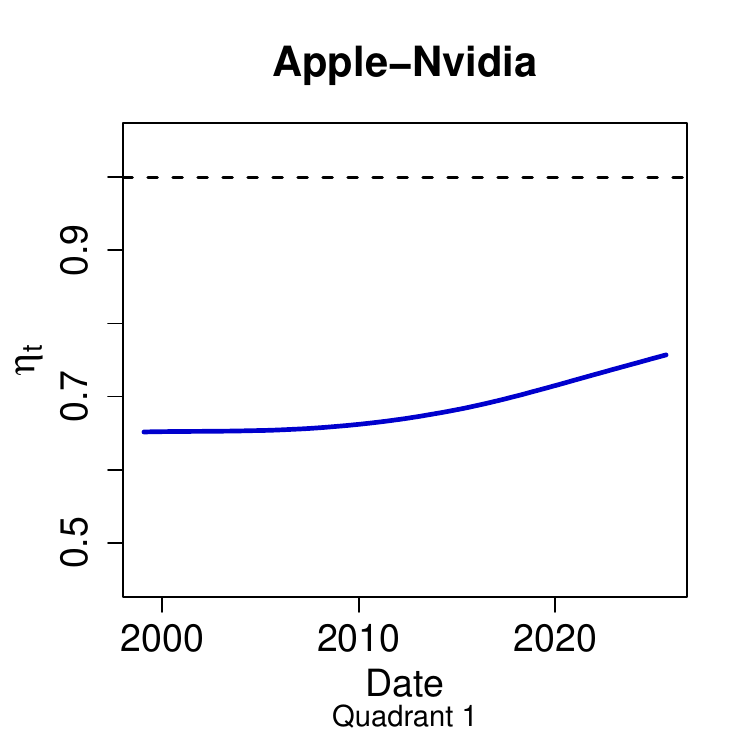}
    \end{subfigure}%

    \begin{subfigure}[b]{.25\textwidth}
        
        \includegraphics[width=\textwidth]{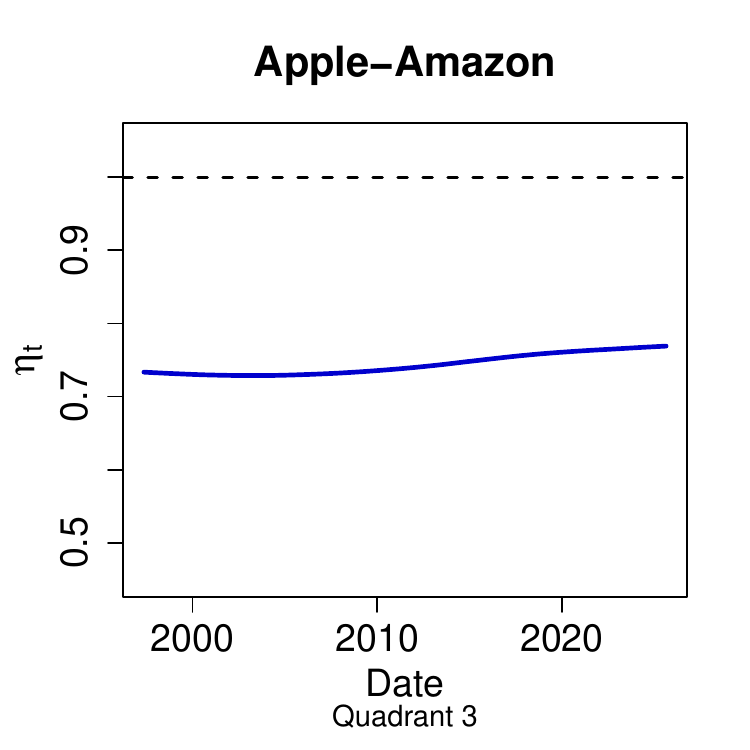}
    \end{subfigure}%
    \begin{subfigure}[b]{.25\textwidth}
        
        \includegraphics[width=\textwidth]{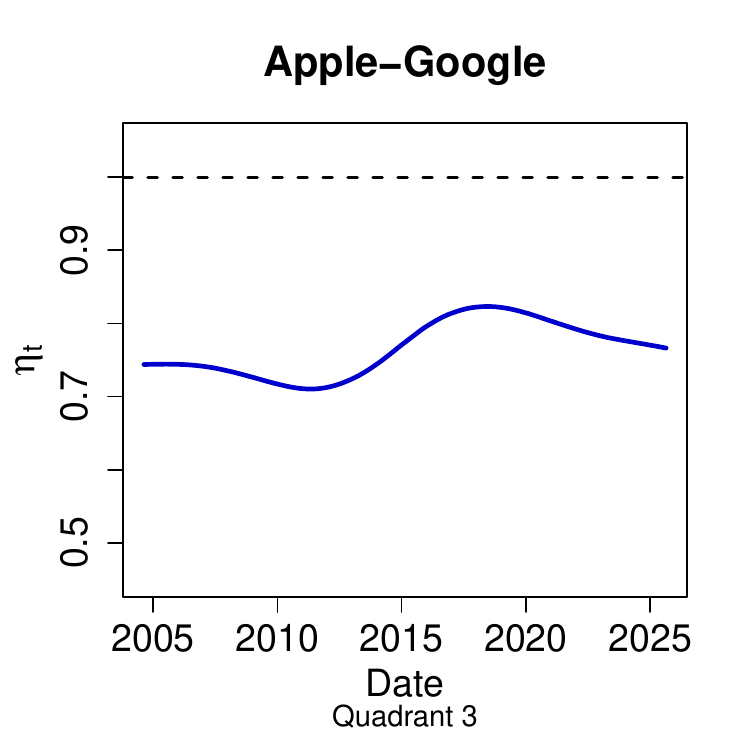}
    \end{subfigure}%
      \begin{subfigure}[b]{.25\textwidth}
        
        \includegraphics[width=\textwidth]{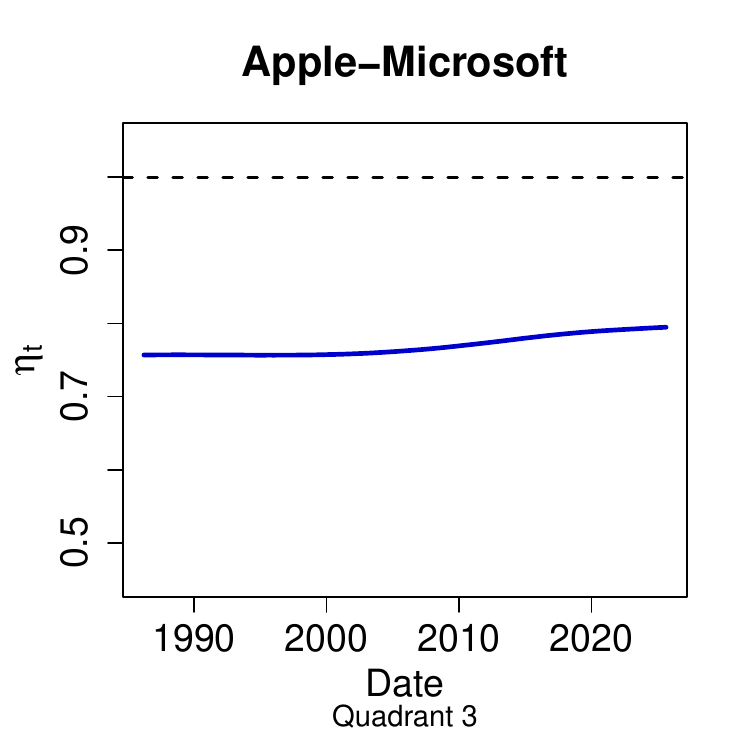}
    \end{subfigure}%
      \begin{subfigure}[b]{.25\textwidth}
        
        \includegraphics[width=\textwidth]{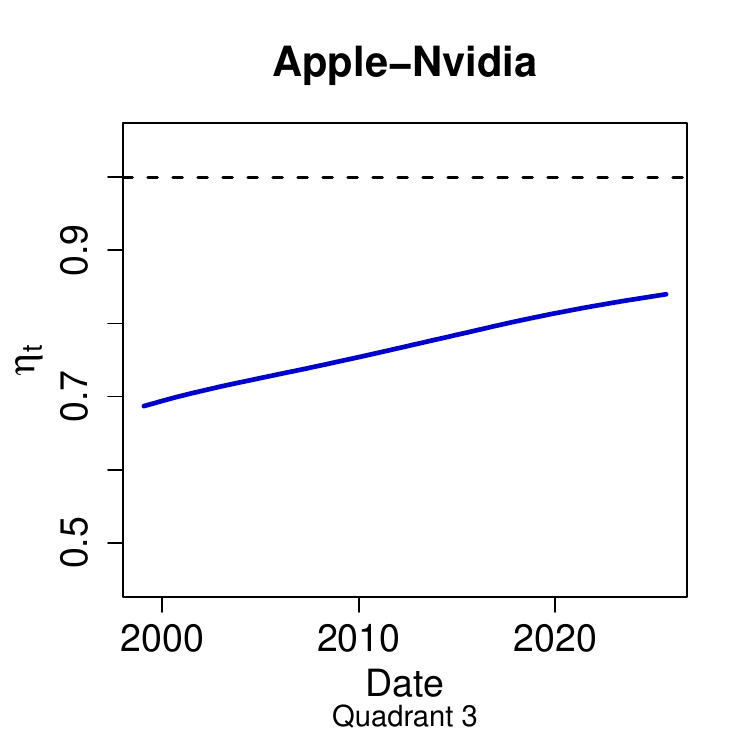}
    \end{subfigure}%
    \caption{The coefficient of tail dependence $\eta_t$ over time for quadrants 1 (top row) and 3 (bottom row). The black dotted line denotes the upper bound for $\eta_t$. }
    \label{fig:eta_ests_quads_1_3}
\end{figure}

\begin{figure}[H]
    \centering
     \begin{subfigure}[b]{.25\textwidth}
        
        \includegraphics[width=\textwidth]{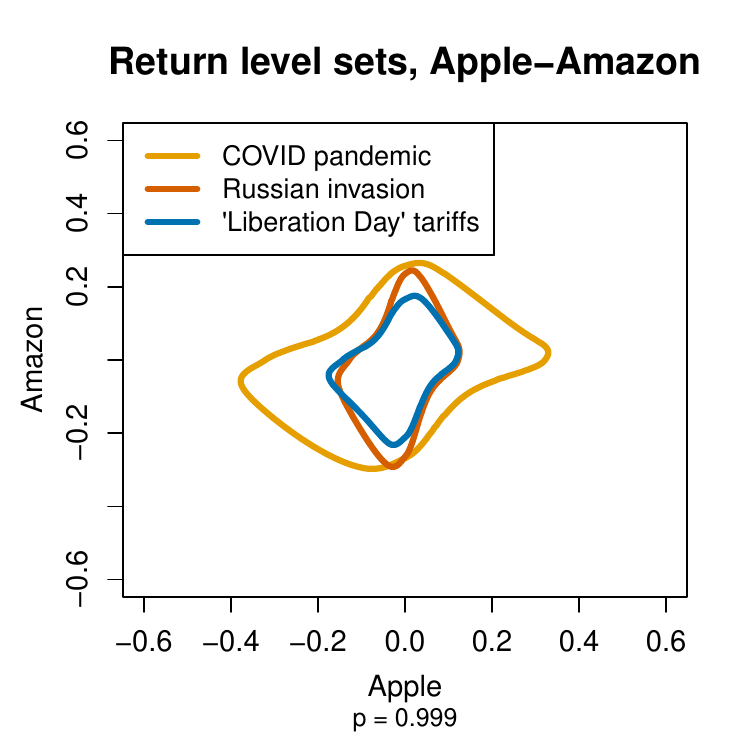}
    \end{subfigure}%
    \begin{subfigure}[b]{.25\textwidth}
        
        \includegraphics[width=\textwidth]{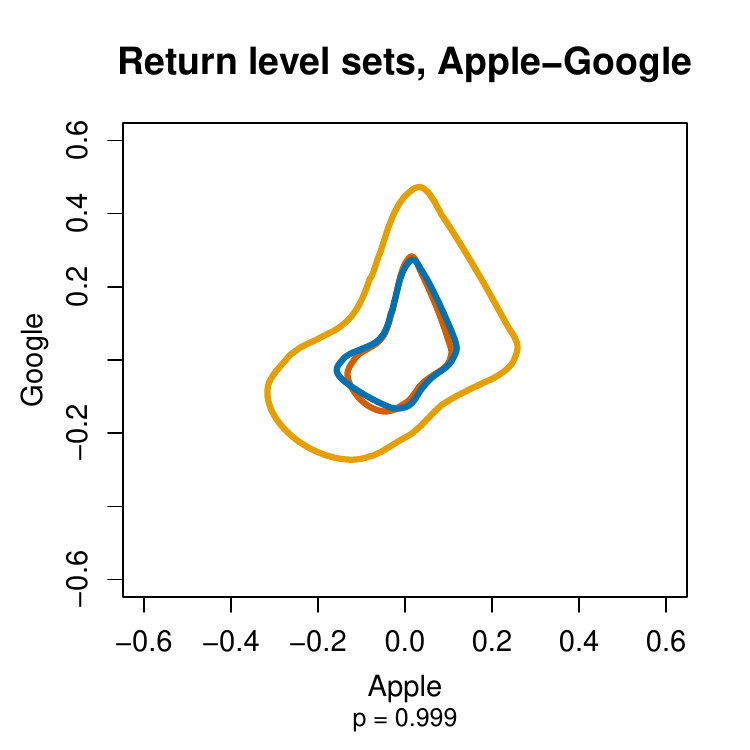}
    \end{subfigure}%
      \begin{subfigure}[b]{.25\textwidth}
        
        \includegraphics[width=\textwidth]{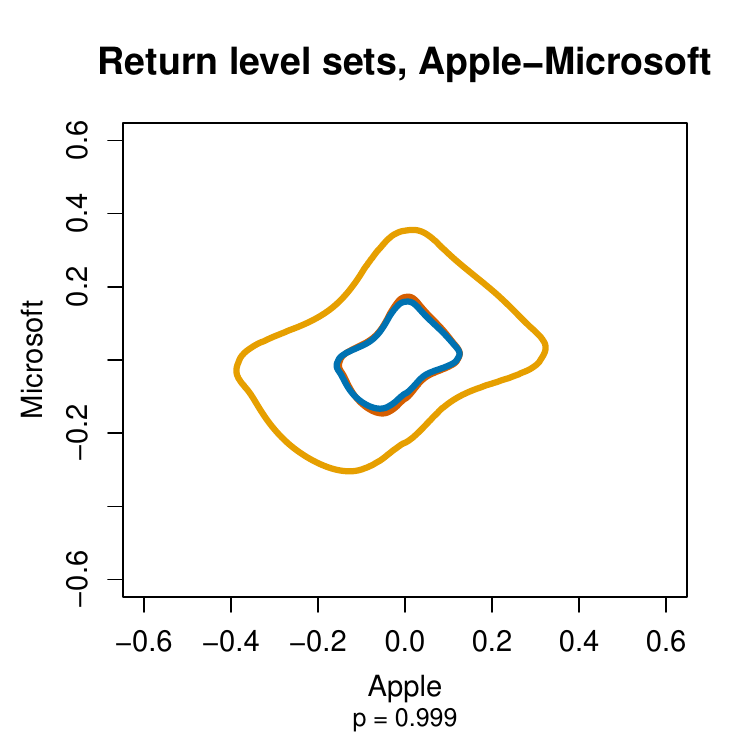}
    \end{subfigure}%
      \begin{subfigure}[b]{.25\textwidth}
        
        \includegraphics[width=\textwidth]{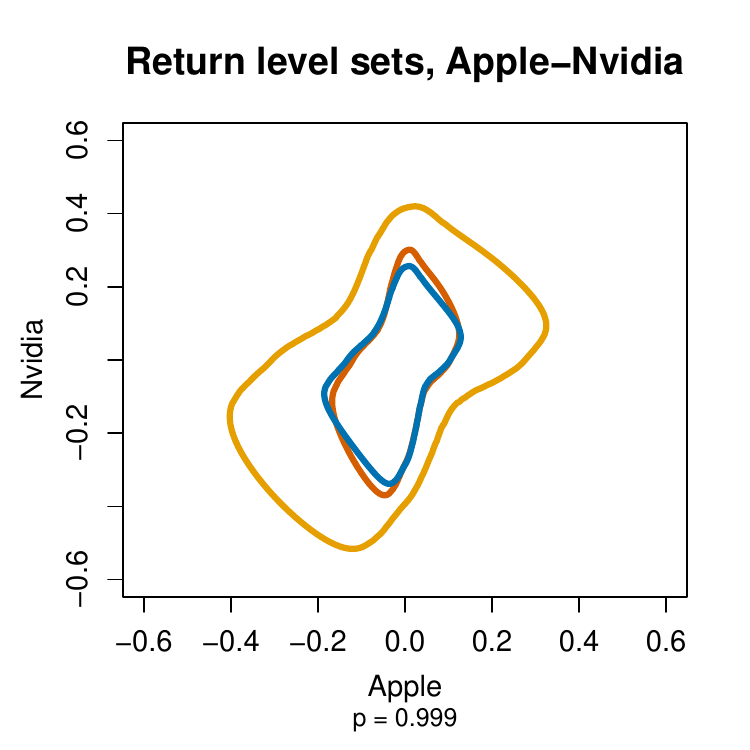}
    \end{subfigure}%
    \caption{Return level set estimates at $p=0.999$ across all pairs for recent dates of significance in the financial context. }
    \label{fig:rl_sets_ests}
\end{figure}

Setting $p = 0.999$, we apply the framework to obtain estimates of return level sets, denoted $\widehat{\mathcal{A}}^p_t$, which can be used as practical risk measures to inform decision making; see \citet{Haselsteiner2019}, \citet{Papastathopoulos2025}, and \citet{Simpson2024} for further discussion. We consider three recent dates of significance in the financial context; namely, the onset of the COVID 2020 pandemic \citep[March 16th, 2020,][]{mazur2021covid}, the beginning of the Russian invasion of Ukraine \citep[February 24th, 2022,][]{lo2022russo}, and the 2025 `Liberation Day' tariffs \citep[April 2nd, 2025,][]{ignatenko2025making}. While the onset of the pandemic cannot be expressed by a single day, stock markets experienced significant volatility in mid-March, and thus this time period is of interest for studying joint tail behaviour. Return level sets for each date were computed across all pairs. The estimated sets were then transformed back to the original scale of the log-returns data, i.e., inverting the modelling procedure described in Section~\ref{subsec:case_marg_models} and effectively `de-filtering' the data. The resulting sets are illustrated in Figure~\ref{fig:rl_sets_ests}. These plots make it clear that the behaviour of this joint risk measure varies noticeably not only between different pairings, but also over different observation dates. We observe significantly more market volatility around the onset of COVID, which helps to explain the difference in magnitudes across the dates. 

\subsubsection*{Generative extreme value modelling}
Finally, we demonstrate how our framework can be used to simulate from the conditional variable $(\boldsymbol{X}_t \mid R_t>r^\tau(\phi,t))$, i.e., the random vector in regions where the radial component is relatively large. Given $t \in \{1,\hdots,T \}$, this involves two steps:
\begin{enumerate}
    \item Simulation from $(\Phi_t \mid R_t>r^\tau(\phi,t))$.
    \item Simulation from $(R_t \mid \Phi_t = \phi, R_t>r^\tau(\phi,t))$. 
\end{enumerate}
Given a simulation $\phi$ from $(\Phi_t \mid R_t>r^\tau(\phi,t))$, step 2) simply involves applying inverse transform sampling using equation~\eqref{eqn:trunc_gamma_assum}. Therefore, all that remains is to specify a model for step 1) from which angular observations can be simulated. For this, we apply a non-parametric kernel density approach known as the Nadaraya-Watson estimator \citep{de2003conditional}. Here, we set $\hat{f}(\phi \mid t) = [\sum_{t^* \in \mathcal{T}^\tau} K^\phi_{h_1}(\phi - \phi_t) K^t_{h_2}(\| t^* -t  \|)]/[\sum_{t \in \mathcal{T}^\tau}K^t_{h_2}(\| t^* -t  \|)]$, where $\mathcal{T}^{\tau} := \{t : \|\boldsymbol{x}_t\| > \hat{r}^\tau(\atantwo(x_{2,t},x_{1,t}),t) \}$, and $K^\phi_{h_1}$ and $K^t_{h_2}$ denote kernel functions with bandwidth parameters $h_1>0$ and $h_2 > 0$ respectively. We use a circular von-Mises kernel for $K^\phi_{h_1}$ and a Gaussian kernel for $K^t_{h_2}$; the former accounts for the periodicity in the angular component. Bandwidth parameters are selected manually by comparing the estimated density functions to local histograms defined on time windows. Numerical integration is then applied to simulate directly from $\hat{f}(\cdot \mid t)$ for any $t$. See \citet{holmes2012fast} and \citet{Castro-Camilo2018} for similar applications of this modelling technique. 

For each time point $t \in \{1, \hdots, T \}$, we can simulate from the entire joint tail of $(\boldsymbol{X}_t \mid R_t>r^\tau(\phi,t))$. This presents a wide range of potential uses, and we present one specific use case here. When managing portfolios, practitioners often wish to assess the (tail) relationships between different stocks \citep{McNeil2015}. Joint risk measures are often used to aid with this assessment. One popular measure is known as the \textit{conditional value-at-risk} (CoVaR) \citep{girardi2013systemic,tobias2016covar}. Given some probability $p$ close to 0, the \textit{downside} CoVaR at level $p$ is defined in the non-stationary, bivariate setting via 
\begin{equation}\label{eqn:low_covar}
\Pr(X_{2,t} \leq \operatorname{CoVaR}_{t,p} \mid X_{1,t} \leq \operatorname{VaR}_{t,p}(X_{1,t}))=p. 
\end{equation}
where $\operatorname{VaR}_{t,p}(X_{1,t}) := F^{-1}_{X_{1,t}}(p)$ is termed the \textit{downside value-at-risk} (VaR) of $X_{1,t}$. One can view $\operatorname{CoVaR}_{t,p}$ as the resulting stress on $X_{2,t}$ given an extreme loss for $X_{1,t}$ \citep{Nolde2021}. However, as noted in Section~\ref{subsec:case_overview}, we are not just interested in extreme losses within our analysis. Thus, we adapt equation~\eqref{eqn:low_covar} to also consider the joint upper tail and define the \textit{upside} CoVaR at level $p$ via
\begin{equation}\label{eqn:upp_covar}
\Pr(X_{2,t} \geq \operatorname{CoVaR}^*_{t,p} \mid X_{1,t} \geq \operatorname{VaR}^*_{t,p}(X_{1,t}))=1-p. 
\end{equation}
where $\operatorname{VaR}^*_{t,p}(X_{1,t}) := F^{-1}_{X_{1,t}}(1-p)$ is termed the \textit{upside} VaR of $X_{1,t}$. Conversely, $\operatorname{CoVaR}^*_{t,p}$ describes the effect on $X_{2,t}$ given an extreme increase for $X_{1,t}$. We note the downside and upside naming conventions follow from similar nomenclature in the literature \citep[e.g.,][]{reboredo2016downside}.

\begin{figure}
    \centering
     \begin{subfigure}[b]{.25\textwidth}
        
        \includegraphics[width=\textwidth]{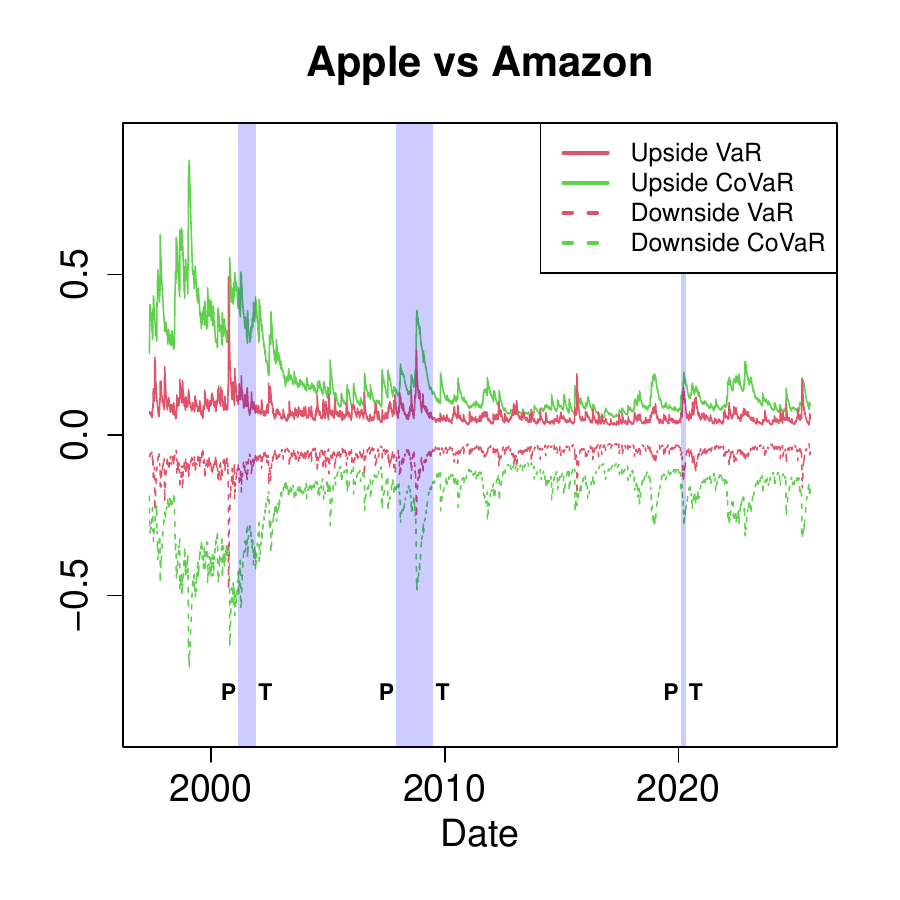}
    \end{subfigure}%
    \begin{subfigure}[b]{.25\textwidth}
        
        \includegraphics[width=\textwidth]{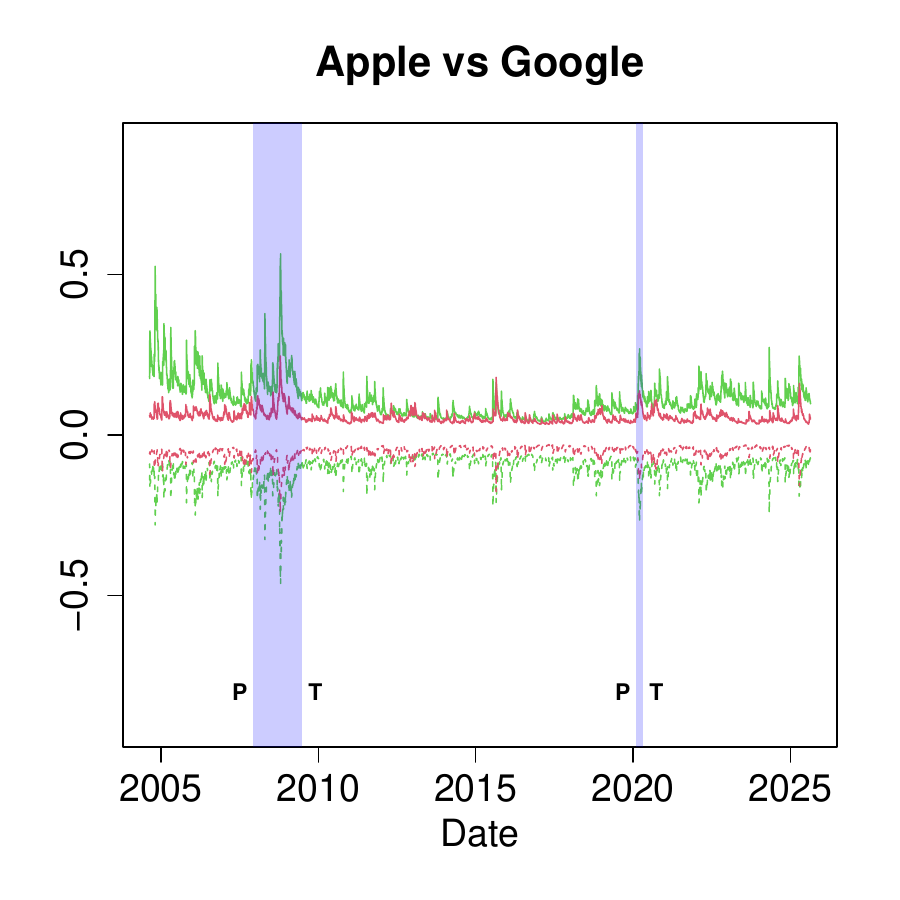}
    \end{subfigure}%
      \begin{subfigure}[b]{.25\textwidth}
        
        \includegraphics[width=\textwidth]{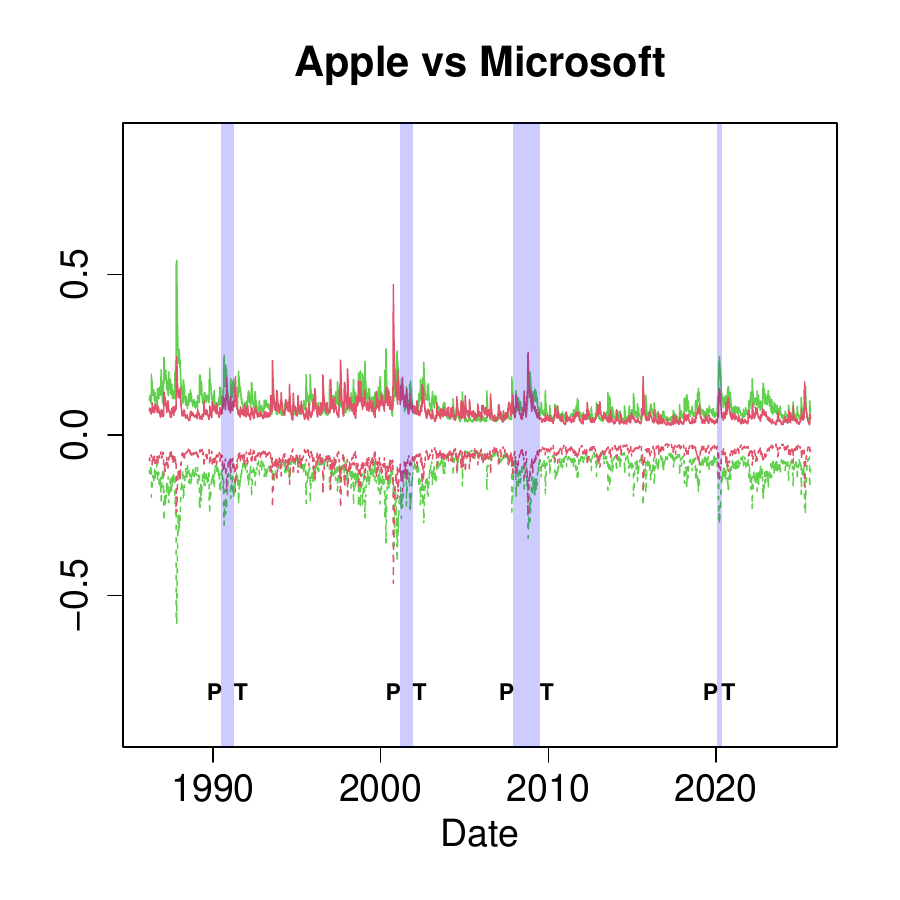}
    \end{subfigure}%
      \begin{subfigure}[b]{.25\textwidth}
        
        \includegraphics[width=\textwidth]{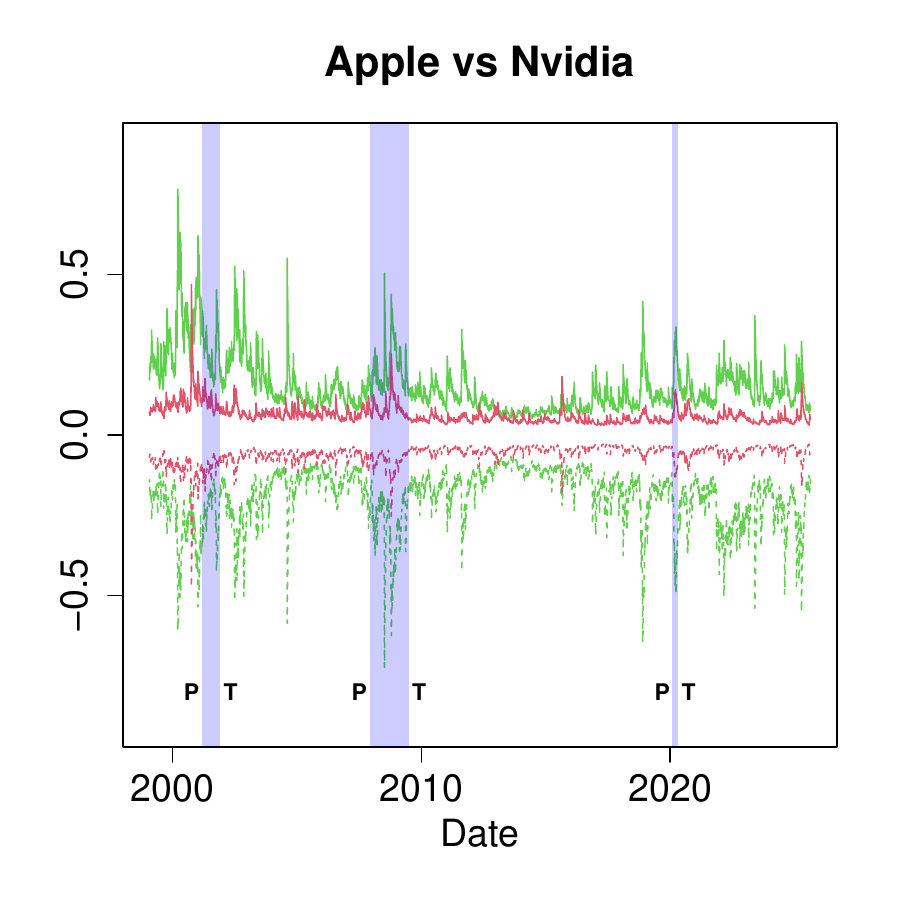}
    \end{subfigure}%
   
    \caption{Downside and upside VaR and CoVaR estimates for all pairs, with the `stress' events defined by the Apple stock in each case. }
    \label{fig:covar_all_mirrors}
\end{figure}

Observe that simulation can be used to approximate both $\operatorname{CoVaR}_{t,p}$ and $\operatorname{CoVaR}^*_{t,p}$ by empirically estimating the probabilities denoted in equations~\eqref{eqn:low_covar} and \eqref{eqn:upp_covar} and using a rootfinder to solve for equality. For each date, we simulate $100,000$ observations in the joint tail and estimate the CoVaR measures at $p = 0.01$ relative to the Apple returns (i.e., $X_{1,t}$ denoting the Apple series). The estimated measures are then transformed from the Laplace to the original observation scale. Figure~\ref{fig:covar_all_mirrors} illustrates the corresponding estimates for each pair of stocks. In these plots, we also illustrate recent US business cycles relevant to the returns data in question. Each cycle corresponds to a period of recession, beginning at the peak (P) and ending at the trough (T) \citep[see, e.g.,][]{de2012tracking}. One can observe that the extremes in the VaR estimates do not always correspond to the extremes in CoVaR, illustrating the time-varying nature of the extremal dependence within the data. This provides evidence that classical multivariate extreme value modelling approaches would not provide adequate flexibility for capturing the structure in this data. Interestingly, we also observe in Figure~\ref{fig:covar_all_mirrors} that both joint extreme gains and losses appear possible during recession periods. Furthermore, certain phenomena are recognisable from the estimates; for example, the collapse of the dot--com bubble which occurred during 2000--2002, leading to spikes in both the upper and lower tails.   

\section{Discussion} \label{sec:discussion}

In this article, we have introduced a novel geometric modelling framework for approximating non-stationary extremal dependence structures. This framework addresses certain limitations from existing works and, crucially, allows for more general and flexible modelling within the non-stationary setting. Our proposed model can capture a wide range of dependence structures, and is not limited to either AD or AI data sets. Furthermore, while we have proposed our framework in the non-stationary setting, the techniques given in Section~\ref{sec:methodology} could be adapted to obtain stationary limit set estimates, adding to a growing literature in this area. As a by-product, our framework can be used to generate random vectors in regions where the radial component is large, thereby contributing to the fast-emerging literature on generative models for multivariate extremes, which aim to simulate data across all joint tail regions; see, e.g., \citet{DeMonte2025} or \citet{Wessel2025}. The ability to generate, or simulate, extremal data in different tail regions is desirable beyond the financial setting; for example, in the environmental setting to assist with flood risk assessment and catastrophe modelling \citep{Keef2013a,Quinn2019}.

We close the paper with some comments on open problems and future work. Firstly, as noted in Section~\ref{subsec:reml}, the boundary set estimates obtained from our approach are not required to satisfy all of the theoretical properties of boundary sets outlined in Section~\ref{sec:intro}; one could in future work explore whether this rescaling could be directly incorporated into the model fitting procedure, as was the case in \citet{Murphy-Barltrop2024d} and \citet{Campbell2024}. 

Currently, the proposed approach can currently only be applied the bivariate setting. Flexibly modelling non-stationary extremal dependence in the general multivariate setting remains an open problem. In the general $d$-dimensional setting, one can still transform to polar angles; however, fitting GAMs over such angles requires more than cyclic basis splines to achieve the correct periodicity properties. Alternatively, one could try to define models directly on the $(d-1)$-sphere, as in \citet{Papastathopoulos2025} and \citet{Murphy-Barltrop2024d}; however, this requires frameworks that can handle the collinearity that exists between the corresponding angular variables, which is not the case for many regression techniques. 

There is no guarantee that our proposed estimators for the non-stationary gauge functions or boundary sets will converge to the corresponding true values. While it would be desirable to have theoretical proofs of consistency, say, such results generally necessitate strict and unreasonable modelling assumptions, which themselves can be difficult to verify. We have therefore opted for a more practical treatment of our proposed estimators, noting that for real data sets, one can only evaluate model performance using diagnostic procedures, such as those presented in Sections~\ref{subsec:model_checking} and \ref{sec:case_study}.

One notable observation from this analysis is that the choice of norm (i.e., the selected coordinate system) greatly affects the shapes of boundary set that can be represented by the modelling framework. In particular, when the boundary set exhibits `pointiness' in certain regions, it may be appropriate to select a norm for which the corresponding unit ball is also pointy in these regions. However, it remains unclear how one can select this norm in practice, and future work could therefore explore whether model selection techniques could help one to perform this selection in a robust and systematic manner. 

As stated in Section~\ref{subsec:case_fitting}, best practices for establishing the existence of time-varying extremal dependence within data are yet to be developed. Constructing tests with a reasonable statistical power is implicitly more complicated in the extreme setting, since one achieves better convergence to asymptotic models as one moves further into the tail, but this vastly reduces the available sample size and hence power. Developing robust and reliable techniques for diagnosing trends in extremal dependence represents an important line for future work. 

Finally, an obvious omission from the case study of Section~\ref{sec:case_study} is the lack of uncertainty quantification for the fitted models and computed statistics. Standard bootstrapping techniques are not applicable here, since resampling such data samples would not preserve a time-varying dependence structure. Uncertainty quantification is further complicated by the fact our modelling framework contains many components, each of which is fitted independently; the quantile function, the gauge function, and the angular model. Jointly quantifying the uncertainty arising from each of these components is non-trivial and would require careful consideration. We defer the quantification of uncertainty for our setup to future work, noting that one could possibly adapt established approaches for non-stationary, univariate time series, such as the local block bootstrap \citep[e.g.,][Chapter 4]{Politis1999}. 

\section*{Declarations}
\subsection*{Ethical Approval}
Not Applicable
\subsection*{Availability of supporting data}
The data sets analysed in Section~\ref{sec:case_study} are available from the corresponding author upon reasonable request. 

\subsection*{Competing interests}
The authors have no relevant financial or non-financial interests to disclose.
\subsection*{Funding details} 
JW gratefully acknowledges support from UK Engineering and Physical Sciences Research Council grant EP/X010449/1. MdC is partially funded by Leverhulme Trust and FCT under Grants UIDB/04106/2020 and UIDP/04106/2020. 

\subsection*{Code availability}
Code for fitting the proposed framework is freely available at \url{https://github.com/callumbarltrop/NSGE}.

\subsection*{Authors’ contribution}
CMB proposed the methodological framework, conducted the simulation study and case analyses, prepared the initial manuscript draft, and led the overall project. JW and MdC provided substantial guidance throughout the development of the work, contributed to the refinement of the methodology, and gave extensive feedback on successive drafts. BY contributed to the implementation and development of the computational code and also provided feedback on the manuscript. All authors approved the final version of the manuscript. 

\bibliographystyle{apalike}

{\small
\bibliography{allrefs.bib}
}

\clearpage

\renewcommand{\theequation}{A.\arabic{equation}}
\renewcommand{\thefigure}{\AlphAlph{\value{figure}}}
\setcounter{figure}{0} 
\renewcommand{\thetable}{\Alph{table}}
\setcounter{table}{0} 
\renewcommand{\thesection}{A\arabic{section}}
\renewcommand\theHtable{Appendix.\thetable}
\renewcommand\theHfigure{Figure.\thefigure}

\setcounter{figure}{0}
\setcounter{table}{0}
\setcounter{equation}{0}
\setcounter{theorem}{0}

\spacingset{1.9}

\begin{appendix}

\section*{Appendix}

\section{Additional illustrative figures} \label{sec:appen_additional_figures}

\begin{figure}[H]
    \centering
    \includegraphics[width=0.38\linewidth]{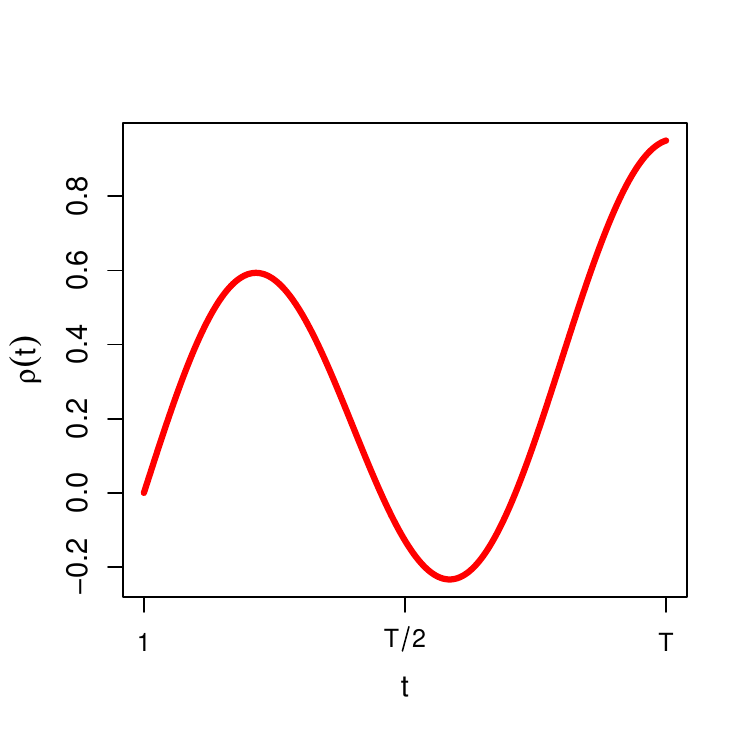}
    \caption{Plot illustrating the correlation parameter function $\rho_2(t)$ over $t$ for the second copula example of Section~\ref{subsec:sim_examples}.}
    \label{fig:rho_func}
\end{figure}

Figure~\ref{fig:rho_func} illustrates the function $\rho_2(t)$ defined in Section~\ref{subsec:sim_examples}.

\begin{figure}[H]
    \centering
    \includegraphics[width=\linewidth]{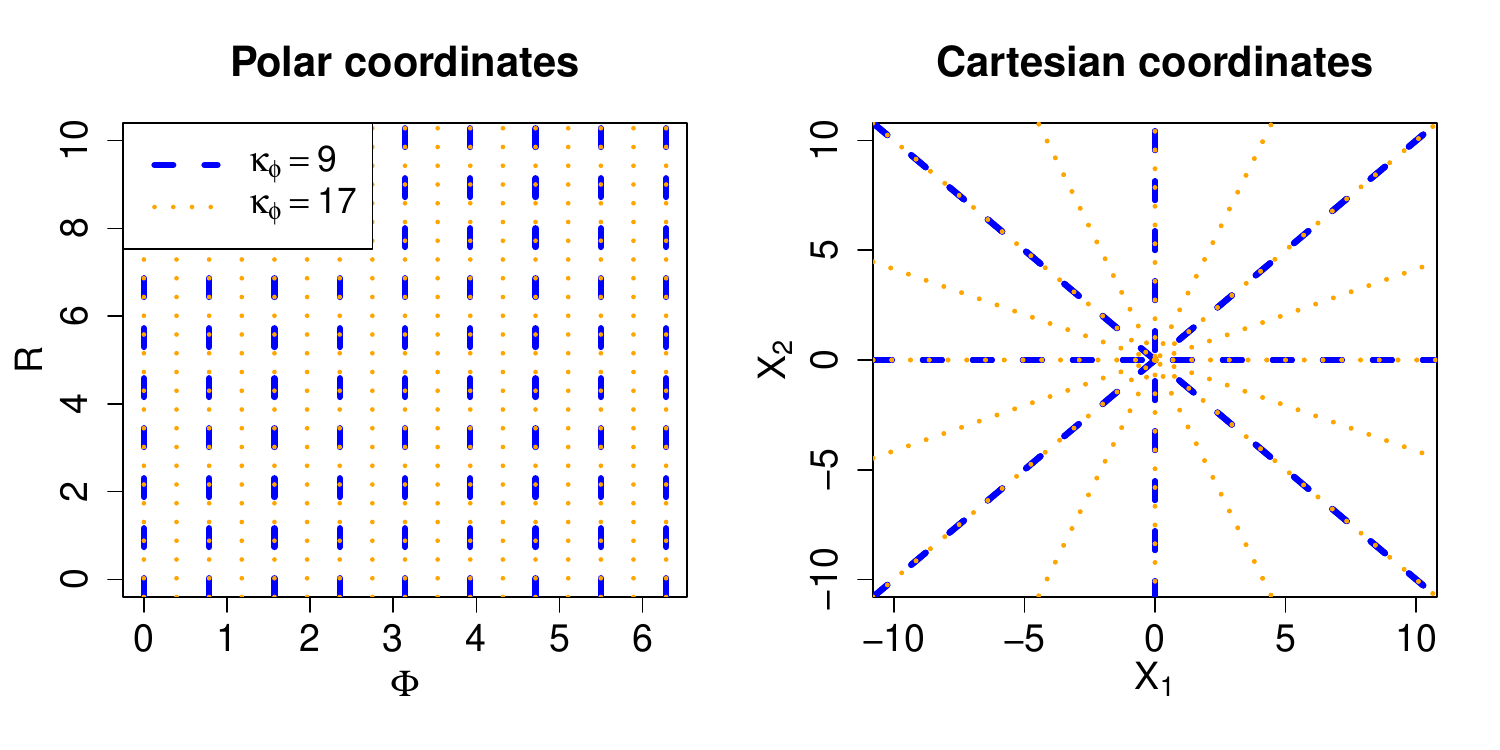}
    \caption{Figure illustrating the two choices of knot locations for $\Phi_t$ in polar (left) and Cartesian (right) coordinates.}
    \label{fig:knot_locations_angle}
\end{figure}

Figure~\ref{fig:knot_locations_angle} illustrates two choices of knot locations for the polar angle $\Phi_t$. Observe that both sets of knots intersect all axes and primary diagonals.

Figure illustrates tensor product basis splines at two pairs of knots for the predictor variables $\phi$ and $t$, with $T = 100$. One can observe the cyclic nature of the splines at the end points of the angular component. 

\begin{figure}[H]
    \centering
    \includegraphics[width=\linewidth]{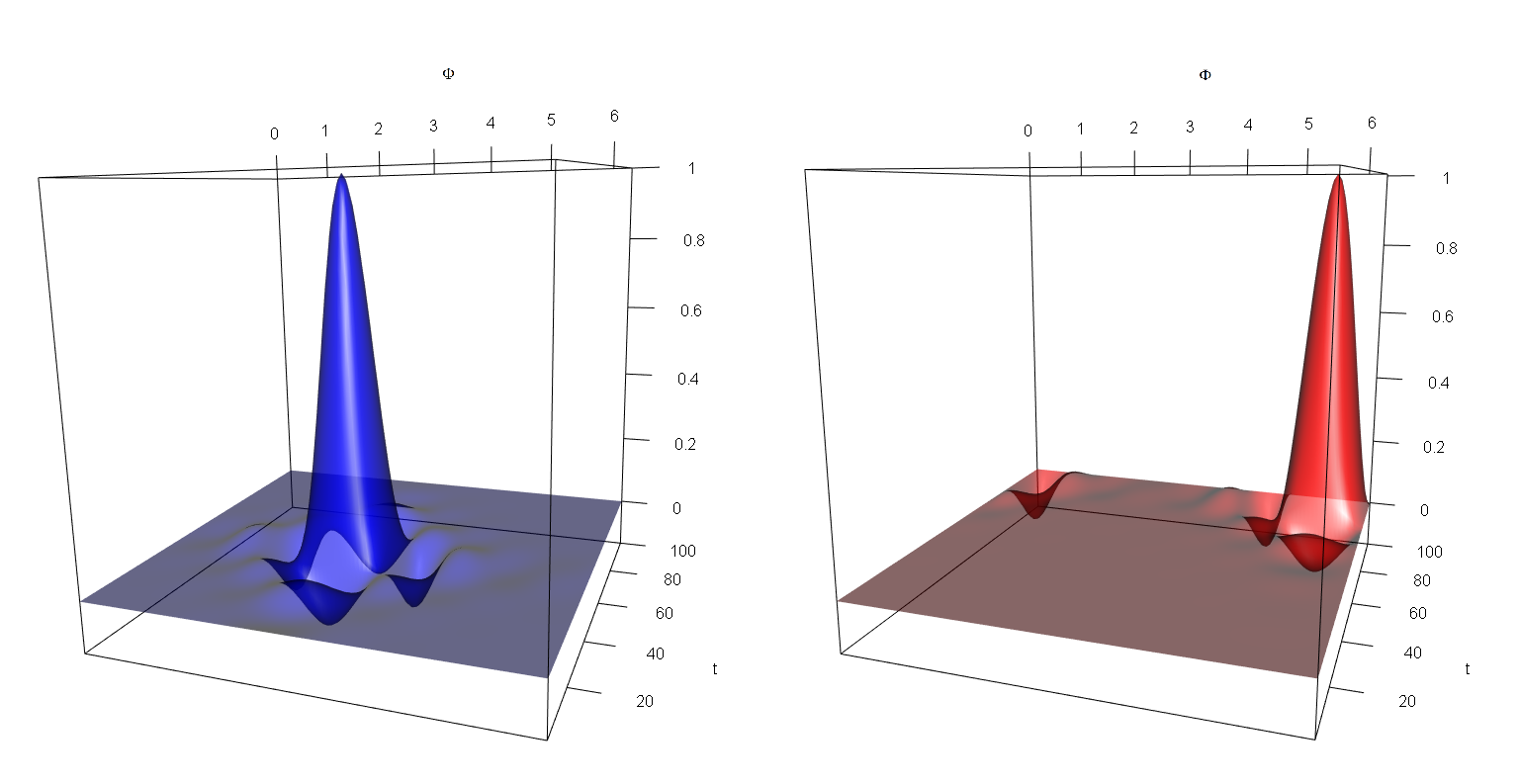}
    \caption{Example tensor product basis functions for $\phi$ and $t$, with $T=100$.}
    \label{fig:tensor_prod}
\end{figure}

\section{The \citet{Ledford1996} model on standard Laplace margins.} \label{sec:appen_led_tawn}

While originally proposed on unit Fr\'echet margins, the \citet{Ledford1996} model can be generalised to other marginal scales \citep[see, e.g.,][]{Simpson2024a,Murphy-Barltrop2024d}. Given a random vector $\boldsymbol{X} = (X_1,X_2)$ on standard Laplace margins, we consider the following modelling formulation: for any $\boldsymbol{o} = (o_1,o_2) \in \{-1,1 \}^2$, we assume
\begin{equation} \label{eqn:led_tawn_ext}
    \Pr( \boldsymbol{o}\boldsymbol{X} > u ) \sim L(e^u)\exp(-u/\eta_{\boldsymbol{o}}) \; \; \text{as} \; u \to \infty,
\end{equation}
where $L(\cdot)$ is a slowly varying function, i.e., $\lim_{u \to \infty}L(c u)/L(u) = 1$ for any constant $c > 0$, and $\eta_{\boldsymbol{o}} \in (0,1]$ is termed the coefficient of tail dependence. These coefficients provide information about the tail dependence structure; for example, setting $\boldsymbol{o} = (1,1)$, asymptotic dependence corresponds to the case when $\eta = 1$ and $\lim_{u \to \infty} L(u) > 0$. Observe also that $\boldsymbol{o} = (1,1)$, $\boldsymbol{o} = (-1,1)$, $\boldsymbol{o} = (-1,-1)$, and $\boldsymbol{o} = (1,-1)$ correspond to the first, second, third, and fourth quadrants, or orthants, respectively. For simplicity, we drop the $(\cdot)_{\boldsymbol{o}}$ notation throughout the article, making it clear in each instance which quadrant we are referring to. 

\clearpage

\section{Simulation study results}

In this section, we provide figures illustrating the results of the simulation study detailed in Section~\ref{sec:sim_study}. We present these figures in several subsections, corresponding to which component of the model formulation we are considering. Note that the legend for each figure is identical to that of Figure~\ref{fig:final_res_bs} in the main text. 

\subsection{Evaluating the effect of quantile level} \label{subsec:appen_tau}

\begin{figure}[H]
    \centering
    \includegraphics[width=.8\linewidth]{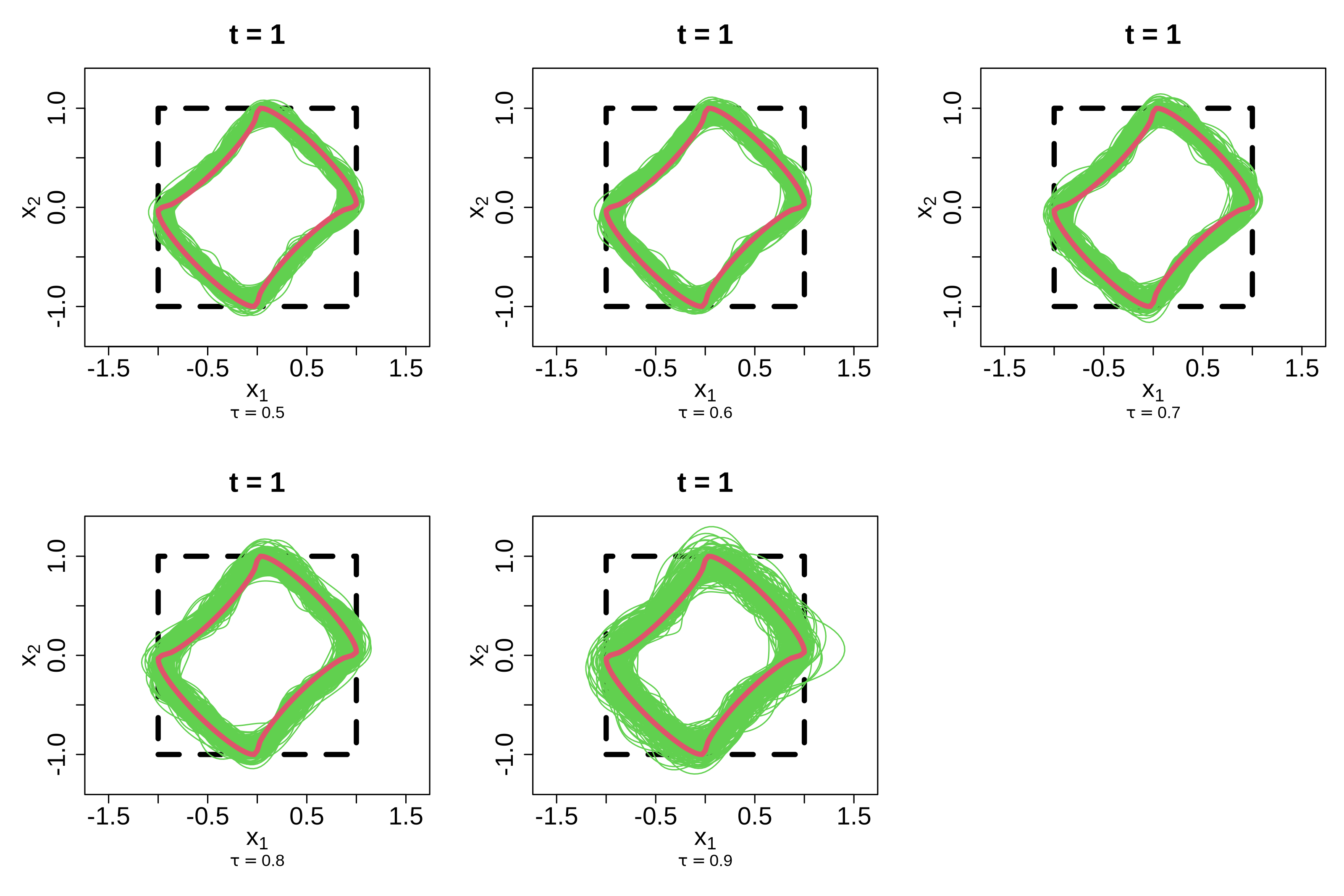}
    \caption{Boundary set estimates as $t = 1$ across $\tau \in \{0.5,0.6,0.7,0.8,0.9\}$ for the first copula example.}
    \label{fig:res_tau_t1_c1}
\end{figure}

\begin{figure}[H]
    \centering
    \includegraphics[width=.8\linewidth]{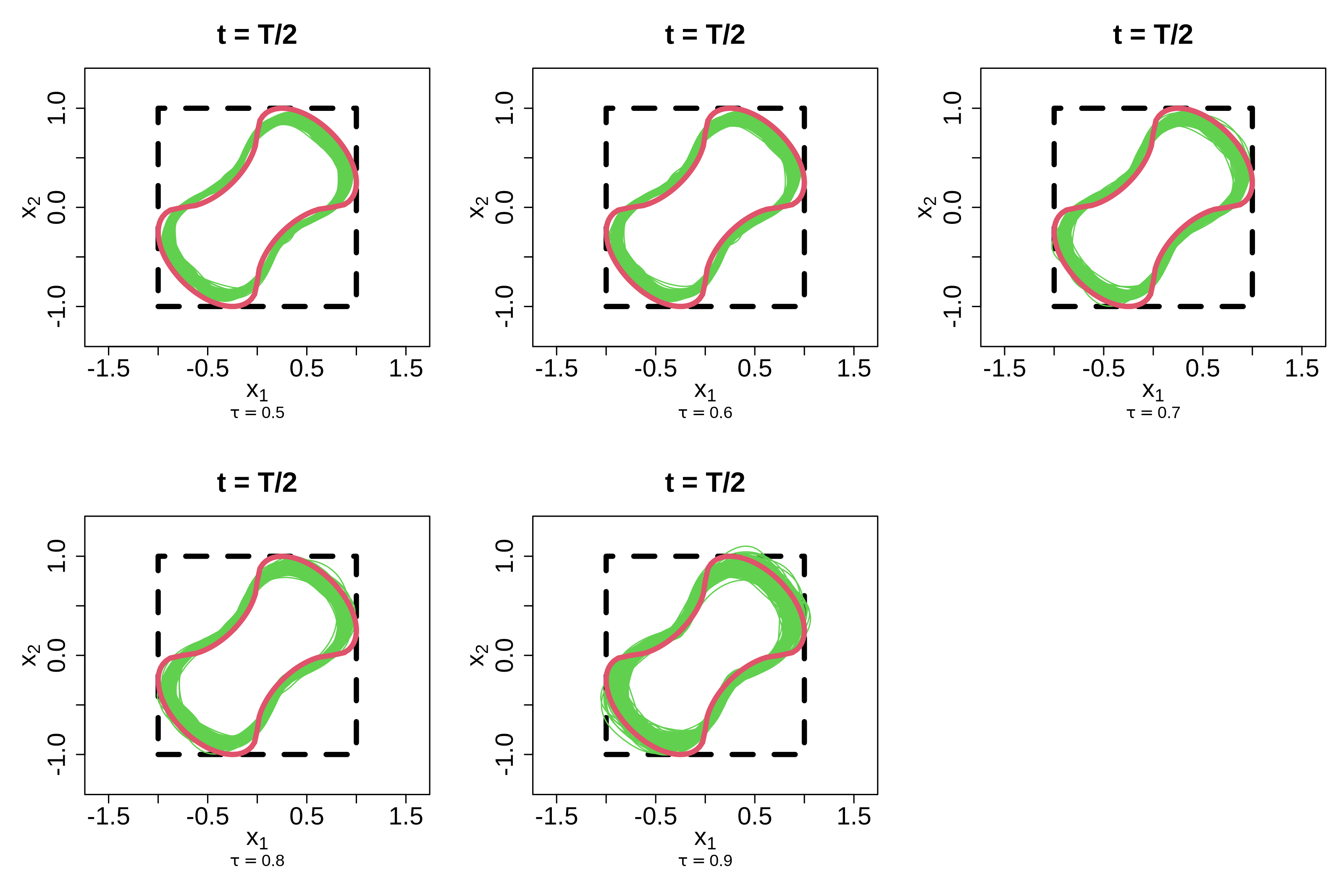}
    \caption{Boundary set estimates as $t = T/2$ across $\tau \in \{0.5,0.6,0.7,0.8,0.9\}$ for the first copula example.}
    \label{fig:res_tau_t2_c1}
\end{figure}

\begin{figure}[H]
    \centering
    \includegraphics[width=.8\linewidth]{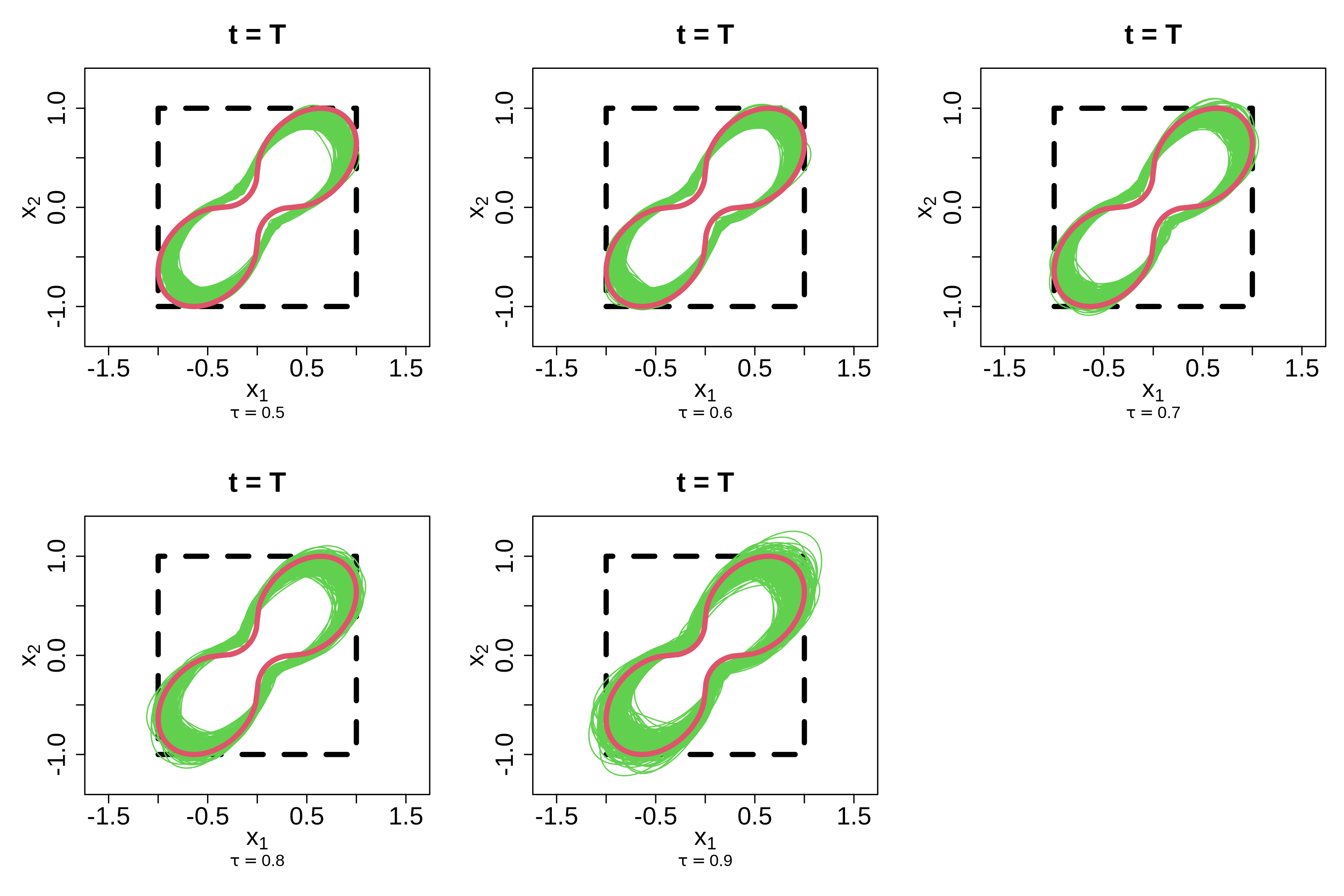}
    \caption{Boundary set estimates at $t = T$ across $\tau \in \{0.5,0.6,0.7,0.8,0.9\}$ for the first copula example.}
    \label{fig:res_tau_t3_c1}
\end{figure}

\begin{figure}[H]
    \centering
    \includegraphics[width=.8\linewidth]{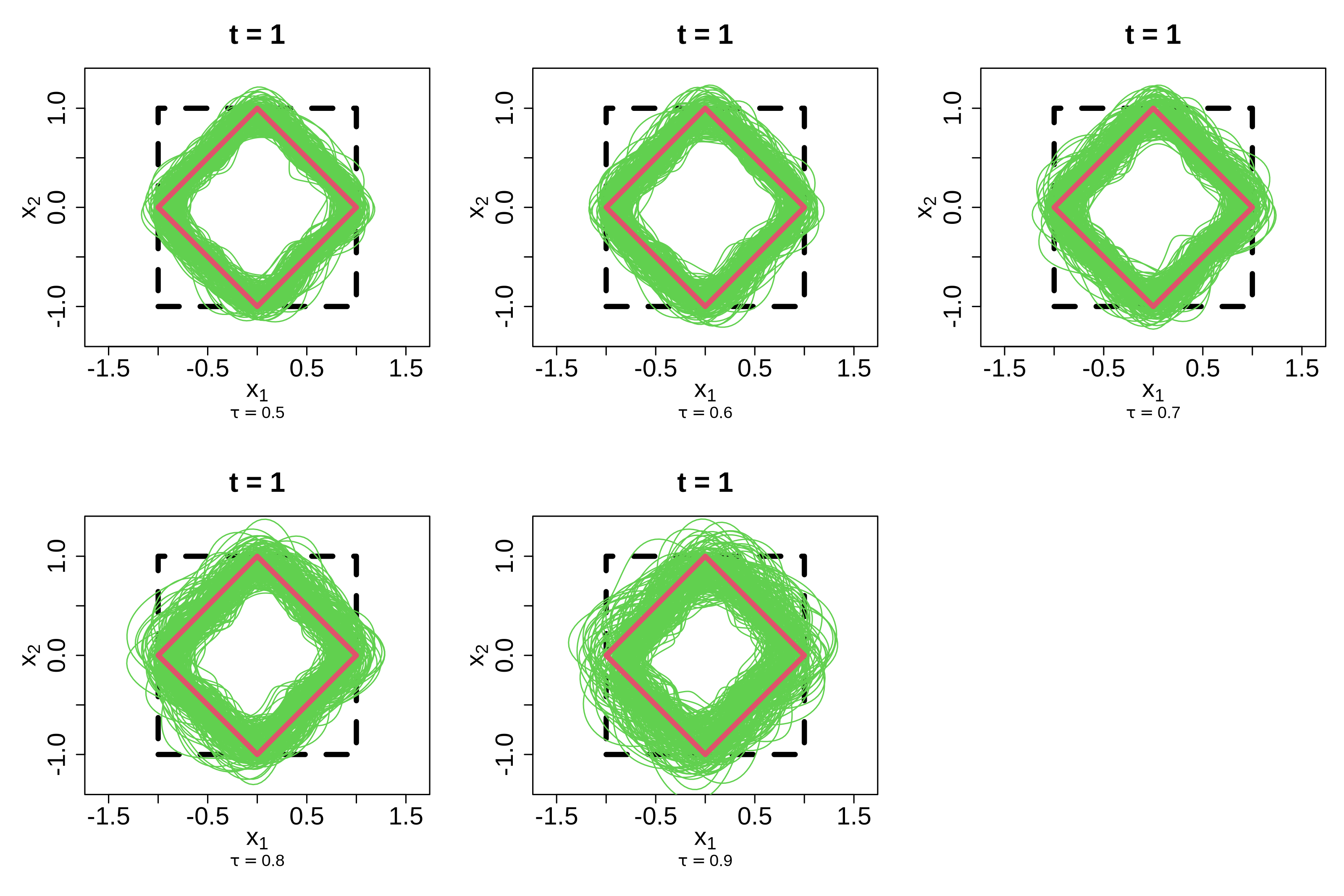}
    \caption{Boundary set estimates as $t = 1$ across $\tau \in \{0.5,0.6,0.7,0.8,0.9\}$ for the second copula example.}
    \label{fig:res_tau_t1_c2}
\end{figure}

\begin{figure}[H]
    \centering
    \includegraphics[width=.8\linewidth]{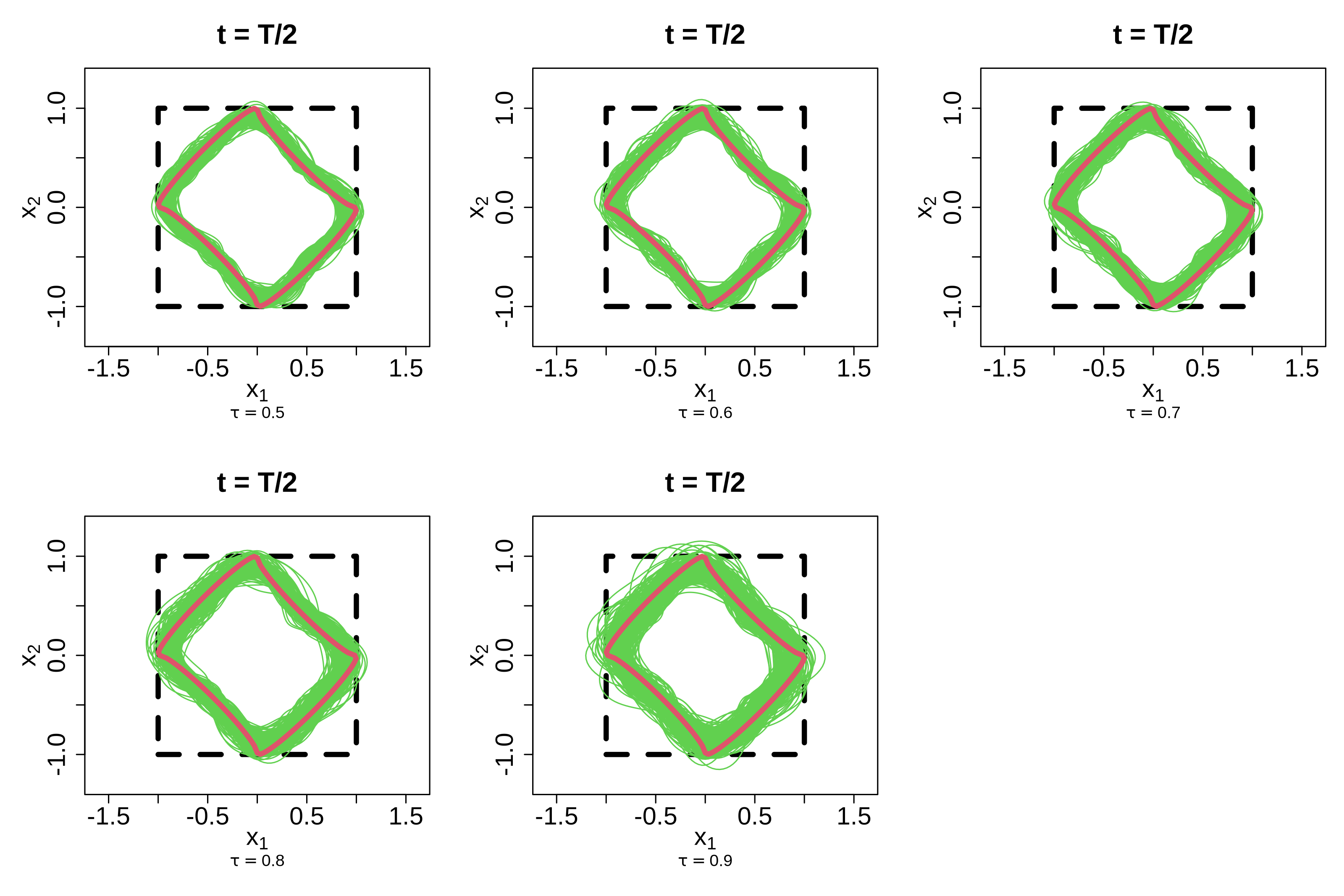}
    \caption{Boundary set estimates as $t = T/2$ across $\tau \in \{0.5,0.6,0.7,0.8,0.9\}$ for the second copula example.}
    \label{fig:res_tau_t2_c2}
\end{figure}

\begin{figure}[H]
    \centering
    \includegraphics[width=.8\linewidth]{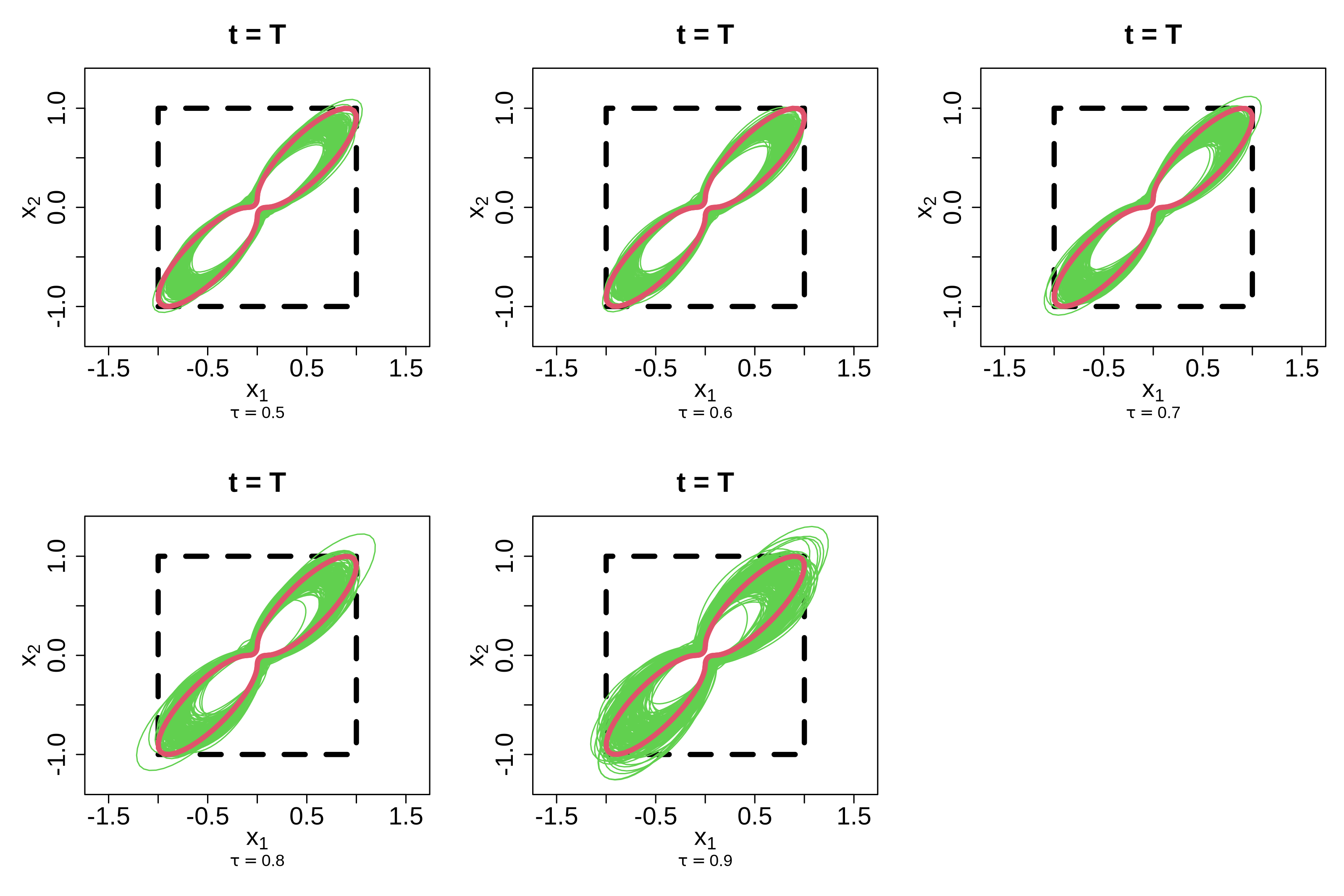}
    \caption{Boundary set estimates at $t = T$ across $\tau \in \{0.5,0.6,0.7,0.8,0.9\}$ for the second copula example.}
    \label{fig:res_tau_t3_c2}
\end{figure}

\begin{figure}[H]
    \centering
    \includegraphics[width=.8\linewidth]{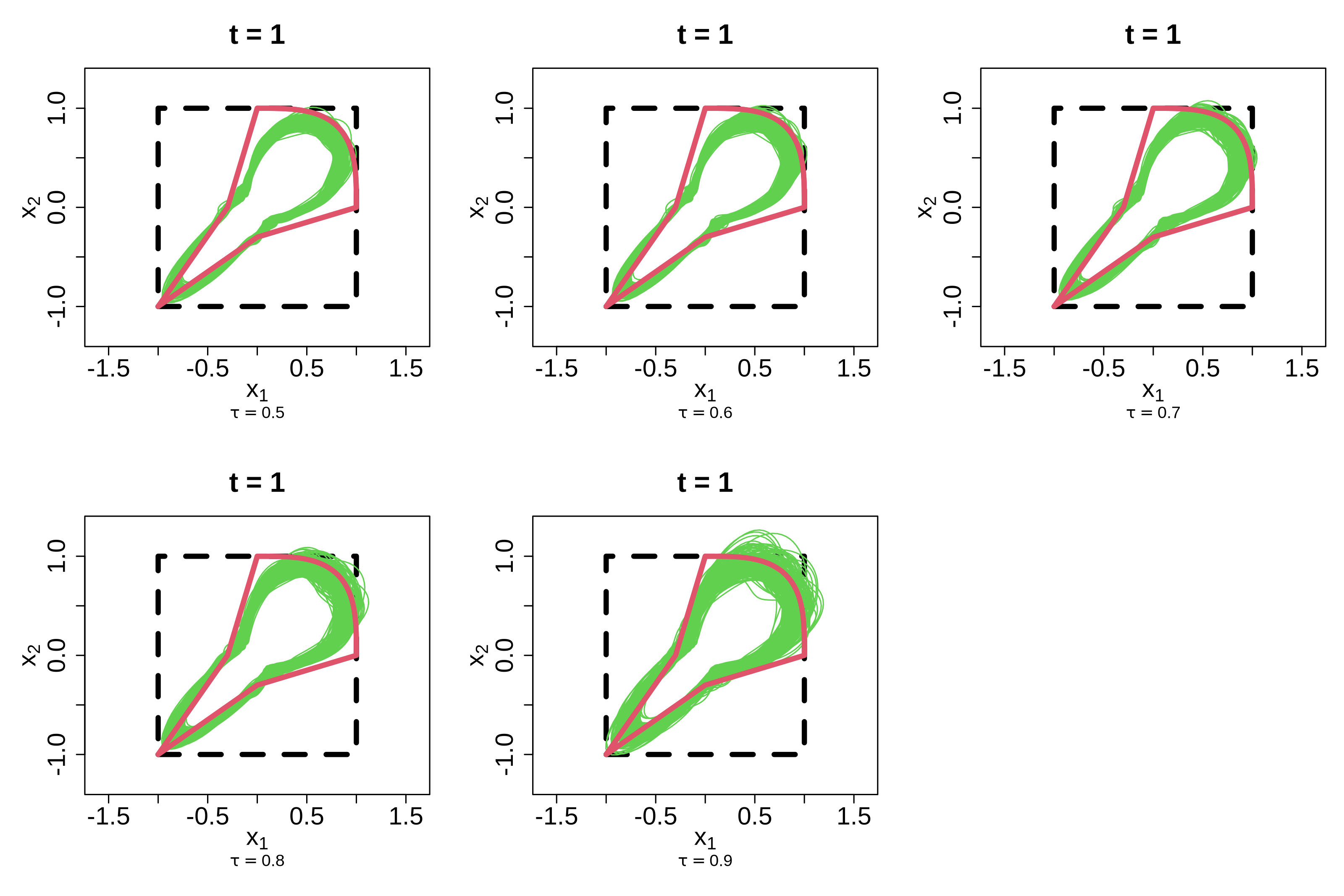}
    \caption{Boundary set estimates as $t = 1$ across $\tau \in \{0.5,0.6,0.7,0.8,0.9\}$ for the third copula example.}
    \label{fig:res_tau_t1_c3}
\end{figure}

\begin{figure}[H]
    \centering
    \includegraphics[width=.8\linewidth]{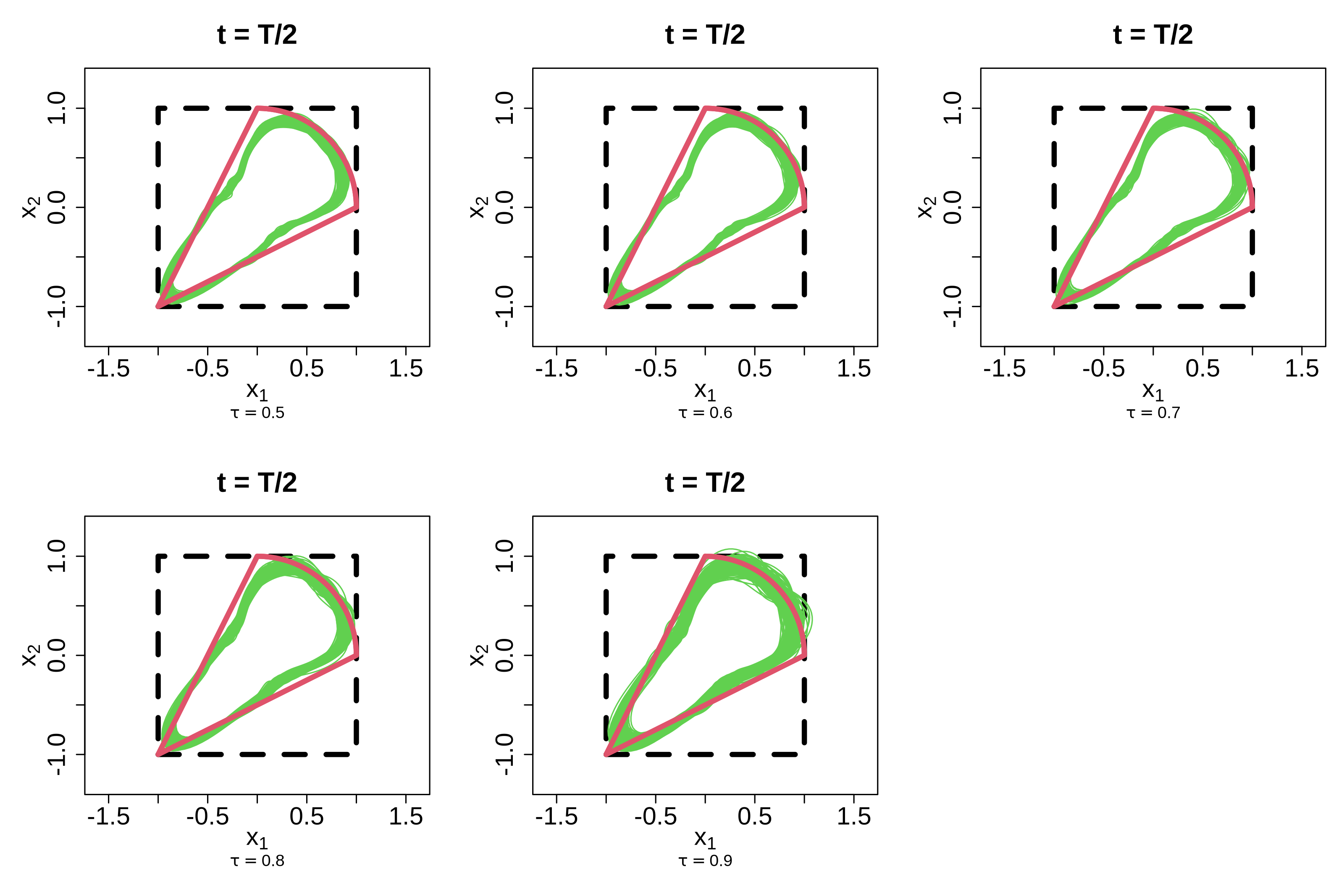}
    \caption{Boundary set estimates as $t = T/2$ across $\tau \in \{0.5,0.6,0.7,0.8,0.9\}$ for the third copula example.}
    \label{fig:res_tau_t2_c3}
\end{figure}

\begin{figure}[H]
    \centering
    \includegraphics[width=.8\linewidth]{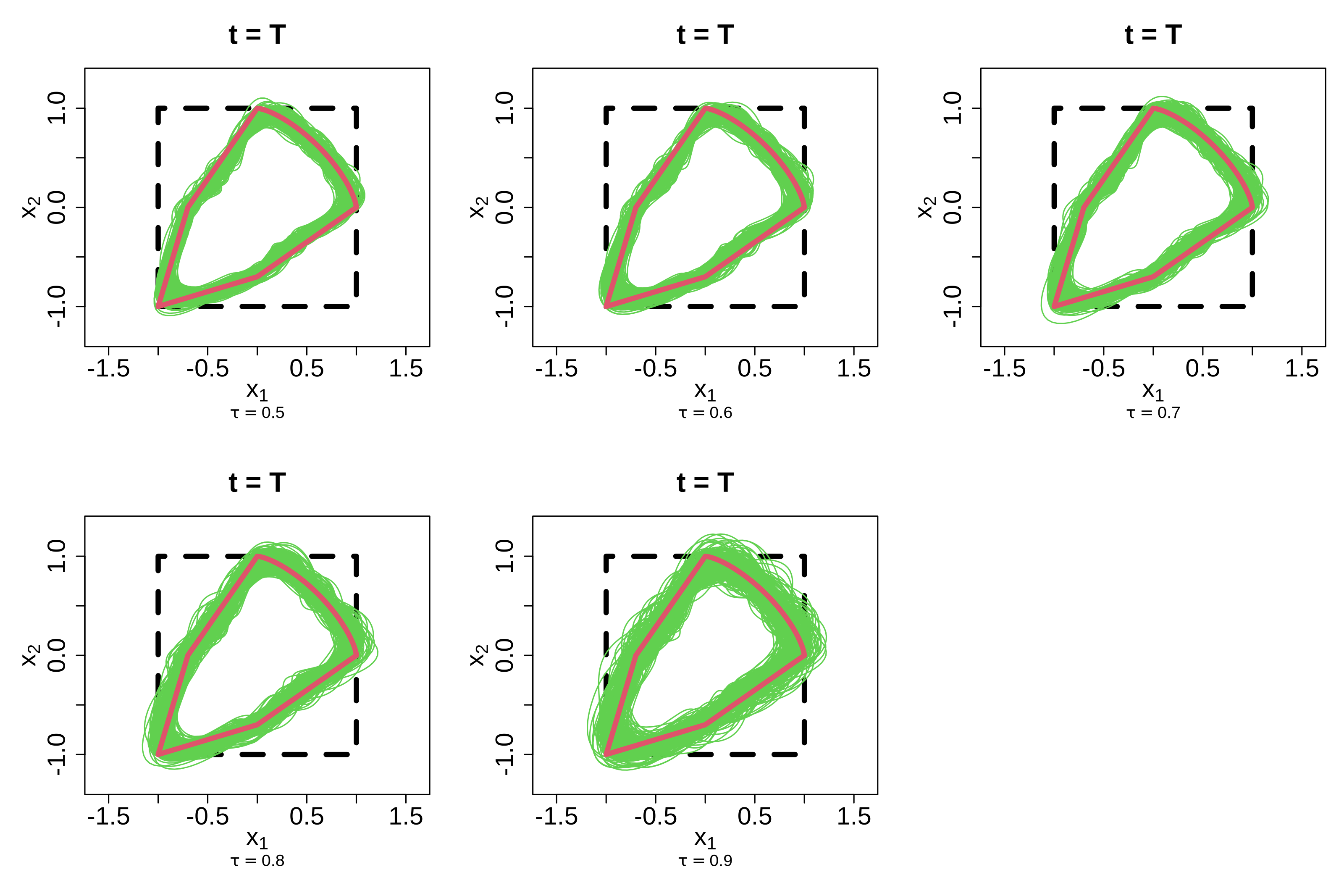}
    \caption{Boundary set estimates at $t = T$ across $\tau \in \{0.5,0.6,0.7,0.8,0.9\}$ for the third copula example.}
    \label{fig:res_tau_t3_c3}
\end{figure}

\begin{figure}[H]
    \centering
    \includegraphics[width=.8\linewidth]{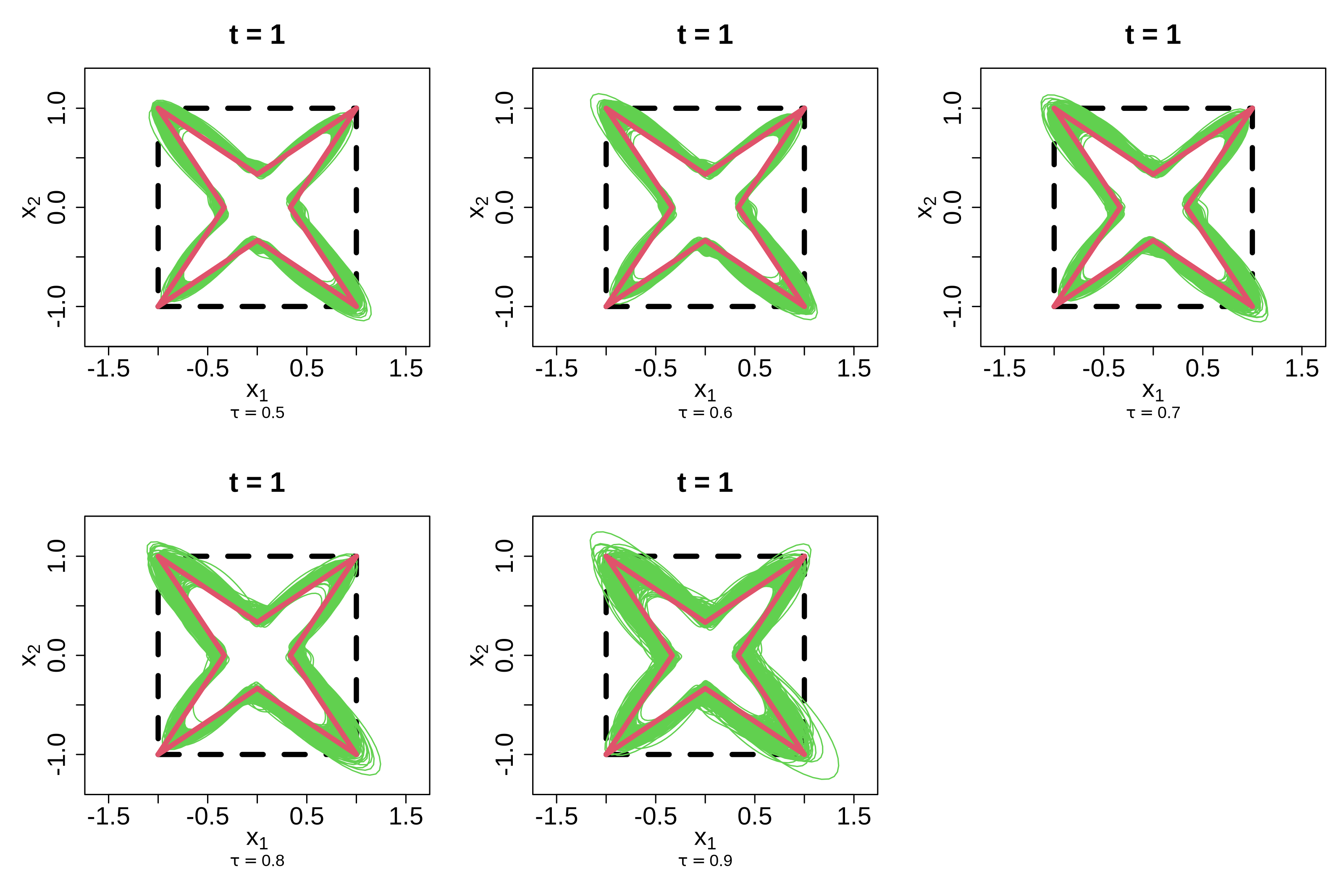}
    \caption{Boundary set estimates as $t = 1$ across $\tau \in \{0.5,0.6,0.7,0.8,0.9\}$ for the fourth copula example.}
    \label{fig:res_tau_t1_c4}
\end{figure}

\begin{figure}[H]
    \centering
    \includegraphics[width=.8\linewidth]{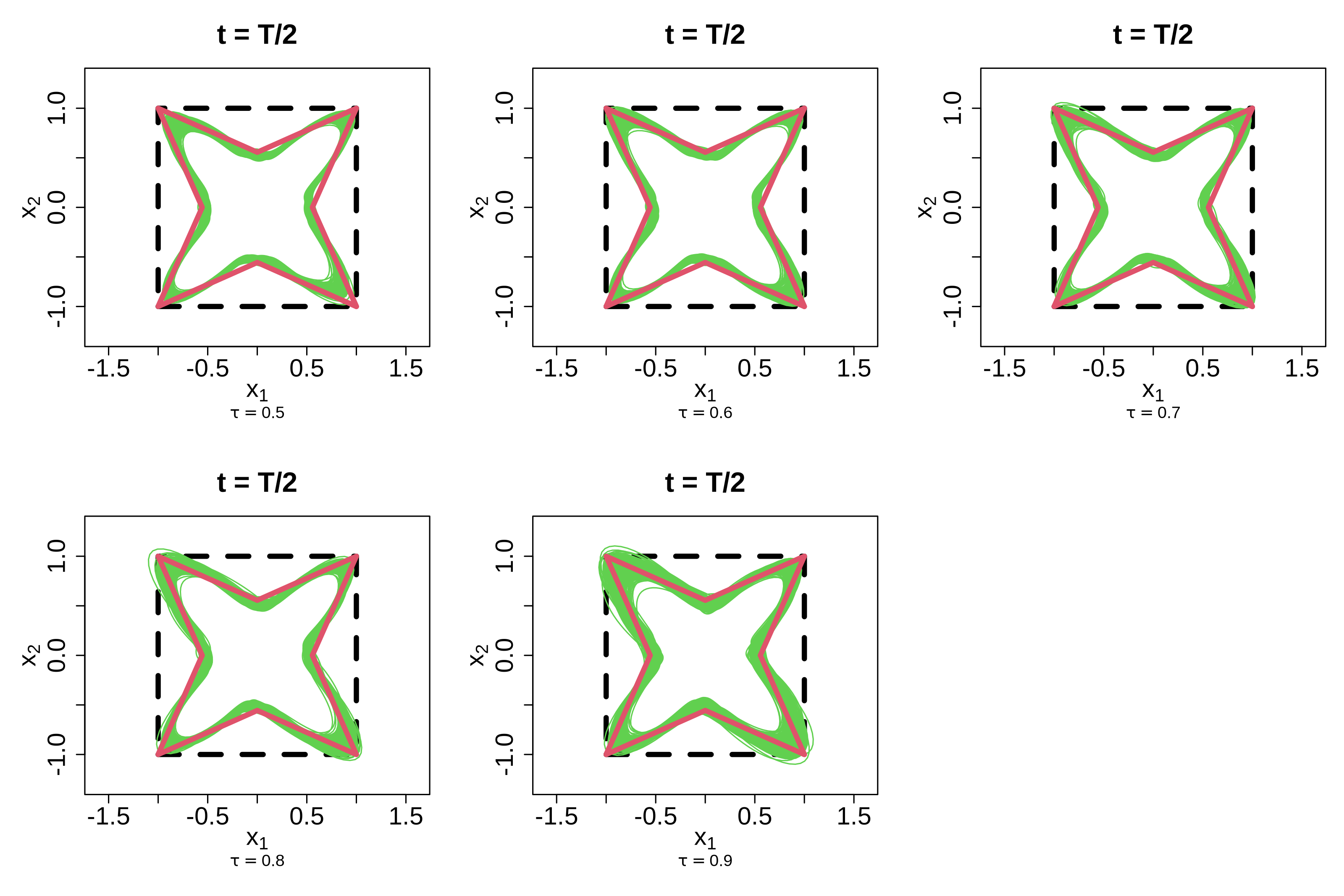}
    \caption{Boundary set estimates as $t = T/2$ across $\tau \in \{0.5,0.6,0.7,0.8,0.9\}$ for the fourth copula example.}
    \label{fig:res_tau_t2_c4}
\end{figure}

\begin{figure}[H]
    \centering
    \includegraphics[width=.8\linewidth]{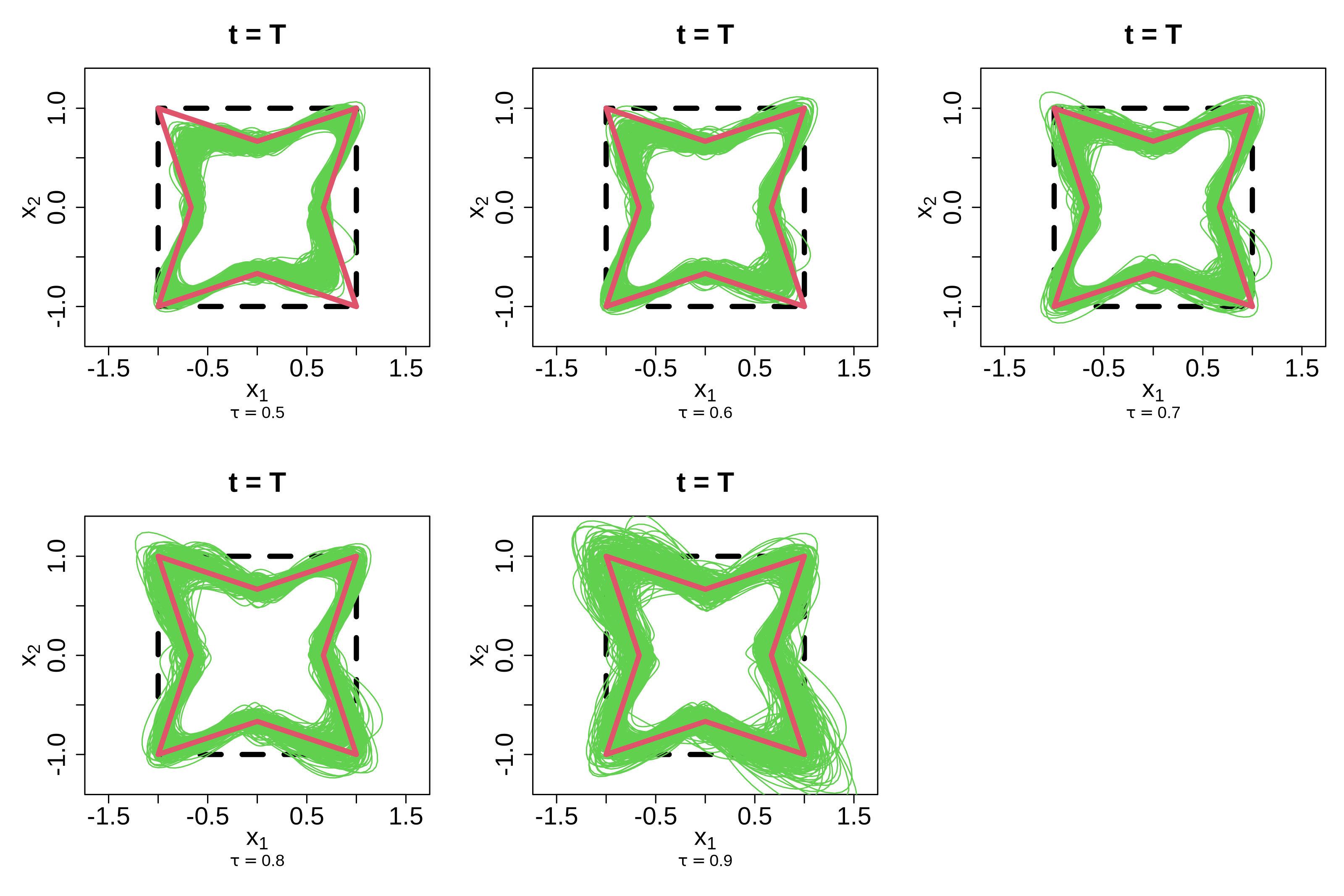}
    \caption{Boundary set estimates at $t = T$ across $\tau \in \{0.5,0.6,0.7,0.8,0.9\}$ for the fourth copula example.}
    \label{fig:res_tau_t3_c4}
\end{figure}

\begin{figure}[H]
    \centering
    \includegraphics[width=.8\linewidth]{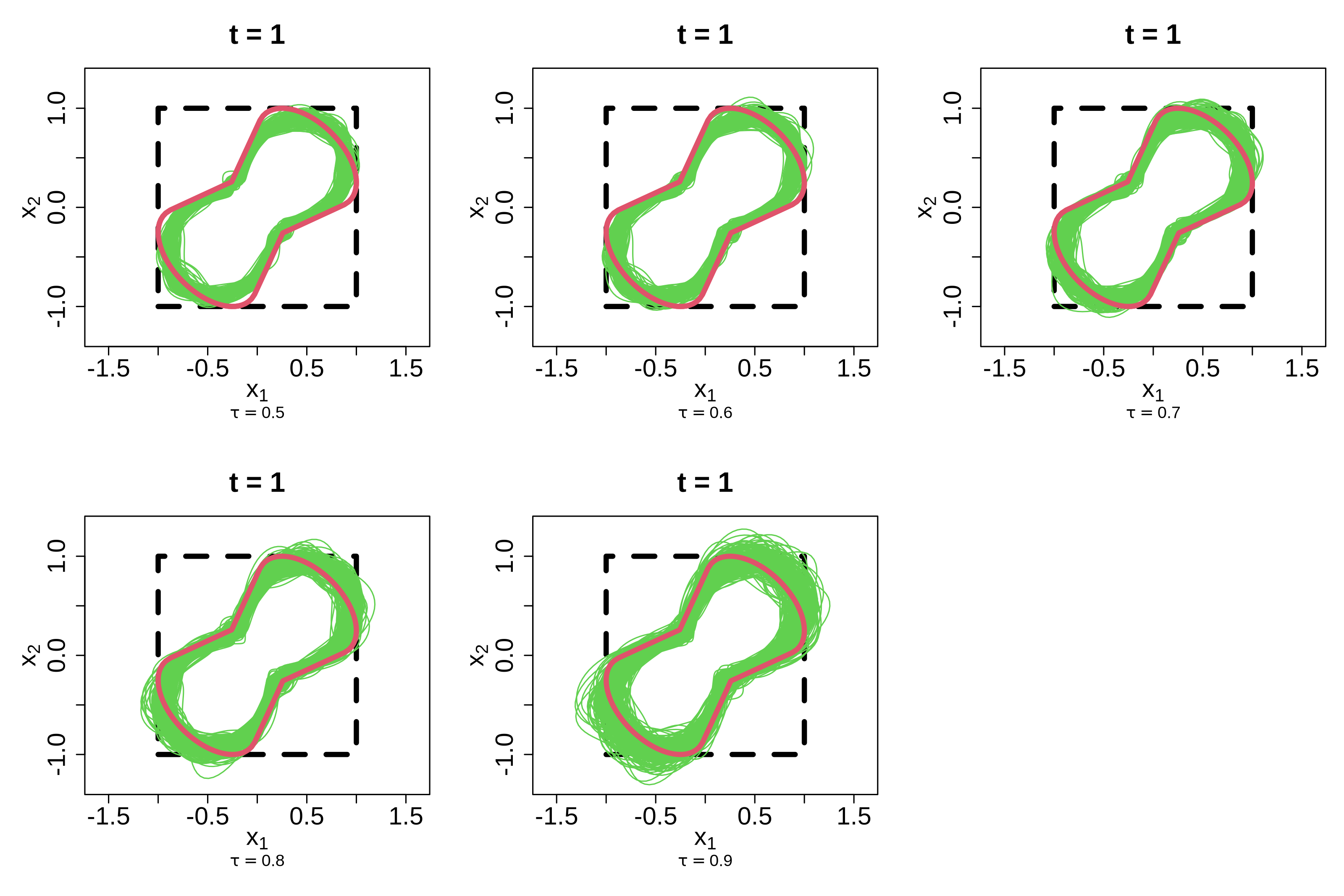}
    \caption{Boundary set estimates as $t = 1$ across $\tau \in \{0.5,0.6,0.7,0.8,0.9\}$ for the fifth copula example.}
    \label{fig:res_tau_t1_c5}
\end{figure}

\begin{figure}[H]
    \centering
    \includegraphics[width=.8\linewidth]{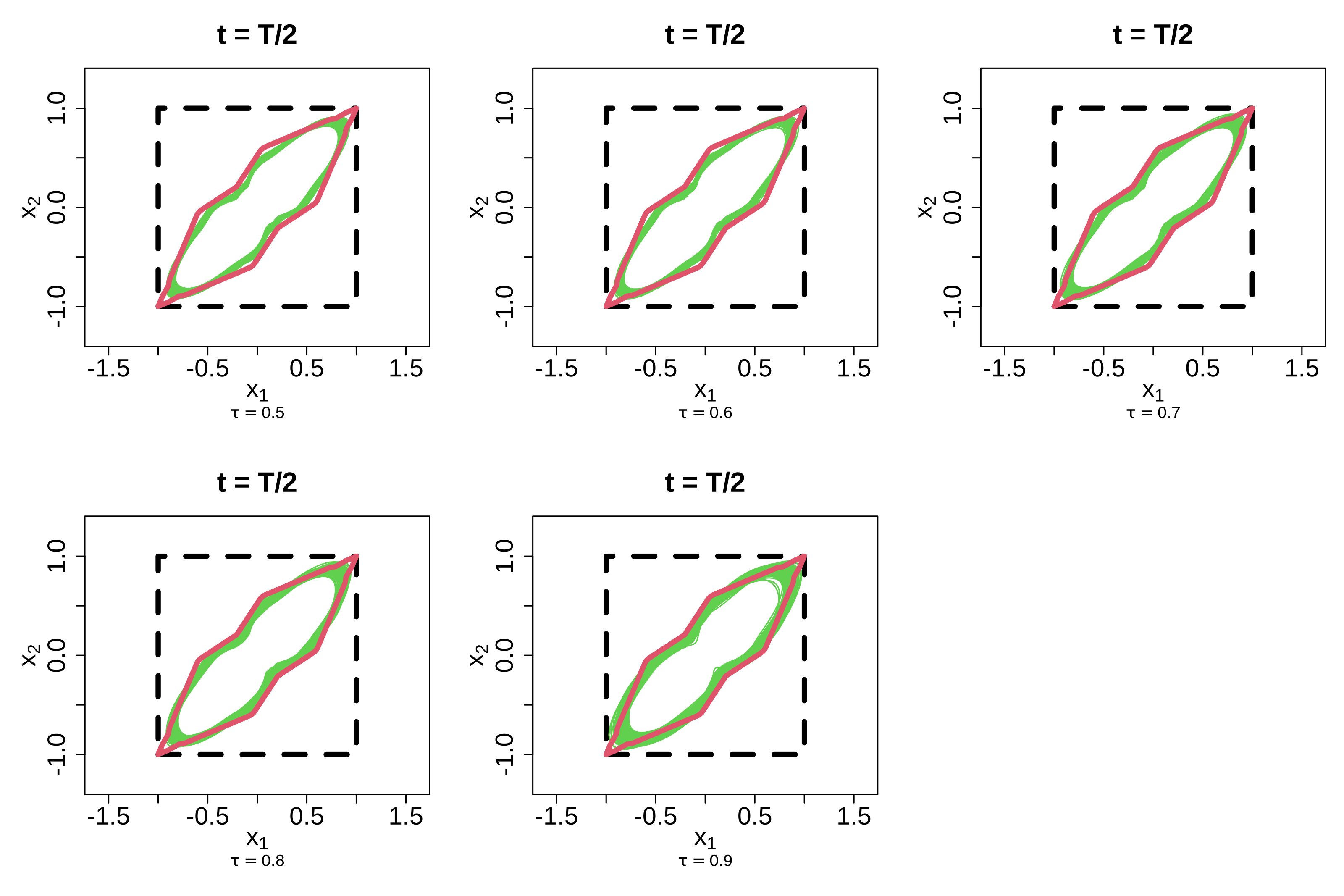}
    \caption{Boundary set estimates as $t = T/2$ across $\tau \in \{0.5,0.6,0.7,0.8,0.9\}$ for the fifth copula example.}
    \label{fig:res_tau_t2_c5}
\end{figure}

\begin{figure}[H]
    \centering
    \includegraphics[width=.8\linewidth]{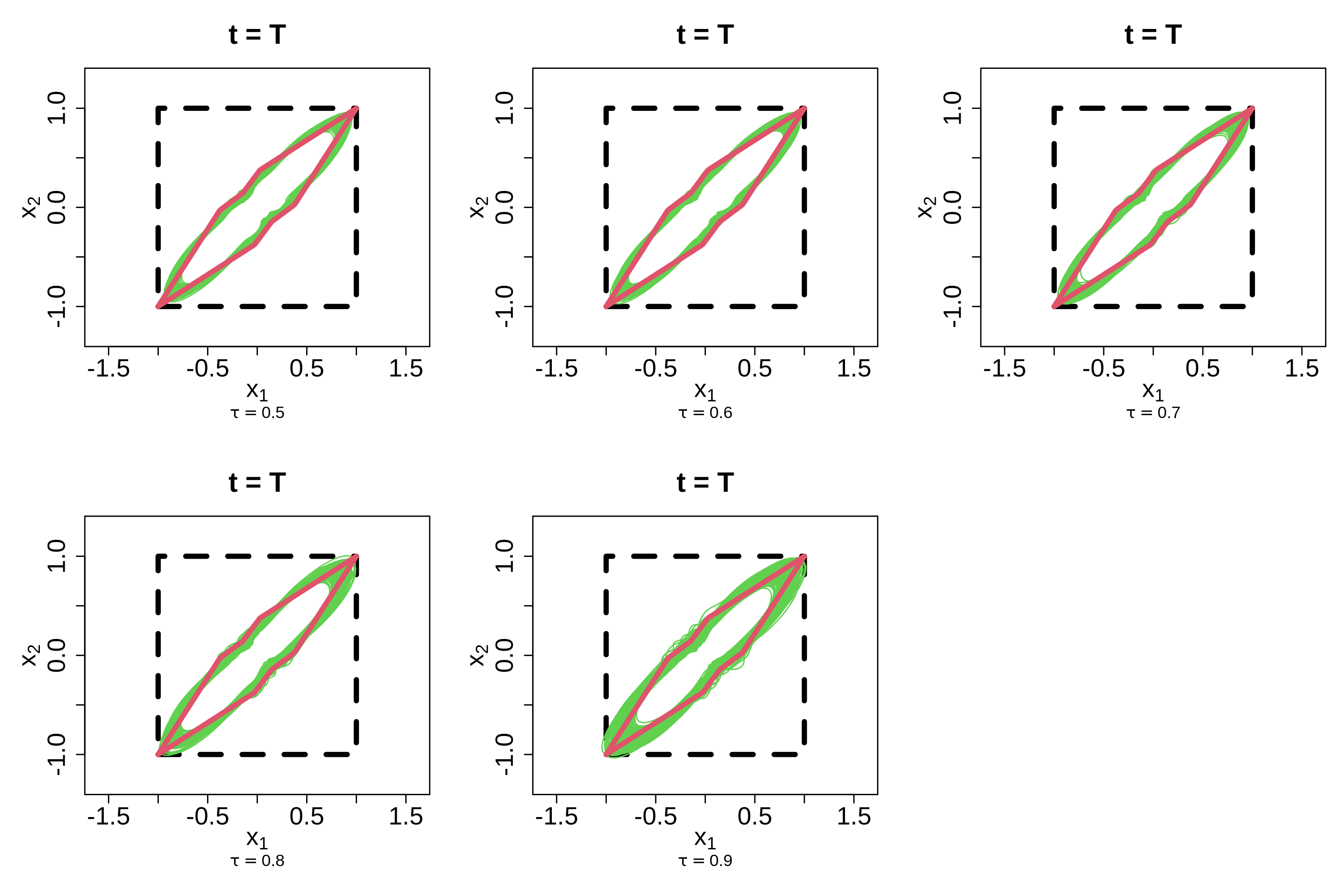}
    \caption{Boundary set estimates at $t = T$ across $\tau \in \{0.5,0.6,0.7,0.8,0.9\}$ for the fifth copula example.}
    \label{fig:res_tau_t3_c5}
\end{figure}

\begin{figure}[H]
    \centering
    \includegraphics[width=.8\linewidth]{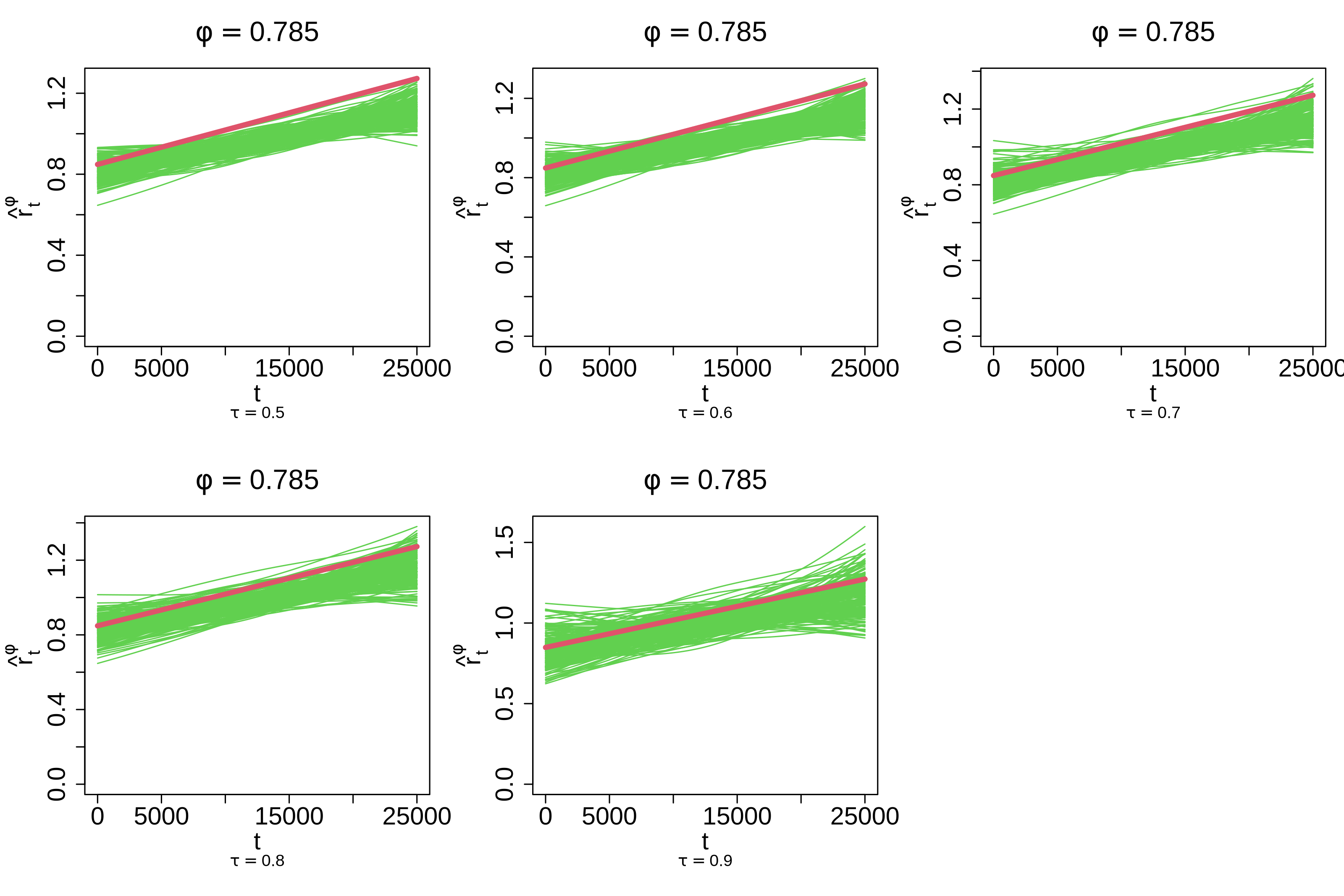}
    \caption{Boundary set radii estimates at $\phi = \pi/4$ across $\tau \in \{0.5,0.6,0.7,0.8,0.9\}$ for the first copula example.}
    \label{fig:res_tau_p1_c1}
\end{figure}

\begin{figure}[H]
    \centering
    \includegraphics[width=.8\linewidth]{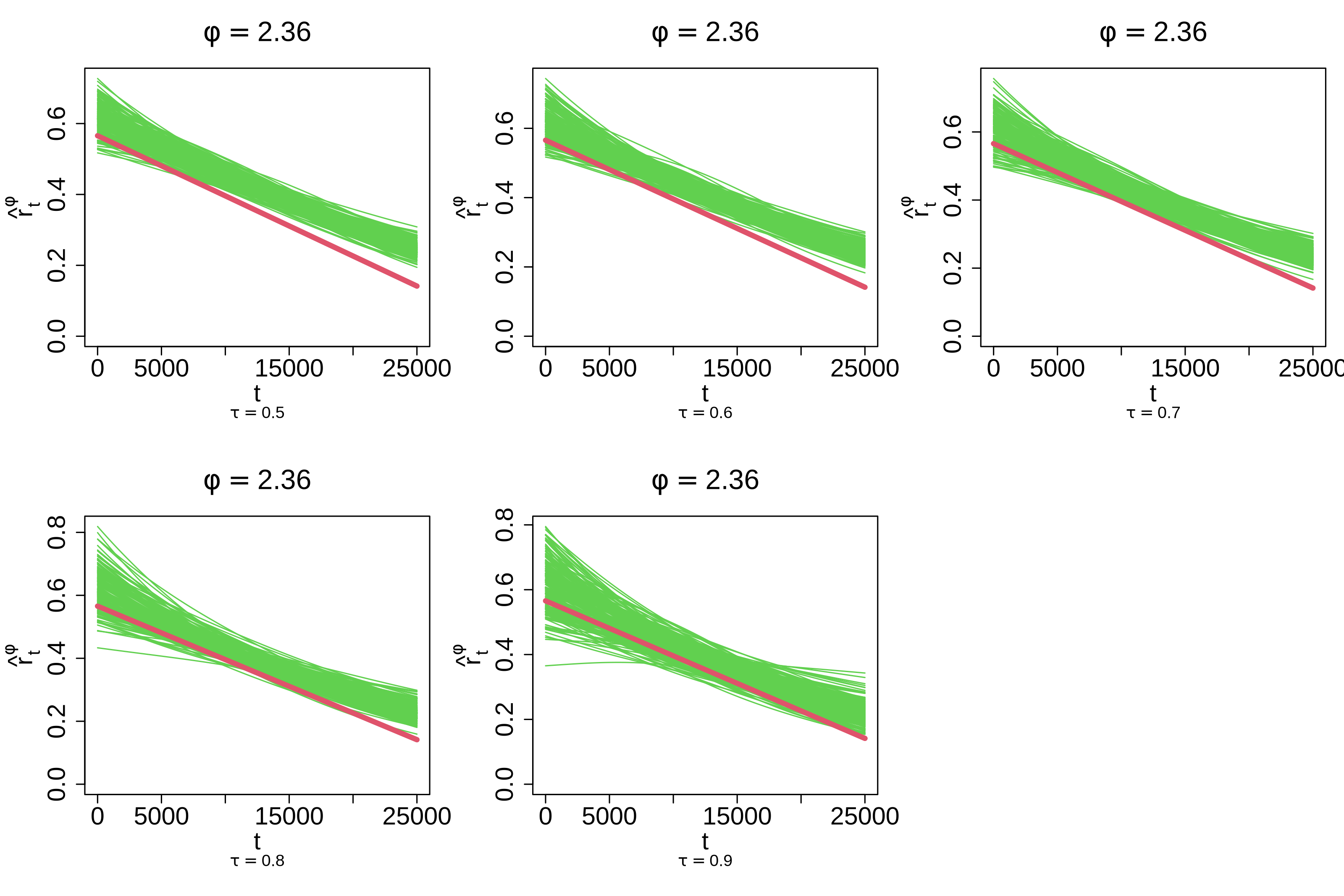}
    \caption{Boundary set radii estimates at $\phi = 3\pi/4$ across $\tau \in \{0.5,0.6,0.7,0.8,0.9\}$ for the first copula example.}
    \label{fig:res_tau_p2_c1}
\end{figure}

\begin{figure}[H]
    \centering
    \includegraphics[width=.8\linewidth]{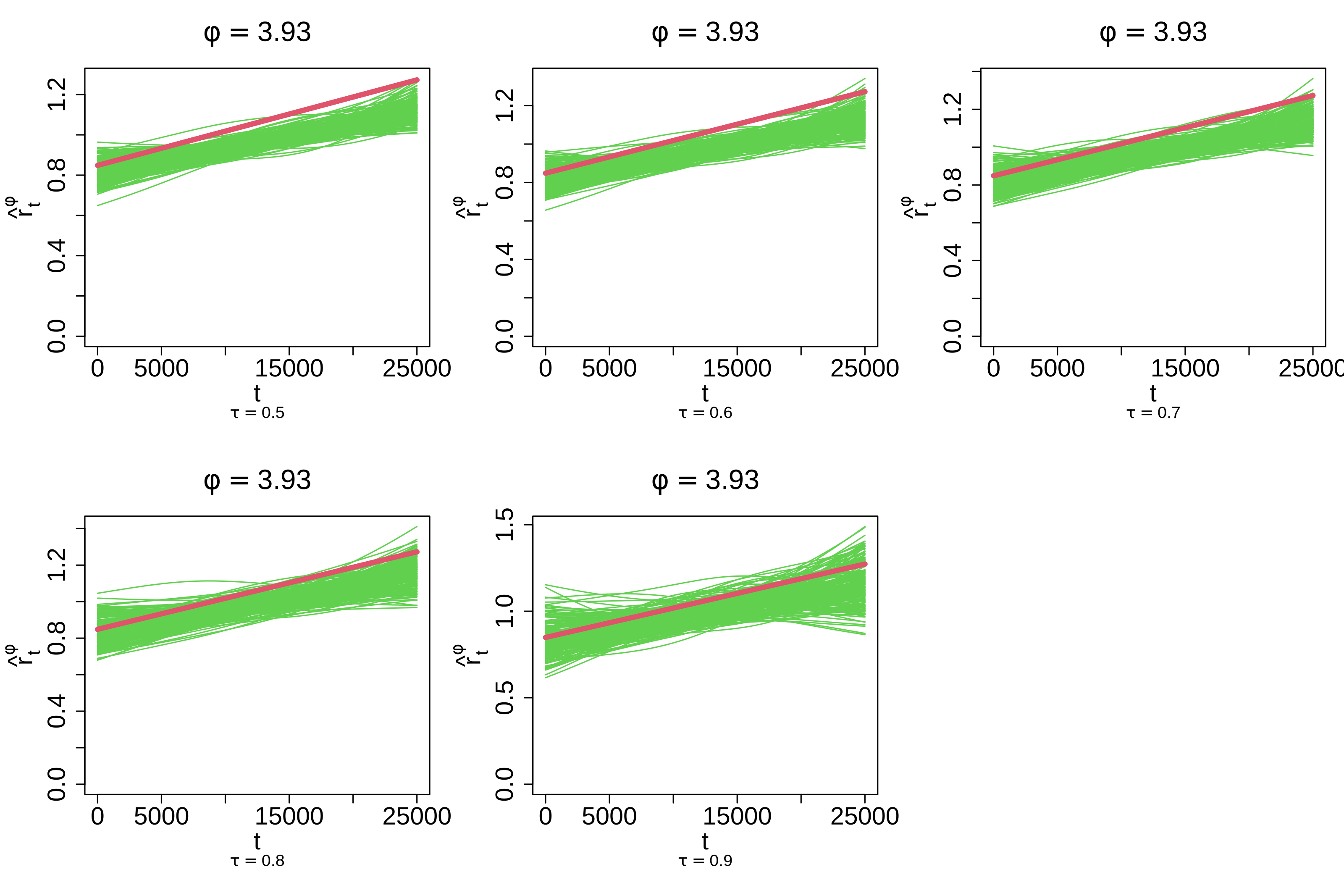}
    \caption{Boundary set radii estimates at $\phi = 5\pi/4$ across $\tau \in \{0.5,0.6,0.7,0.8,0.9\}$ for the first copula example.}
    \label{fig:res_tau_p3_c1}
\end{figure}

\begin{figure}[H]
    \centering
    \includegraphics[width=.8\linewidth]{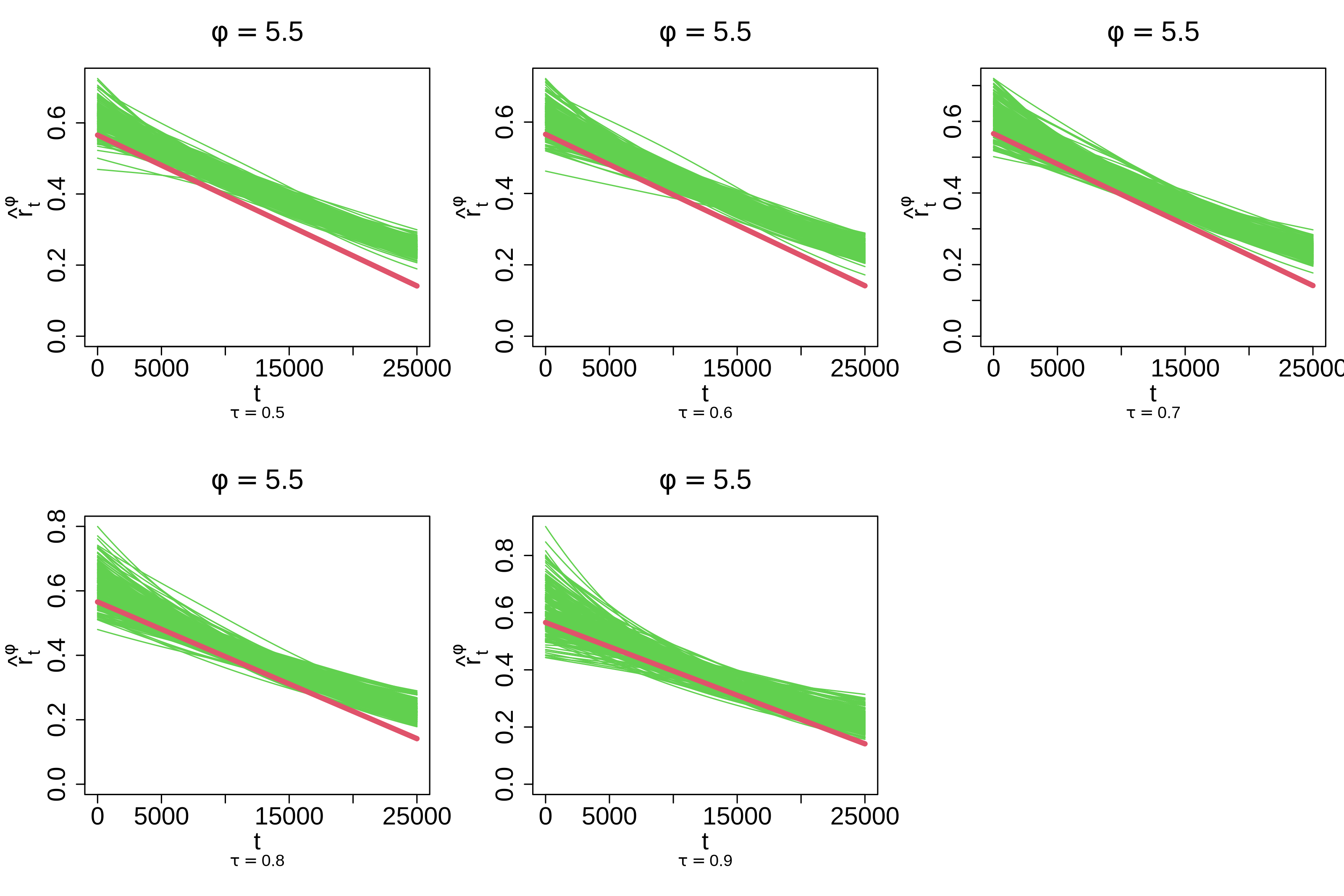}
    \caption{Boundary set radii estimates at $\phi = 7\pi/4$ across $\tau \in \{0.5,0.6,0.7,0.8,0.9\}$ for the first copula example.}
    \label{fig:res_tau_p4_c1}
\end{figure}

\begin{figure}[H]
    \centering
    \includegraphics[width=.8\linewidth]{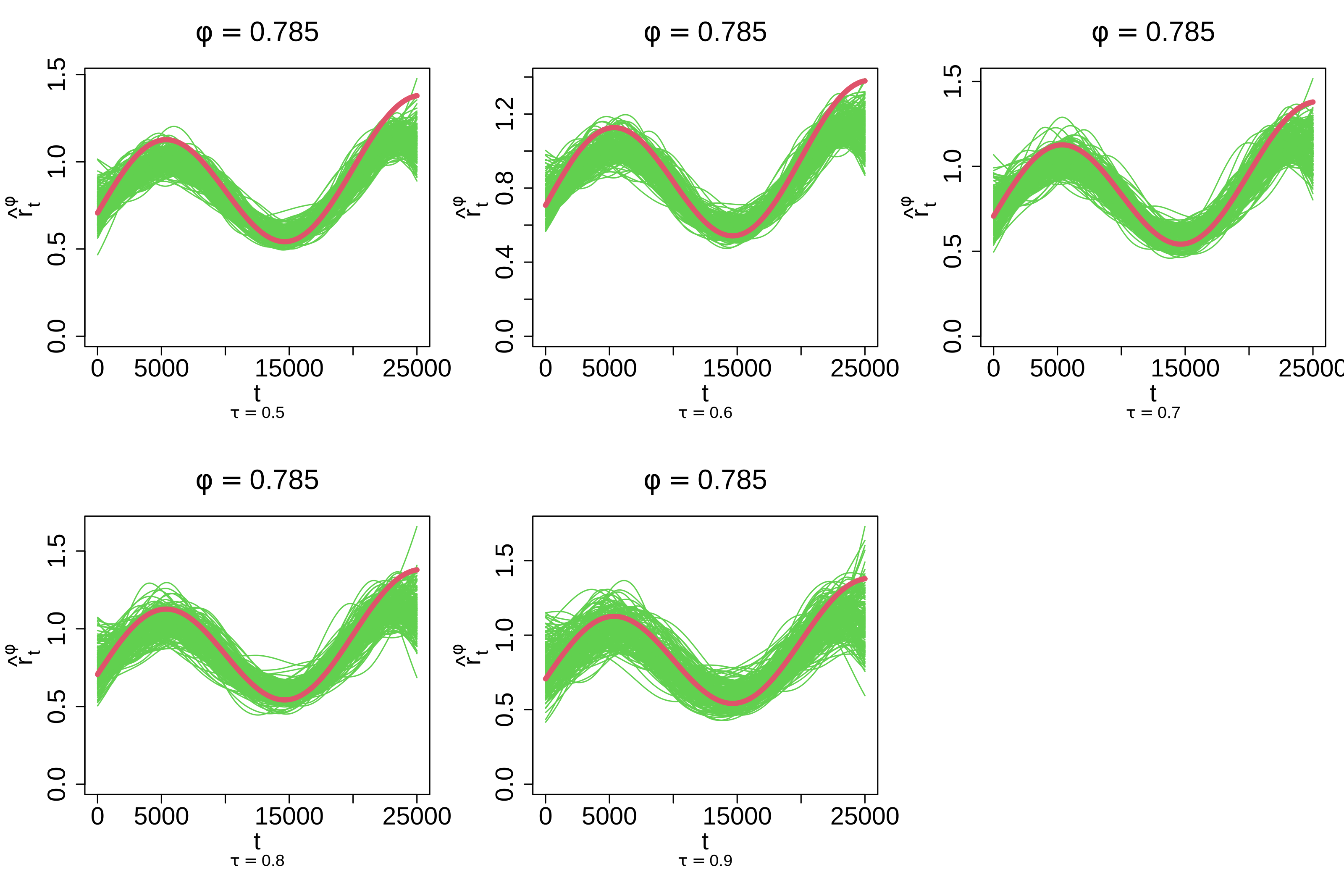}
    \caption{Boundary set radii estimates at $\phi = \pi/4$ across $\tau \in \{0.5,0.6,0.7,0.8,0.9\}$ for the second copula example.}
    \label{fig:res_tau_p1_c2}
\end{figure}

\begin{figure}[H]
    \centering
    \includegraphics[width=.8\linewidth]{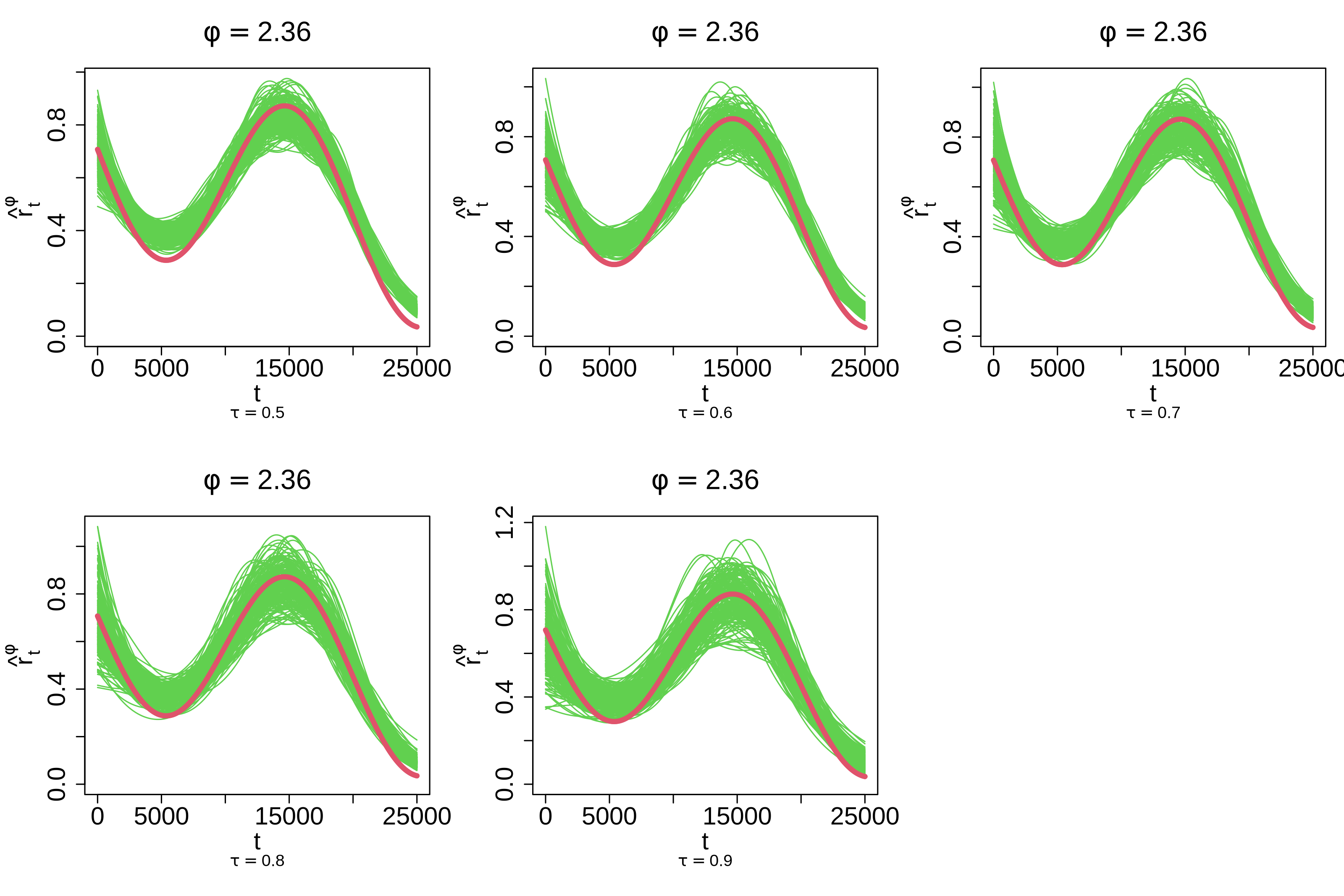}
    \caption{Boundary set radii estimates at $\phi = 3\pi/4$ across $\tau \in \{0.5,0.6,0.7,0.8,0.9\}$ for the second copula example.}
    \label{fig:res_tau_p2_c2}
\end{figure}

\begin{figure}[H]
    \centering
    \includegraphics[width=.8\linewidth]{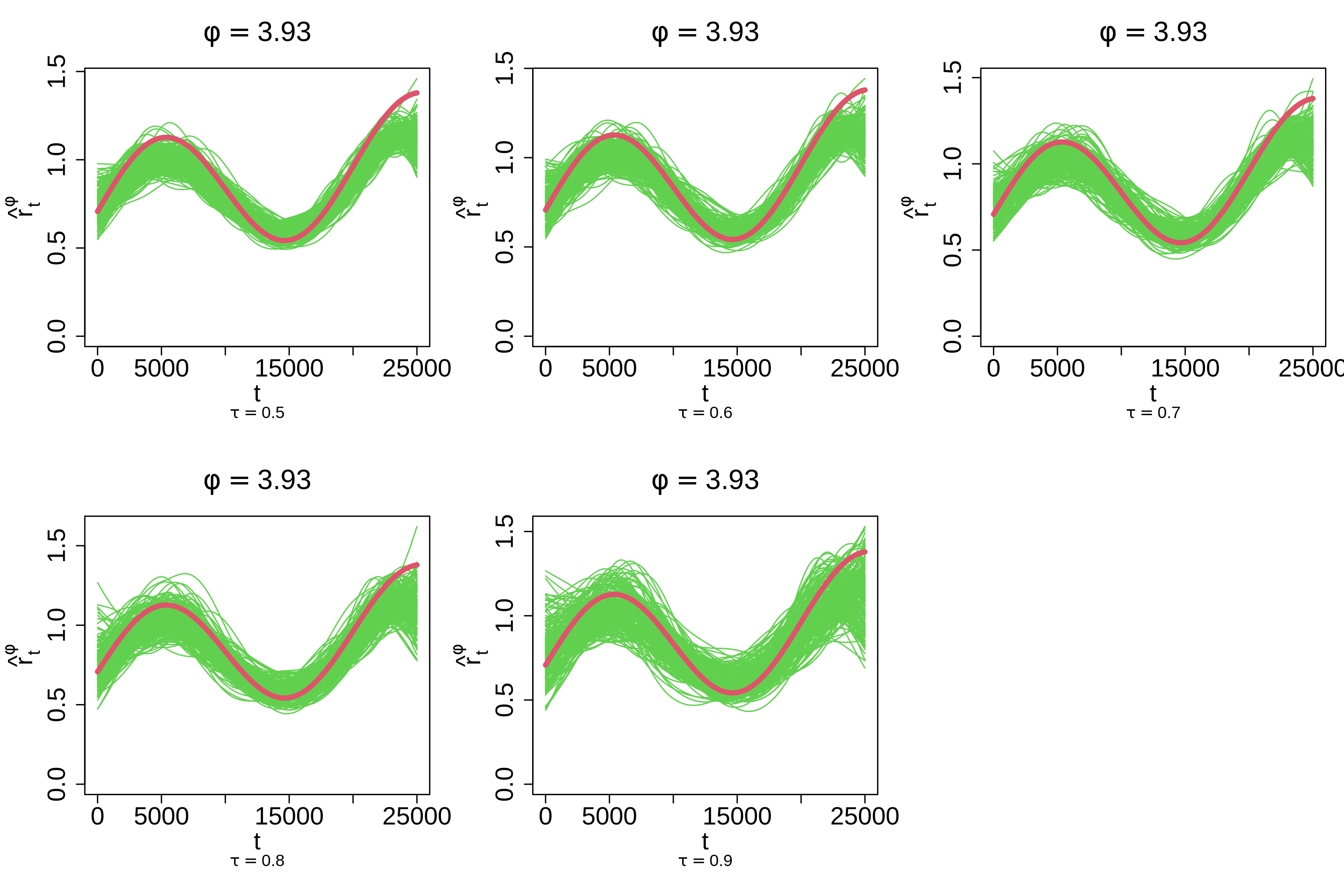}
    \caption{Boundary set radii estimates at $\phi = 5\pi/4$ across $\tau \in \{0.5,0.6,0.7,0.8,0.9\}$ for the second copula example.}
    \label{fig:res_tau_p3_c2}
\end{figure}

\begin{figure}[H]
    \centering
    \includegraphics[width=.8\linewidth]{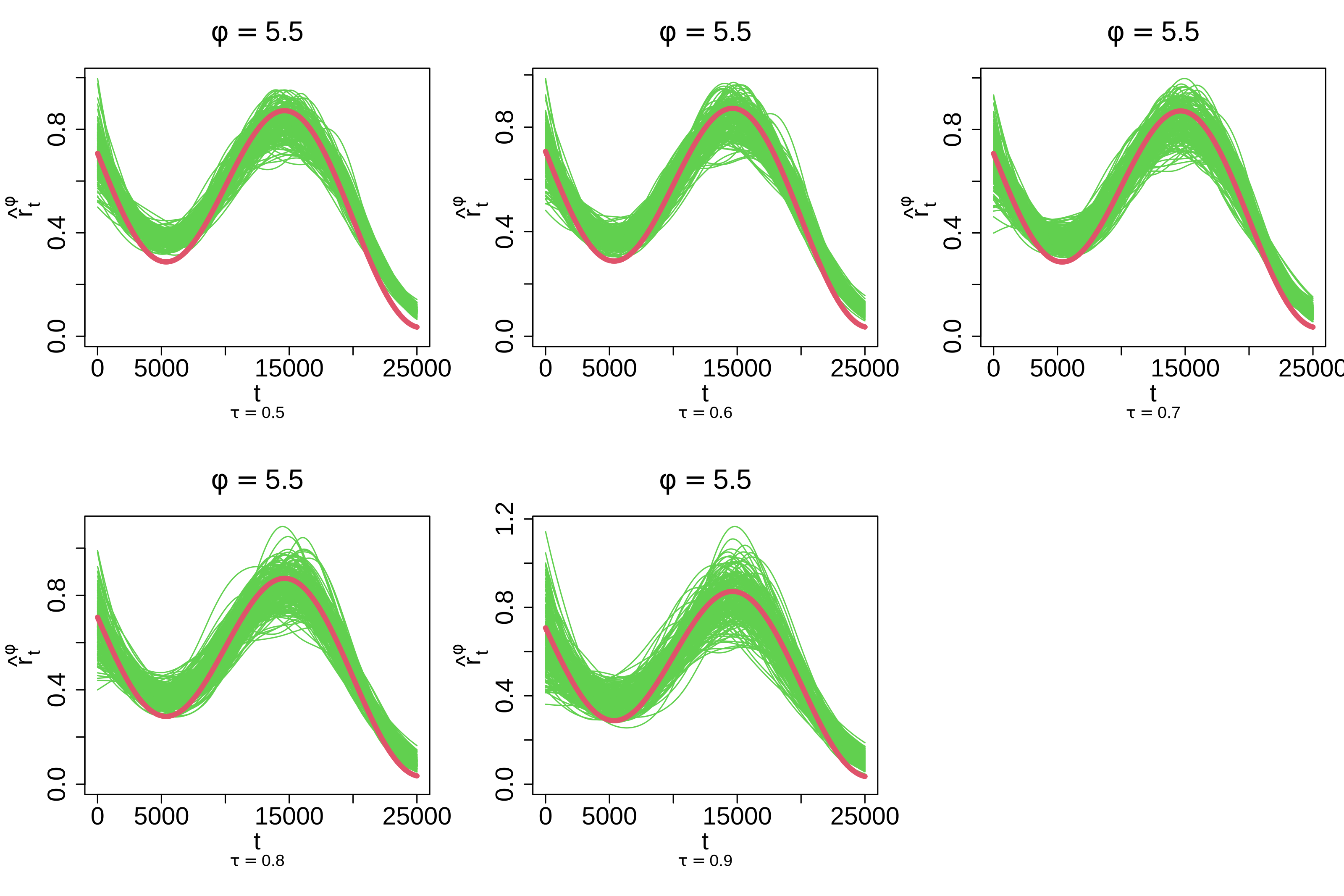}
    \caption{Boundary set radii estimates at $\phi = 7\pi/4$ across $\tau \in \{0.5,0.6,0.7,0.8,0.9\}$ for the second copula example.}
    \label{fig:res_tau_p4_c2}
\end{figure}

\begin{figure}[H]
    \centering
    \includegraphics[width=.8\linewidth]{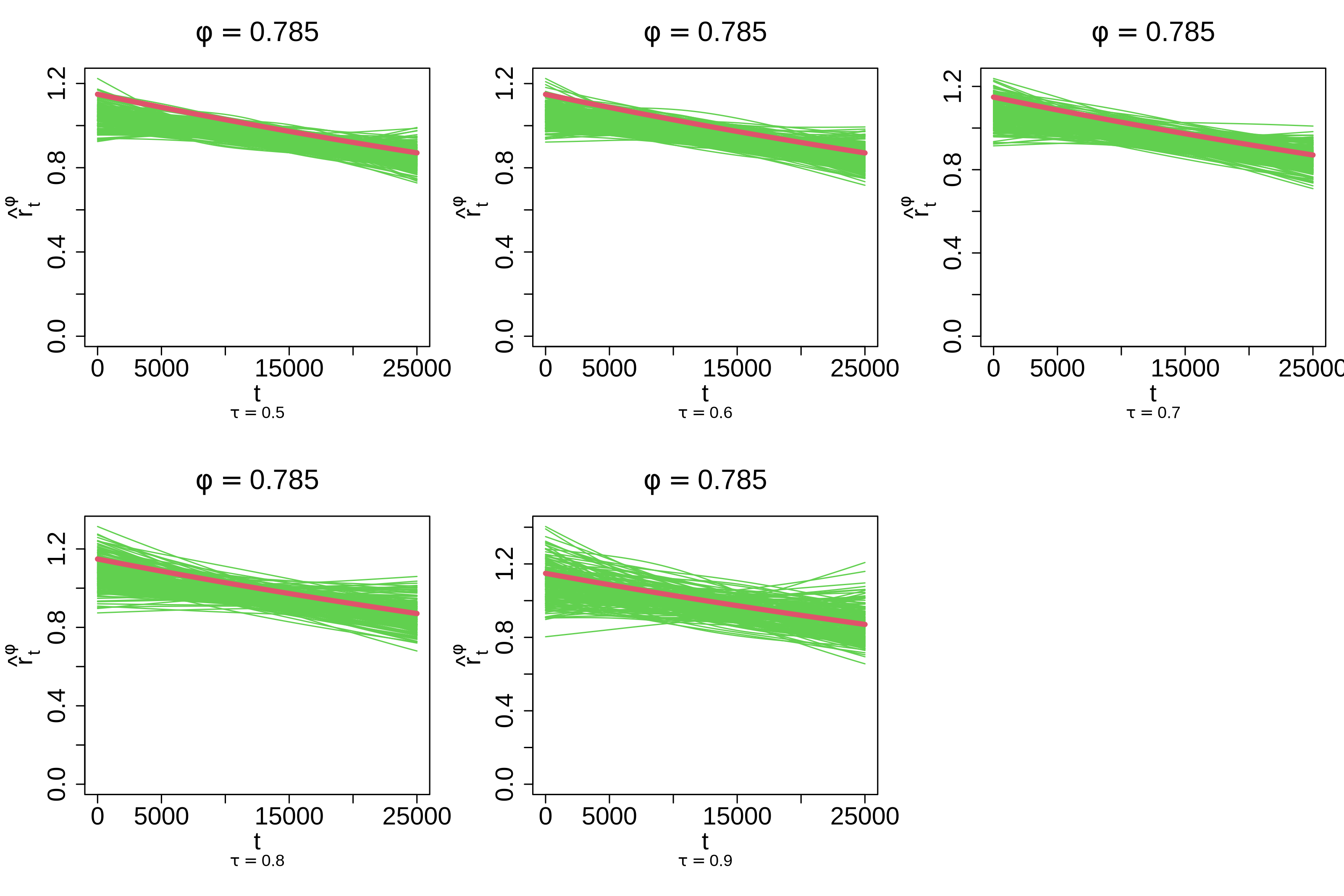}
    \caption{Boundary set radii estimates at $\phi = \pi/4$ across $\tau \in \{0.5,0.6,0.7,0.8,0.9\}$ for the third copula example.}
    \label{fig:res_tau_p1_c3}
\end{figure}

\begin{figure}[H]
    \centering
    \includegraphics[width=.8\linewidth]{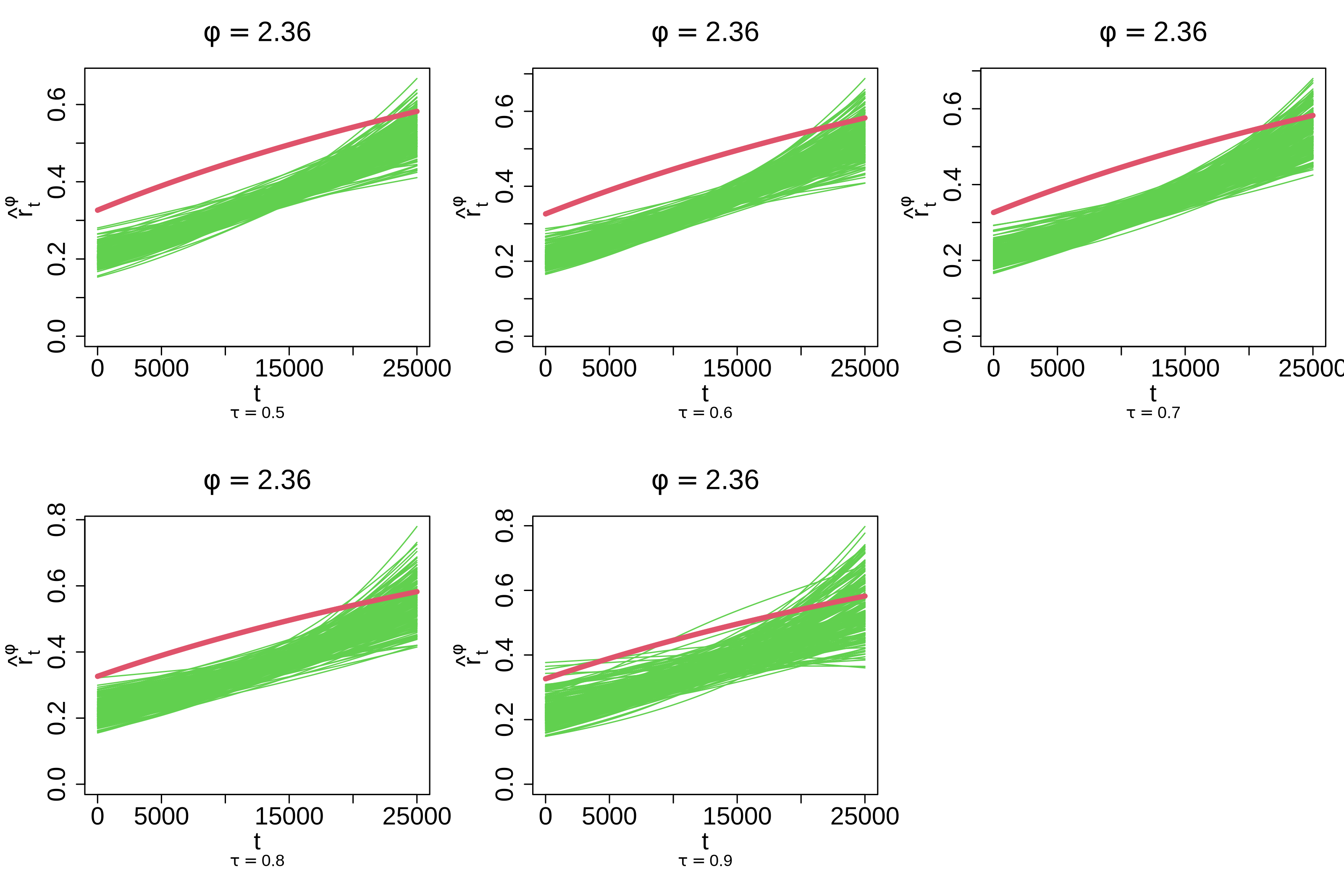}
    \caption{Boundary set radii estimates at $\phi = 3\pi/4$ across $\tau \in \{0.5,0.6,0.7,0.8,0.9\}$ for the third copula example.}
    \label{fig:res_tau_p2_c3}
\end{figure}

\begin{figure}[H]
    \centering
    \includegraphics[width=.8\linewidth]{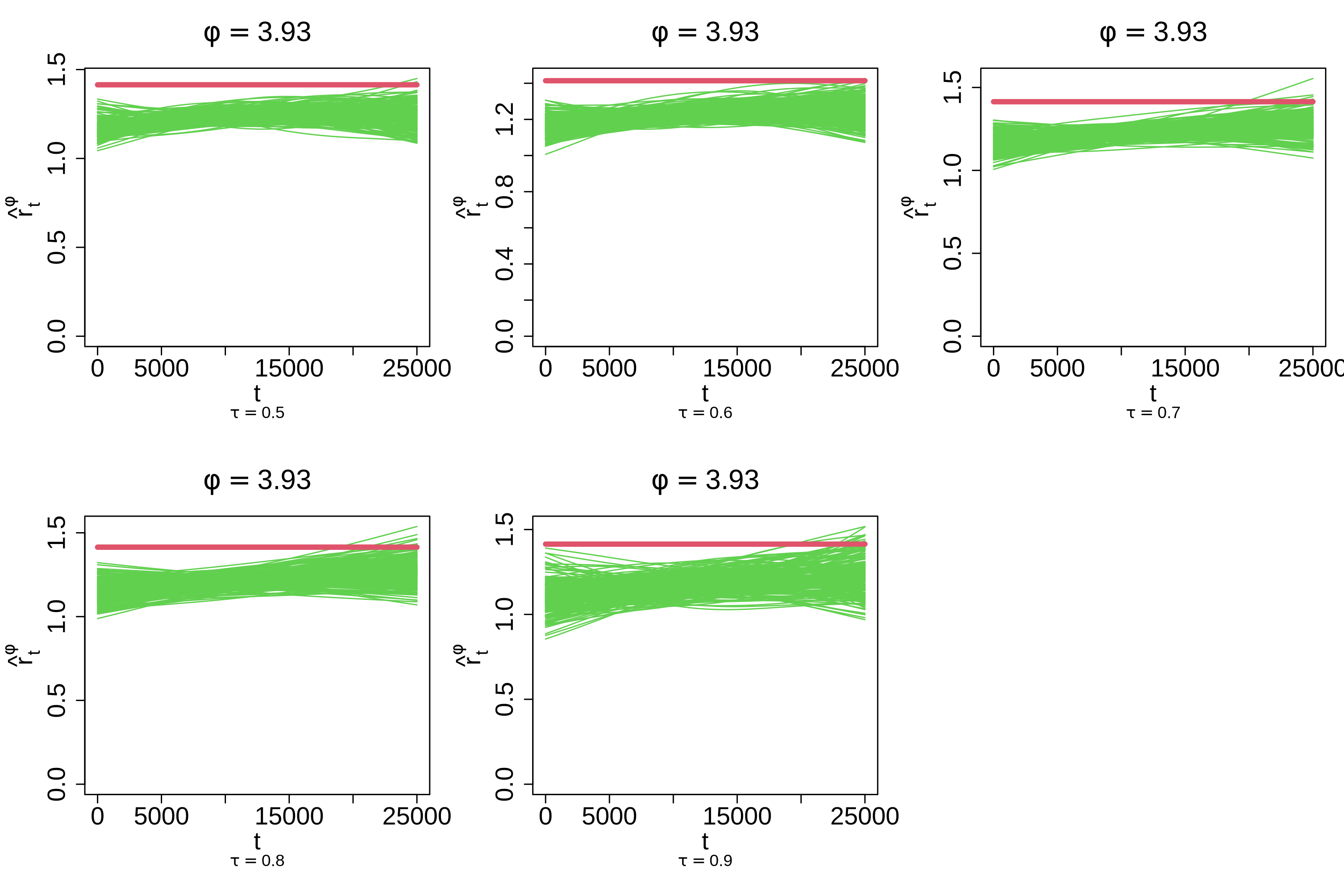}
    \caption{Boundary set radii estimates at $\phi = 5\pi/4$ across $\tau \in \{0.5,0.6,0.7,0.8,0.9\}$ for the third copula example.}
    \label{fig:res_tau_p3_c3}
\end{figure}

\begin{figure}[H]
    \centering
    \includegraphics[width=.8\linewidth]{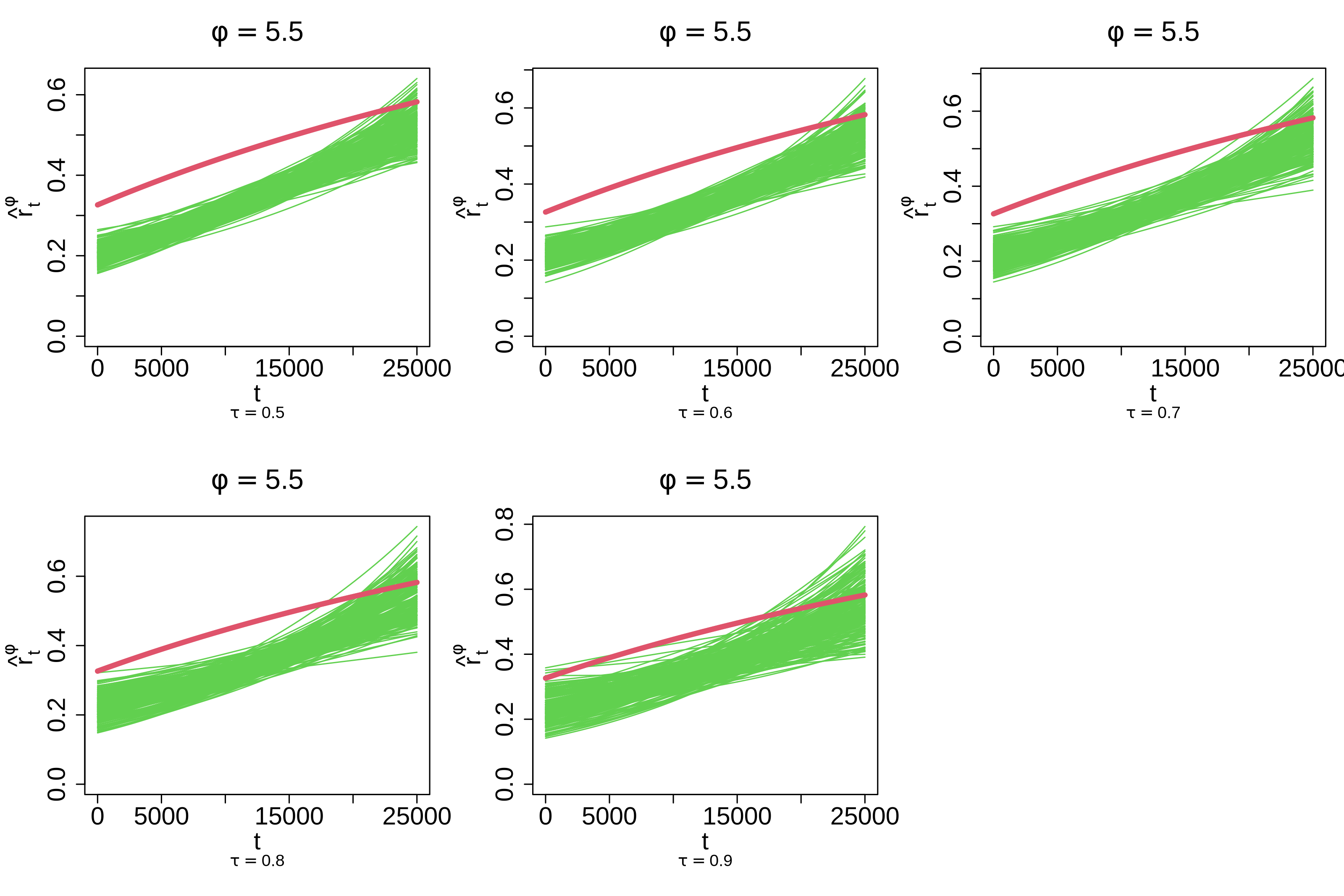}
    \caption{Boundary set radii estimates at $\phi = 7\pi/4$ across $\tau \in \{0.5,0.6,0.7,0.8,0.9\}$ for the third copula example.}
    \label{fig:res_tau_p4_c3}
\end{figure}

\begin{figure}[H]
    \centering
    \includegraphics[width=.8\linewidth]{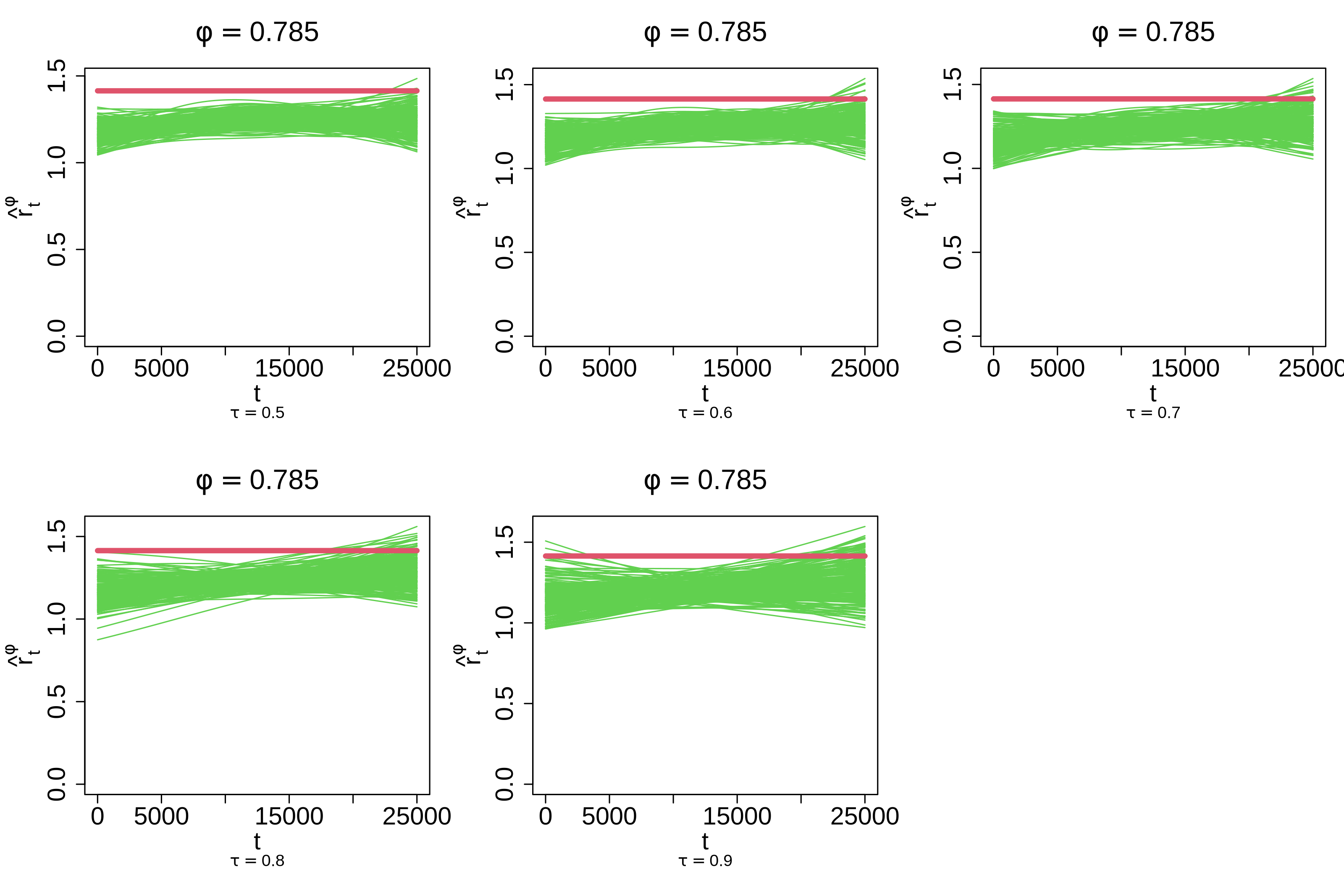}
    \caption{Boundary set radii estimates at $\phi = \pi/4$ across $\tau \in \{0.5,0.6,0.7,0.8,0.9\}$ for the fourth copula example.}
    \label{fig:res_tau_p1_c4}
\end{figure}

\begin{figure}[H]
    \centering
    \includegraphics[width=.8\linewidth]{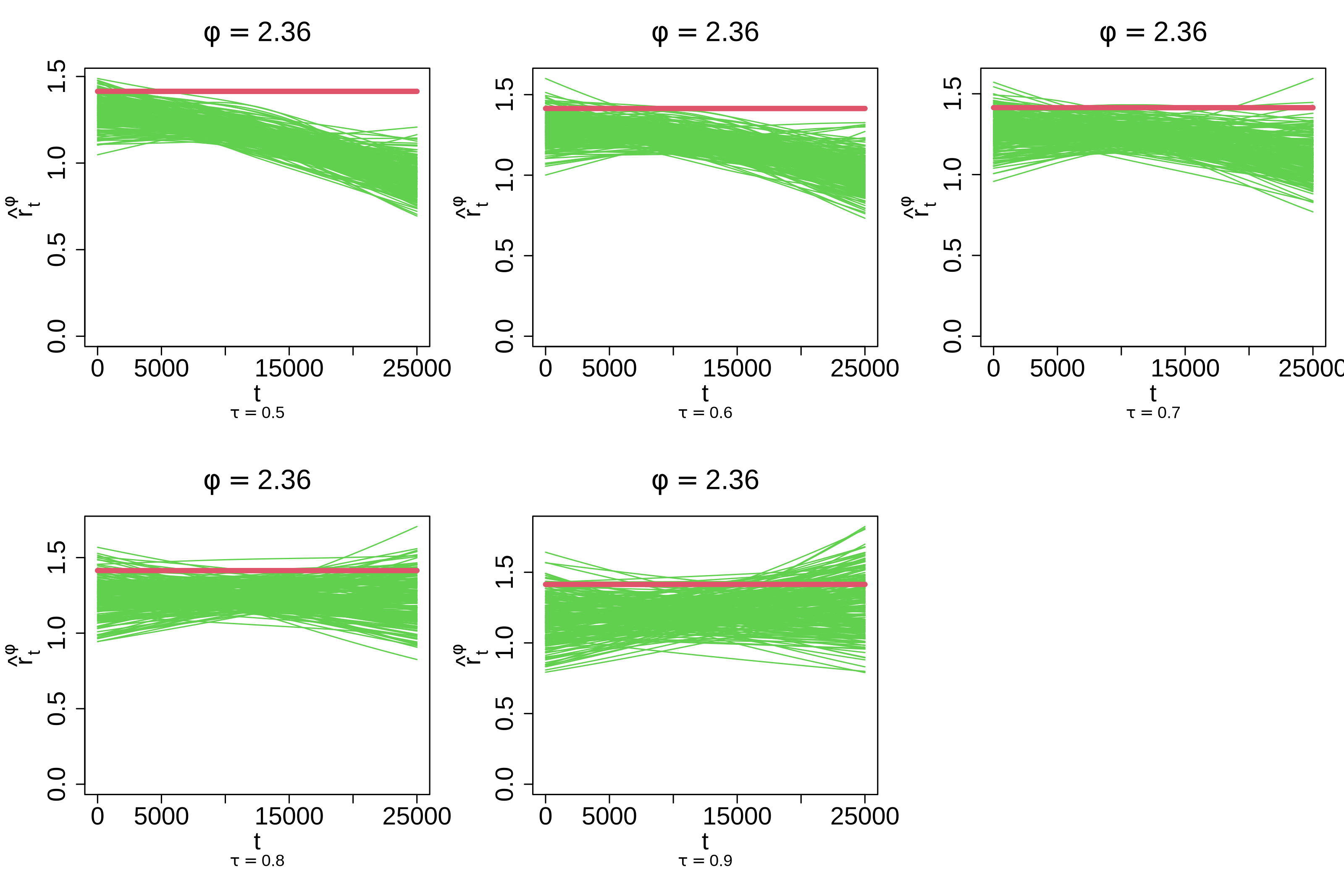}
    \caption{Boundary set radii estimates at $\phi = 3\pi/4$ across $\tau \in \{0.5,0.6,0.7,0.8,0.9\}$ for the fourth copula example.}
    \label{fig:res_tau_p2_c4}
\end{figure}

\begin{figure}[H]
    \centering
    \includegraphics[width=.8\linewidth]{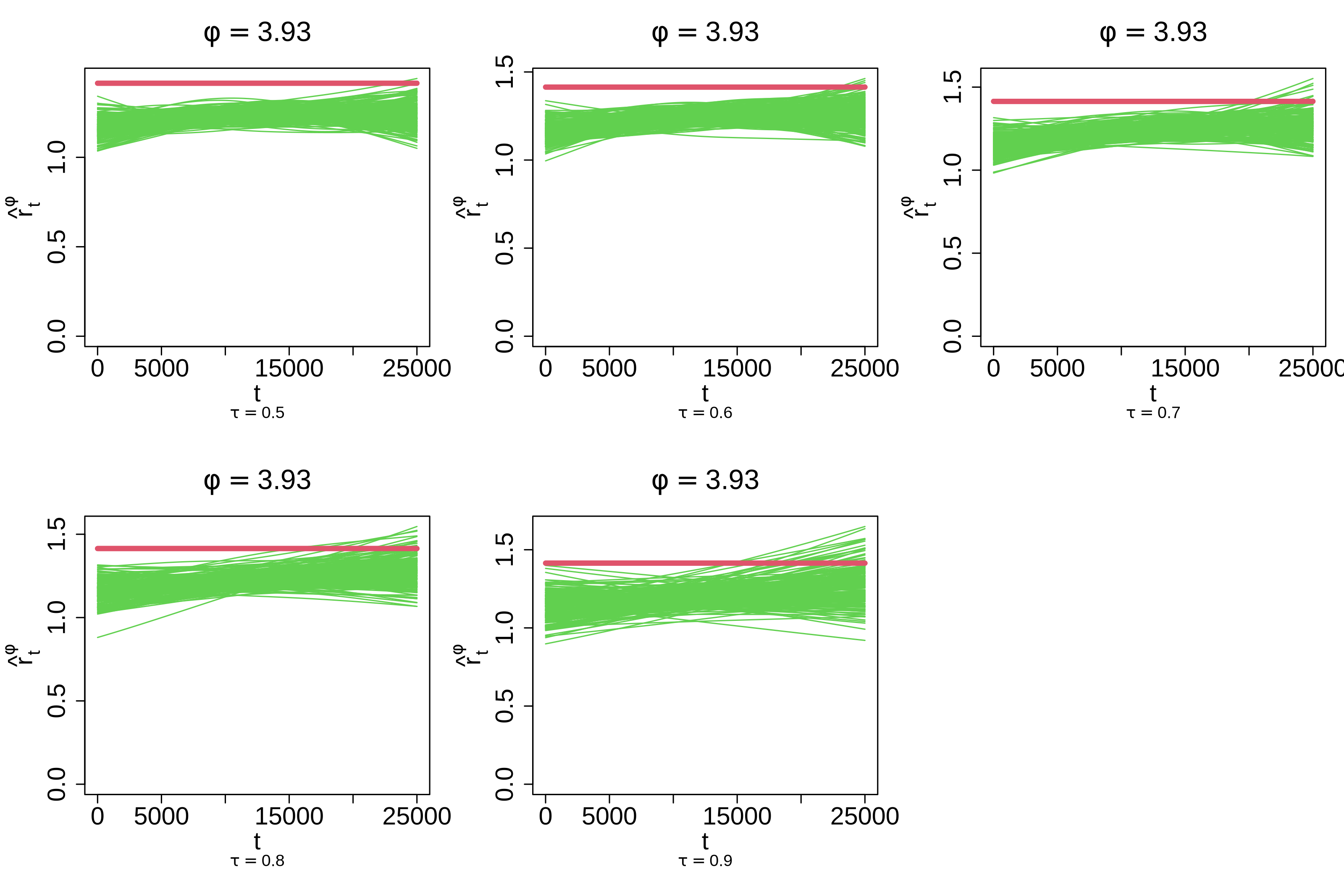}
    \caption{Boundary set radii estimates at $\phi = 5\pi/4$ across $\tau \in \{0.5,0.6,0.7,0.8,0.9\}$ for the fourth copula example.}
    \label{fig:res_tau_p3_c4}
\end{figure}

\begin{figure}[H]
    \centering
    \includegraphics[width=.8\linewidth]{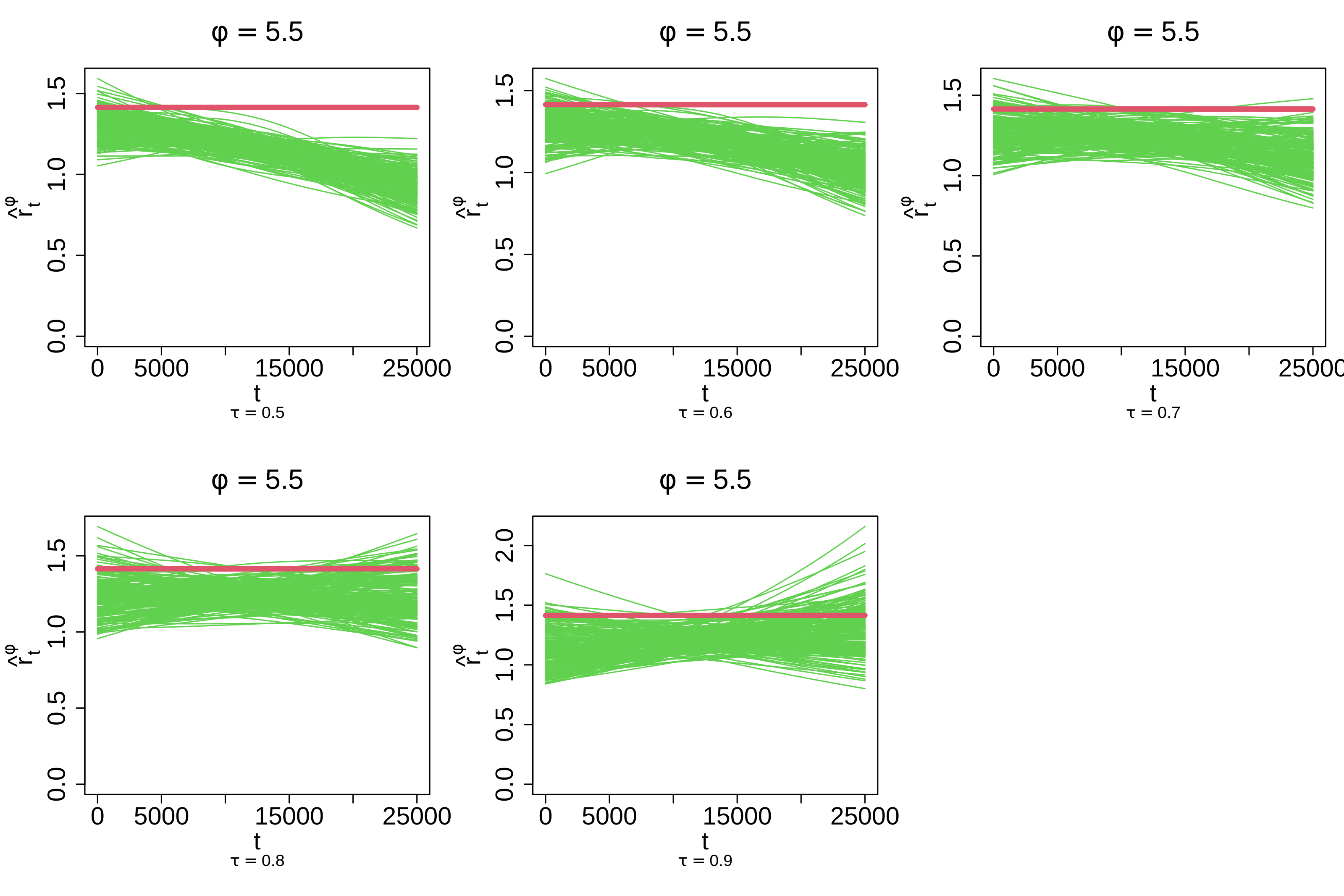}
    \caption{Boundary set radii estimates at $\phi = 7\pi/4$ across $\tau \in \{0.5,0.6,0.7,0.8,0.9\}$ for the fourth copula example.}
    \label{fig:res_tau_p4_c4}
\end{figure}

\begin{figure}[H]
    \centering
    \includegraphics[width=.8\linewidth]{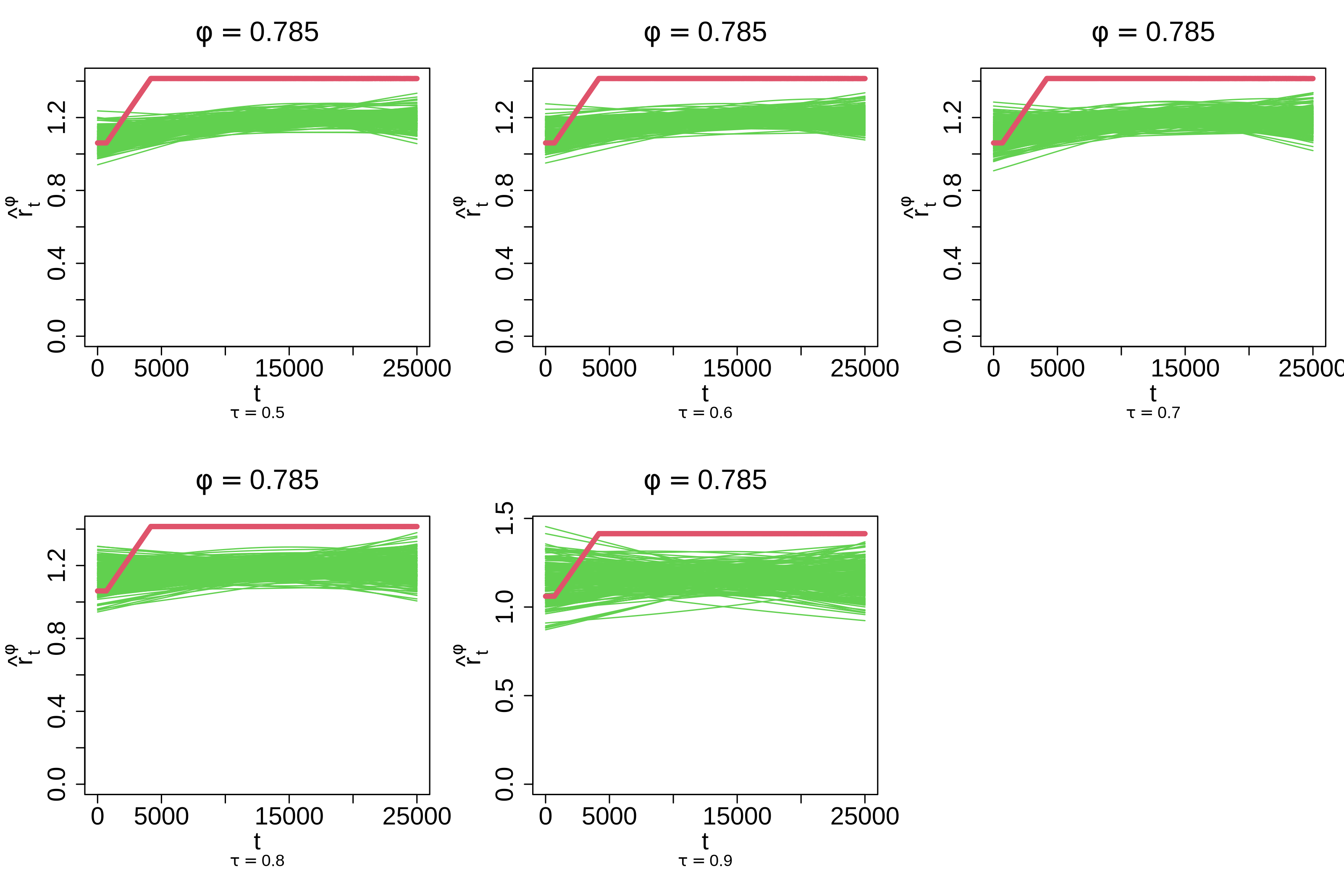}
    \caption{Boundary set radii estimates at $\phi = \pi/4$ across $\tau \in \{0.5,0.6,0.7,0.8,0.9\}$ for the fifth copula example.}
    \label{fig:res_tau_p1_c5}
\end{figure}

\begin{figure}[H]
    \centering
    \includegraphics[width=.8\linewidth]{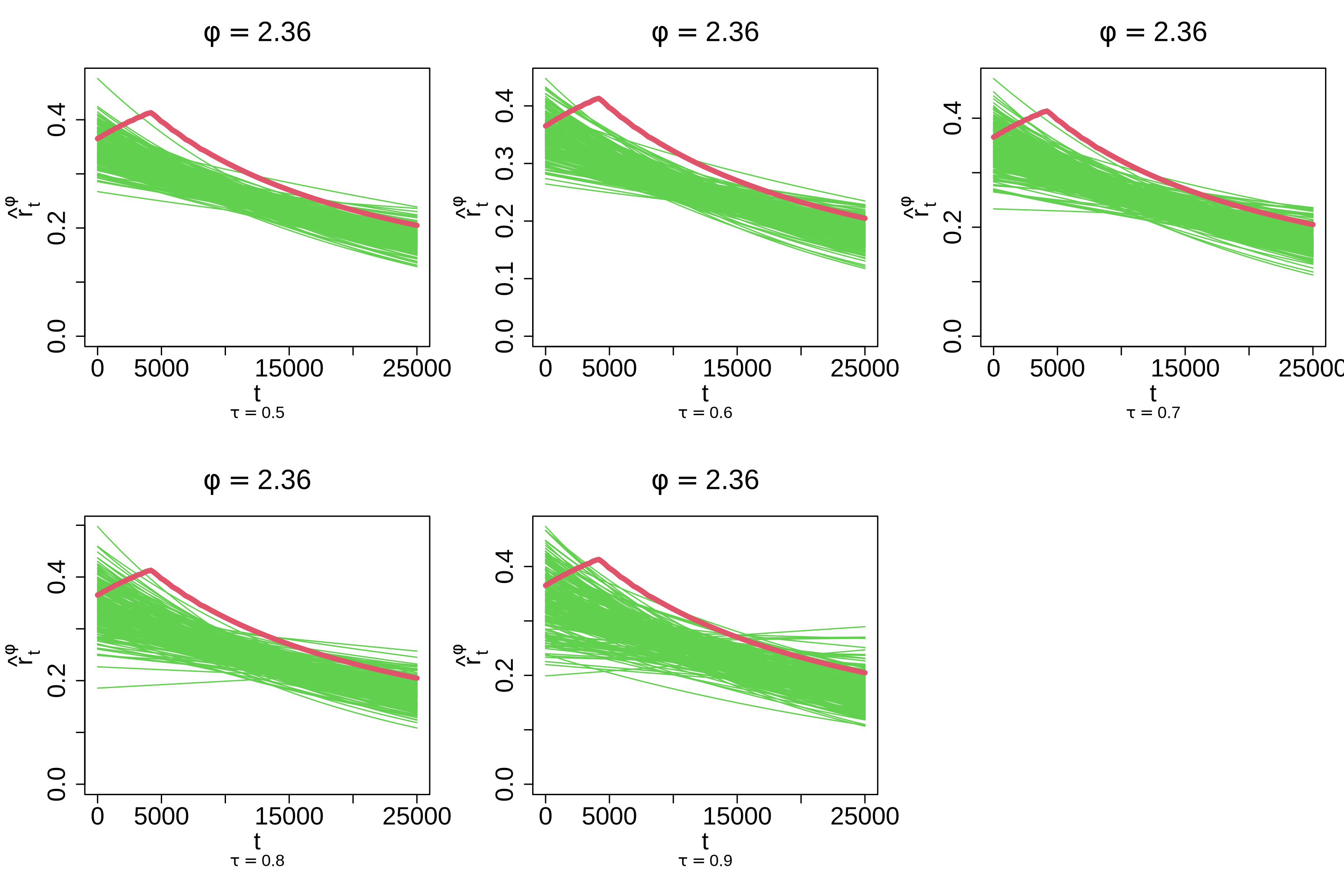}
    \caption{Boundary set radii estimates at $\phi = 3\pi/4$ across $\tau \in \{0.5,0.6,0.7,0.8,0.9\}$ for the fifth copula example.}
    \label{fig:res_tau_p2_c5}
\end{figure}

\begin{figure}[H]
    \centering
    \includegraphics[width=.8\linewidth]{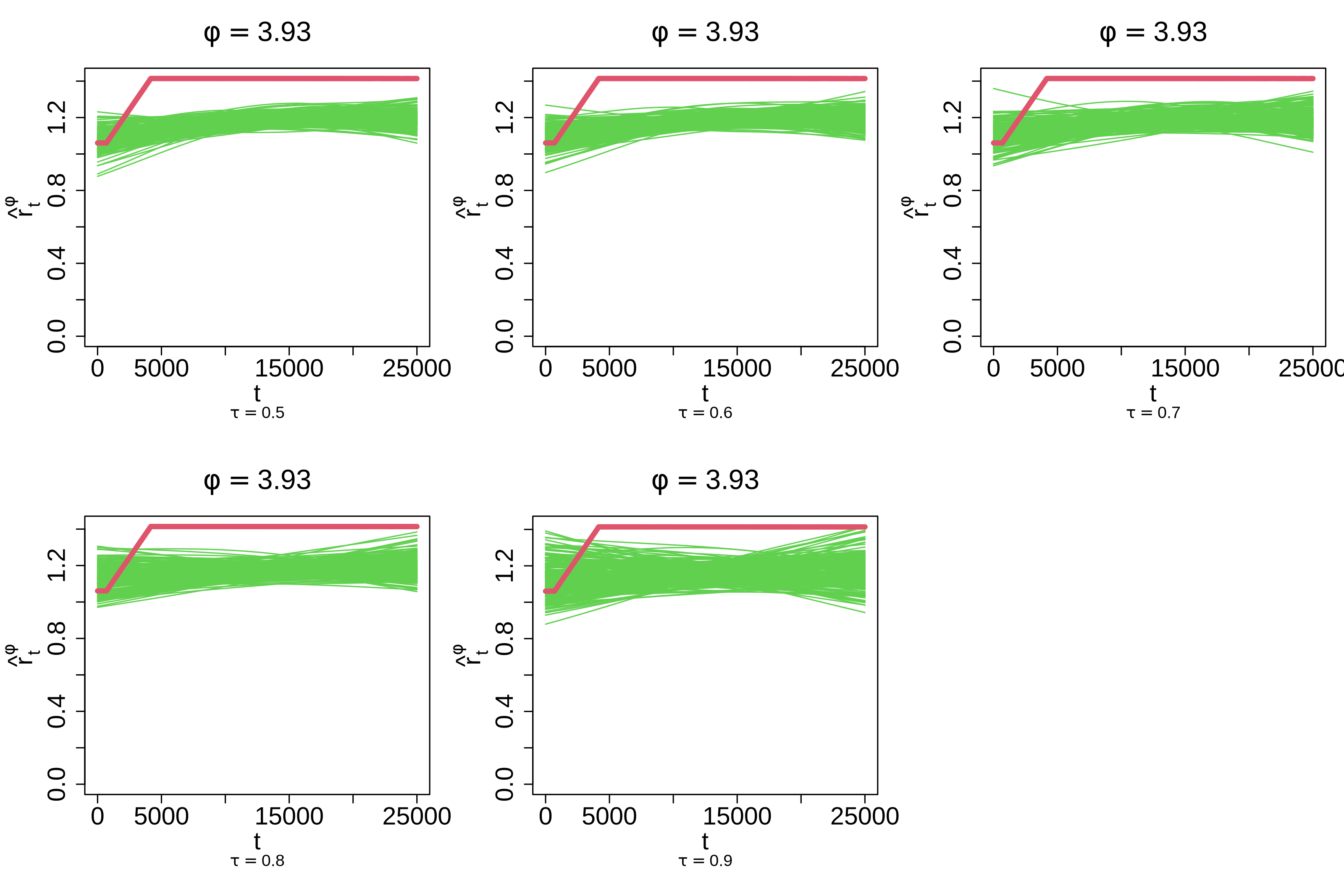}
    \caption{Boundary set radii estimates at $\phi = 5\pi/4$ across $\tau \in \{0.5,0.6,0.7,0.8,0.9\}$ for the fifth copula example.}
    \label{fig:res_tau_p3_c5}
\end{figure}

\begin{figure}[H]
    \centering
    \includegraphics[width=.8\linewidth]{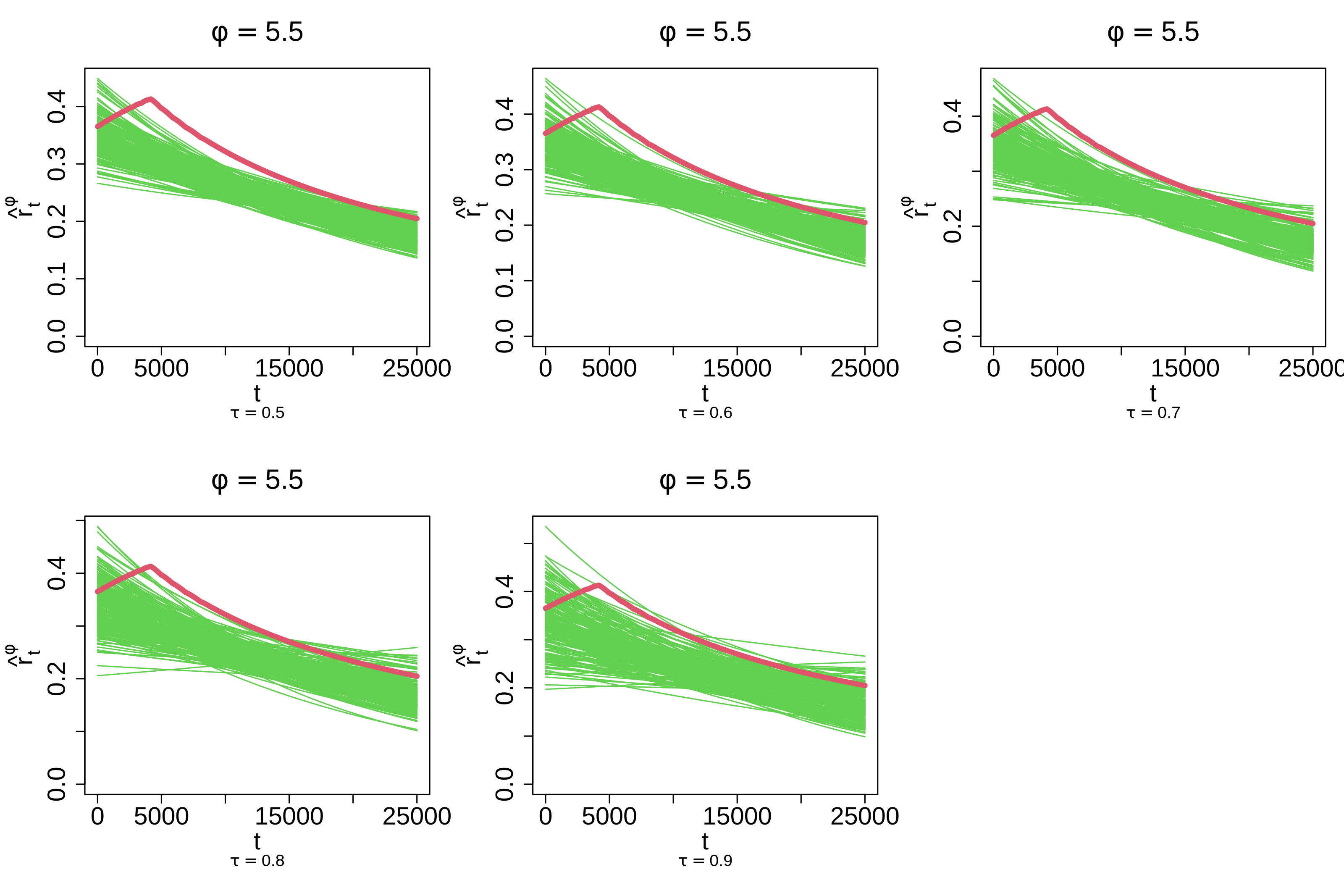}
    \caption{Boundary set radii estimates at $\phi = 7\pi/4$ across $\tau \in \{0.5,0.6,0.7,0.8,0.9\}$ for the fifth copula example.}
    \label{fig:res_tau_p4_c5}
\end{figure}

Figures~\ref{fig:res_tau_t1_c1}-\ref{fig:res_tau_p4_c5} illustrate the effect of the quantile level $\tau$ on the boundary set estimates. 

\subsection{Evaluating the effect of basis dimension} \label{subsec:appen_basis_dimension}

\begin{figure}[H]
    \centering
    \includegraphics[width=.8\linewidth]{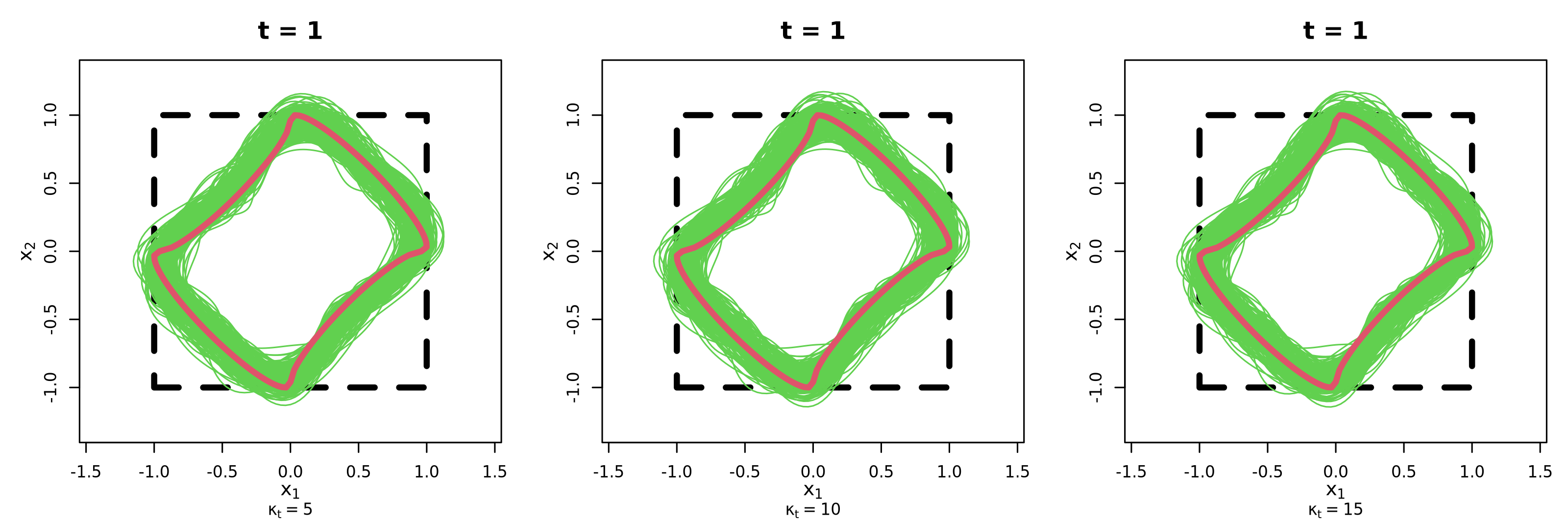}
    \caption{Boundary set estimates as $t = 1$ across $\kappa_t \in \{5,10,15\}$ for the first copula example.}
    \label{fig:res_kappa_t_t1_c1}
\end{figure}

\begin{figure}[H]
    \centering
    \includegraphics[width=.8\linewidth]{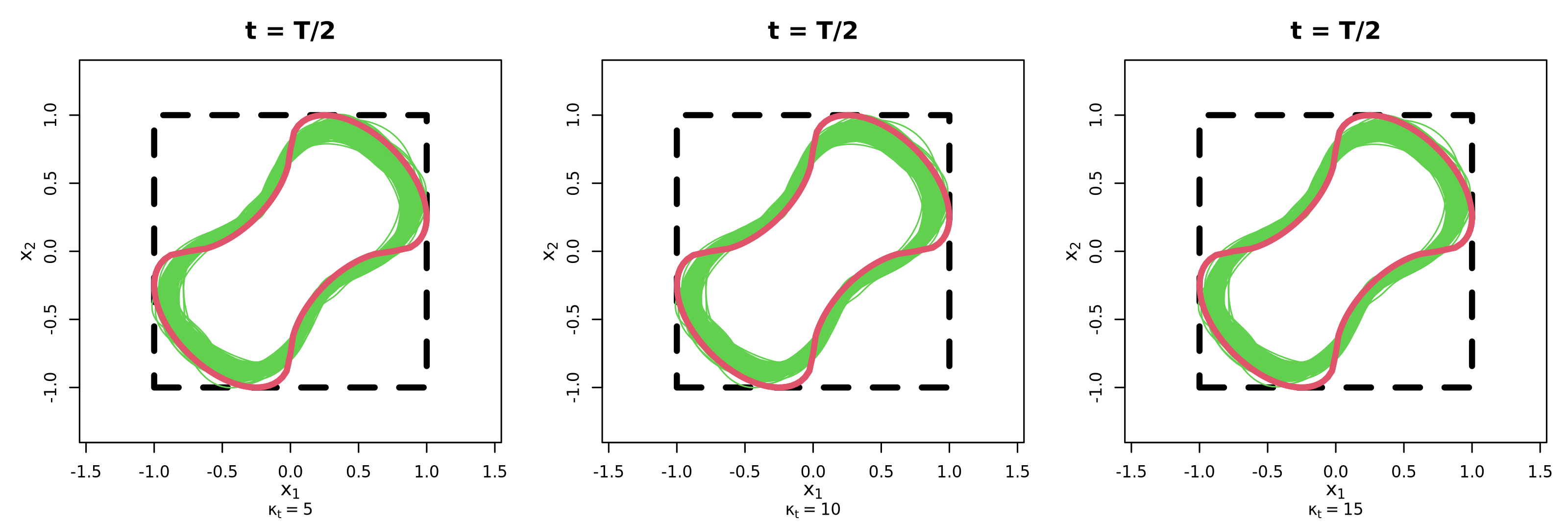}
    \caption{Boundary set estimates as $t = T/2$ across $\kappa_t \in \{5,10,15\}$ for the first copula example.}
    \label{fig:res_kappa_t_t2_c1}
\end{figure}

\begin{figure}[H]
    \centering
    \includegraphics[width=.8\linewidth]{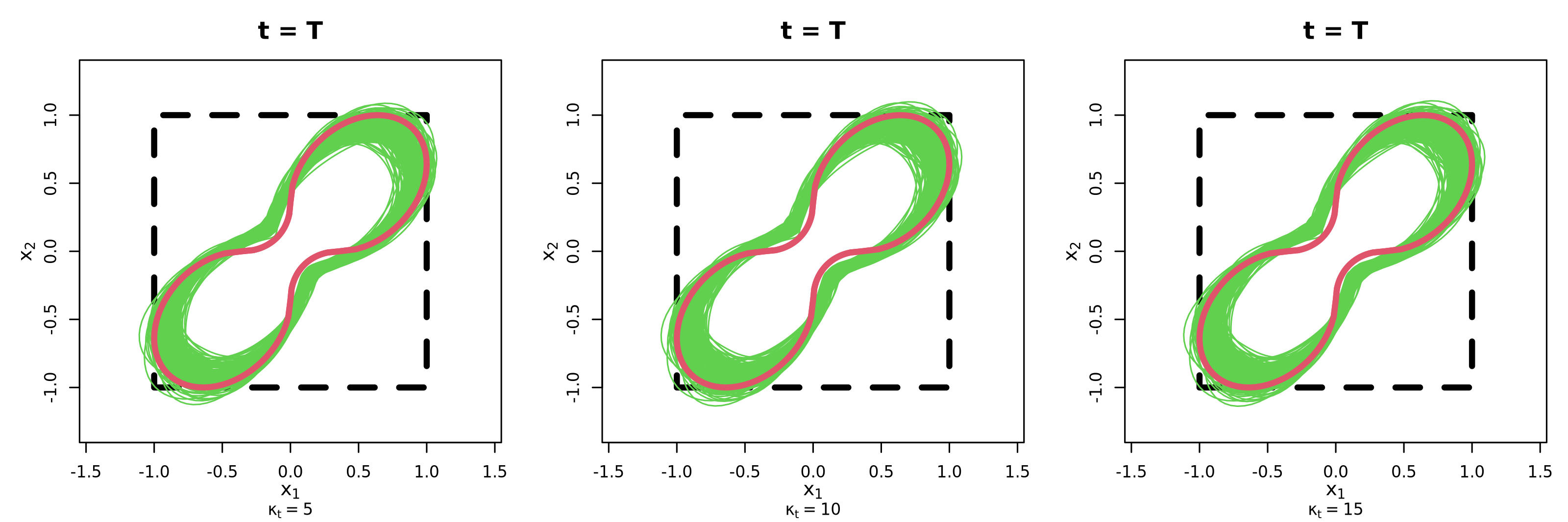}
    \caption{Boundary set estimates at $t = T$ across $\kappa_t \in \{5,10,15\}$ for the first copula example.}
    \label{fig:res_kappa_t_t3_c1}
\end{figure}

\begin{figure}[H]
    \centering
    \includegraphics[width=.8\linewidth]{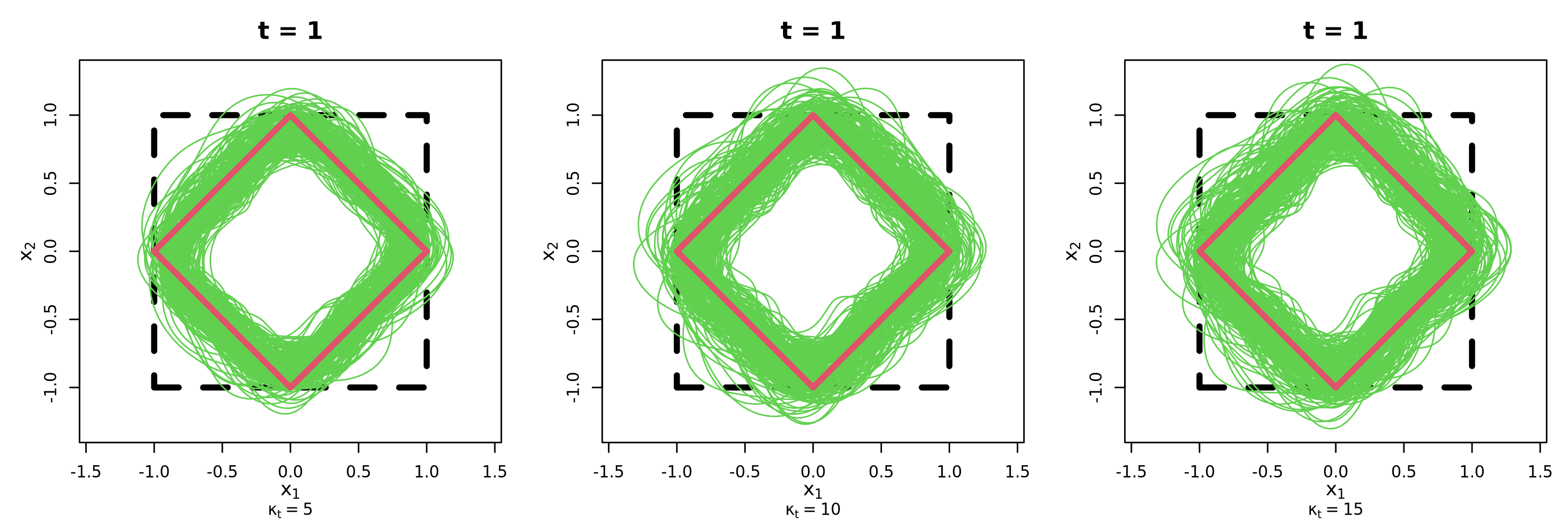}
    \caption{Boundary set estimates as $t = 1$ across $\kappa_t \in \{5,10,15\}$ for the second copula example.}
    \label{fig:res_kappa_t_t1_c2}
\end{figure}

\begin{figure}[H]
    \centering
    \includegraphics[width=.8\linewidth]{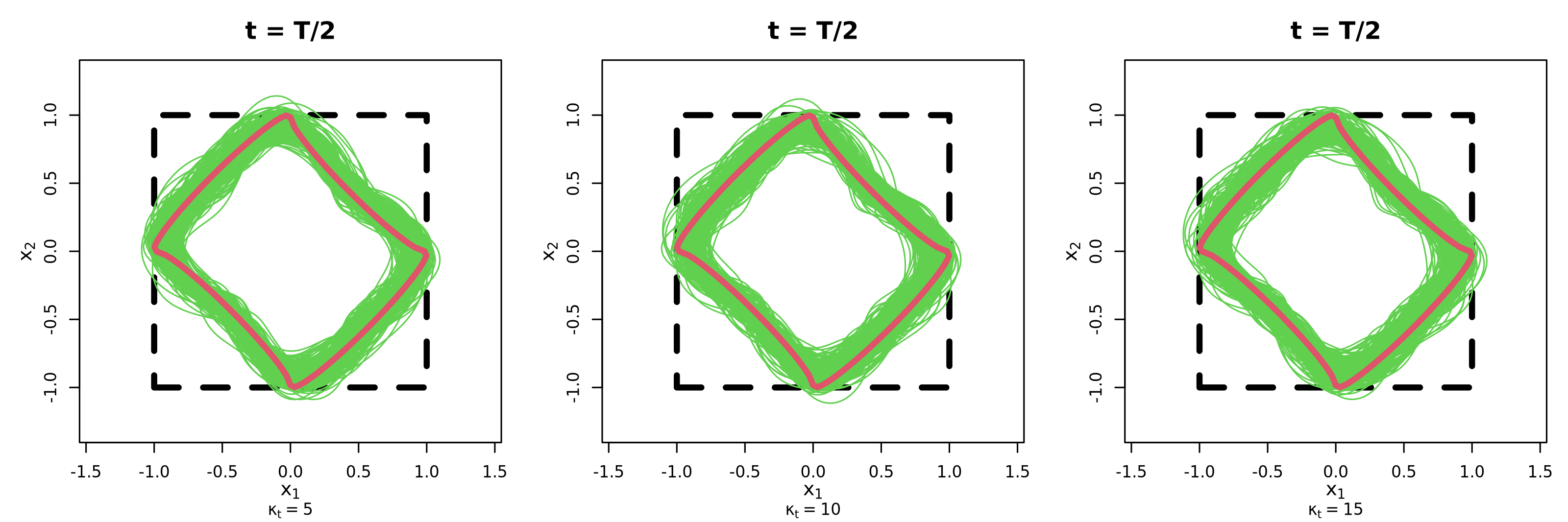}
    \caption{Boundary set estimates as $t = T/2$ across $\kappa_t \in \{5,10,15\}$ for the second copula example.}
    \label{fig:res_kappa_t_t2_c2}
\end{figure}

\begin{figure}[H]
    \centering
    \includegraphics[width=.8\linewidth]{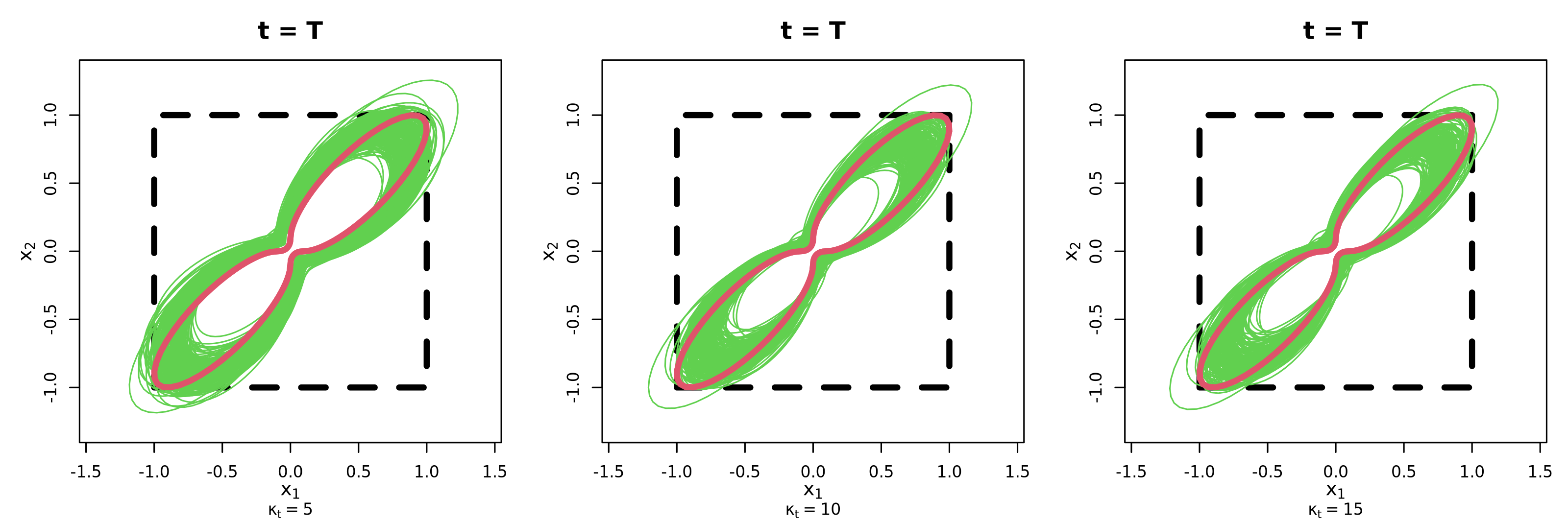}
    \caption{Boundary set estimates at $t = T$ across $\kappa_t \in \{5,10,15\}$ for the second copula example.}
    \label{fig:res_kappa_t_t3_c2}
\end{figure}

\begin{figure}[H]
    \centering
    \includegraphics[width=.8\linewidth]{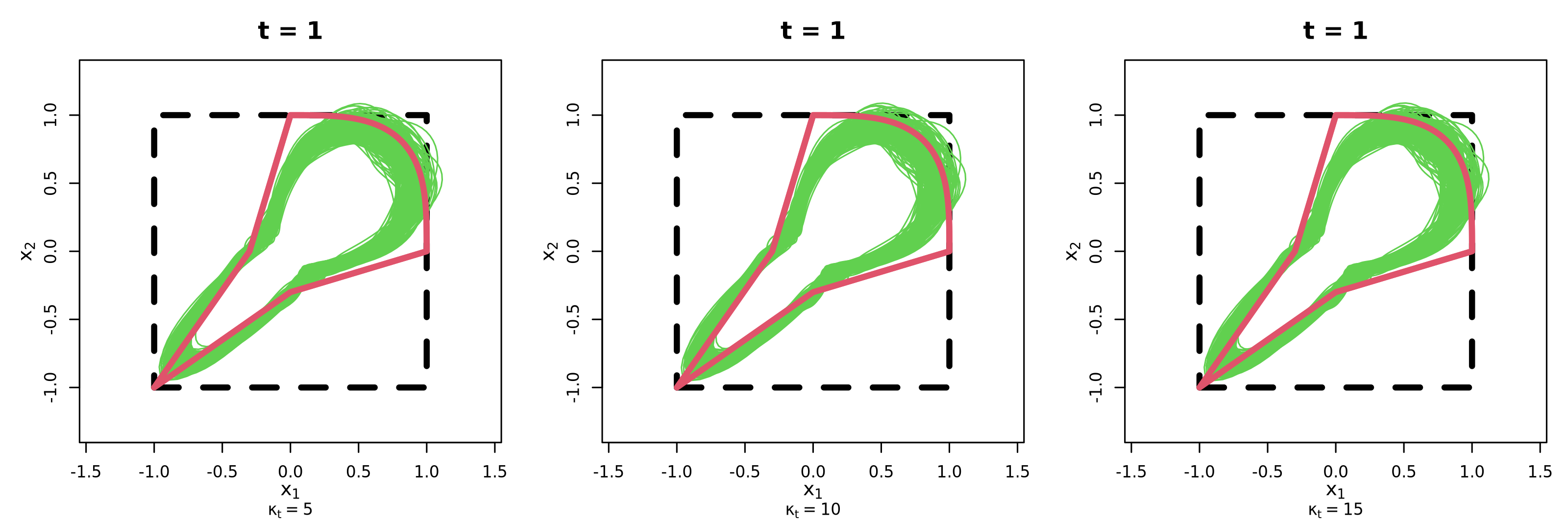}
    \caption{Boundary set estimates as $t = 1$ across $\kappa_t \in \{5,10,15\}$ for the third copula example.}
    \label{fig:res_kappa_t_t1_c3}
\end{figure}

\begin{figure}[H]
    \centering
    \includegraphics[width=.8\linewidth]{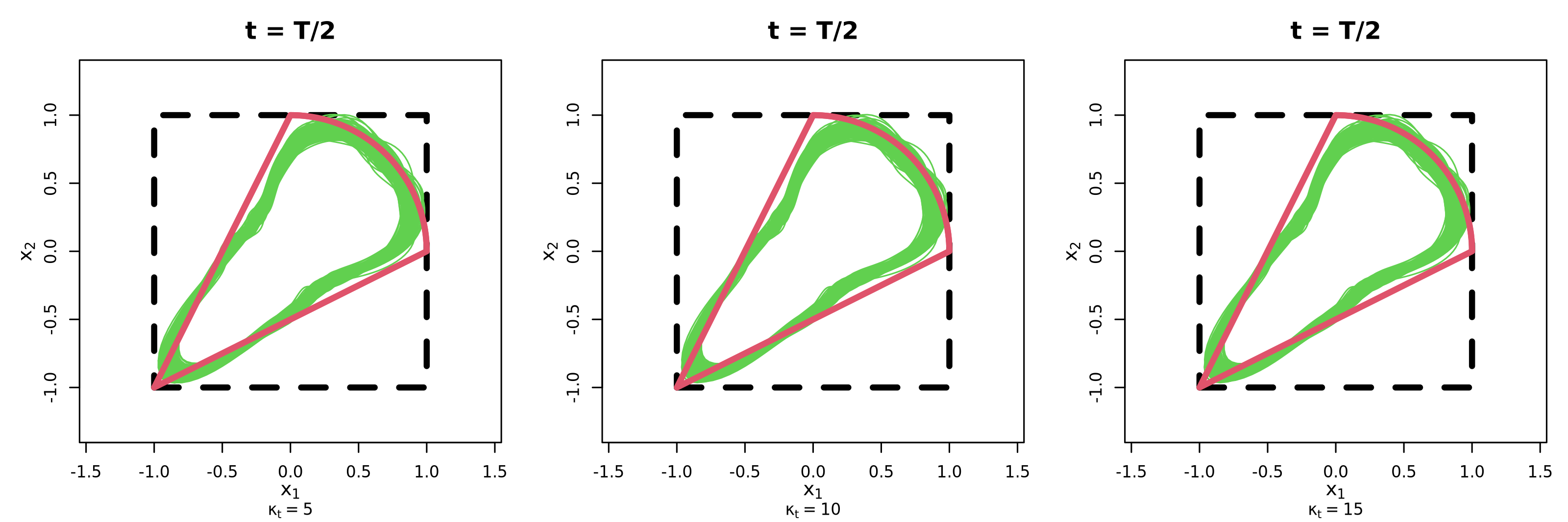}
    \caption{Boundary set estimates as $t = T/2$ across $\kappa_t \in \{5,10,15\}$ for the third copula example.}
    \label{fig:res_kappa_t_t2_c3}
\end{figure}

\begin{figure}[H]
    \centering
    \includegraphics[width=.8\linewidth]{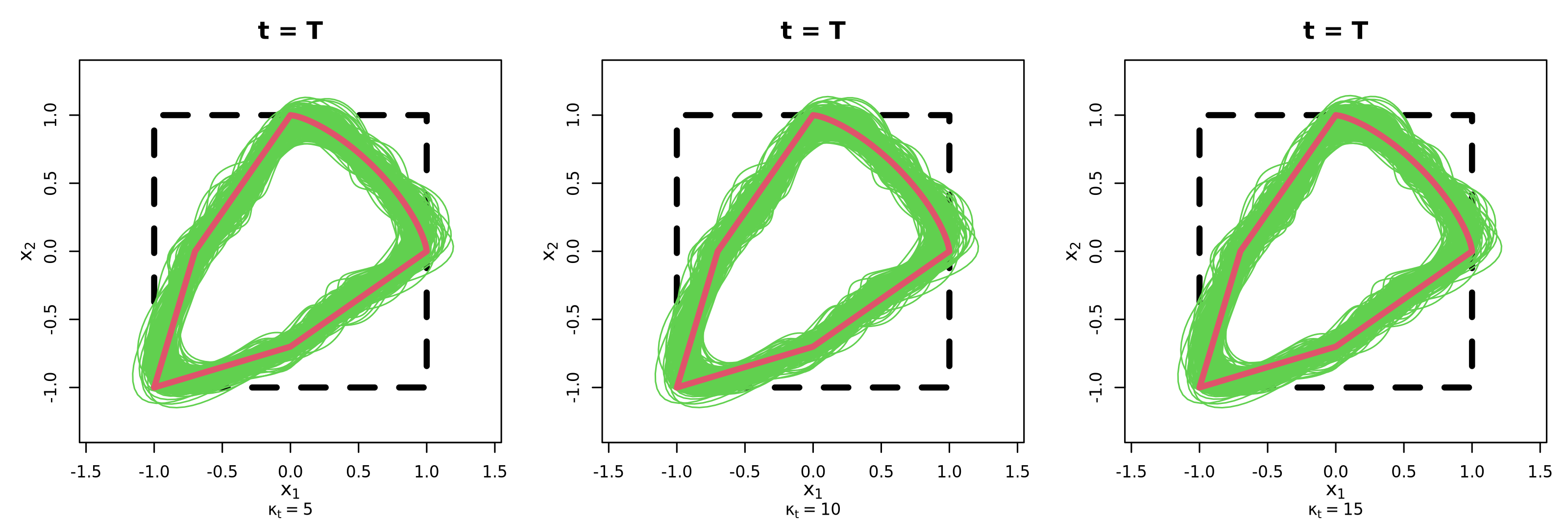}
    \caption{Boundary set estimates at $t = T$ across $\kappa_t \in \{5,10,15\}$ for the third copula example.}
    \label{fig:res_kappa_t_t3_c3}
\end{figure}

\begin{figure}[H]
    \centering
    \includegraphics[width=.8\linewidth]{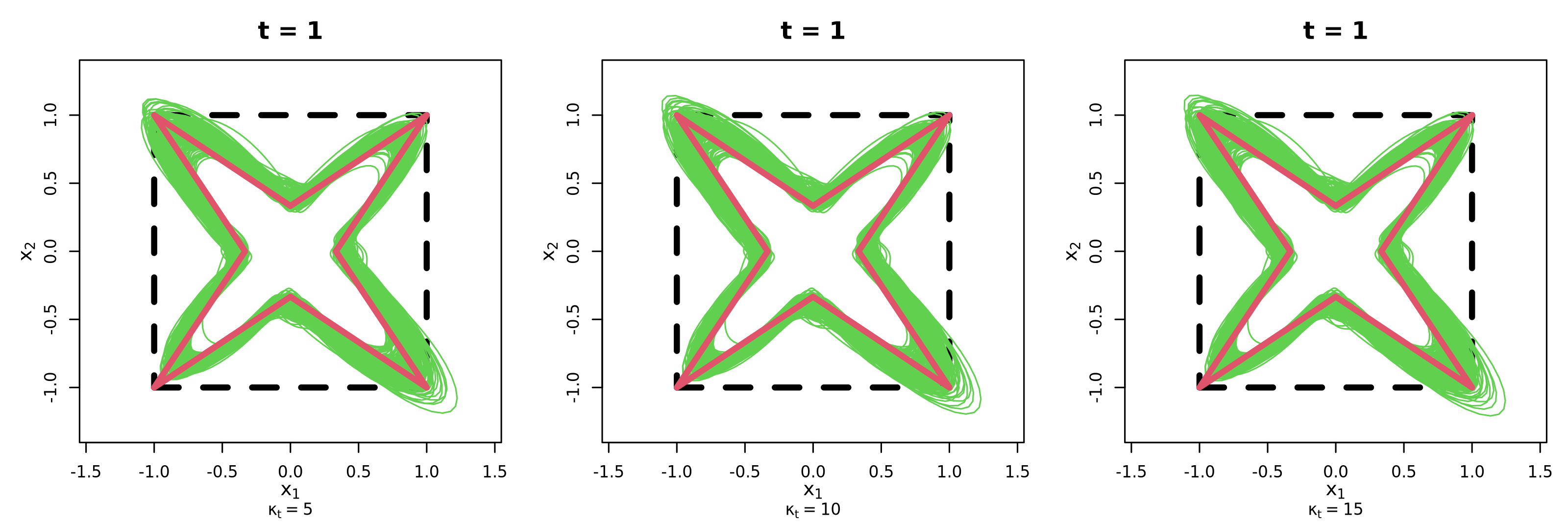}
    \caption{Boundary set estimates as $t = 1$ across $\kappa_t \in \{5,10,15\}$ for the fourth copula example.}
    \label{fig:res_kappa_t_t1_c4}
\end{figure}

\begin{figure}[H]
    \centering
    \includegraphics[width=.8\linewidth]{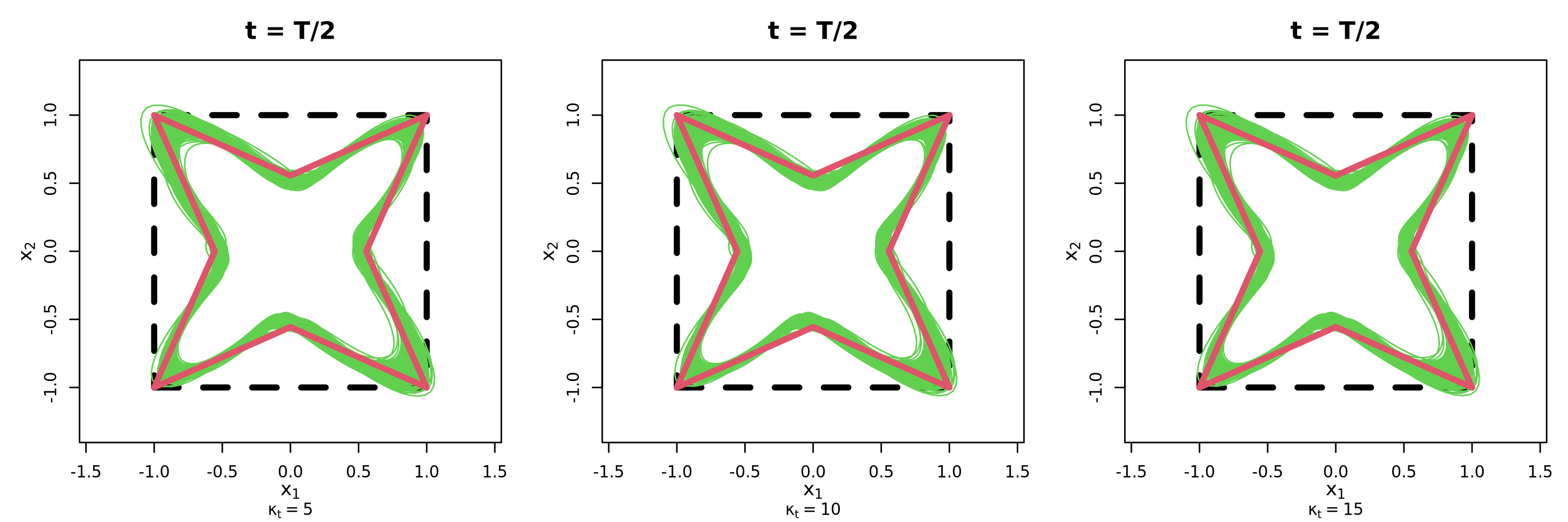}
    \caption{Boundary set estimates as $t = T/2$ across $\kappa_t \in \{5,10,15\}$ for the fourth copula example.}
    \label{fig:res_kappa_t_t2_c4}
\end{figure}

\begin{figure}[H]
    \centering
    \includegraphics[width=.8\linewidth]{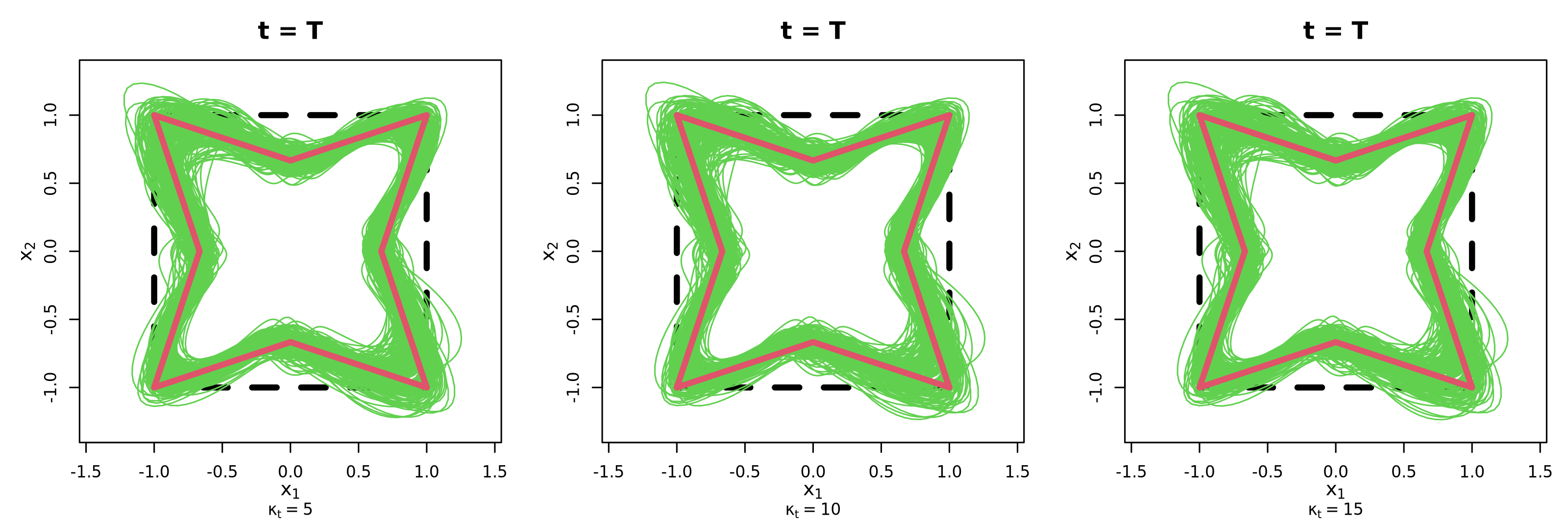}
    \caption{Boundary set estimates at $t = T$ across $\kappa_t \in \{5,10,15\}$ for the fourth copula example.}
    \label{fig:res_kappa_t_t3_c4}
\end{figure}

\begin{figure}[H]
    \centering
    \includegraphics[width=.8\linewidth]{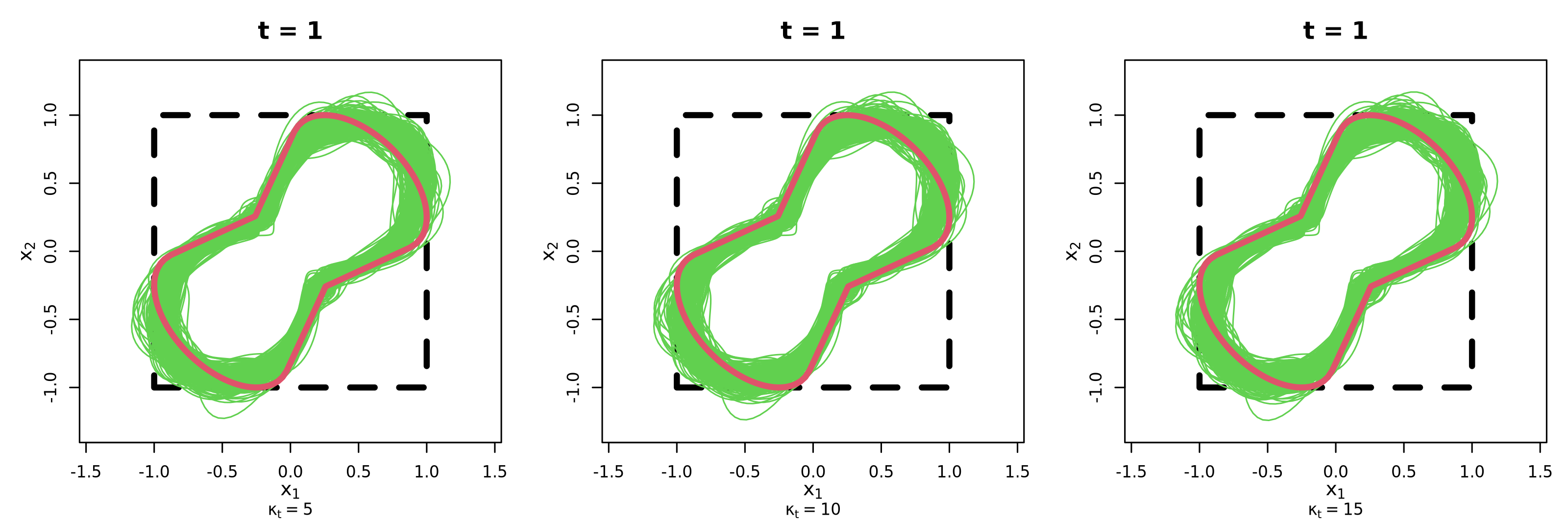}
    \caption{Boundary set estimates as $t = 1$ across $\kappa_t \in \{5,10,15\}$ for the fifth copula example.}
    \label{fig:res_kappa_t_t1_c5}
\end{figure}

\begin{figure}[H]
    \centering
    \includegraphics[width=.8\linewidth]{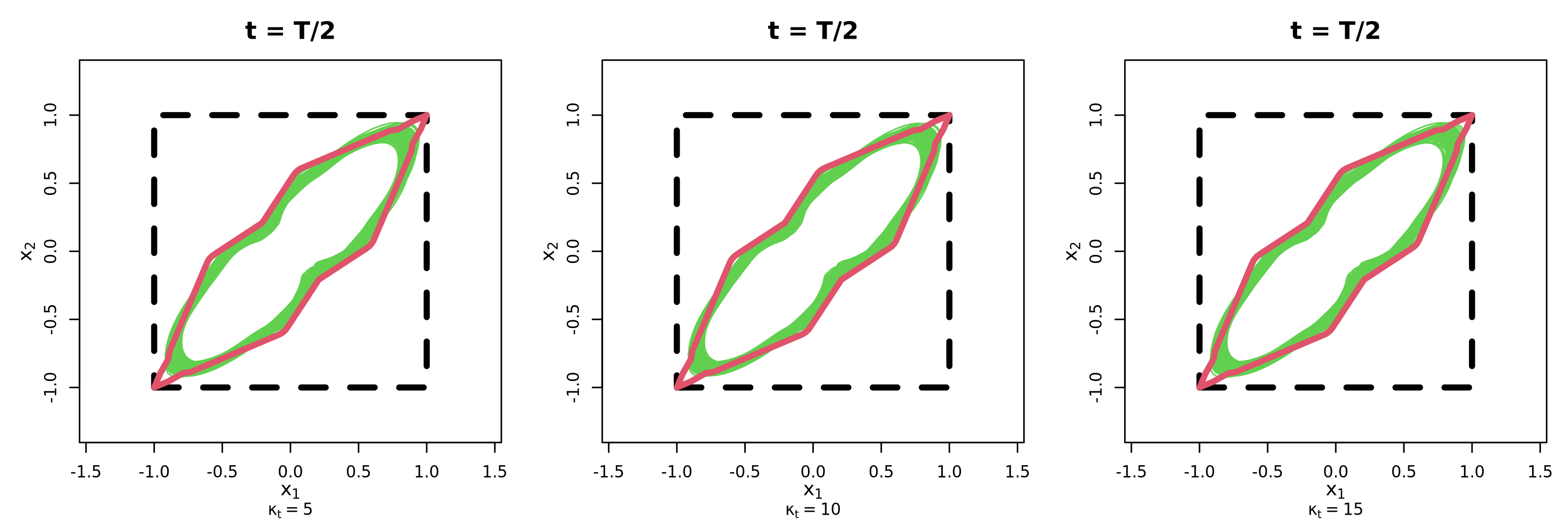}
    \caption{Boundary set estimates as $t = T/2$ across $\kappa_t \in \{5,10,15\}$ for the fifth copula example.}
    \label{fig:res_kappa_t_t2_c5}
\end{figure}

\begin{figure}[H]
    \centering
    \includegraphics[width=.8\linewidth]{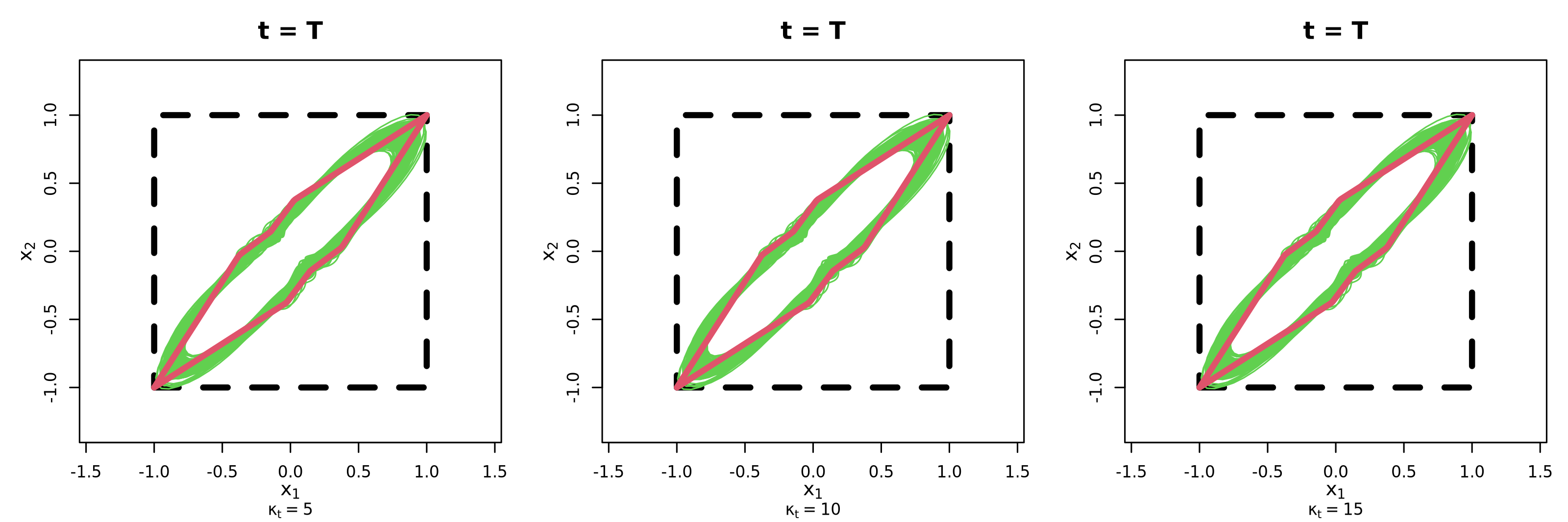}
    \caption{Boundary set estimates at $t = T$ across $\kappa_t \in \{5,10,15\}$ for the fifth copula example.}
    \label{fig:res_kappa_t_t3_c5}
\end{figure}

\begin{figure}[H]
    \centering
    \includegraphics[width=.8\linewidth]{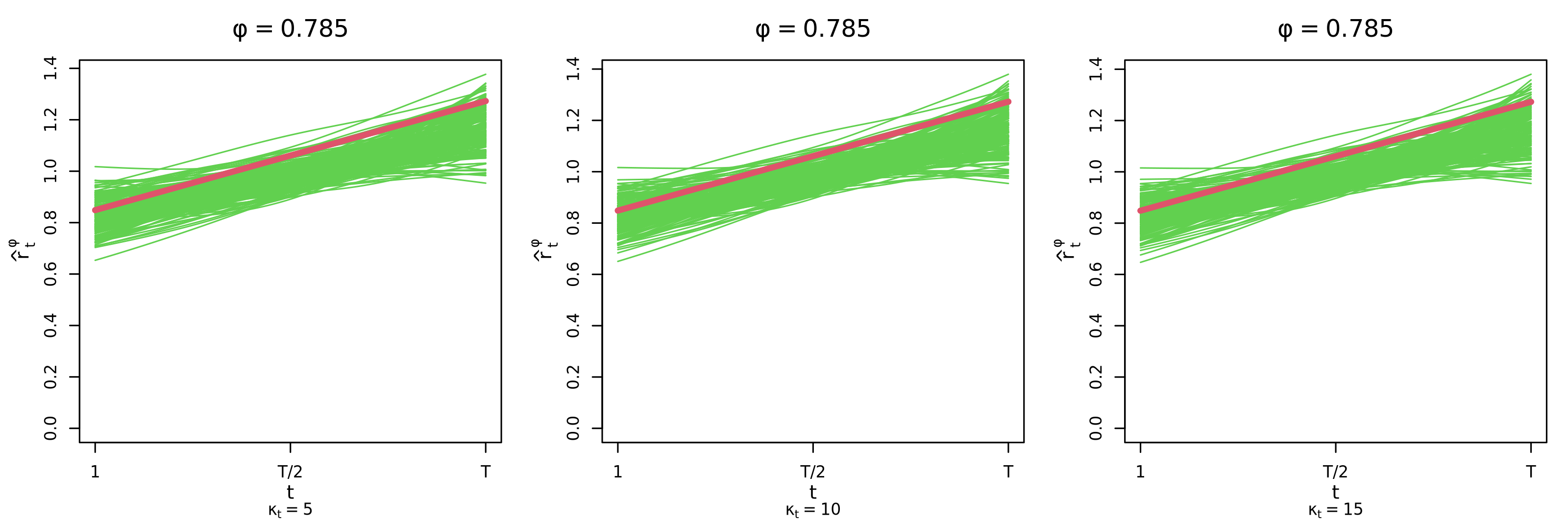}
    \caption{Boundary set radii estimates at $\phi = \pi/4$ across $\kappa_t \in \{5,10,15\}$ for the first copula example.}
    \label{fig:res_kappa_t_p1_c1}
\end{figure}

\begin{figure}[H]
    \centering
    \includegraphics[width=.8\linewidth]{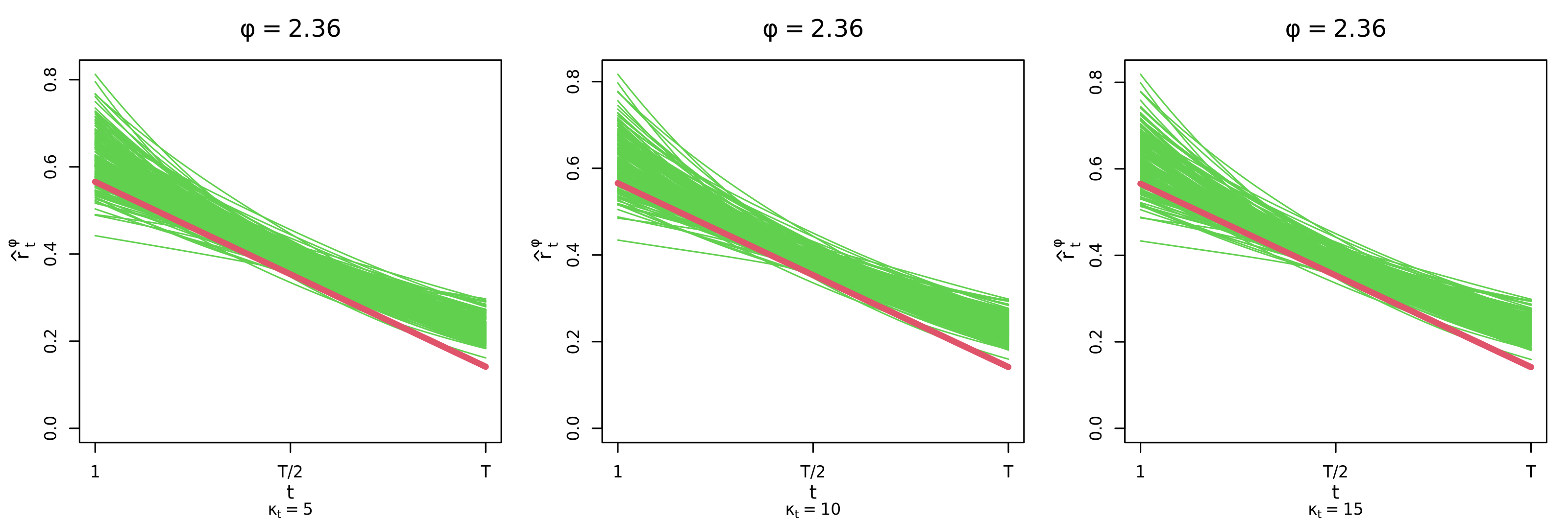}
    \caption{Boundary set radii estimates at $\phi = 3\pi/4$ across $\kappa_t \in \{5,10,15\}$ for the first copula example.}
    \label{fig:res_kappa_t_p2_c1}
\end{figure}

\begin{figure}[H]
    \centering
    \includegraphics[width=.8\linewidth]{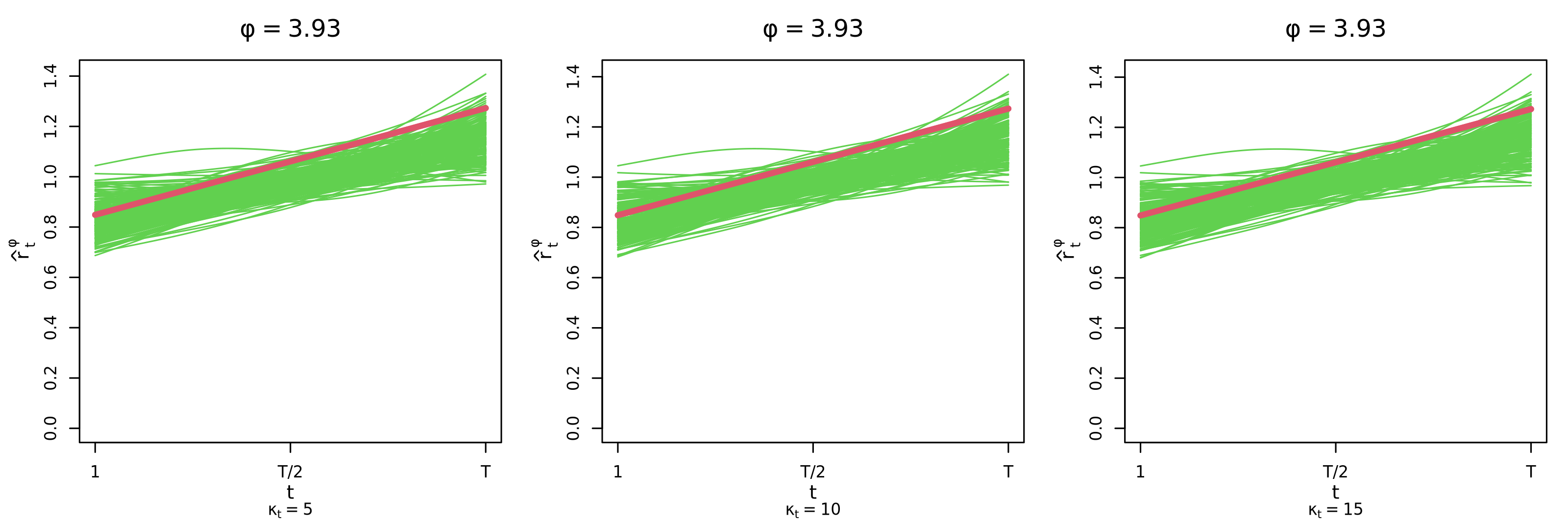}
    \caption{Boundary set radii estimates at $\phi = 5\pi/4$ across $\kappa_t \in \{5,10,15\}$ for the first copula example.}
    \label{fig:res_kappa_t_p3_c1}
\end{figure}

\begin{figure}[H]
    \centering
    \includegraphics[width=.8\linewidth]{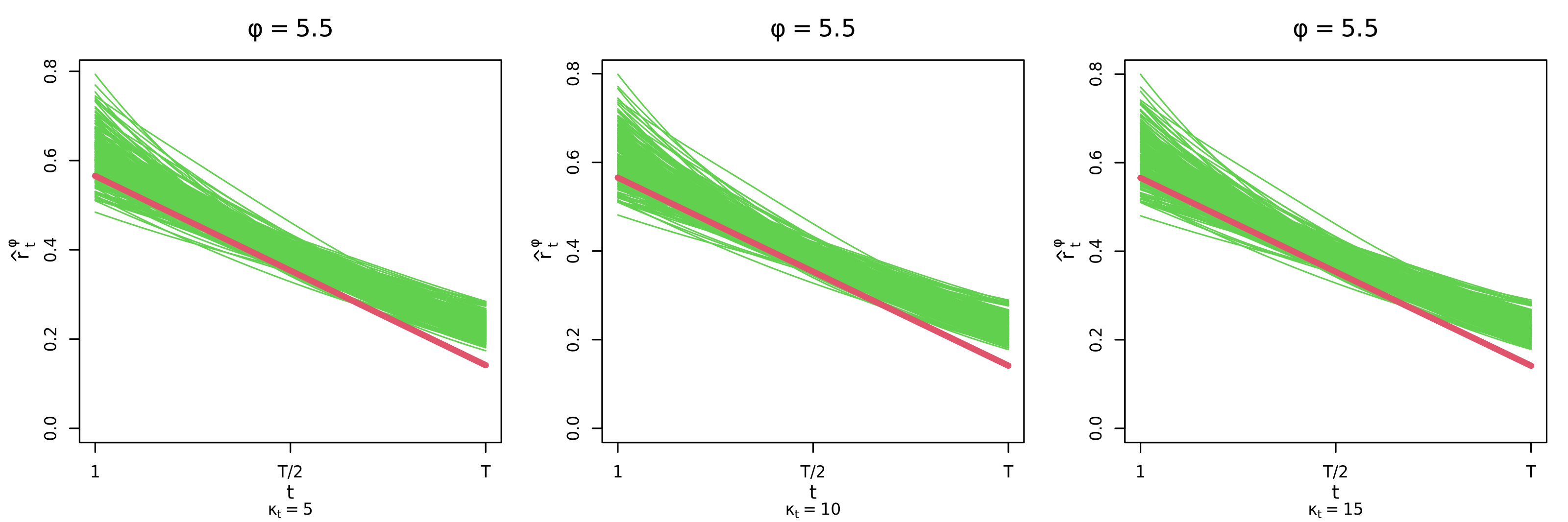}
    \caption{Boundary set radii estimates at $\phi = 7\pi/4$ across $\kappa_t \in \{5,10,15\}$ for the first copula example.}
    \label{fig:res_kappa_t_p4_c1}
\end{figure}

\begin{figure}[H]
    \centering
    \includegraphics[width=.8\linewidth]{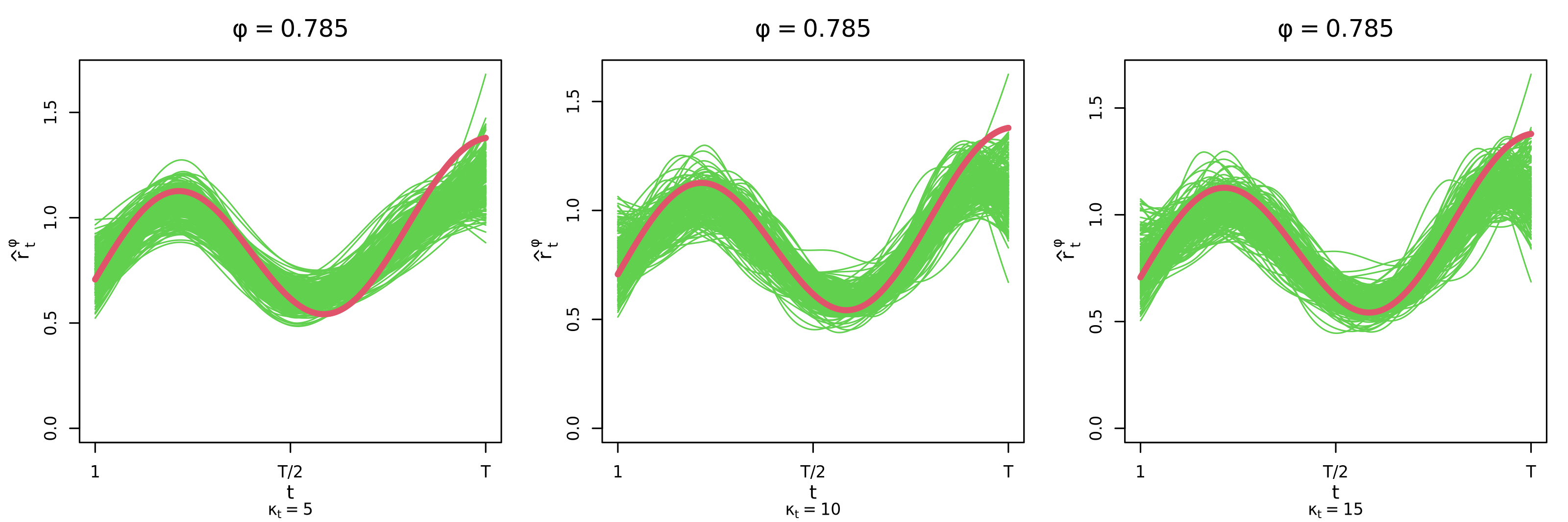}
    \caption{Boundary set radii estimates at $\phi = \pi/4$ across $\kappa_t \in \{5,10,15\}$ for the second copula example.}
    \label{fig:res_kappa_t_p1_c2}
\end{figure}

\begin{figure}[H]
    \centering
    \includegraphics[width=.8\linewidth]{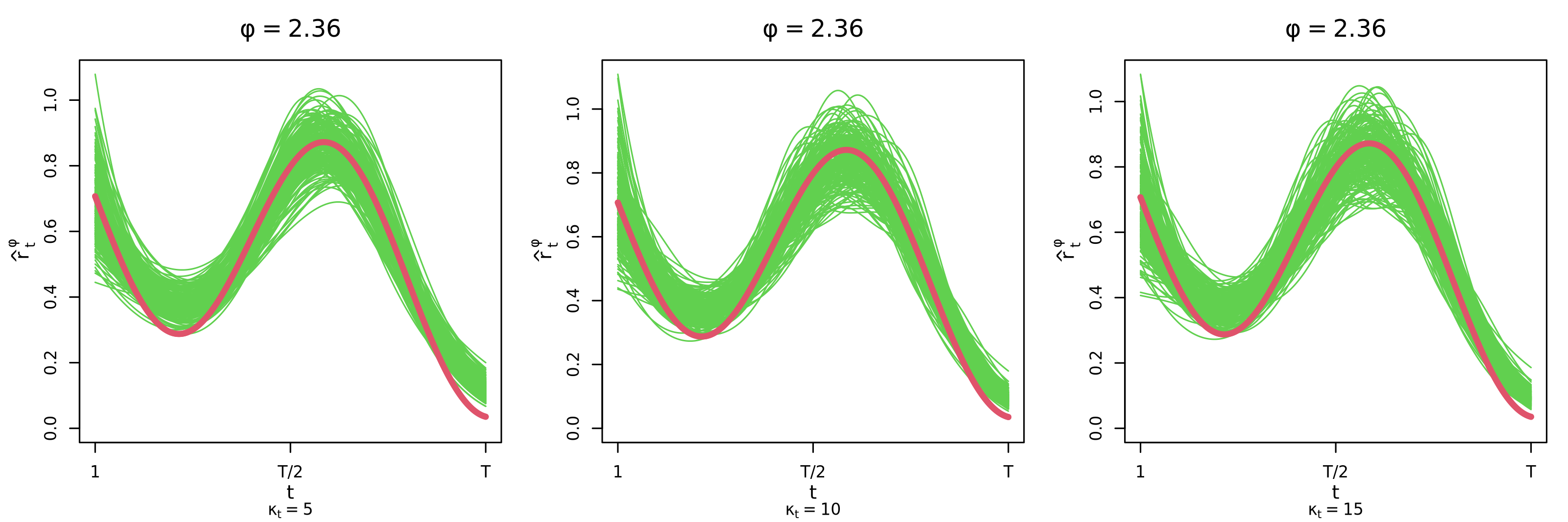}
    \caption{Boundary set radii estimates at $\phi = 3\pi/4$ across $\kappa_t \in \{5,10,15\}$ for the second copula example.}
    \label{fig:res_kappa_t_p2_c2}
\end{figure}

\begin{figure}[H]
    \centering
    \includegraphics[width=.8\linewidth]{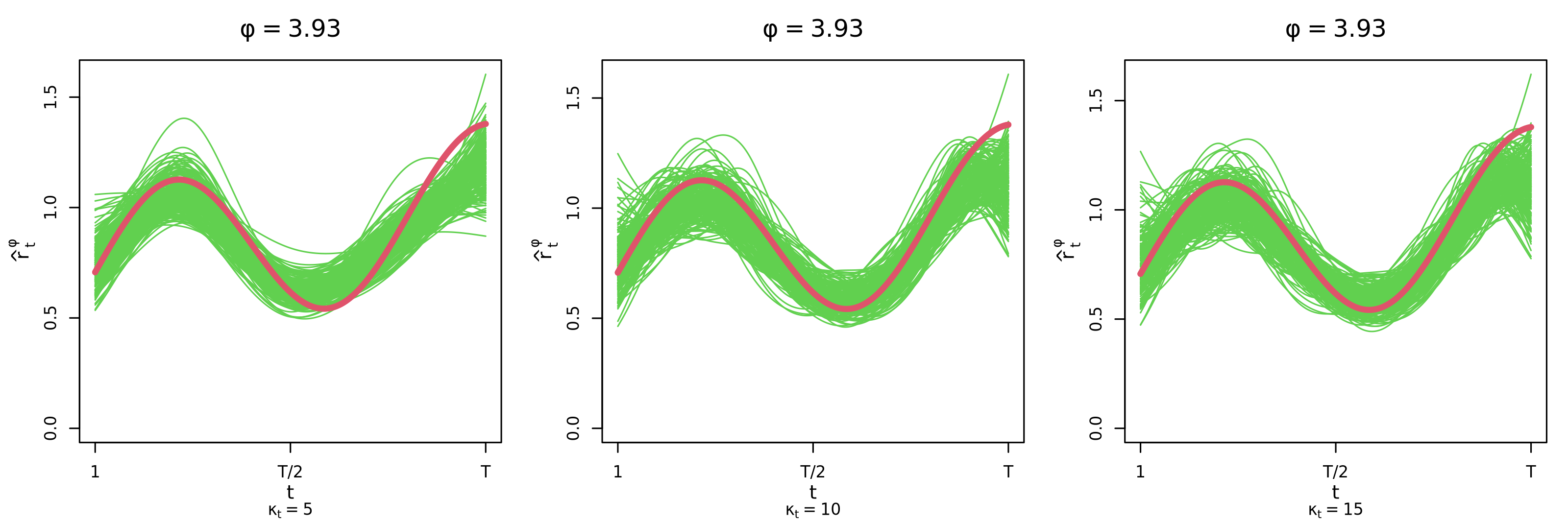}
    \caption{Boundary set radii estimates at $\phi = 5\pi/4$ across $\kappa_t \in \{5,10,15\}$ for the second copula example.}
    \label{fig:res_kappa_t_p3_c2}
\end{figure}

\begin{figure}[H]
    \centering
    \includegraphics[width=.8\linewidth]{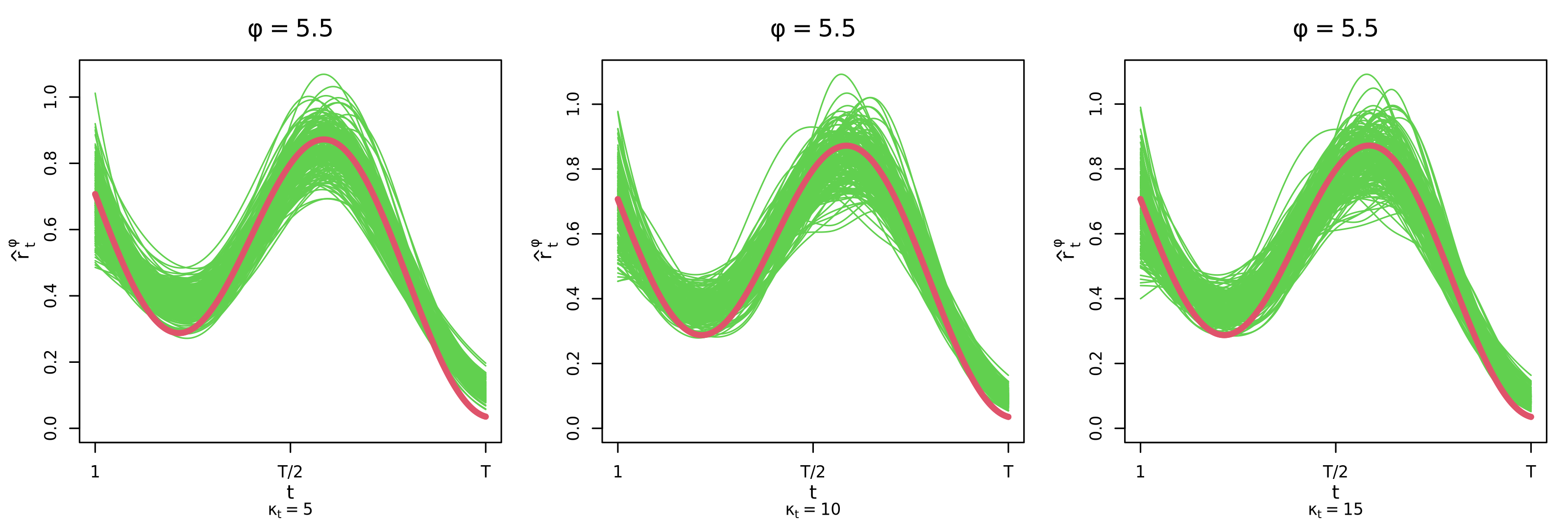}
    \caption{Boundary set radii estimates at $\phi = 7\pi/4$ across $\kappa_t \in \{5,10,15\}$ for the second copula example.}
    \label{fig:res_kappa_t_p4_c2}
\end{figure}

\begin{figure}[H]
    \centering
    \includegraphics[width=.8\linewidth]{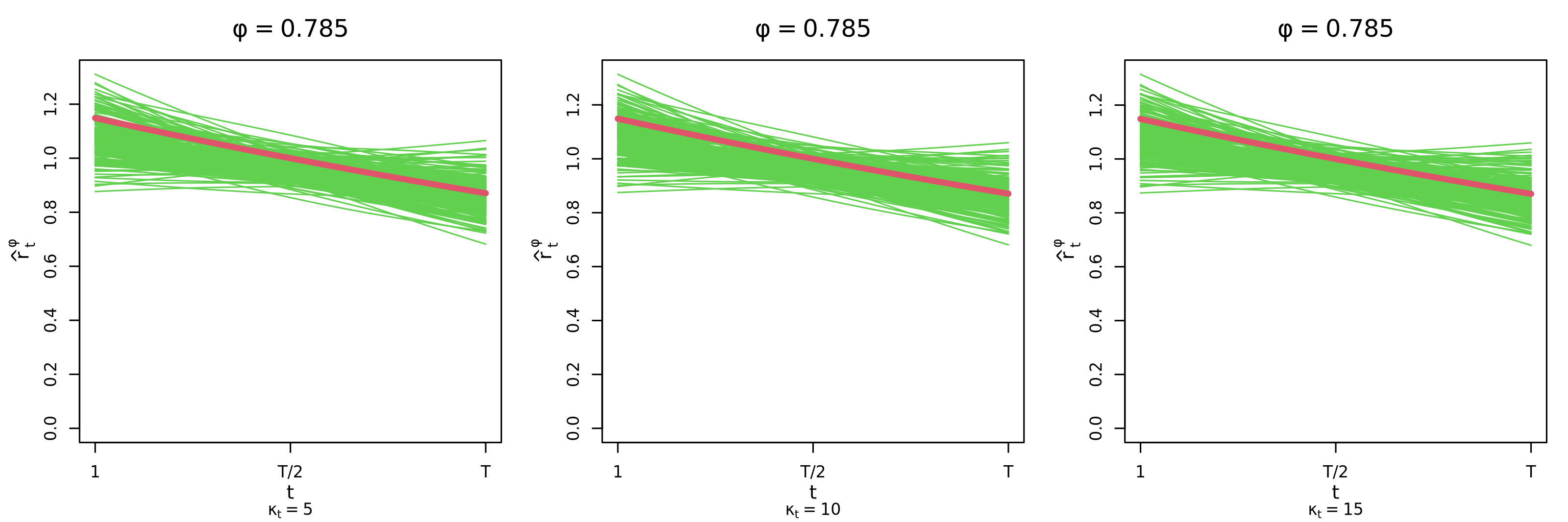}
    \caption{Boundary set radii estimates at $\phi = \pi/4$ across $\kappa_t \in \{5,10,15\}$ for the third copula example.}
    \label{fig:res_kappa_t_p1_c3}
\end{figure}

\begin{figure}[H]
    \centering
    \includegraphics[width=.8\linewidth]{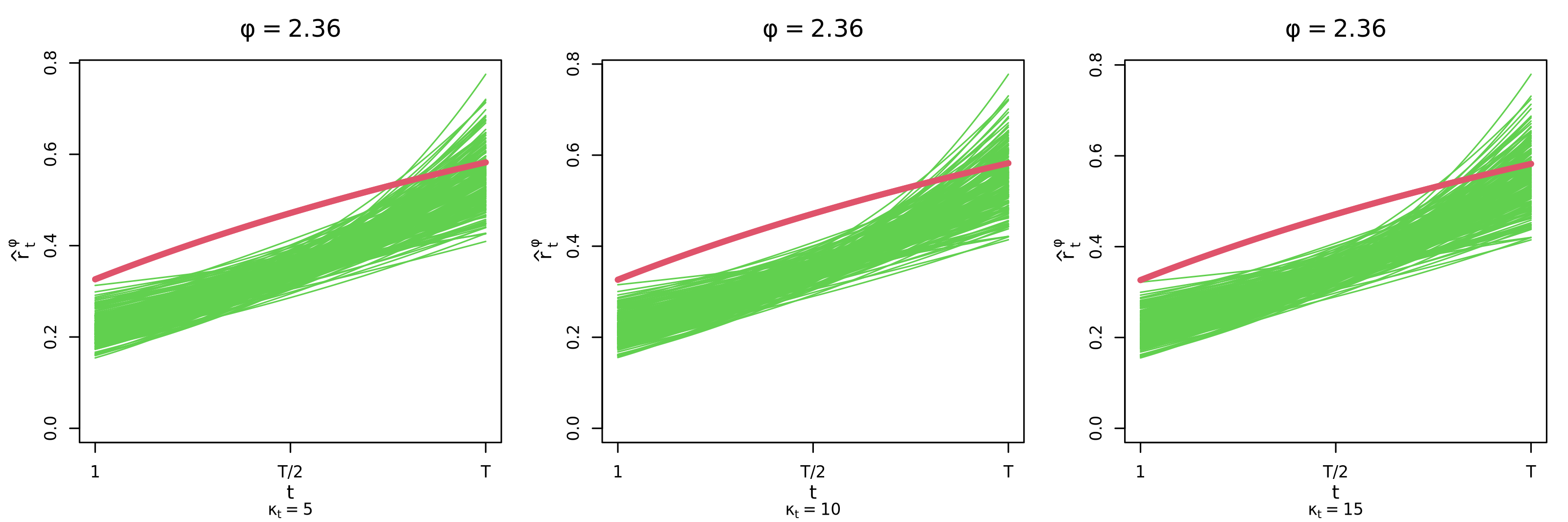}
    \caption{Boundary set radii estimates at $\phi = 3\pi/4$ across $\kappa_t \in \{5,10,15\}$ for the third copula example.}
    \label{fig:res_kappa_t_p2_c3}
\end{figure}

\begin{figure}[H]
    \centering
    \includegraphics[width=.8\linewidth]{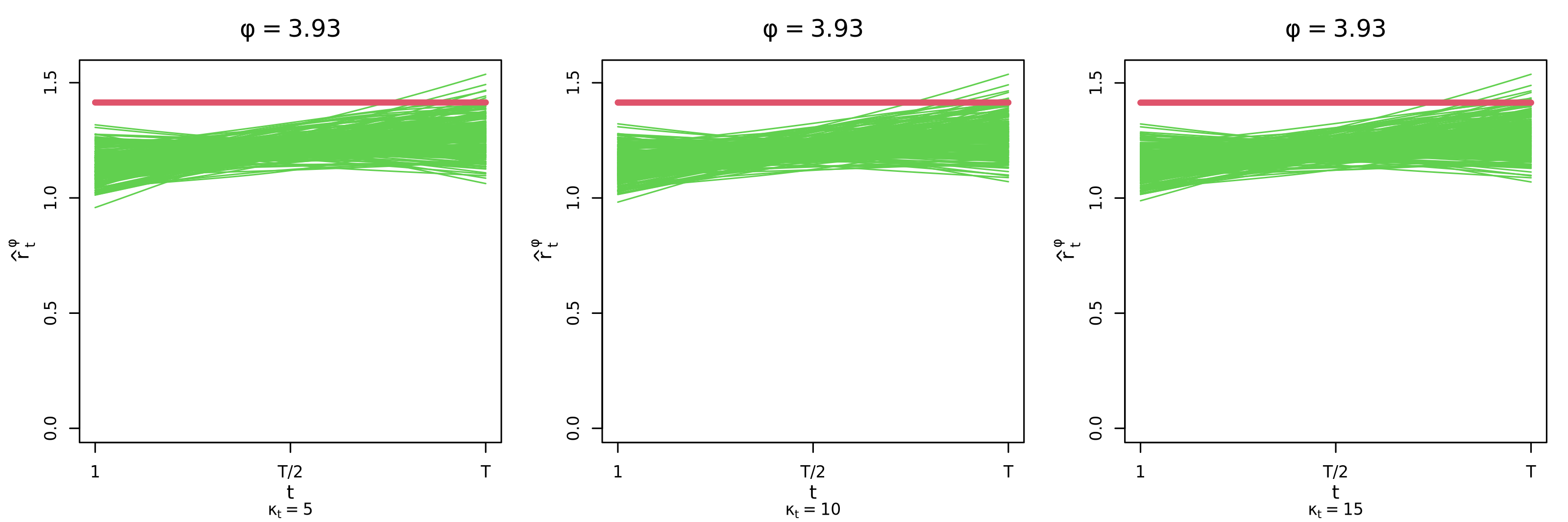}
    \caption{Boundary set radii estimates at $\phi = 5\pi/4$ across $\kappa_t \in \{5,10,15\}$ for the third copula example.}
    \label{fig:res_kappa_t_p3_c3}
\end{figure}

\begin{figure}[H]
    \centering
    \includegraphics[width=.8\linewidth]{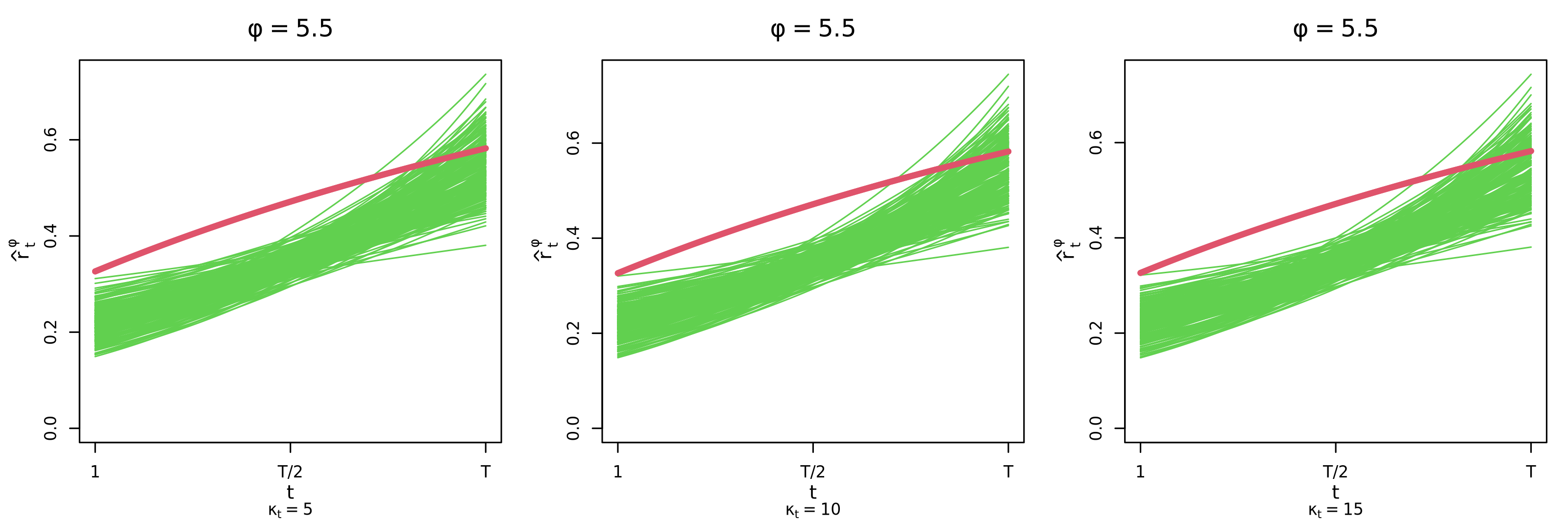}
    \caption{Boundary set radii estimates at $\phi = 7\pi/4$ across $\kappa_t \in \{5,10,15\}$ for the third copula example.}
    \label{fig:res_kappa_t_p4_c3}
\end{figure}

\begin{figure}[H]
    \centering
    \includegraphics[width=.8\linewidth]{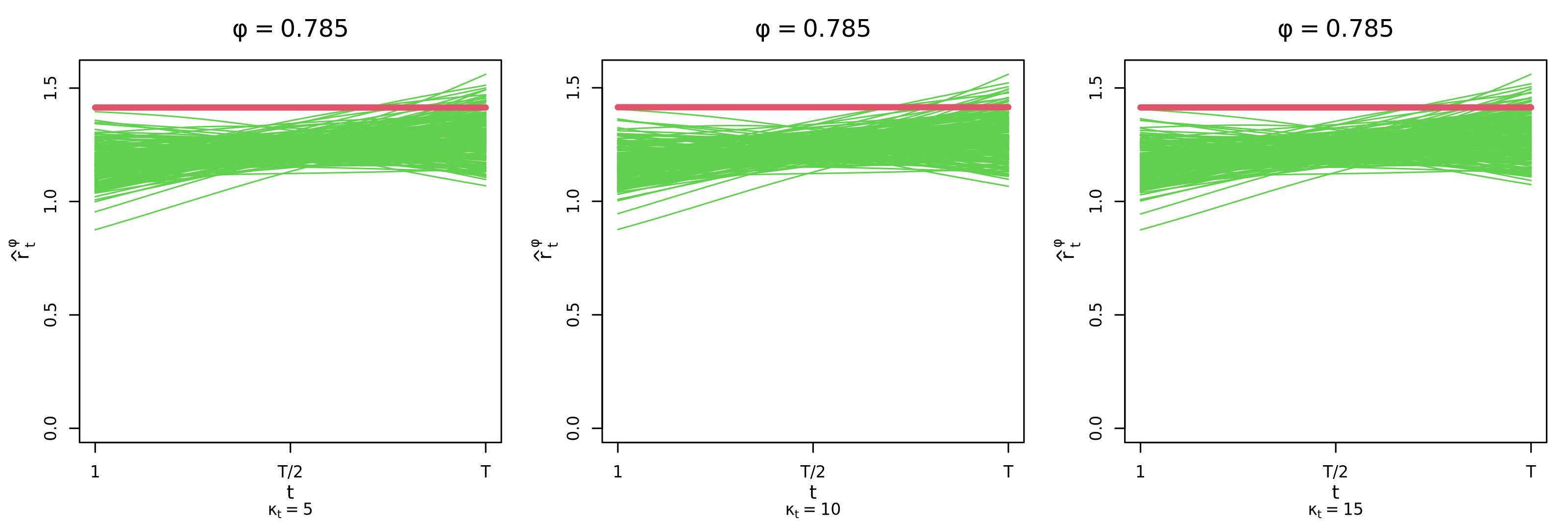}
    \caption{Boundary set radii estimates at $\phi = \pi/4$ across $\kappa_t \in \{5,10,15\}$ for the fourth copula example.}
    \label{fig:res_kappa_t_p1_c4}
\end{figure}

\begin{figure}[H]
    \centering
    \includegraphics[width=.8\linewidth]{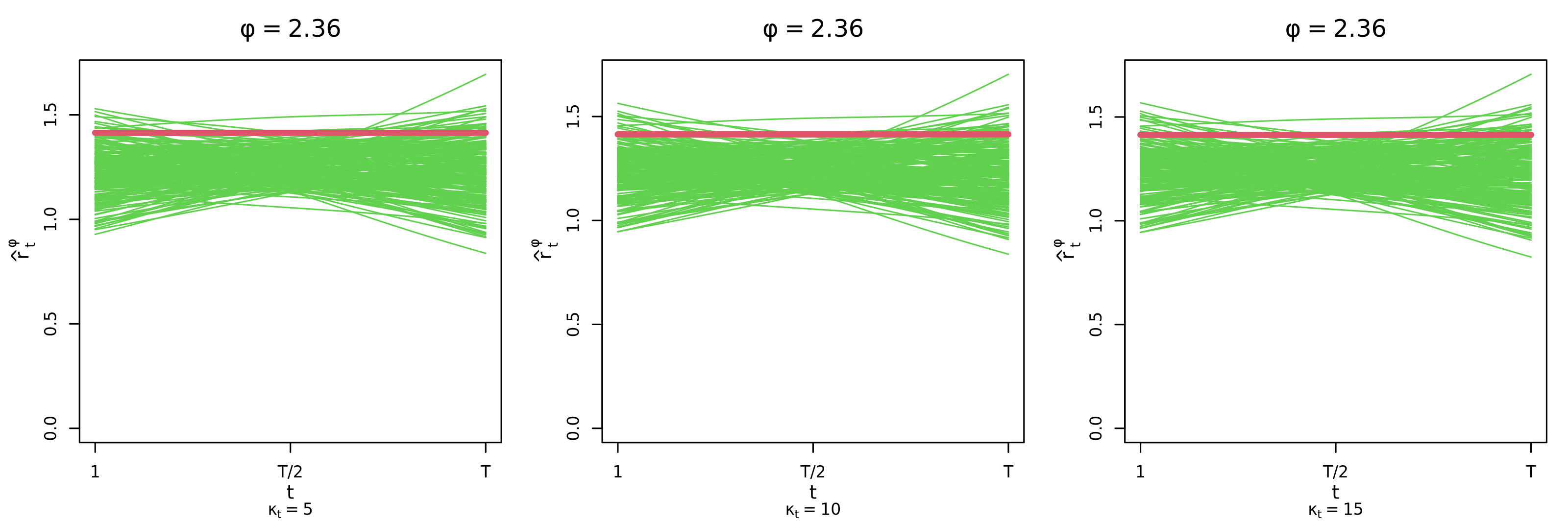}
    \caption{Boundary set radii estimates at $\phi = 3\pi/4$ across $\kappa_t \in \{5,10,15\}$ for the fourth copula example.}
    \label{fig:res_kappa_t_p2_c4}
\end{figure}

\begin{figure}[H]
    \centering
    \includegraphics[width=.8\linewidth]{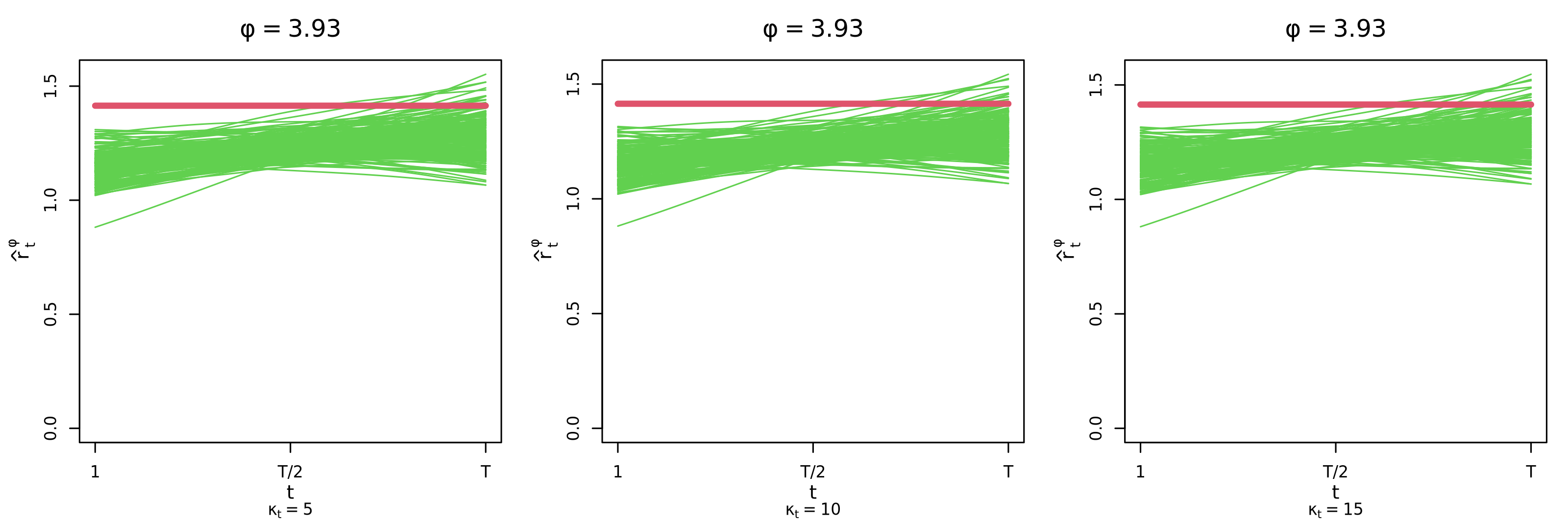}
    \caption{Boundary set radii estimates at $\phi = 5\pi/4$ across $\kappa_t \in \{5,10,15\}$ for the fourth copula example.}
    \label{fig:res_kappa_t_p3_c4}
\end{figure}

\begin{figure}[H]
    \centering
    \includegraphics[width=.8\linewidth]{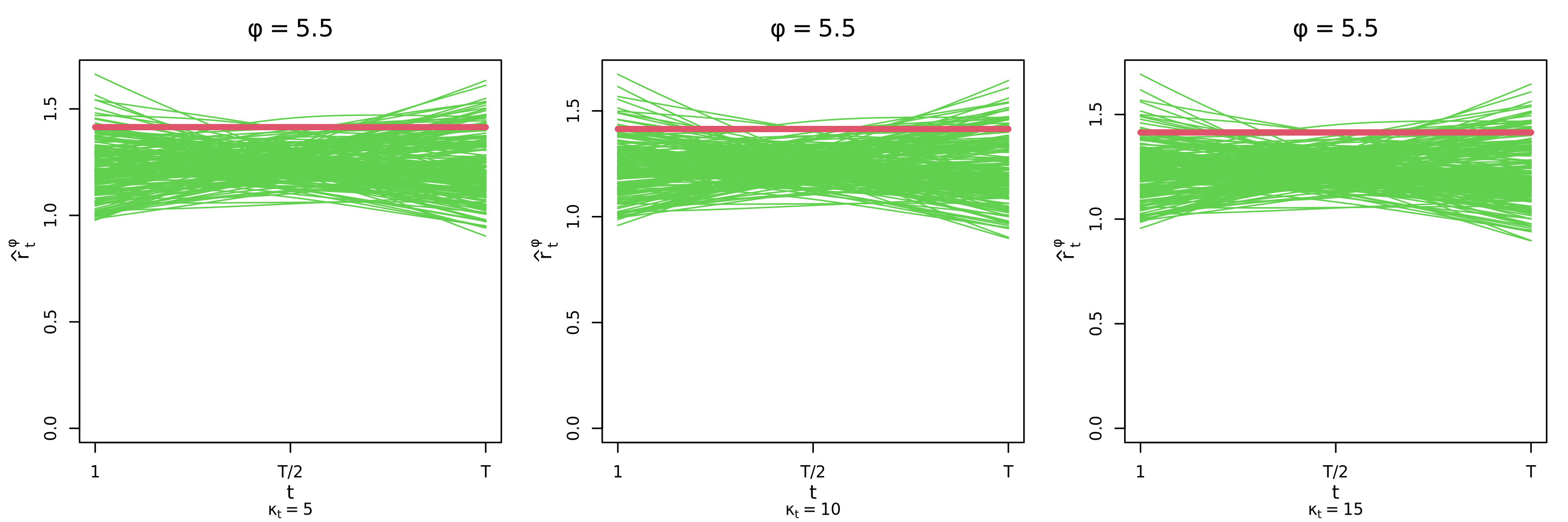}
    \caption{Boundary set radii estimates at $\phi = 7\pi/4$ across $\kappa_t \in \{5,10,15\}$ for the fourth copula example.}
    \label{fig:res_kappa_t_p4_c4}
\end{figure}

\begin{figure}[H]
    \centering
    \includegraphics[width=.8\linewidth]{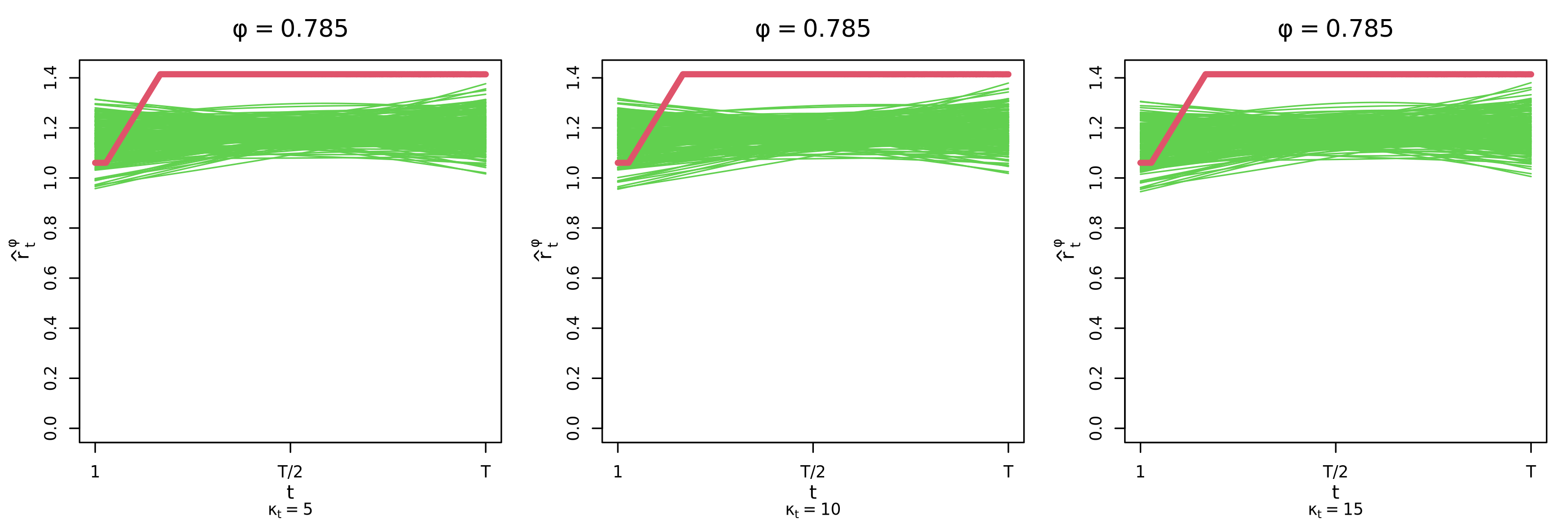}
    \caption{Boundary set radii estimates at $\phi = \pi/4$ across $\kappa_t \in \{5,10,15\}$ for the fifth copula example.}
    \label{fig:res_kappa_t_p1_c5}
\end{figure}

\begin{figure}[H]
    \centering
    \includegraphics[width=.8\linewidth]{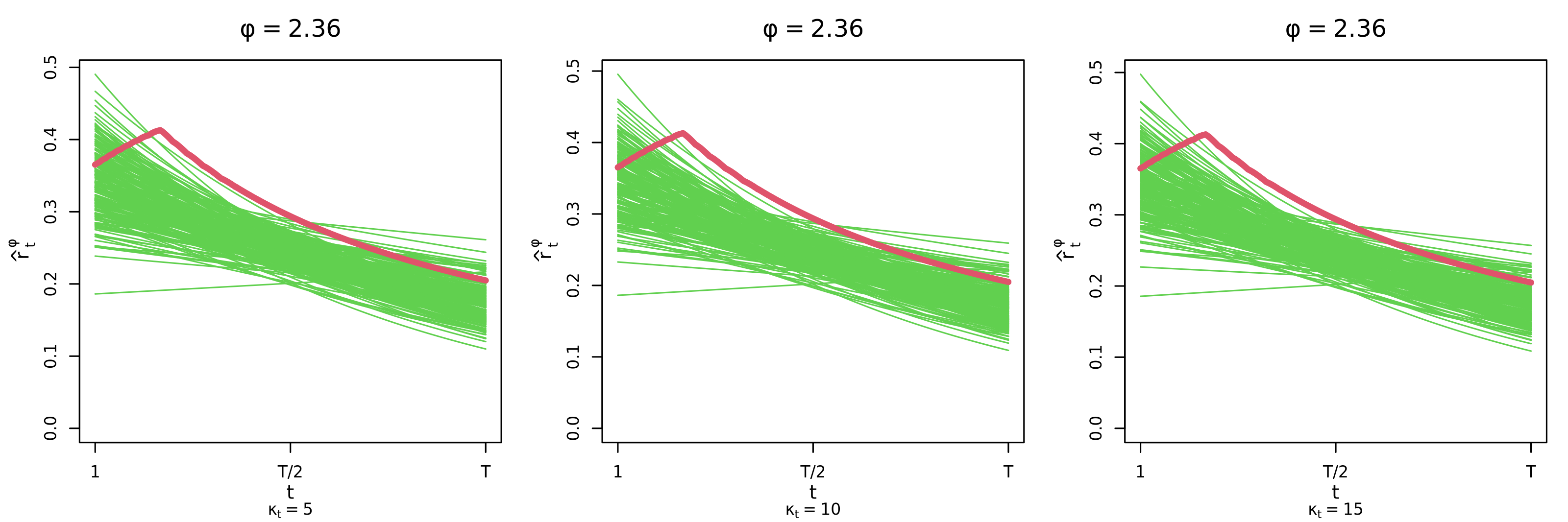}
    \caption{Boundary set radii estimates at $\phi = 3\pi/4$ across $\kappa_t \in \{5,10,15\}$ for the fifth copula example.}
    \label{fig:res_kappa_t_p2_c5}
\end{figure}

\begin{figure}[H]
    \centering
    \includegraphics[width=.8\linewidth]{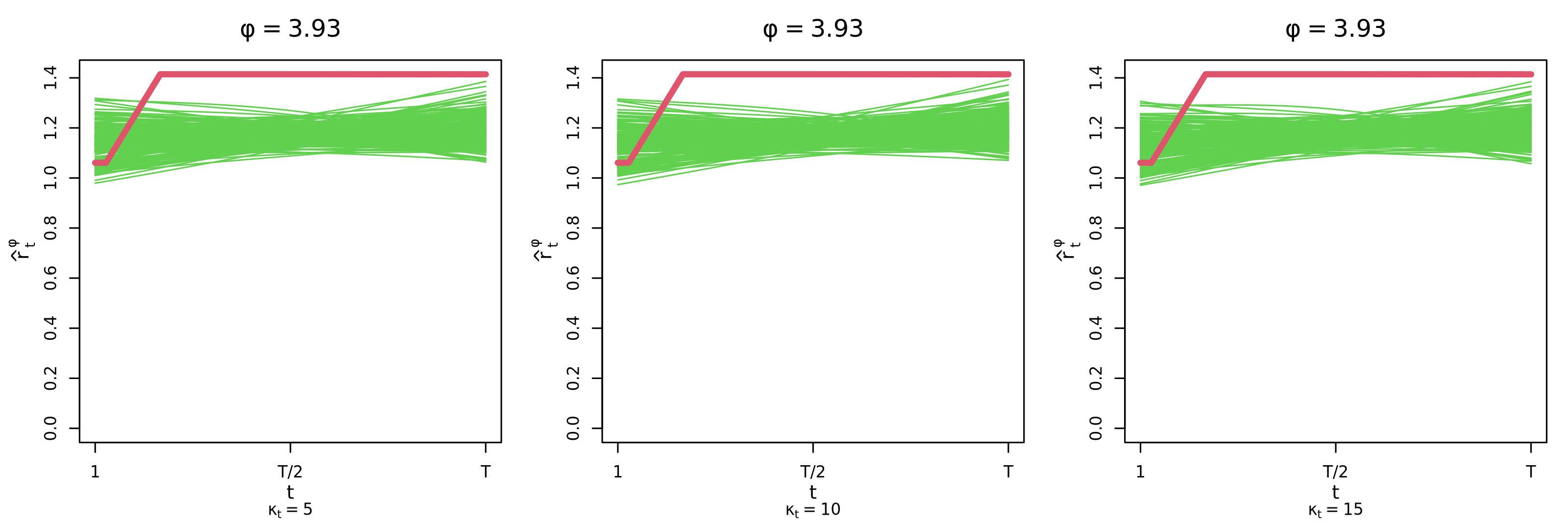}
    \caption{Boundary set radii estimates at $\phi = 5\pi/4$ across $\kappa_t \in \{5,10,15\}$ for the fifth copula example.}
    \label{fig:res_kappa_t_p3_c5}
\end{figure}

\begin{figure}[H]
    \centering
    \includegraphics[width=.8\linewidth]{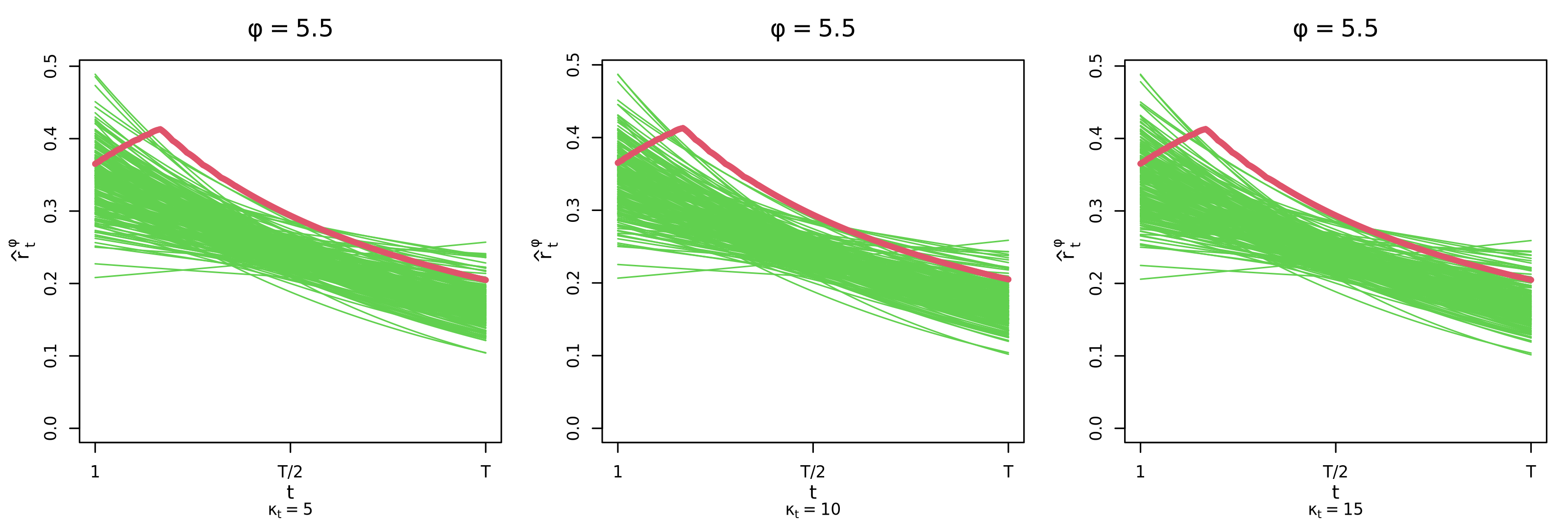}
    \caption{Boundary set radii estimates at $\phi = 7\pi/4$ across $\kappa_t \in \{5,10,15\}$ for the fifth copula example.}
    \label{fig:res_kappa_t_p4_c5}
\end{figure}

\begin{figure}[H]
    \centering
    \includegraphics[width=.6\linewidth]{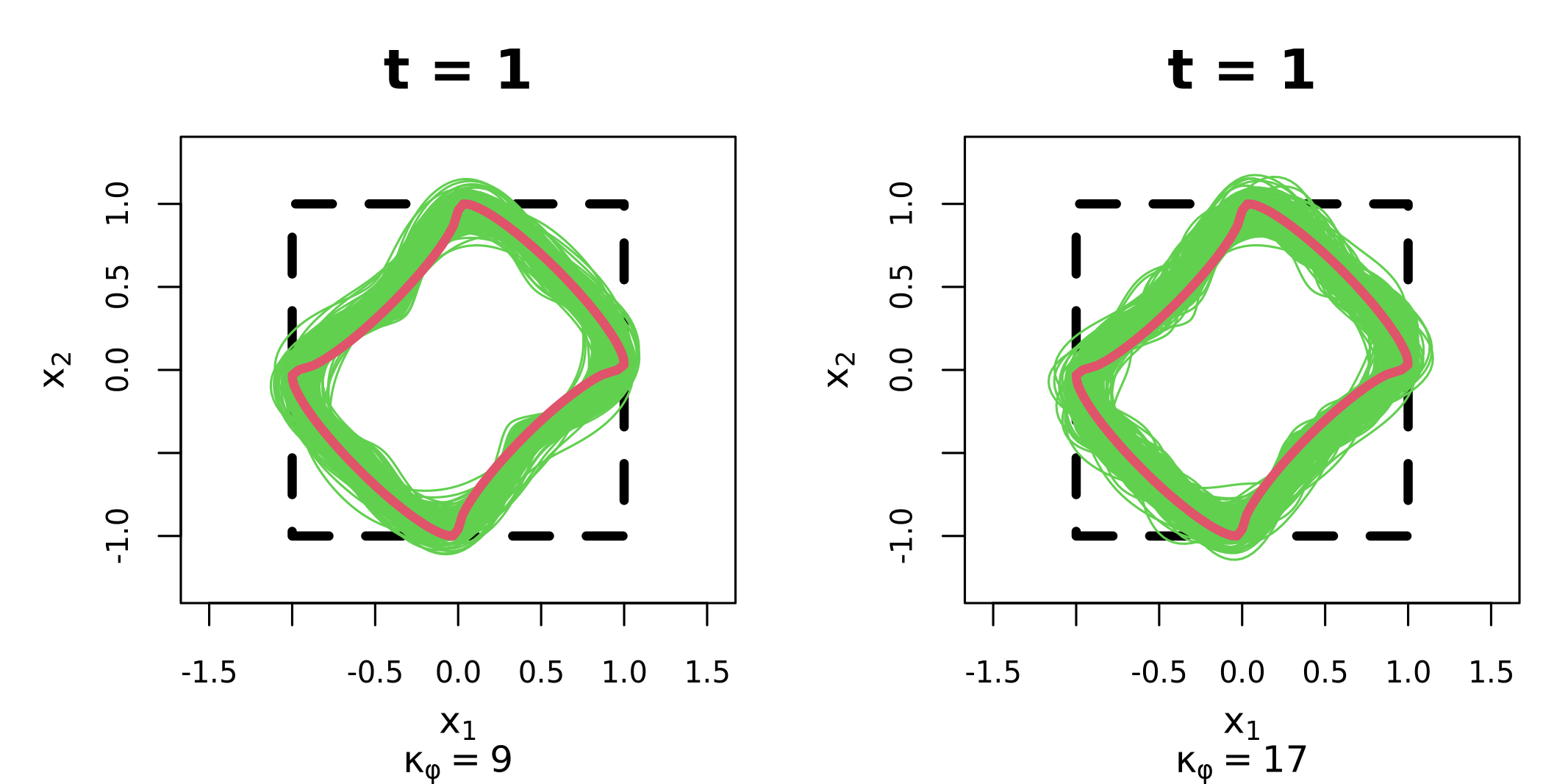}
    \caption{Boundary set estimates as $t = 1$ across $\kappa_{\phi} \in \{9,17\}$ for the first copula example.}
    \label{fig:res_kappa_phi_t1_c1}
\end{figure}

\begin{figure}[H]
    \centering
    \includegraphics[width=.6\linewidth]{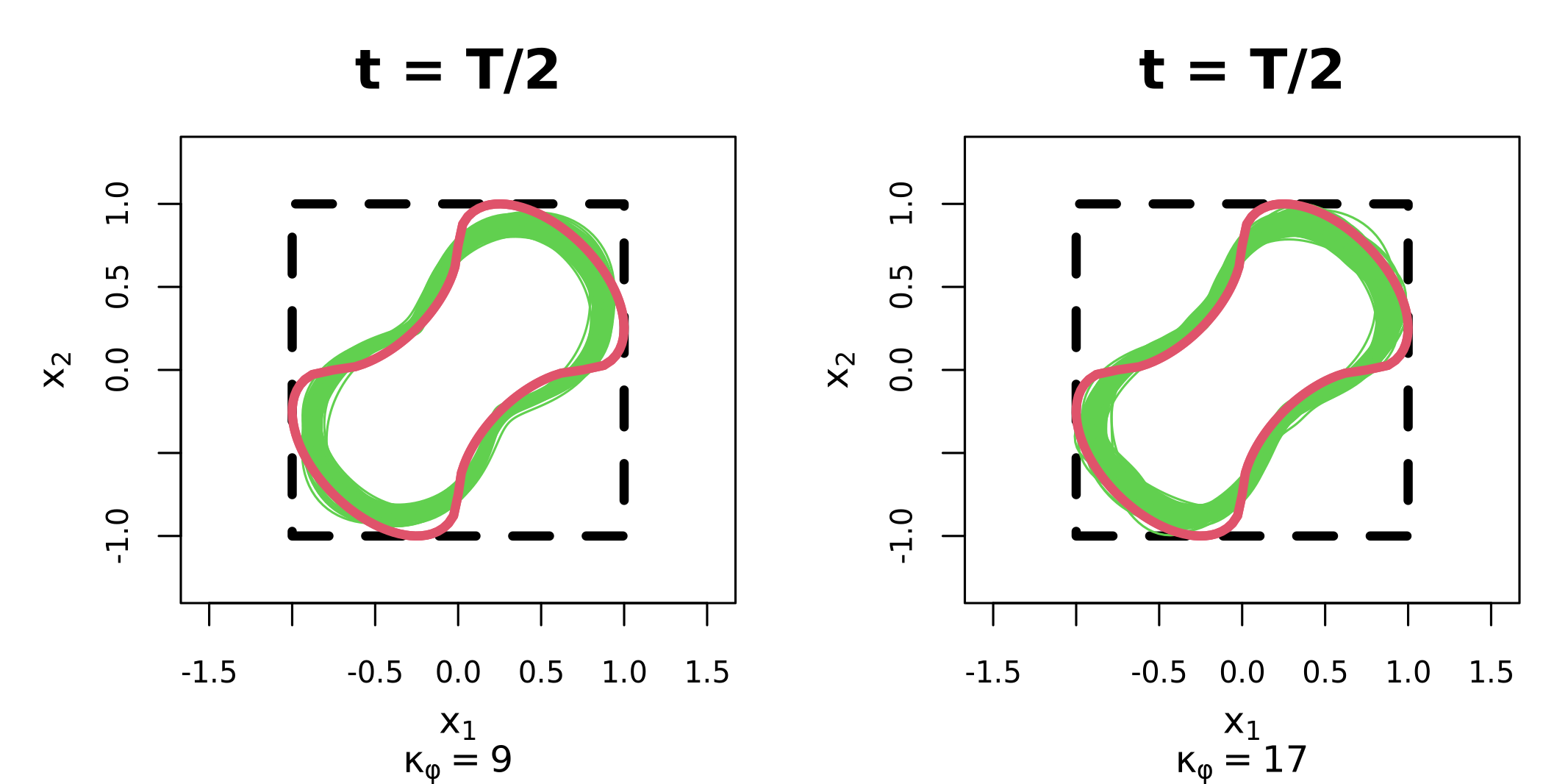}
    \caption{Boundary set estimates as $t = T/2$ across $\kappa_{\phi} \in \{9,17\}$ for the first copula example.}
    \label{fig:res_kappa_phi_t2_c1}
\end{figure}

\begin{figure}[H]
    \centering
    \includegraphics[width=.6\linewidth]{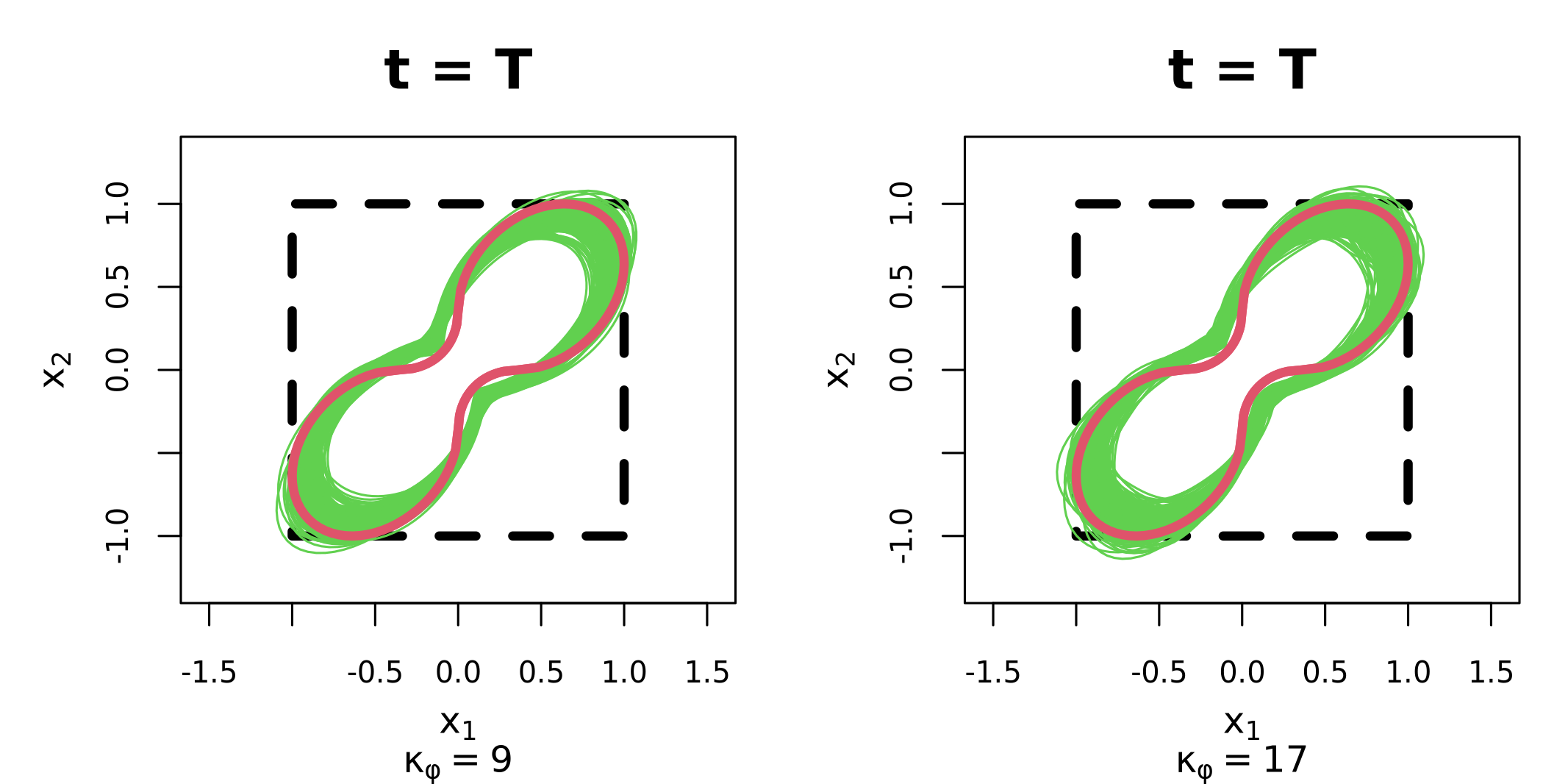}
    \caption{Boundary set estimates at $t = T$ across $\kappa_{\phi} \in \{9,17\}$ for the first copula example.}
    \label{fig:res_kappa_phi_t3_c1}
\end{figure}

\begin{figure}[H]
    \centering
    \includegraphics[width=.6\linewidth]{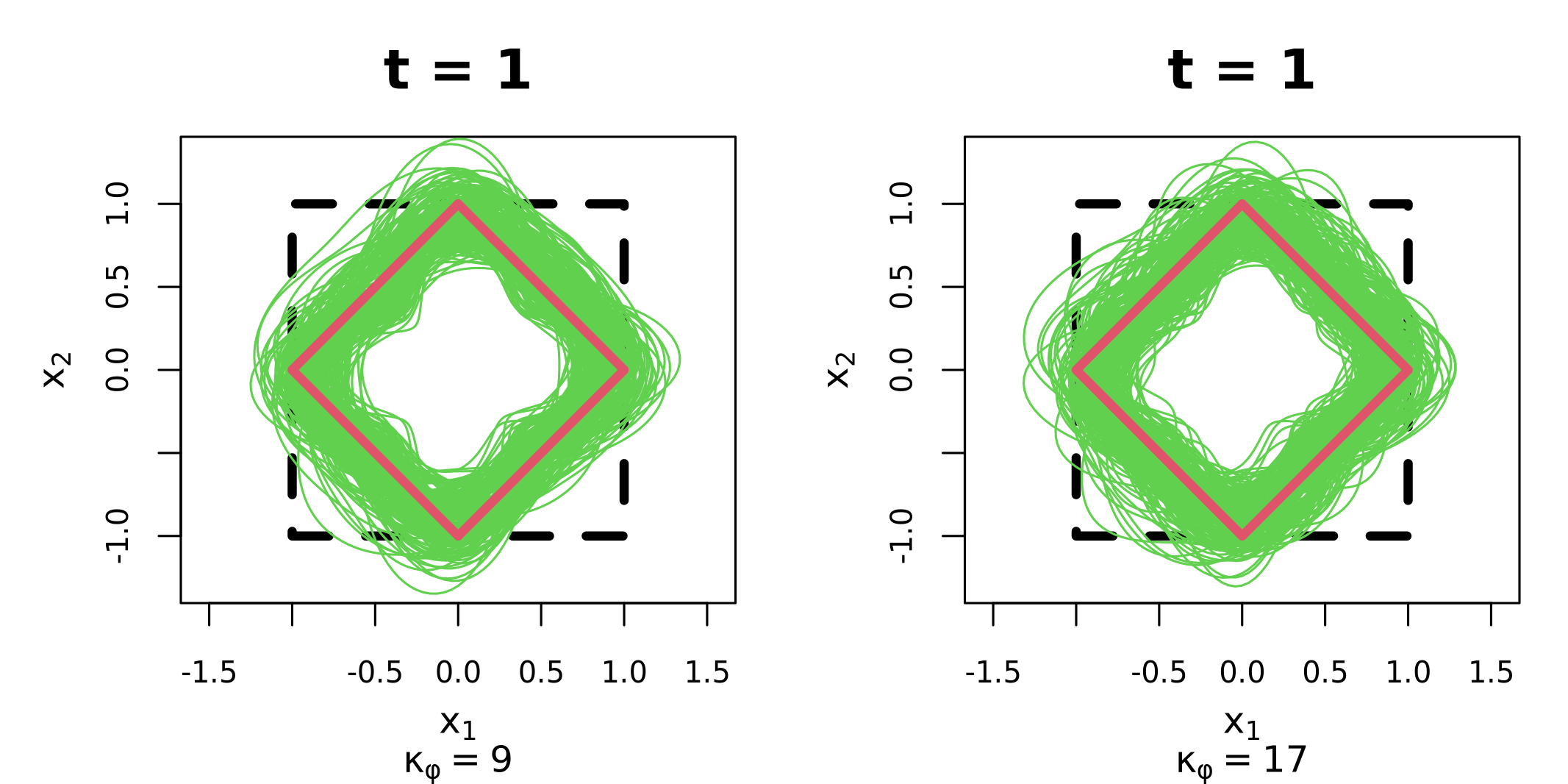}
    \caption{Boundary set estimates as $t = 1$ across $\kappa_{\phi} \in \{9,17\}$ for the second copula example.}
    \label{fig:res_kappa_phi_t1_c2}
\end{figure}

\begin{figure}[H]
    \centering
    \includegraphics[width=.6\linewidth]{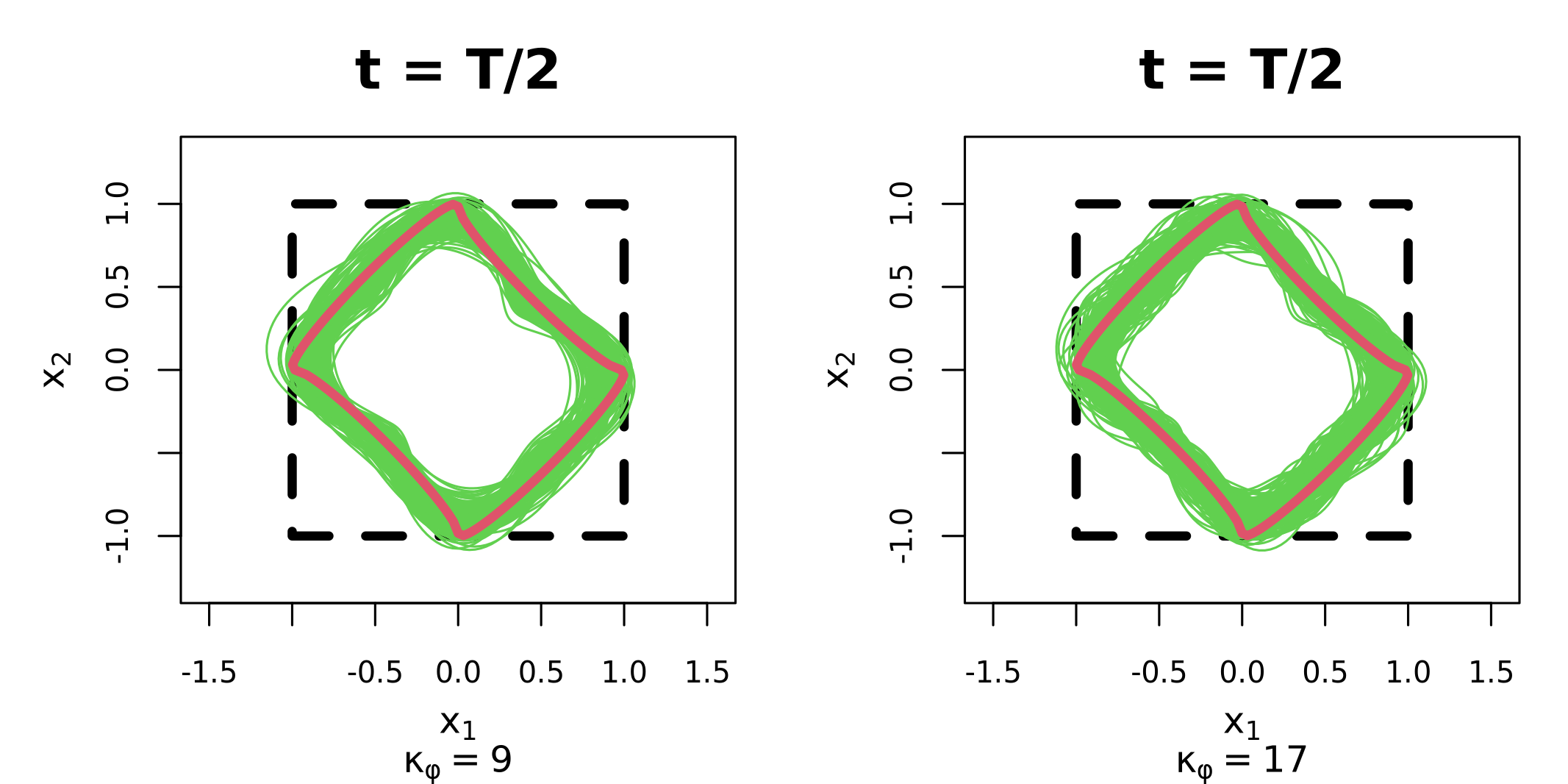}
    \caption{Boundary set estimates as $t = T/2$ across $\kappa_{\phi} \in \{9,17\}$ for the second copula example.}
    \label{fig:res_kappa_phi_t2_c2}
\end{figure}

\begin{figure}[H]
    \centering
    \includegraphics[width=.6\linewidth]{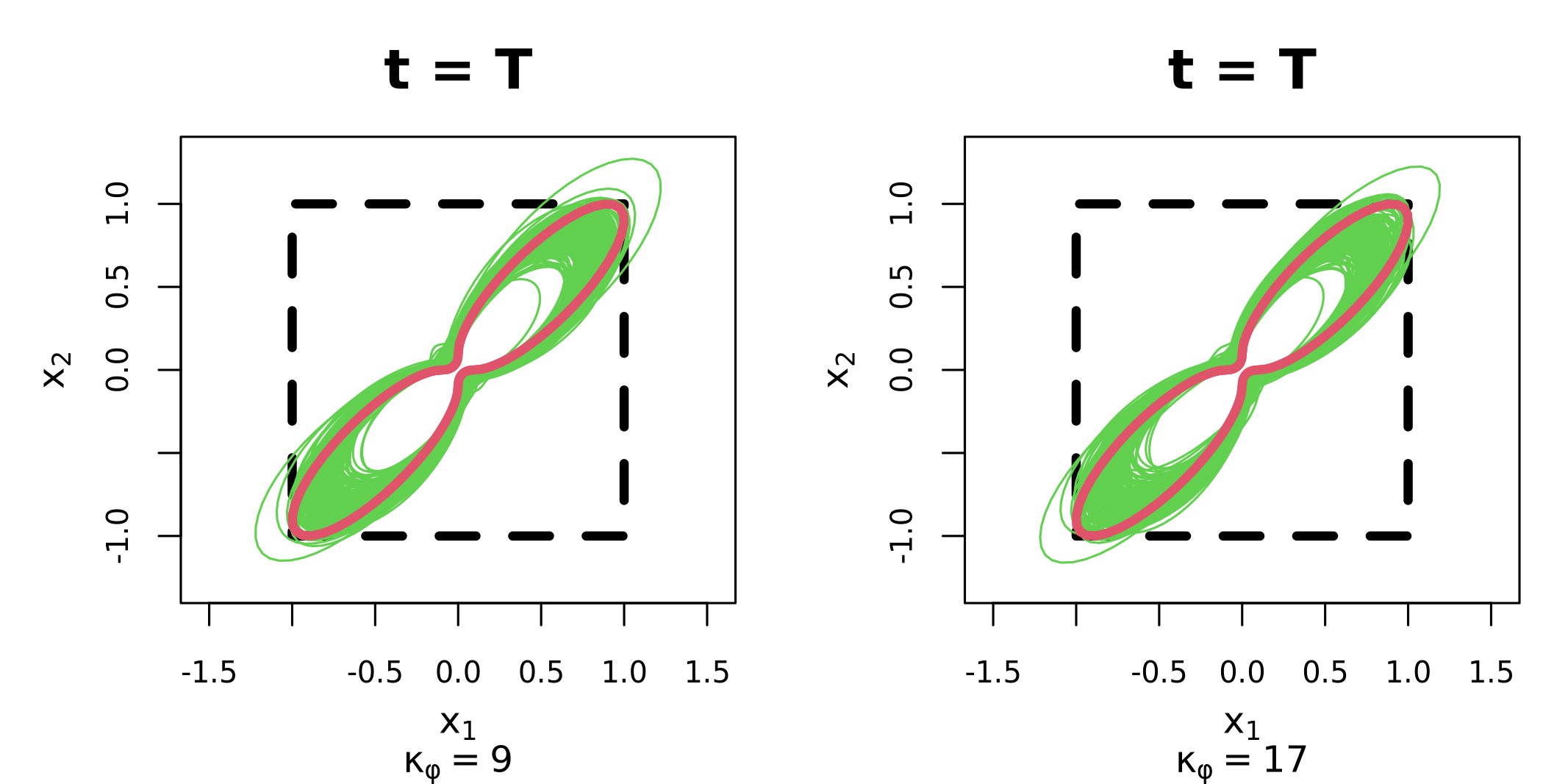}
    \caption{Boundary set estimates at $t = T$ across $\kappa_{\phi} \in \{9,17\}$ for the second copula example.}
    \label{fig:res_kappa_phi_t3_c2}
\end{figure}

\begin{figure}[H]
    \centering
    \includegraphics[width=.6\linewidth]{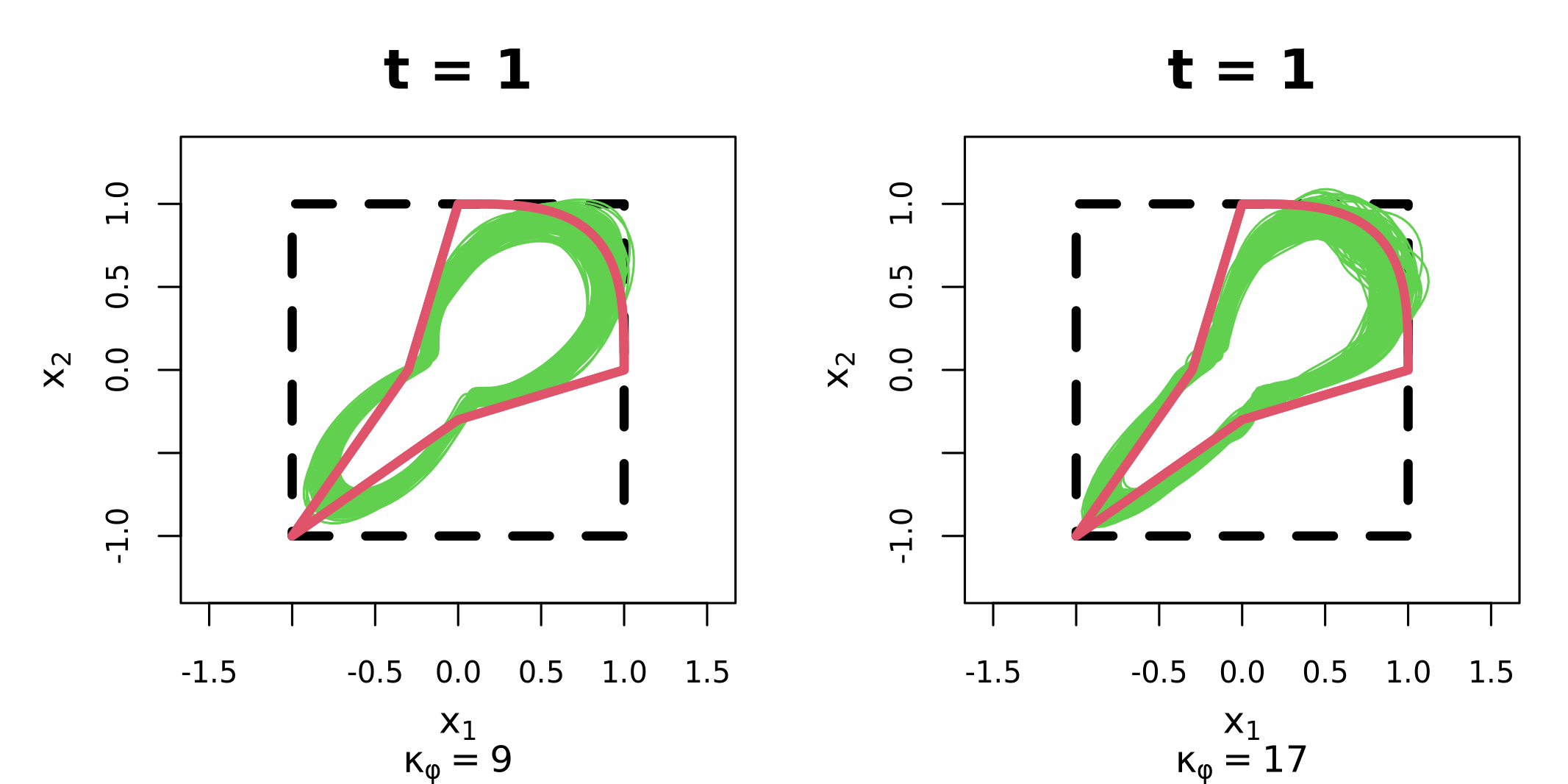}
    \caption{Boundary set estimates as $t = 1$ across $\kappa_{\phi} \in \{9,17\}$ for the third copula example.}
    \label{fig:res_kappa_phi_t1_c3}
\end{figure}

\begin{figure}[H]
    \centering
    \includegraphics[width=.6\linewidth]{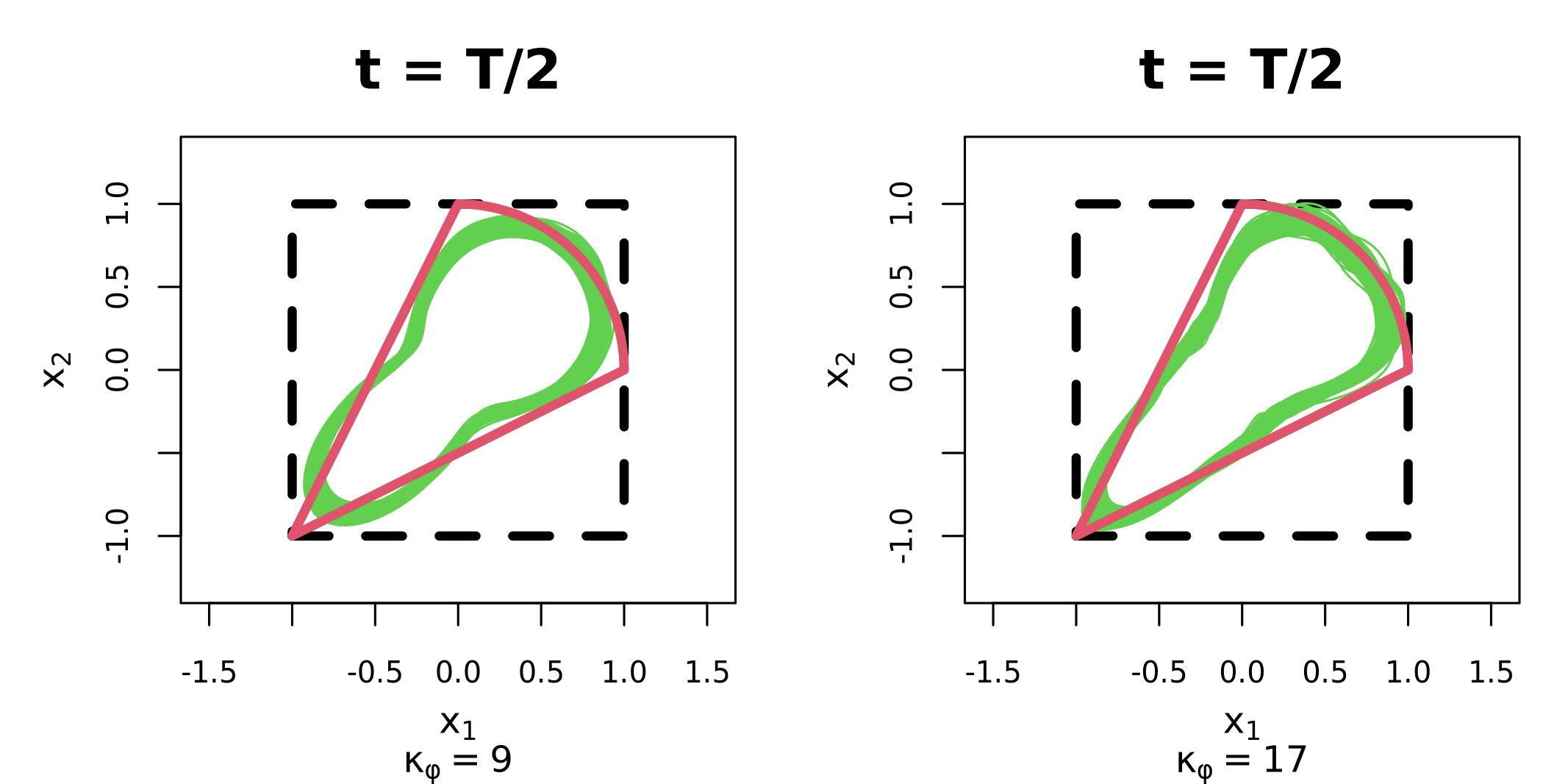}
    \caption{Boundary set estimates as $t = T/2$ across $\kappa_{\phi} \in \{9,17\}$ for the third copula example.}
    \label{fig:res_kappa_phi_t2_c3}
\end{figure}

\begin{figure}[H]
    \centering
    \includegraphics[width=.6\linewidth]{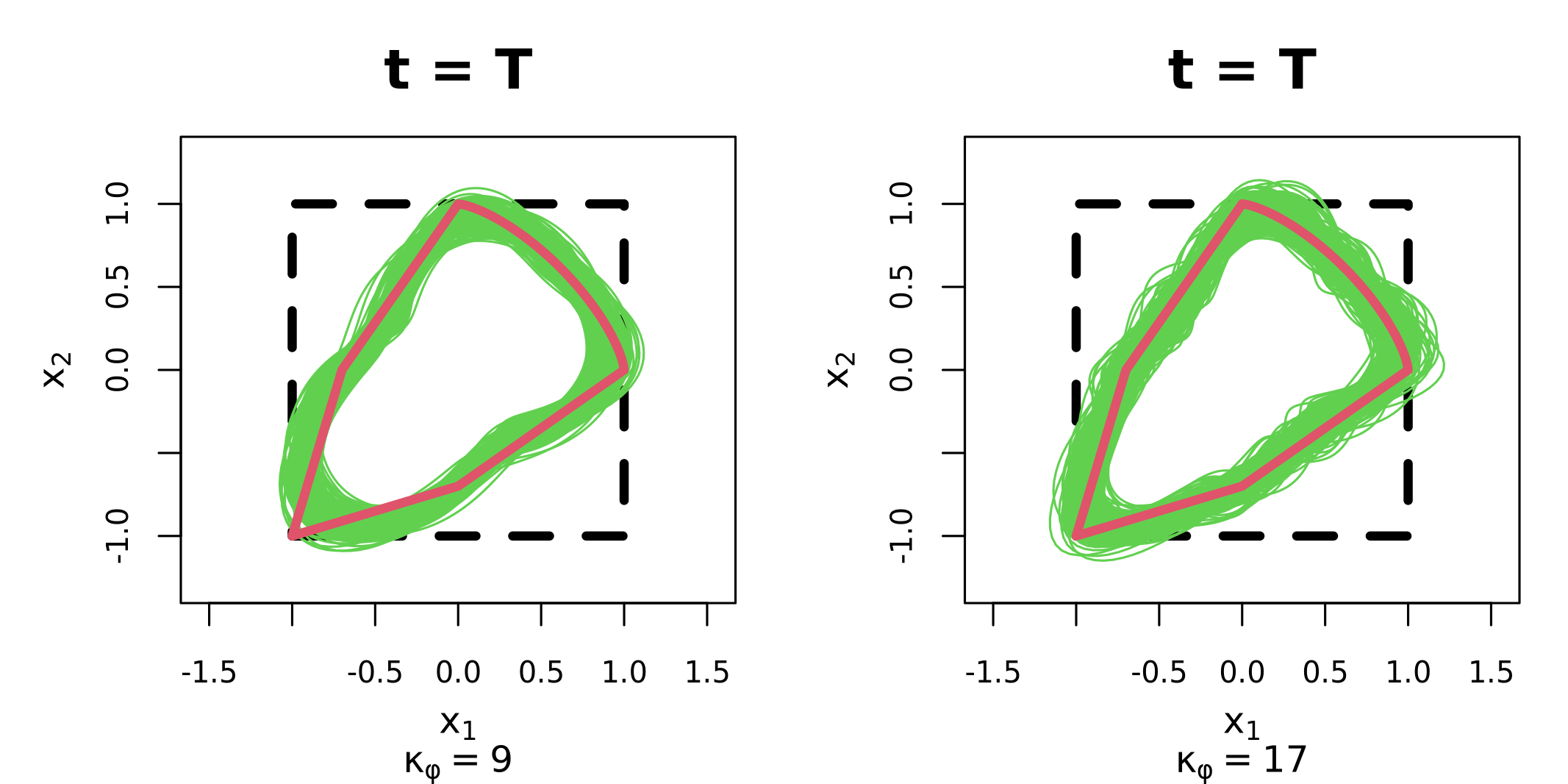}
    \caption{Boundary set estimates at $t = T$ across $\kappa_{\phi} \in \{9,17\}$ for the third copula example.}
    \label{fig:res_kappa_phi_t3_c3}
\end{figure}

\begin{figure}[H]
    \centering
    \includegraphics[width=.6\linewidth]{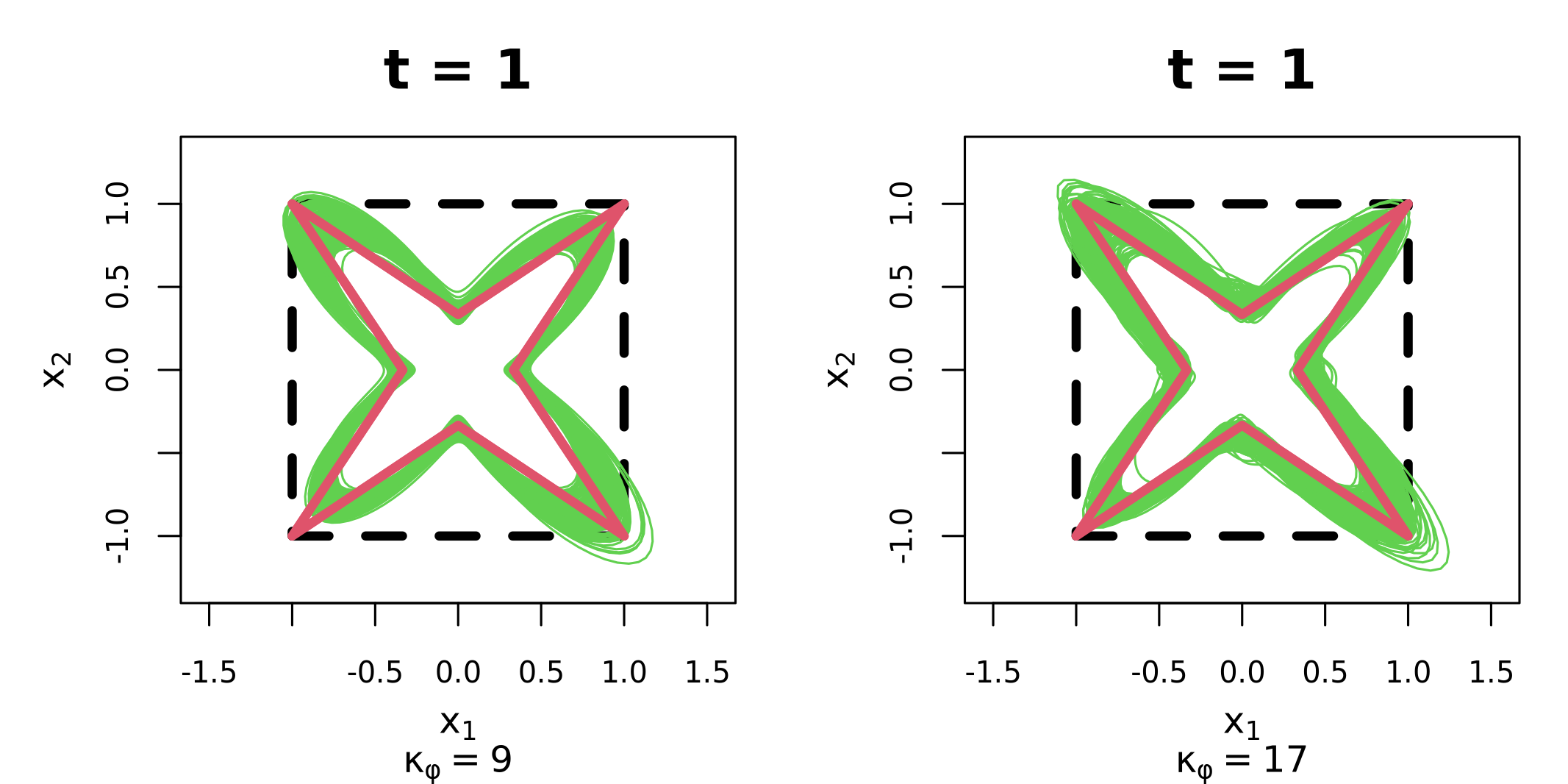}
    \caption{Boundary set estimates as $t = 1$ across $\kappa_{\phi} \in \{9,17\}$ for the fourth copula example.}
    \label{fig:res_kappa_phi_t1_c4}
\end{figure}

\begin{figure}[H]
    \centering
    \includegraphics[width=.6\linewidth]{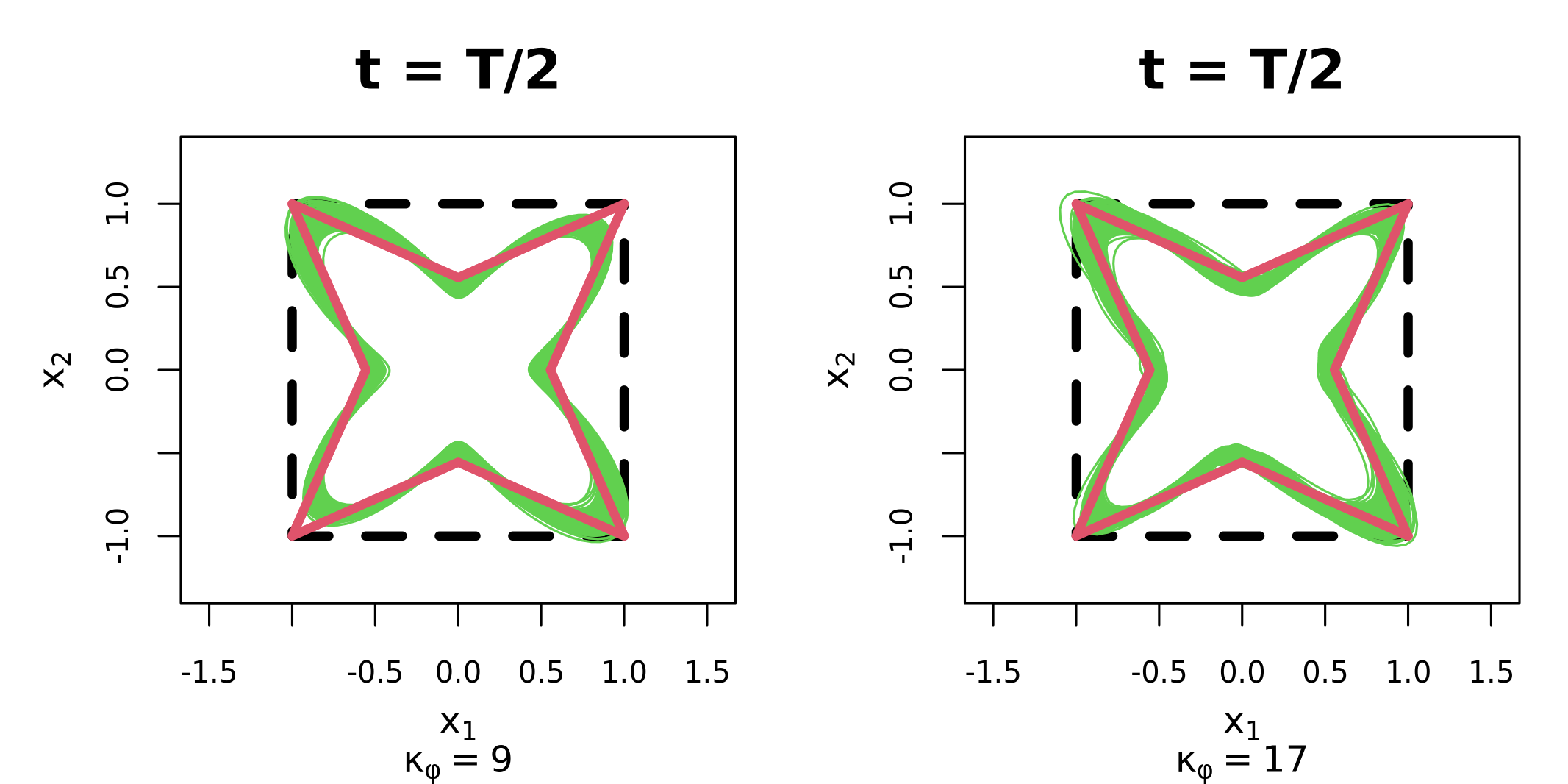}
    \caption{Boundary set estimates as $t = T/2$ across $\kappa_{\phi} \in \{9,17\}$ for the fourth copula example.}
    \label{fig:res_kappa_phi_t2_c4}
\end{figure}

\begin{figure}[H]
    \centering
    \includegraphics[width=.6\linewidth]{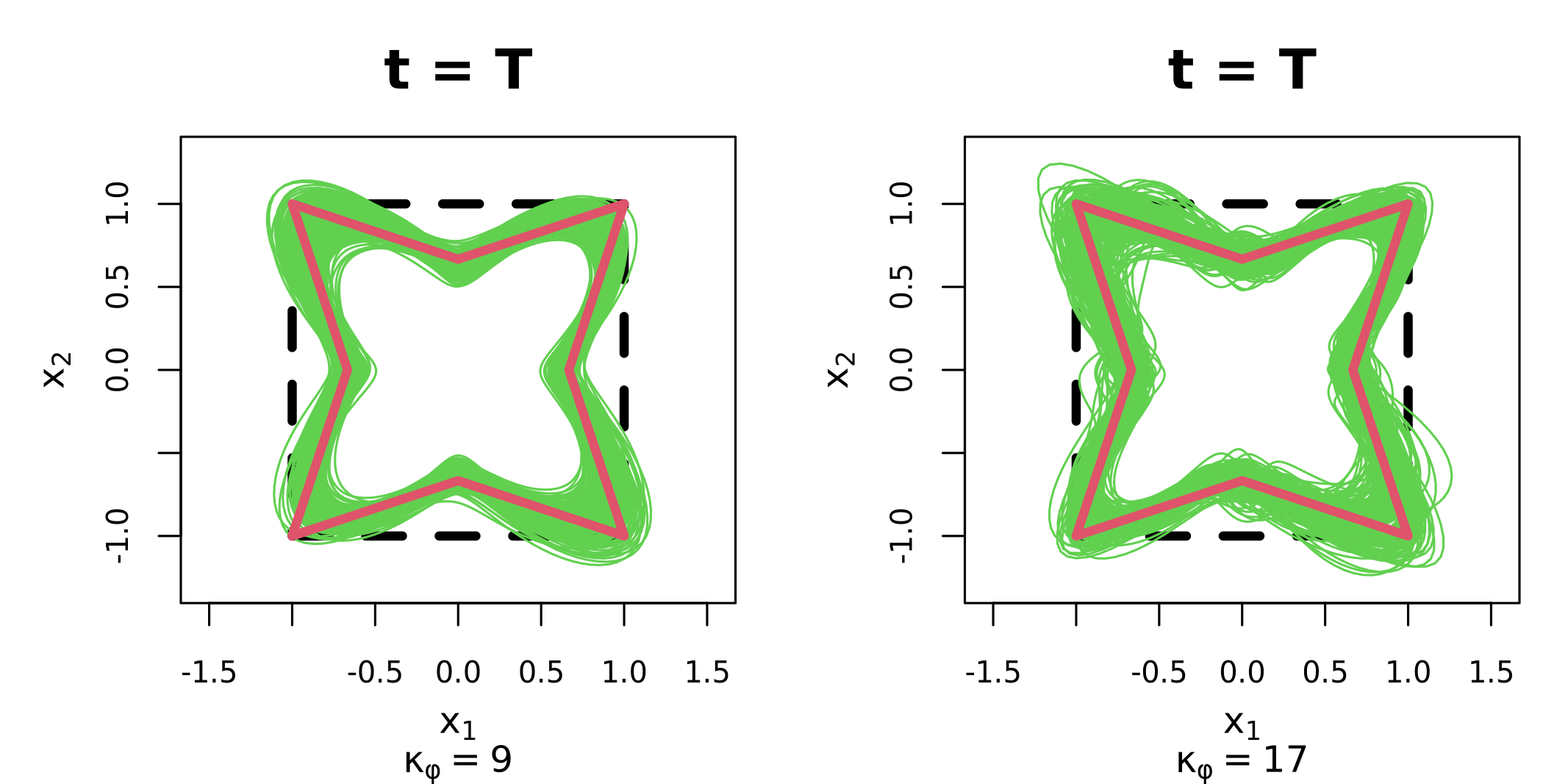}
    \caption{Boundary set estimates at $t = T$ across $\kappa_{\phi} \in \{9,17\}$ for the fourth copula example.}
    \label{fig:res_kappa_phi_t3_c4}
\end{figure}

\begin{figure}[H]
    \centering
    \includegraphics[width=.6\linewidth]{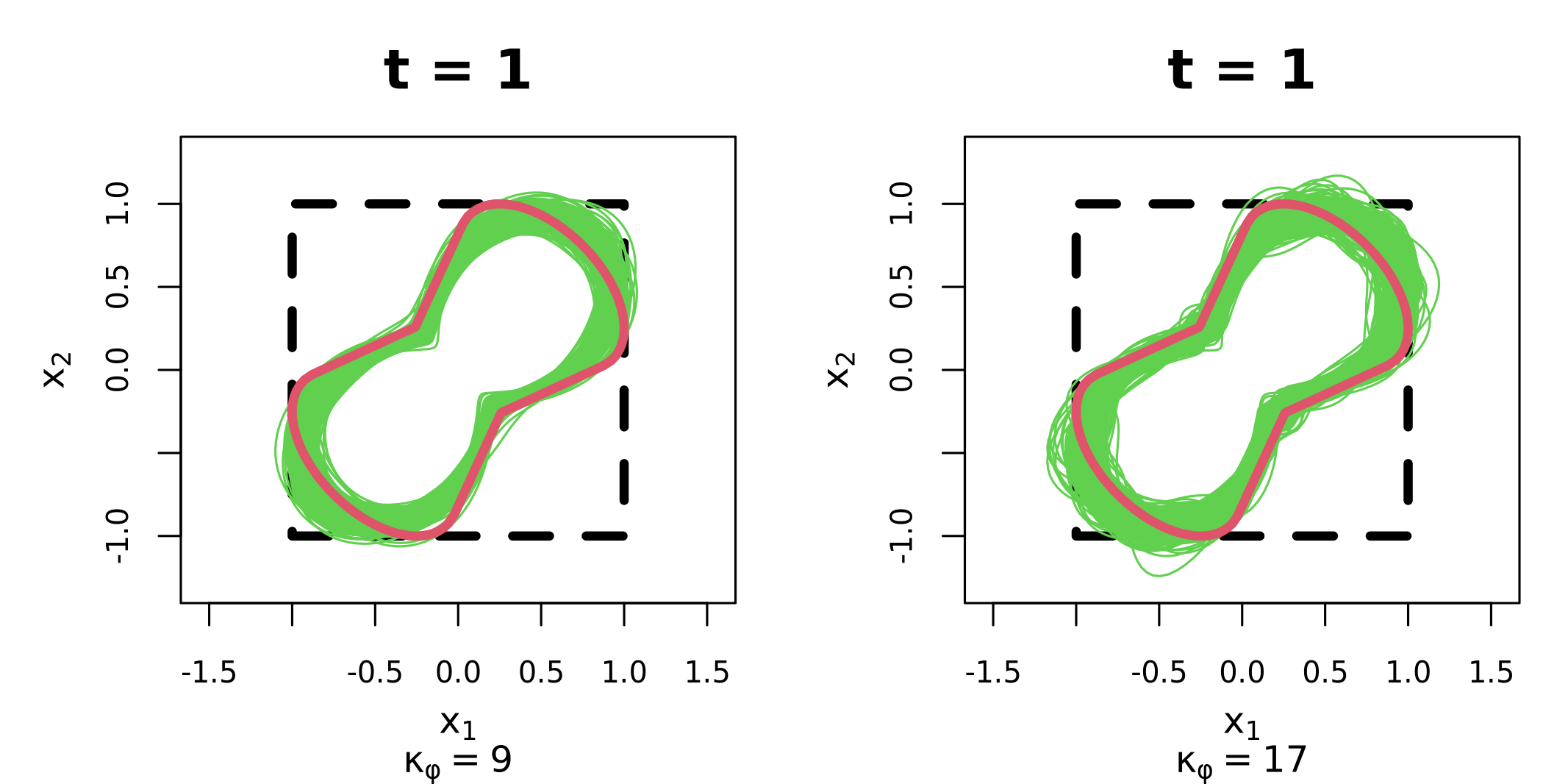}
    \caption{Boundary set estimates as $t = 1$ across $\kappa_{\phi} \in \{9,17\}$ for the fifth copula example.}
    \label{fig:res_kappa_phi_t1_c5}
\end{figure}

\begin{figure}[H]
    \centering
    \includegraphics[width=.6\linewidth]{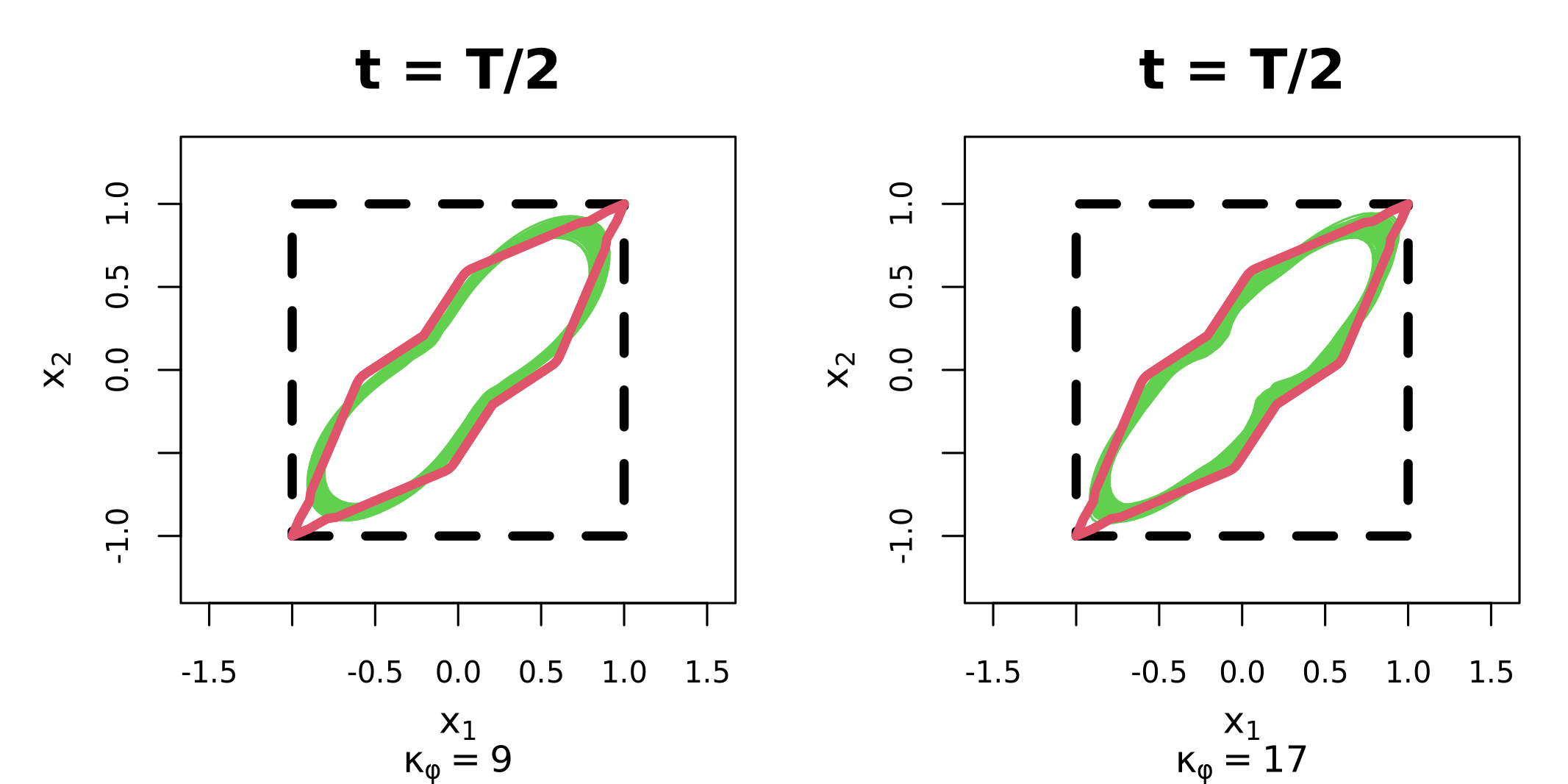}
    \caption{Boundary set estimates as $t = T/2$ across $\kappa_{\phi} \in \{9,17\}$ for the fifth copula example.}
    \label{fig:res_kappa_phi_t2_c5}
\end{figure}

\begin{figure}[H]
    \centering
    \includegraphics[width=.6\linewidth]{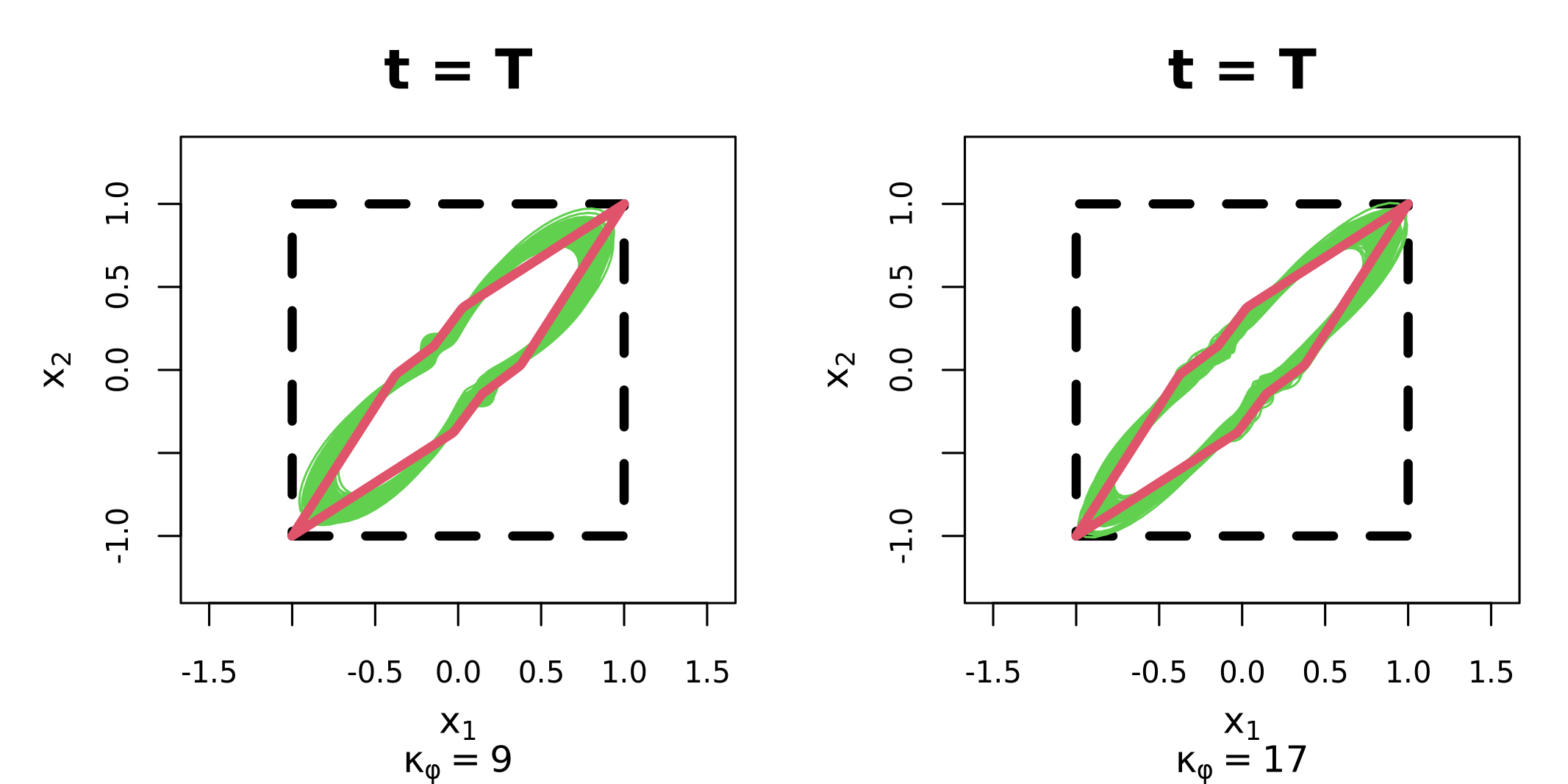}
    \caption{Boundary set estimates at $t = T$ across $\kappa_{\phi} \in \{9,17\}$ for the fifth copula example.}
    \label{fig:res_kappa_phi_t3_c5}
\end{figure}

\begin{figure}[H]
    \centering
    \includegraphics[width=.6\linewidth]{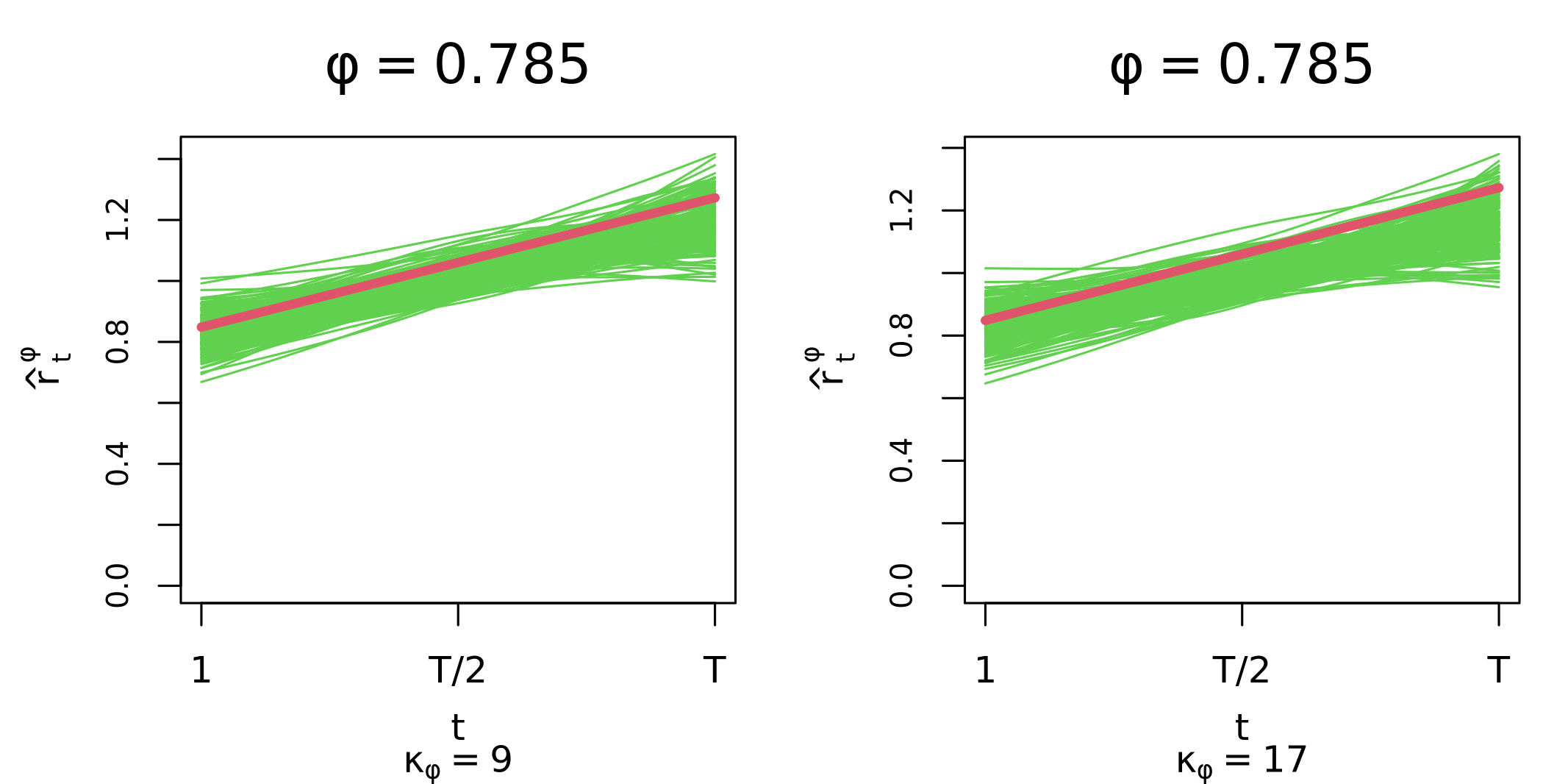}
    \caption{Boundary set radii estimates at $\phi = \pi/4$ across $\kappa_{\phi} \in \{9,17\}$ for the first copula example.}
    \label{fig:res_kappa_phi_p1_c1}
\end{figure}

\begin{figure}[H]
    \centering
    \includegraphics[width=.6\linewidth]{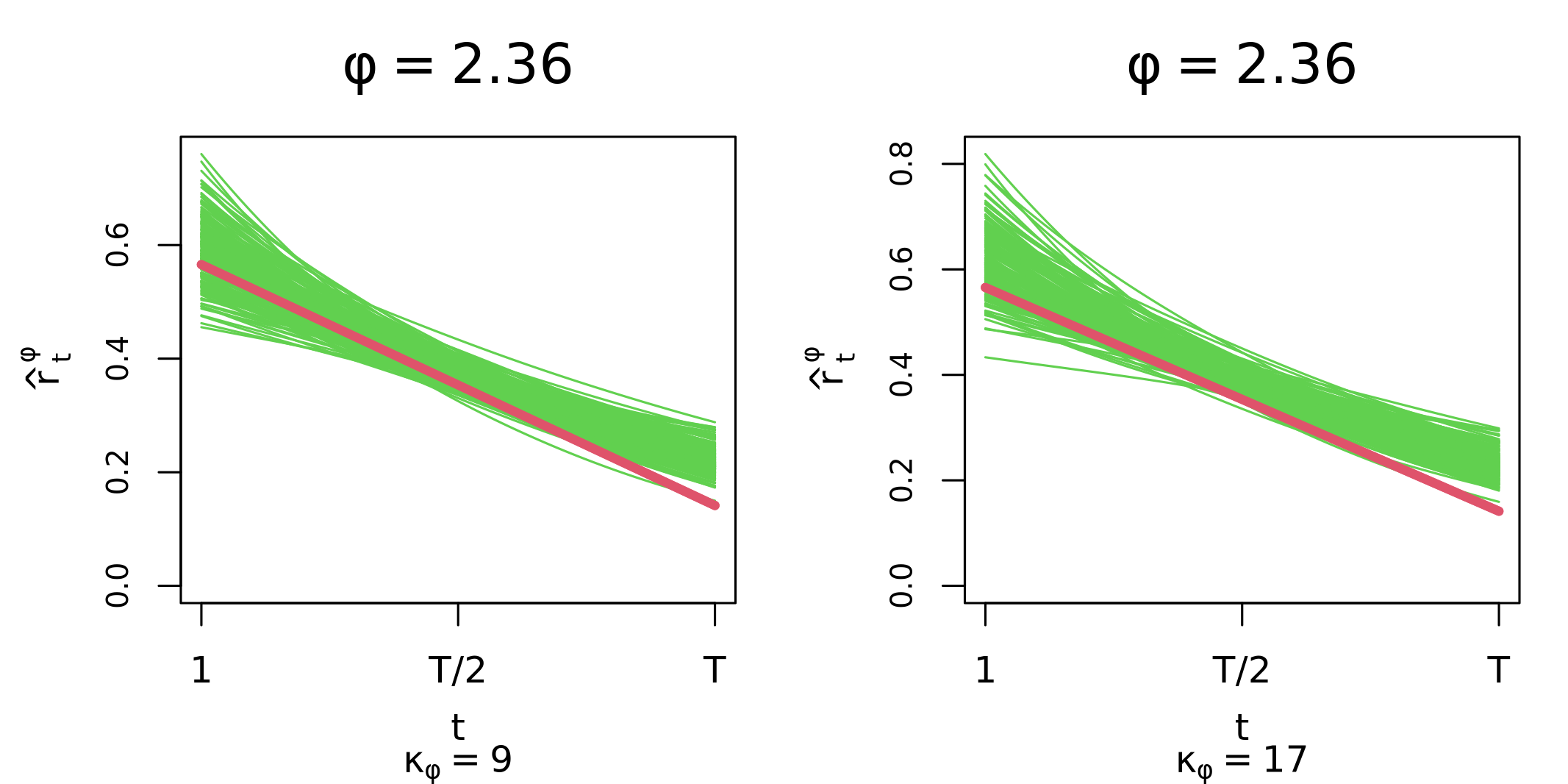}
    \caption{Boundary set radii estimates at $\phi = 3\pi/4$ across $\kappa_{\phi} \in \{9,17\}$ for the first copula example.}
    \label{fig:res_kappa_phi_p2_c1}
\end{figure}

\begin{figure}[H]
    \centering
    \includegraphics[width=.6\linewidth]{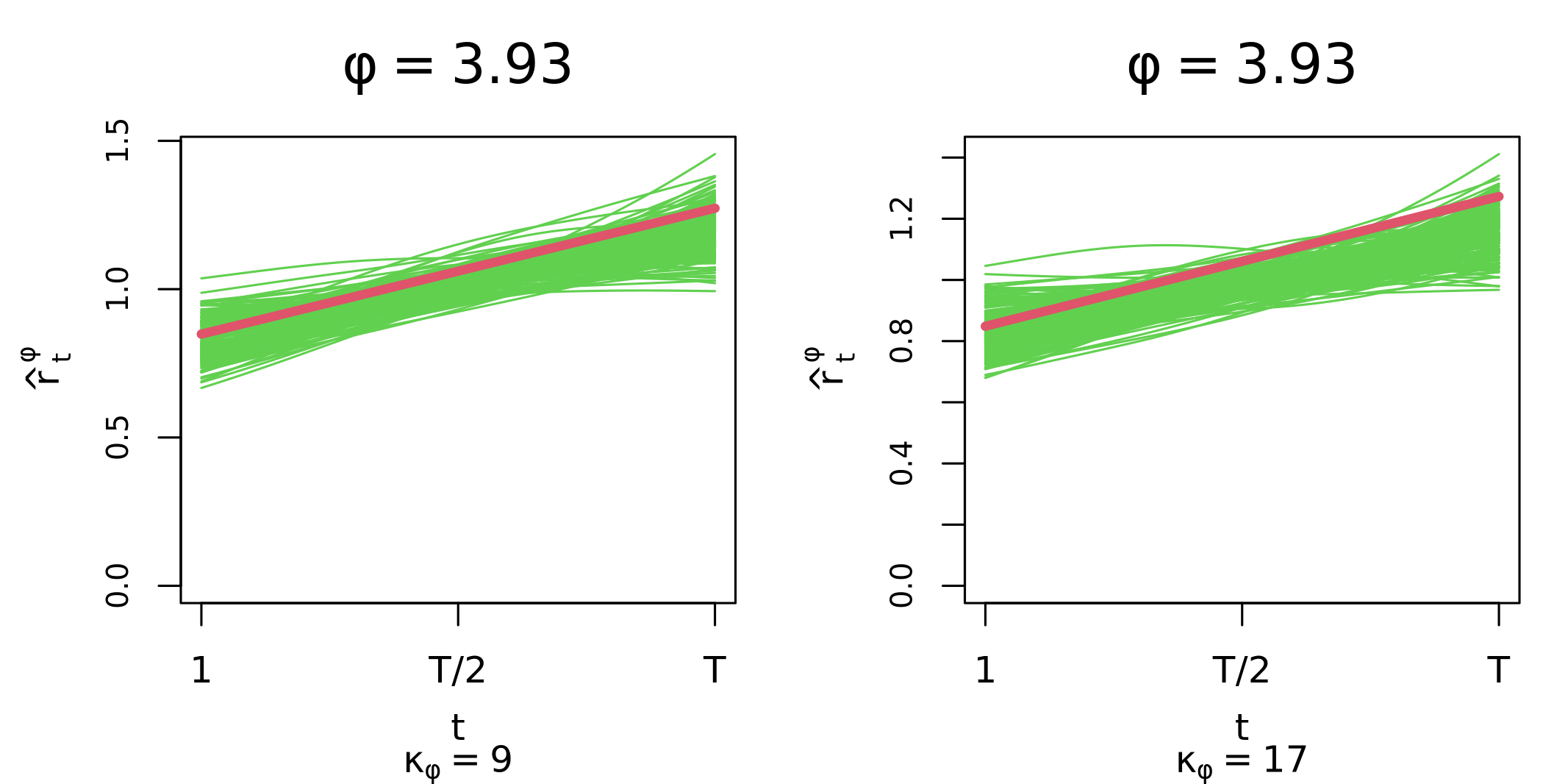}
    \caption{Boundary set radii estimates at $\phi = 5\pi/4$ across $\kappa_{\phi} \in \{9,17\}$ for the first copula example.}
    \label{fig:res_kappa_phi_p3_c1}
\end{figure}

\begin{figure}[H]
    \centering
    \includegraphics[width=.6\linewidth]{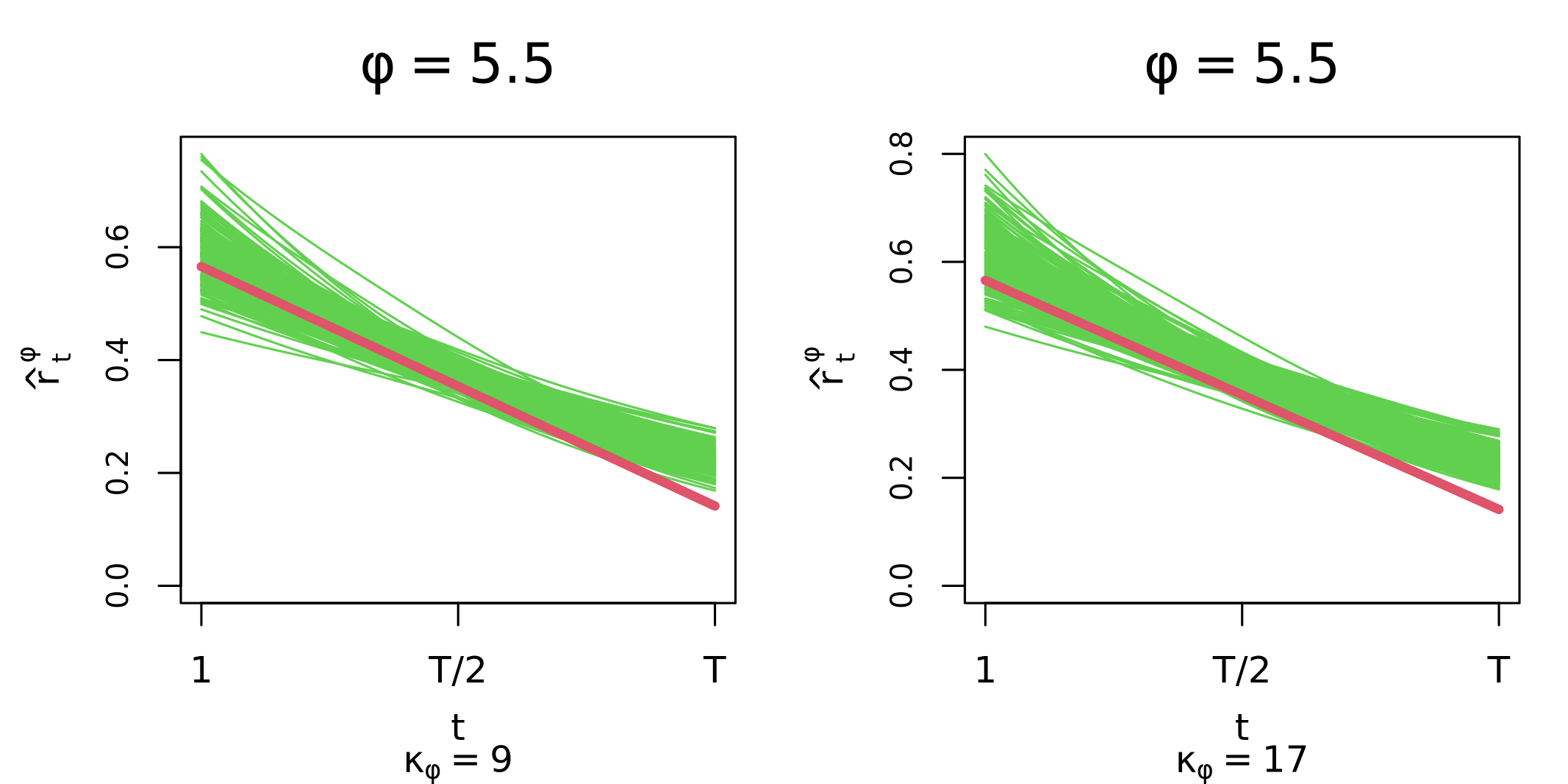}
    \caption{Boundary set radii estimates at $\phi = 7\pi/4$ across $\kappa_{\phi} \in \{9,17\}$ for the first copula example.}
    \label{fig:res_kappa_phi_p4_c1}
\end{figure}

\begin{figure}[H]
    \centering
    \includegraphics[width=.6\linewidth]{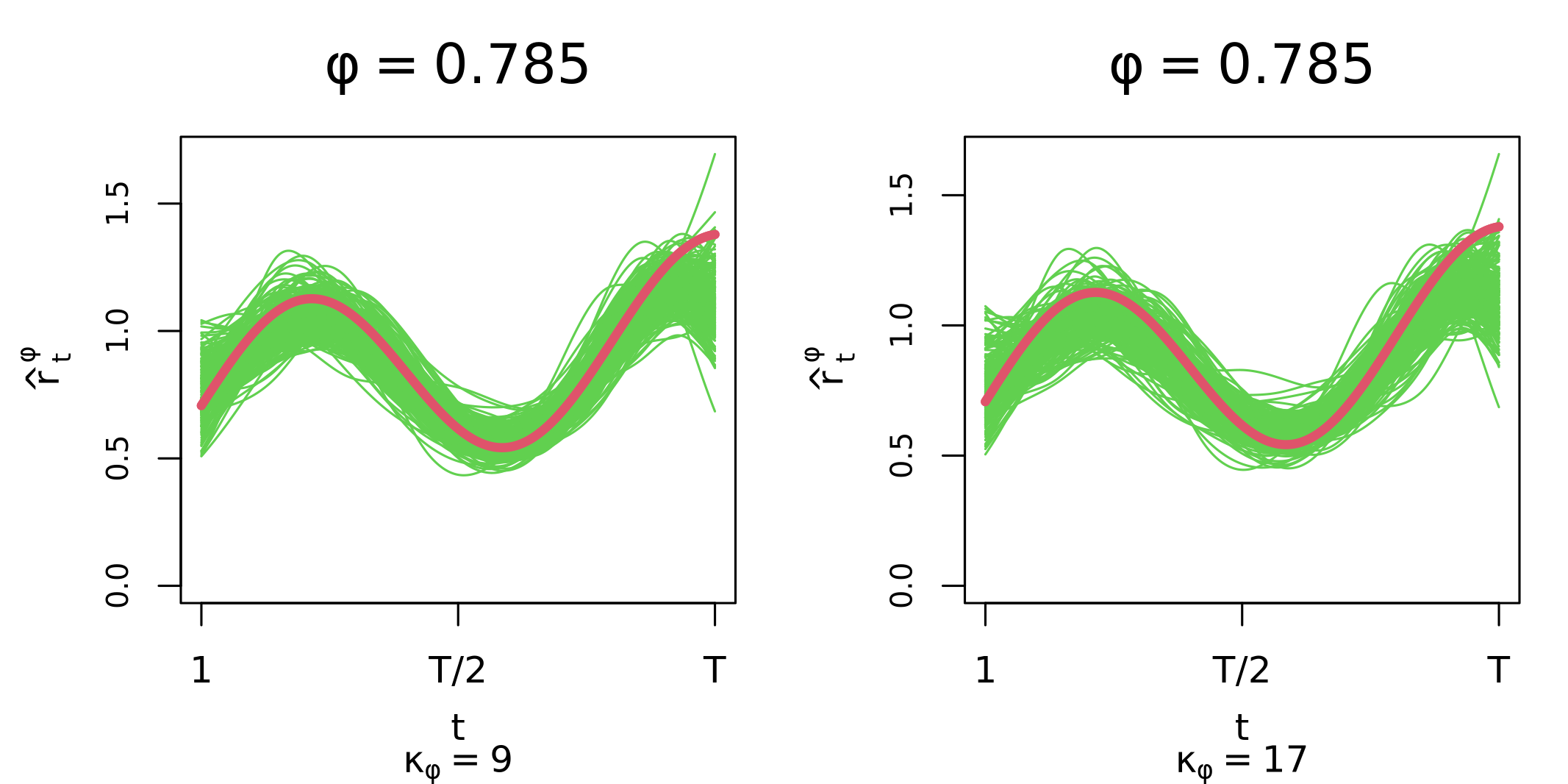}
    \caption{Boundary set radii estimates at $\phi = \pi/4$ across $\kappa_{\phi} \in \{9,17\}$ for the second copula example.}
    \label{fig:res_kappa_phi_p1_c2}
\end{figure}

\begin{figure}[H]
    \centering
    \includegraphics[width=.6\linewidth]{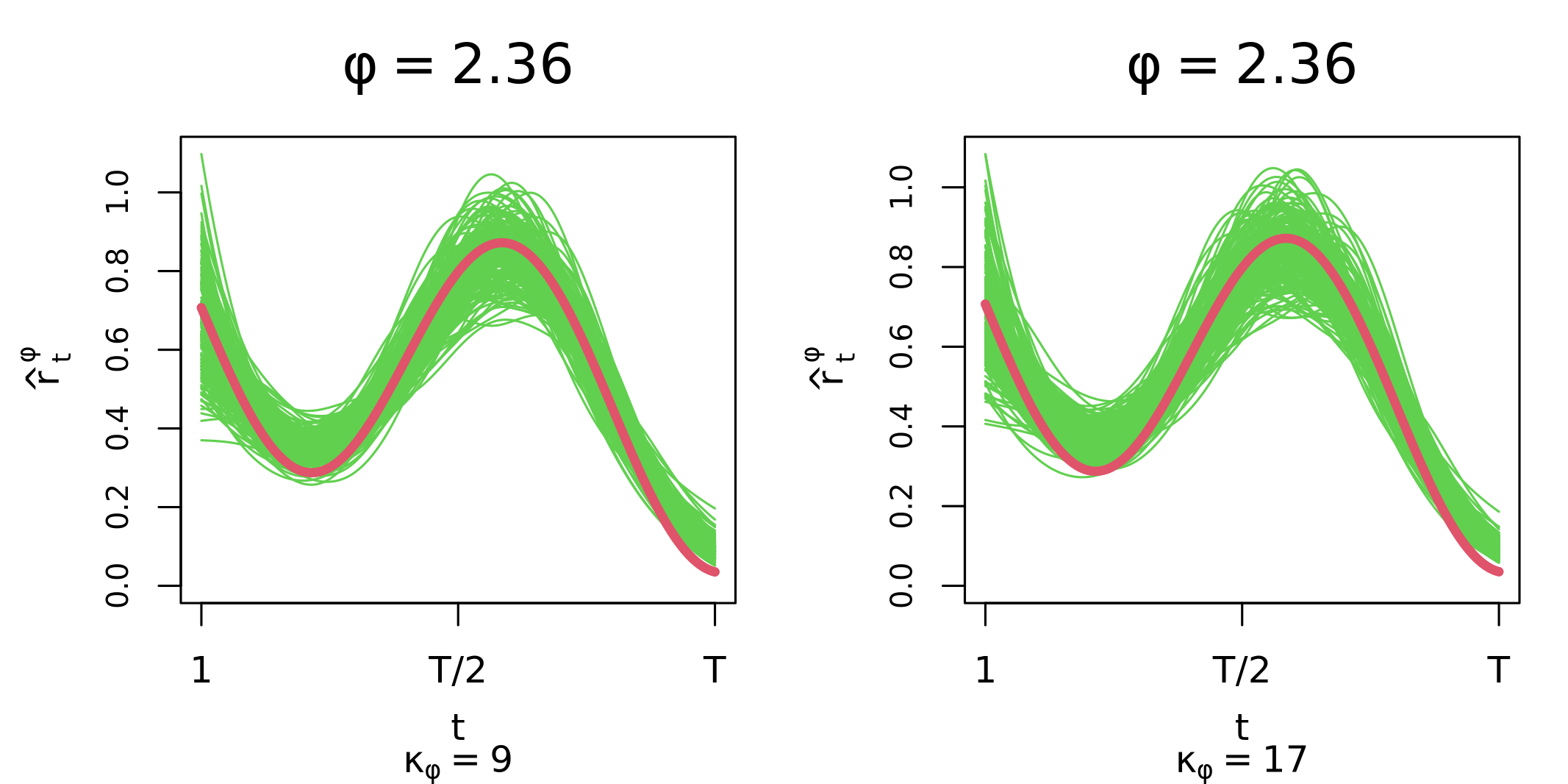}
    \caption{Boundary set radii estimates at $\phi = 3\pi/4$ across $\kappa_{\phi} \in \{9,17\}$ for the second copula example.}
    \label{fig:res_kappa_phi_p2_c2}
\end{figure}

\begin{figure}[H]
    \centering
    \includegraphics[width=.6\linewidth]{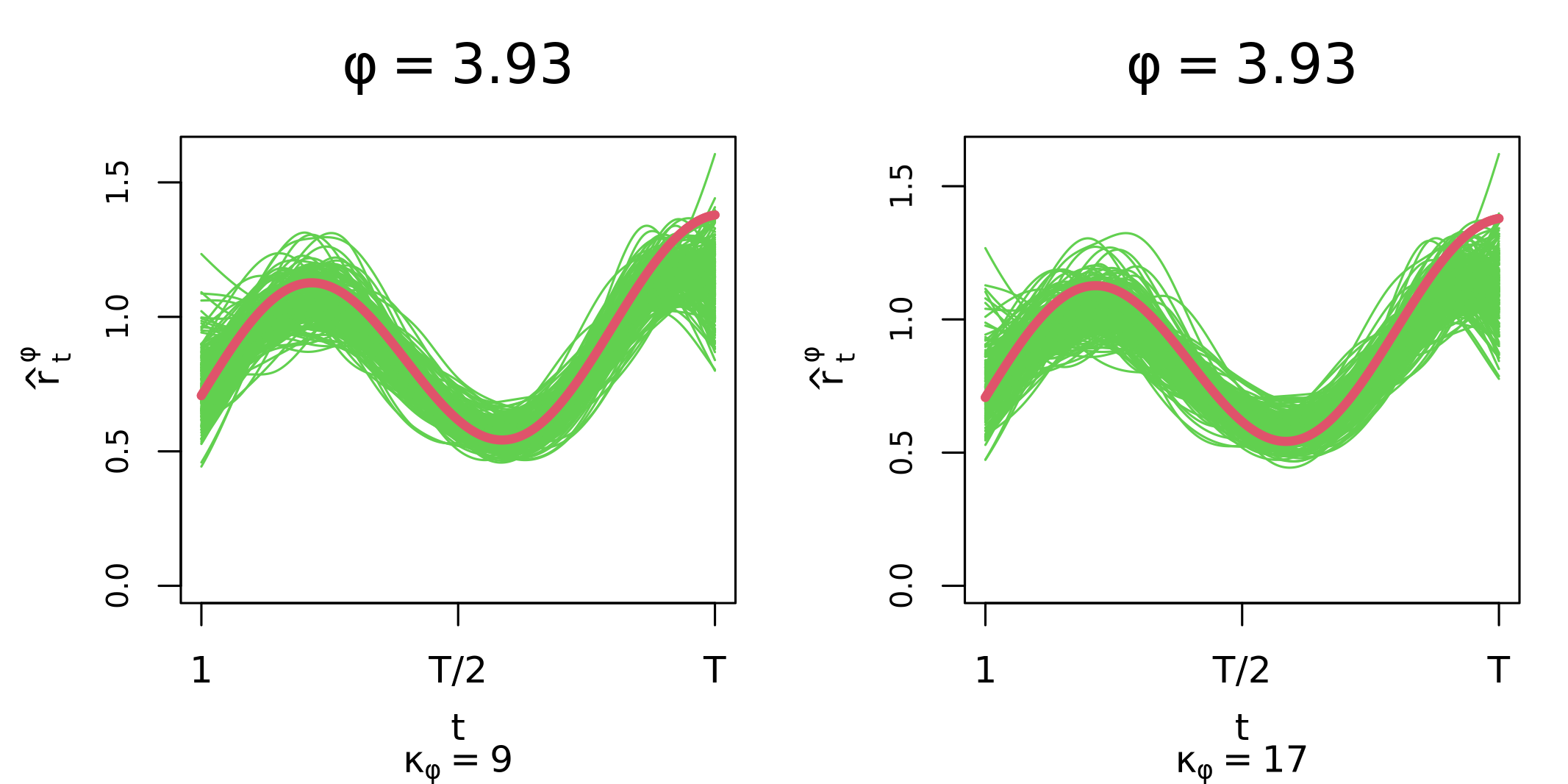}
    \caption{Boundary set radii estimates at $\phi = 5\pi/4$ across $\kappa_{\phi} \in \{9,17\}$ for the second copula example.}
    \label{fig:res_kappa_phi_p3_c2}
\end{figure}

\begin{figure}[H]
    \centering
    \includegraphics[width=.6\linewidth]{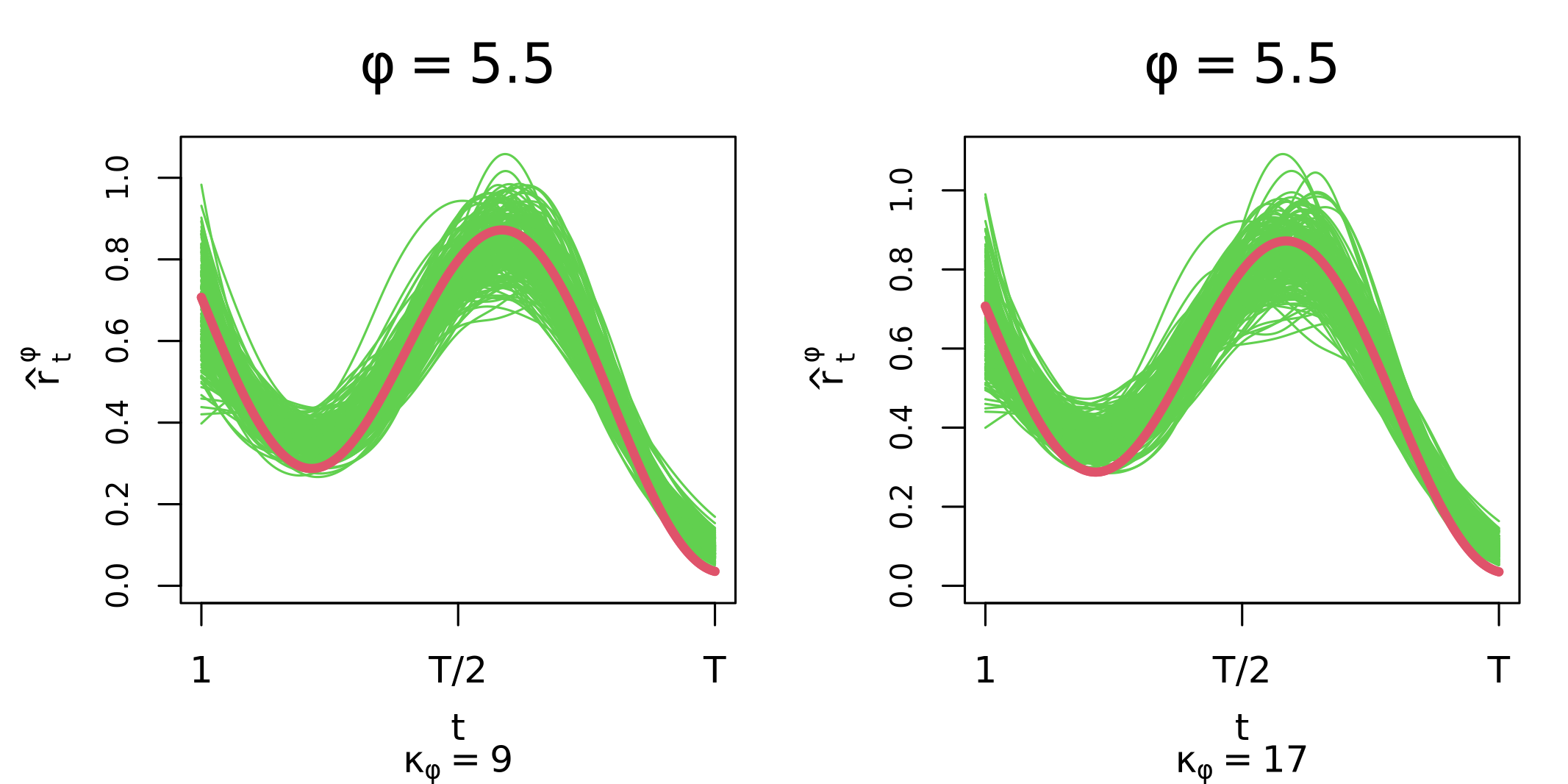}
    \caption{Boundary set radii estimates at $\phi = 7\pi/4$ across $\kappa_{\phi} \in \{9,17\}$ for the second copula example.}
    \label{fig:res_kappa_phi_p4_c2}
\end{figure}

\begin{figure}[H]
    \centering
    \includegraphics[width=.6\linewidth]{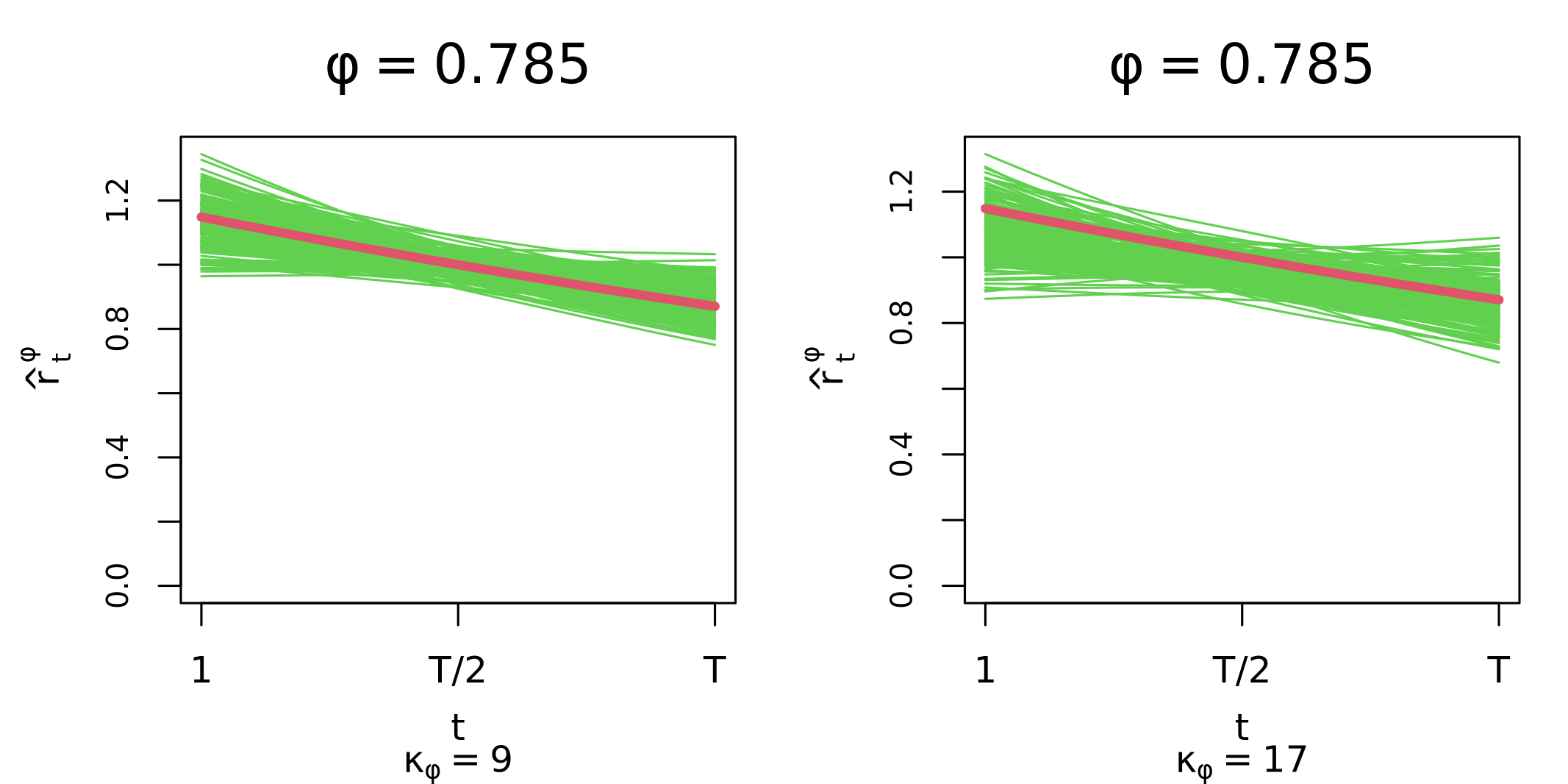}
    \caption{Boundary set radii estimates at $\phi = \pi/4$ across $\kappa_{\phi} \in \{9,17\}$ for the third copula example.}
    \label{fig:res_kappa_phi_p1_c3}
\end{figure}

\begin{figure}[H]
    \centering
    \includegraphics[width=.6\linewidth]{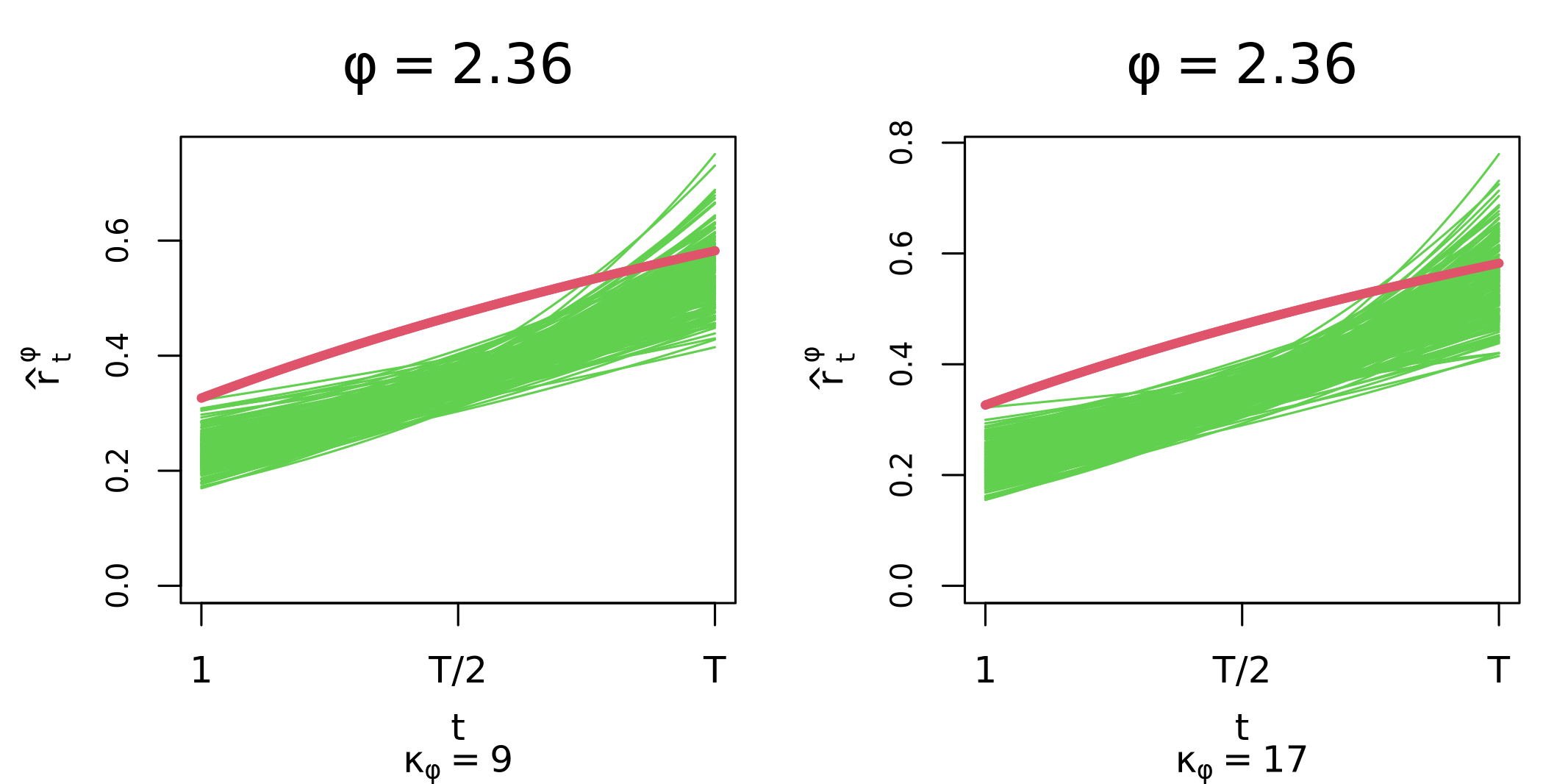}
    \caption{Boundary set radii estimates at $\phi = 3\pi/4$ across $\kappa_{\phi} \in \{9,17\}$ for the third copula example.}
    \label{fig:res_kappa_phi_p2_c3}
\end{figure}

\begin{figure}[H]
    \centering
    \includegraphics[width=.6\linewidth]{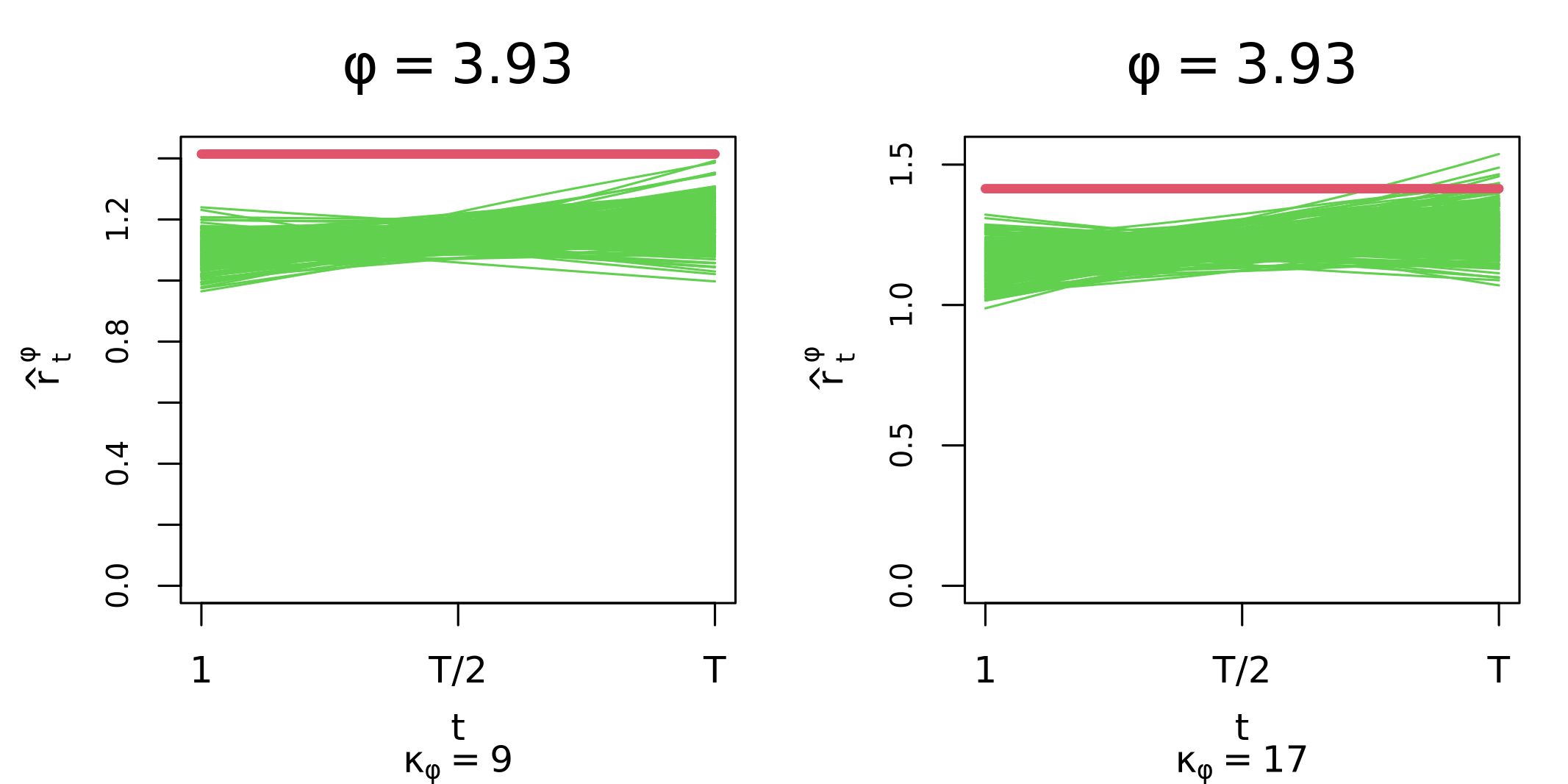}
    \caption{Boundary set radii estimates at $\phi = 5\pi/4$ across $\kappa_{\phi} \in \{9,17\}$ for the third copula example.}
    \label{fig:res_kappa_phi_p3_c3}
\end{figure}

\begin{figure}[H]
    \centering
    \includegraphics[width=.6\linewidth]{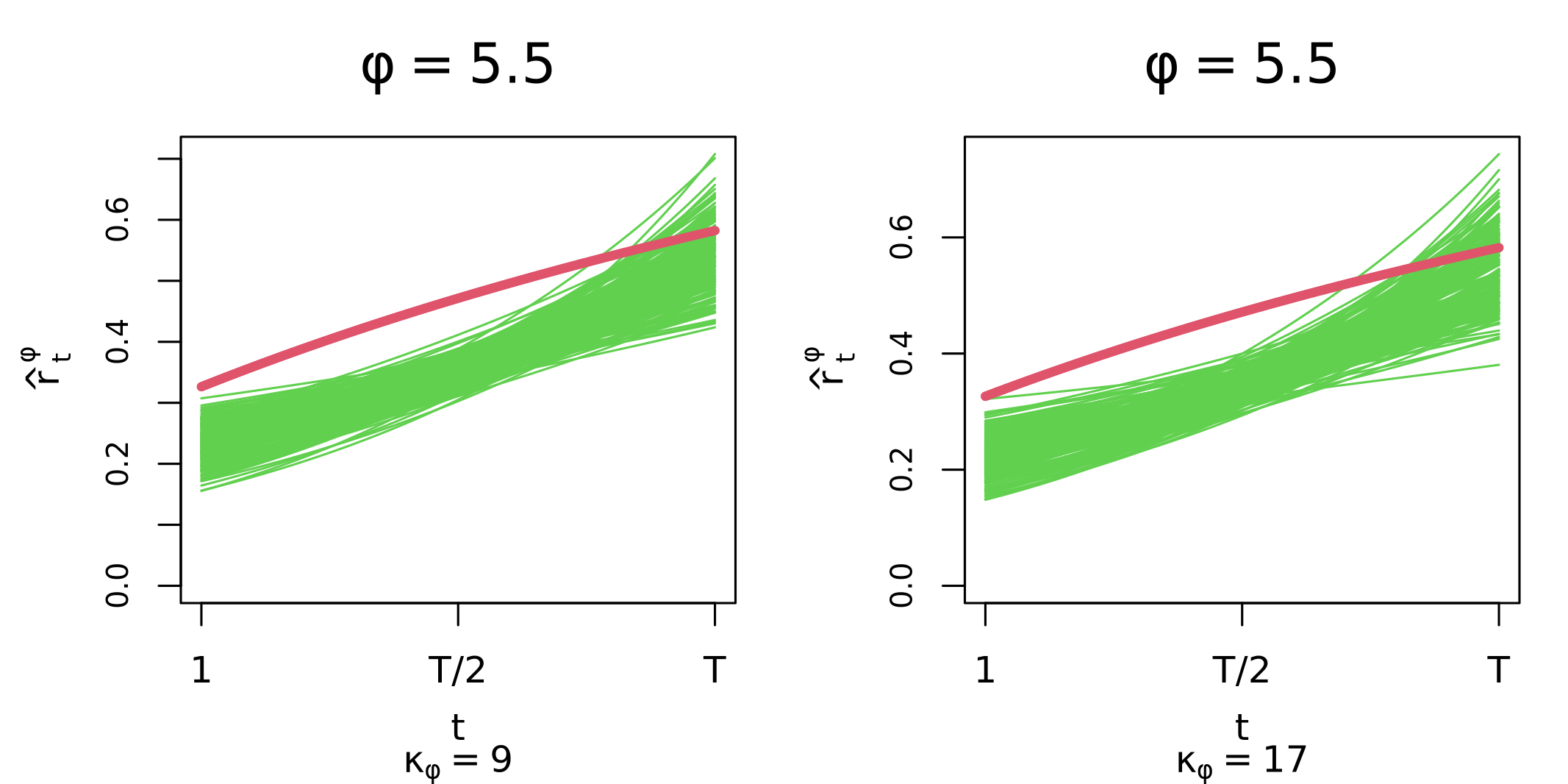}
    \caption{Boundary set radii estimates at $\phi = 7\pi/4$ across $\kappa_{\phi} \in \{9,17\}$ for the third copula example.}
    \label{fig:res_kappa_phi_p4_c3}
\end{figure}

\begin{figure}[H]
    \centering
    \includegraphics[width=.6\linewidth]{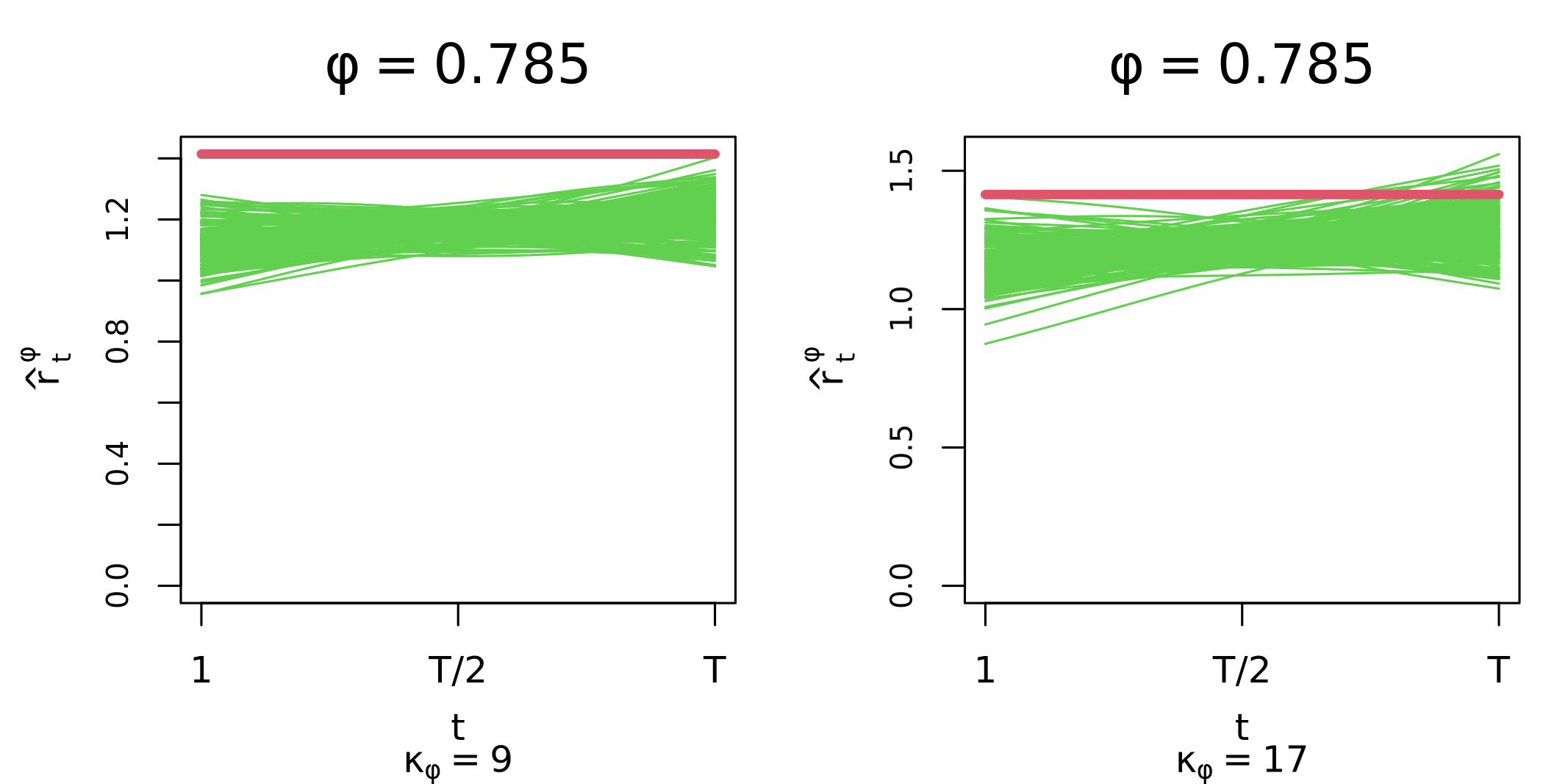}
    \caption{Boundary set radii estimates at $\phi = \pi/4$ across $\kappa_{\phi} \in \{9,17\}$ for the fourth copula example.}
    \label{fig:res_kappa_phi_p1_c4}
\end{figure}

\begin{figure}[H]
    \centering
    \includegraphics[width=.6\linewidth]{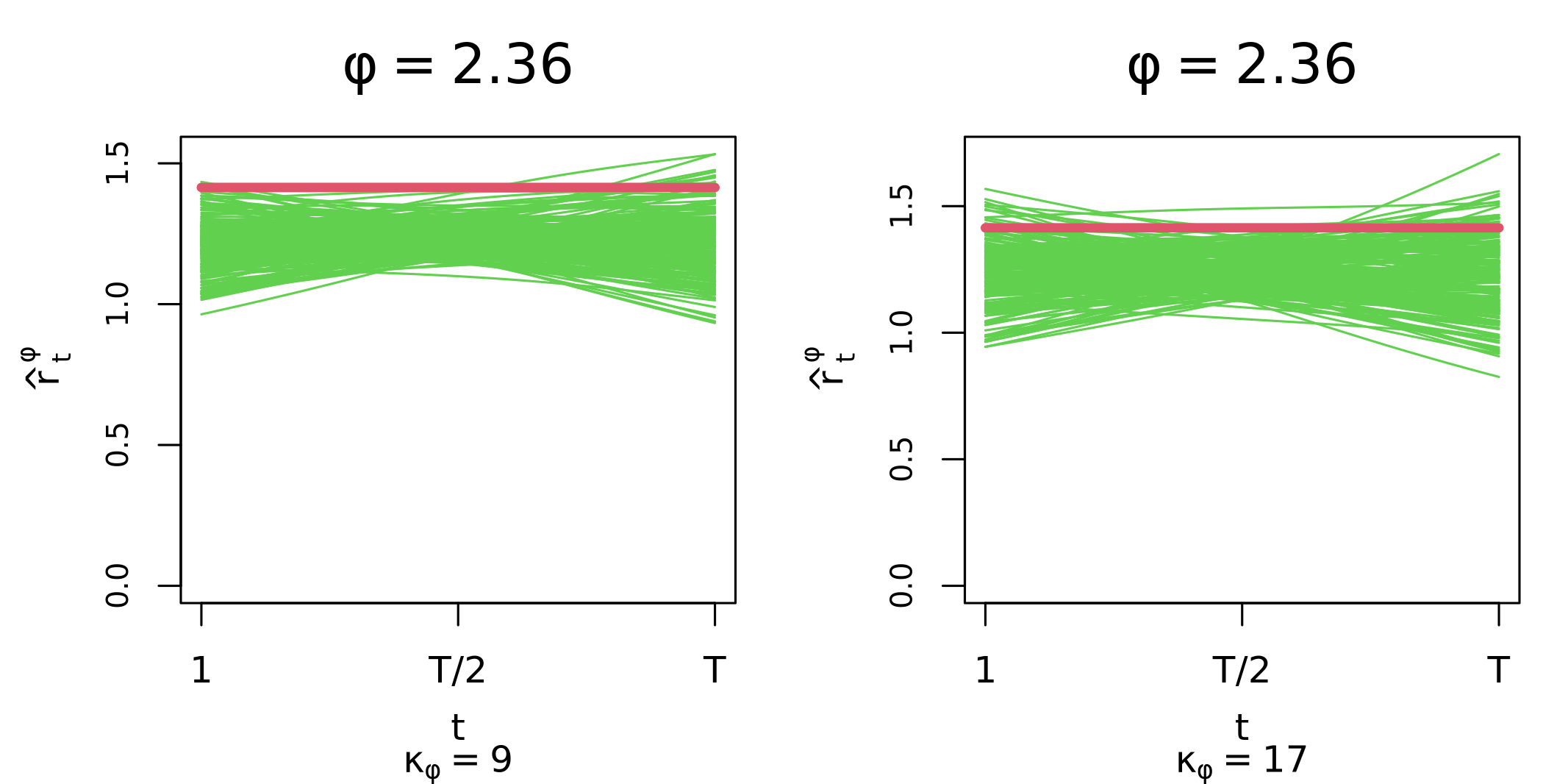}
    \caption{Boundary set radii estimates at $\phi = 3\pi/4$ across $\kappa_{\phi} \in \{9,17\}$ for the fourth copula example.}
    \label{fig:res_kappa_phi_p2_c4}
\end{figure}

\begin{figure}[H]
    \centering
    \includegraphics[width=.6\linewidth]{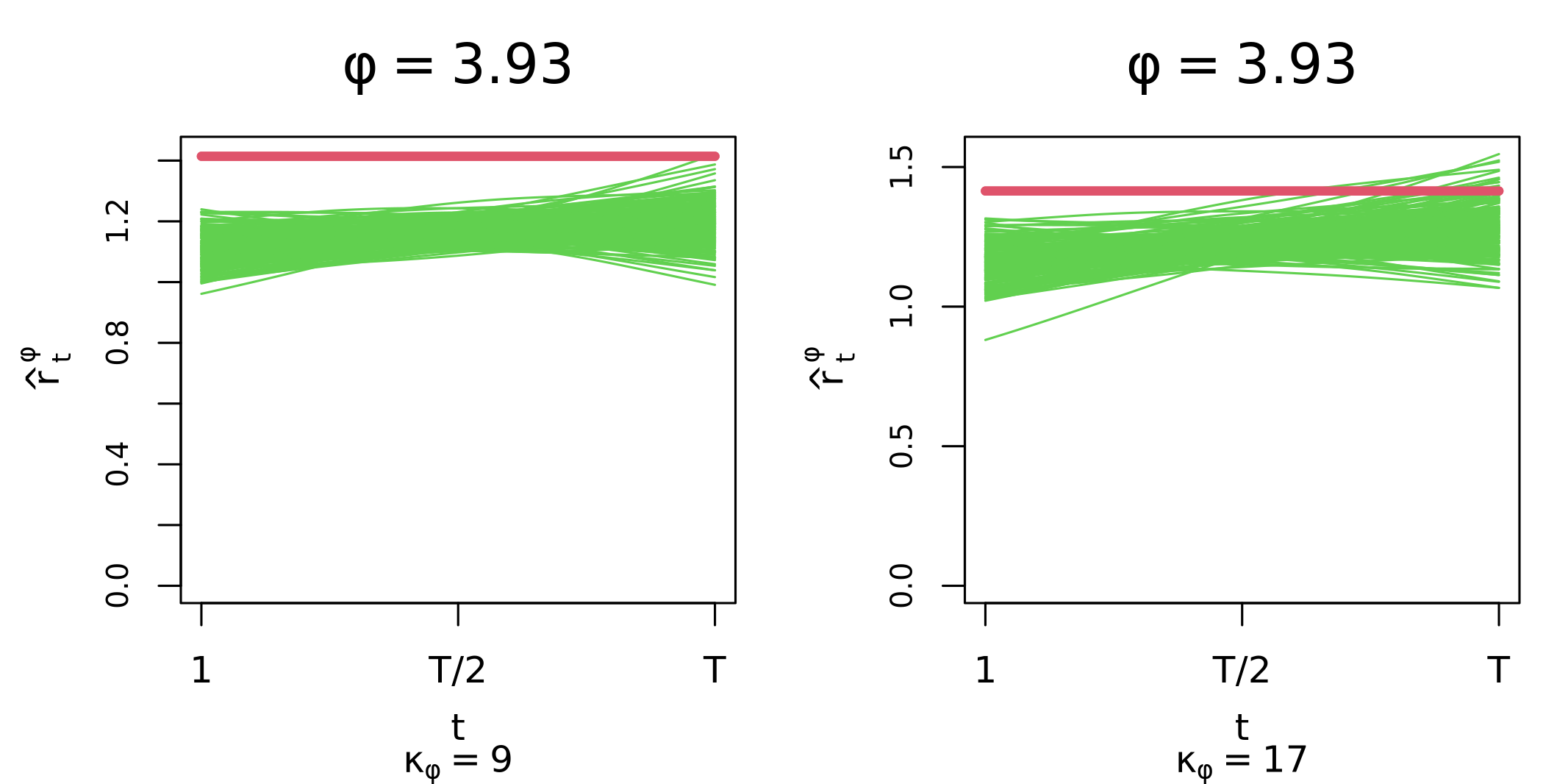}
    \caption{Boundary set radii estimates at $\phi = 5\pi/4$ across $\kappa_{\phi} \in \{9,17\}$ for the fourth copula example.}
    \label{fig:res_kappa_phi_p3_c4}
\end{figure}

\begin{figure}[H]
    \centering
    \includegraphics[width=.6\linewidth]{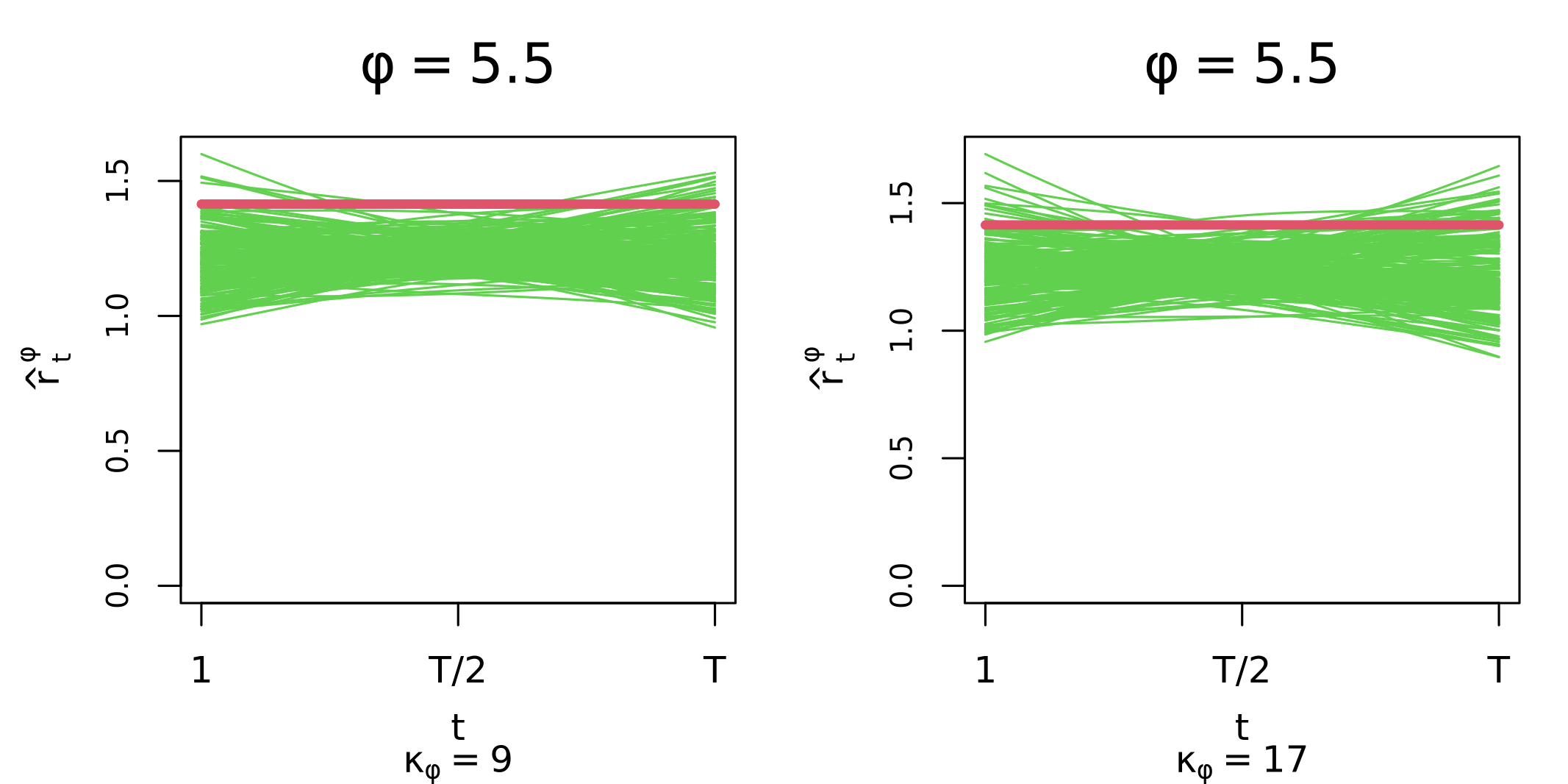}
    \caption{Boundary set radii estimates at $\phi = 7\pi/4$ across $\kappa_{\phi} \in \{9,17\}$ for the fourth copula example.}
    \label{fig:res_kappa_phi_p4_c4}
\end{figure}

\begin{figure}[H]
    \centering
    \includegraphics[width=.6\linewidth]{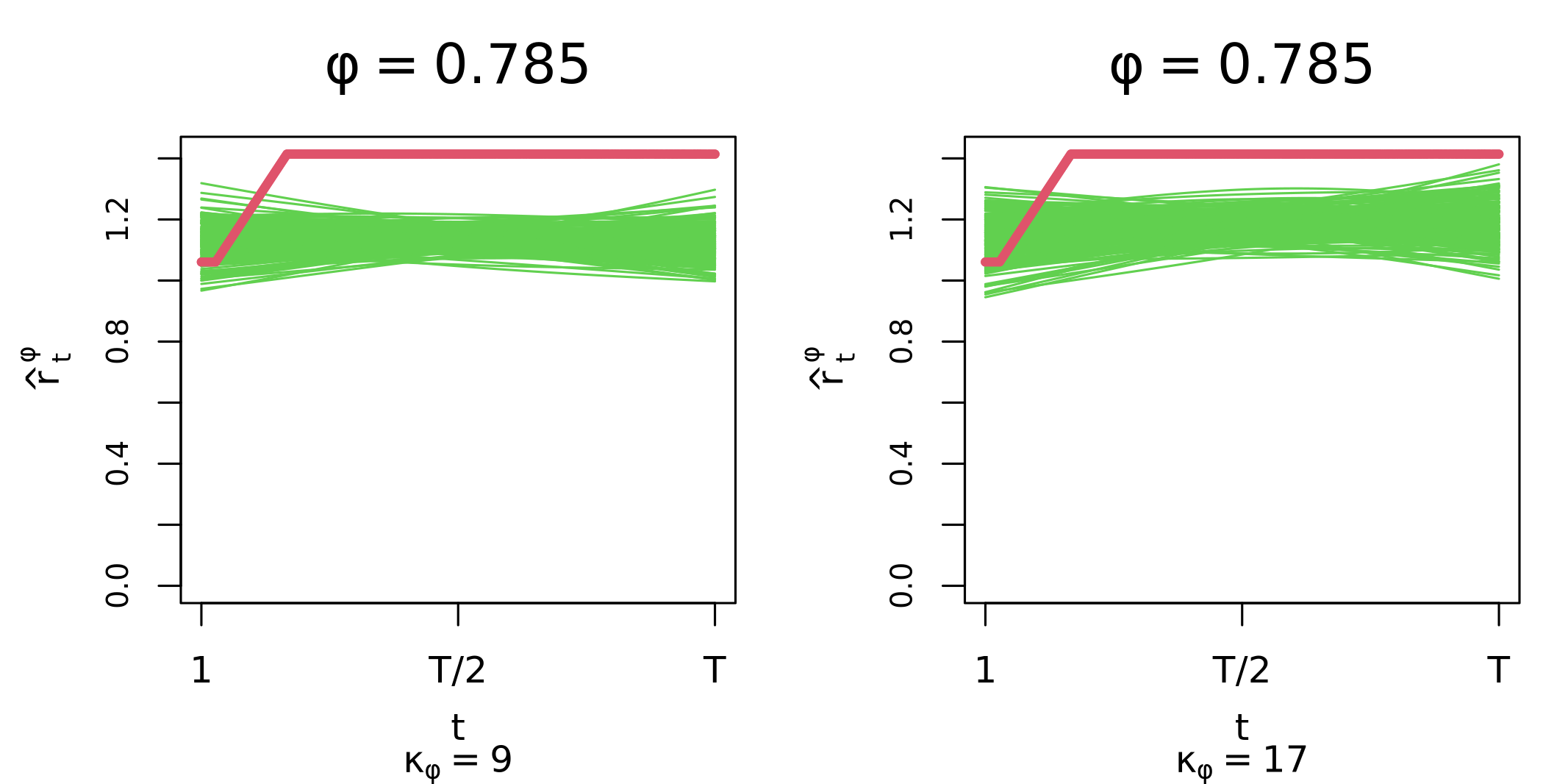}
    \caption{Boundary set radii estimates at $\phi = \pi/4$ across $\kappa_{\phi} \in \{9,17\}$ for the fifth copula example.}
    \label{fig:res_kappa_phi_p1_c5}
\end{figure}

\begin{figure}[H]
    \centering
    \includegraphics[width=.6\linewidth]{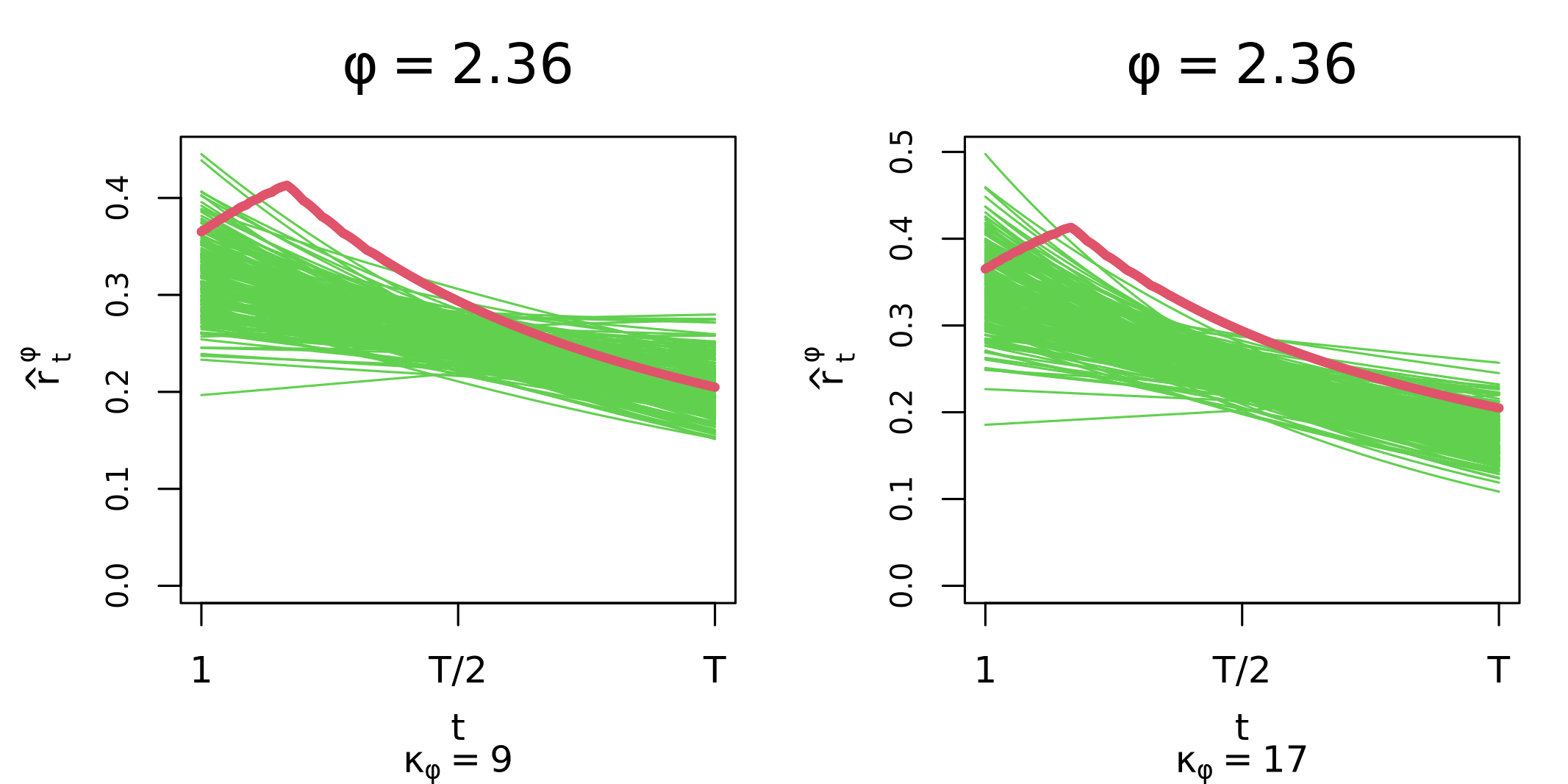}
    \caption{Boundary set radii estimates at $\phi = 3\pi/4$ across $\kappa_{\phi} \in \{9,17\}$ for the fifth copula example.}
    \label{fig:res_kappa_phi_p2_c5}
\end{figure}

\begin{figure}[H]
    \centering
    \includegraphics[width=.6\linewidth]{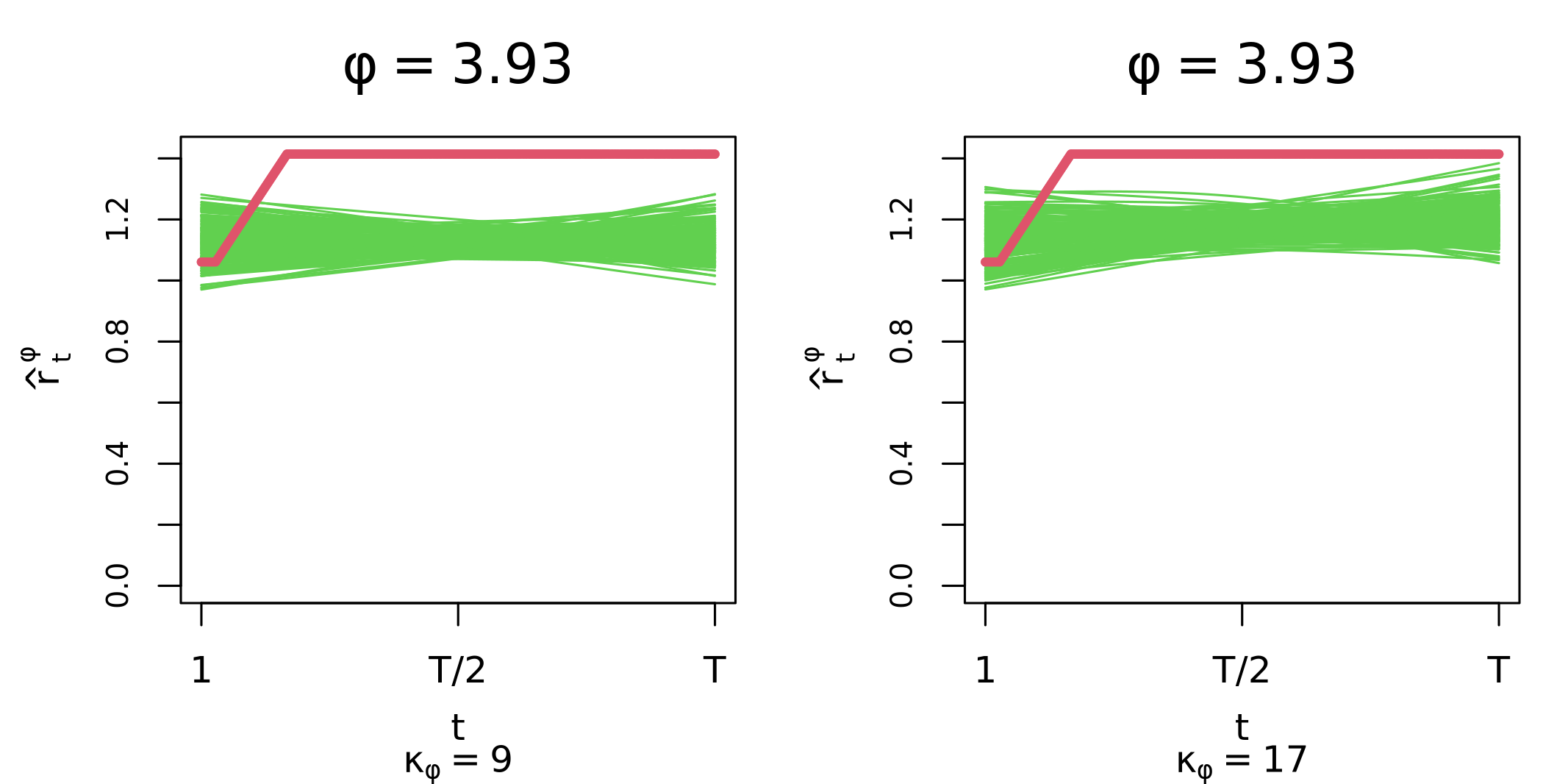}
    \caption{Boundary set radii estimates at $\phi = 5\pi/4$ across $\kappa_{\phi} \in \{9,17\}$ for the fifth copula example.}
    \label{fig:res_kappa_phi_p3_c5}
\end{figure}

\begin{figure}[H]
    \centering
    \includegraphics[width=.6\linewidth]{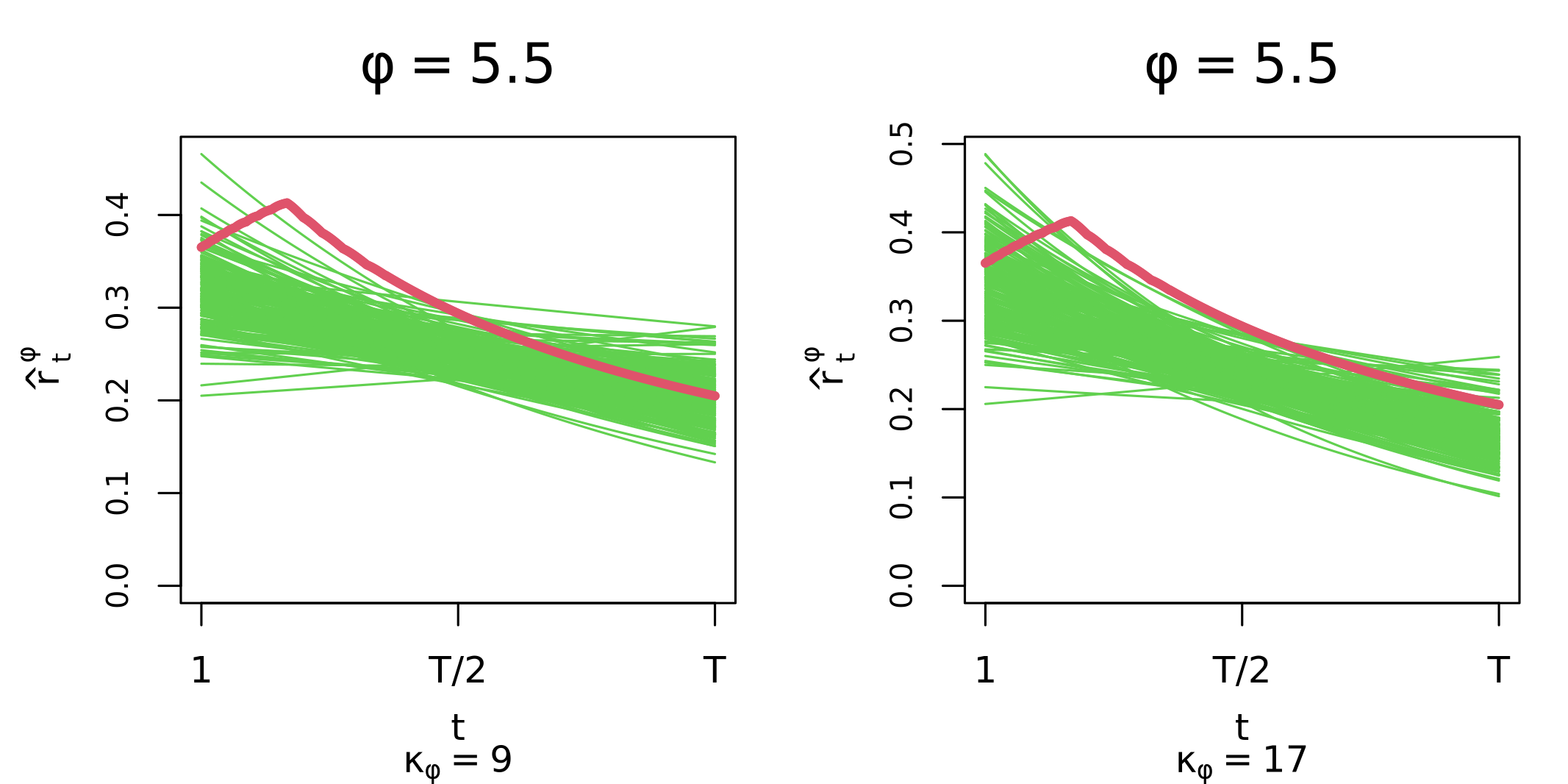}
    \caption{Boundary set radii estimates at $\phi = 7\pi/4$ across $\kappa_{\phi} \in \{9,17\}$ for the fifth copula example.}
    \label{fig:res_kappa_phi_p4_c5}
\end{figure}

Figures~\ref{fig:res_kappa_t_t1_c1}-\ref{fig:res_kappa_t_p4_c5} illustrate the effect of the basis dimension $\kappa_t$ on the boundary set estimates, while Figures~\ref{fig:res_kappa_phi_t1_c1}-\ref{fig:res_kappa_phi_p4_c5} illustrate the effect of the basis dimension $\kappa_\phi$.  

\subsection{Evaluating the effect of the norm choice} \label{subsec:appen_norm}

\begin{figure}[H]
    \centering
    \includegraphics[width=.8\linewidth]{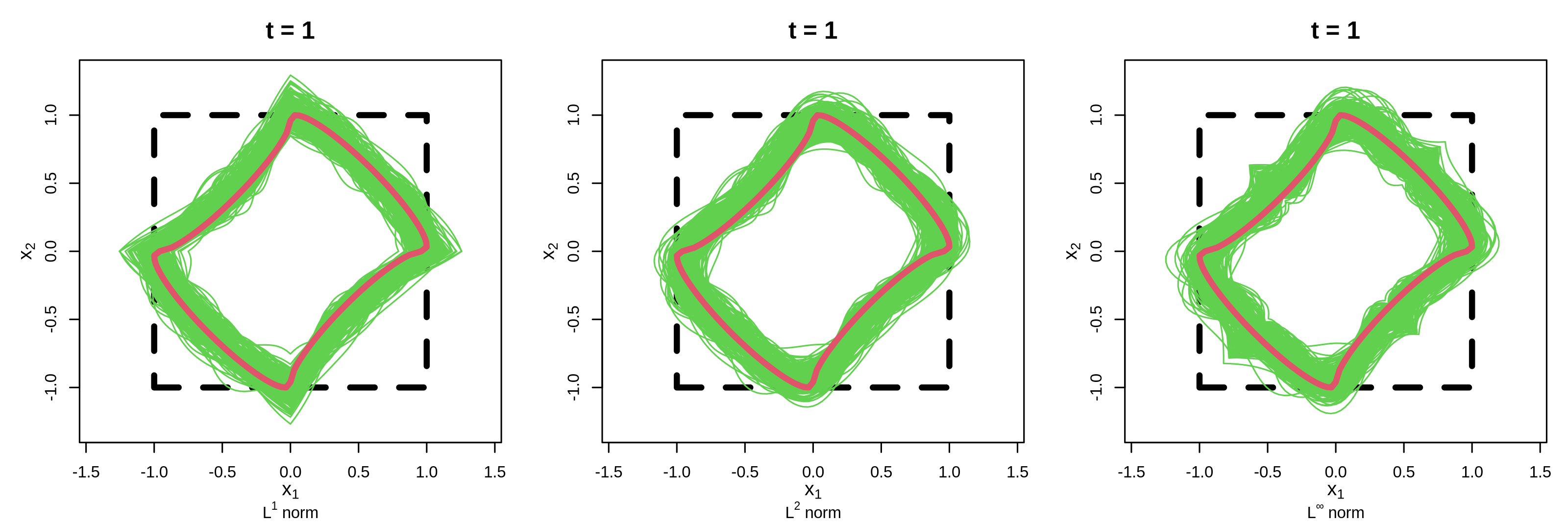}
    \caption{Boundary set estimates as $t = 1$ across $\|\cdot\| \in \{\|\cdot \|_1 ,\|\cdot \|_2,\|\cdot \|_{\infty}\}$ for the first copula example.}
    \label{fig:res_norm_t1_c1}
\end{figure}

\begin{figure}[H]
    \centering
    \includegraphics[width=.8\linewidth]{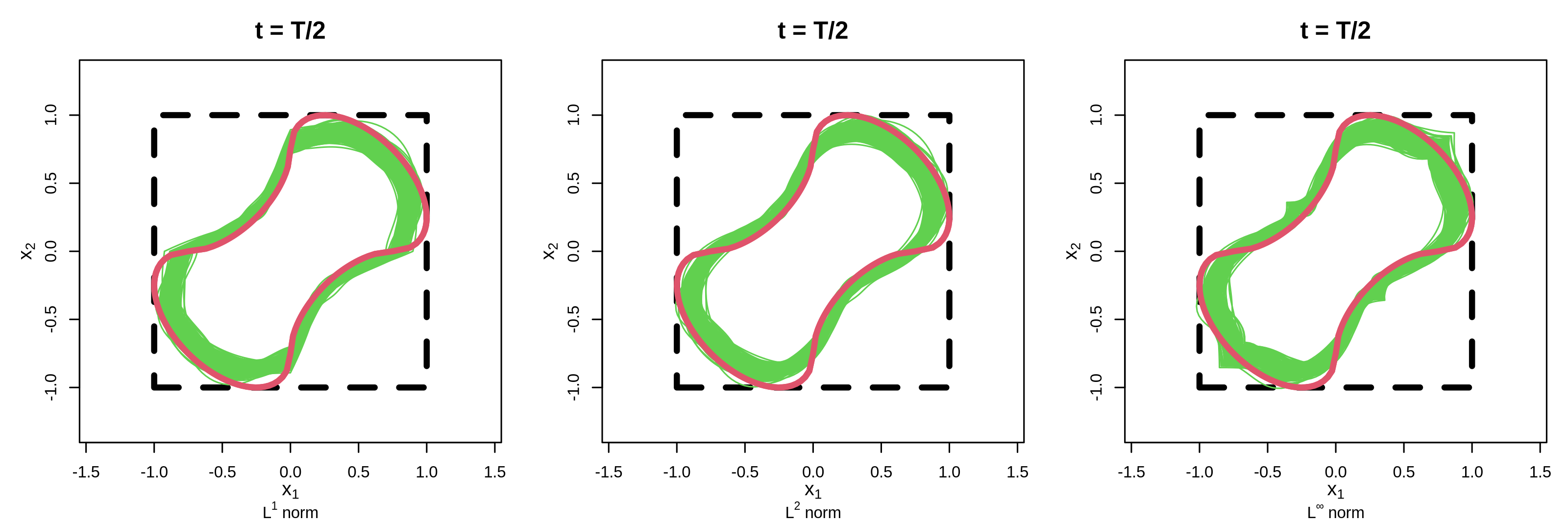}
    \caption{Boundary set estimates as $t = T/2$ across $\|\cdot\| \in \{\|\cdot \|_1 ,\|\cdot \|_2,\|\cdot \|_{\infty}\}$ for the first copula example.}
    \label{fig:res_norm_t2_c1}
\end{figure}

\begin{figure}[H]
    \centering
    \includegraphics[width=.8\linewidth]{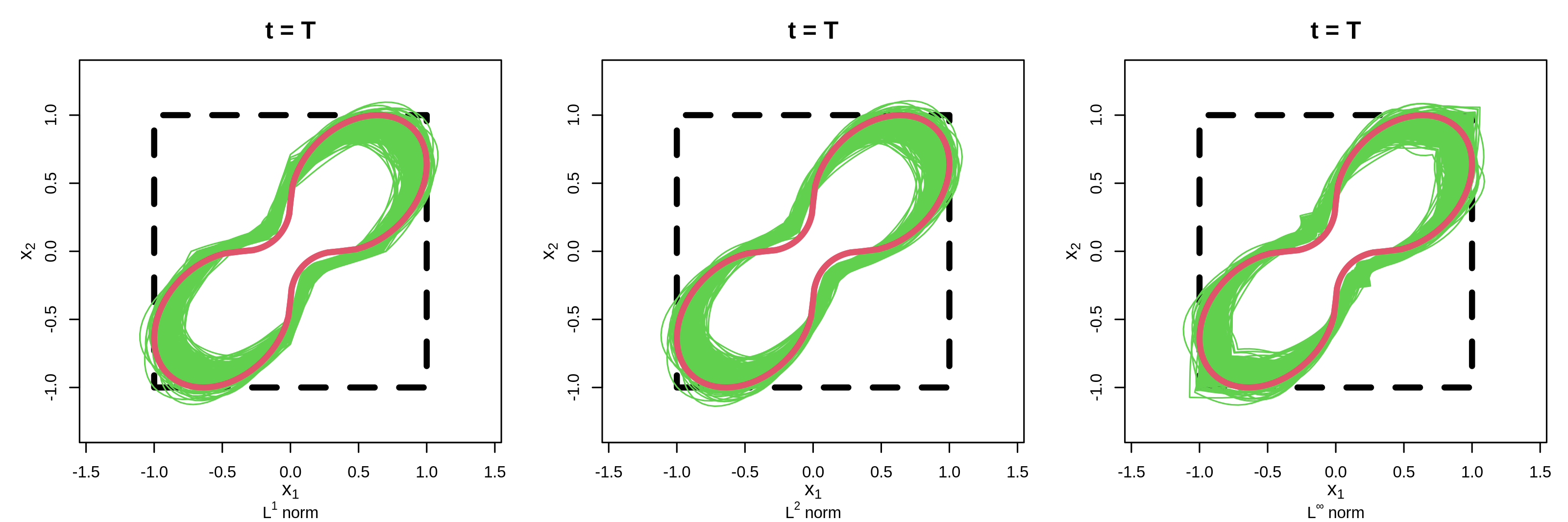}
    \caption{Boundary set estimates at $t = T$ across $\|\cdot\| \in \{\|\cdot \|_1 ,\|\cdot \|_2,\|\cdot \|_{\infty}\}$ for the first copula example.}
    \label{fig:res_norm_t3_c1}
\end{figure}

\begin{figure}[H]
    \centering
    \includegraphics[width=.8\linewidth]{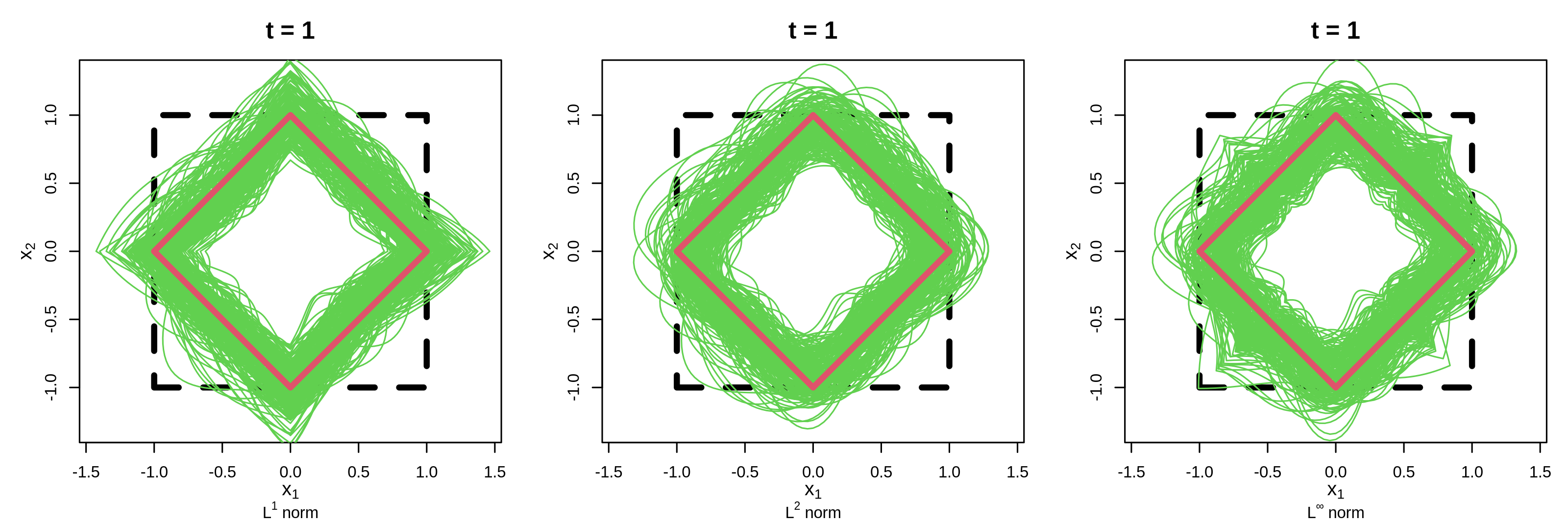}
    \caption{Boundary set estimates as $t = 1$ across $\|\cdot\| \in \{\|\cdot \|_1 ,\|\cdot \|_2,\|\cdot \|_{\infty}\}$ for the second copula example.}
    \label{fig:res_norm_t1_c2}
\end{figure}

\begin{figure}[H]
    \centering
    \includegraphics[width=.8\linewidth]{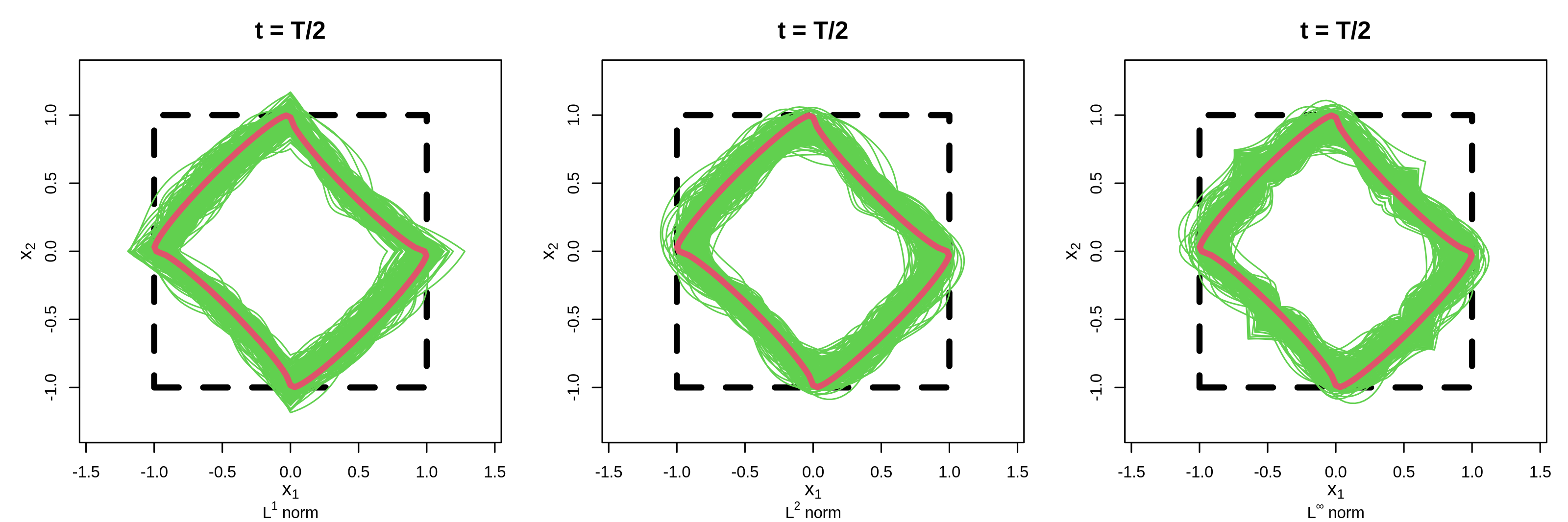}
    \caption{Boundary set estimates as $t = T/2$ across $\|\cdot\| \in \{\|\cdot \|_1 ,\|\cdot \|_2,\|\cdot \|_{\infty}\}$ for the second copula example.}
    \label{fig:res_norm_t2_c2}
\end{figure}

\begin{figure}[H]
    \centering
    \includegraphics[width=.8\linewidth]{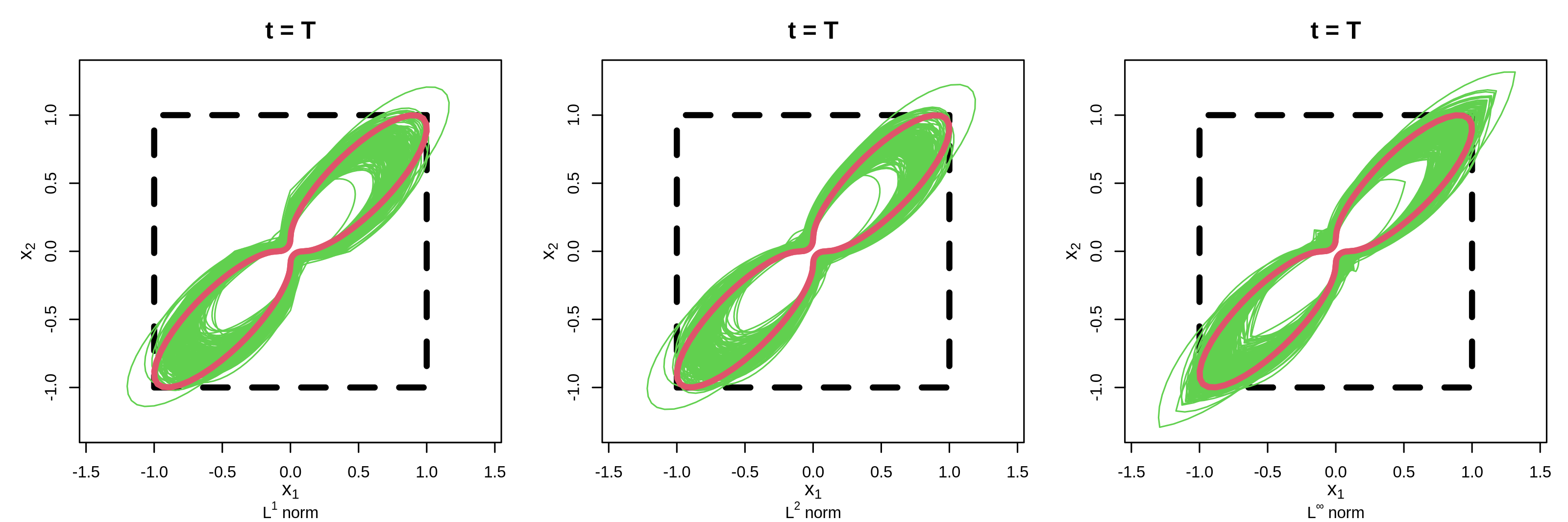}
    \caption{Boundary set estimates at $t = T$ across $\|\cdot\| \in \{\|\cdot \|_1 ,\|\cdot \|_2,\|\cdot \|_{\infty}\}$ for the second copula example.}
    \label{fig:res_norm_t3_c2}
\end{figure}

\begin{figure}[H]
    \centering
    \includegraphics[width=.8\linewidth]{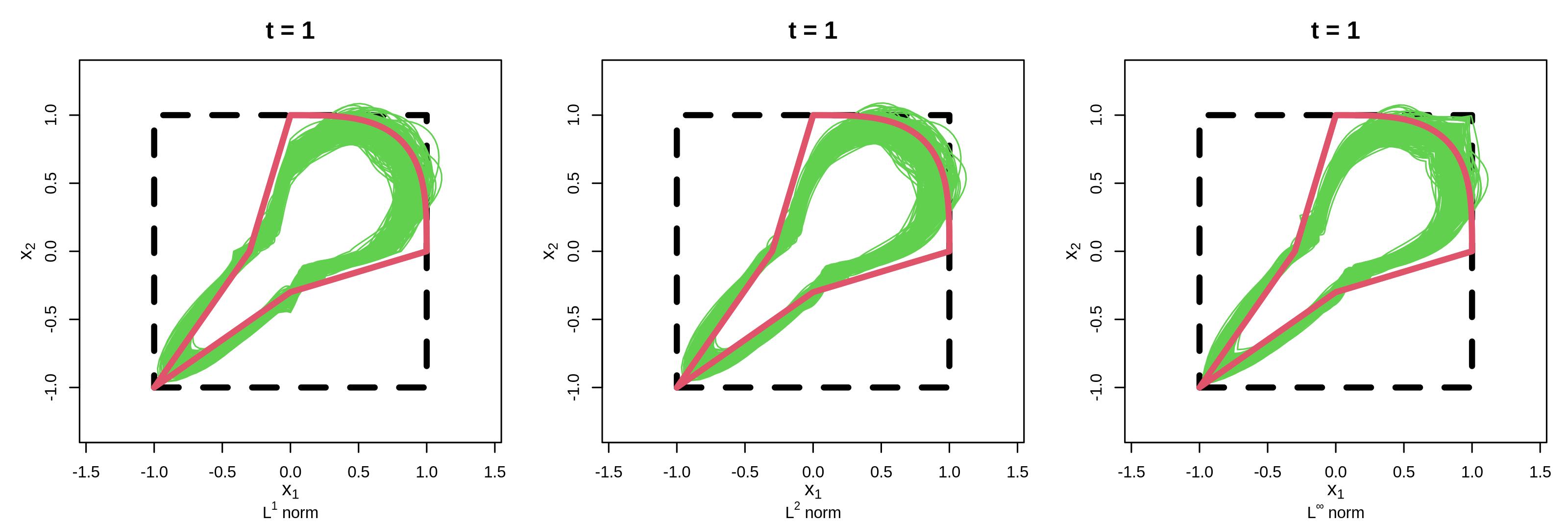}
    \caption{Boundary set estimates as $t = 1$ across $\|\cdot\| \in \{\|\cdot \|_1 ,\|\cdot \|_2,\|\cdot \|_{\infty}\}$ for the third copula example.}
    \label{fig:res_norm_t1_c3}
\end{figure}

\begin{figure}[H]
    \centering
    \includegraphics[width=.8\linewidth]{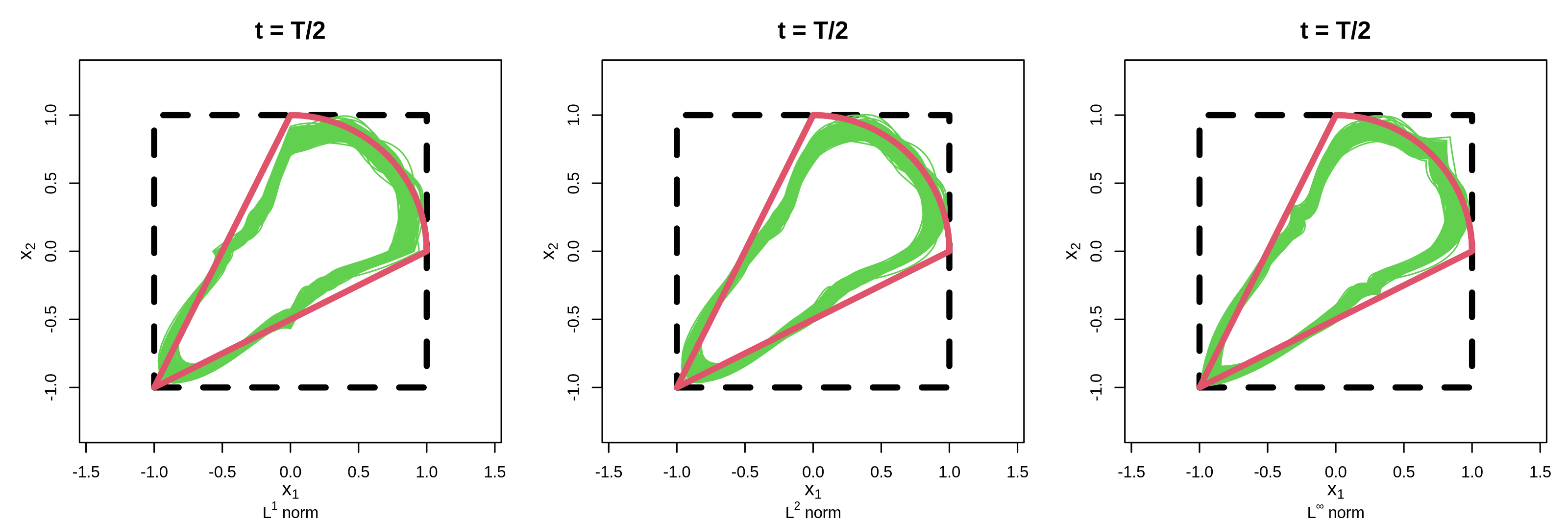}
    \caption{Boundary set estimates as $t = T/2$ across $\|\cdot\| \in \{\|\cdot \|_1 ,\|\cdot \|_2,\|\cdot \|_{\infty}\}$ for the third copula example.}
    \label{fig:res_norm_t2_c3}
\end{figure}

\begin{figure}[H]
    \centering
    \includegraphics[width=.8\linewidth]{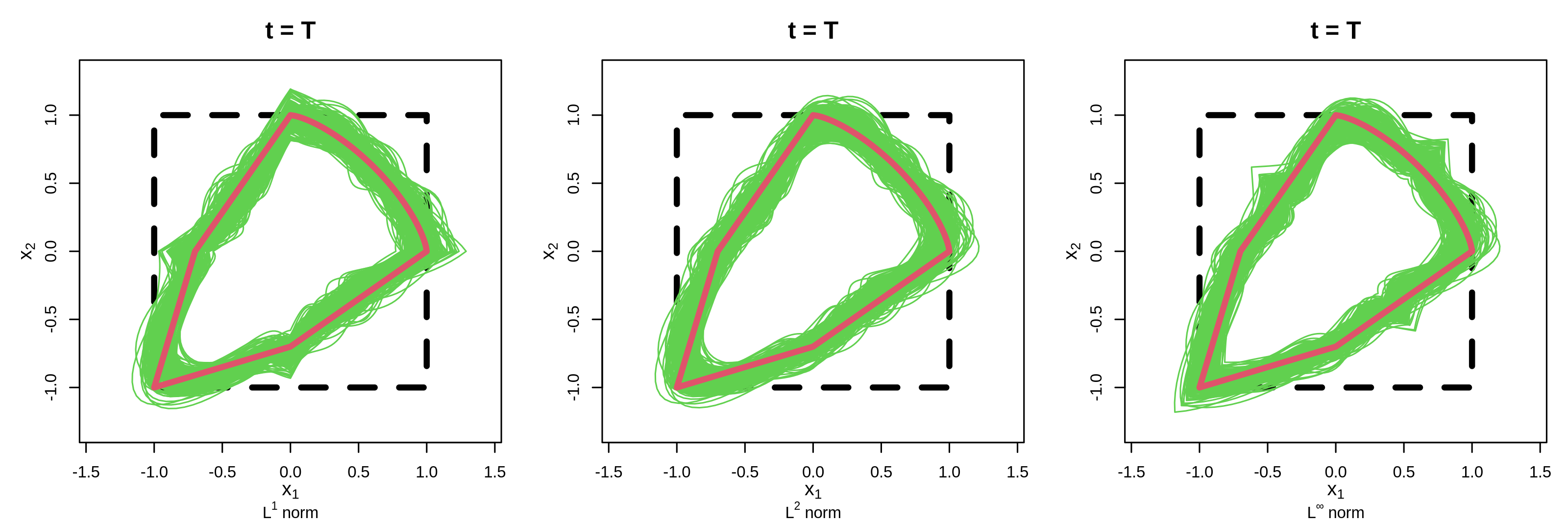}
    \caption{Boundary set estimates at $t = T$ across $\|\cdot\| \in \{\|\cdot \|_1 ,\|\cdot \|_2,\|\cdot \|_{\infty}\}$ for the third copula example.}
    \label{fig:res_norm_t3_c3}
\end{figure}

\begin{figure}[H]
    \centering
    \includegraphics[width=.8\linewidth]{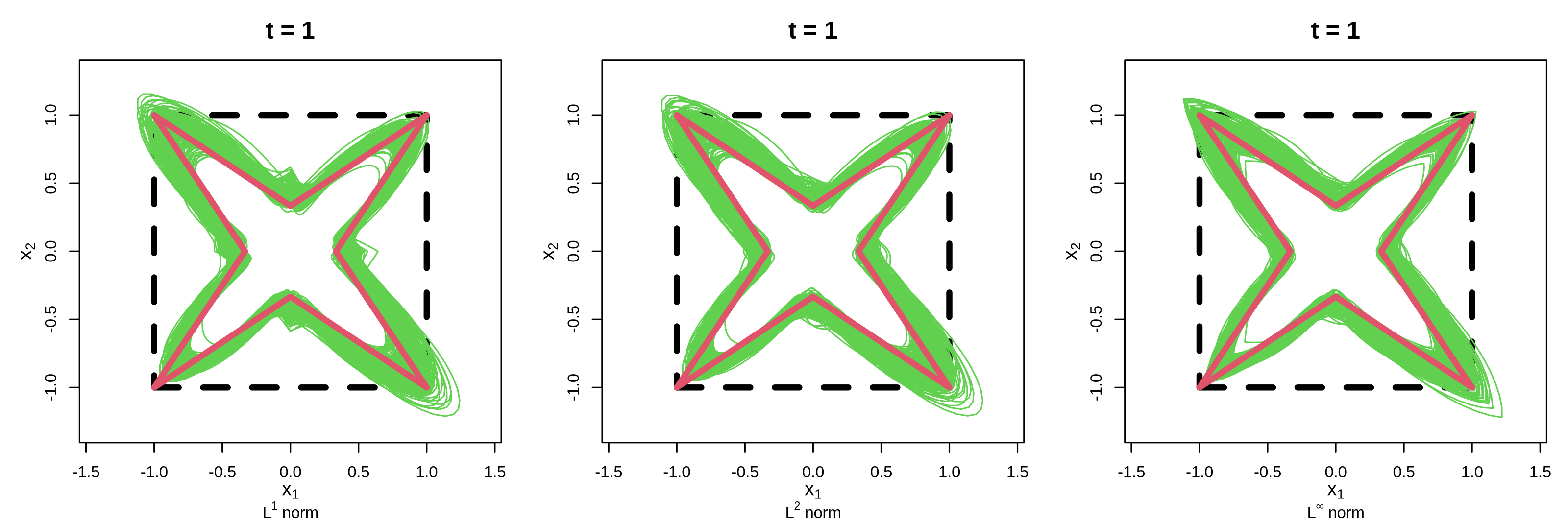}
    \caption{Boundary set estimates as $t = 1$ across $\|\cdot\| \in \{\|\cdot \|_1 ,\|\cdot \|_2,\|\cdot \|_{\infty}\}$ for the fourth copula example.}
    \label{fig:res_norm_t1_c4}
\end{figure}

\begin{figure}[H]
    \centering
    \includegraphics[width=.8\linewidth]{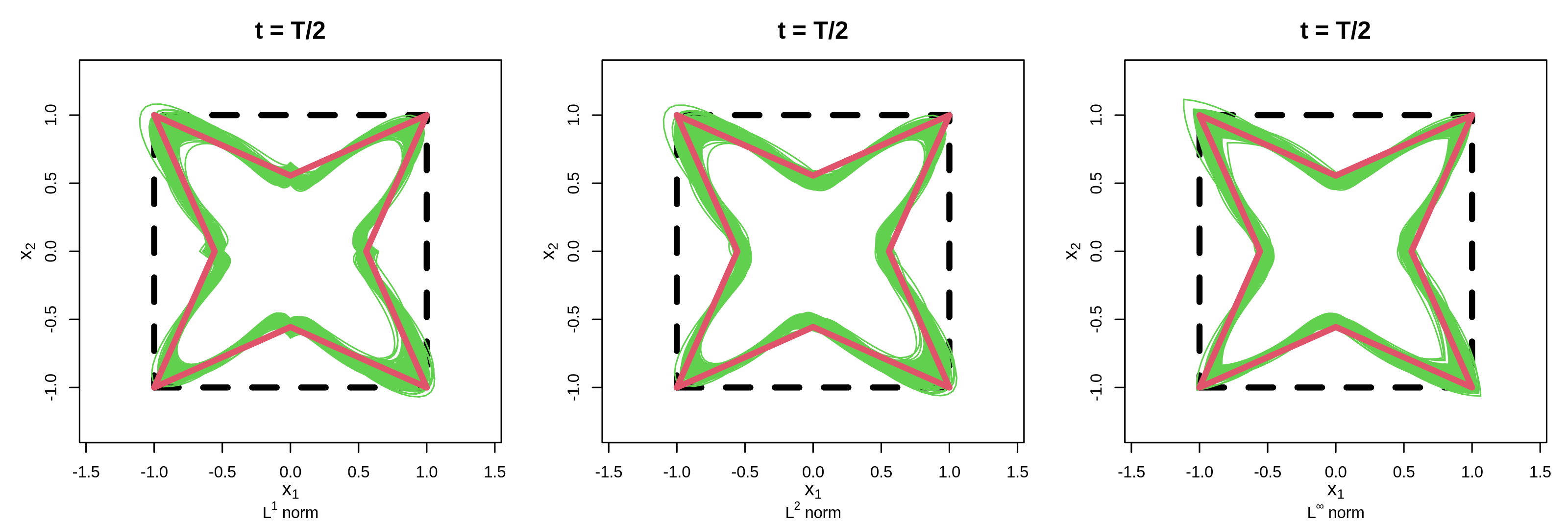}
    \caption{Boundary set estimates as $t = T/2$ across $\|\cdot\| \in \{\|\cdot \|_1 ,\|\cdot \|_2,\|\cdot \|_{\infty}\}$ for the fourth copula example.}
    \label{fig:res_norm_t2_c4}
\end{figure}

\begin{figure}[H]
    \centering
    \includegraphics[width=.8\linewidth]{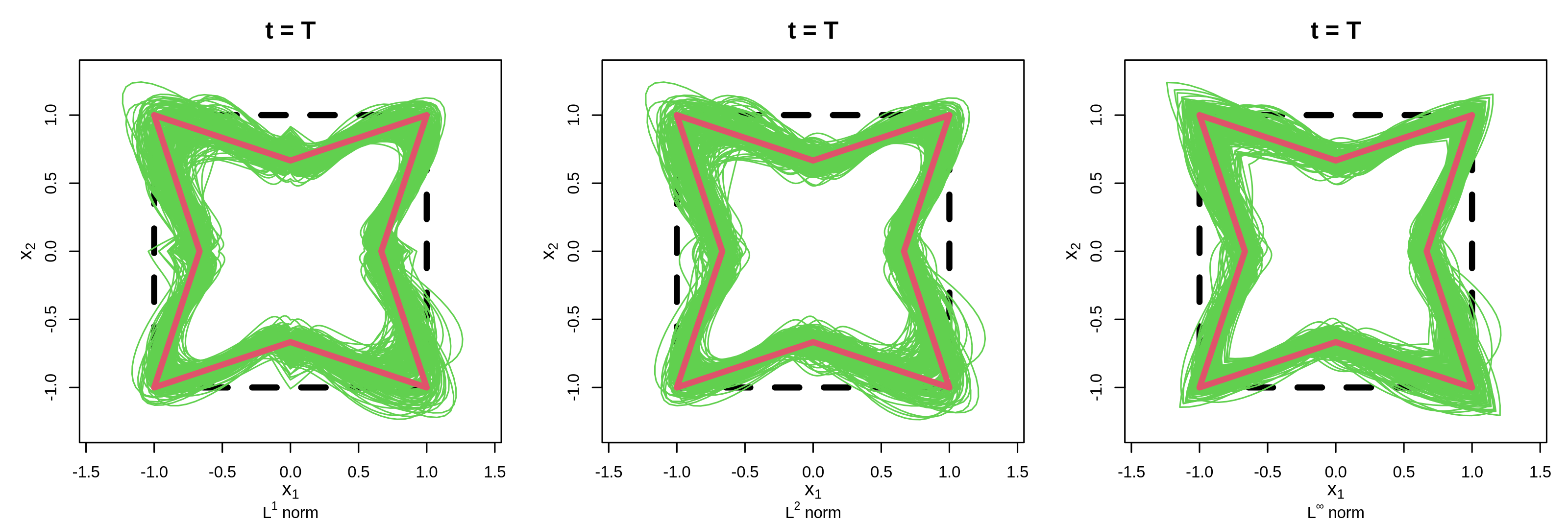}
    \caption{Boundary set estimates at $t = T$ across $\|\cdot\| \in \{\|\cdot \|_1 ,\|\cdot \|_2,\|\cdot \|_{\infty}\}$ for the fourth copula example.}
    \label{fig:res_norm_t3_c4}
\end{figure}

\begin{figure}[H]
    \centering
    \includegraphics[width=.8\linewidth]{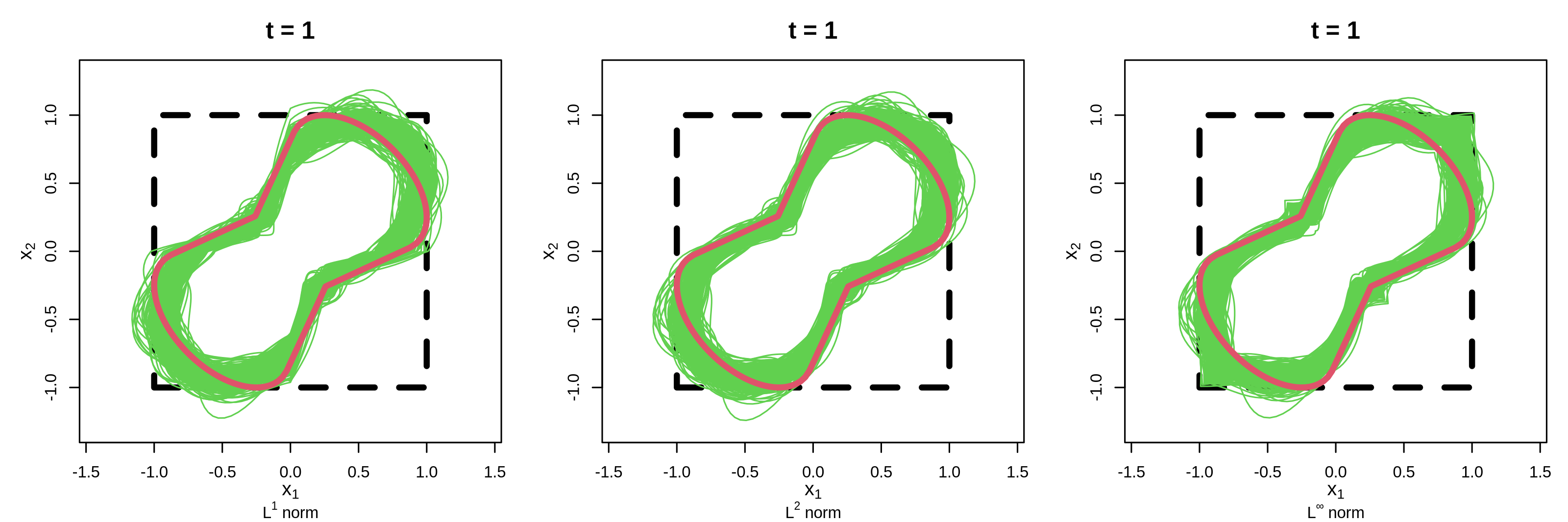}
    \caption{Boundary set estimates as $t = 1$ across $\|\cdot\| \in \{\|\cdot \|_1 ,\|\cdot \|_2,\|\cdot \|_{\infty}\}$ for the fifth copula example.}
    \label{fig:res_norm_t1_c5}
\end{figure}

\begin{figure}[H]
    \centering
    \includegraphics[width=.8\linewidth]{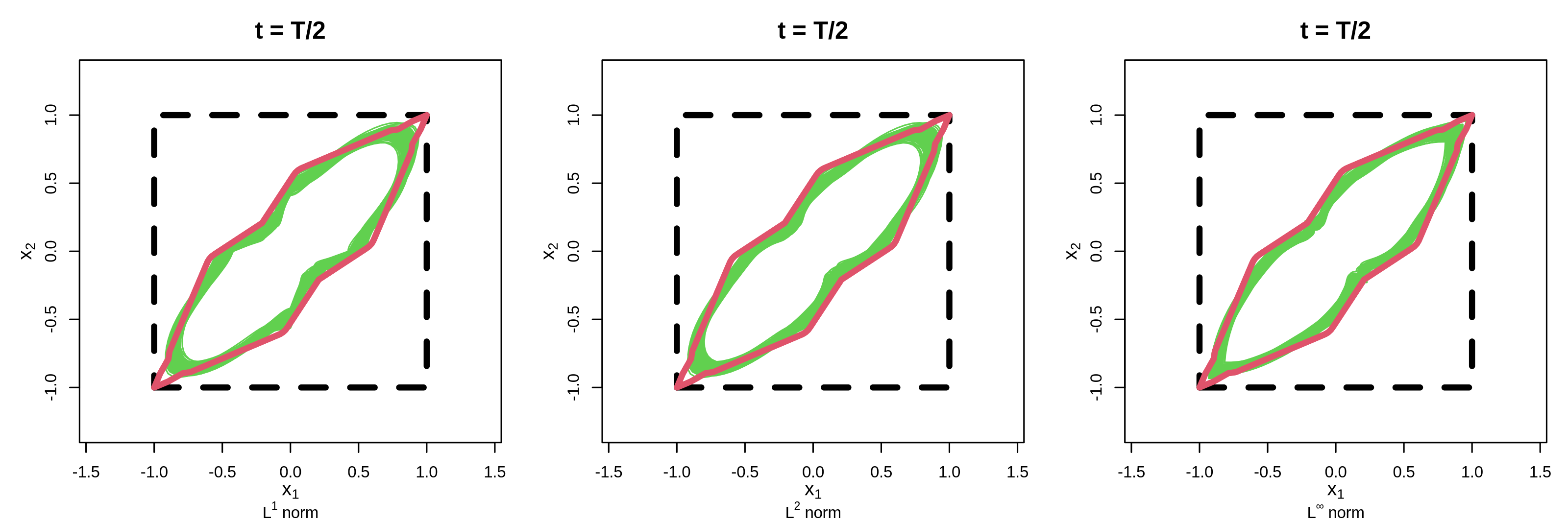}
    \caption{Boundary set estimates as $t = T/2$ across $\|\cdot\| \in \{\|\cdot \|_1 ,\|\cdot \|_2,\|\cdot \|_{\infty}\}$ for the fifth copula example.}
    \label{fig:res_norm_t2_c5}
\end{figure}

\begin{figure}[H]
    \centering
    \includegraphics[width=.8\linewidth]{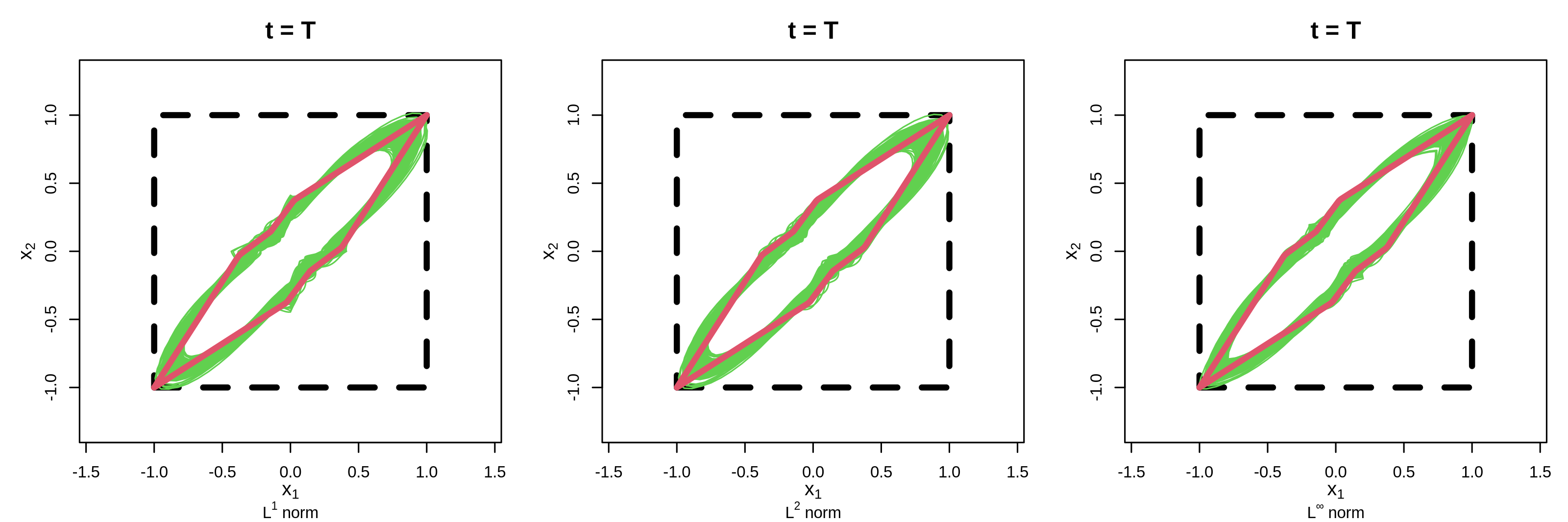}
    \caption{Boundary set estimates at $t = T$ across $\|\cdot\| \in \{\|\cdot \|_1 ,\|\cdot \|_2,\|\cdot \|_{\infty}\}$ for the fifth copula example.}
    \label{fig:res_norm_t3_c5}
\end{figure}

\begin{figure}[H]
    \centering
    \includegraphics[width=.8\linewidth]{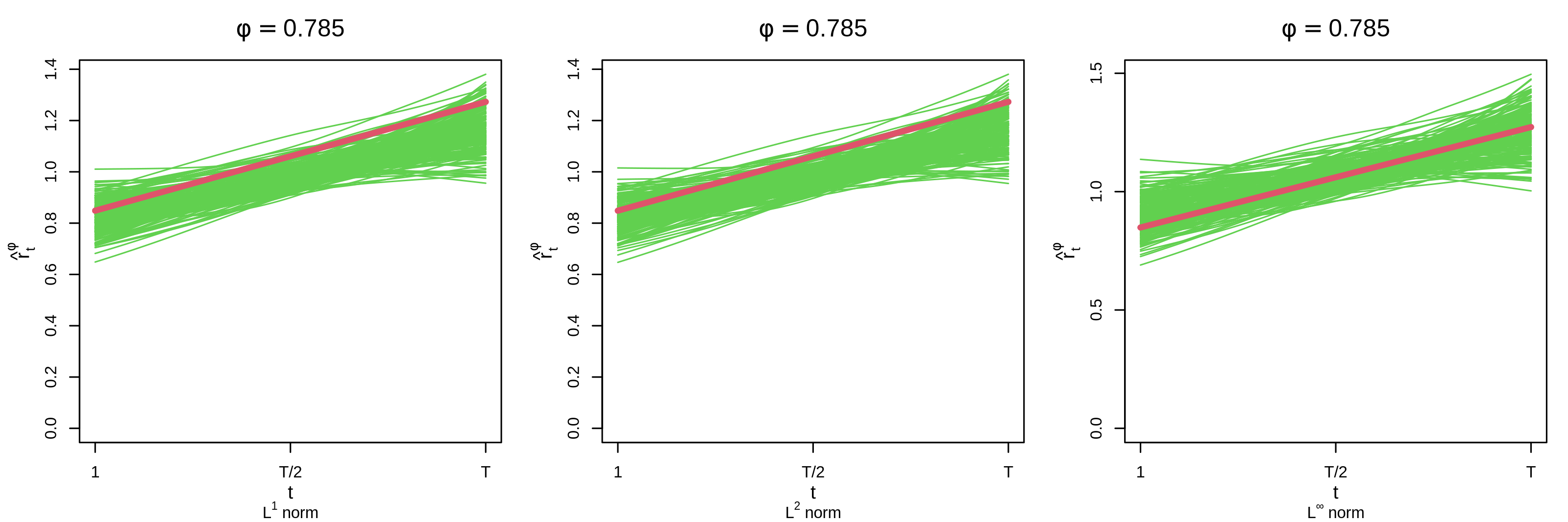}
    \caption{Boundary set radii estimates at $\phi = \pi/4$ across $\|\cdot\| \in \{\|\cdot \|_1 ,\|\cdot \|_2,\|\cdot \|_{\infty}\}$ for the first copula example.}
    \label{fig:res_norm_p1_c1}
\end{figure}

\begin{figure}[H]
    \centering
    \includegraphics[width=.8\linewidth]{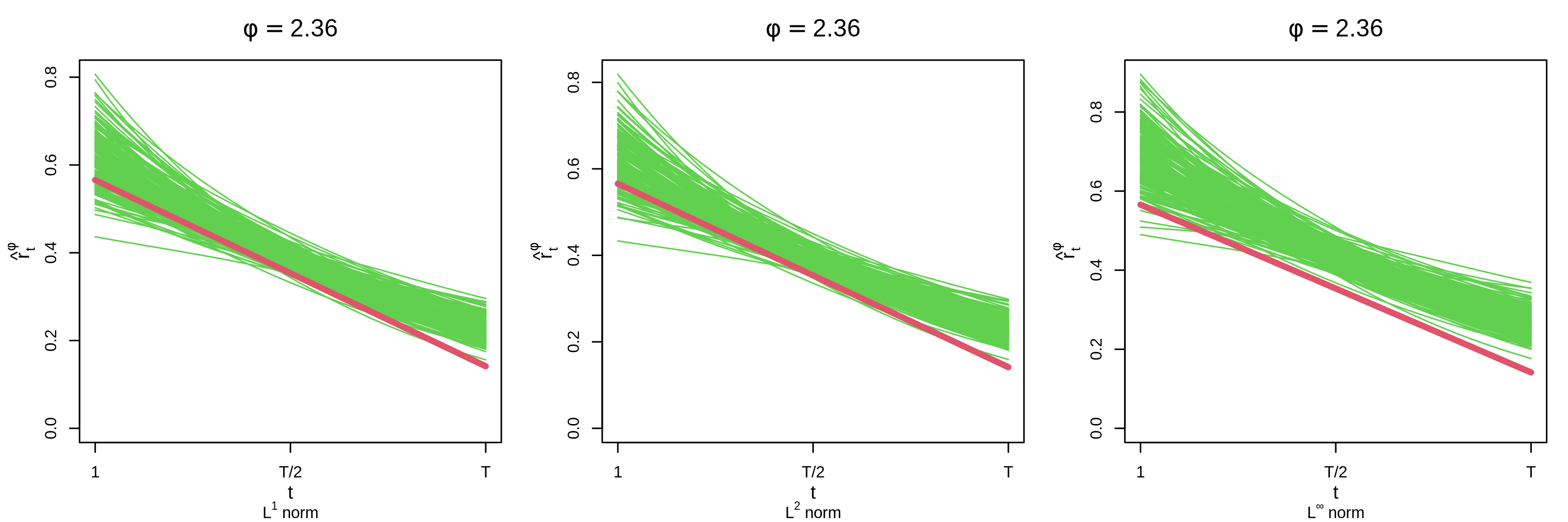}
    \caption{Boundary set radii estimates at $\phi = 3\pi/4$ across $\|\cdot\| \in \{\|\cdot \|_1 ,\|\cdot \|_2,\|\cdot \|_{\infty}\}$ for the first copula example.}
    \label{fig:res_norm_p2_c1}
\end{figure}

\begin{figure}[H]
    \centering
    \includegraphics[width=.8\linewidth]{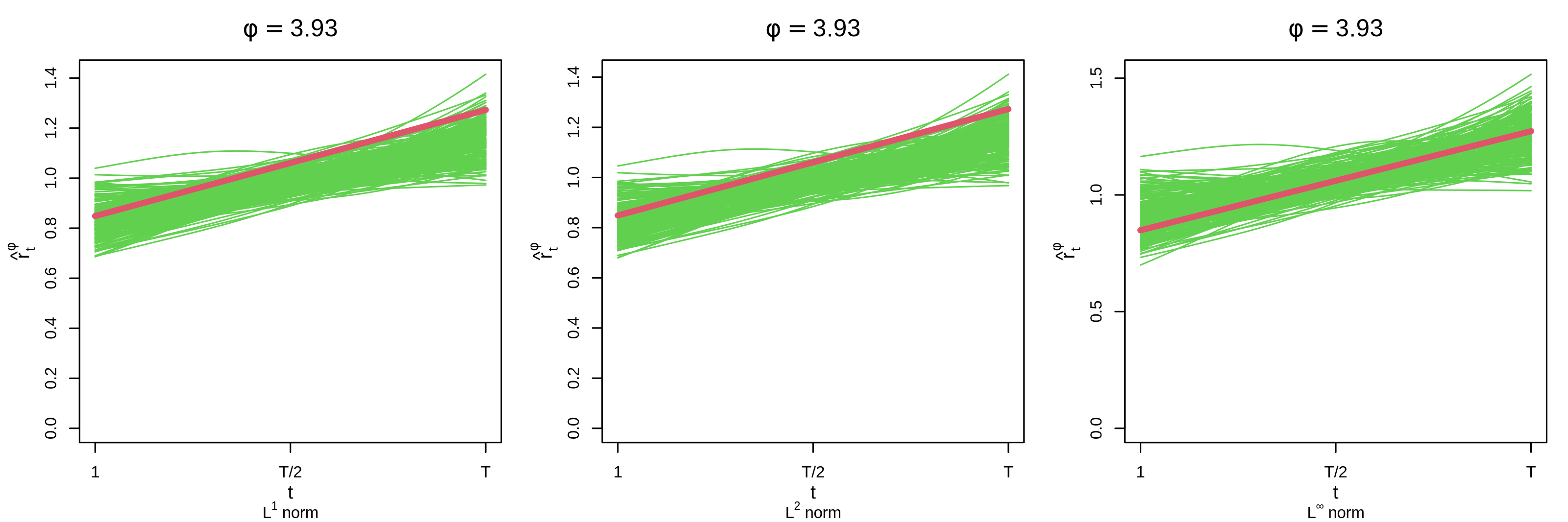}
    \caption{Boundary set radii estimates at $\phi = 5\pi/4$ across $\|\cdot\| \in \{\|\cdot \|_1 ,\|\cdot \|_2,\|\cdot \|_{\infty}\}$ for the first copula example.}
    \label{fig:res_norm_p3_c1}
\end{figure}

\begin{figure}[H]
    \centering
    \includegraphics[width=.8\linewidth]{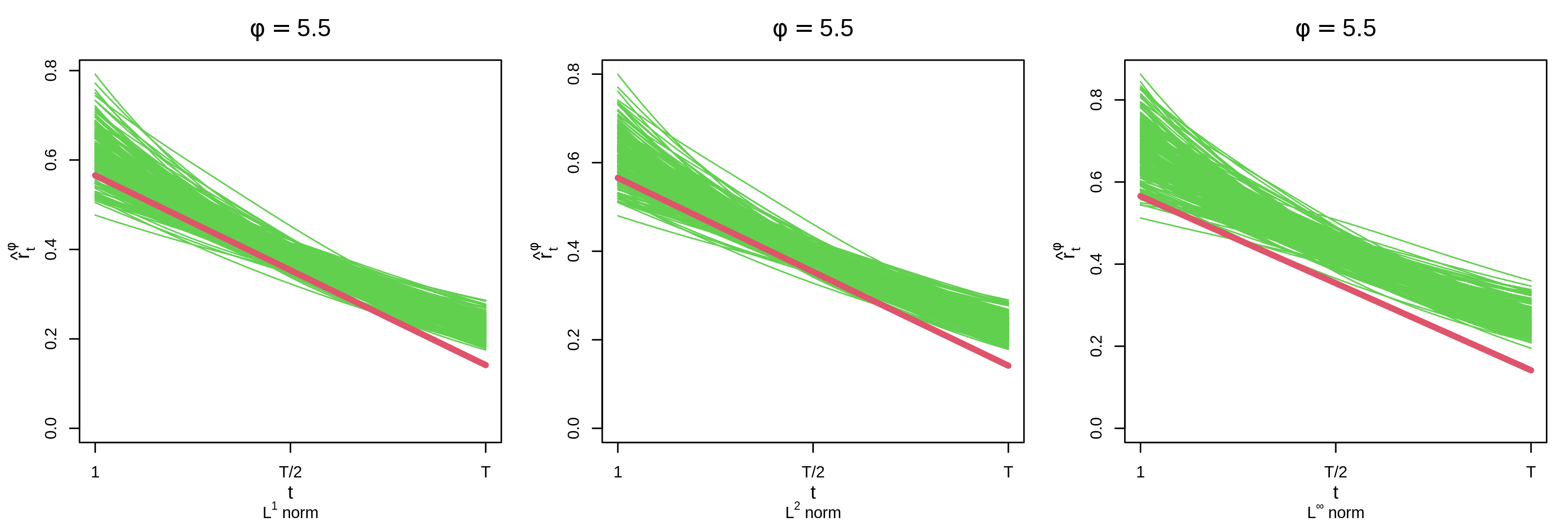}
    \caption{Boundary set radii estimates at $\phi = 7\pi/4$ across $\|\cdot\| \in \{\|\cdot \|_1 ,\|\cdot \|_2,\|\cdot \|_{\infty}\}$ for the first copula example.}
    \label{fig:res_norm_p4_c1}
\end{figure}

\begin{figure}[H]
    \centering
    \includegraphics[width=.8\linewidth]{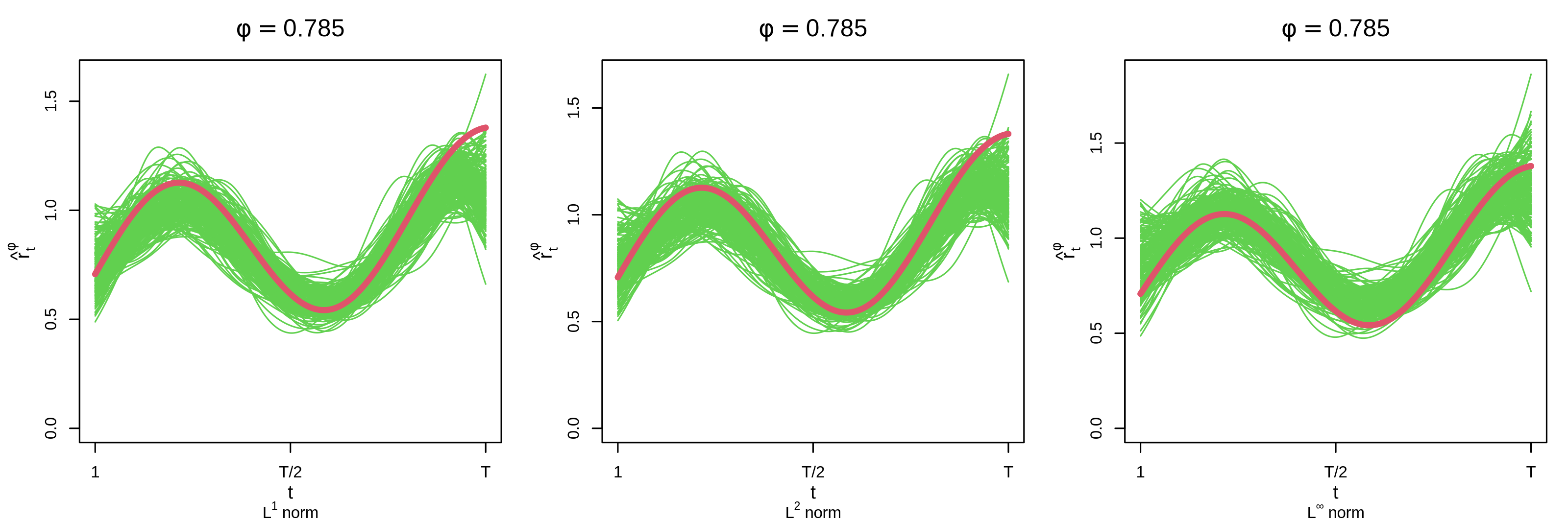}
    \caption{Boundary set radii estimates at $\phi = \pi/4$ across $\|\cdot\| \in \{\|\cdot \|_1 ,\|\cdot \|_2,\|\cdot \|_{\infty}\}$ for the second copula example.}
    \label{fig:res_norm_p1_c2}
\end{figure}

\begin{figure}[H]
    \centering
    \includegraphics[width=.8\linewidth]{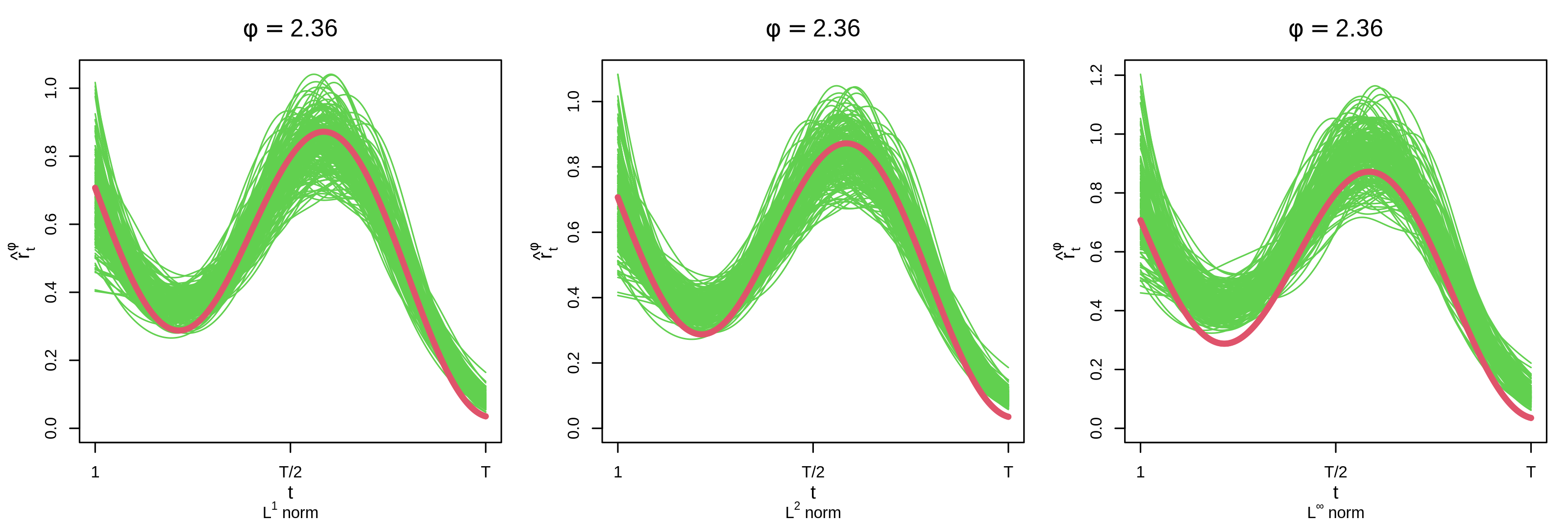}
    \caption{Boundary set radii estimates at $\phi = 3\pi/4$ across $\|\cdot\| \in \{\|\cdot \|_1 ,\|\cdot \|_2,\|\cdot \|_{\infty}\}$ for the second copula example.}
    \label{fig:res_norm_p2_c2}
\end{figure}

\begin{figure}[H]
    \centering
    \includegraphics[width=.8\linewidth]{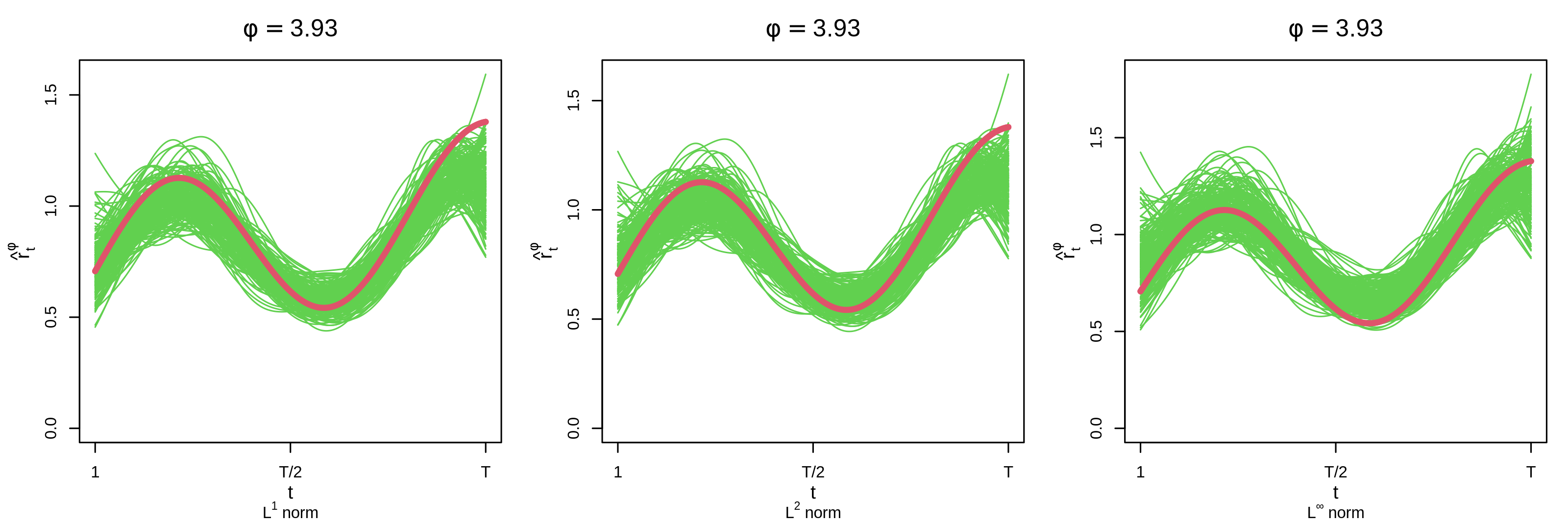}
    \caption{Boundary set radii estimates at $\phi = 5\pi/4$ across $\|\cdot\| \in \{\|\cdot \|_1 ,\|\cdot \|_2,\|\cdot \|_{\infty}\}$ for the second copula example.}
    \label{fig:res_norm_p3_c2}
\end{figure}

\begin{figure}[H]
    \centering
    \includegraphics[width=.8\linewidth]{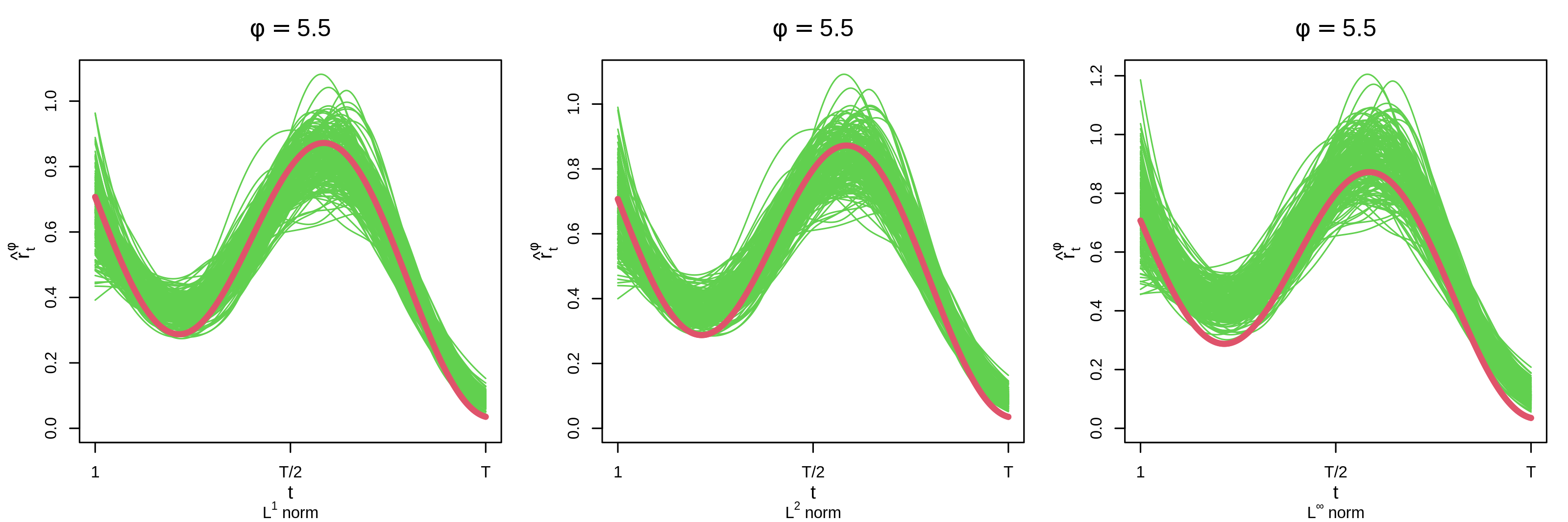}
    \caption{Boundary set radii estimates at $\phi = 7\pi/4$ across $\|\cdot\| \in \{\|\cdot \|_1 ,\|\cdot \|_2,\|\cdot \|_{\infty}\}$ for the second copula example.}
    \label{fig:res_norm_p4_c2}
\end{figure}

\begin{figure}[H]
    \centering
    \includegraphics[width=.8\linewidth]{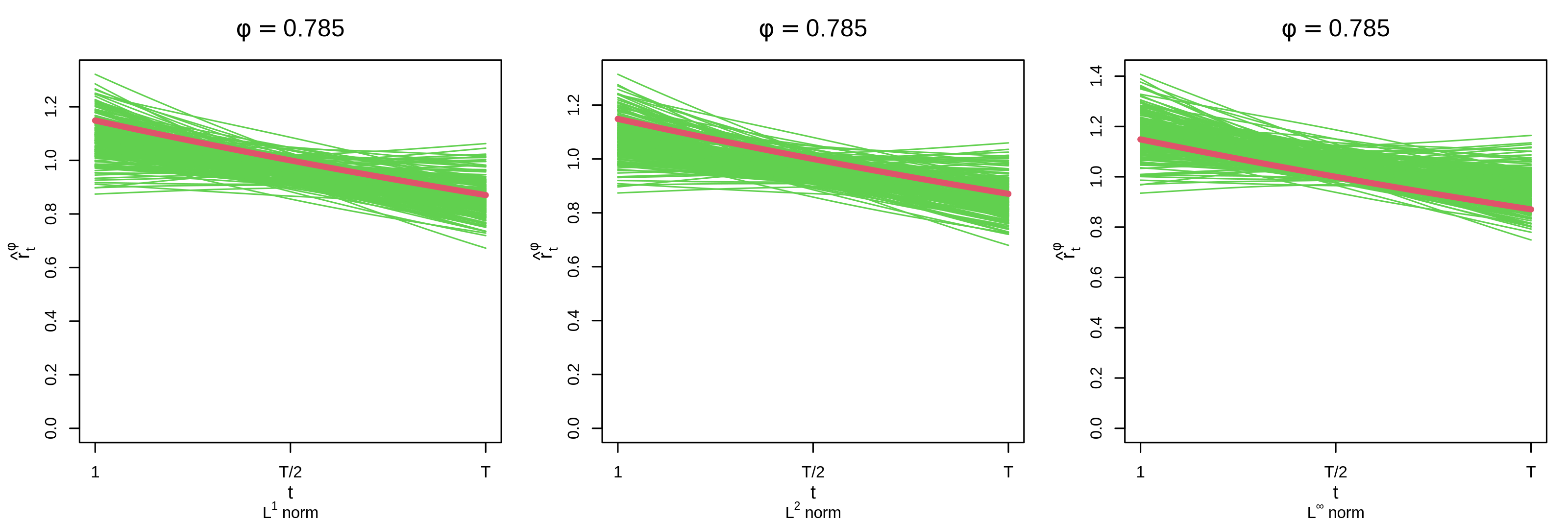}
    \caption{Boundary set radii estimates at $\phi = \pi/4$ across $\|\cdot\| \in \{\|\cdot \|_1 ,\|\cdot \|_2,\|\cdot \|_{\infty}\}$ for the third copula example.}
    \label{fig:res_norm_p1_c3}
\end{figure}

\begin{figure}[H]
    \centering
    \includegraphics[width=.8\linewidth]{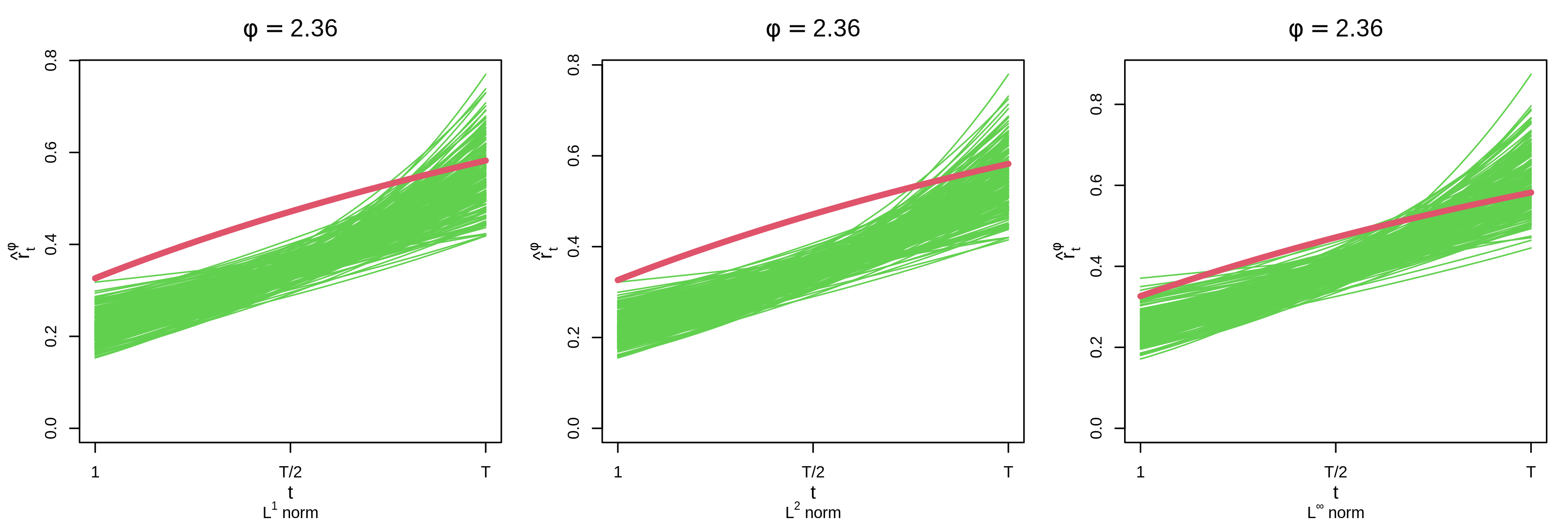}
    \caption{Boundary set radii estimates at $\phi = 3\pi/4$ across $\|\cdot\| \in \{\|\cdot \|_1 ,\|\cdot \|_2,\|\cdot \|_{\infty}\}$ for the third copula example.}
    \label{fig:res_norm_p2_c3}
\end{figure}

\begin{figure}[H]
    \centering
    \includegraphics[width=.8\linewidth]{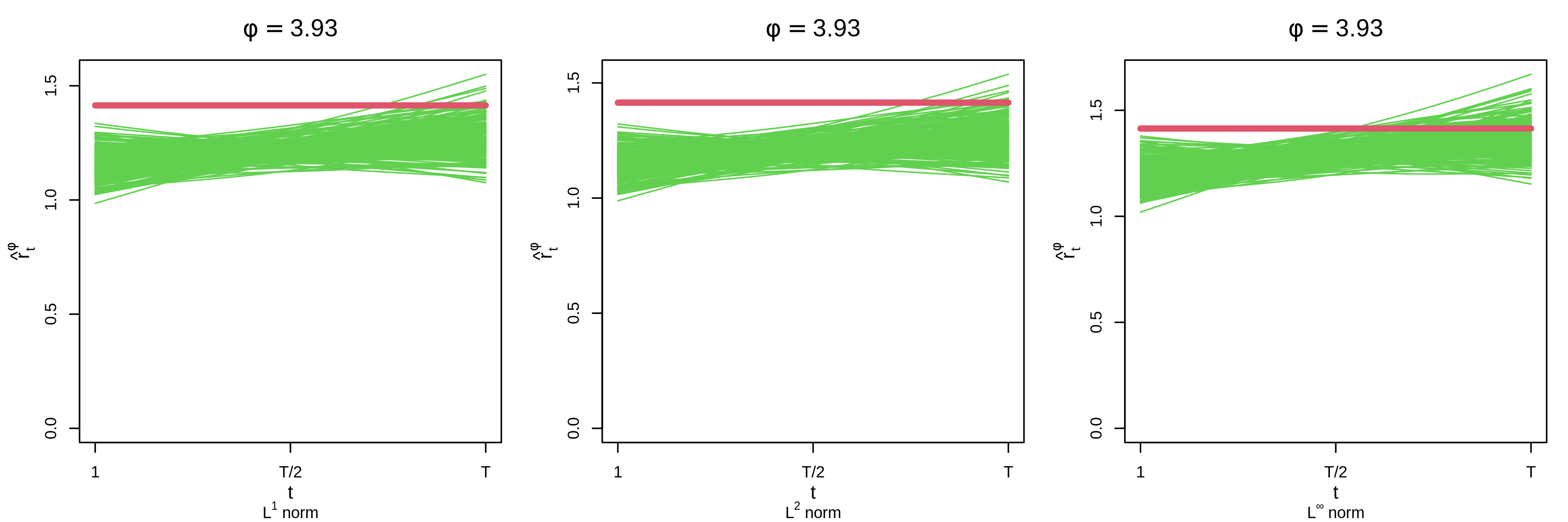}
    \caption{Boundary set radii estimates at $\phi = 5\pi/4$ across $\|\cdot\| \in \{\|\cdot \|_1 ,\|\cdot \|_2,\|\cdot \|_{\infty}\}$ for the third copula example.}
    \label{fig:res_norm_p3_c3}
\end{figure}

\begin{figure}[H]
    \centering
    \includegraphics[width=.8\linewidth]{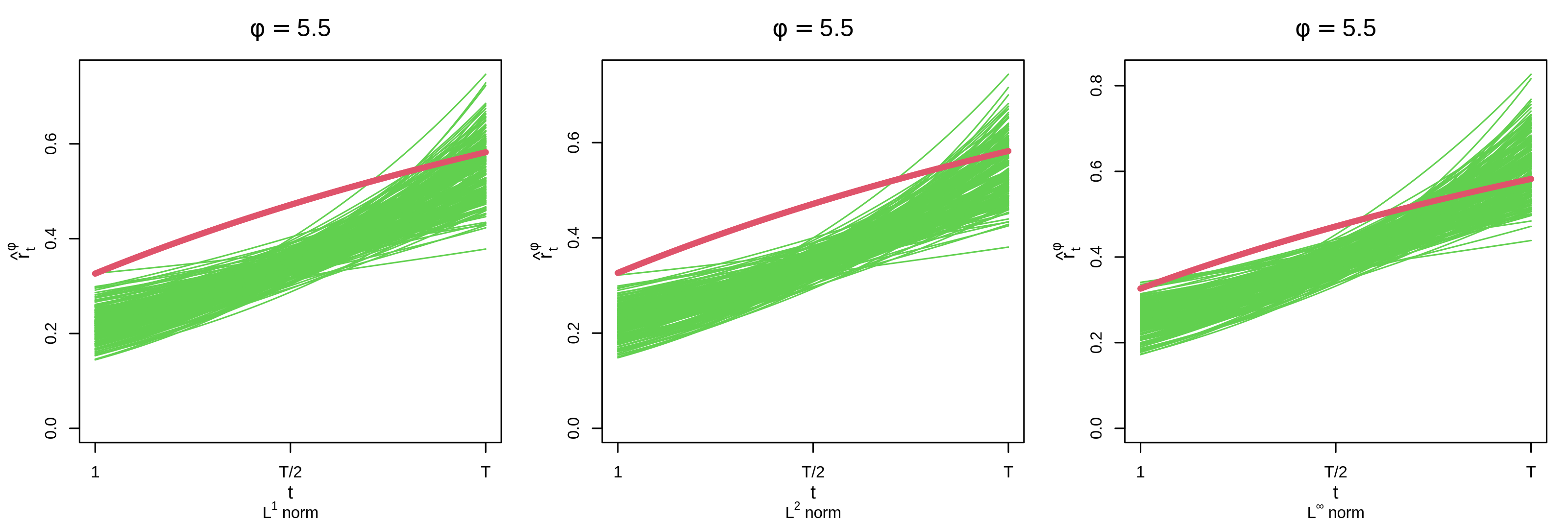}
    \caption{Boundary set radii estimates at $\phi = 7\pi/4$ across $\|\cdot\| \in \{\|\cdot \|_1 ,\|\cdot \|_2,\|\cdot \|_{\infty}\}$ for the third copula example.}
    \label{fig:res_norm_p4_c3}
\end{figure}

\begin{figure}[H]
    \centering
    \includegraphics[width=.8\linewidth]{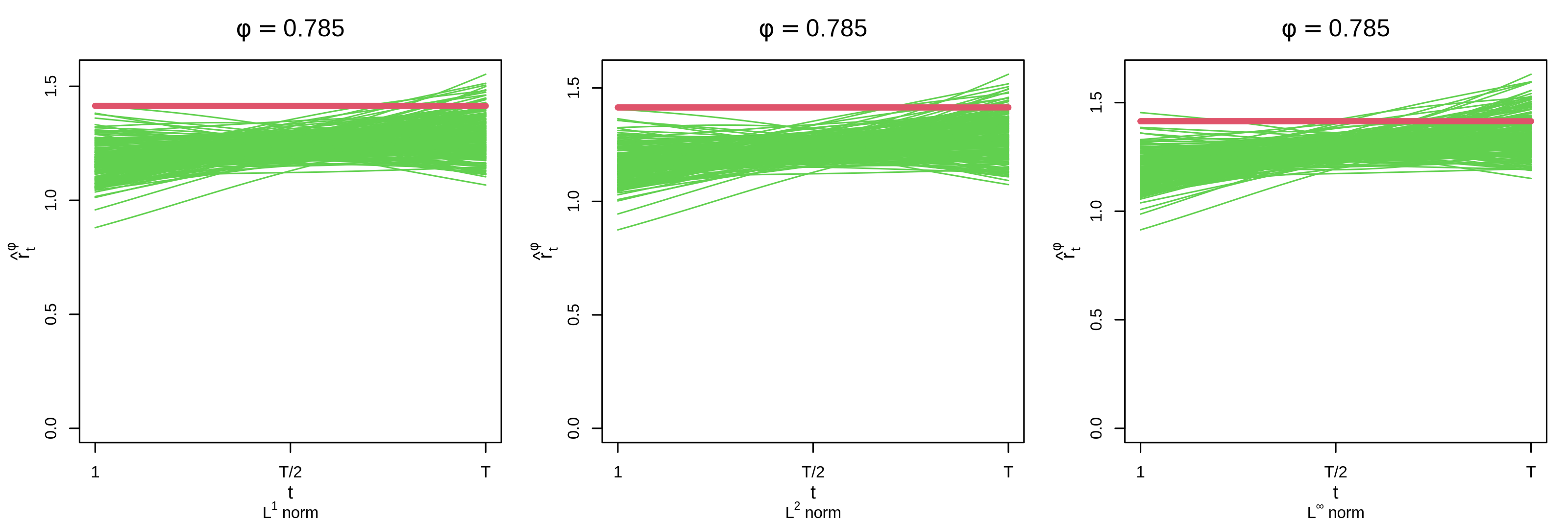}
    \caption{Boundary set radii estimates at $\phi = \pi/4$ across $\|\cdot\| \in \{\|\cdot \|_1 ,\|\cdot \|_2,\|\cdot \|_{\infty}\}$ for the fourth copula example.}
    \label{fig:res_norm_p1_c4}
\end{figure}

\begin{figure}[H]
    \centering
    \includegraphics[width=.8\linewidth]{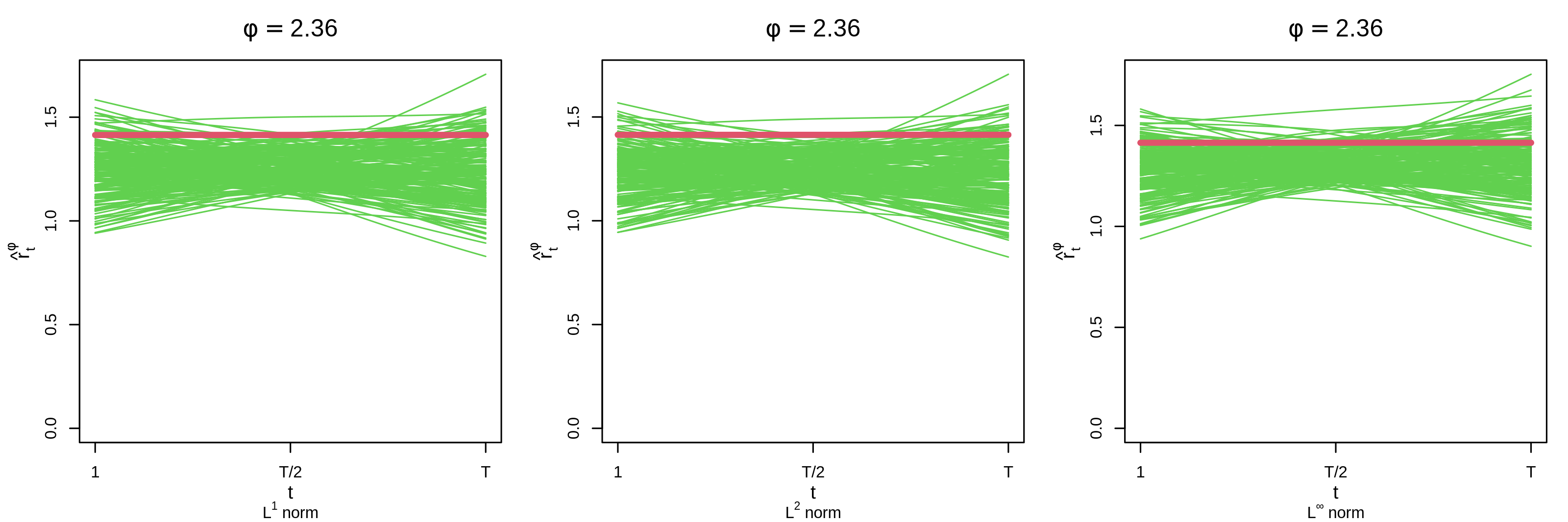}
    \caption{Boundary set radii estimates at $\phi = 3\pi/4$ across $\|\cdot\| \in \{\|\cdot \|_1 ,\|\cdot \|_2,\|\cdot \|_{\infty}\}$ for the fourth copula example.}
    \label{fig:res_norm_p2_c4}
\end{figure}

\begin{figure}[H]
    \centering
    \includegraphics[width=.8\linewidth]{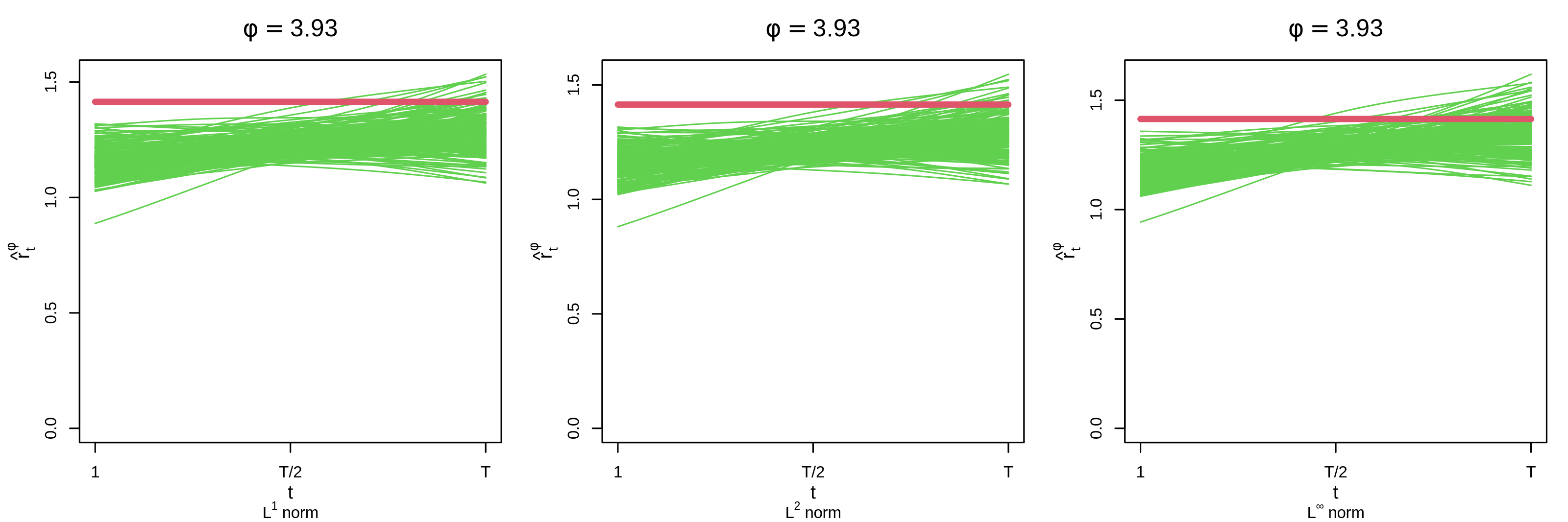}
    \caption{Boundary set radii estimates at $\phi = 5\pi/4$ across $\|\cdot\| \in \{\|\cdot \|_1 ,\|\cdot \|_2,\|\cdot \|_{\infty}\}$ for the fourth copula example.}
    \label{fig:res_norm_p3_c4}
\end{figure}

\begin{figure}[H]
    \centering
    \includegraphics[width=.8\linewidth]{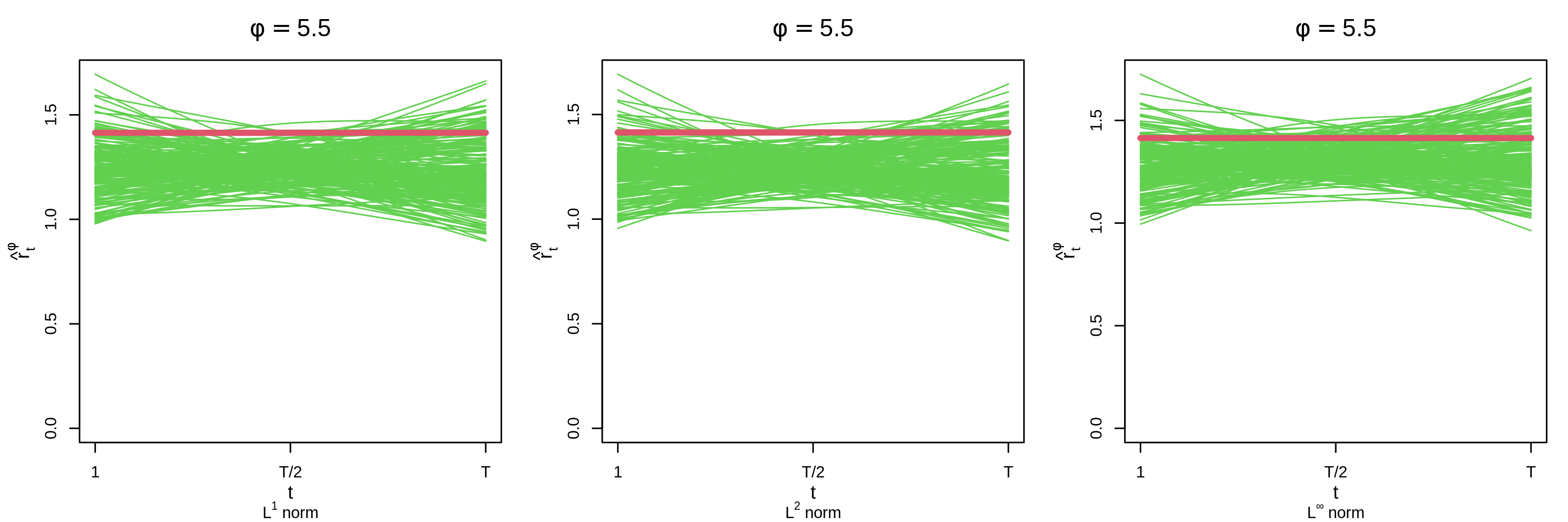}
    \caption{Boundary set radii estimates at $\phi = 7\pi/4$ across $\|\cdot\| \in \{\|\cdot \|_1 ,\|\cdot \|_2,\|\cdot \|_{\infty}\}$ for the fourth copula example.}
    \label{fig:res_norm_p4_c4}
\end{figure}

\begin{figure}[H]
    \centering
    \includegraphics[width=.8\linewidth]{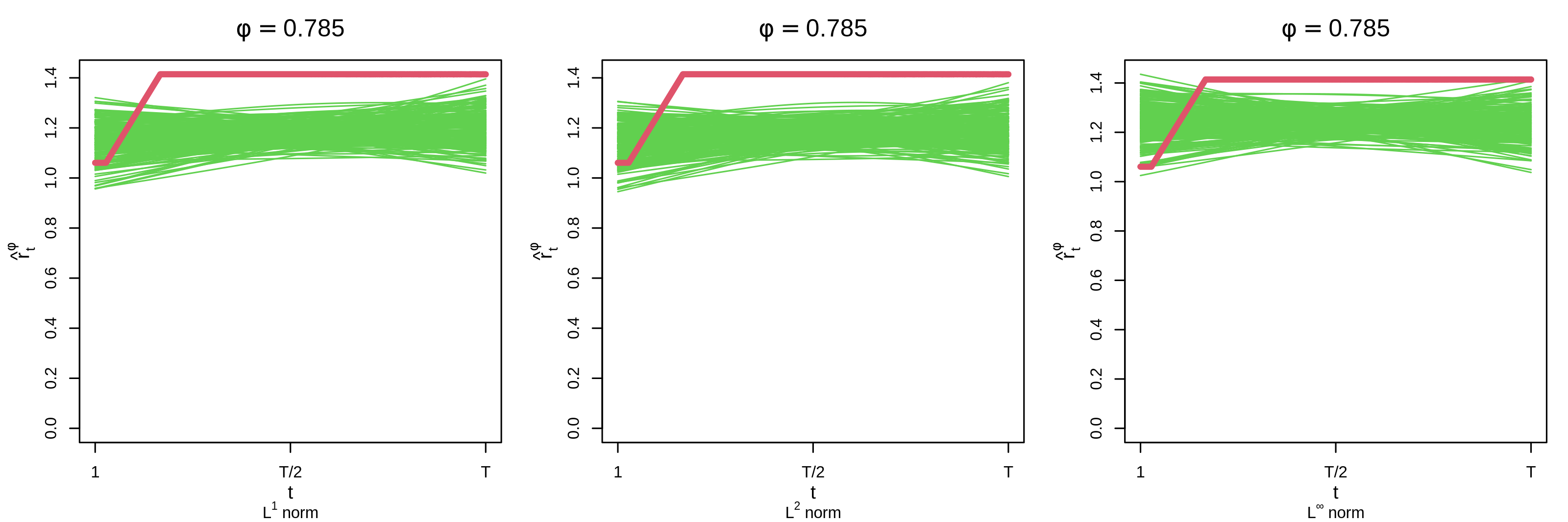}
    \caption{Boundary set radii estimates at $\phi = \pi/4$ across $\|\cdot\| \in \{\|\cdot \|_1 ,\|\cdot \|_2,\|\cdot \|_{\infty}\}$ for the fifth copula example.}
    \label{fig:res_norm_p1_c5}
\end{figure}

\begin{figure}[H]
    \centering
    \includegraphics[width=.8\linewidth]{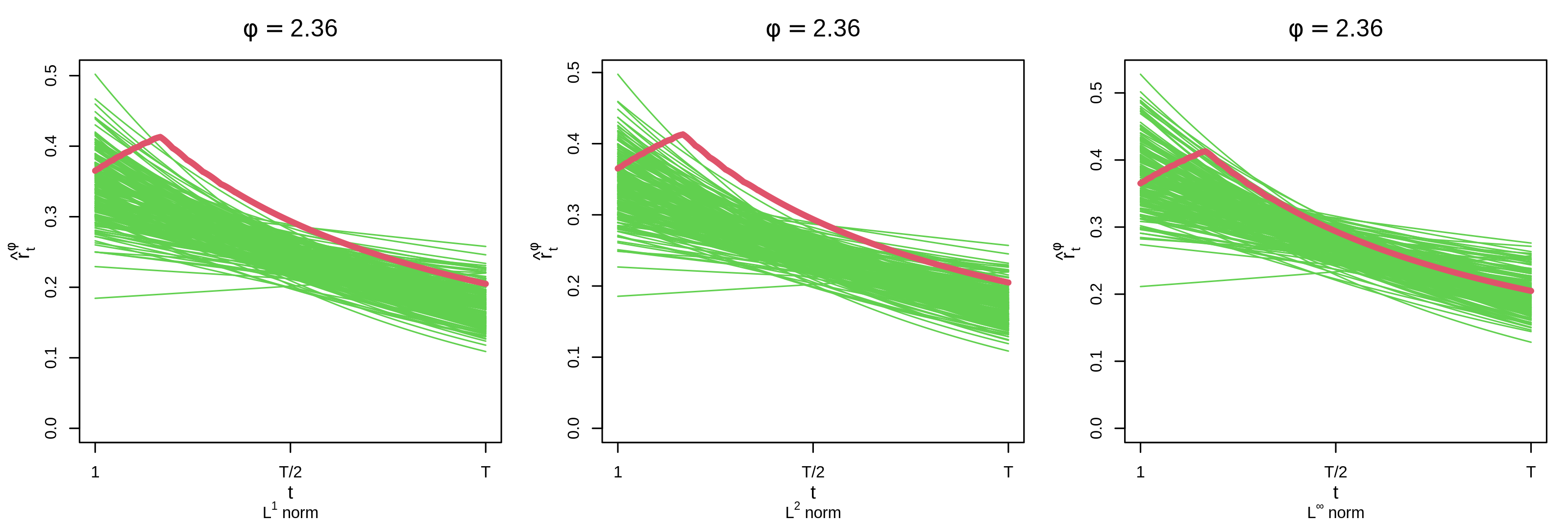}
    \caption{Boundary set radii estimates at $\phi = 3\pi/4$ across $\|\cdot\| \in \{\|\cdot \|_1 ,\|\cdot \|_2,\|\cdot \|_{\infty}\}$ for the fifth copula example.}
    \label{fig:res_norm_p2_c5}
\end{figure}

\begin{figure}[H]
    \centering
    \includegraphics[width=.8\linewidth]{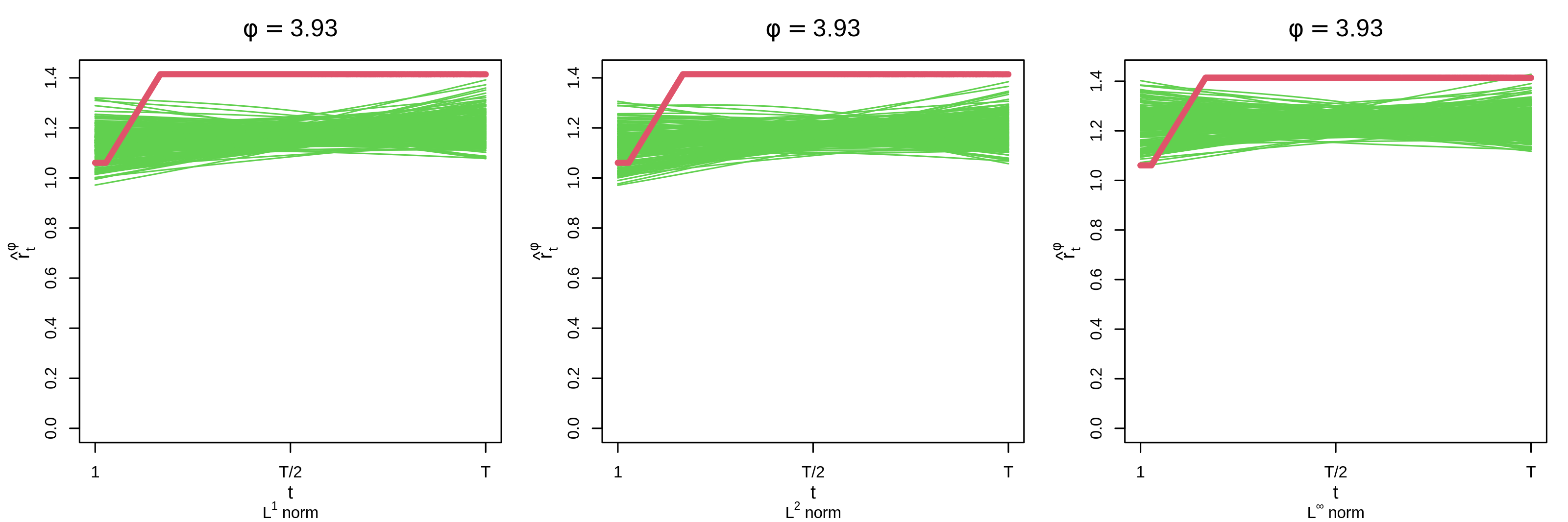}
    \caption{Boundary set radii estimates at $\phi = 5\pi/4$ across $\|\cdot\| \in \{\|\cdot \|_1 ,\|\cdot \|_2,\|\cdot \|_{\infty}\}$ for the fifth copula example.}
    \label{fig:res_norm_p3_c5}
\end{figure}

\begin{figure}[H]
    \centering
    \includegraphics[width=.8\linewidth]{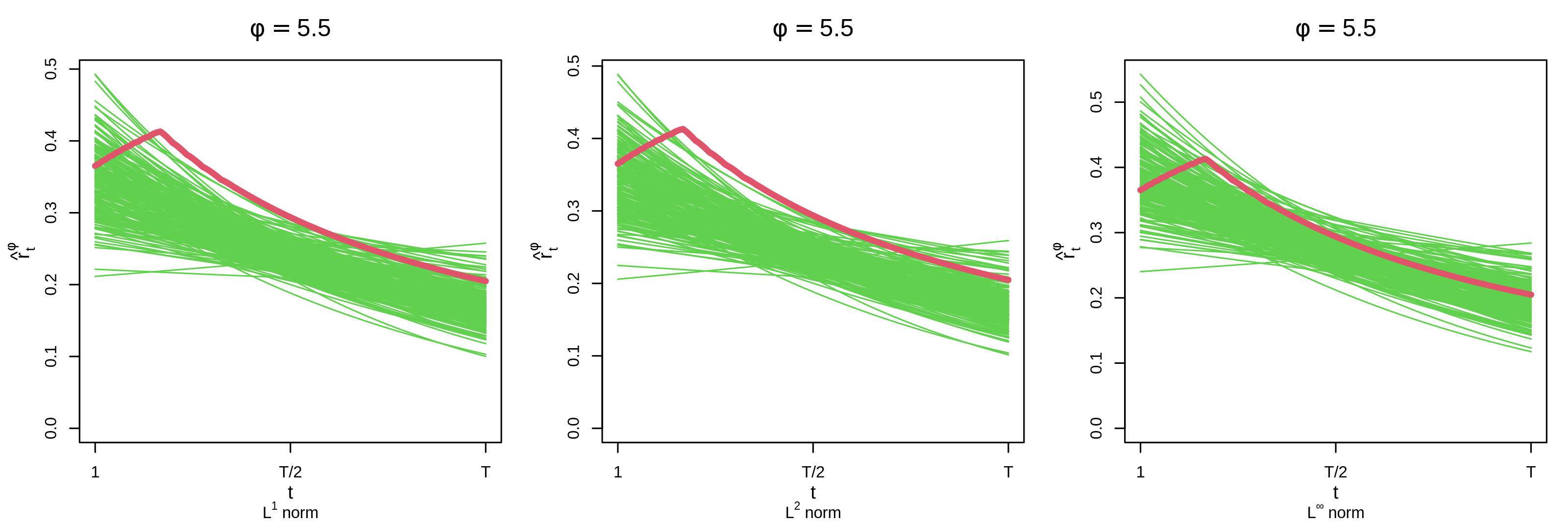}
    \caption{Boundary set radii estimates at $\phi = 7\pi/4$ across $\|\cdot\| \in \{\|\cdot \|_1 ,\|\cdot \|_2,\|\cdot \|_{\infty}\}$ for the fifth copula example.}
    \label{fig:res_norm_p4_c5}
\end{figure}

Figures~\ref{fig:res_norm_t1_c1}-\ref{fig:res_norm_p4_c5} illustrate the effect of the choice of norm $\| \cdot\|$ on the boundary set estimates. 

\subsection{Additional results} \label{subsec:appen_final_results}

Figures~\ref{fig:res_ss_t1_c1}-\ref{fig:res_ss_t3_c5} illustrate additional results for the model formulation selected in Section~\ref{subsec:model_formulation}. 

\begin{figure}[H]
    \centering
    \begin{subfigure}{0.32\textwidth}
        \includegraphics[width=\linewidth]{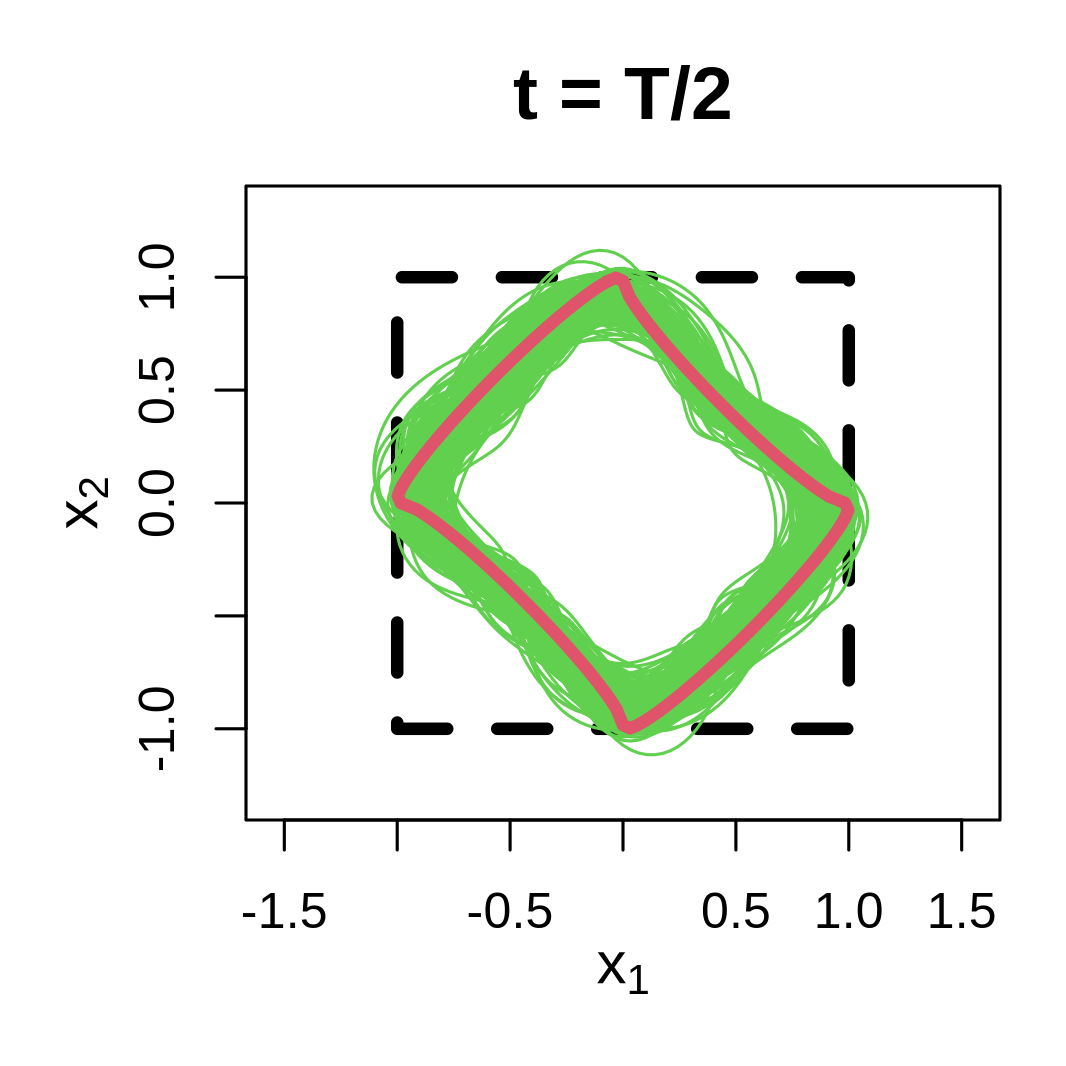}
    \end{subfigure}
    \begin{subfigure}{0.32\textwidth}
        \includegraphics[width=\linewidth]{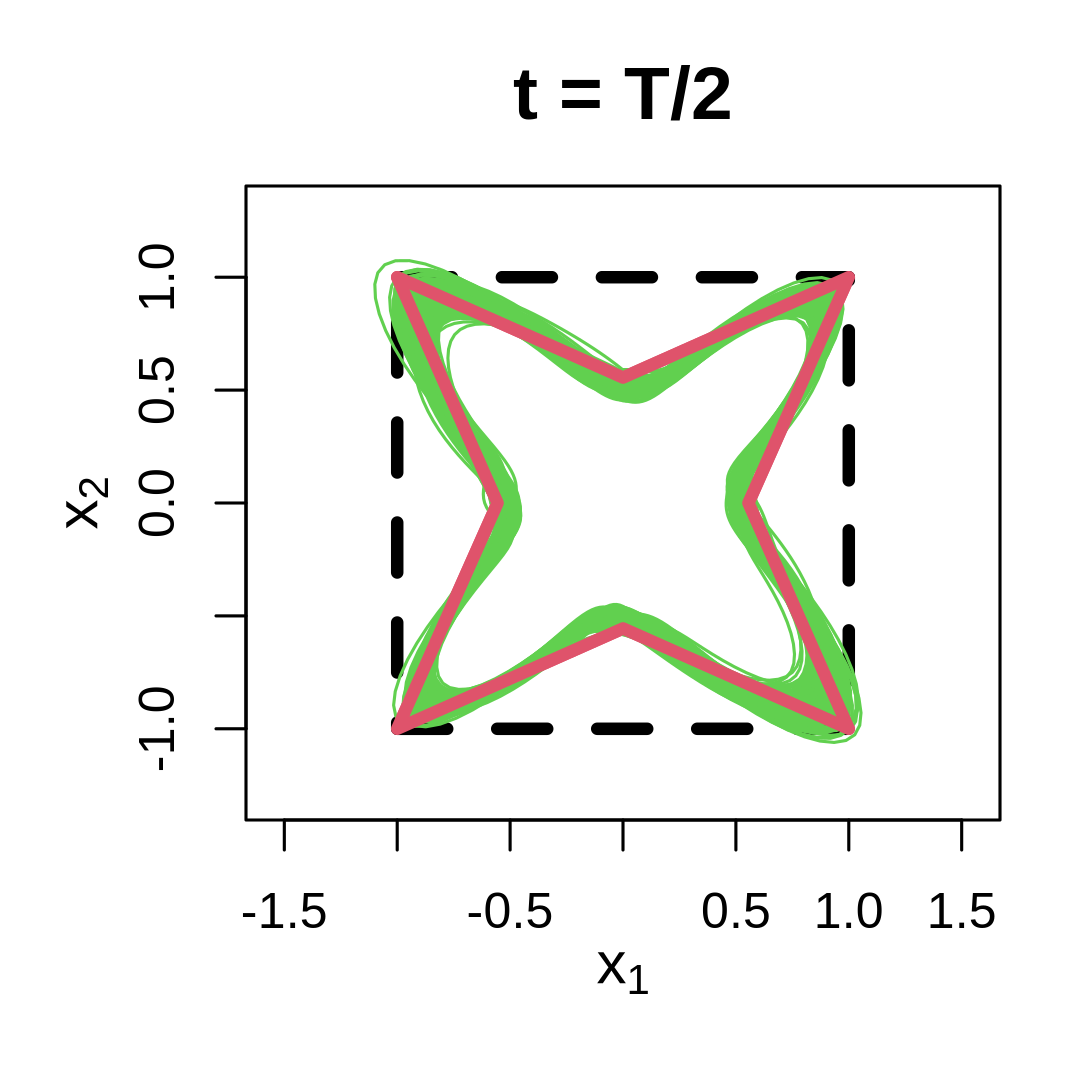}
    \end{subfigure}
    \begin{subfigure}{0.32\textwidth}
        \includegraphics[width=\linewidth]{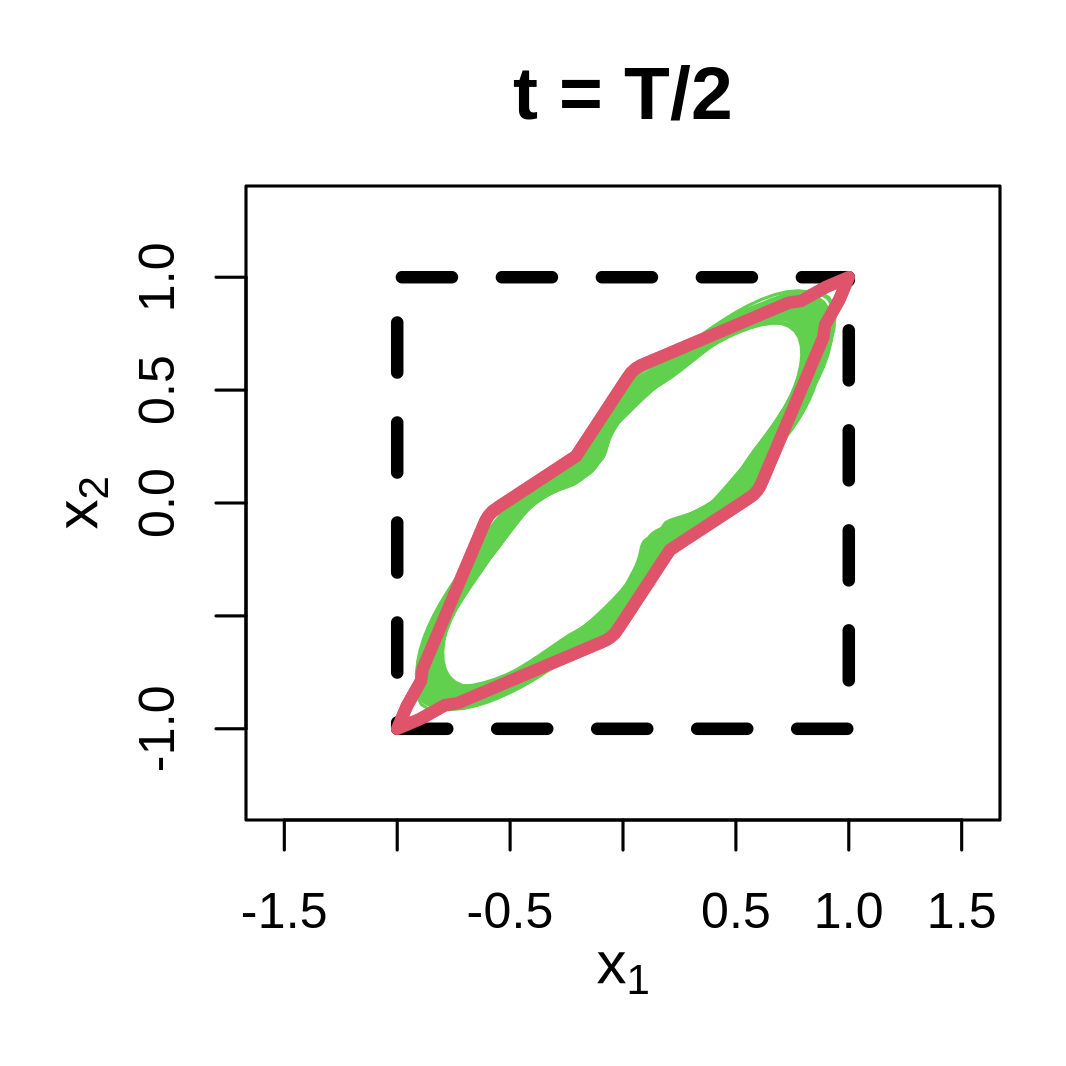}
    \end{subfigure}
    
    \caption{Boundary set estimates for $t=T/2$. The left, centre and right columns correspond to the second, fourth and fifth copula examples, respectively.}
    \label{fig:final_bs_1}
\end{figure}

\begin{figure}[H]
    \centering
    \begin{subfigure}{0.32\textwidth}
        \includegraphics[width=\linewidth]{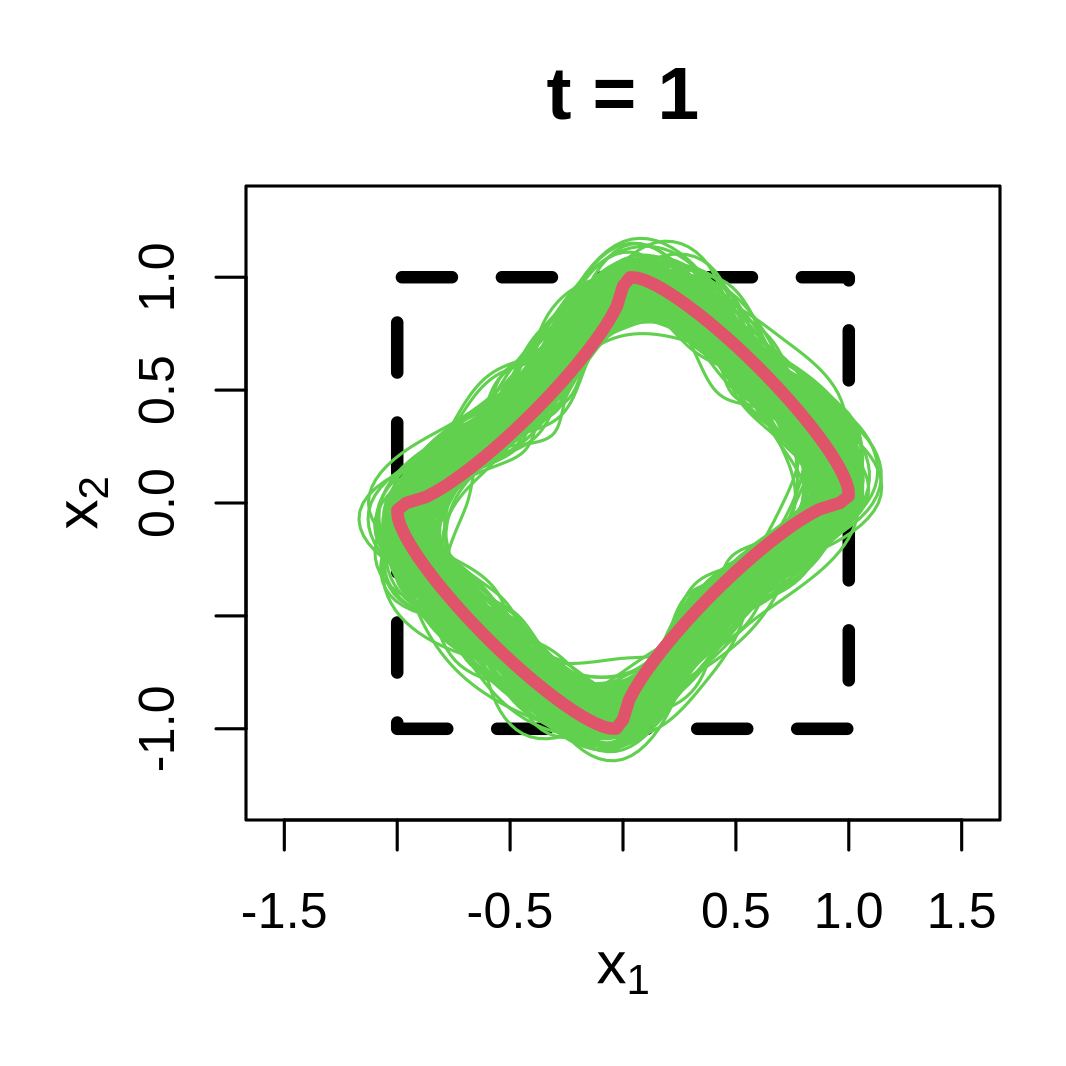}
    \end{subfigure}
    \begin{subfigure}{0.32\textwidth}
        \includegraphics[width=\linewidth]{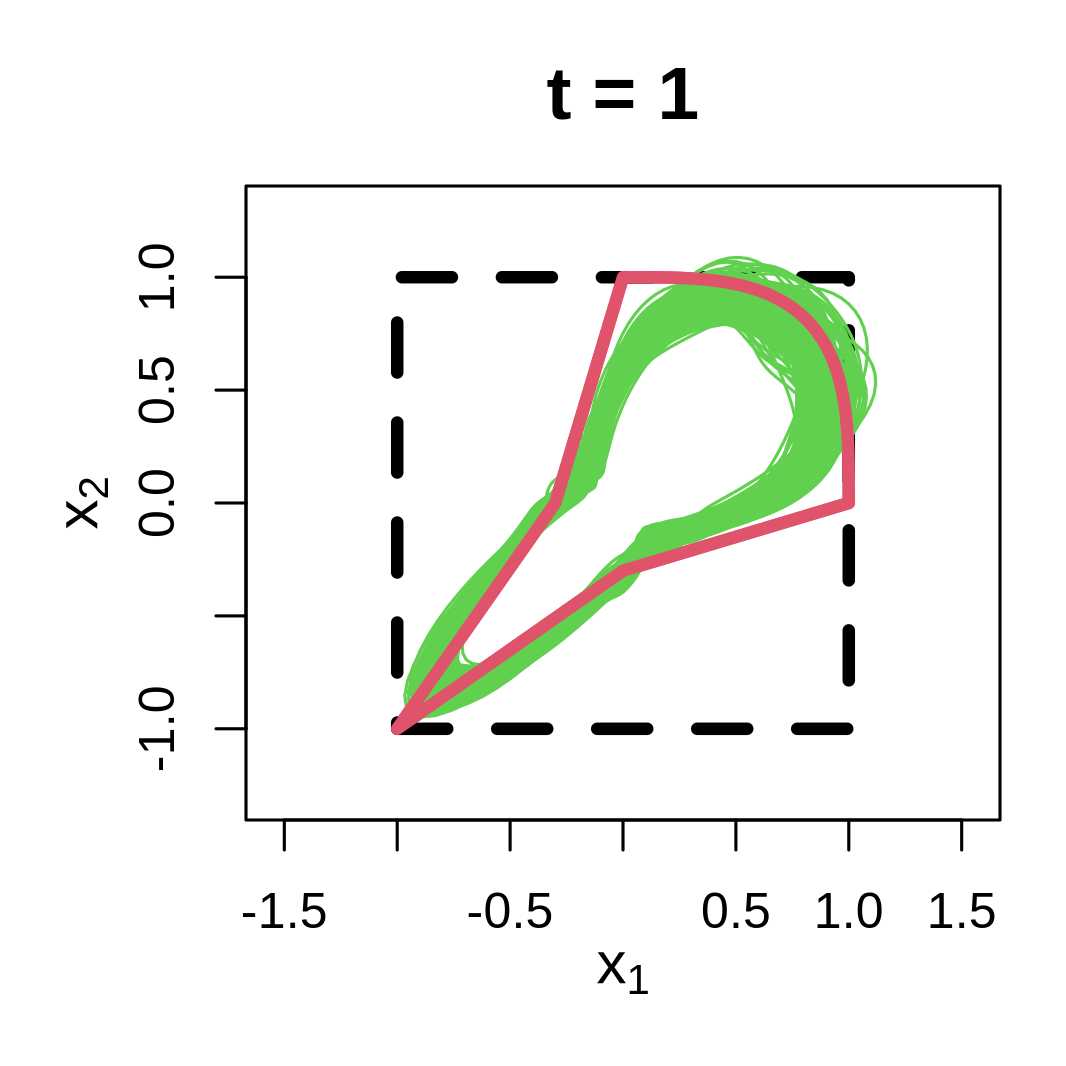}
    \end{subfigure}

        \begin{subfigure}{0.32\textwidth}
        \includegraphics[width=\linewidth]{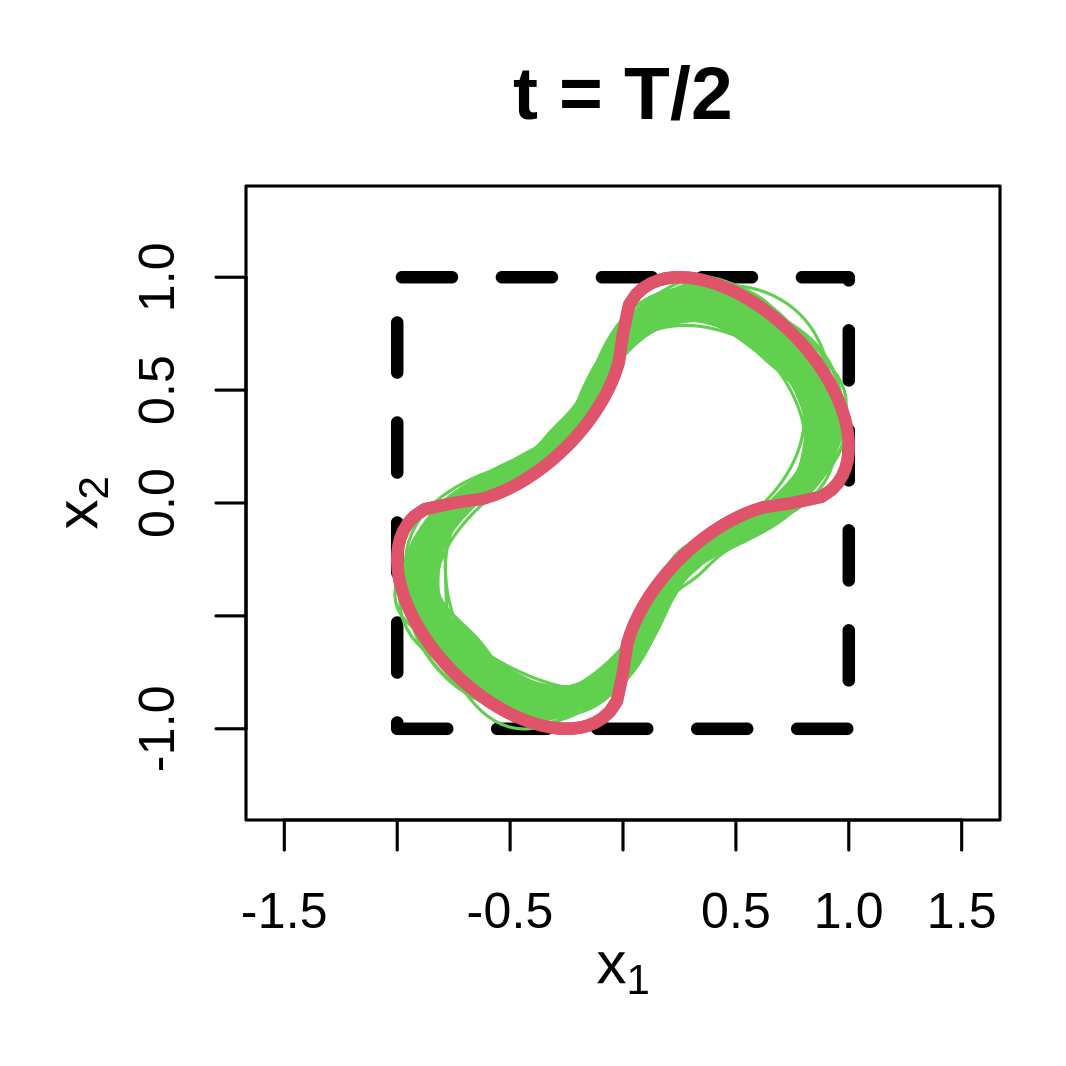}
    \end{subfigure}
    \begin{subfigure}{0.32\textwidth}
        \includegraphics[width=\linewidth]{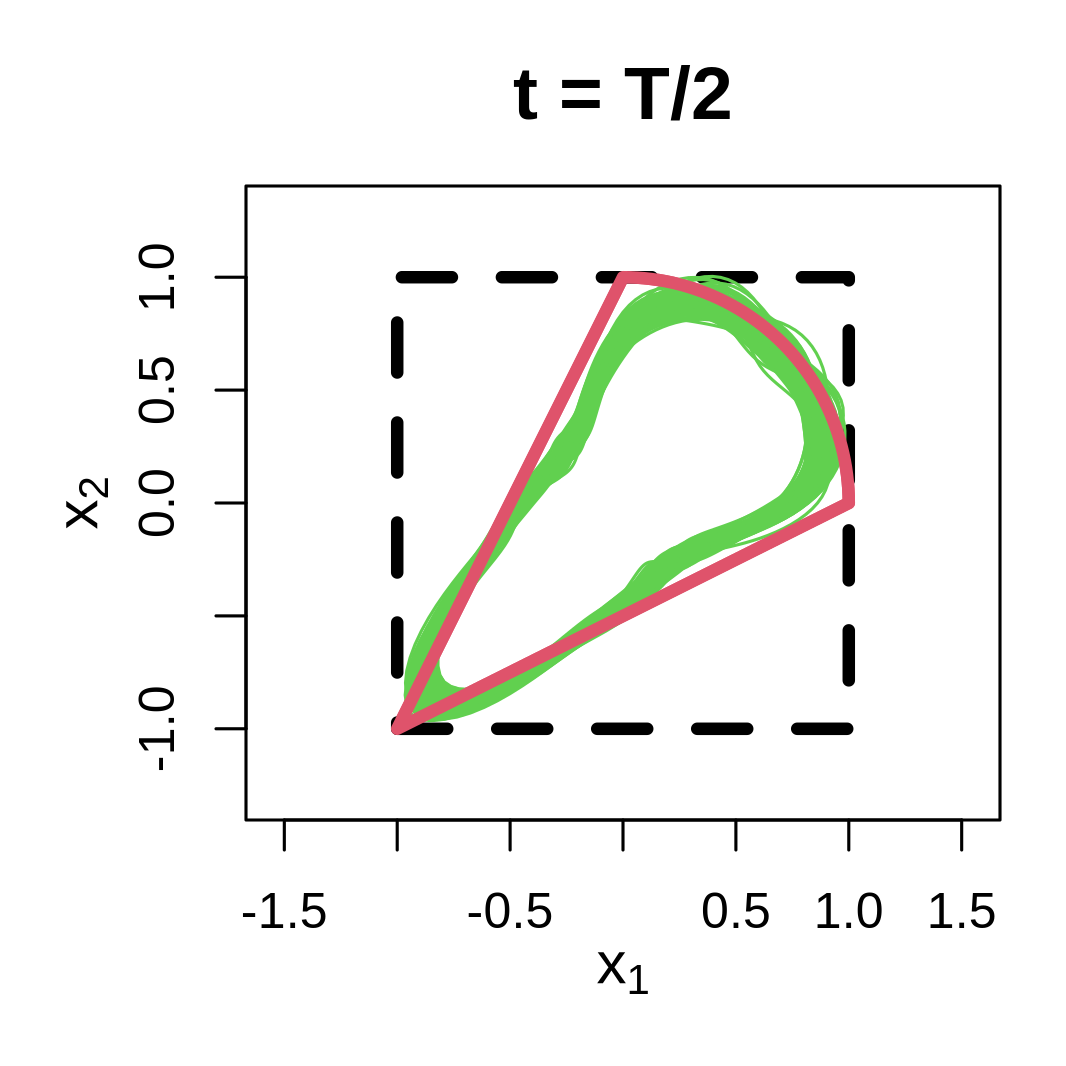}
    \end{subfigure}

        \begin{subfigure}{0.32\textwidth}
        \includegraphics[width=\linewidth]{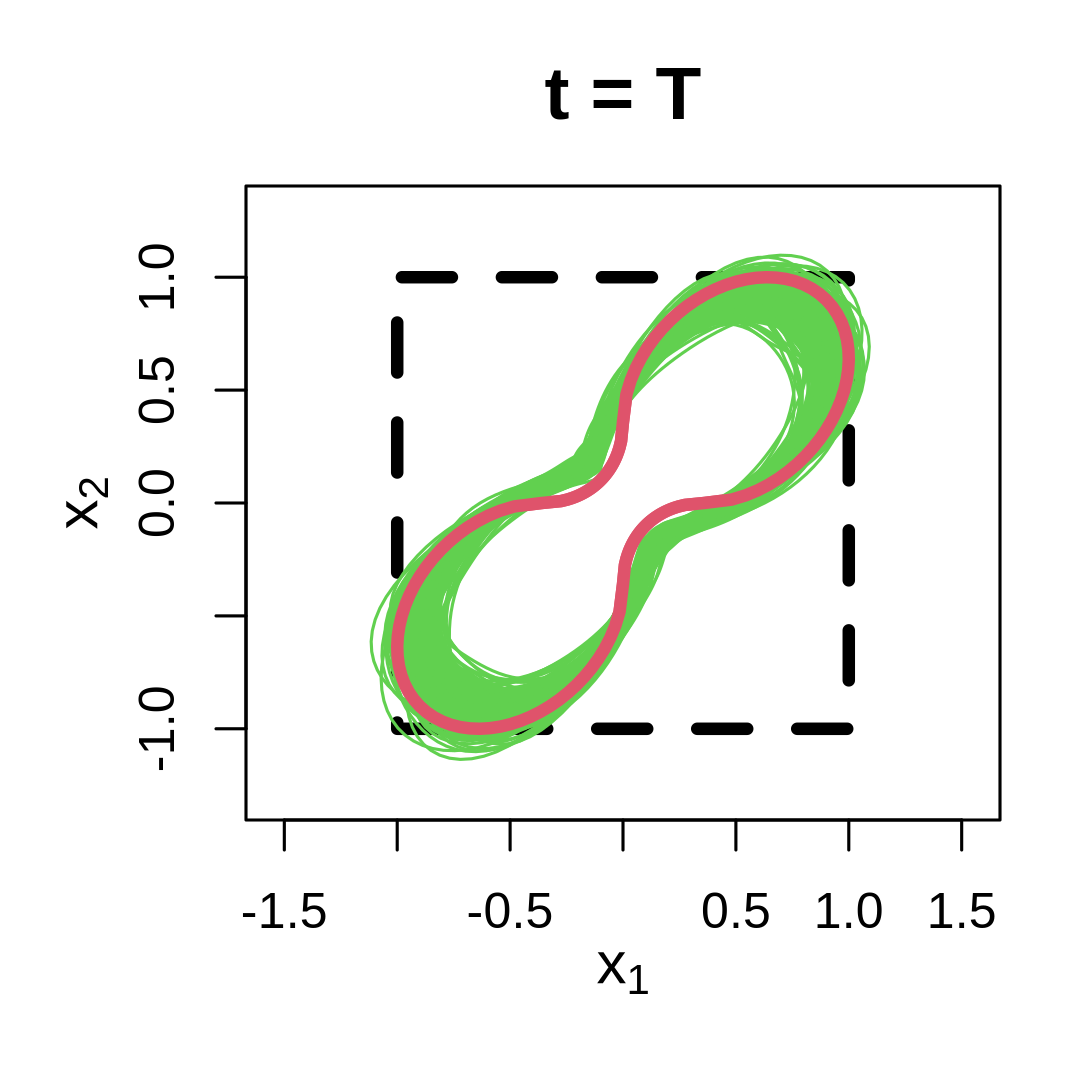}
    \end{subfigure}
    \begin{subfigure}{0.32\textwidth}
        \includegraphics[width=\linewidth]{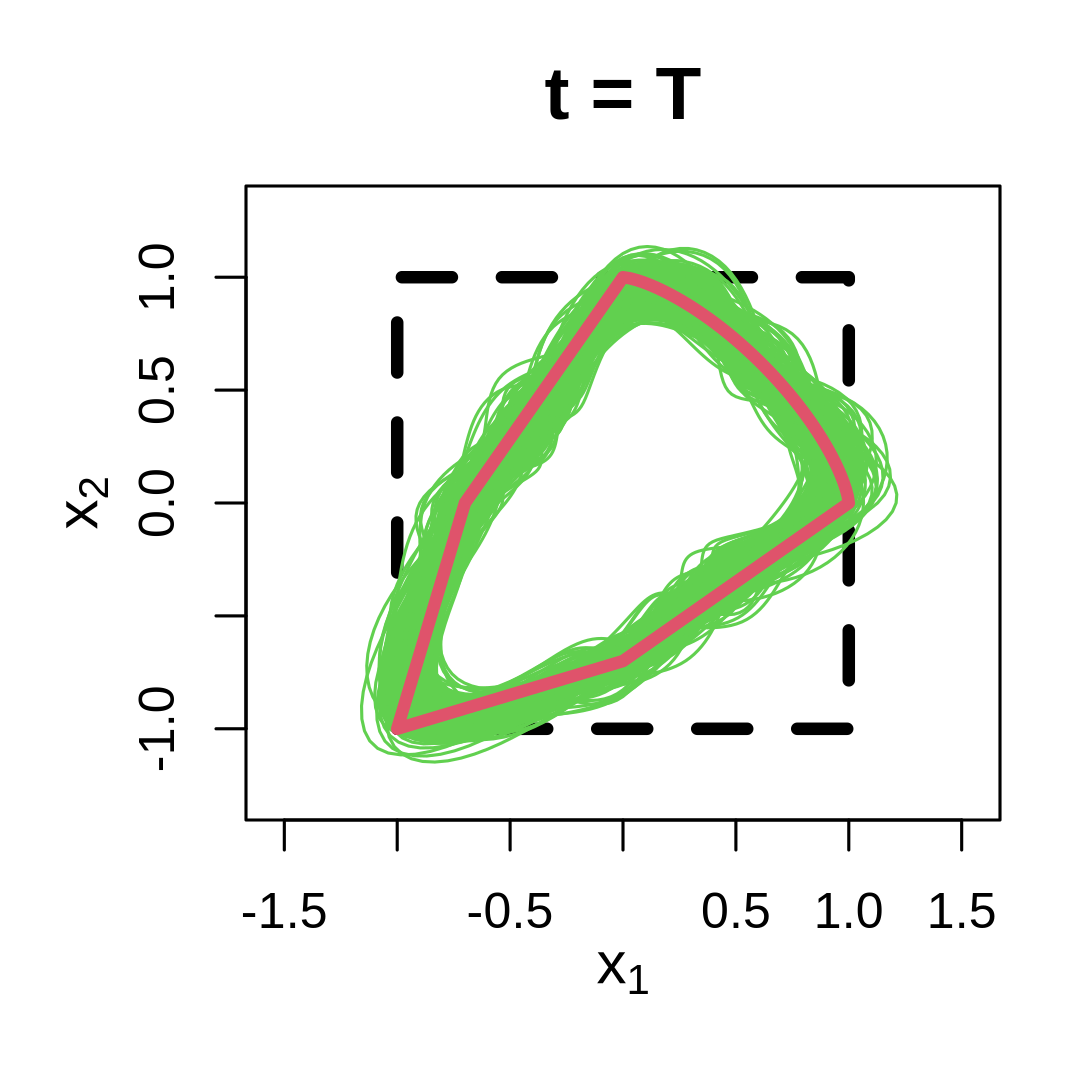}
    \end{subfigure}
    
    \caption{Boundary set estimates for $t=1$ (top row), $t = T/2$ (middle row) and $t=T$ (bottom row). The left, centre and right columns correspond to the first and third copula examples, respectively.}
    \label{fig:final_bs_2}
\end{figure}

\begin{figure}[H]
    \centering
    \begin{subfigure}{0.32\textwidth}
        \includegraphics[width=\linewidth]{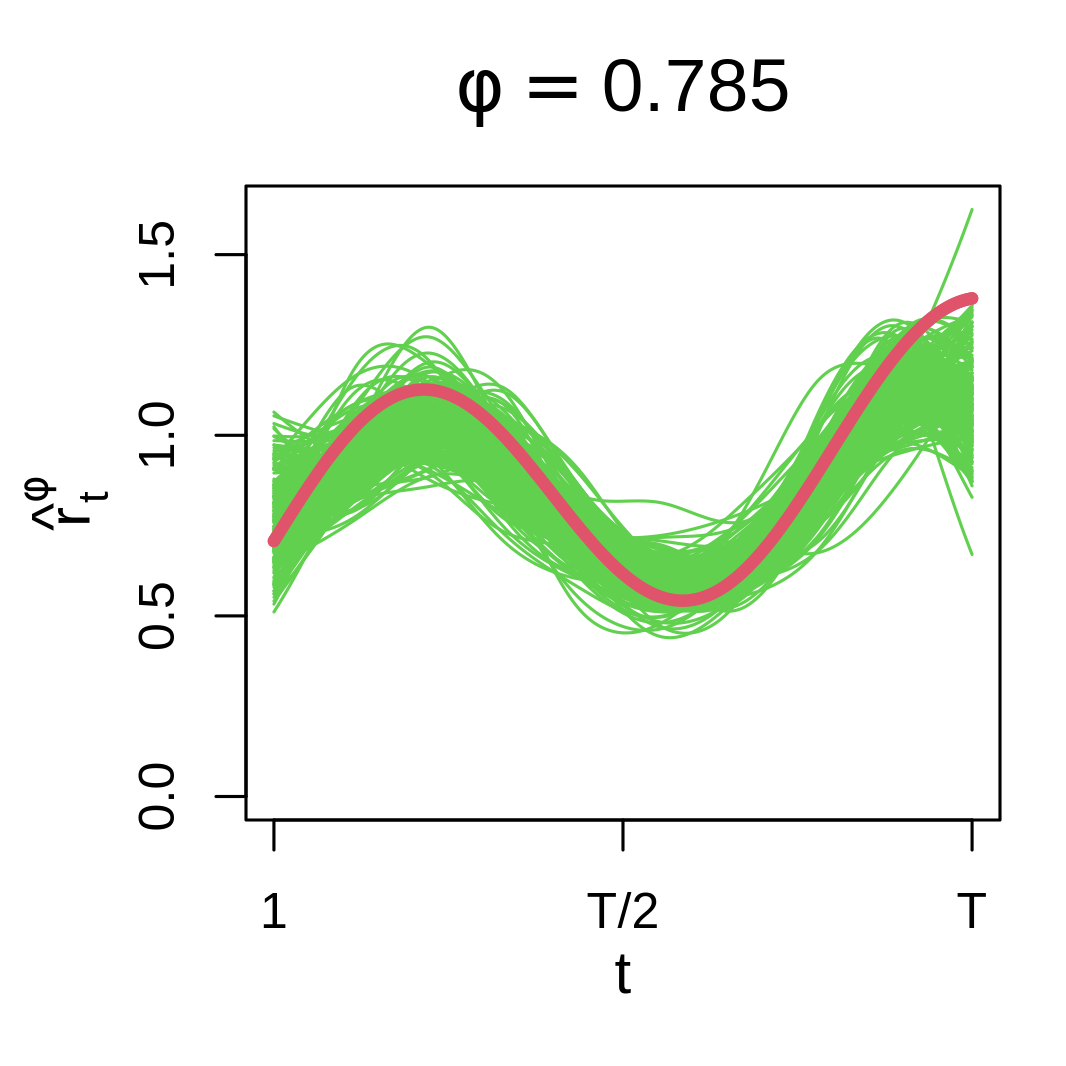}
    \end{subfigure}
    \begin{subfigure}{0.32\textwidth}
        \includegraphics[width=\linewidth]{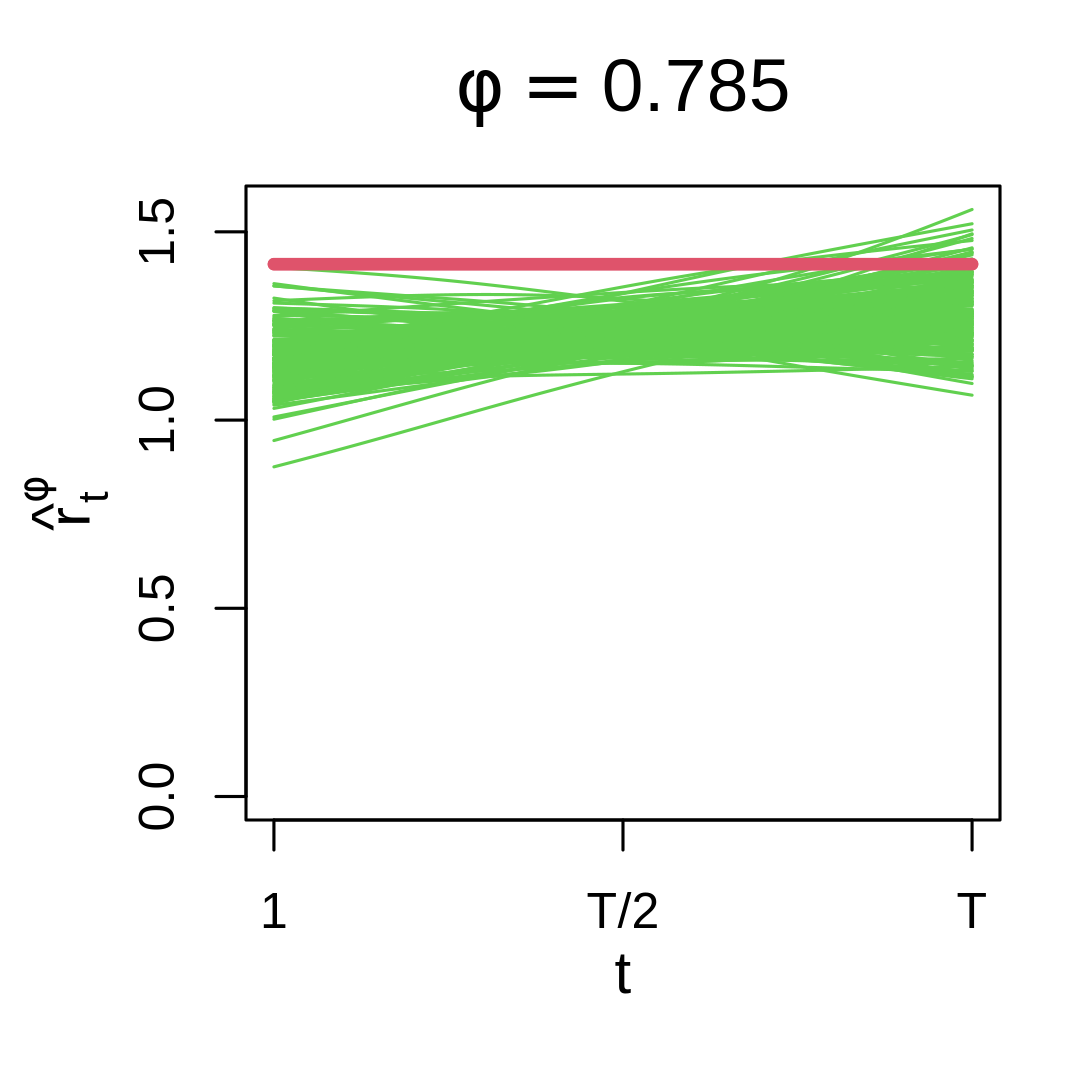}
    \end{subfigure}
    \begin{subfigure}{0.32\textwidth}
        \includegraphics[width=\linewidth]{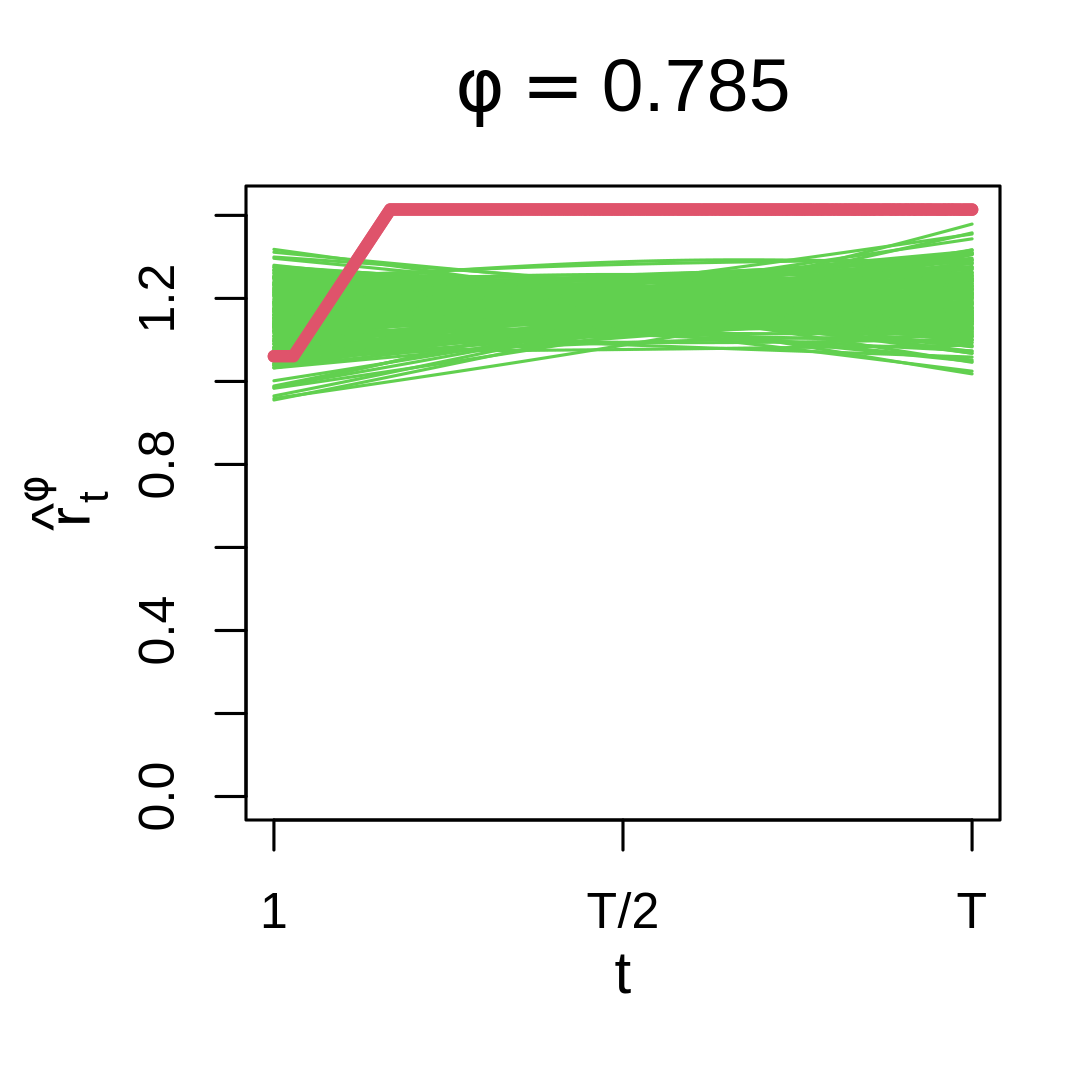}
    \end{subfigure}

    \begin{subfigure}{0.32\textwidth}
        \includegraphics[width=\linewidth]{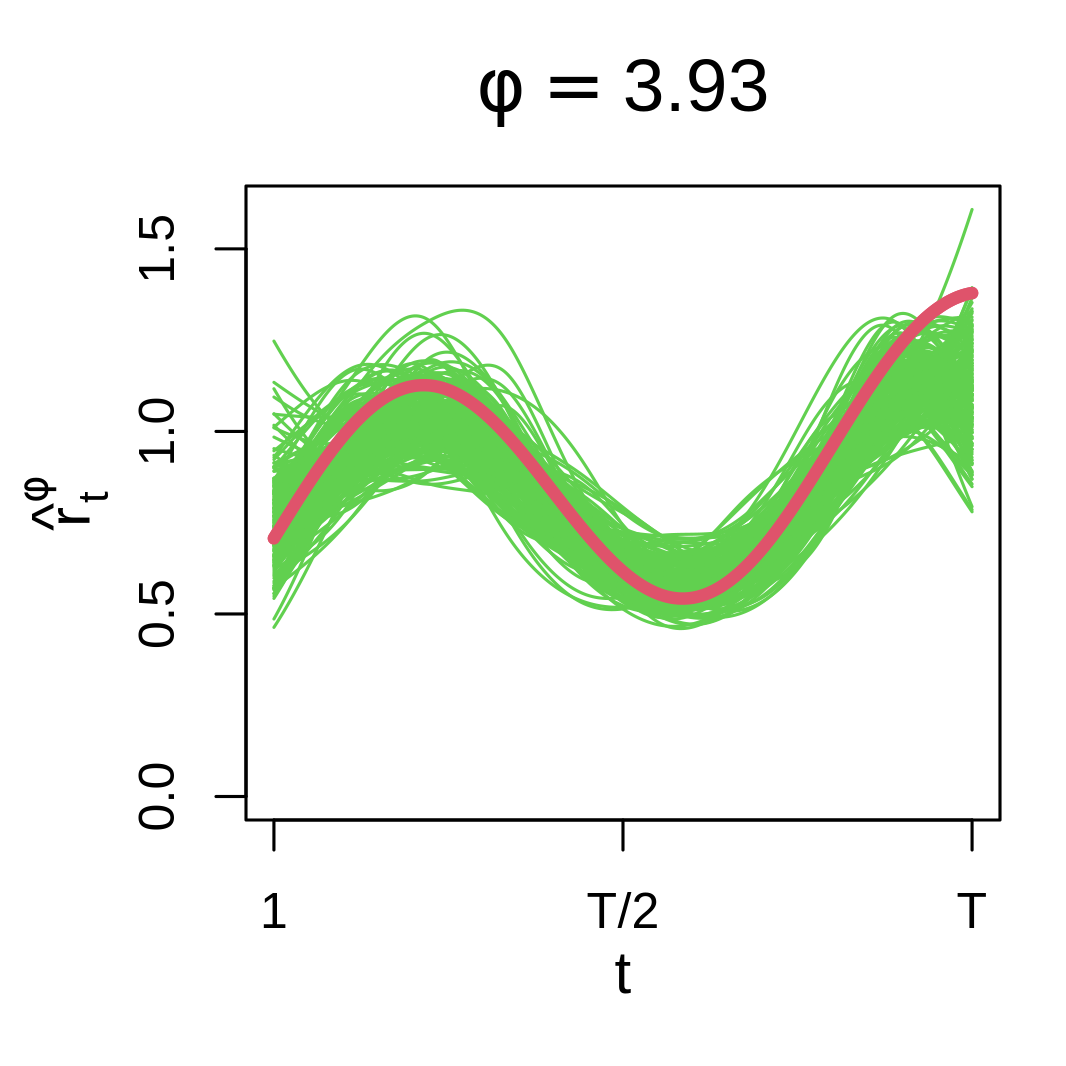}
    \end{subfigure}
    \begin{subfigure}{0.32\textwidth}
        \includegraphics[width=\linewidth]{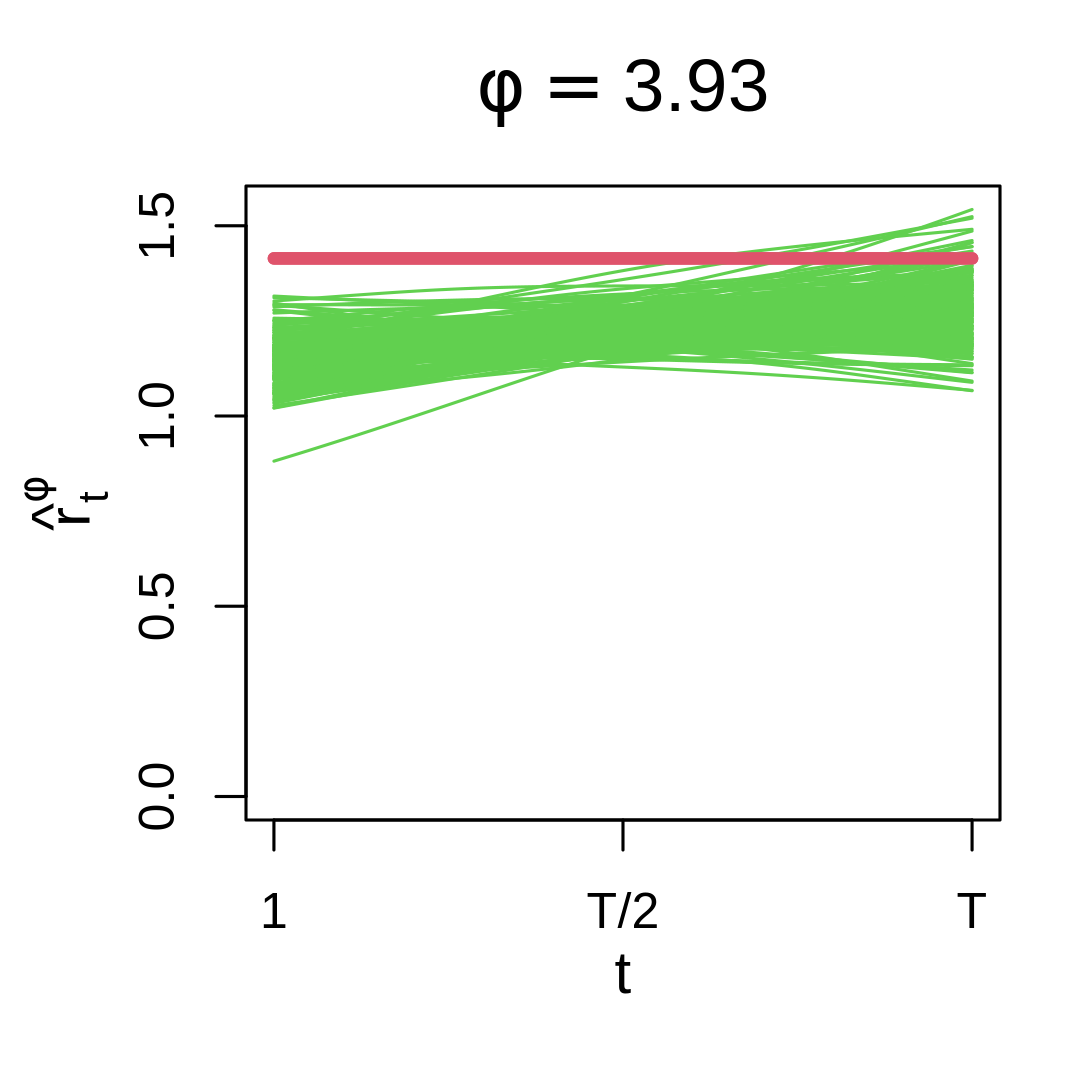}
    \end{subfigure}
    \begin{subfigure}{0.32\textwidth}
        \includegraphics[width=\linewidth]{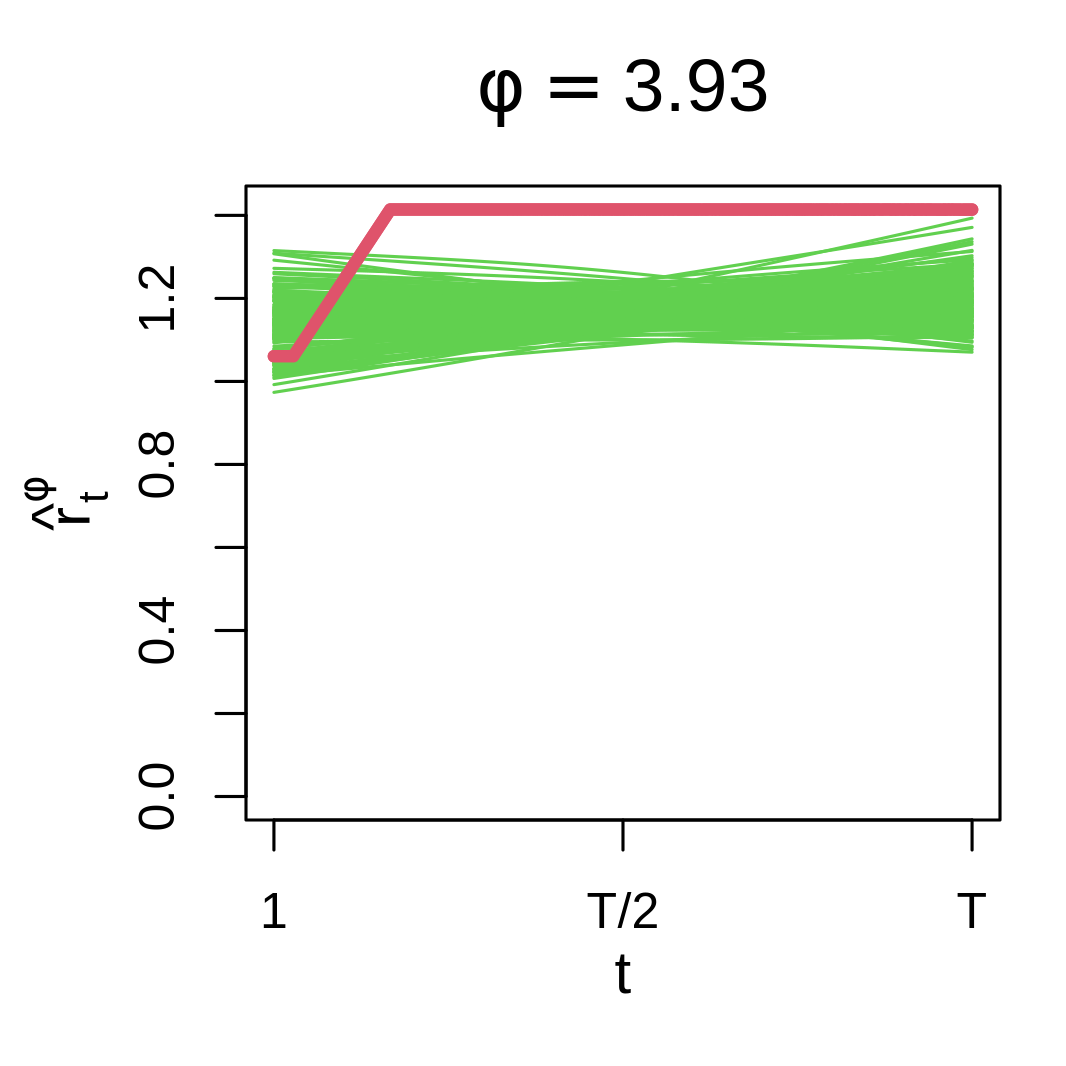}
    \end{subfigure}
    
    \caption{Boundary set radii estimates over time for $\phi=5\pi/4$ (top row) and $\phi = 7\pi/4$ (bottom row). The left, centre and right columns correspond to the second, fourth and fifth copula examples, respectively.}
    \label{fig:final_radii_1}
\end{figure}

\begin{figure}[H]
    \centering
    \begin{subfigure}{0.32\textwidth}
        \includegraphics[width=\linewidth]{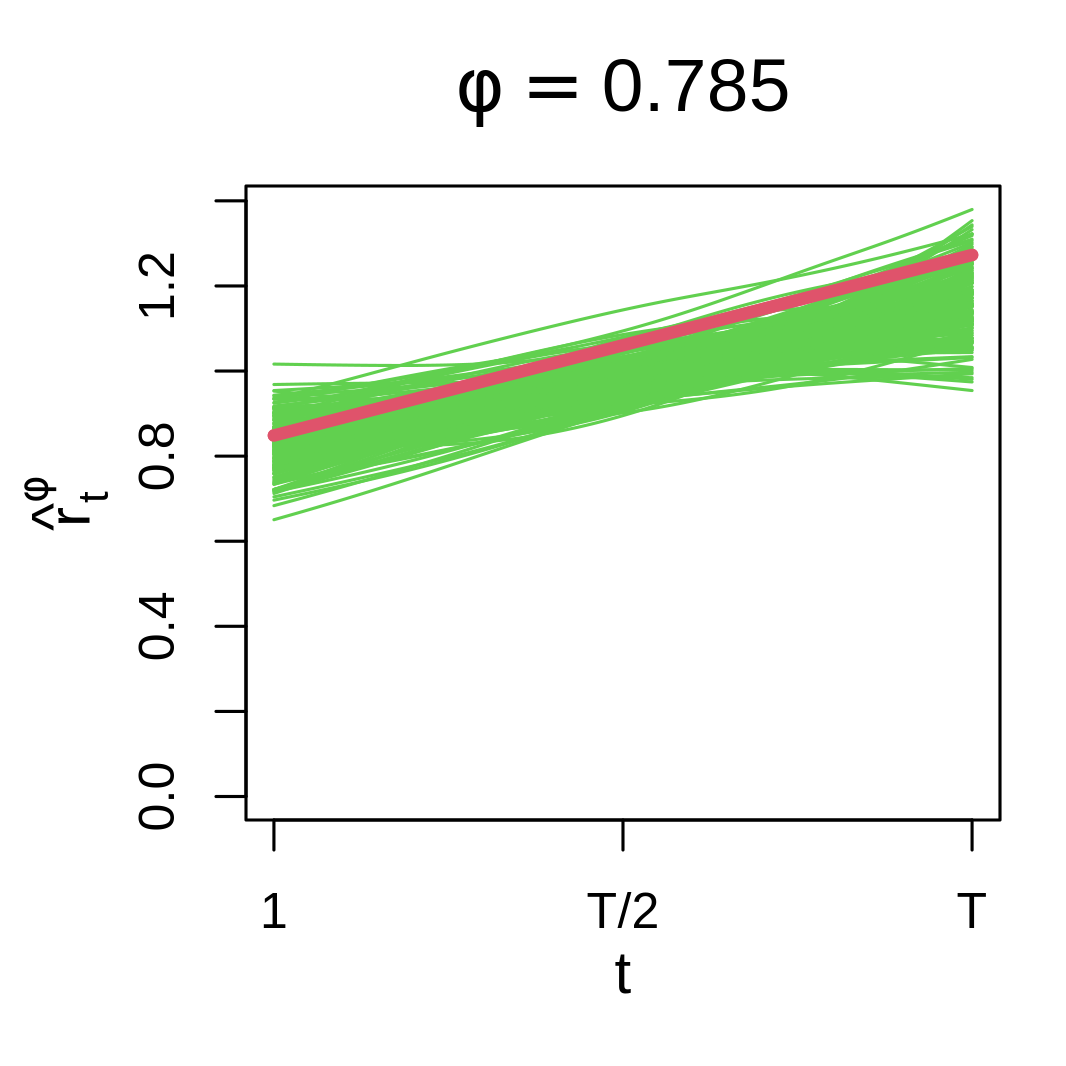}
    \end{subfigure}
    \begin{subfigure}{0.32\textwidth}
        \includegraphics[width=\linewidth]{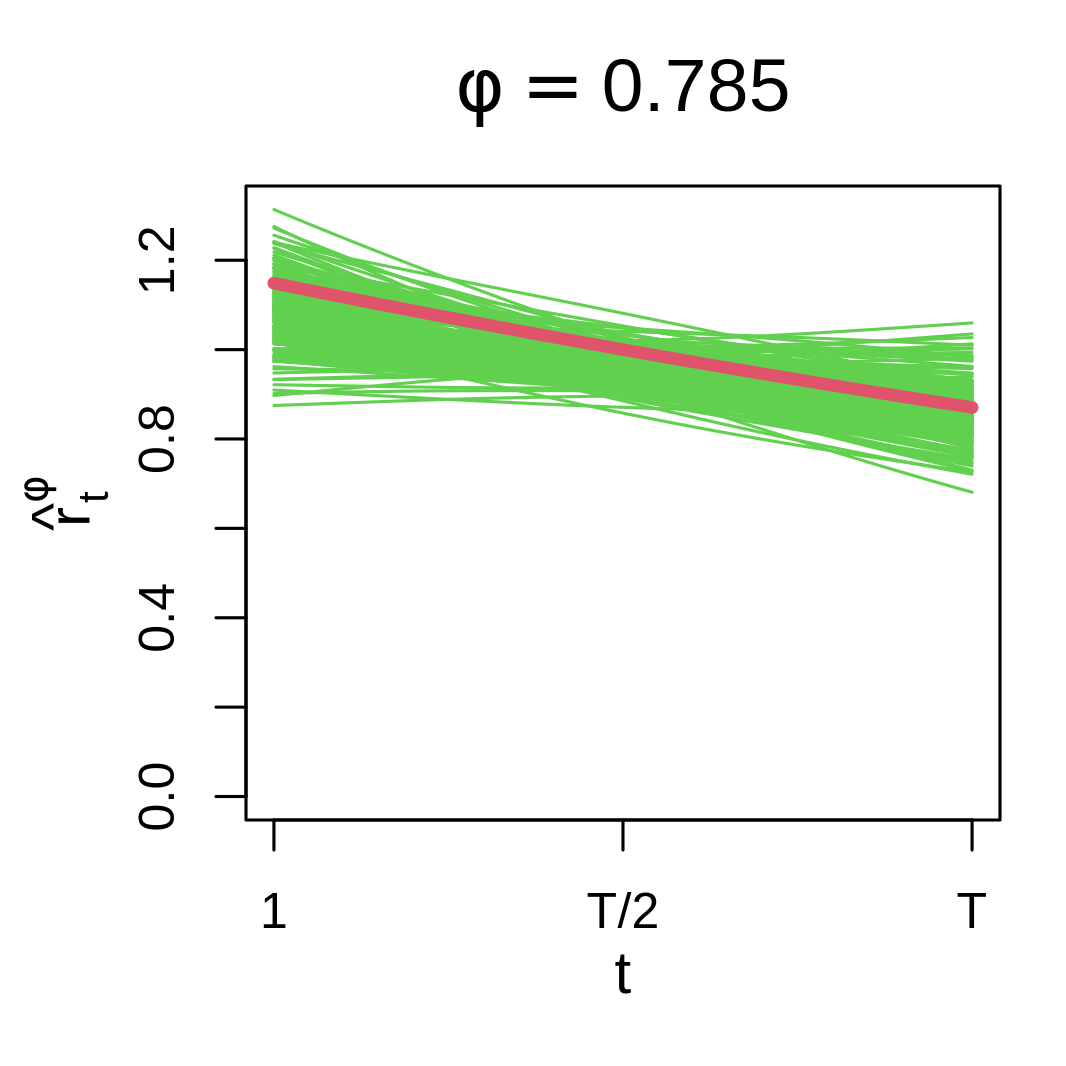}
    \end{subfigure}

    \begin{subfigure}{0.32\textwidth}
        \includegraphics[width=\linewidth]{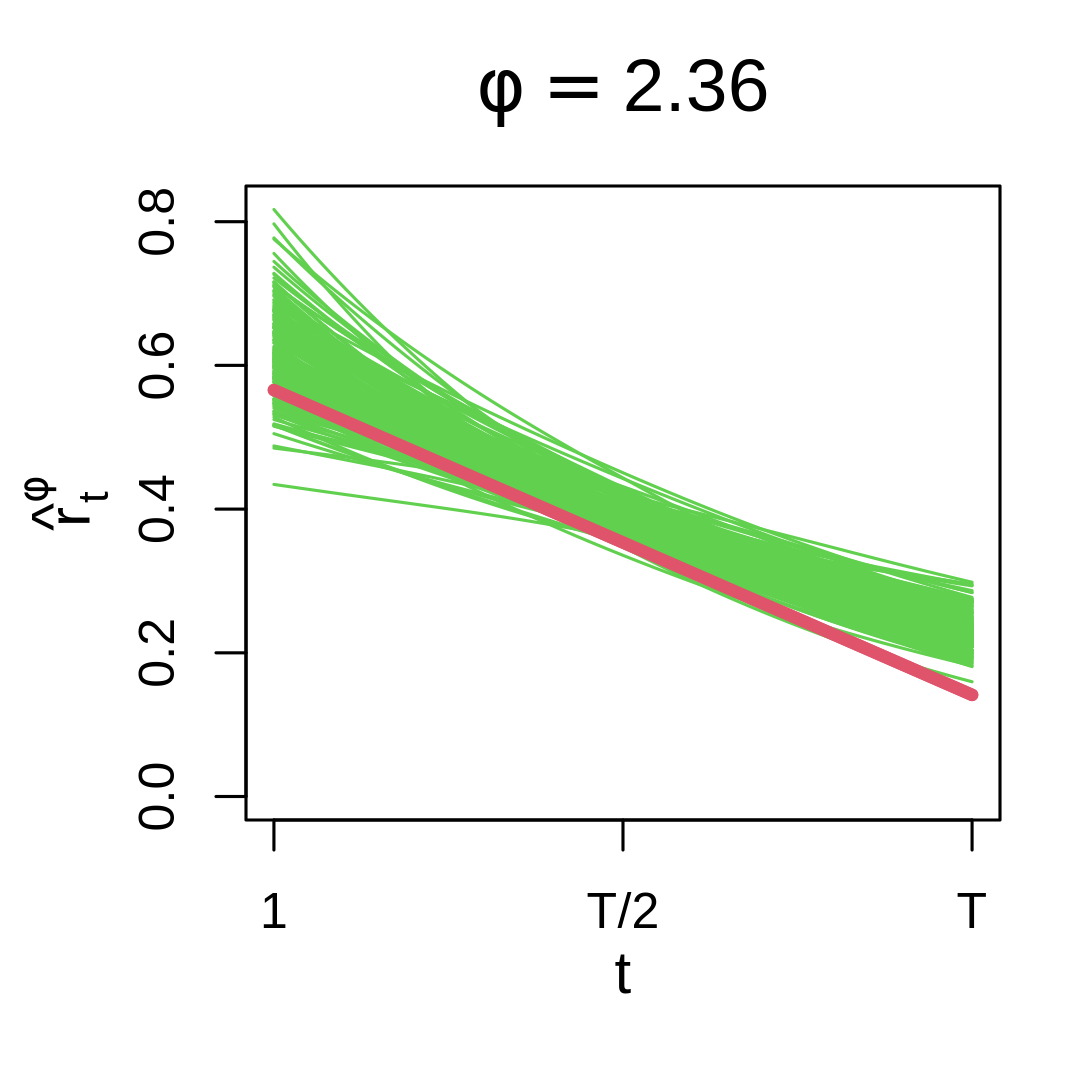}
    \end{subfigure}
    \begin{subfigure}{0.32\textwidth}
        \includegraphics[width=\linewidth]{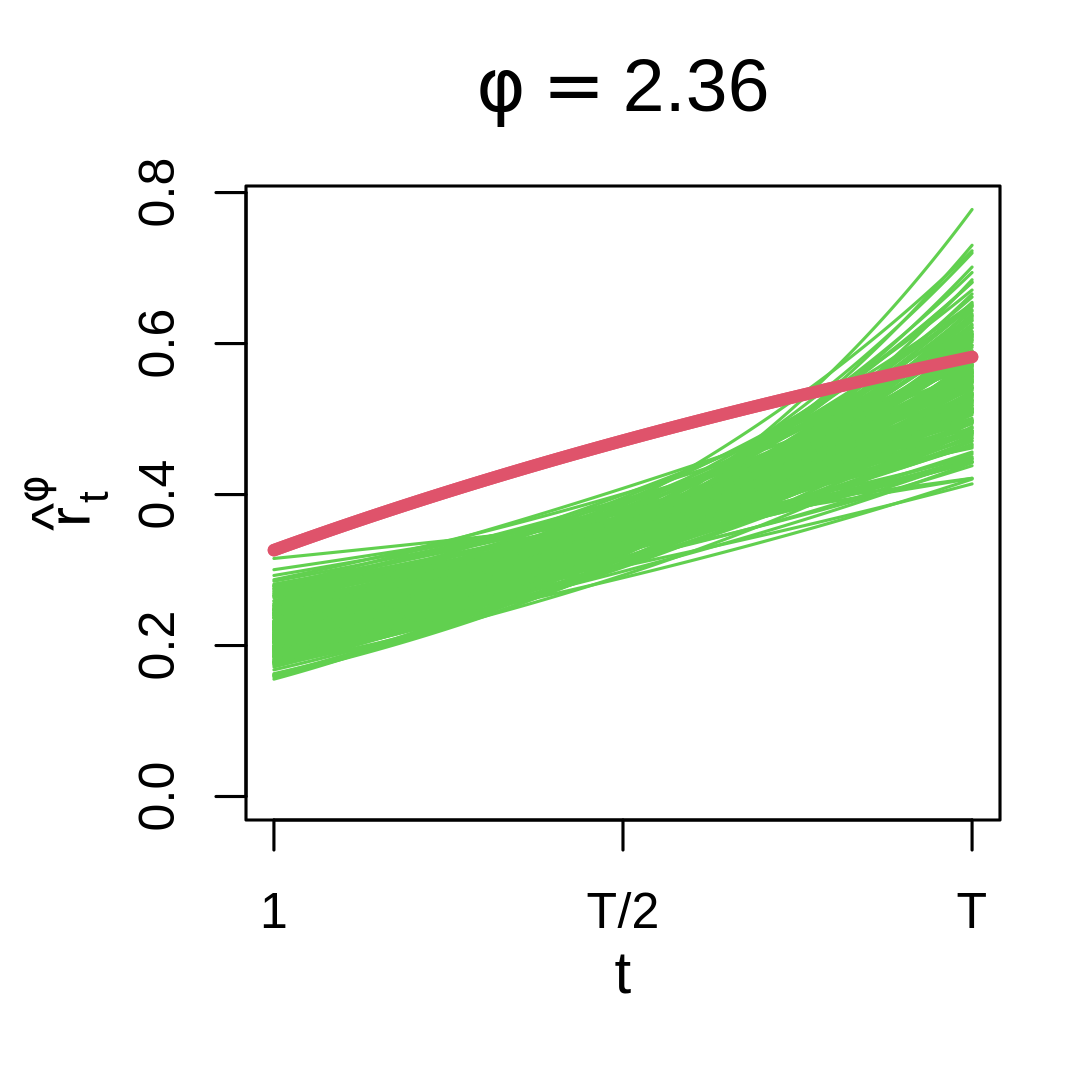}
    \end{subfigure}

    \begin{subfigure}{0.32\textwidth}
        \includegraphics[width=\linewidth]{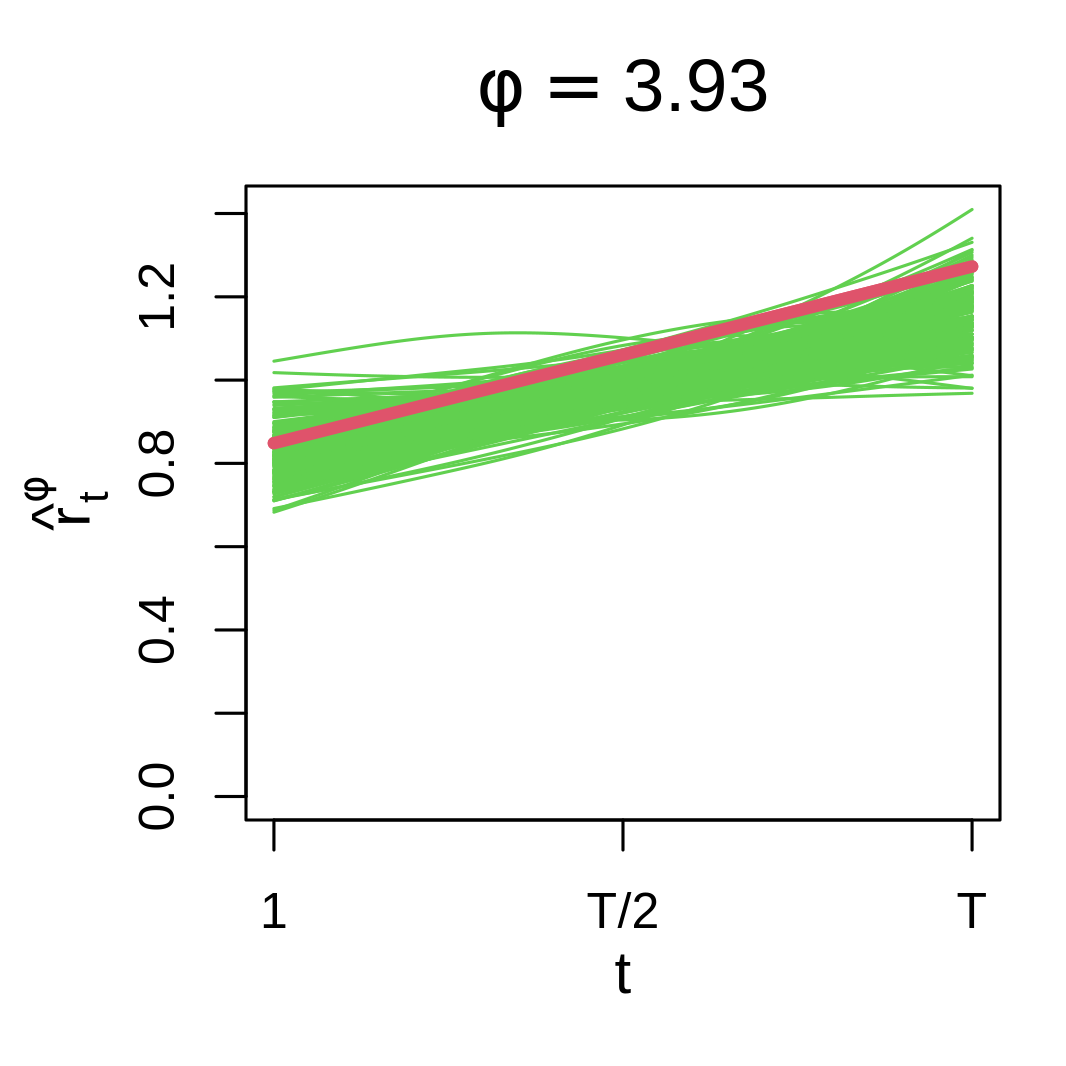}
    \end{subfigure}
    \begin{subfigure}{0.32\textwidth}
        \includegraphics[width=\linewidth]{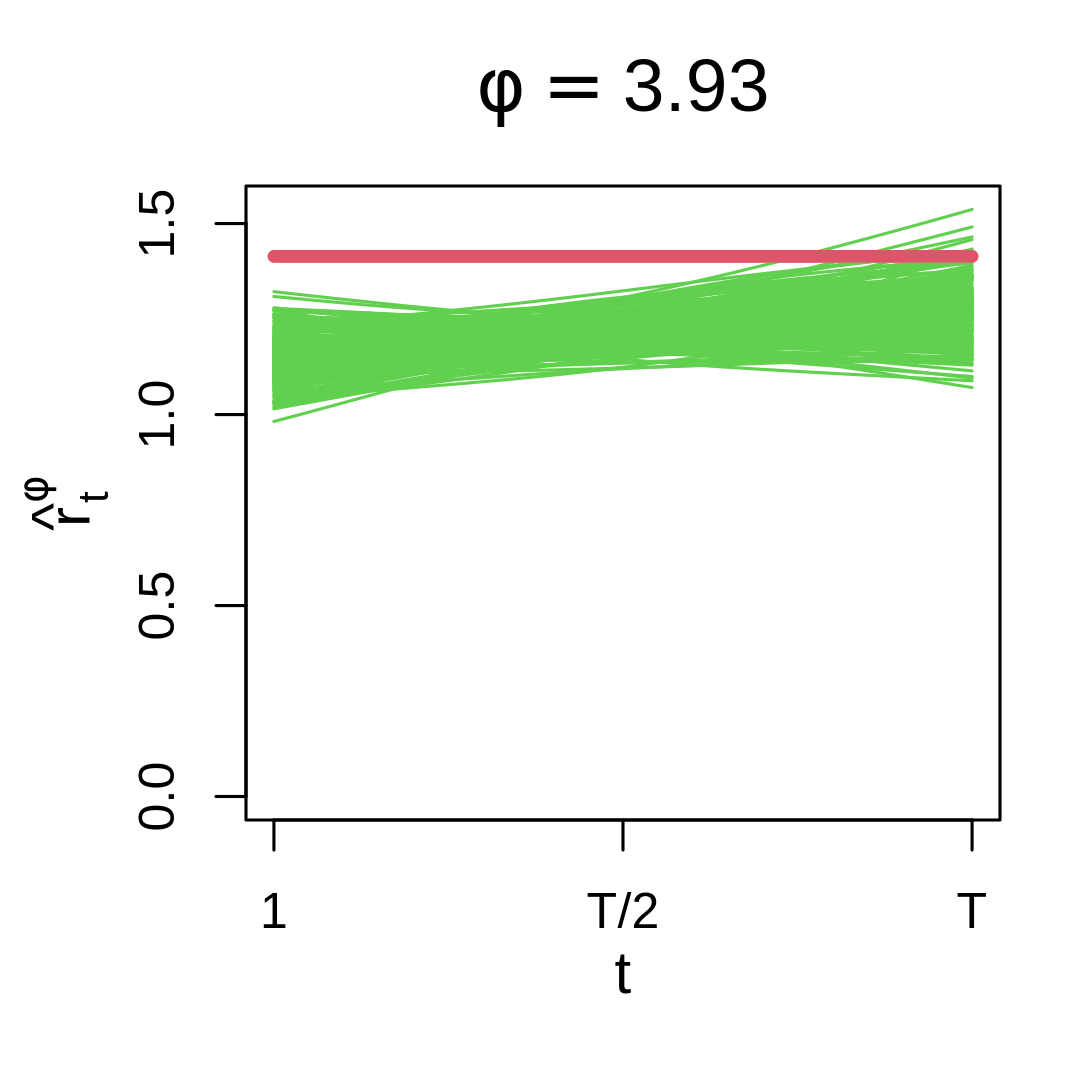}
    \end{subfigure}

    \begin{subfigure}{0.32\textwidth}
        \includegraphics[width=\linewidth]{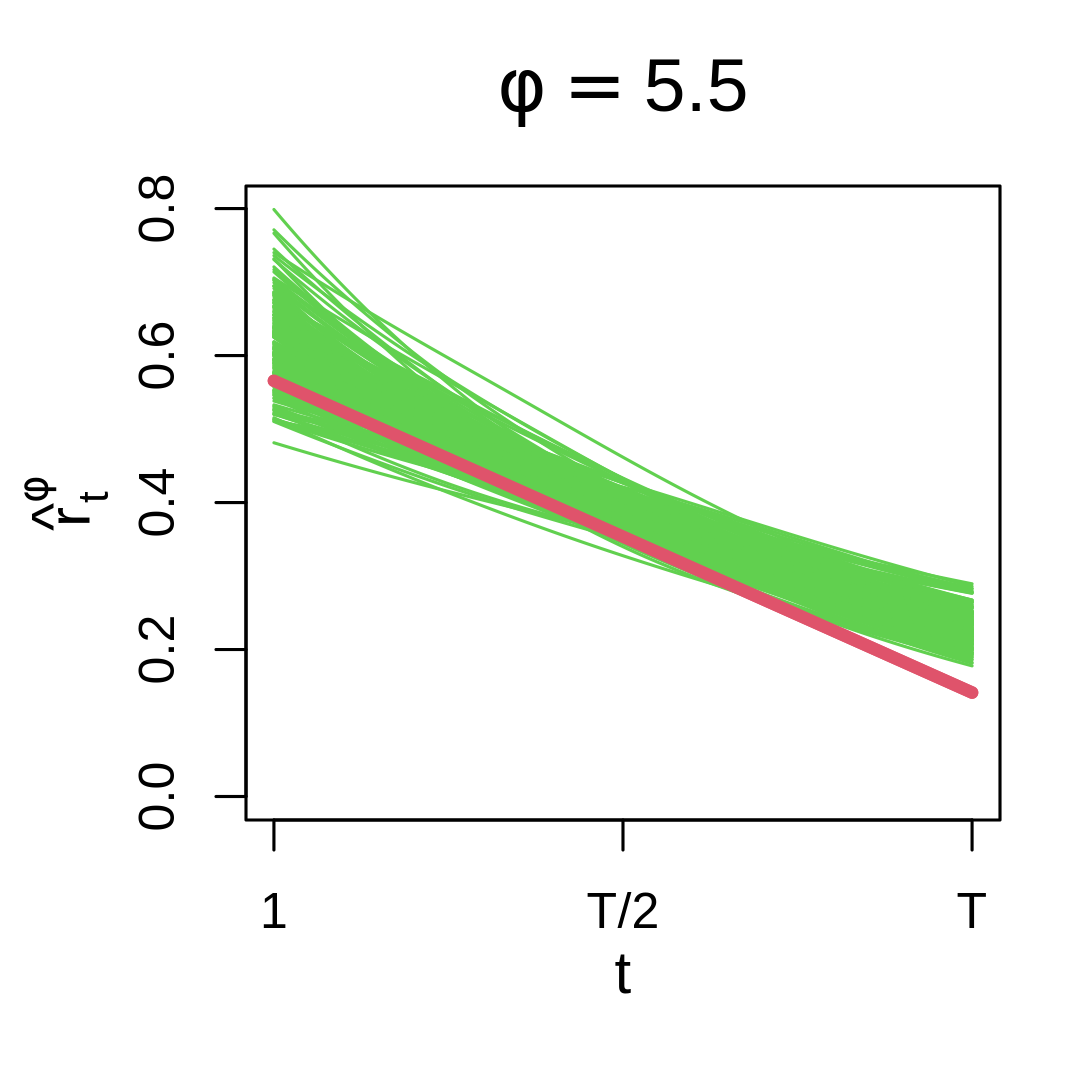}
    \end{subfigure}
    \begin{subfigure}{0.32\textwidth}
        \includegraphics[width=\linewidth]{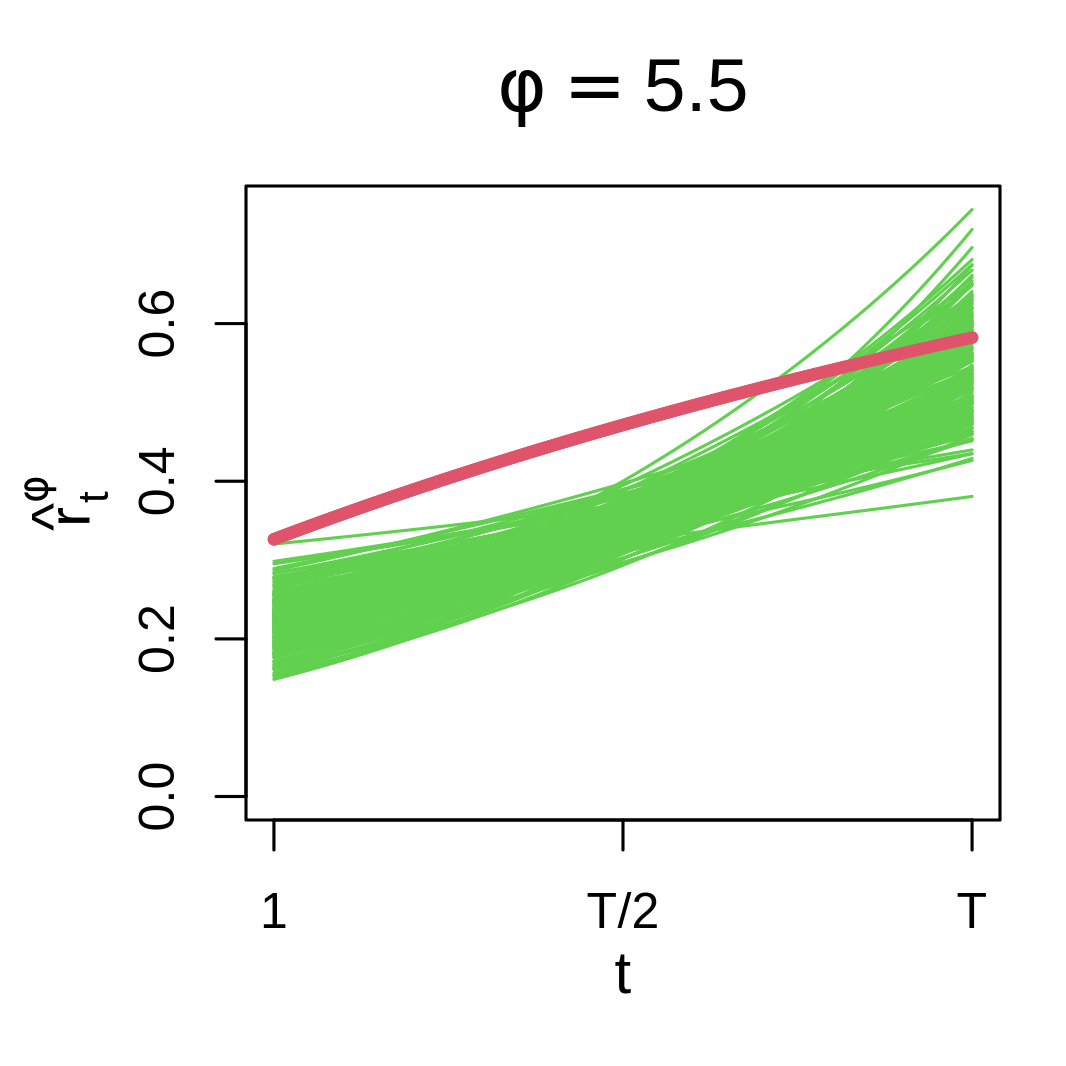}
    \end{subfigure}
    
    \caption{Boundary set radii estimates over time for $\phi=\pi/4$ (top row), $\phi = 3\pi/4$ (centre-top row), $\phi=5\pi/4$ (centre-bottom row) and $\phi = 7\pi/4$ (bottom row). The left and right columns correspond to the first and third copula examples, respectively.}
    \label{fig:final_radii_2}
\end{figure}

\subsection{Evaluating the effect of sample size} \label{subsec:appen_ss}

\begin{figure}[H]
    \centering
    \includegraphics[width=.8\linewidth]{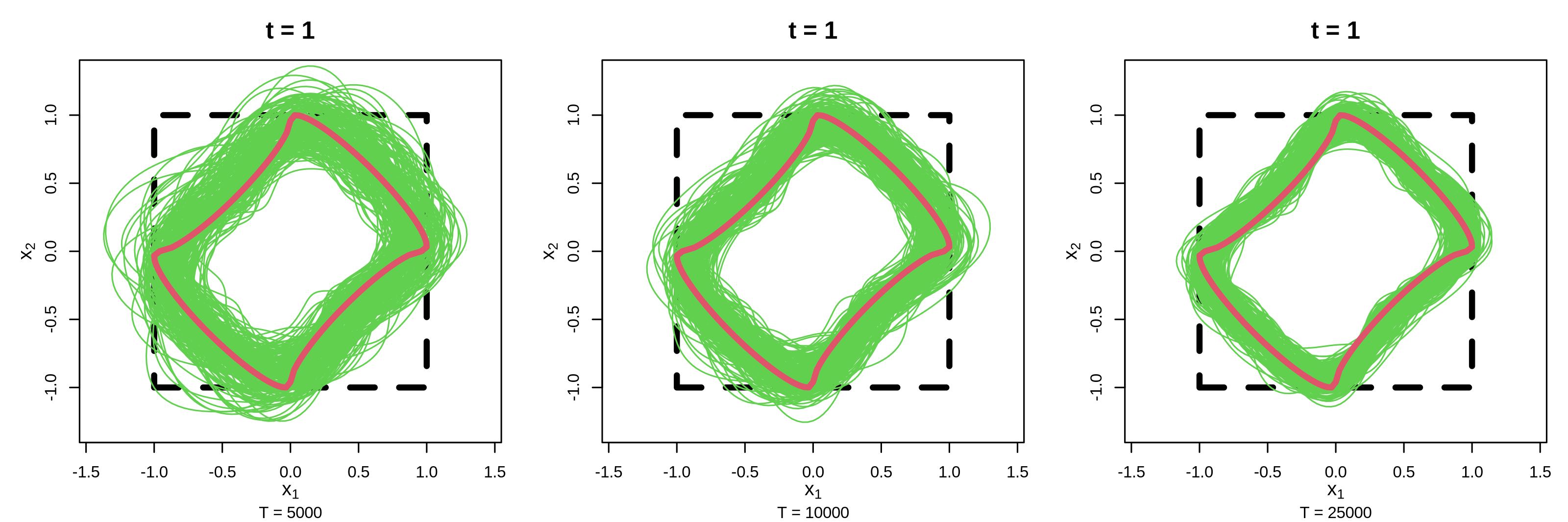}
    \caption{Boundary set estimates as $t = 1$ across $T \in \{5,000, \; 10,000, \; 25,000 \}$ for the first copula example.}
    \label{fig:res_ss_t1_c1}
\end{figure}

\begin{figure}[H]
    \centering
    \includegraphics[width=.8\linewidth]{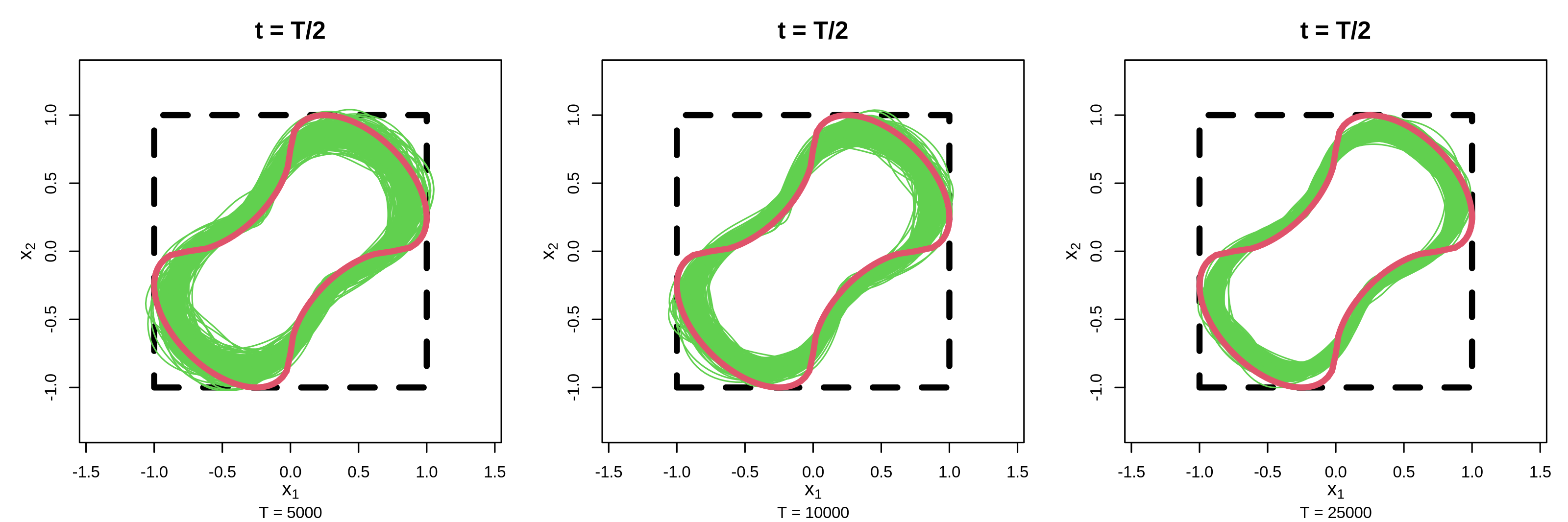}
    \caption{Boundary set estimates as $t = T/2$ across $T \in \{5,000, \; 10,000, \; 25,000 \}$ for the first copula example.}
    \label{fig:res_ss_t2_c1}
\end{figure}

\begin{figure}[H]
    \centering
    \includegraphics[width=.8\linewidth]{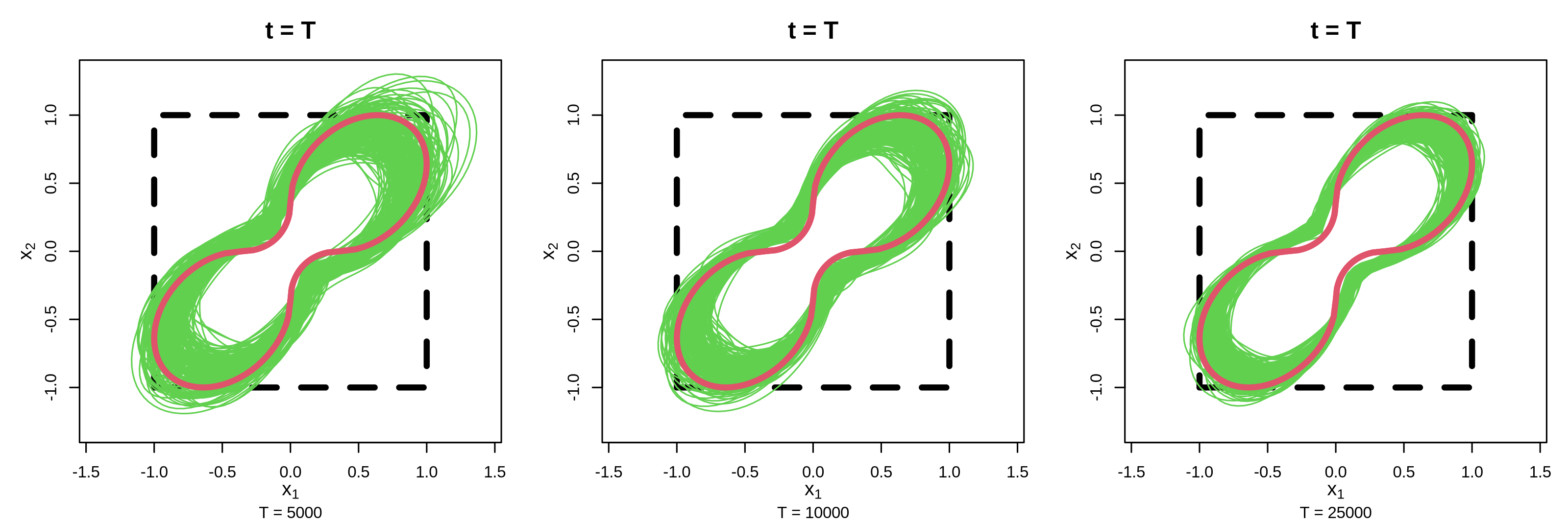}
    \caption{Boundary set estimates at $t = T$ across $T \in \{5,000, \; 10,000, \; 25,000 \}$ for the first copula example.}
    \label{fig:res_ss_t3_c1}
\end{figure}

\begin{figure}[H]
    \centering
    \includegraphics[width=.8\linewidth]{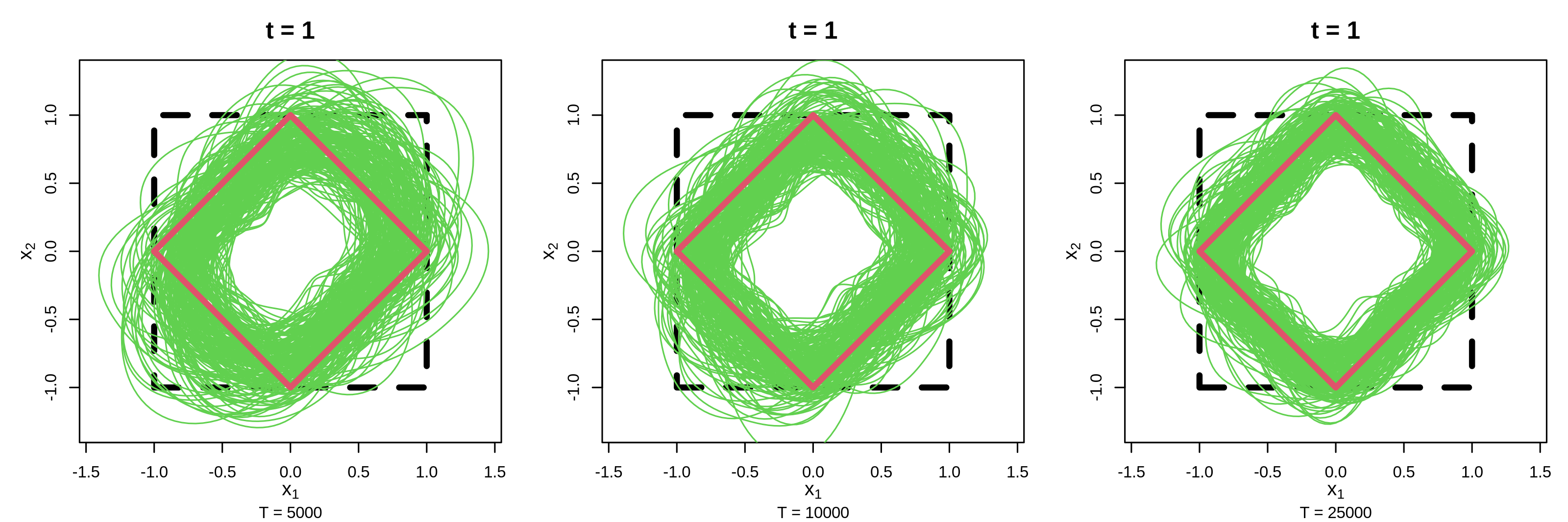}
    \caption{Boundary set estimates as $t = 1$ across $T \in \{5,000, \; 10,000, \; 25,000 \}$ for the second copula example.}
    \label{fig:res_ss_t1_c2}
\end{figure}

\begin{figure}[H]
    \centering
    \includegraphics[width=.8\linewidth]{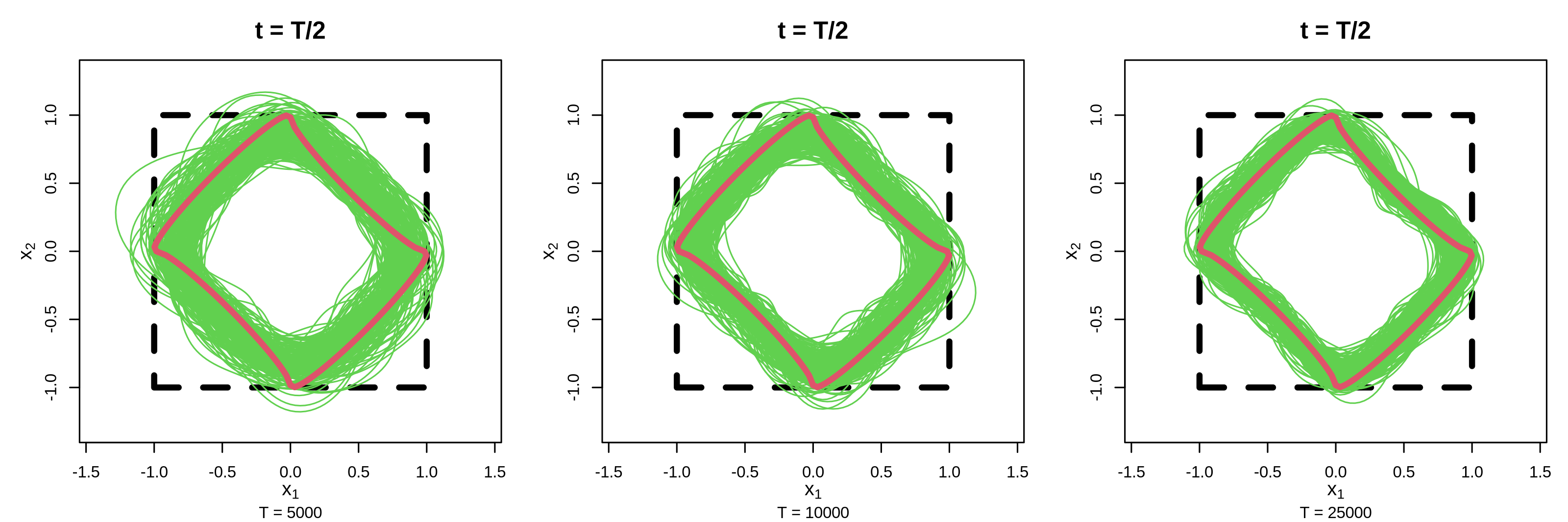}
    \caption{Boundary set estimates as $t = T/2$ across $T \in \{5,000, \; 10,000, \; 25,000 \}$ for the second copula example.}
    \label{fig:res_ss_t2_c2}
\end{figure}

\begin{figure}[H]
    \centering
    \includegraphics[width=.8\linewidth]{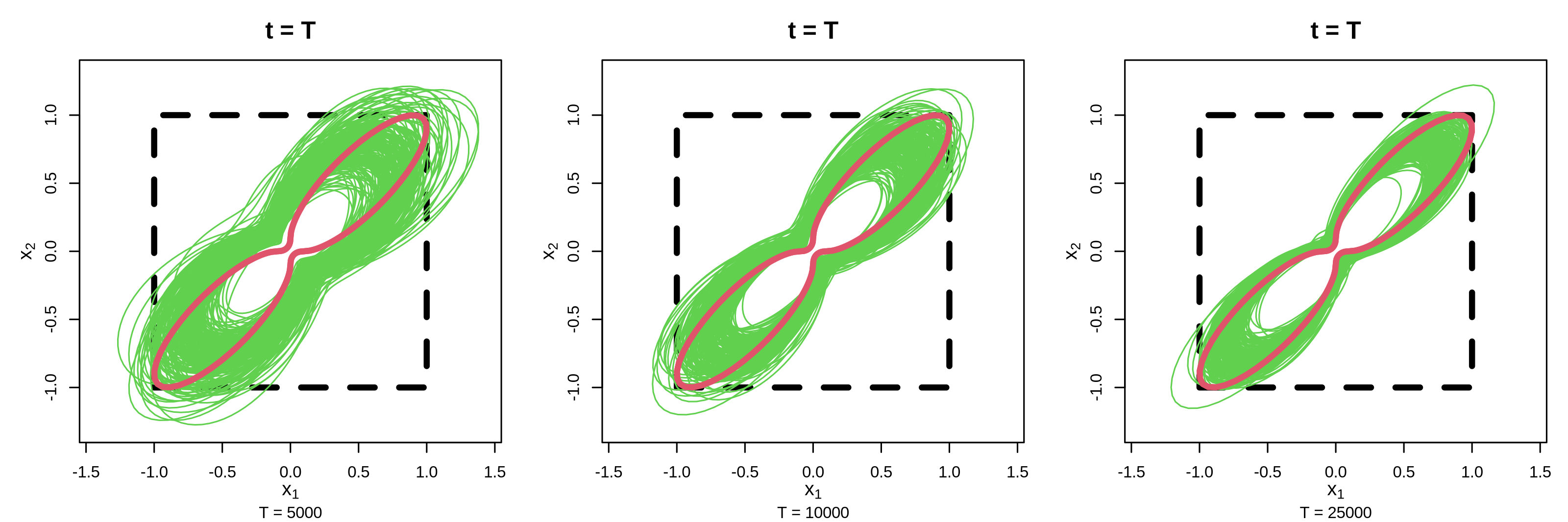}
    \caption{Boundary set estimates at $t = T$ across $T \in \{5,000, \; 10,000, \; 25,000 \}$ for the second copula example.}
    \label{fig:res_ss_t3_c2}
\end{figure}

\begin{figure}[H]
    \centering
    \includegraphics[width=.8\linewidth]{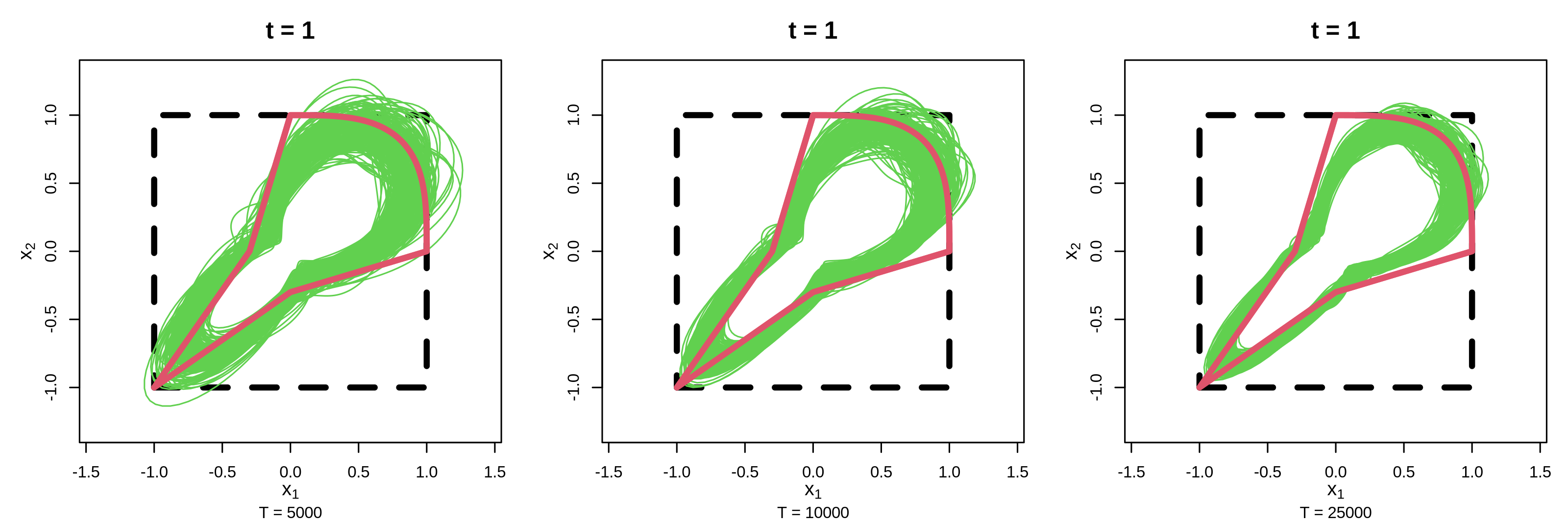}
    \caption{Boundary set estimates as $t = 1$ across $T \in \{5,000, \; 10,000, \; 25,000 \}$ for the third copula example.}
    \label{fig:res_ss_t1_c3}
\end{figure}

\begin{figure}[H]
    \centering
    \includegraphics[width=.8\linewidth]{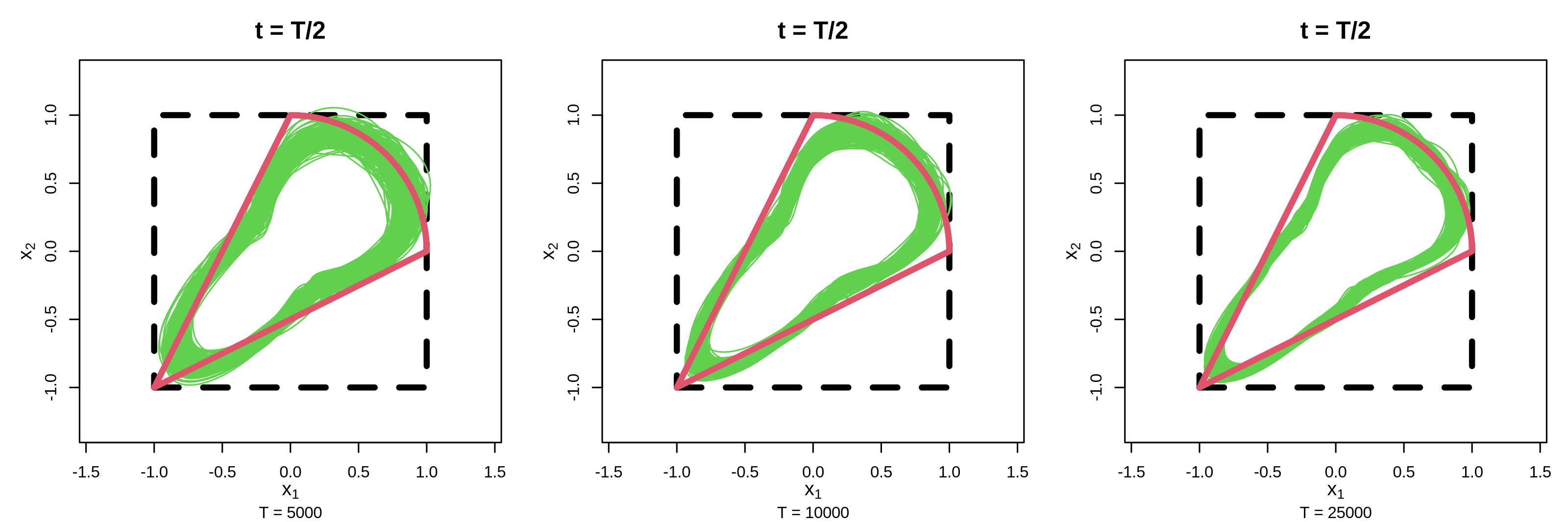}
    \caption{Boundary set estimates as $t = T/2$ across $T \in \{5,000, \; 10,000, \; 25,000 \}$ for the third copula example.}
    \label{fig:res_ss_t2_c3}
\end{figure}

\begin{figure}[H]
    \centering
    \includegraphics[width=.8\linewidth]{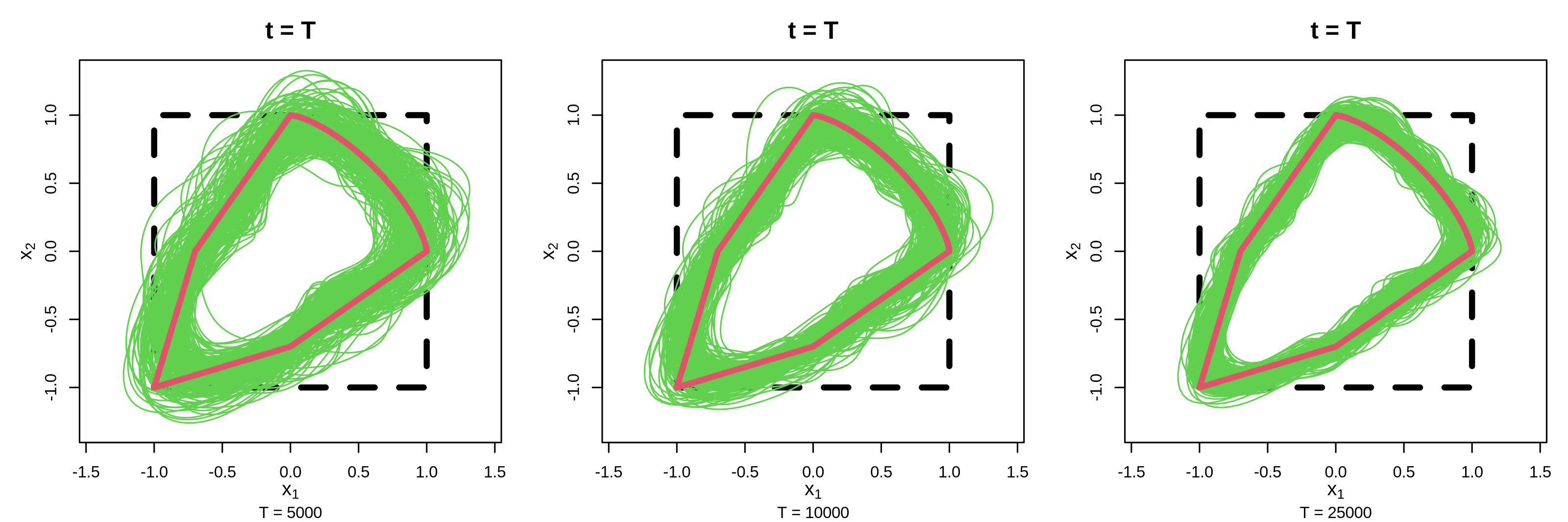}
    \caption{Boundary set estimates at $t = T$ across $T \in \{5,000, \; 10,000, \; 25,000 \}$ for the third copula example.}
    \label{fig:res_ss_t3_c3}
\end{figure}

\begin{figure}[H]
    \centering
    \includegraphics[width=.8\linewidth]{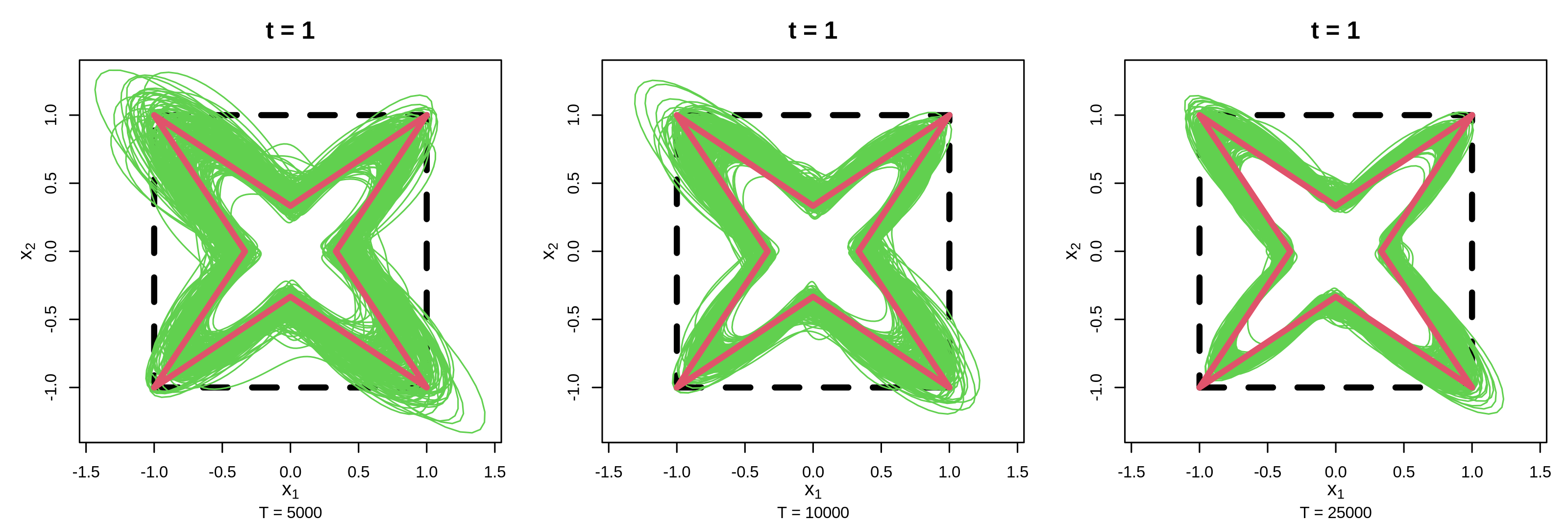}
    \caption{Boundary set estimates as $t = 1$ across $T \in \{5,000, \; 10,000, \; 25,000 \}$ for the fourth copula example.}
    \label{fig:res_ss_t1_c4}
\end{figure}

\begin{figure}[H]
    \centering
    \includegraphics[width=.8\linewidth]{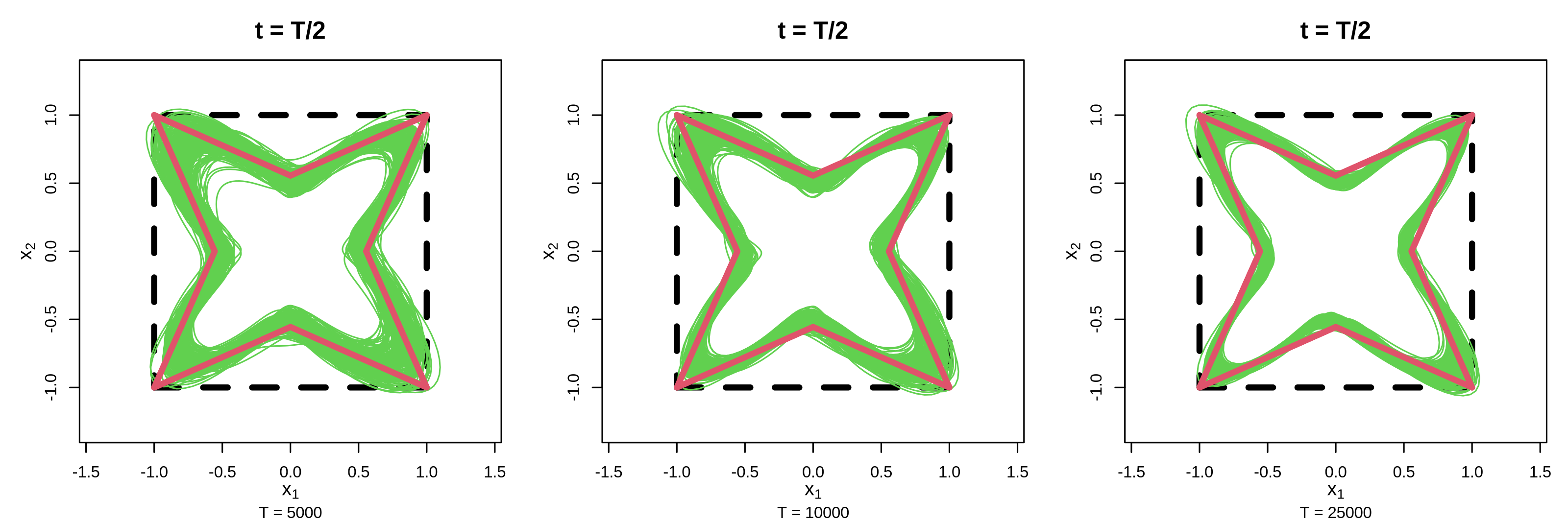}
    \caption{Boundary set estimates as $t = T/2$ across $T \in \{5,000, \; 10,000, \; 25,000 \}$ for the fourth copula example.}
    \label{fig:res_ss_t2_c4}
\end{figure}

\begin{figure}[H]
    \centering
    \includegraphics[width=.8\linewidth]{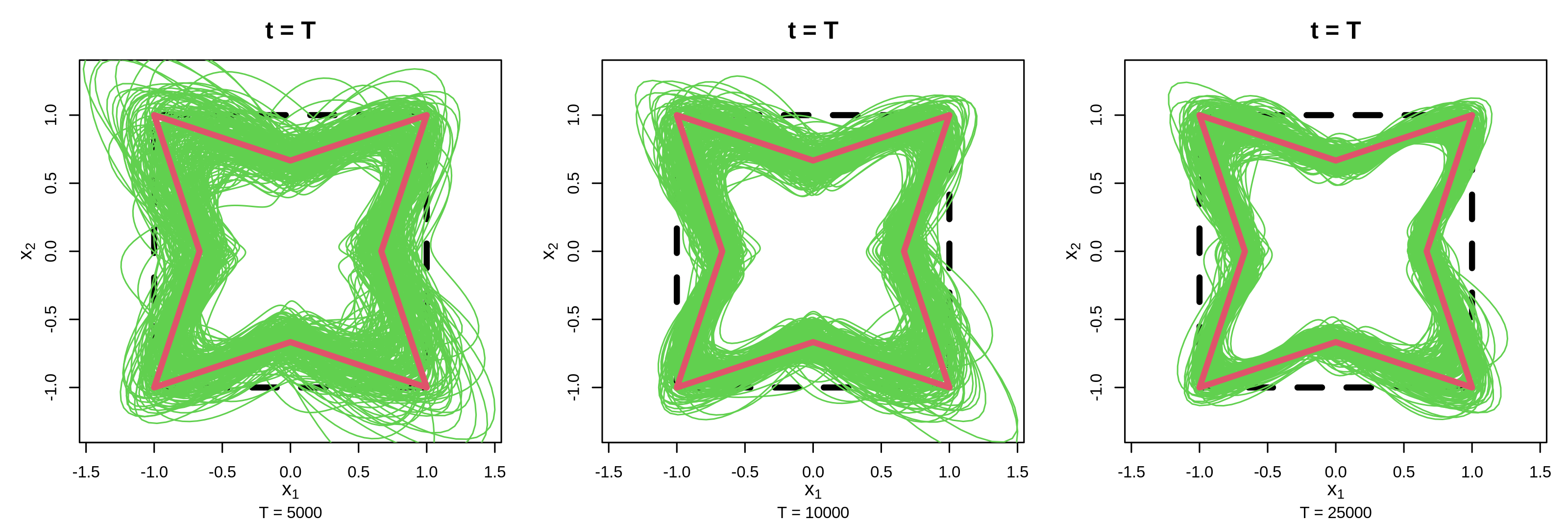}
    \caption{Boundary set estimates at $t = T$ across $T \in \{5,000, \; 10,000, \; 25,000 \}$ for the fourth copula example.}
    \label{fig:res_ss_t3_c4}
\end{figure}

\begin{figure}[H]
    \centering
    \includegraphics[width=.8\linewidth]{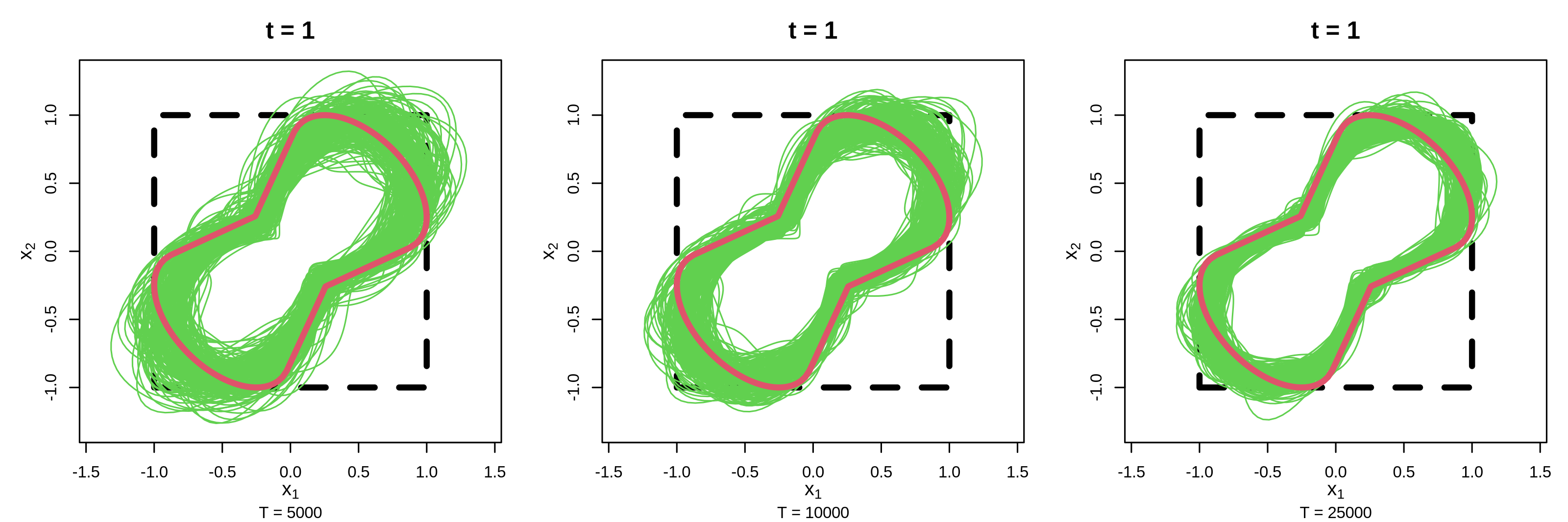}
    \caption{Boundary set estimates as $t = 1$ across $T \in \{5,000, \; 10,000, \; 25,000 \}$ for the fifth copula example.}
    \label{fig:res_ss_t1_c5}
\end{figure}

\begin{figure}[H]
    \centering
    \includegraphics[width=.8\linewidth]{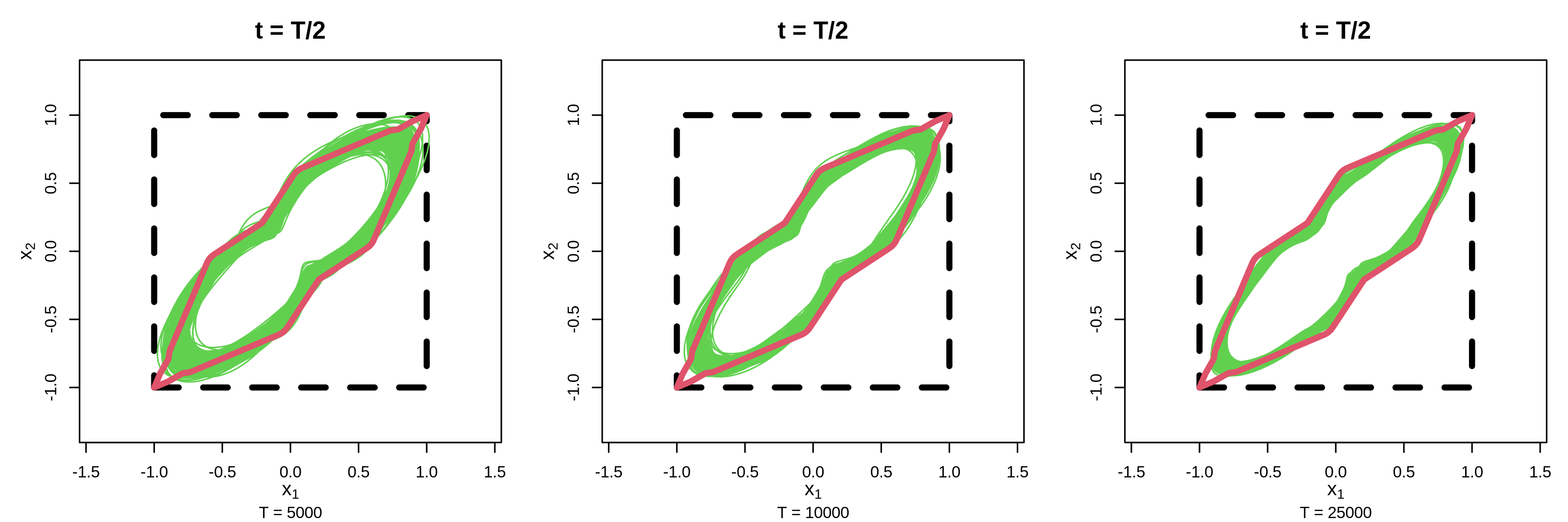}
    \caption{Boundary set estimates as $t = T/2$ across $T \in \{5,000, \; 10,000, \; 25,000 \}$ for the fifth copula example.}
    \label{fig:res_ss_t2_c5}
\end{figure}

\begin{figure}[H]
    \centering
    \includegraphics[width=.8\linewidth]{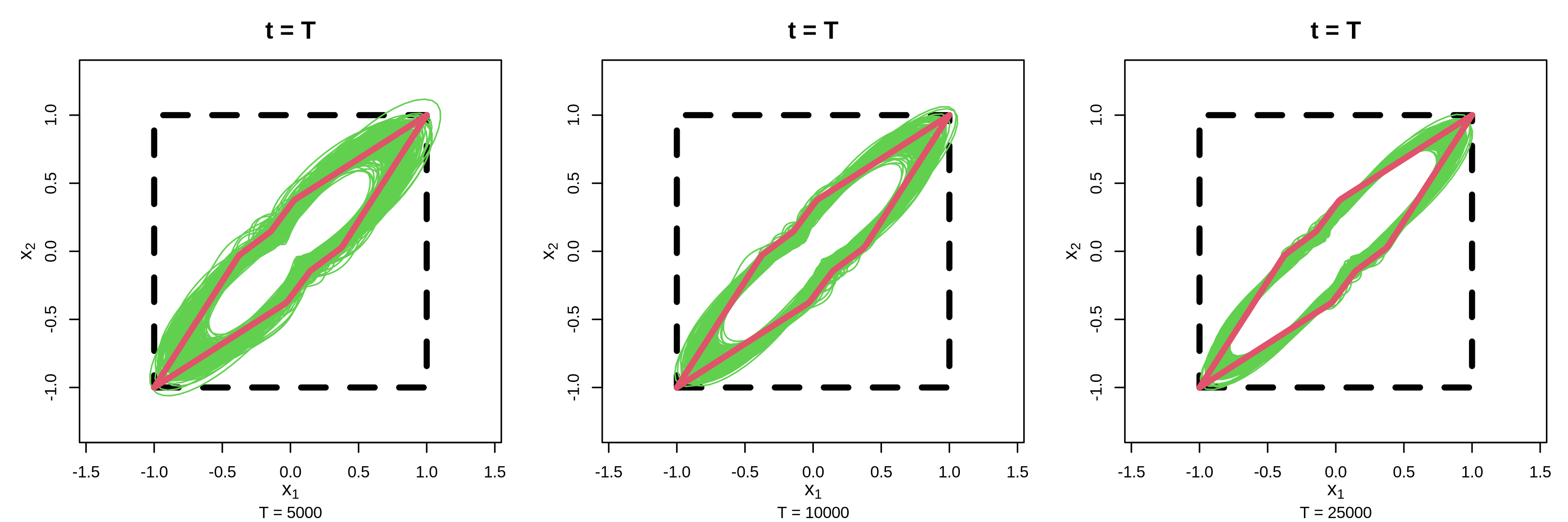}
    \caption{Boundary set estimates at $t = T$ across $T \in \{5,000, \; 10,000, \; 25,000 \}$ for the fifth copula example.}
    \label{fig:res_ss_t3_c5}
\end{figure}

Figures~\ref{fig:res_ss_t1_c1}-\ref{fig:res_ss_t3_c5} illustrate the effect of the sample size $T$ on the boundary set estimates. 

\section{Additional case study figures and tables} \label{sec:appen_add_case_figs}

\begin{figure}[H]
    \centering
    \includegraphics[width=\linewidth]{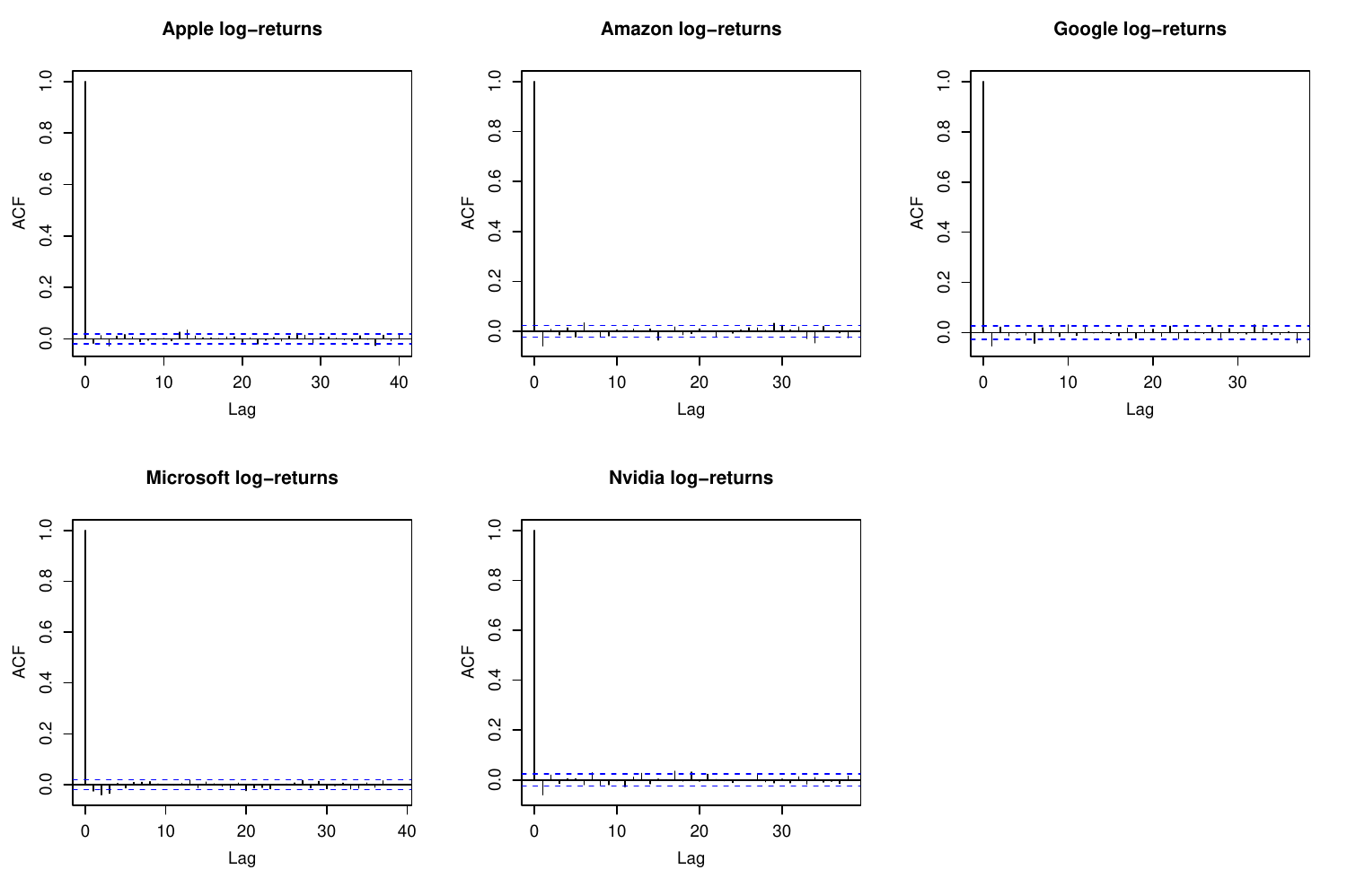}
    \caption{Log-returns ACF plots for each stock. }
    \label{fig:returns_acf}
\end{figure}

\begin{figure}[H]
    \centering
    \includegraphics[width=\linewidth]{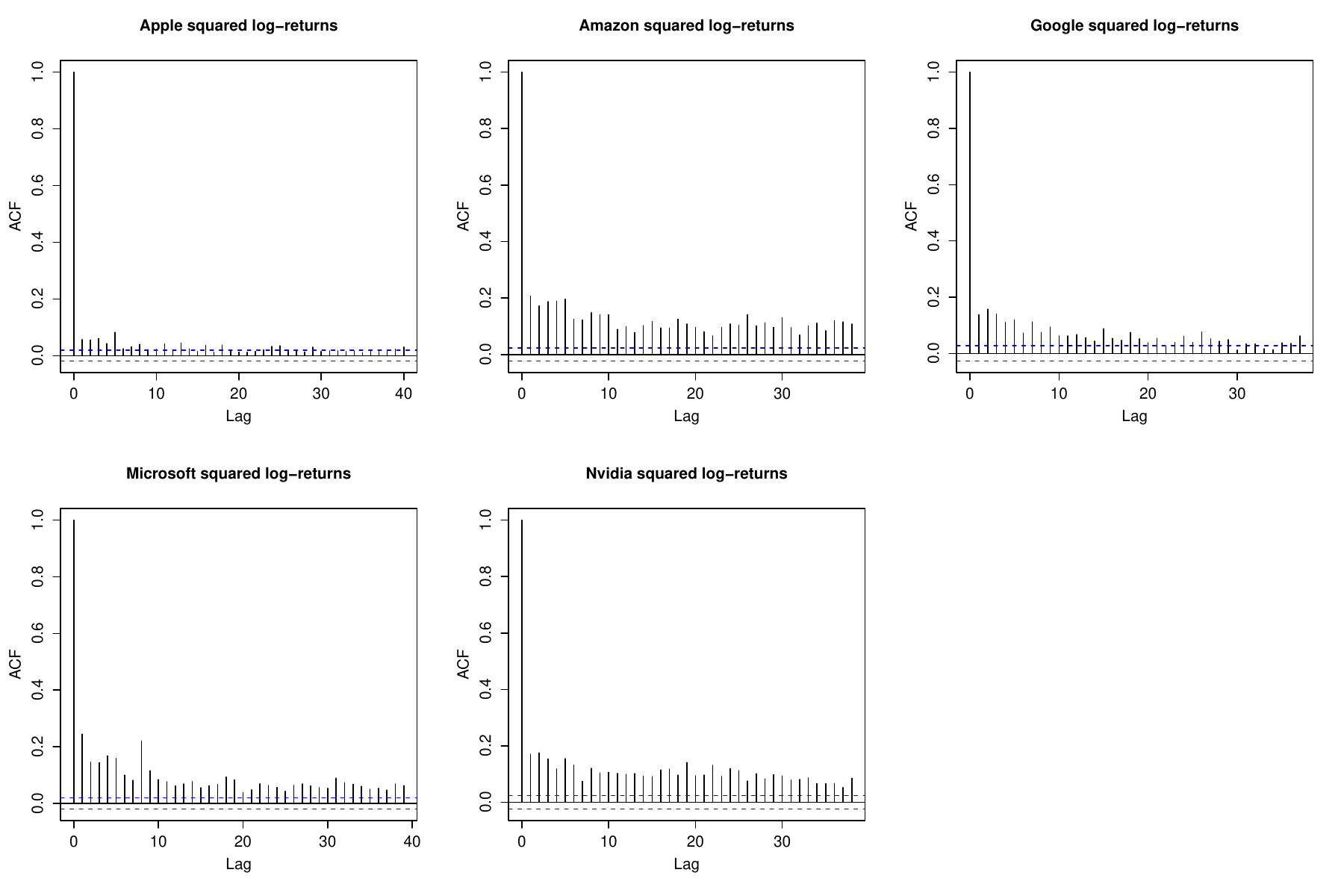}
    \caption{Squared log-returns ACF plots for each stock.}
    \label{fig:vol_acf}
\end{figure}

Figures~\ref{fig:returns_acf} and \ref{fig:vol_acf} illustrate ACF plots for $Q_t$ and $Q_t^2$, respectively, suggesting an independence assumption is unreasonable for each of the time series. 

\begin{figure}[H]
    \centering
    \includegraphics[width=\linewidth]{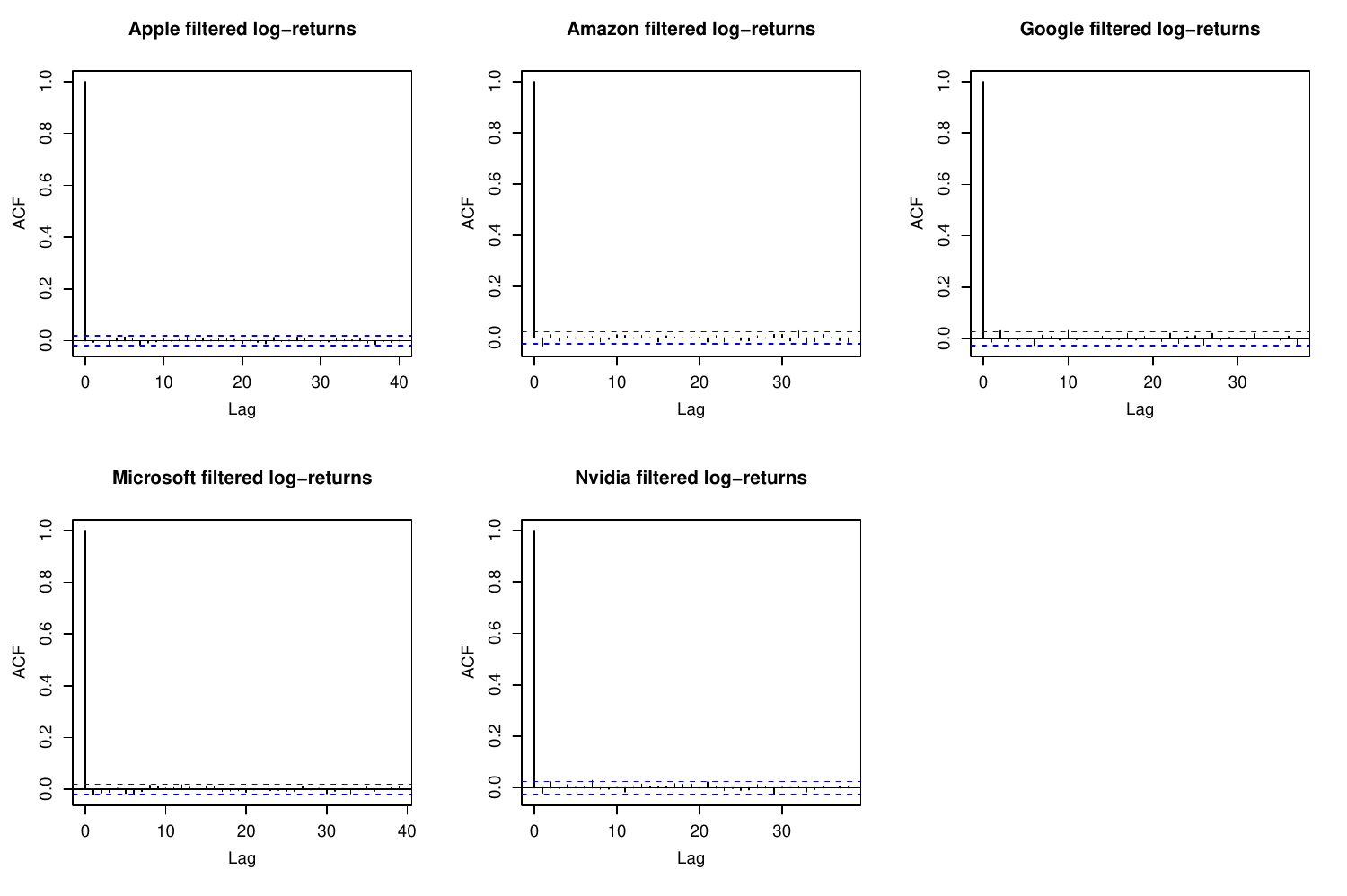}
    \caption{Filtered log-returns ACF plots for each stock. }
    \label{fig:fil_returns_acf}
\end{figure}

\begin{figure}[H]
    \centering
    \includegraphics[width=\linewidth]{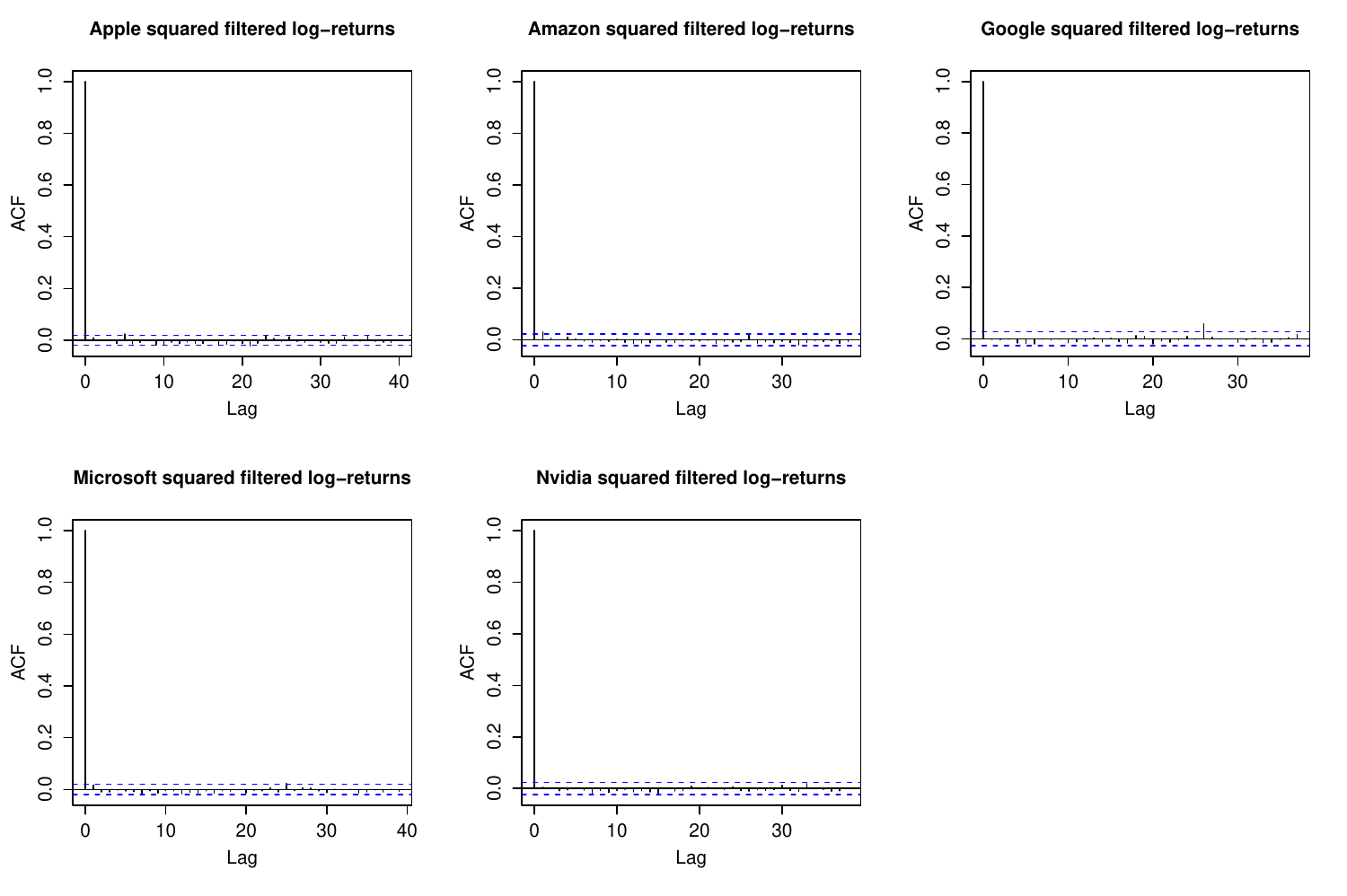}
    \caption{Squared filtered log-returns ACF plots for each stock.}
    \label{fig:fil_vol_acf}
\end{figure}

Figures~\ref{fig:fil_returns_acf} and \ref{fig:fil_vol_acf} illustrate ACF plots for the filtered and squared filtered log-returns, $\hat{\varepsilon}_t$ and $\hat{\varepsilon}_t^2$, respectively. These plots indicate an independence assumption is reasonable for the residual process. 

\begin{figure}[H]
    \centering
    \includegraphics[width=\linewidth]{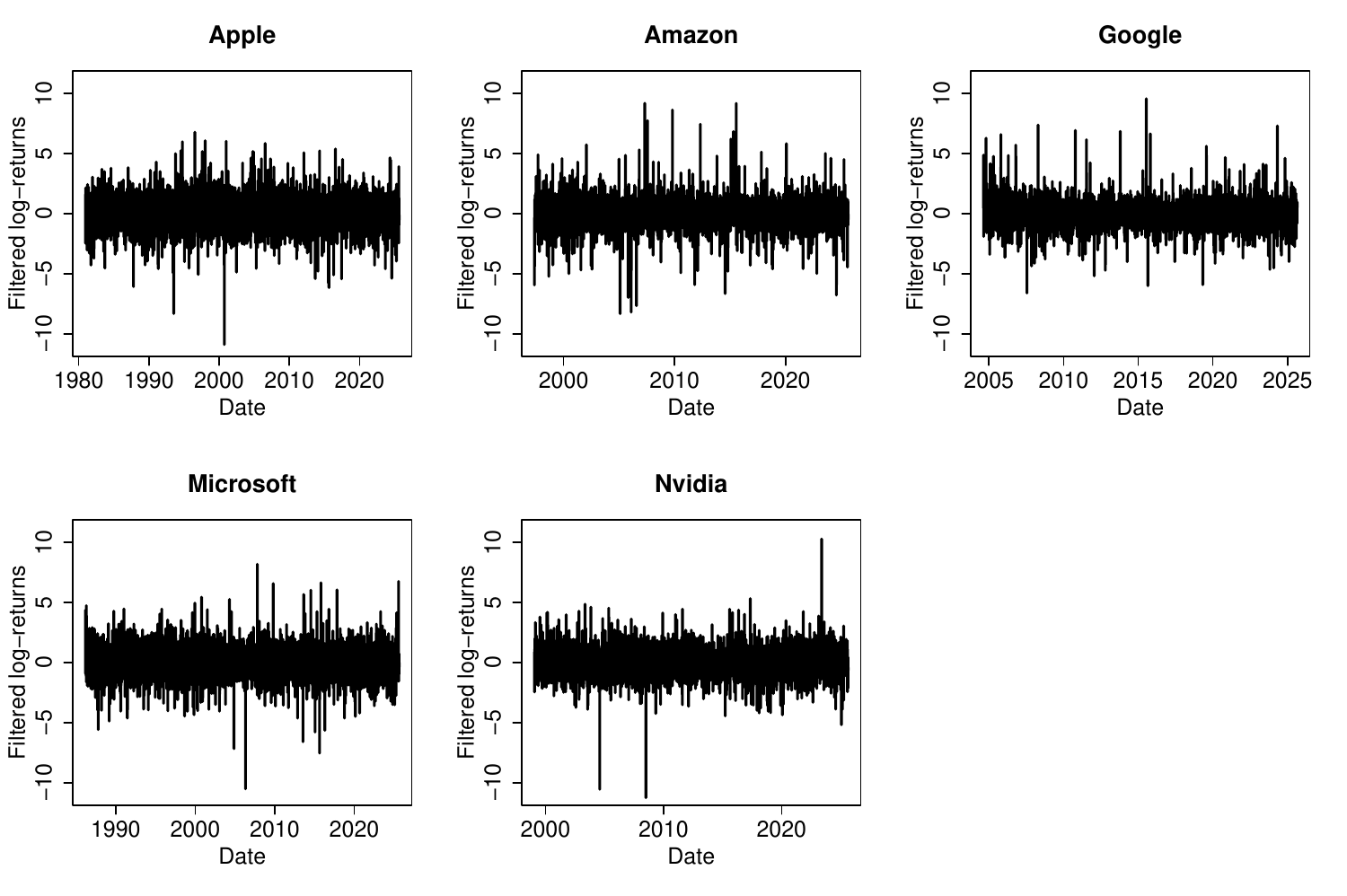}
    \caption{Filtered log-returns time series for each stock. }
    \label{fig:fil_returns_ts}
\end{figure}

Figures~\ref{fig:fil_returns_ts} illustrate the individual time series for the filtered log-returns. These plots display no obvious periodicity or long term trends, suggesting a stationarity assumption is reasonable for the filtered data. 

\begin{figure}[H]
    \centering
     \begin{subfigure}[b]{\textwidth}
        \centering
        \includegraphics[width=\textwidth]{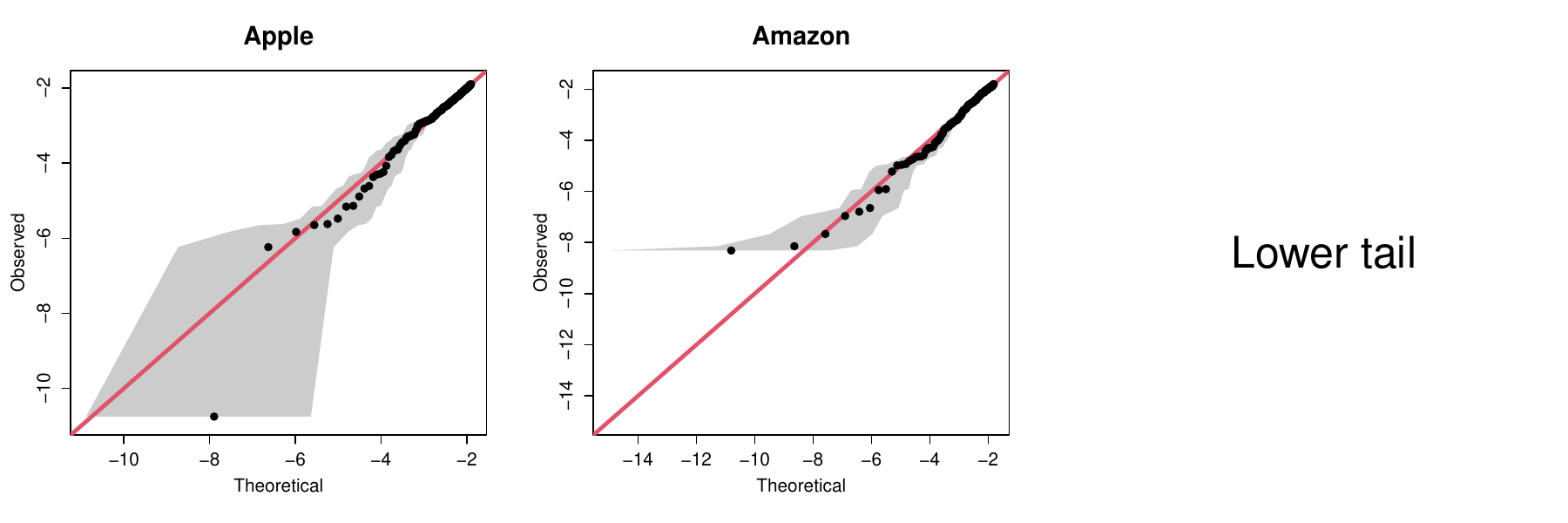}
    \end{subfigure}%
    \\
    \begin{subfigure}[b]{\textwidth}
        \centering
        \includegraphics[width=\textwidth]{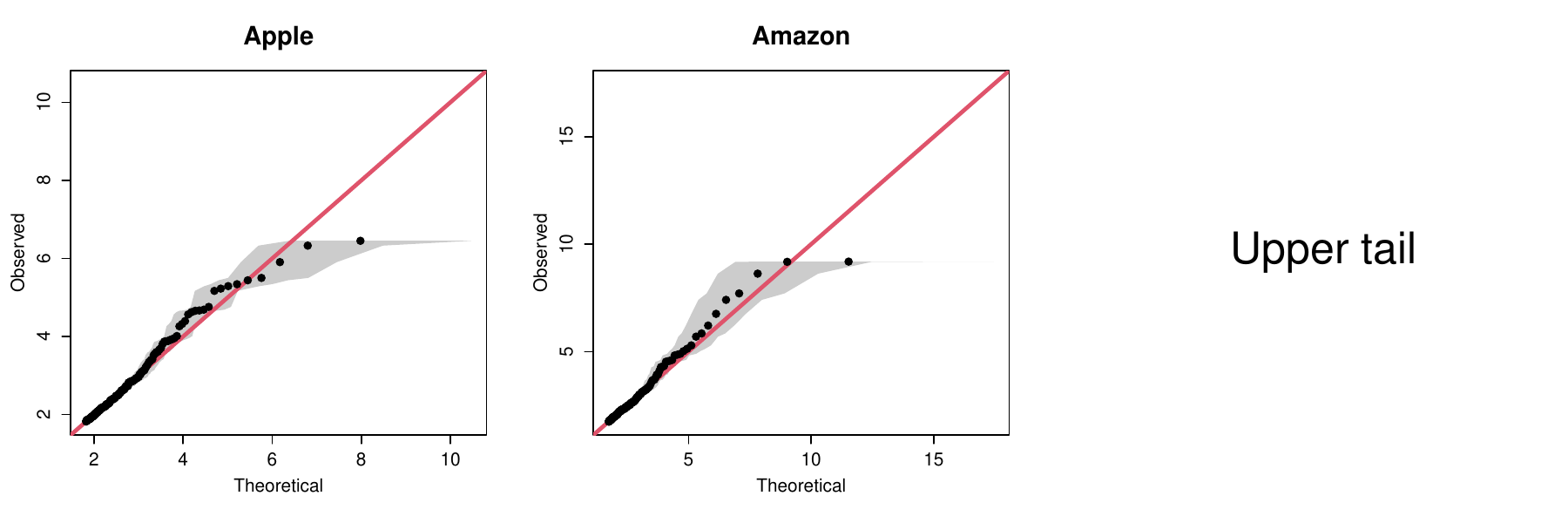}
    \end{subfigure}%
   
    \caption{GPD QQ plots for the lower (top row) and upper (bottom row) tails of the Apple and Amazon series. }
    \label{fig:gpd_qq_AMZN}
\end{figure}

\begin{figure}[H]
    \centering
     \begin{subfigure}[b]{\textwidth}
        \centering
        \includegraphics[width=\textwidth]{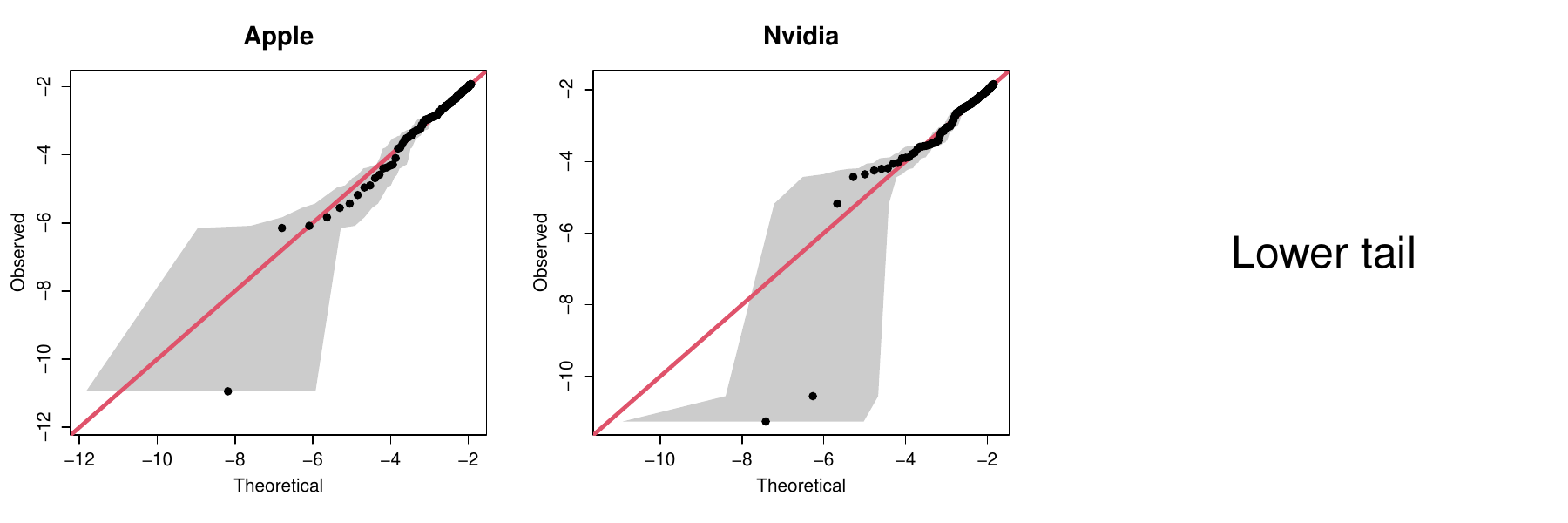}
    \end{subfigure}%
    \\
    \begin{subfigure}[b]{\textwidth}
        \centering
        \includegraphics[width=\textwidth]{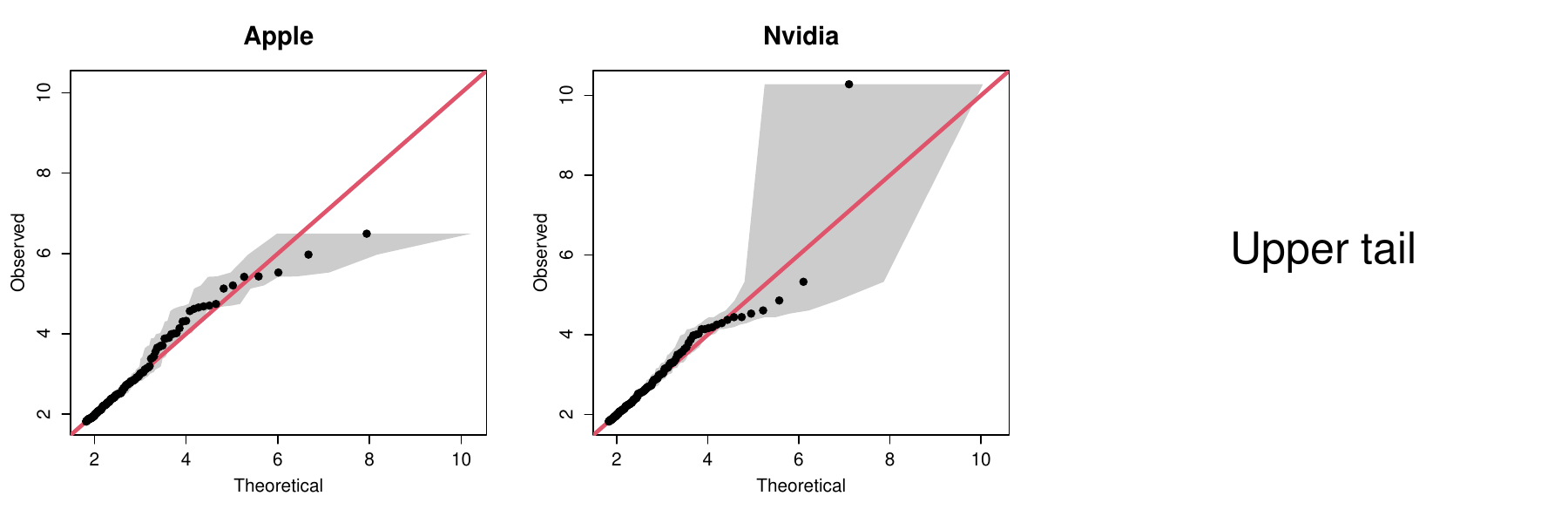}
    \end{subfigure}%
   
    \caption{GPD QQ plots for the lower (top row) and upper (bottom row) tails of the Apple and Nvidia series. }
    \label{fig:gpd_qq_NVDA}
\end{figure}

Figures~\ref{fig:gpd_qq_AMZN} and \ref{fig:gpd_qq_NVDA} give the QQ plots obtained from the GPD model fits on the lower and upper tails of the filtered log-returns time series. Each plot illustrates model quantile estimates, alongside 95\% parametric bootstrapped confidence intervals. One can observe generally good agreement between the observed and estimated quantiles, with only mild disagreements at the most extreme observations for some stocks (e.g., Nvidia), suggesting the proposed marginal model is reasonable for capturing both tails. Note that we fit the GPD model to the Apple data multiple times, due to the fact observation lengths and zero movement days vary significantly between different pairs of stocks, meaning the Apple time series changes for each pairing.  

\begin{figure}[H]
    \centering
     \begin{subfigure}[b]{\textwidth}
        \centering
        \includegraphics[width=\textwidth]{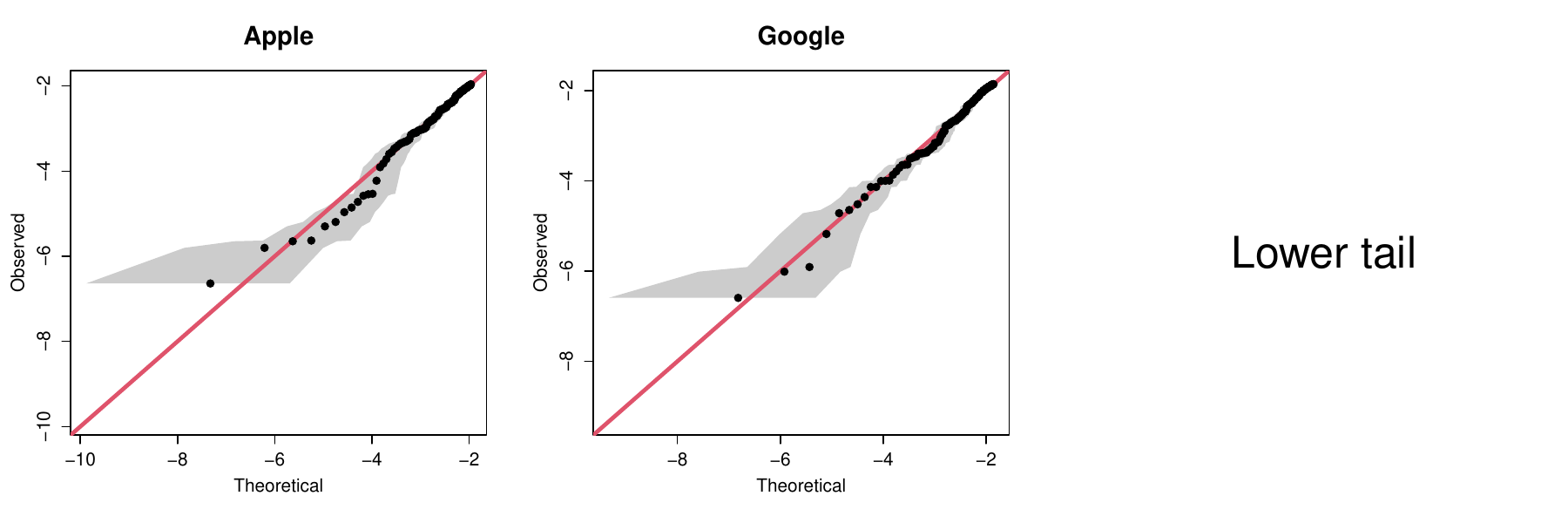}
    \end{subfigure}%
    \\
    \begin{subfigure}[b]{\textwidth}
        \centering
        \includegraphics[width=\textwidth]{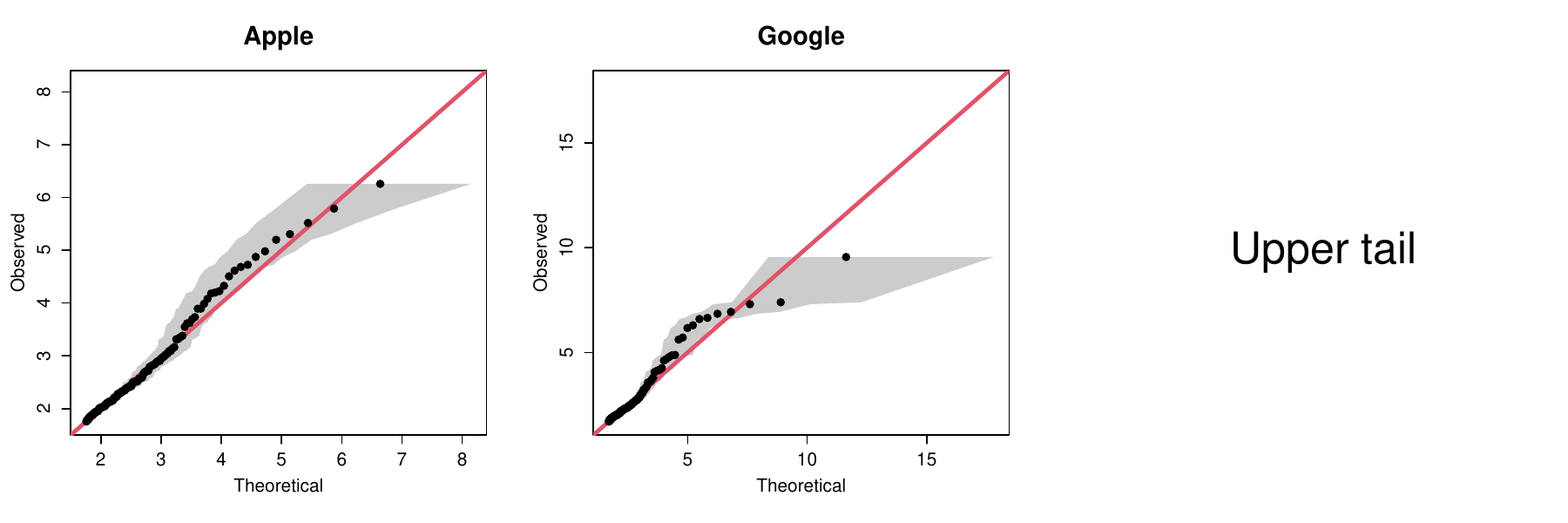}
    \end{subfigure}%
   
    \caption{GPD QQ plots for the lower (top row) and upper (bottom row) tails of the Apple and Google series. }
    \label{fig:gpd_qq_GOOGL}
\end{figure}

\begin{figure}[H]
    \centering
     \begin{subfigure}[b]{\textwidth}
        \centering
        \includegraphics[width=\textwidth]{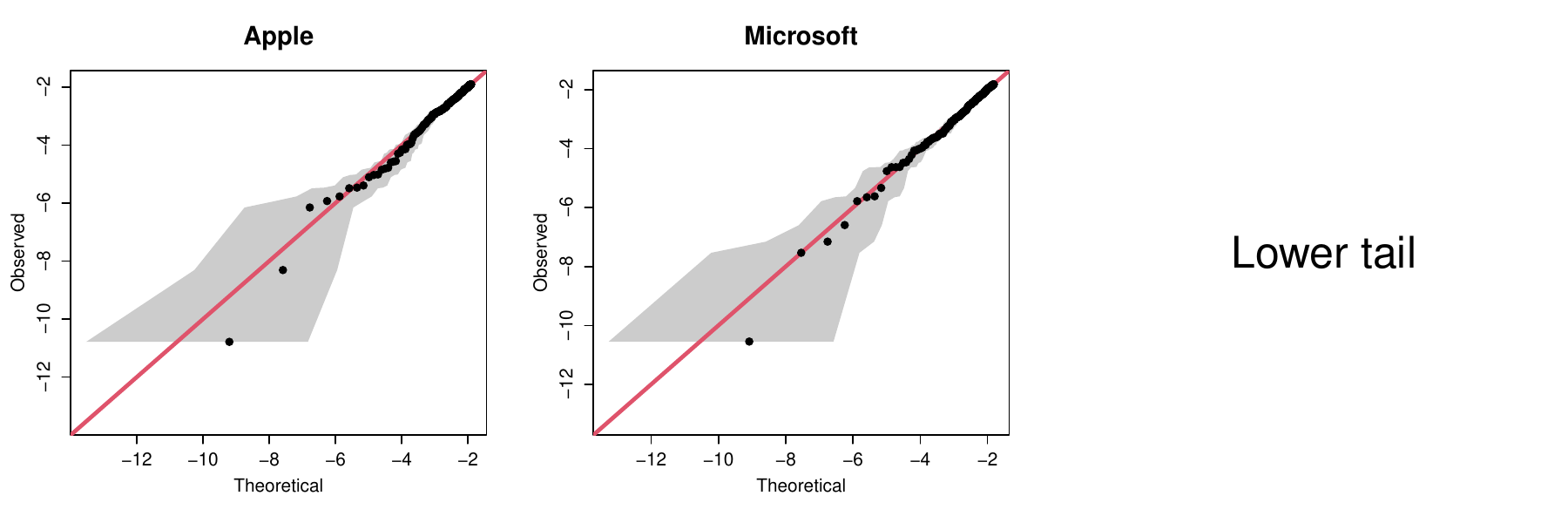}
    \end{subfigure}%
    \\
    \begin{subfigure}[b]{\textwidth}
        \centering
        \includegraphics[width=\textwidth]{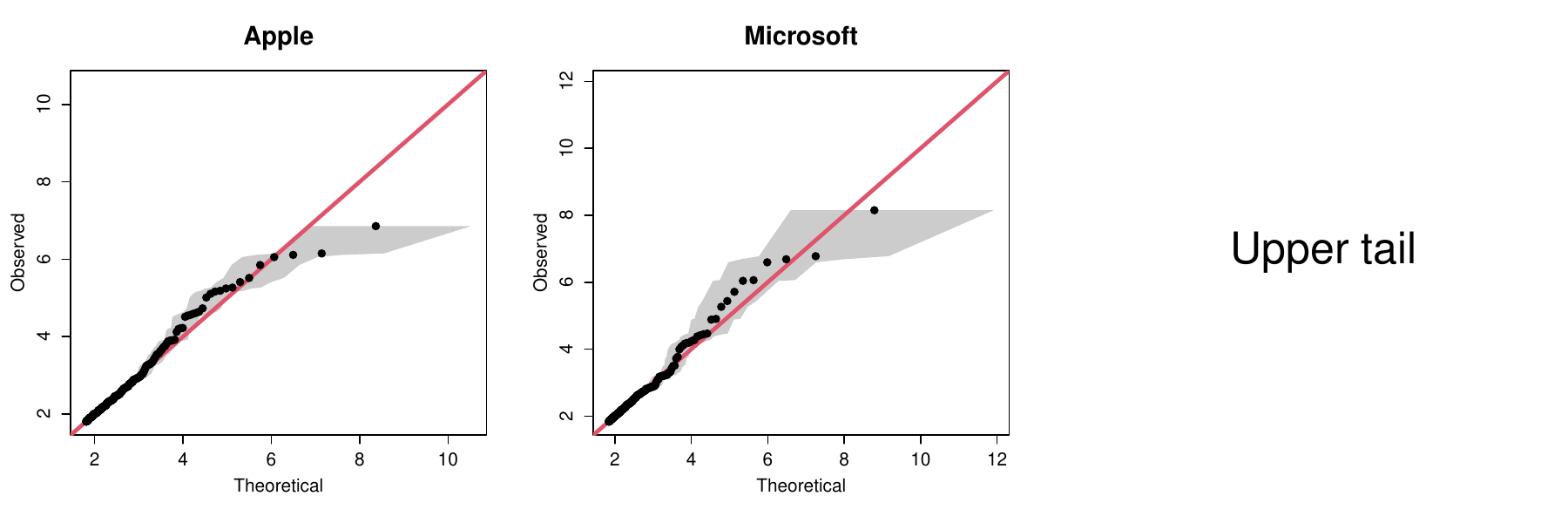}
    \end{subfigure}%
   
    \caption{GPD QQ plots for the lower (top row) and upper (bottom row) tails of the Apple and Microsoft series. }
    \label{fig:gpd_qq_MSFT}
\end{figure}

\begin{figure}[H]
    \centering
    \includegraphics[width=\linewidth]{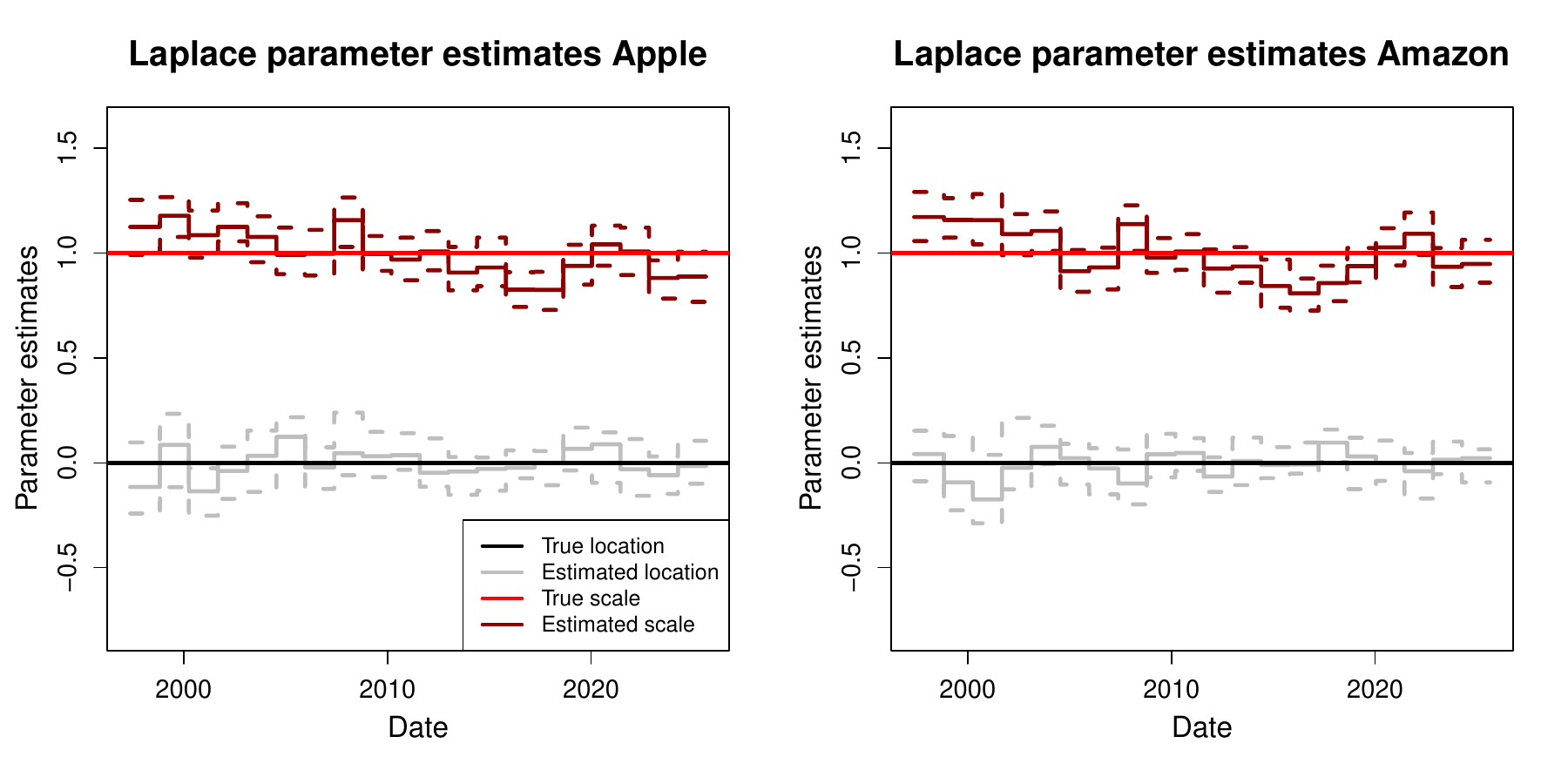}
    \caption{Rolling window Laplace parameter estimates for the transformed Apple and Amazon series. }
    \label{fig:laplace_est_AMZN}
\end{figure}

\begin{figure}[H]
    \centering
    \includegraphics[width=\linewidth]{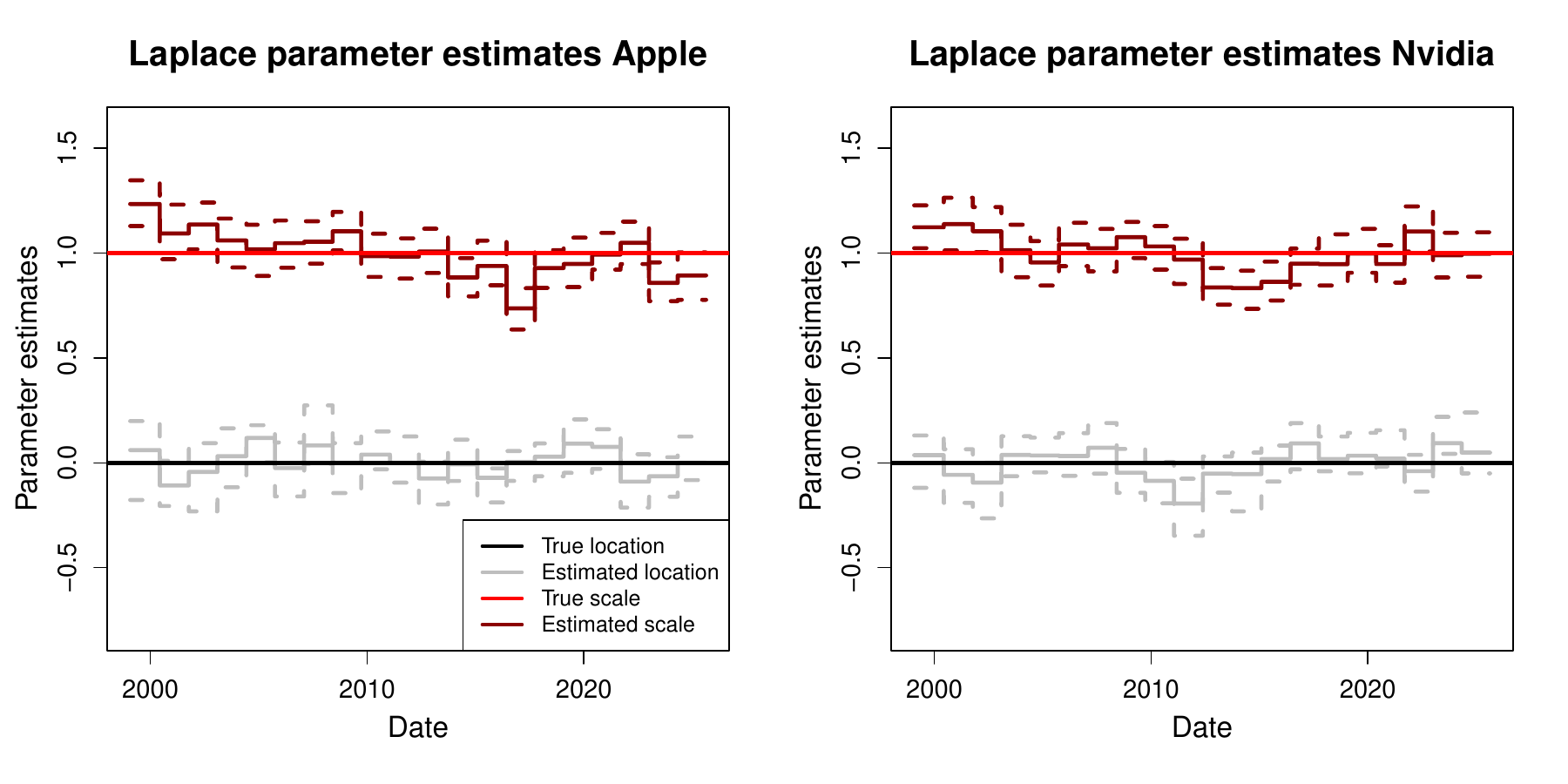}
    \caption{Rolling window Laplace parameter estimates for the transformed Apple and Nvidia series. }
    \label{fig:laplace_est_NVDA}
\end{figure}

Figures~\ref{fig:laplace_est_AMZN} and \ref{fig:laplace_est_NVDA} illustrate rolling window Laplace distribution parameter estimates for each pair of stocks post marginal transformation. In all cases, the rolling window estimates remain close to the `true' standard Laplace parameters, which are also shown in the panels. 

\begin{figure}[H]
    \centering
    \includegraphics[width=\linewidth]{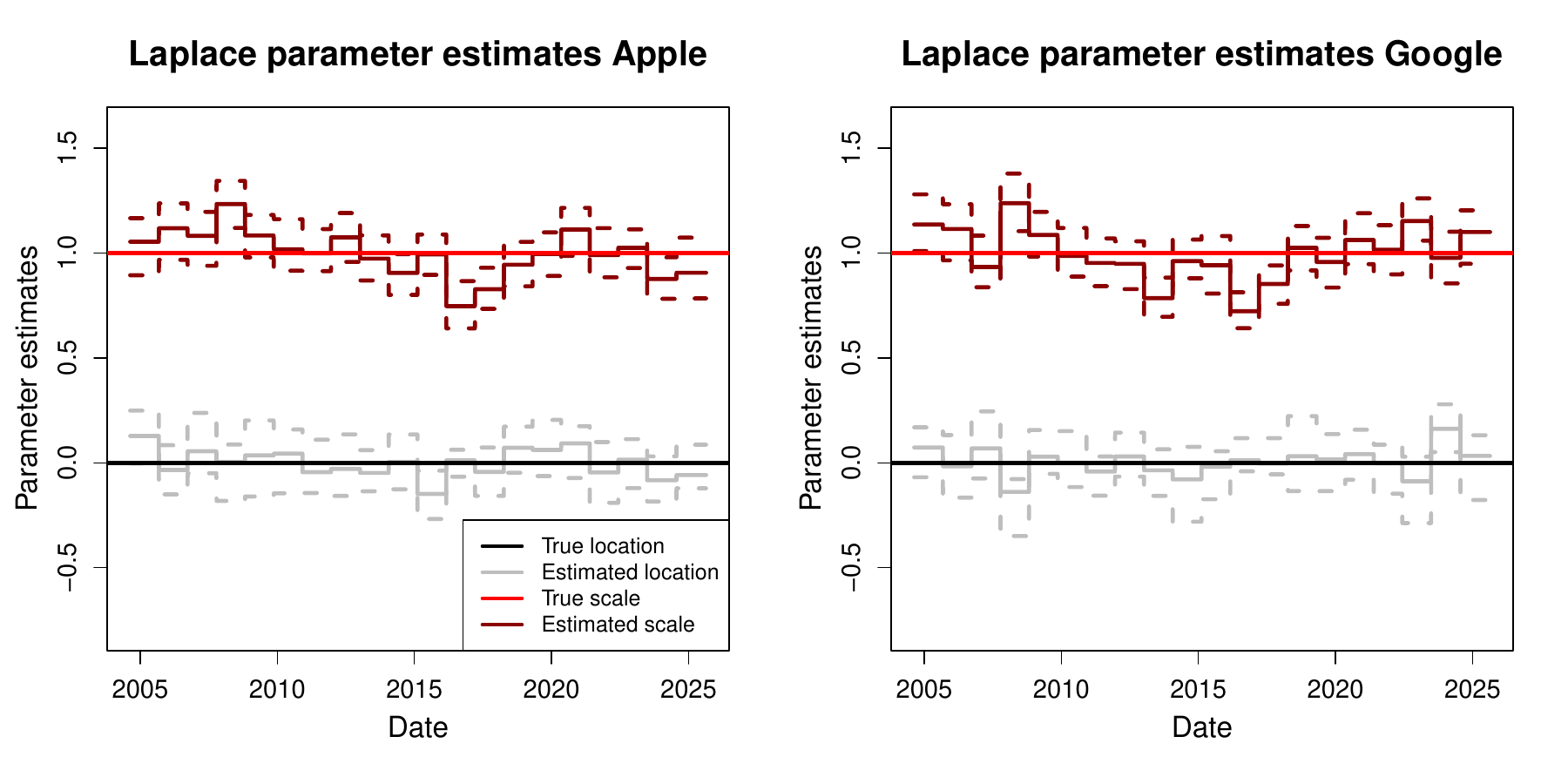}
    \caption{Rolling window Laplace parameter estimates for the transformed Apple and Google series. }
    \label{fig:laplace_est_GOOGL}
\end{figure}

\begin{figure}[H]
    \centering
    \includegraphics[width=\linewidth]{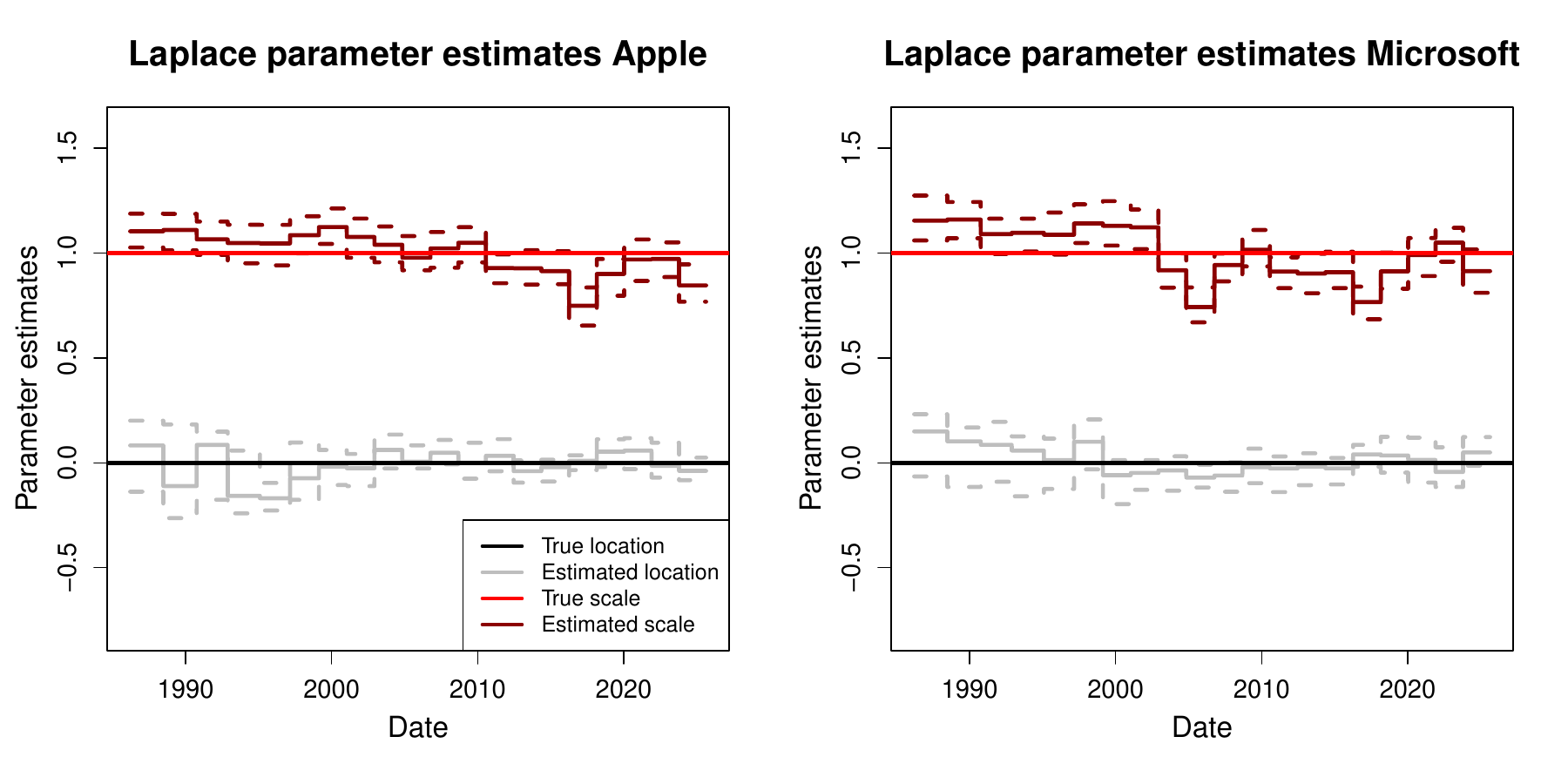}
    \caption{Rolling window Laplace parameter estimates for the transformed Apple and Microsoft series. }
    \label{fig:laplace_est_MSFT}
\end{figure}

\begin{figure}[H]
    \centering
     \begin{subfigure}[b]{.25\textwidth}
        
        \includegraphics[width=\textwidth]{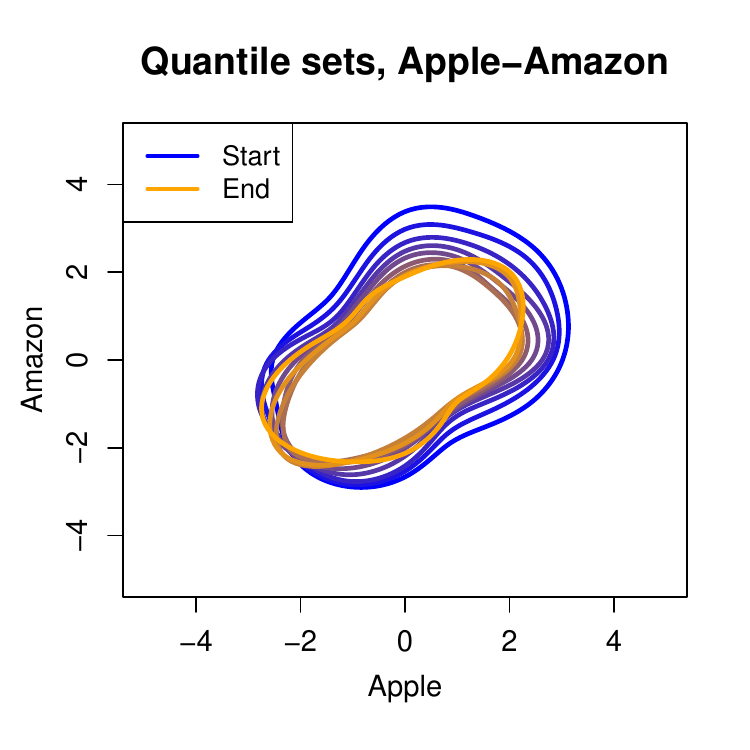}
    \end{subfigure}%
    \begin{subfigure}[b]{.25\textwidth}
        
        \includegraphics[width=\textwidth]{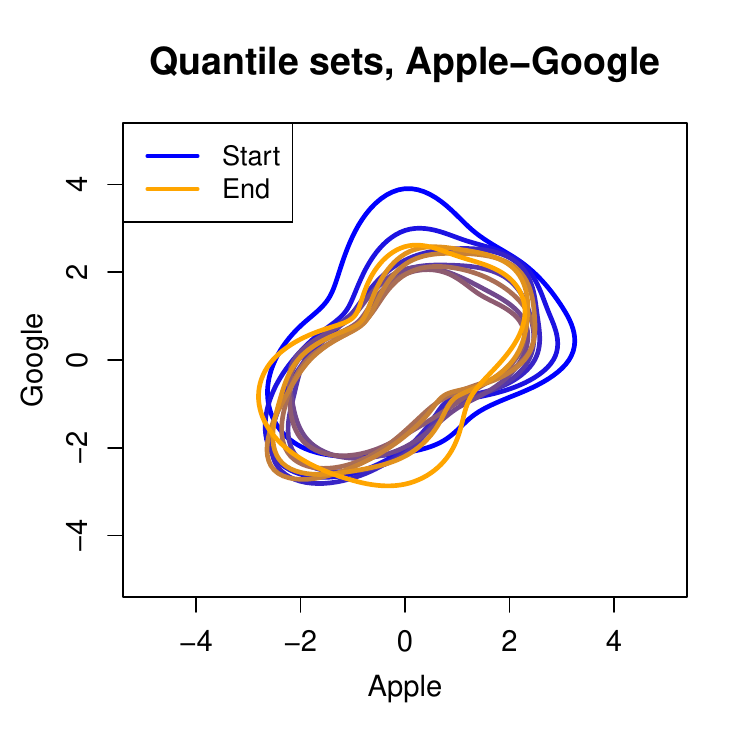}
    \end{subfigure}%
      \begin{subfigure}[b]{.25\textwidth}
        
        \includegraphics[width=\textwidth]{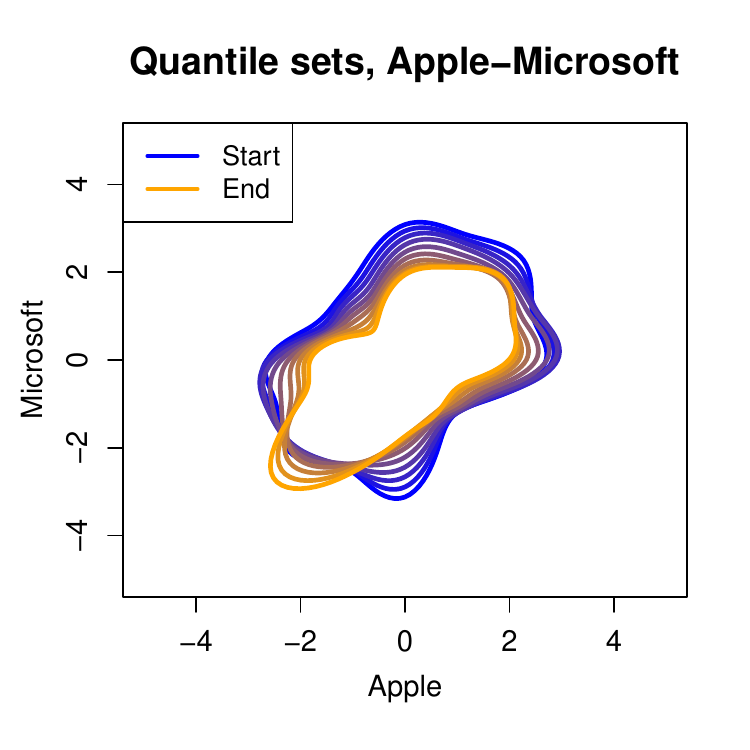}
    \end{subfigure}%
      \begin{subfigure}[b]{.25\textwidth}
        
        \includegraphics[width=\textwidth]{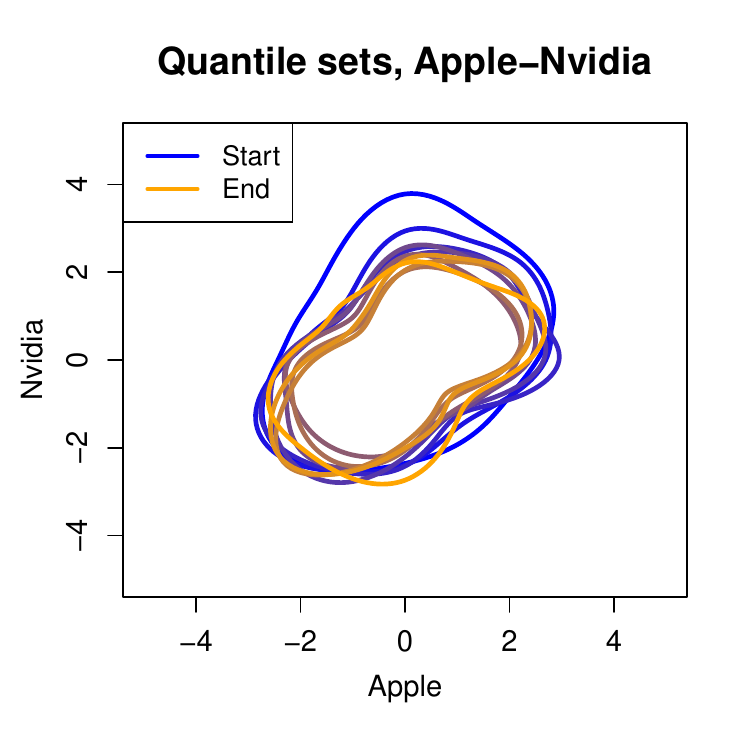}
    \end{subfigure}%
   
    \caption{Threshold quantile set estimates $\widehat{\mathcal{R}}^{\tau}_t$ at $\tau = 0.8$ over time for each pair of stocks. The colour scale is used to illustrate the variation over time, with the blue and orange sets corresponding to the start and end of the observation periods, respectively.}
    \label{fig:all_threshold_sets}
\end{figure}

Figure~\ref{fig:all_threshold_sets} illustrates the time-varying threshold quantile sets for each pair of filtered log-returns. Such sets can be seen at the `threshold level' above which we fit the truncated gamma model. 

\begin{figure}[H]
    \centering
     \begin{subfigure}[b]{.25\textwidth}
        
        \includegraphics[width=\textwidth]{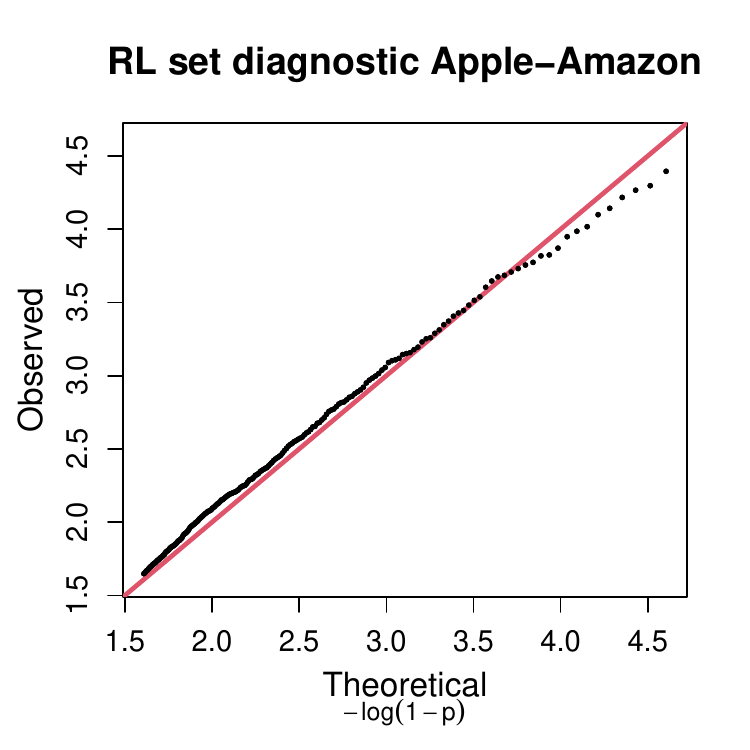}
    \end{subfigure}%
    \begin{subfigure}[b]{.25\textwidth}
        
        \includegraphics[width=\textwidth]{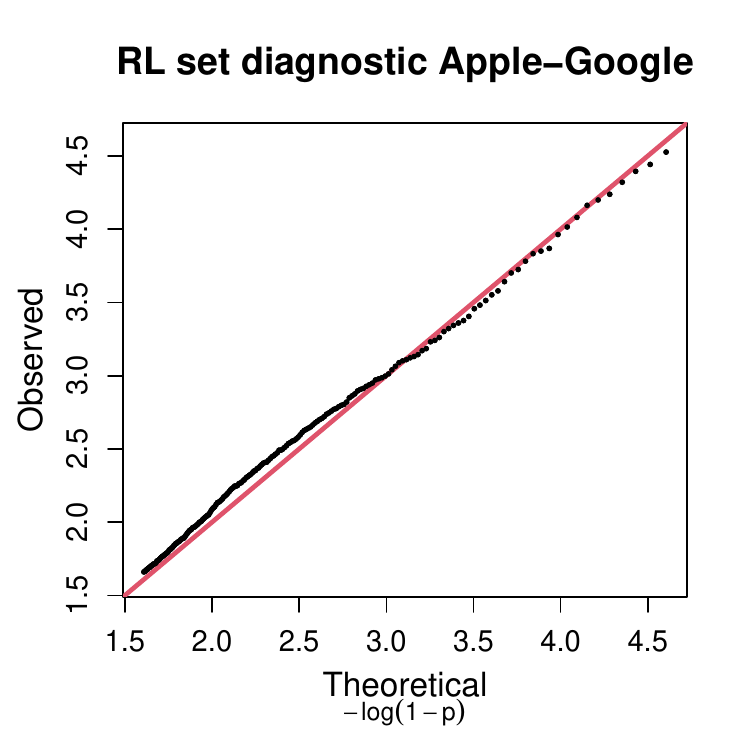}
    \end{subfigure}%
      \begin{subfigure}[b]{.25\textwidth}
        
        \includegraphics[width=\textwidth]{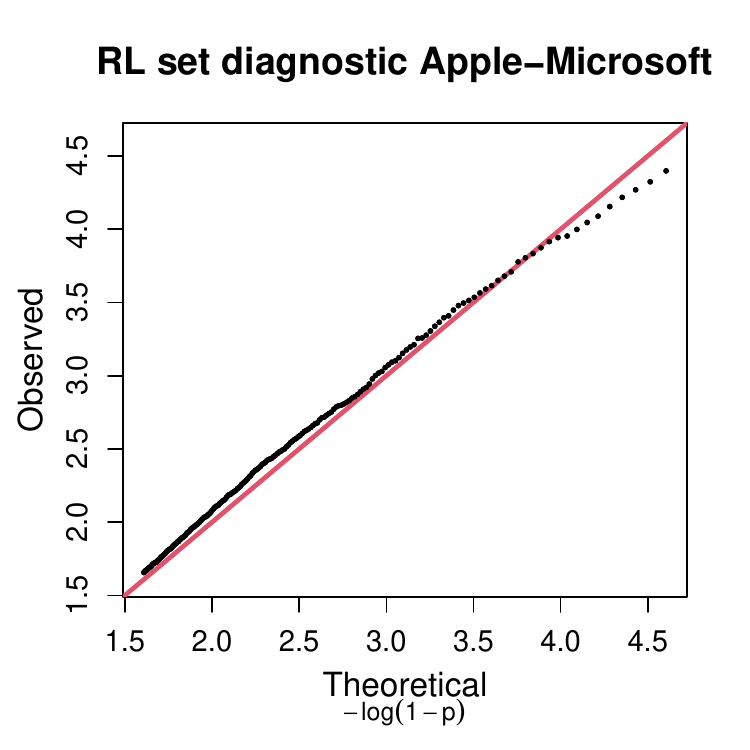}
    \end{subfigure}%
      \begin{subfigure}[b]{.25\textwidth}
        
        \includegraphics[width=\textwidth]{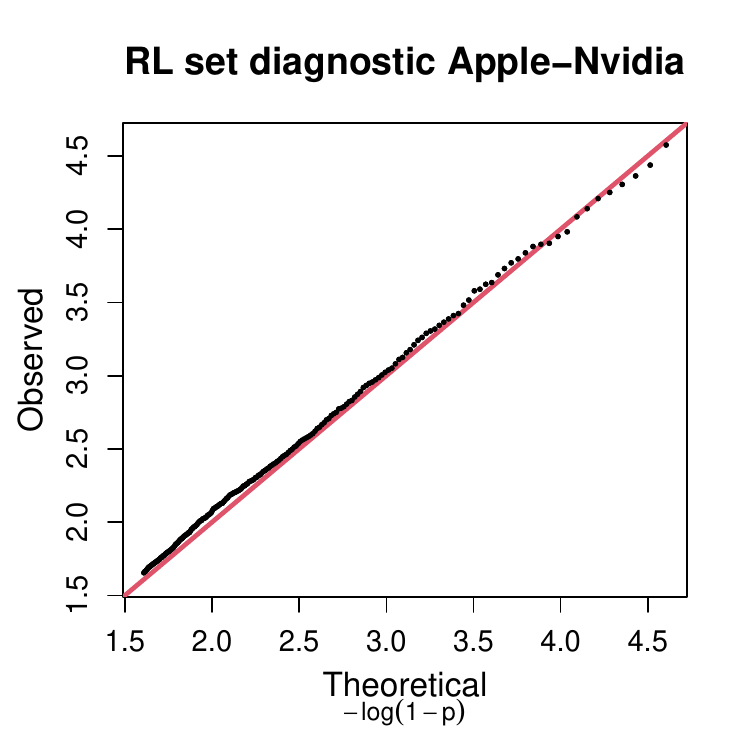}
    \end{subfigure}%
   
    \caption{Return level set diagnostic for each pair of stocks. The red lines denote equity. }
    \label{fig:rl_set_diag_all}
\end{figure}

Figure~\ref{fig:rl_set_diag_all} illustrates the return level set probabilities diagnostic for each pair of stocks. One can observe good agreement between the tuples in every case.

\begin{figure}[H]
    \centering
     \begin{subfigure}[b]{.25\textwidth}
        
        \includegraphics[width=\textwidth]{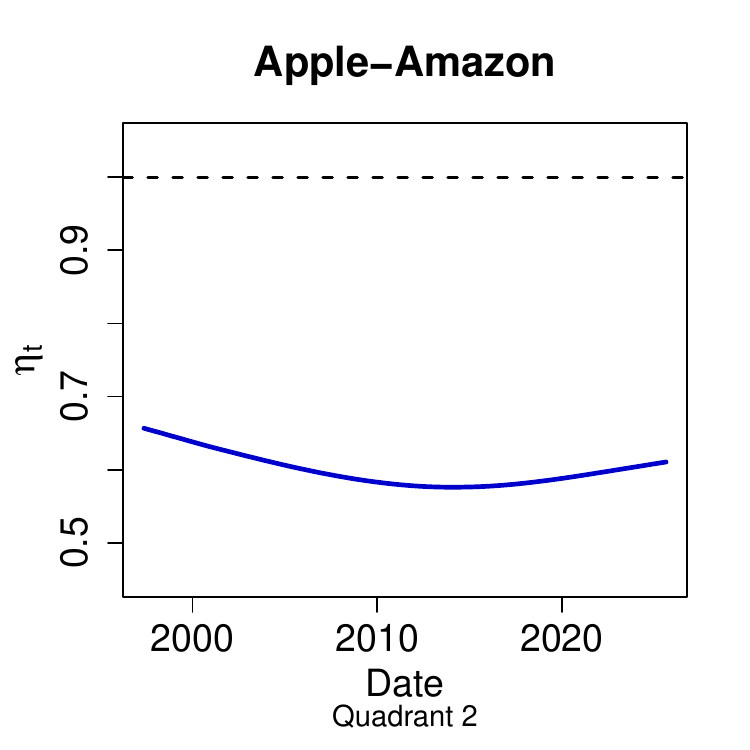}
    \end{subfigure}%
    \begin{subfigure}[b]{.25\textwidth}
        
        \includegraphics[width=\textwidth]{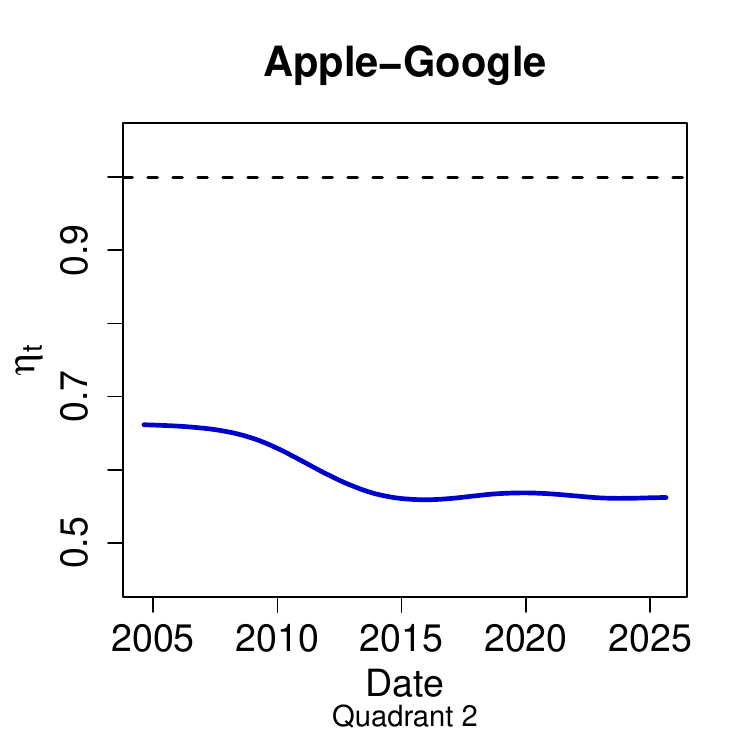}
    \end{subfigure}%
      \begin{subfigure}[b]{.25\textwidth}
        
        \includegraphics[width=\textwidth]{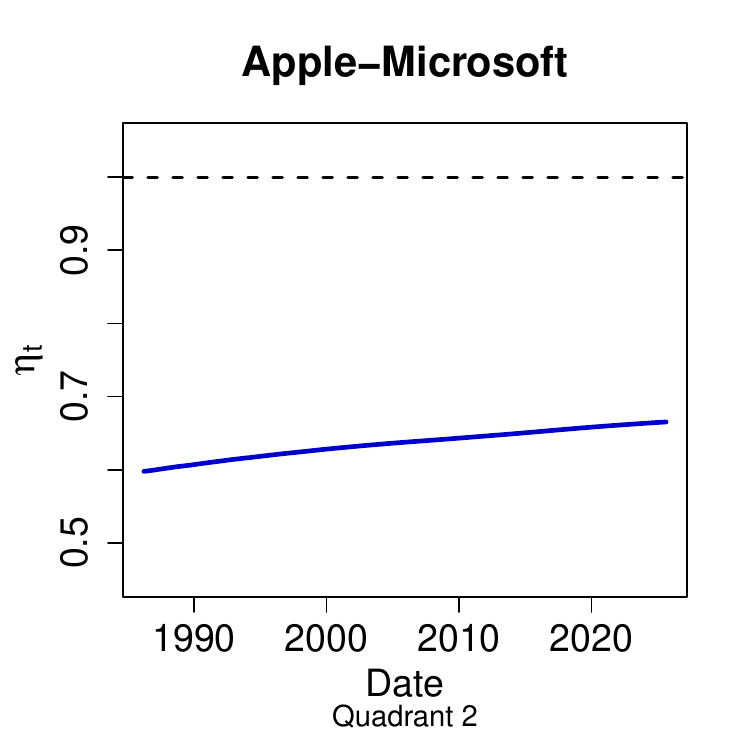}
    \end{subfigure}%
      \begin{subfigure}[b]{.25\textwidth}
        
        \includegraphics[width=\textwidth]{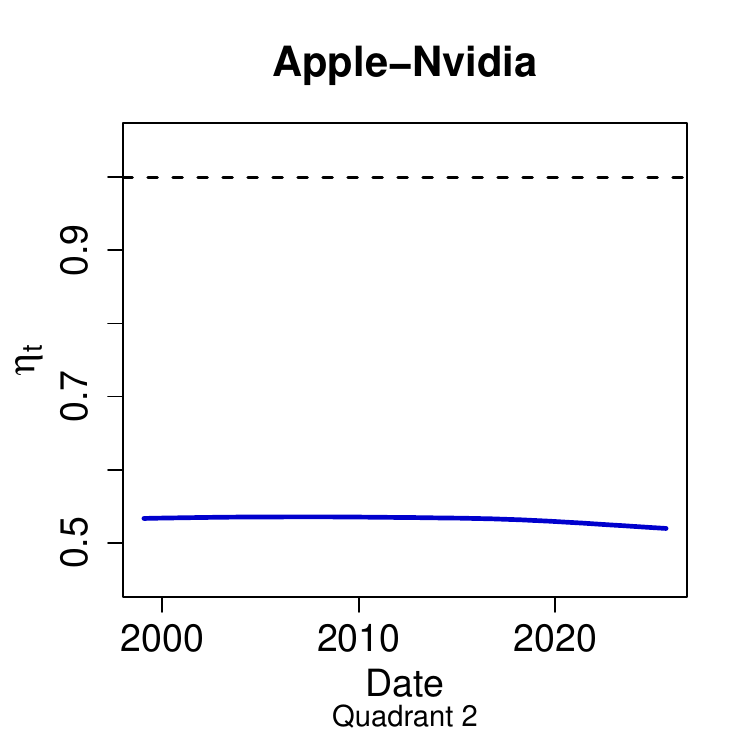}
    \end{subfigure}%

    \begin{subfigure}[b]{.25\textwidth}
        
        \includegraphics[width=\textwidth]{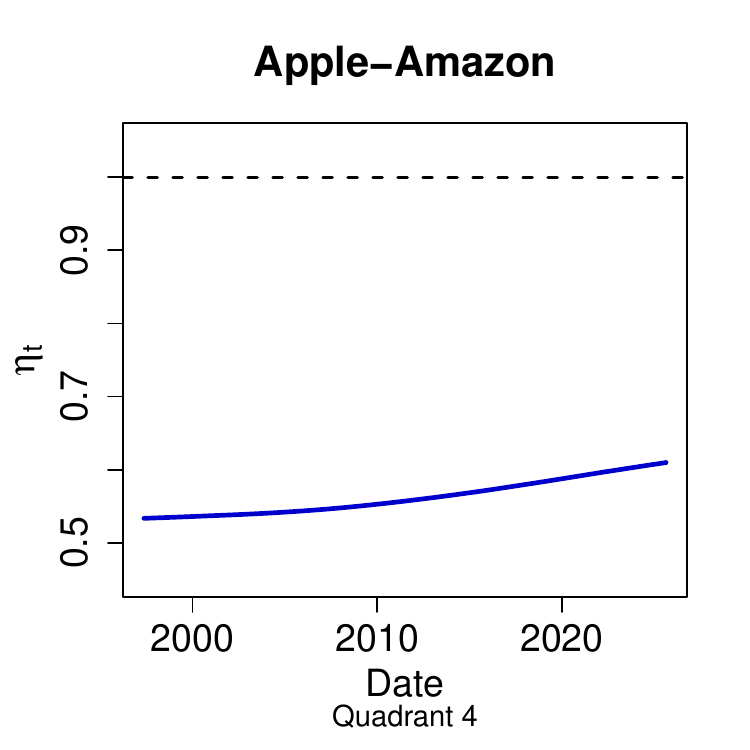}
    \end{subfigure}%
    \begin{subfigure}[b]{.25\textwidth}
        
        \includegraphics[width=\textwidth]{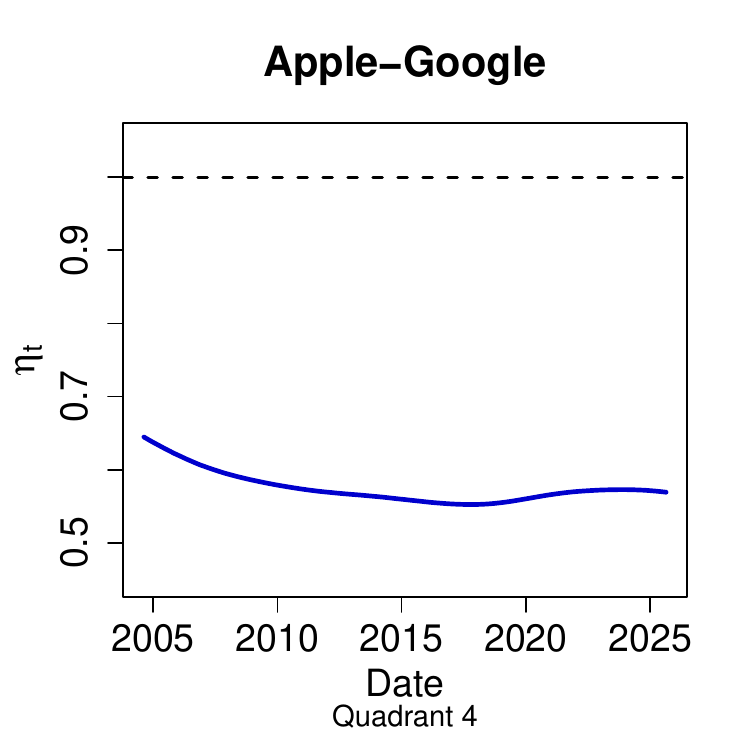}
    \end{subfigure}%
      \begin{subfigure}[b]{.25\textwidth}
        
        \includegraphics[width=\textwidth]{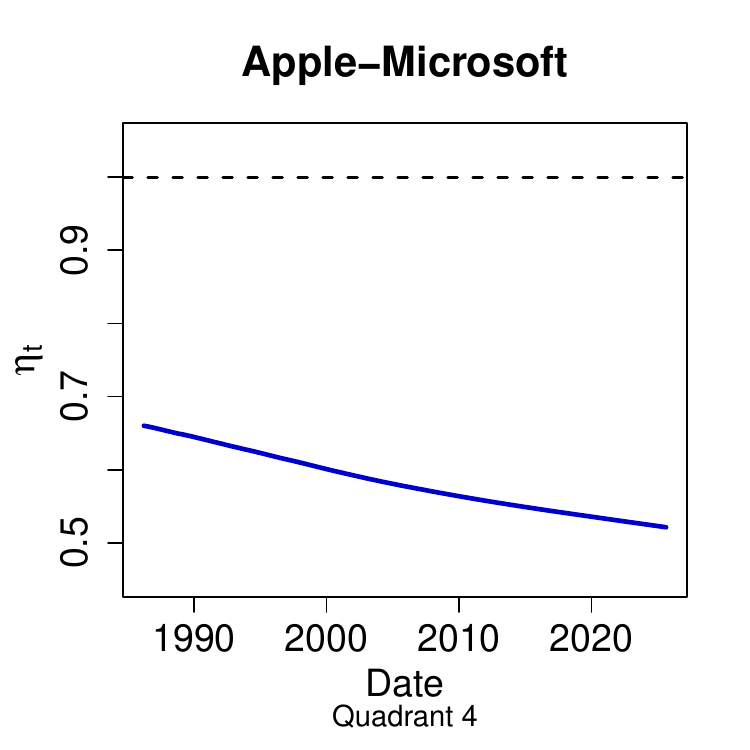}
    \end{subfigure}%
      \begin{subfigure}[b]{.25\textwidth}
        
        \includegraphics[width=\textwidth]{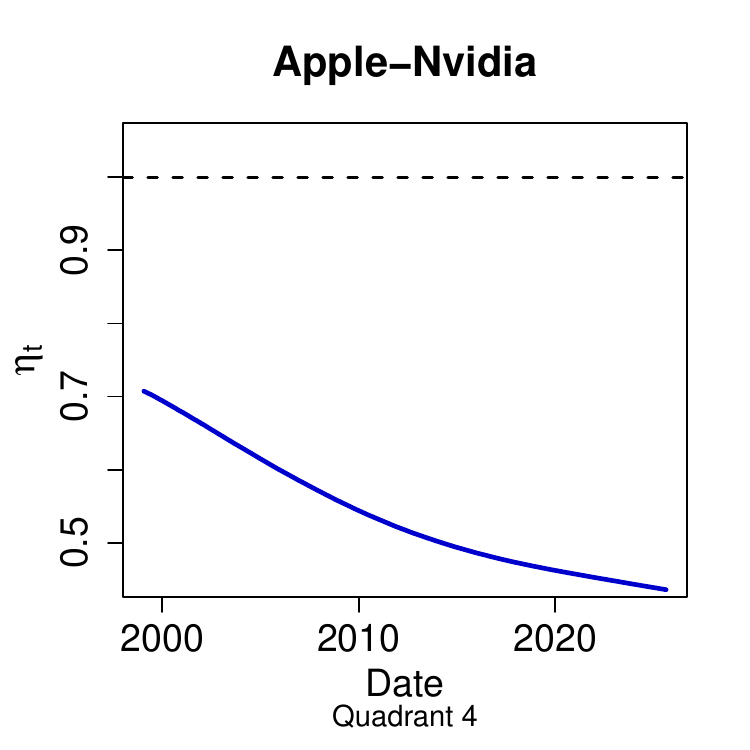}
    \end{subfigure}%
    \caption{The coefficient of tail dependence $\eta_t$ over time for quadrants 2 (top row) and 4 (bottom row). The black dotted line denotes the upper bound for $\eta_t$. }
    \label{fig:eta_ests_quads_2_4}
\end{figure}

The time-varying $\eta$ estimates for each pair in quadrants 2 and 4 are shown in Figure~\ref{fig:eta_ests_quads_2_4}. As in Section~\ref{subsec:case_sim_ret_level}, we observe a range of complex dependence trends across the different pairs.  

\end{appendix}

\end{document}